\documentclass[twoside,10pt]{report}
\usepackage{myunmdiss}
\usepackage{epsfig}
\usepackage{rotating}
\input{epsf}

\newcommand{\beqn}{\begin{eqnarray}}
\newcommand{\eeqn}{\end{eqnarray}}
\newcommand{\be}{\begin{eqnarray}}
\newcommand{\ee}{\end{eqnarray}}
\newcommand{\tr}{\mbox{tr}}
\newcommand{\la}{\langle}
\newcommand{\ra}{\rangle}

\newcommand{\ket}[1]{|#1\ra}
\newcommand{\bra}[1]{\la #1|}
\newcommand{\Sch}{\mbox{Sch}}
\newcommand{\mattwoc}[4]{\left[ \begin{array}{cc} #1 & #2 \\ #3 & #4 \end{array} \right]}
\newcommand{\evop}{{\cal E}}

\newcommand{\ex}{{\bf E}}
\newcommand{\qed}{{\bf QED}}

\newtheorem{theorem}{Theorem}
\newtheorem{lemma}{Lemma}
\newtheorem{corollary}{Corollary}
\newenvironment{proof}{{\bf Proof}}{\qed}

\phdthesis
\leftchapter
\renewcommand{\baselinestretch}{1.0}

\begin{document}

\newpage
\thispagestyle{empty}

\begin{center}
\textbf{\large Quantum Information Theory \\
\vspace{3.7cm}
By \\
\vspace{1cm}
Michael Aaron Nielsen} \\
\vspace{1cm}
B.Sc., University of Queensland, 1993 \\
B.Sc. (First Class Honours), Mathematics, University of Queensland,
1994 \\
M.Sc., Physics, University of Queensland, 1998 \\
\vspace{2.5cm}
{\large DISSERTATION} \\
\vspace{0.3cm}
Submitted in Partial Fulfillment of the \\
Requirements for the Degree of \\
\vspace{0.3cm}
\textbf{Doctor of Philosophy \\ Physics} \\
\vspace{0.3cm} The University of New Mexico \\
Albuquerque, New Mexico, USA \\
\vspace{0.5cm}
December 1998
\end{center}
\newpage

\thesiscopyrightpage
\thesisdedicationpage

\begin{thesisacknowledgments}

\begin{spacing}{1}
\vspace{-1cm} It is a great pleasure to thank the many people who have
contributed to this Dissertation. My deepest thanks goes to my friends
and family, especially my parents, Howard and Wendy, for their support
and encouragement.  Warm thanks also to the many other people who have
contributed to this Dissertation, especially Carl~Caves, who has been
a terrific mentor, colleague, and friend; to Gerard~Milburn, who got
me started in physics, in research, and in quantum information; and to
Ben~Schumacher, whose boundless enthusiasm and encouragement has
provided so much inspiration for my research.  Throughout my graduate
career I have had the pleasure of many enjoyable and helpful
discussions with Howard~Barnum, Ike~Chuang, Chris~Fuchs,
Raymond~Laflamme, Manny~Knill, Mark~Tracy, and Wojtek~Zurek.  In
particular, Howard~Barnum, Carl~Caves, Chris~Fuchs, Manny~Knill, and
Ben~Schumacher helped me learn much of what I know about quantum
operations, entropy, and distance measures for quantum
information. The material reviewed in chapters 3 through 5 I learnt in
no small measure from these people. Many other friends and colleagues
have contributed to this Dissertation. A partial list includes
Chris Adami, Dorit Aharonov, James Anglin, David Beckman,
Paul Benioff, Charles Bennett, Sam Braunstein, Nicholas Cerf,
Ignacio Cirac, Richard Cleve, John Cortese, Vageli Coutsias,
Wim van Dam, Ivan Deutsch, David DiVincenzo, Rob Duncan,
Mark Ettinger, Betty Fry, Jack Glassman, Daniel Gottesman,
Bob Griffiths, Gary Herling, Tom Hess, Alexander Holevo,
Peter H{\o}yer, Richard Hughes, Chris Jarzynski, Brad Johnson,
Jeff Kimble, Alexei Kitaev, Andrew Landahl, Debbie Leung, Seth Lloyd,
Hoi-Kwong Lo, Hideo Mabuchi, Eleanor Maes, Norine Meyer, Cesar Miquel,
Juan Pablo Paz, Mitch Porter, John Preskill, Peter Shor, John Smolin,
Alain Tapp, Barbara Terhal, Amnon Ta-Shma, Mike Westmoreland,
Barby Woods, Bill Wootters, Christof Zalka, and Peter Zoller. My
especial thanks also to those creative people without whom the physics
of information would not exist as a field of research; especial
influences on my thinking were the writings of Paul~Benioff,
Charles~Bennett, Gilles~Brassard, David~Deutsch, Artur~Ekert,
Richard~Feynman, Alexander~Holevo, Richard~Jozsa, Rolf~Landauer,
G\"oran~Lindblad, Seth~Lloyd, Asher~Peres, Ben~Schumacher, Peter~Shor,
John~Wheeler, Stephen~Wiesner, Bill~Wootters, and Wojtek~Zurek.
\end{spacing}

\end{thesisacknowledgments}

\newpage
\begin{center}
\textbf{\large Quantum Information Theory \\
\vspace{5cm}
By \\
\vspace{1cm}
Michael Aaron Nielsen} \\
\vspace{4.5cm}
{\large Abstract of Dissertation} \\
\vspace{0.3cm}
Submitted in Partial Fulfillment of the \\
Requirements for the Degree of \\
\vspace{0.3cm}
\textbf{Doctor of Philosophy \\ Physics} \\
\vspace{0.3cm} The University of New Mexico \\
Albuquerque, New Mexico, USA \\
\vspace{0.5cm}
December 1998
\end{center}
\newpage

\begin{center}
\textbf{Quantum Information Theory \\ \vspace{0.5cm}
By \\ \vspace{0.5cm}
Michael Aaron Nielsen} \\ \vspace{0.3cm}
B.Sc., University of Queensland, 1993 \\
B.Sc. (First Class Honours), Mathematics, University of Queensland,
1994 \\
M.Sc., Physics, University of Queensland, 1998
\end{center}

\vspace{1cm}

\begin{center}
\textbf{Abstract}
\end{center}
Quantum information theory is the study of the
achievable limits of information processing within quantum mechanics.
Many different types of information can be accommodated within quantum
mechanics, including {\em classical information}, {\em coherent
quantum information}, and {\em entanglement}. Exploring the rich
variety of capabilities allowed by these types of information is the
subject of quantum information theory, and of this Dissertation. In
particular, I demonstrate several novel limits to the information
processing ability of quantum mechanics.  Results of especial interest
include: the demonstration of limitations to the class of measurements
which may be performed in quantum mechanics; a capacity theorem giving
achievable limits to the transmission of classical information through
a two-way noiseless quantum channel; resource bounds on distributed
quantum computation; a new proof of the quantum noiseless channel
coding theorem; an information-theoretic characterization of the
conditions under which quantum error-correction may be achieved; an
analysis of the thermodynamic limits to quantum error-correction, and
new bounds on channel capacity for noisy quantum channels.

\tableofcontents
%\listoftables
\listoffigures

\newpage
\addcontentsline{toc}{chapter}{{\bf Nomenclature and notation}}
\noindent
{\bf \huge Nomenclature and notation}

\vspace{0.5cm}
There are several items of nomenclature and notation which have two or
more meanings in common use in the field of quantum information
theory. To prevent confusion from arising, this section collects many
of the more frequently used of these items, together with the
conventions that will be adhered to in this Dissertation.

As befits good information theorists, logarithms are {\em always}
taken to base two, unless otherwise noted.

A {\em positive} operator $A$ is one for which $\la \psi|A|\psi\ra
\geq 0$ for all $|\psi\ra$. A {\em positive definite} operator $A$ is
one for which $\la \psi|A|\psi\ra > 0$ for all $|\psi\ra \neq 0$.

The {\em relative entropy} of a positive operator $A$ with respect to
a positive operator $B$ is defined by
\beqn
S(A||B) \equiv \tr(A\log A)-\tr(A\log B). \eeqn

Conventionally, most researchers use $|\psi\ra$ to represent a pure
state of a quantum system, and $\rho$ to represent a mixed state.  We
will use this notation on occasion, but we will also make use of a
different notation.  Suppose we are dealing with a composite quantum
system with component parts labeled $R$ and $Q$.  Then we will use
$R$, $Q$, and $RQ$ to denote the quantum states associated with those
systems, in addition to their use as labels for the systems.  When one
or more of these systems is known to be in a pure state we will use
the notation $|R\ra, |Q\ra$, and $|RQ\ra$, as appropriate.

A {\em purification} of a mixed state, $Q$, of some quantum system
$Q$, is a pure state $|RQ\ra$ of some larger system $RQ$, such that
when the system $R$ is traced out, the state $Q$ is recovered,
\beqn
Q = \tr_R(|RQ\ra \la RQ|). \eeqn

The {\em support} of an operator is defined to be the vector space
orthogonal to its kernel. For a Hermitian operator, this means the
vector space spanned by eigenvectors of the operator with non-zero
eigenvalues.

The term {\em probability distribution} is used to refer to a finite
set of real numbers, $p_x$, such that $p_x \geq 0$ and $\sum_x p_x =
1$.

All Hilbert spaces are assumed to be finite dimensional.  In many
instances this restriction is unnecessary, or can be removed with some
additional technical work, but making the restriction globally makes
the presentation more easily comprehensible, and doesn't detract much
from many of the intended applications of the results.  Furthermore,
in some instances, extension of a result to general Hilbert spaces is
beyond my technical expertise!

\clearpage
\pagestyle{headings}
\pagenumbering{arabic}

\part{Fundamentals of quantum information}

\chapter{The physics of information}
 \label{chap:intro}

\begin{quote}
Information is physical.
\end{quote}
-- Rolf Landauer \cite{Landauer91a}
\index{physics of information}

\section{A collision of ideas}

{\em What is discovered when the laws of physics are used as the
foundation for investigations of information processing and
computation?} This Dissertation is an attempt to provide a partial
answer to this question. To understand why the attempt should be
fruitful, it is useful to remind ourselves of what it is that a
physicist or computer scientist does.

What is physics? Physics is a messy human endeavour, so any answer to
this question is somewhat inaccurate. Nevertheless, an examination of
the history and current state of physics reveals at least two
overarching themes within physics.  The first theme is that physics
studies {\em universal} properties of nature. We expect that black
holes in the cores of galaxies a billion light years away obey the
same laws of general relativity that govern the motion of planets in
our own solar system, or the motion of a ball through the
air. Likewise, we expect that the structure of matter based upon
quarks and leptons is the same throughout the universe.

A second overarching theme of physics is the {\em reduction of
phenomena}. This theme has two aspects. One aspect is the ongoing
search for simplified, unified frameworks in which it is possible to
understand more complicated phenomena.  For example, there is the
current search for a unified description of the particles and fields
of nature \cite{Wilczek97a}, or attempts to understand the principles
underlying pattern formation in physics
\cite{Langer97a}. The second aspect of this theme is the discovery and
explanation of phenomena in terms of simple frameworks. For example,
there is the remarkable Bardeen-Cooper-Schrieffer theory of
superconductivity
\cite{Bardeen57a}, based upon the principles of quantum mechanics, or
the current search for gravitational waves \cite{Thorne97a},
potentially one of the most useful consequences of the general theory
of relativity.

Note that both these themes are somewhat gray. There are differing
degrees of universality, and physics does not concern itself with the
reduction of all phenomena to fundamentals. It leaves many phenomena
-- the human body, climate patterns, computer design -- to other
disciplines. Here too, universality plays a role, with physics being
primarily interested in relatively simple phenomena, such as
superfluidity, which do not have an especially detailed historical
dependence such as may be found, for example, in the functioning of a
cell, and can therefore be relatively easily reproduced by a variety
of means, in many locations.

The themes of universality and reduction both have strong parallels
within computer science\footnote{I am using ``computer science'' as a
pseudonym for all those fields of science concerned with information
processing, including, for example, computer science, information
theory, signal processing, and many others.}. Traditionally, computer
science is based upon a small number of {\em universal} models that
are each supposed to capture the essence of some aspect of information
processing. For example, the majority of work done on algorithm design
has been framed within the well known Turing machine model
\cite{Turing36a} of computation, or one of its equivalents. Shannon's
model \cite{Shannon49a} of a communications channel is the foundation
for modern work in information theory.

Computer science is also concerned with the reduction of phenomena,
but in a different way than is often the case in physics. Reduction in
physics often concerns the explanation of phenomena discovered without
specific intent, such as superconductivity. In computer science, it is
more typical to set a specific information processing goal -- ``I
would like my computer to sort this list of names for me in such and
such an amount of time'' -- and then to attempt to meet that goal
within an existing model of information processing.

What is the origin of the fundamental models used as the basis for
further progress in computer science? Examination of the original
papers shows that the founders used systems existing in the real world
as inspiration and justification for the models of computation they
proposed. For example, Turing analyzed the set of operations which a
mathematician could perform with pen and paper, in order to help
justify the claim that his model of computation was truly universal.

It is a key insight of the last thirty years that these pseudophysical
justifications for the fundamental models of computation may be
carried much further. For example, a theory of computation which has
its foundations in quantum mechanics has been formulated
\cite{Ekert96a}. {\em Information is physical}, as Landauer
reminds us \cite{Landauer91a}. That is, any real information
processing system relies for its implementation upon systems whose
behaviour is completely described by the laws of physics.

Remarkable progress has been achieved by acting on this insight,
re-examining and reformulating the fundamental models of information
based upon physical principles. The hope, which has been fulfilled, is
that such a reformulation will reveal information processing
capabilities that go beyond what was thought to be possible in the old
models of computation.

The field of science which studies these fundamental connections
between physics and information processing has come to be known as the
{\em physics of information}. The connection between physics and
information processing is a two way street, with potential benefits
for both computer science and physics.

Computer science benefits from physics by the introduction of new
models of information processing. {\em Any physical theory may be
regarded as the basis for a theory of information processing.} We may,
for example, enquire about the computational power of Einstein's
general theory of relativity, or about the computational power of a
quantum field theory. The hope is that these new models of information
processing may give rise to capabilities not present in existing
models of information processing.  In this Dissertation we will
primarily be concerned with the information processing power of
quantum mechanics.  The other possible implication for computer
science is more ominous: there may be unphysical elements in existing
theories of information processing which need to be rooted out if
those theories are to accurately reflect reality.

Physics benefits in at least four ways from computer science. First,
computer science may act as a stimulus for the development of new
techniques which assist in the pursuit of the fundamental goals of
physics. For example, inspired by coding theory, error correction
methods to protect against noise in quantum mechanics have been
developed.  One of the chief obstacles to precision measurement is, of
course, the presence of noise, so error correcting codes to reduce the
effects of that noise are welcome. They are doubly useful, however, as
a diagnostic tool, since error correcting codes can be used to
determine what types of noise occur in a system. 

The second way physics benefits from computer science is via
simulation. Computational physics has allowed us to investigate
physical theories in regimes that were not previously accessible. Such
investigations can lead to interesting new questions about those
theories, and yield important insights into the predictions made by
our physical theories.

The third way physics benefits from computer science is that computers
enable us to perform experiments that would once have been impossible
or, at the least, much more difficult and expensive.  Computer-based
methods for obtaining, analysing, and presenting data have opened up
new experimental realms.  For example, computers enormously simplify
the analysis of data taken in particle accelerators, in which only a
miniscule fraction of the events detected in a given experimental run
may be of direct interest.  Automated sifting of the data and
identification of the relevant events is performed in an instant using
powerful computers, rather than the time of years or more that it
would take a human being to achieve the same results.

The fourth way physics benefits from computer science is more difficult
to describe or justify. My experience has been that computer science
is a great inspiration for fundamental questions about physics, and
can sometimes suggest useful approaches to take in the solution of
physics problems. This will be apparent several times during the main
body of this Dissertation. I can not yet say precisely why this should
be the case, although as we have seen, both physics and computer
science involve the development of tools to reduce phenomena involving
complex interacting systems to certain fundamental models, as well as
continual questioning and refinement of those models.  Perhaps it is
not so surprising that each field should have much to teach the other.

\section{What observables are realizable as quantum measurements?}
\label{sec:turing}

This Dissertation is concerned principally with a special subfield of
the physics of information, {\em quantum information}, in which the
fundamental models for information processing are based upon the laws
of quantum mechanics. The earlier formulation of the question
investigated by this Dissertation may thus be refined: {\em What is
discovered when the laws of quantum mechanics are used as the
foundation for investigations of information processing and
computation?}

To better understand the subject of quantum information, it is useful
to have a concrete example in hand. This section presents a simple
example which illustrates many of the basic themes of quantum
information. The example is also interesting in its own right, as it
takes us straight to the edge of what is known, posing a fundamental
question about quantum mechanics, inspired by the methods of computer
science.

The example concerns the question of {\em what properties of a quantum
mechanical system may be measured?} In the 1920s, Heisenberg and other
researchers formulated the notion of a {\em quantum mechanical
observable}. Observables were introduced into quantum mechanics as a
means of describing what properties of a quantum system may be
measured. For example, a particle's position is regarded as an
observable in quantum mechanics.

Mathematically, the concept of an observable is usually formulated as
follows. An observable is any Hermitian operator acting on the state
space of a physical system, where by ``state space'' we shall mean the
usual Hilbert space associated with a physical system. Recall from
elementary quantum mechanics that the measurement postulate of quantum
mechanics as usually formulated has the following consequences: To
each measurable quantity of a quantum mechanical system there is
associated a mathematical object, an {\em observable}, which is a
Hermitian operator acting on the state space of the quantum system.
% Recall that in the usual formulation of quantum
%measurement theory,
The possible outcomes of the measurement are given
by the spectrum of the observable.
% If $m$ is such an outcome, and
% $P_m$ is the projector onto the corresponding eigenspace of the
% observable, then the probability of obtaining the measurement outcome
% $m$ is $\tr(P_m \rho)$, where $\rho$ is the state of the quantum
% system immediately before the measurement is performed.
% The state
%after the measurement is $P_m \rho P_m / \tr(P_m \rho)$.
If the state of the quantum system immediately before the system is
observed is an eigenstate of the observable then, with certainty, the
outcome of the measurement is the corresponding eigenvalue, $m$.

One of the most remarkable discoveries of quantum mechanics is that
the theory implies limits to the class of measurements which may be
performed on a physical system.  The most famous example of this is
the {\em Heisenberg uncertainty principle}, which establishes
fundamental limits upon our ability to perform simultaneous
measurements of position and momentum.  Given the shock caused by
Heisenberg's result that there are limits, in principle, to our
ability to make observations on a physical system, it is natural to
ask for a precise characterization of what properties of a system may
be measured. For example, Dirac's influential text (\cite{Dirac58a},
page 37) makes the following assertion on the subject:

{\em The question now presents itself -- Can every observable be
measured? The answer theoretically is yes. In practice it may be very
awkward, or perhaps even beyond the ingenuity of the experimenter, to
devise an apparatus which could measure some particular observable, but
the theory always allows one to imagine that the measurement can be
made.}

That is, Dirac is asserting that given any observable for a reasonable
quantum system, it is possible in principle to build a measuring
device that makes the measurement corresponding to that observable.
Dirac leaves his discussion of the subject at that, making no attempt
to further justify his claims. Later, Wigner \cite{Wigner52a}
investigated the problem, and discovered that conservation laws do, in
fact, impose interesting physical constraints upon what properties of
a system may be measured. This work was subsequently extended by Araki
and Yanase \cite{Araki60a}, resulting in what Peres \cite{Peres93a}
terms the Wigner-Araki-Yanase or WAY theorem.  To my knowledge, there
has been remarkably little other work done on the fundamental question
of what observables may be measured in quantum mechanics.

Not long after Heisenberg, Dirac and others were laying the
foundations for the new quantum mechanics, a revolution of similar
magnitude was underway in computer science. The remarkable English
mathematician Alan Turing laid out the foundations for modern computer
science in a paper written in 1936 \cite{Turing36a}\footnote{It is
worth noting that many other researchers arrived at similar results
around the same time, notably Church and Post.  However it is my
opinion that it is Turing's grand vision that has ultimately proved to
be the deepest and most influential.}.

Turing's work was motivated, in part, by a challenge set down by the
great mathematician David Hilbert at the International Congress of
Mathematicians held in Bologna in 1928. Hilbert's problem, the {\em
Entscheidungsproblem}, was to find an algorithm by which all
mathematical questions could be decided. Remarkably, Turing was able
to show that there is {\em no such procedure}.  Turing demonstrated
this by giving an explicit example of an interesting mathematical
question whose answer could not be decided by algorithmic means. In
order to do this, Turing had to formalize our intuitive notion of what
it means to perform some task by algorithmic means.

To do this, Turing invented what is now known as the {\em universal
Turing machine}. Essentially, a universal Turing machine behaves like
an idealized modern computer, with an infinite memory. Turing's
computer was capable of being {\em programmed}, in much the same sense
as a modern computer may be programmed. Turing's programs computed
mathematical functions: the machine would take a number as input, and
return a number as output, with the function computed by the machine
in this way being determined by the program being run on the
machine. In addition, it was possible that programs would fail to
halt, continuing to execute forever, never giving a definite output.

The most important assertion in Turing's paper has come to be known as
the {\em Church-Turing thesis}\index{Church-Turing thesis}. Roughly
speaking this thesis states that {\em any function which may be
computed by what we intuitively regard as an algorithm may be computed
by a program running on a universal Turing machine, and vice
versa}. The reason this thesis is so important is because it asserts
the equivalence of an intuitive concept -- that of an algorithm --
with the rigorously defined mathematical concept of a program running
on a universal Turing machine. The validity of the Church-Turing
thesis has been repeatedly tested and verified inductively since
Turing's original paper, and it is this continuing success that
ensures that Turing's model of computation, and others equivalent to
it, remain the foundation of theoretical work in computer science.

One observation made by Turing was that the programs for his universal
machine could be numbered, $0,1,2,\ldots$. This led him to pose the
{\em halting problem}\index{halting problem}: does program number $x$
halt on input of the value $x$, or does it continue forever? Turing
showed that this apparently innocuous question has no solution by
algorithmic means.  In fact, it is now known that in some sense
``most'' questions admit no algorithmic solution. The way Turing
demonstrated the unsolvability of the halting problem was to note that
it is equivalent to being able to compute the {\em halting function},
\beqn h(x) \equiv \left\{ \begin{array}{ll} 1 & \mbox{if program } x
\mbox{ halts on input } x \\ 0 & \mbox{if program } x \mbox{ does not
halt on input } x, \end{array} \right. \eeqn by algorithmic means.

In Chapter \ref{chap:fundamentals} we review the proof that there is
no algorithm which can compute the halting function, establishing
Turing's great result. For now, we will assume that this remarkable
result is correct.

Turing's result paves the way for an interesting quantum mechanical
construction. Suppose we consider a quantum mechanical system whose
state space is spanned by orthonormal states $|0\ra,|1\ra,\ldots$,
such as the quantum mechanical simple harmonic oscillator. We use the
halting function to define a Hermitian operator, $\hat h$, by the
formula: \beqn \hat h \equiv \sum_{x=0}^{\infty} h(x) |x\ra \la
x|. \eeqn This operator is clearly Hermitian, and thus represents a
quantum mechanical observable, which we call the {\em halting
observable}\index{halting observable}. Notice that it has two
eigenvalues, $0$ and $1$. The eigenspace corresponding to the
eigenvalue $1$ is spanned by those states $|x\ra$ for which $h(x) =
1$, while the eigenspace corresponding to the eigenvalue $0$ is
spanned by those states $|x\ra$ for which $h(x) = 0$.

Is the halting observable a measurable property of the quantum
mechanical system? More precisely, is it possible to construct a
measuring device which performs a measurement of the halting
observable? There are two possibilities:
\begin{enumerate}
\item It is possible, in principle, to construct a measuring device
which can measure the halting observable. In this case, we can give a
physical algorithm for solving the halting problem: to evaluate
$h(x)$, build the device to measure the halting observable, prepare
the quantum system in the state $|x\ra$, and perform a measurement of
the halting observable. By the quantum measurement postulate, the
result of the measurement is, with certainty, the correct value of
$h(x)$.
\item It is not possible, in principle to construct a measuring device
which can measure the halting observable.
\end{enumerate}

If the first possibility is correct, we are forced to conclude that
Turing's model of computation is insufficient to describe all possible
algorithms, and thus the Church-Turing thesis needs to be
re-evaluated. If, on the other hand, the second possibility is
correct, then we are left to ponder the problem of determining the
fundamental limits to measurement in quantum mechanics.

A resolution of this dichotomy, which I posed in
\cite{Nielsen97d}\footnote{Benioff (private communication) has
independently constructed related examples, with similar ends in
mind.}, is not presently known. It is instructive to note several
features of the problem posed. First, it is a problem concerning the
ultimate limits to our ability to perform a particular ``information
processing'' task, in this case, the performance of a quantum
measurement. We are interested in finding limitations on what is
possible, and also in constructive techniques for certain physical
tasks. In the present example of measurement theory, I think it is
fair to say that we do not yet fully understand either the limits or
the possibilities available in the measurement process. Second, it is
interesting to note the fruitful interplay between physics and
computation taking place here: a fundamental question from computer
science has been translated into physical terms, and gives rise to an
interesting fundamental question about physics. Both these features
are repeated many times through the course of this Dissertation, and
throughout quantum information in general.

With this concrete example in hand, we now return to understand in
more detail what the subject of quantum information is about. In
Chapter \ref{chap:fundamentals} we return to study the problems posed
by the halting observable and similar constructions in greater depth.

\section{Overview of the field of quantum information}
\label{sec:overview_qinfo}
\index{quantum information}

Quantum information may be defined as {\em the study of the achievable
limits to information processing possible within quantum mechanics}.
Thus, the field of quantum information has two tasks.

First, it aims to determine limits on the class of information
processing tasks which are possible in quantum mechanics.  For
example, one might be interested in limitations on the class of
measurements that may be performed on a quantum system -- if it is
impossible to measure the halting observable, then that would be an
interesting fact to know, and explore in greater detail. Another
example which will be examined in this Dissertation is the question of
determining bounds to how much information may be stored using given
quantum resources.

The second task of quantum information theory is to provide
constructive means for achieving information processing tasks. For
example, it would be extremely useful to have a means for implementing
any desired measurement in quantum mechanics. Another example, where
this goal of constructive success has to some extent been achieved, is
in the development of unbreakable schemes for doing {\em
cryptography}, based upon the principles of quantum mechanics
\cite{Wiesner69a,Bennett84a,Hughes95a}. This is an especially interesting
example, as Shannon used the tools of classical information theory to
``prove'' that the task accomplished by quantum cryptography was not
possible \cite{Shannon49b}. Of course, the flaw in Shannon's proof is
that he assumed a model of communication that did not include the
possibilities afforded by quantum mechanics.

Ideally, these two tasks would dovetail perfectly; for each limit to
information processing that we prove, we would find a constructive
procedure for achieving that limit. Alas, that ideal is often not
achieved, although it remains a central goal of all investigations
into quantum information processing.

We have been rather vague about what is meant by the term ``quantum
information''. What sorts of entities qualify as quantum information?
The answer to this question will evolve as we proceed through the
Dissertation, however it is useful to look ahead at what is in store,
by taking a historical tour to enlighten us as to the tasks which may
be performed using quantum information.

Quantum information really began to get going during the 1960s and
1970s. For example, several researchers began to ask and answer
questions about what communications tasks could be accomplished using
quantum states as intermediary resources. The inputs and outputs to
the processes considered were usually classical information, with the
novelty coming from the use of quantum resources during the process to
aid in the accomplishment of the task. The questions being asked about
these processes were framed in terms of classical information
processing. Much of this early work is reviewed in the inspiring books
of Holevo \cite{Holevo82a} and Helstrom \cite{Helstrom76a}.

A little later came the invention of quantum cryptography
\cite{Wiesner69a,Bennett82a} and the quantum computer
\cite{Benioff80a,Feynman82a,Deutsch85a}. In these applications the
role of quantum mechanics is rather more subtle.  Both possibilities,
most decisively quantum cryptography, enable the performance of
information processing tasks which are considered ``impossible'' in
classical information theory. These new information processing
capabilities acted as a great motivator for the idea that essentially
new types of ``information'' were being used to perform these tasks --
quantum information. Moreover, in both applications it is necessary to
take into account the effect of noise on quantum states, and if
possible minimize the effect of that noise. That problem is strongly
reminiscent of the problem of protecting against noise which arises in
classical information theory, yet without any classical
``information'' apparently involved in the process.

More recently, the pioneering work of Schumacher \cite{Schumacher95a}
on quantum data compression, by Shor \cite{Shor95a} and Steane
\cite{Steane96a} on quantum error correcting codes, and by Wootters
and coworkers \cite{Bennett96b,Bennett96a,Hill97a,Wootters98a} on
measures of entanglement has made this idea of essentially new types
of information much more precise. Schumacher quantified the physical
resources necessary to store the quantum states being emitted by a
``quantum source''. Shor and Steane showed how to protect quantum
states and entanglement against the effects of noise. Finally,
Wootters and coworkers have emphasized the use of quantum entanglement
as a resource that may be useful in the solution of many information
processing problems, and have characterized entanglement by its
efficiency as an aid in those problems.

Perhaps, then, we may distill the following heuristic definition of
information from this historical tour: {\em (Quantum) Information is
any physical resource which may be of assistance in the performance of
an interesting (quantum) information processing task}. Of course, this
simply moves the definitional difficulty elsewhere, but speaking for
myself, I believe that I have a better intuitive feel for what
constitutes an information processing task than for the more ethereal
question of what information is.

Reflecting on this historical tour we see that quantum information
comes in many different types. Some of the types of information of
interest include classical information, entanglement, and actual
quantum states. This is in contrast to the classical theory of
information processing which is largely focused on information types
derived from a single structure: the bit\footnote{There is a well
developed theory of analogue computation, which, however, appears to
be equivalent to the theory based on bits, when physically realistic
assumptions about the presence of noise are made.}. The greater
variety of information structures available in quantum information
necessitate a broader range of tools for understanding the different
information types, and open up a richer range of information
processing possibilities for exploration.

We conclude from the history that quantum information is an evolving
concept, and it seems likely that we are yet a long way from grasping
all the subtleties of the different kinds of quantum
information. Indeed, in the future we may discover new quantum
resources whose importance is not yet glimpsed, but which will one day
be seen as a crucial part of quantum information theory.

\section{Overview of the Dissertation}

The primary purpose of the Dissertation is {\em to develop theoretical
bounds on our ability to perform information processing tasks in
quantum mechanics.}

Two aspects of this purpose deserve special comment. First, it is to
be emphasized that the purpose is to find bounds on our ability to
process information. We will not always be able to determine whether
the bounds we discover are achievable. Nevertheless, it is still of
considerable interest to understand limits to what is in principle
possible. Second, the focus of the Dissertation is theoretical,
although Chapter \ref{chap:fundamentals} does contain an overview of
the experimental state of the field, and the results of a simple
experiment in quantum information. However, a full exposition of the
experimental state of the field is beyond the scope of this
Dissertation.

The Dissertation is structured into three parts.

The first part of the Dissertation, ``Fundamentals of quantum
information'', provides an introductory overview of quantum
information, and develops tools for the study of quantum information.
Part I consists of Chapters \ref{chap:intro} through
\ref{chap:distance}. The primary purpose of Part I is to provide a
pedagogical introduction and reference for concepts in the
field. While Part I contains a substantial amount of original research
material, the presentation of that material is ancillary to the main
goal, which is to provide a solid basis for the understanding of the
quantum information-theoretic problems investigated in Part II of the
Dissertation, which is primarily oriented towards original research
results.

The following is a brief summary of the contents of Part I.

Chapter \ref{chap:fundamentals} provides an introduction to many of
the most basic notions used in quantum information, such as quantum
states, dynamics, quantum gates, quantum measurements and the notion
of a quantum computer. These notions are illustrated using a number of
simple examples, most notably quantum teleportation and superdense
coding. We revisit in greater detail the question of what measurements
may be performed in a quantum system. The Chapter concludes with a
summary of some of the challenges facing experimental quantum
information. Notable original features of the Chapter include a
discussion of realizable measurements in quantum mechanics, and the
description of an experimental implementation of quantum teleportation
using nuclear magnetic resonance.

Chapter \ref{chap:qops} is a review of the {\em quantum operations}
formalism, used to describe state changes in quantum systems. This
formalism includes as special cases the unitary evolution generated by
the Schr\"odinger equation, quantum measurements, and noise processes
such as phase decoherence and dissipation. Notable original features
of the Chapter include a discussion of {\em quantum process
tomography}, a procedure by which the dynamics of a quantum system may
be experimentally determined, and a formulation of quantum
teleportation within the quantum operations formalism.

Chapter \ref{chap:entropy} reviews the concepts of {\em entropy} and
{\em information} that underpin much of quantum information. Entropic
measures often arise naturally in the study of resource problems in
quantum information, which are usually of the form {\em how much of
physical resource X do I need to accomplish task Y?} Much of Part II
of the Dissertation is concerned with such resource problems, so it is
crucial that we obtain a solid understanding of the basic facts about
entropy. A notable feature of the Chapter is the inclusion of several
inequalities relating von Neumann entropies which I believe to be new.

Chapter \ref{chap:distance} reviews {\em distance measures for quantum
information}. A distance measure provides a means for determining the
similarity of two items of quantum information. For example, we may be
interested in the question of what it means for two quantum states to
be ``close'' to one another. Many different measures of distance may
be proposed, motivated by different physical questions one may ask
about quantum information. This Chapter reviews the motivation for
many of these definitions, and attempts to relate some of the
definitions that have been proposed. This Chapter contains the most
original material of any Chapter in Part I, including many new
properties of the various measures of distance investigated, and some
new relationships between the distance measures that have been
proposed.

%different 

This concludes the summary of the contents of Part I.

% part II

Part II of the Dissertation, ``Bounds on quantum information
transmission'', poses a number of questions about information
transmission, and provides bounds on the answers to those
questions. Part II consists of Chapters \ref{chap:qcomm} through
\ref{chap:capacity} of the Dissertation. In Part II, the tools
developed in Part I are employed in the investigation of several
substantive questions in quantum information theory. Part II is
largely devoted to the presentation of original research results.

The following is a brief summary of the contents of Part II.

Chapter \ref{chap:qcomm} studies quantum communication
complexity. Quantum communication complexity is concerned with the
communication cost incurred during the performance of some distributed
computation, if quantum resources are employed for the
communication. Recently, several remarkable results have been proved
showing that in some cases the use of quantum resources may provide a
substantial saving over the communication cost required to solve a
problem in distributed computation with classical resources. The
Chapter begins with an explanation of Holevo's theorem, which is a
fundamental bound on the ability to perform quantum
communication. This bound is then applied to give a new capacity
theorem which precisely quantifies the resources required to send
classical information over a two way quantum noiseless channel. This
capacity theorem is applied to demonstrate a significant new negative
result in quantum information: that there exist problems of
distributed computation for which the use of quantum resources can
provide no improvement over the situation in which only classical
resources are used. Next, we turn our attention to the following
problem: what communication resources are required to compute a {\em
quantum function} -- a unitary evolution -- if that function is
distributed over two or more parties? To my knowledge all previous
work on quantum communication complexity has focused on distributed
computation of {\em classical functions}. In addition to posing this
problem for the first time, this Chapter contains the first
non-trivial lower-bound on such a problem, the communication
complexity for computation of the quantum Fourier transform by two
parties, as well as a general lower bound for the communication
complexity of an arbitrary unitary operator.

Chapter \ref{chap:data_compress} studies the problem of quantum data
compression. It is well known that it is often possible to compress
classical information so that it uses up fewer physical resources. For
example, there are many widely used programs which can be used to
compress computer files so that they take up less disk space. It turns
out that it is possible to compress quantum states along somewhat
similar lines, so that they may be stored using fewer physical
resources. This Chapter provides a new proof of the fundamental
theorem of quantum data compression, substantially simplifying earlier
proofs. Furthermore, the Chapter reports results on \emph{universal
data compression}, which allows the compression of a quantum source
whose characteristics are not completely known.

Chapter \ref{chap:ent} studies the fundamental problem of providing
quantitative measures of the entanglement between two quantum systems.
More than any other resource, it appears to be quantum entanglement
which enables the most striking departures of quantum information
processing from classical information processing, and it is to be
hoped that developing quantitative measures of entanglement will
enable us to better understand the nature of this resource.  Several
measures of entanglement are reviewed, and many new bounds on these
measures and relationships between the measures are proved. I discuss
the insights into quantum information which are given by these bounds,
emphasizing connections with other problems studied in the
Dissertation.  This Chapter does not contain any results which are
especially striking in their own right; rather it proves several new
results and examples which provide insight into the results in other
Chapters of Part II.

Chapter \ref{chap:qec} describes the methods that have been developed
for the performance of quantum error correction. New
information-theoretic conditions for quantum error-correction are
developed, together with other information-theoretic constraints upon
the error-correction process.  The Chapter concludes with an original
analysis of the thermodynamic cost of quantum error correction.

Chapter \ref{chap:capacity} studies the {\em capacity} of a noisy
quantum channel. The capacity is a measure of how much quantum
information can be transferred over a noisy quantum communications
channel with high reliability. Unfortunately, at present the quantum
channel capacity is still rather poorly understood. This Chapter
presents several new bounds on the quantum channel capacity, and
emphasizes the differences between classical and quantum information
which make the quantum channel capacity especially interesting. The
Chapter concludes with the presentation of a new problem in quantum
information theory, that of determining the quantum channel capacity
of a noisy quantum channel in which partial classical access to the
channel environment is allowed. New expressions upper bounding the
capacity in this instance are proved.

This concludes the summary of the contents of Part II.

Part III of the Dissertation, ``Conclusion'', consists of a single
Chapter, Chapter \ref{chap:conc}, which summarizes the results of the
Dissertation, and sketches out some directions for future
work. Chapter \ref{chap:conc} begins with a brief summary of the
results of the Dissertation, highlighting specific questions raised in
the Dissertation which deserve further investigation. The Chapter and
main body of the Dissertation concludes by taking a broader look at
the future directions available to quantum information theory,
outlining a number of possible research programs that might be
pursued.

Some miscellaneous remarks on the style and structure of the
Dissertation:

The front matter of the Dissertation contains a detailed table of
contents, which I encourage you to read, as well as a list of figures
with their associated captions. There is also a guide to nomenclature
and notation, which contains notes to assist the reader in translation
between the often incompatible conventions used by different authors
in the field of quantum information.

\begin{figure}[htbp]
\begin{center}
\mbox{\psfig{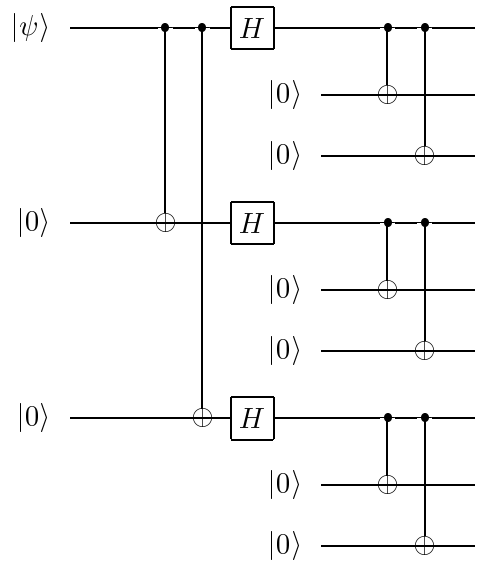}}
\end{center}
\caption{Encoding circuit for the Shor nine qubit code. \label{fig: Shor code}}.
\end{figure}

Each Chapter in Parts I and II of the Dissertation begins with an
overview of the problems to be addressed in the Chapter, and concludes
with a boxed summary of the main results of the
Chapter. Collaborations with other researchers are indicated where
appropriate, usually at the beginning of a Chapter or section. In
addition, I have tried whenever possible to give credit for prior work
in the field, with citations pointing to the extensive bibliography
which may be found at the end of the Dissertation. My apologies to any
researcher whose work I have inadvertently omitted.  Ike Chuang
supplied figures \ref{fig: venn}, \ref{fig: three qubit encoding} and
\ref{fig: Shor code}.

The end matter of the Dissertation contains a single Appendix, a
Bibliography, and an Index.

%\ref{app:mixed}
The Appendix contains material which I felt was outside the main
thrust of the Dissertation, but nevertheless is sufficiently
interesting and useful to warrant inclusion.  It discusses the Schmidt
decomposition, a structural theorem useful for the study of composite
quantum systems. A new generalization of the Schmidt decomposition is
proved in the Appendix, and related concepts such as purifications of
mixed states are discussed.
%Appendix \ref{app:asymptotic} contains a brief review of
%the asymptotic notation used by computer scientists to quantify the
%uses required to solve some information processing problem problem.
% Appendix B contains a brief discussion of the
% information processing power available in physical theories which
% resemble quantum mechanics but allow non-linear dynamics for states.

The Bibliography contains a listing of all reference materials cited
in the text of the Dissertation, ordered alphabetically by the family
name of the first author.

The Index references the most important occurrences of technical terms
and results appearing in this Dissertation.  Only subjects are
indexed, not names.

Finally, I note that the Dissertation has been written in the first
person. When ``I'' appears, it indicates my opinion, or something for
which I claim responsibility. ``We'' indicates occasions where I hope
you, the reader, and I, the author, can fully agree.

\section{Quantum information, science, and technology}

What is the broader relationship of quantum information with science
and technology? This question is well beyond my ability to answer in
full, however based upon what we now know it is interesting to essay
some possible answers.

Let us start with science. Predicting the future impact of quantum
information on science is obviously impossible in detail (although see
Chapter \ref{chap:conc} for an attempt in this direction). Instead, we
will attempt to relate the existing goals and achievements of quantum
information to other areas of science, and science as a whole.

% impact on QM

An area in which quantum information theory has already had a
substantial impact is on physicists' understanding of quantum
mechanics. Quantum mechanics is legendary for the counter-intuitive
nature of its predictions. One way to lift the veil of mystery
surrounding quantum mechanics is to develop a toolkit containing
simple tools on which we can rely to help us navigate quantum
mechanics. The development of such a toolkit is one of the primary
aims of quantum information, and is the central goal of Part I of this
Dissertation.

One consequence of this tool building is the development of many
equivalent ways of formulating fundamental physical principles. For
example, Westmoreland and Schumacher \cite{Westmoreland98a} have
recently argued that the physical prohibition against superluminal
communication can be deduced from elementary quantum mechanics, via
the no-cloning theorem \cite{Wootters82a,Dieks82a}.  Feynman
\cite{Feynman65d} has argued that such development of new ways of
looking at physical principles has great value for fundamental
research.
% {\bf (insert Feynman's nice quote on this, once I track it
%down)}.
As we do not yet have a complete fundamental physical theory of the
world \cite{Wilczek97a,Witten97a}, new perspectives on old theories
such as quantum mechanics may be extremely useful in the search for a
more complete theory of the world.

% Collective phenomena

A second area in which I expect the physics of information to
eventually have a great impact is in the study of statistical physics
and collective phenomena. Collective phenomena involve large numbers
of systems interacting to produce some interesting, complicated
behaviour. The investigation of computer science and collective
phenomena both involve the study of complicated behaviours emerging
from simple systems following simple rules. This is particularly so in
models of computation such as cellular automata or object oriented
programming, in which the programs being executed do not have a
natural sequential structure, but rather involve the parallel
interaction of many relatively simple systems. The hope is that
connections between the two fields can be found, based upon the
analogy in the tasks the two fields attempt to accomplish. Indeed,
some connections between the two fields are already known at the
classical level (see for example \cite{Wolfram94a} and references
therein). However, little work investigating possible connections
seems to have been done in the quantum case. We will return to this
problem in Chapter \ref{chap:conc} with some concrete proposals for
investigation of the connections between these two areas.

% Fundamental physics

\index{fundamental physics}
I have repeatedly stressed the impact that physics has on the
foundations of computer science, as it causes us to re-evaluate the
fundamental models used in the study of information processing. Does
the physics of information, especially quantum information, have a
similar impact on fundamental physics?

The term {\em fundamental physics} itself has been the subject of
considerable debate in recent years. On occasion, it appears merely to
mean ``my research is more important than yours''.  Two particularly
strongly argued cases for what it means for a phenomenon to be
fundamental have been presented by Anderson
\cite{Anderson72a} and Weinberg \cite{Weinberg93a}.

Anderson's article, entitled ``More is different'', argues that
essentially new principles appear at higher levels of complexity in
physical systems, that cannot be deduced from the constituent parts
alone. Anderson argues that the study of such phenomena is as
fundamental as the study of particle physics or cosmology, which are
traditionally regarded as the most fundamental parts of physics.

Weinberg takes a very different tack. He introduces what might be
called ``arrow diagrams'' relating different realms of science to one
another. There is an arrow from one field to another if the first
field depends critically upon the second. For example, physical
chemistry ``points'' to quantum mechanics, because the interactions of
atoms and molecules are determined by the rules of quantum
mechanics. Weinberg argues that fundamental phenomena are those which
can not be reduced to some simpler level; they have arrows pointing
towards them, but none pointing away from them. Up to this point I'm
with him, and believe his point of view dovetails nicely with
Anderson's. Weinberg then goes on to assert, without any evidence that
I can see, that particle physics and cosmology are the unique branches
of science which have this property of irreducibility.

What seems to me to be going on here is a confusion of two separate
issues. First is the question of whether or not a phenomenon is {\em
universal}. Particle physics and cosmology study phenomena which are,
without a doubt, universal. The second question is whether or not a
phenomenon is {\em reducible} to a simpler theoretical level. For
example, the energy levels of the Hydrogen atom can be explained quite
well using simple quantum mechanics.

It seems to me that the term {\em fundamental} refers primarily to
whether or not a phenomenon is reducible to some simpler level or
not. If it is not, then our task as scientists must surely be to
explain that phenomenon on its own terms. 

It is instructive to consider the concrete example of
thermodynamics. It is expected by many people that the principles of
thermodynamics should be reducible to mechanics, yet despite decades
of hard work such a reduction has never been generally achieved. It
may be that such a reduction is in principle impossible. It is known,
for example, that many behavioural properties of certain types of
cellular automata can not be deduced merely by knowing their starting
configurations and dynamics without performing a full simulation of
the entire process \cite{Wolfram94a}.

What if such a situation were to obtain in the study of real
phenomena: that the behaviour of those phenomena could not be deduced
from their starting configurations and a detailed knowledge of their
microscopic dynamics, by any means short of observing the actual
ensuing dynamics? Would we give up attempts at explanation of those
phenomena?  Of course not! Our task then would be to discern higher
level principles governing the behaviour of those systems, and to
subject those principles to the same thorough empirical scrutiny which
has been our wont at the microscopic level throughout the history of
physics. This does not imply that we should give up the search for
reductions of one theory to another, but rather, that we should
acknowledge that such attempts may not always be successful, nor need
they be possible, {\em even in principle}.

What then is the role of quantum information in fundamental science,
especially fundamental physics?  First, I believe it can be used to
aid in the reduction of mesoscopic quantum phenomena to the level of
elementary quantum mechanics. It is difficult to point to many
situations where this has yet occurred, but I believe that is
primarily because much of the field has been focused inward, on the
development of basic tools. Recently there has been some indicators
that this is occurring, such as the work of Huelga {\em et al} on
using concepts from quantum information to develop better frequency
standards \cite{Huelga97a}.

Second, quantum information can directly assist the process of
research into fundamental physics. One way of doing this is by
throwing new light on old quantum principles which, as suggested
earlier, is potentially a major stimulant of further progress in
fundamental research. Another way in which quantum information can
inform the progress of fundamental physics is to act as a source and
catalyst for fundamental questions, such as the questions about the
class of realizable measurements raised earlier in this Chapter, or to
suggest new methods of approaching existing questions, such as
Preskill's recent suggestion \cite{Preskill97a} that quantum error
correction could provide a missing link between Hawking's claim
\cite{Hawking76a} that at the fundamental level nature may be
non-unitary, and the unitarity which appears to be the rule in all
experimental work done to date.  More precisely, Preskill has proposed
that this apparent contradiction may be caused by some sort of
``natural'' quantum error correction, in which nature is non-unitary
at very small length scales, but this non-unitarity gradually becomes
less important at longer length scales.

Obviously, it is not possible to say with any degree of certainty how
quantum information will affect fundamental physics in years to
come. Yet I hope to have convinced you that quantum information is a
subject worth thinking about in connection with fundamental physics,
and that it has already resulted in some interesting work of a
fundamental nature.

% Landauer-Wheeler meaning circuit.

\begin{figure}
\begin{center}
\unitlength 1cm
\begin{picture}(5,6)(0,0)
\put(0,0){\framebox(5,2){Computation}}
\put(0,4){\framebox(5,2){Physics}}
\put(1,2.2){\vector(0,1){1.6}}
\put(1.2,2.97){?}
\put(4,3.8){\vector(0,-1){1.6}}
\end{picture}
\caption{Adaptation of the meaning circuits proposed by Wheeler
\cite{Wheeler86a} and Landauer \cite{Landauer86a}. \label{fig:Landauer circuit}}
\end{center}
\end{figure}
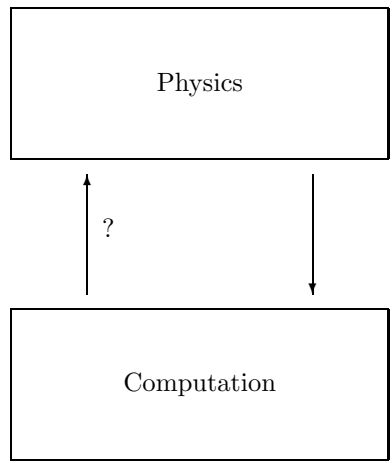

\index{meaning circuits}
There is an interesting and related question one can ask about the
comprehensibility of physical laws, which to my knowledge was first
raised by Wheeler and Landauer, two of the first researchers to
appreciate the deep connections between physics and computer
science. In
\cite{Wheeler86a} and \cite{Landauer86a} they each proposed a
``meaning circuit'' to represent the connections between physics and
computation. An adapted version of these circuits is shown in figure
\ref{fig:Landauer circuit}.

The circuit illustrates two connections between physics and
computation. One is the observation that the laws of physics determine
the scope of possible computational processes. This is an observation
that we have discussed at length in this introductory Chapter, and is
the founding insight for the entire Dissertation. I don't believe we
can reject this part of the circuit without rejecting the founding
principle of physics, namely that the world is essentially orderly,
being governed by some set of laws.

The second part of the meaning circuit may be encapsulated in a
question: {\em Are the consequences of the fundamental laws of physics
computable?} The answer to this question depends on what is
computable. That, as we have seen, depends on what the laws of physics
are, so the question has an interesting self-referential
nature. Another way of stating the question is: {\em Do the laws of
physics allow the existence of structures capable of comprehending
those laws?}

To make progress in physics, it is necessary to assume that the answer
to this question is yes, at least in some limited domain. However,
{\em a priori} there does not seem to be any especially good reason
why the answer to this question ought to be yes, despite the empirical
fact that a good deal about the universe {\em does} appear to be
comprehensible. As Einstein noted, ``{\em The most incomprehensible
thing about the world is that it is comprehensible.}''

In my day to day work I, of course, assume that the laws of physics
do allow the existence of structures capable of comprehending those
laws; there wouldn't be much point to my work otherwise. Nevertheless,
there are some interesting, amusing, and possibly even fruitful
speculations one can engage in, by questioning this assumption.

One observation is that human beings may not naturally exploit the
full information processing power provided by the laws of physics. In
particular, it appears as though information processing devices based
upon quantum mechanics may be intrinsically much more powerful than
devices which process information according to the principles of
classical computer science. It is usually assumed, with considerable
supporting evidence, that the human mind processes information in ways
adequately modeled by classical models of information processing;
this is is sometimes known as the {\em computational hypothesis} of
cognitive science
\cite{Newell76a,vanGelder98a}. Might it be possible that quantum
mechanical devices for computation may exploit their additional
processing power to achieve a more complete comprehension of the world
than is possible using classical computational devices?

This line of speculation may be restated in the language of computer
science. Roughly speaking, the class of problems known as $NP$ in
computer science are those for which a {\em solution} to the problem
can be checked efficiently, that is, in polynomial time on a classical
computer. For example, solutions to the well-known traveling salesman
problem can be be checked quickly on a classical computer: given a
potential solution, one simply checks whether that path length is less
than the desired bound on path lengths. At the 1997 DIMACS Quantum
Computing Tutorial and Workshop, Peter Shor commented on one possible
relationship between classical and quantum complexity classes during
his talk. Shor indicated a suspicion that there may be problems which
can be solved efficiently on a quantum computer, for which even the
solution may not be efficiently checkable on a classical computer.
This would be truly remarkable if correct, and would open up the
possibility of having a quantum oracle that can tell you what the
answer to certain problems are, but not give you a proof of the answer
which you can efficiently check. Might it be that the laws of physics
can be comprehended by intelligences which do quantum information
processing, yet some aspects of those laws remain uncomprehended by
beings which only utilize classical models of computation?

% impact on technology

Let us return from the far edges of speculation to the more practical
concern of understanding the effect quantum information will have on
technology.

% trend to minituarize
\index{Moore's law}

In 1965, Gordon Moore presented a now-famous talk\footnote{To my
surprise, I have not been able to determine where this talk was given,
or whether a written record of the talk exists.} in which he made a
variety of predictions about how computer power would behave over the
coming years. ``Moore's Law'' is quoted in a number of different
forms; perhaps the most famous form is the economic form, that
computer power will double for constant cost every two years or
so\footnote{The quoted doubling time varies quite a bit depending from
source to source, with the usual range of figures being one to two
years.}. We are more interested in the closely related physical forms
of Moore's law, which state, for example, that the number of atoms
needed to represent one bit of information should halve every two
years or so. This prediction has, indeed, been borne out
\cite{Intel98a,Williams98a} over the past thirty years.

%\begin{figure}
%\begin{center}
%\vspace{3cm}
%\caption{The number of transistors per chip. \label{fig:bit_density}}
%\end{center}
%\end{figure}

Extrapolating this trend, it has been predicted \cite{Williams98a} that
around the year 2020, bits will be stored in individual atoms. At that
level, we would certainly expect quantum mechanical effects to become
very important. Techniques for quantum control\footnote{Depending on
what one counts as quantum control, this is already a huge field.
\cite{Mabuchi98a,Viola98a,Lloyd97b} are references on quantum control
that are directly motivated by concerns of quantum information
processing, and which provide an entry into the wider literature.},
motivated in part by potential applications to quantum information
processing, will be necessary to build components at that scale, even
if quantum effects are not harnessed in the information processing
model being implemented.

There is a second physical aspect to Moore's law. The amount of heat
that may be dissipated by a chip with a given surface area per unit
time is roughly a constant, without making use of elaborate
refrigeration techniques. Thus, if the number of components being
squeezed onto the chip is increasing, and the speed of logical
operations on the chip is increasing, then the amount of heat that is
being dissipated per logical operation must necessarily be
decreasing. Once again, heat dissipation per logical operation has
been following its own version of Moore's Law, halving approximately
every twelve months \cite{Intel98a,Williams98a}.

%\begin{figure}
%\begin{center}
%\vspace{3cm}
%\caption{Heat dissipation per logical operation. \label{fig:heat_dissipation}}
%\end{center}
%\end{figure}

\index{Landauer limit}
If current trends continue, by 2020, the amount of heat being
dissipated per logical operation will be $kT$, which Landauer
\cite{Landauer61a} has shown is the fundamental limit for irreversible
logical operations. All modern computers are based upon such
irreversible operations, so without radical changes in the
architecture being used, we will hit a limit to computation set by
heat dissipation requirements. Fortunately, a radical alternative does
exist: reversible computation. Lecerf \cite{Lecerf63a} and Bennett
\cite{Bennett73a} have shown that it is possible to do universal
computation when restricted to reversible logical operations, with
negligible cost in terms of computational resources.

\index{reversible computation} One way of accomplishing the big switch
to reversible computation would be to move to quantum
computation. Ideally, quantum computers are reversible devices at the
atomic level. As we have seen, this is exactly what will be required
for further progress, if current trends continue for another twenty
years. A potential problem with this solution is that both quantum and
classical reversible computers will require error correction
techniques. As we discuss in Chapter \ref{chap:qec}, error correction
itself necessarily dissipates energy, at a rate determined by the
fundamental error rate in the information processing components. It
seems likely that reversible classical information processing
components will have a much lower fundamental error rate than their
quantum components, which would make it much easier to switch to
reversible computing rather than fully quantum computing. On the other
hand, quantum computers have significant advantages over classical
reversible computers, so it may make more sense to ameliorate the cost
of switching to quantum computers by absorbing the funds that would
have been necessary in any case to make the switch to classical
reversible computation.

Until now, we have concentrated on the eventual {\em necessity} of
taking quantum effects into account, if present trends in computer
hardware are to continue. There is substantial economic incentive for
the trends to continue, so it does not seem unreasonable to conclude
that large semiconductor companies may eventually put serious effort
into understanding and harnessing quantum effects.

Of course, the great promise of quantum information processing is to
enable information processing tasks that are either intractably
difficult, or downright impossible, using classical information
processing techniques. At present, the most exciting potential
applications known for quantum information are the ability to factor
large composite numbers \cite{Shor97a,Shor94a}, which in principle
enables many currently popular cryptographic systems to be broken, and
quantum cryptography \cite{Hughes95a,Bennett84a}, which ironically
enables unbreakable cryptographic systems. The attractiveness of both
is derived from the widespread interest in private communications, for
example, for financial transactions. I do not believe that either
factoring or quantum cryptography is a truly ``killer application''
which makes the development of large scale quantum information
processing imperative. One of the chief unknowns in the future of
quantum information is whether such killer applications are possible,
and if so, how they are to be found. This is a subject we will return
to and discuss in more detail in Chapter \ref{chap:conc}.

% Side effect on nanotech:
% - nanotech: biotech,
% - diagnostic tool.

What of the effect of quantum information on other fields? It is
difficult to assess that effect until it becomes more clear what
diagnostic use the tools of quantum information are. It is possible
that techniques developed within the field of quantum information such
as quantum error correction and quantum process tomography will
provide much more precise information about the noise processes taking
place at the atomic level than is currently known. If this hope is
fulfilled, quantum information may have an enabling effect for fields
such as biotechnology \cite{Grace97a}, quantum electronics
\cite{Yariv89a}, and molecular nanotechnology
\cite{Drexler92a}, whose eventual effectiveness depends, to some
extent, on a detailed understanding of the processes taking place at
the atomic level.

%
% Moore's law
%

\index{Moore's law!quantum corollary to} Let us conclude the Chapter
by noting an amusing and optimistic ``quantum corollary'' to Moore's
law. In general, it is very difficult to simulate a quantum system on
a classical computer. The difficulty of doing so rises exponentially
with the number of qubits\footnote{The base two logarithm of the
number of Hilbert space dimensions in non-qubit systems.} in the
system being simulated. In practice, what this means is that at least
twice as much classical computer power is required to simulate a
quantum system containing one more qubit. In fact, the problem may
even be worse than that. The amount of memory required to store the
quantum state at least doubles with each additional qubit, however the
time required to do the simulation also rises because many more
computational paths must be accounted for.

On a quantum computer, however, it is known how to efficiently
simulate a wide class of quantum systems
\cite{Lloyd96a,Abrams97a,Zalka98a}. Very roughly, we may say that for
this class of problems, quantum computers are keeping pace with
classical computers provided a {\em single qubit} can be added to the
quantum computer for every classical doubling period, according to
Moore.  This corollary should not be taken too seriously, as the exact
nature of the gain, if any, of quantum computation over classical will
not be clear until some major problems in computational complexity are
resolved.  Nevertheless, this is an amusing heuristic statement that
helps convey why we should be interested in quantum computers, and
hopeful that they will one day be able to outperform the most powerful
classical computers, at least for some applications.

\vspace{1cm}
\begin{center}
\fbox{\parbox{14cm}{
\begin{center} {\bf Summary of Chapter \ref{chap:intro}: The physics 
of information} \end{center}

\begin{itemize}
\item Information is physical. Each physical theory may be treated as
the basis for a theory of information processing, with possible
differences in resulting computational power.

\item Quantum information is the study of the achievable limits to
information processing possible within quantum mechanics.

\item There is more than one type of information available to be processed
in quantum mechanics, including {\em coherent quantum states,
classical information}, and {\em quantum entanglement}.

\item The promise of quantum information is to reveal new information
processing capabilities beyond what is possible in traditional models
of information processing, and to inform us as to the limits of
quantum mechanics as a means for information processing.

\item Quantum corollary to Moore's law: for certain applications,
quantum computers need only increase in size by one qubit every two
years, in order to keep pace with classical computers.
\end{itemize}

}}
\end{center}

\chapter{Quantum information: fundamentals}
\label{chap:fundamentals}

This Chapter introduces many of the fundamental notions of quantum
information, emphasizing notation, terminology, and simple
examples. It is assumed that you are familiar with elementary quantum
mechanics, in particular, the bra-ket notation for state vectors. The
Chapter begins with an introduction to the fundamental unit of quantum
information, the quantum bit, or {\em qubit}, and then gives two
important examples of quantum information processing -- superdense
coding and quantum teleportation.  With these concrete examples in
hand, we introduce a general model of quantum information processing,
the {\em quantum circuit}, or {\em quantum computing} model. This
model is an attempt to formulate a general framework for the
description of quantum information processing, and will be used as a
tool in the description of many of the information processing tasks we
discuss in the Dissertation. The Chapter concludes with an overview of
the challenges facing an experimentalist wishing to do quantum
information processing in the laboratory, a brief description of some
of the technologies for quantum information processing which have been
proposed or implemented, and a description of a new experimental
implementation of quantum teleportation using liquid state nuclear
magnetic resonance.

Before we begin the Chapter proper, a remark about notation. State
vectors will be written in the standard bra-ket notation. Often,
however, we will have occasion to use density operators. The standard
notation for density operators is sometimes inappropriate when
discussing composite systems. For example, a composite system
consisting of two parts, $A$ and $B$, will have three density
operators associated to it and its various parts, $\rho^A, \rho^B$ and
$\rho^{AB}$. In addition, we will often wish to compare two or more
different density operators on the same system, and may be interested
in the states of the various system at different times.  All this adds
up to a mess of notation, with primes, subscripts and
superscripts. For that reason, where it is clear, I often drop the
$\rho$, and simply write $A$, $B$, and $AB$ to indicate the density
operators associated with the corresponding systems.

\section{Quantum bits}
\label{sec:qubits}
\index{qubit}

The simplest quantum mechanical system has a two dimensional complex
state space. Suppose we single out an orthonormal basis set in the
state space of such a system, and label the basis vectors $|0\ra$ and
$|1\ra$. Then an arbitrary pure state of the system has the form
\beqn
|\psi\ra = \alpha |0\ra + \beta |1\ra, \eeqn where $\alpha$ and
$\beta$ are complex numbers which must satisfy the condition
$|\alpha|^2+|\beta|^2 = 1$ in order for $|\psi\ra$ to be correctly
normalized.

This two dimensional quantum system is known as the {\em quantum bit}
or {\em qubit} \cite{Schumacher95a}, by analogy with the {\em bit},
the fundamental unit of information in the classical theory of
information processing. The states $|0\ra$ and $|1\ra$ are known as
the {\em computational basis states}\index{computational basis
states}. They are merely reference states; it does not matter how they
are chosen, just that we agree upon which states they are. In abstract
discussions of quantum information processing, the computational basis
states are no more than a fixed reference set of orthonormal basis
states. In discussions of real physical systems implementing qubits,
it is usual to pick the computational basis states so that they
correspond to some other physically interesting pair of states. For
example, in nuclear magnetic resonance implementations of a single
qubit on a nucleus of spin $1/2$, it is usual to identify the $|0\ra$
and $|1\ra$ states with the magnetic eigenstates of the spin
corresponding to the large constant applied magnetic field.

It is instructive to compare bits and qubits. A bit can be in one of
two states, $0$ or $1$. A qubit can be in a continuum of states,
described by the complex numbers $\alpha$ and $\beta$. It is possible,
in principle, to distinguish the $0$ and $1$ states of a bit. It is
not possible, in general, to distinguish non-orthogonal states of a
quantum system. For example, if we prepare a qubit in one of the two
states $|0\ra$ and $(|0\ra+|1\ra)/\sqrt 2$, it can be shown that it is
not possible to perform a measurement on that system which will
reliably tell which of these two states was prepared\footnote{See
section \ref{sec:Holevo} for further discussion of this point.}.

By contrast, if we ensure that a qubit is always kept in the $|0\ra$
or $|1\ra$ state, then it is always possible to determine which state
the system is in. Indeed, it turns out that all the information
processing tasks which can be done with bits can also be done with
qubits, provided the qubit remains in one of the two states
$|0\ra$ or $|1\ra$. Thus, information processing models based upon the
qubit are {\em at least} as powerful as models based upon the
bit. 

There are several items of terminology related to qubits which we
ought to agree upon now.  Four standard operators acting on a single
qubit are the {\em Pauli sigma operators}\index{Pauli
sigmas matrices}\index{sigma matrices}, defined by \beqn I \equiv \sigma_0
\equiv \left[ \begin{array}{cc} 1 & 0 \\ 0 & 1 \end{array} \right];
\,\,\,\, X \equiv \sigma_1 \equiv \left[ \begin{array}{cc} 0 & 1 \\ 1
& 0
\end{array} \right]; \\
Y \equiv \sigma_2  \equiv  \left[ \begin{array}{cc} 0 & -i \\ i & 0
\end{array} \right]; \,\,\,\, 
Z \equiv \sigma_3  \equiv  \left[ \begin{array}{cc} 1 & 0 \\ 0 & -1
\end{array} \right], \eeqn
where these matrices are written in the computational basis $|0\ra,
|1\ra$.  The standard notation for the Pauli operators is $\sigma_i$;
we will more often omit the redundant $\sigma$, and just write $I, X,
Y$ or $Z$ instead.

\index{Bloch sphere}
\index{Bloch vector}

The Pauli operators form a basis set for the vector space of operators
on a single qubit.  In particular, an arbitrary operator $A$ acting on
a single qubit can be written uniquely in the \emph{Bloch
representation},
\be
A = \sum_{i=0}^3 a_i \sigma_i. \ee
The Bloch representation has a particularly attractive form for
density operators of a single qubit.  Such a density operator can be
written in the form
\be
\rho = \frac{I+\vec \lambda \cdot \vec \sigma}{2}, \ee
where $\vec \lambda = (\lambda_x,\lambda_y,\lambda_z)$ is the
\emph{Bloch vector} for the state, characterized by the requirement
that the vector is real and satisfies $\| \vec \lambda \| \leq 1$.

The Pauli operators are our first examples of {\em quantum
gates}\index{quantum gates}. Classical information processing is
accomplished by various {\em logic gates} which act on the bits being
processed. Similarly, quantum information processing is accomplished
by quantum gates. Quantum gates are operations acting on a fixed
number of qubits. For example, the Pauli operators represent unitary
evolutions which may take place on a single qubit. The $X$ Pauli
operator is often known as the {\em quantum not gate}\index{not gate},
as it flips the computational basis states, $X|0\ra = |1\ra$ and
$X|1\ra = |0\ra$, much as the classical not gate interchanges $0$ and
$1$. The $Z$ Pauli operator is often known as the {\em phase flip
gate}\index{phase flip operator}, as it flips the relative phase of the
computational basis states, $Z|0\ra = |0\ra$ and $Z|1\ra = -|1\ra$. At
present there is no widely accepted term for the $Y$ operator. $I$ is,
of course, the identity gate\index{identity gate}.

Two more quantum gates which are of great importance are the {\em
Hadamard} and {\em phase shift}
gates\label{defn:Hadamard}\index{Hadamard gate}\index{phase shift
operator}. These gates are defined, respectively, as follows: \beqn H & =
& \frac{1}{\sqrt 2} \left[ \begin{array}{cc} 1 & 1 \\ 1 & -1
\end{array} \right] \\
S & = & \left[ \begin{array}{cc} 1 & 0 \\ 0 & e^{i\pi/4} \end{array}
\right]. \eeqn Note that $S^4 = Z$, $HZH = X$, and $ZX = Y$, up to a
global phase\footnote{More precisely, $ZX = iY$.  In quantum
mechanics, global phase factors such as the $i$ can be ignored.}, so
the Hadamard gate and phase shift together can be used to generate any
of the Pauli operators. Later, we will introduce a two qubit gate, the
controlled not gate. The controlled not gate, the Hadamard gate and
the phase shift gate together form a {\em universal
set}\index{universal set of gates} -- any unitary operation can be
approximated arbitrarily well making use of only these gates.

There are many reasons the qubit is regarded as the fundamental unit
of quantum information. It is the simplest quantum mechanical system,
and is quite easily analyzed. Moreover, the state space of any finite
dimensional quantum system can be understood to be composed of a
number of qubits. In this respect, the qubit closely resembles the
classical bit. It is possible to formulate classical information
processing in terms of {\em trits}, for example, which are classical
systems taking the three values $0, 1$ and $2$. In certain systems, it
may even be more natural to do the analysis this way. However, little
is lost from the theoretical point of view by regarding a trit as
being composed of two bits, in which only the three states $00, 01$
and $10$ are accessible. Similarly, a three dimensional quantum system
can be regarded as essentially identical to a pair of qubits in which
the state is guaranteed to be in the space spanned by the states
$|00\ra, |01\ra$ and $|10\ra$. For all these reasons, and others which
will become apparent as we move deeper into quantum information, the
qubit is regarded as the fundamental unit of quantum information.

\section{Superdense coding}
\label{sec:superdense}

There is a simple but important example of quantum information
processing known as {\em superdense coding}\index{superdense coding}
\cite{Bennett92a} which is explained in this section. This example
shows that there are information processing tasks which can be
performed with qubits which do not have natural analogues in terms of
bits.

Superdense coding involves two parties, conventionally known as
``Alice'' and ``Bob'', who are a long way away from one another. Their
goal is to transmit some classical information from Alice to
Bob. Suppose Alice is in possession of two classical bits of
information which she wishes to send Bob, but is only allowed to send
a single qubit to Bob. Can she achieve her goal?

Superdense coding tells us that the answer to this question is
yes. Suppose Alice and Bob initially share a pair of qubits in the
entangled state
\beqn
	|\psi\ra =\frac{|00\ra+|11\ra}{\sqrt 2}. 
\eeqn
Alice is initially in possession of the first qubit, while Bob has
possession of the second qubit. Note that this is a fixed state; there
is no need for Alice to have sent Bob any qubits in order to prepare
this state. Instead, some third party may prepare the entangled state
ahead of time, sending one of the qubits to Alice, and the other to
Bob.

By sending her single qubit to Bob, it turns out that Alice can
communicate two bits of classical information to Bob.  Here is the
procedure she uses. If she wishes to send the bit string ``00'' to Bob
then she does nothing at all to her state. If she wishes to send
``01'' then she applies the quantum not gate, $X$, to her qubit. If
she wishes to send ``10'' then she applies the phase flip, $Z$, to her
qubit. If she wishes to send ``11'' then she applies the $iY$ gate to
her qubit. The four resulting states are
\beqn \label{eq:superdense1}
00: |\psi\ra & \rightarrow & \frac{|00\ra+|11\ra}{\sqrt 2} \\
01: |\psi\ra & \rightarrow & \frac{|10\ra+|01\ra}{\sqrt 2} \\
10: |\psi\ra & \rightarrow & \frac{|00\ra-|11\ra}{\sqrt 2} \\
\label{eq:superdense4}
11: |\psi\ra & \rightarrow & \frac{|01\ra-|10\ra}{\sqrt 2}.  \eeqn
These four states are known as the {\em Bell basis}\index{Bell basis},
after John Bell, who did so much to emphasize the importance of
entanglement \cite{Bell89a}. Notice that the Bell states form an
orthonormal basis, and can therefore be distinguished by an
appropriate quantum measurement. Alice now sends her qubit to Bob,
giving Bob possession of both qubits. By doing a measurement in the
Bell basis Bob can determine which of the four bit strings Alice sent.

This remarkable prediction of quantum mechanics has been given a
partial experimental validation by Mattle {\em et al} using entangled
photon pairs \cite{Mattle96a}. In the experiment, a trit of classical
information was sent using photon polarization as the qubit. It was
only possible to send a trit, rather than two bits, because with the
measurement scheme used, the experimentalists were unable to
distinguish between the states corresponding to $00$ and $01$, above.

It is surprising enough that a two level quantum system can be used to
transmit two bits of classical information, however there is another
remarkable aspect to this procedure. Suppose Alice sends her qubit to
Bob, but the qubit is intercepted on the way by a third party, Eve.
Examining the four states
(\ref{eq:superdense1})--(\ref{eq:superdense4}), we see that in each
case, the reduced density operator associated with the first qubit is
the same, the completely mixed state $I/2$. Because the reduced states
are the same regardless of which state was prepared, Eve can infer
nothing about the information Alice is trying to send by examining the
qubit she has intercepted. The intercepted qubit contains essentially
no classical information; rather, the classical information is
contained jointly by the two qubits.

Superdense coding is an example of how quantum and classical
information can be combined in an interesting way. In Chapter
\ref{chap:qcomm} we will return to study the limits to superdense
coding in a much more detailed fashion, along the way to some results
about the efficiency of distributed computations in quantum mechanics.

\section{Quantum teleportation}
\label{sec:teleportation} \index{teleportation}

\begin{quote}
The medium is the message.
\end{quote}
-- Marshall McLuhan \cite{McLuhan64a}

Superdense coding shows that quantum information may be used in an
interesting way as the medium for transmission of classical
information. An even more remarkable effect, quantum teleportation
\cite{Bennett93a}, shows that classical information and entanglement
can be used as the medium for transmission of a quantum state.

Suppose Alice is living in London and wishes to send a single qubit to
Bob, who is living in New York.  There are many different ways Alice
could do this. One method is for Alice to send a description of her
state to Bob, who can then create that state in New York. This method
has two major disadvantages. First, quantum states are specified using
sets of complex numbers. For quantum systems of many qubits it
requires a huge number of classical bits to specify the state to
reasonable accuracy. The cost of transmitting these classical bits may
be considerable. Second, suppose the state of Alice's system is not
known to Alice. The situation then is even worse, because it is not
possible, even in principle, for Alice to determine the state of her
system. There is no way she can send her system to Bob by sending Bob
a classical description.

A second method is to physically move the quantum system from London
to New York. For example, a photon could be sent down a highly
idealized fiber optic from London to New York.  This method also
suffers from two major disadvantages. First, it may simply be very
difficult to reliably send qubits from London to New York. The channel
used to do so may degrade over time, or it might be unreliable to begin
with. Second, if the qubit being sent was carrying information that
information could be intercepted by a malevolent third party.

Quantum teleportation is a method for moving quantum states from one
location to another which suffers from none of these problems. Suppose
Alice and Bob share a pair of qubits which are initially in the
entangled state $(|00\ra+|11\ra)/\sqrt 2$.  In addition, Alice has a
system which is in some potentially unknown state
$|\psi\ra$. The total state of the system is therefore
\beqn
|\psi\ra\left( \frac{|00\ra+|11\ra}{\sqrt 2} \right). \eeqn
By writing the state $|\psi\ra$ as $\alpha|0\ra+\beta |1\ra$ and doing
some simple algebra, we see that the initial state can be rewritten as
\beqn \label{eq:teleport}
(|00\ra+|11\ra) |\psi\ra + (|00\ra-|11\ra)
	Z|\psi\ra + (|01\ra+|10\ra) X|\psi\ra +
	(|01\ra-|10\ra) XZ |\psi\ra. \nonumber \\
 {} \eeqn
Here and throughout the remainder of this section we omit
normalization factors from the description of quantum states.

Suppose Alice performs a measurement on the two qubits in her
possession, in the Bell basis, consisting of the
four orthogonal vectors, $|00\ra+|11\ra,
|00\ra-|11\ra,|01\ra+|10\ra,|01\ra-|10\ra$, with corresponding
measurement outcomes which we label $00, 01, 10$ and $11$. From the
previous equation, we see that Bob's state, conditioned on the
respective measurement outcomes, is given by \beqn 00: |\psi\ra;
\,\,\,\, 01: X|\psi\ra; \,\,\,\, 10: Z |\psi\ra; \,\,\,\, 11: XZ
|\psi\ra. \eeqn

Therefore, if Alice transmits the two classical bits of information
she obtains from the measurement to Bob, it is possible for Bob to
recover the original state $|\psi\ra$ by applying unitary operators
inverse to the identity, $X$, $Z$ and $XZ$, respectively. More
explicitly, if Bob receives $00$, he knows his state is $|\psi\ra$, if
he receives $01$ then applying an $X$ gate will cause him to recover
$|\psi\ra$, if he receives $10$ then applying a $Z$ gate will cause
him to recover $|\psi\ra$, and if he receives $11$ then applying an
$X$ gate followed by a $Z$ gate will enable him to recover $|\psi\ra$.

This completes the teleportation process.

It is interesting to note that teleportation involves the transmission
of only {\em two bits} of classical information. This is despite the
fact that in general it takes an infinite amount of classical
information to describe the state to be teleported. Furthermore, the
success of the teleportation procedure did not in any way depend upon
Alice knowing anything about the quantum state she was sending. Even
more remarkably, we see from equation (\ref{eq:teleport}) that each of
the four Bell states appears with equal weight in the superposition
making up the initial state. Thus, the four measurement results have
equal probabilities $1/4$, {\em independent of the initial state}
$|\psi\ra$. Because the probability is independent of the state
$|\psi\ra$, neither Alice nor anybody else can infer anything about
the identity of the state being teleported from the measurement
outcome.

\begin{figure}[t]
\begin{centering}
\unitlength 1.2cm
\begin{picture}(7.1,3)(-0.8,0)
%\put(-0.8,2.2){\framebox(1,2.6){\parbox{0.8cm}{Alice \\ data $\,
%	\psi$}}}
\put(-0.6,2.45){data}
%\put(-0.49,2.25){(C2)}
\put(-0.75,1.45){ancilla}
%\put(-0.46,1.25){(C1)}
\put(-0.70,0.45){target}
%\put(-0.42,0.25){(H)}
\put(0.25,2.45){$\psi$}
\put(0.25,1.45){$0$}
\put(0.25,0.45){$0$}
\put(0.5,0.5){\line(1,0){0.3}}
\put(0.5,1.5){\line(1,0){2.65}}
\put(0.5,2.5){\line(1,0){1.8}}
\put(0.8,0.2){\framebox(0.6,0.6){$R_y$}}	% first y 90 rotation.
\put(1.4,0.5){\line(1,0){2.9}}
\put(1.7,1.6){\line(0,-1){1.1}}			% controlled not from
\put(1.7,1.5){\circle{0.2}}			% target to ancilla
\put(1.7,0.5){\circle*{0.1}}
\put(2.0,2.5){\line(0,-1){1.1}}			% controlled not from
\put(2.0,1.5){\circle{0.2}}			% data to ancilla
\put(2.0,2.5){\circle*{0.1}}
\put(2.3,2.2){\framebox(0.6,0.6){$R_{-y}$}}
\put(2.9,2.5){\line(1,0){0.25}}
\put(3.15,1.4){\framebox(1.2,1.2){Measure}}	% Measurement box
\put(4.35,2.5){\line(1,0){1.75}}
\put(4.35,1.5){\line(1,0){1.75}}
\put(4.3,0.2){\framebox(0.6,0.6){$X$}}
\put(4.9,0.5){\line(1,0){0.3}}
\put(5.2,0.2){\framebox(0.6,0.6){$Z$}}
\put(5.8,0.5){\line(1,0){0.3}}
\put(4.6,1.5){\line(0,-1){0.7}}
\put(4.6,1.5){\circle*{0.1}}
\put(5.5,2.5){\line(0,-1){1.7}}
\put(5.5,2.5){\circle*{0.1}}
\end{picture}
\end{centering}
\caption{Circuit for quantum teleportation. The measurement is in the
computational basis, leaving the measurement result stored in the
data and ancilla qubits. $R_y$ and $R_{-y}$ denote rotations of $90$
degrees about the $y$ and $-y$ axes on the Bloch sphere. 
\label{fig: Brassard}}
\end{figure}
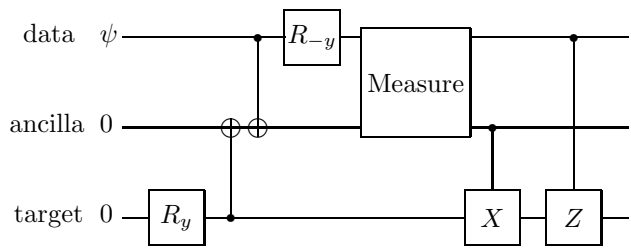

Quantum teleportation can be recast in the language of quantum gates
which we met briefly earlier in this Chapter.  A {\em quantum circuit}
implementing teleportation is shown in figure \ref{fig: Brassard}
\cite{Brassard96b}. \index{quantum circuit!for teleportation} The
three lines traversing the circuit from left to right represent the
three qubits involved in teleportation. The top line represents the
initial state which Alice wishes to teleport. We shall refer to it as
the {\em data qubit}.  The second line represents the qubit which
Alice uses to share the initial entanglement with Bob, which we shall
call the {\em ancilla qubit}.  The third line represents Bob's qubit,
which we shall call the {\em target qubit}.

The quantum circuit is read from left to right. The input state to the
quantum circuit is assumed to be the product state $|\psi\ra|00\ra$,
and the first two gates in the circuit are used to create the
entanglement between Alice and Bob. The very first gate is a $90$
degree rotation about the $y$ axis on the Bloch sphere\footnote{In
general, a rotation by $\theta$ degrees about the $\vec n$ axis on the
Bloch sphere is defined to be $\exp(-i\theta \vec n \cdot \vec \sigma
/2)$, where $\vec n$ is a unit vector, in this case $(0,1,0)$, and
$\vec \sigma = (X,Y,Z)$ is a vector whose entries are the Pauli sigma
operators. Therefore, a $90$ degree rotation about the $y$ axis
corresponds to the operator $(I-iY)/\sqrt 2$.}.  This takes the state
$|0\ra$ to the state $(|0\ra+|1\ra)/\sqrt{2}$. The second gate is
known as the controlled not gate. As can be seen from the figure, the
controlled not gate involves two qubits, which we shall refer to as
the {\em control} and {\em data}\footnote{The data qubit for a
controlled not gate is sometimes known as the target qubit, not to be
confused with the target qubit of our circuit!} qubits. In this
particular gate, the control qubit is on the bottom line, and the data
qubit is on the second line.

\index{controlled not gate} The controlled not gate is a unitary gate
whose action is to flip the data qubit if the control qubit is set to
$|1\ra$, and to leave the data qubit alone if the control qubit is set
to $|0\ra$. Symbolically, \beqn |a\ra|b\ra
\stackrel{\mbox{c-not}}{\longrightarrow} |a\ra X^{a} |b\ra, \eeqn
where $a$ and $b$ are $0$ or $1$. This defines the action of the
controlled not on a basis, and thus on all states.

After the controlled not is applied, the state of the system is seen
to be
\beqn
|\psi\ra \frac{|00\ra+|11\ra}{\sqrt 2}. \eeqn That is, the first two
gates create the necessary entanglement between the target and ancilla
qubits.

The next step of teleportation is to perform a measurement on the data
and ancilla qubits in the Bell basis. The way this is accomplished is
to do two gates which rotate the Bell states into the computational
basis, and then to perform a measurement in the computational
basis. The rotation is accomplished by performing a controlled not
from the data qubit to the ancilla qubit, followed by a rotation by
$-90$ degrees about the $y$ axis on the data qubit. The effect of
these transformations is as follows: \beqn \frac{|00\ra+|11\ra}{\sqrt
2} \rightarrow \frac{|00\ra +|10\ra}{\sqrt 2} & & \rightarrow |00\ra
\\ \frac{|00\ra-|11\ra}{\sqrt 2} \rightarrow \frac{|00\ra
-|10\ra}{\sqrt 2} & & \rightarrow -|10\ra \\
\frac{|01\ra+|10\ra}{\sqrt 2} \rightarrow \frac{|01\ra +|11\ra}{\sqrt
2} & & \rightarrow |01\ra \\ \frac{|01\ra-|10\ra}{\sqrt 2} \rightarrow
\frac{|01\ra-|11\ra}{\sqrt 2} & & \rightarrow -|11\ra. \eeqn Thus, a
measurement in the computational basis will give the result
$00,01,10$, or $11$, corresponding to one of the four Bell
states. Moreover, the state of the data and ancilla qubits after the
measurement is a computational basis state whose value records the
result of the measurement; that is, which Bell state was measured.

Making use of this fact, we apply operations to the target qubit,
conditional upon the measurement result.  First, we apply an $X$ gate
to the target qubit, conditional on the ancilla qubit being set, then
a $Z$ gate to the target qubit, conditional on the data qubit being
set. Thus, the four possible outcomes are: if $00$ is measured, then
the identity transformation is applied to the target; if $01$ is
measured then $X$ is applied; if $10$ is measured then $Z$ is applied
to the target; and if $11$ is measured then $ZX$ is applied to the
target. Notice that this sequence of operations corresponds exactly to
the sequence of operations necessary for quantum teleportation.  It is
interesting to note further that the measurement step can be removed
from the circuit and the state of the data qubit will still be
transferred to the target qubit.  However, this is much less
impressive than the full teleportation operation, in which our
intuition incorrectly tells us that the quantum state initially on the
data qubit is irreversibly destroyed by the measurement process.  This
completes our description of quantum teleportation in the language of
quantum circuits.

Quantum teleportation is an important elementary demonstration of
quantum information theory. Later in this Chapter we discuss the
experimental implementation of quantum teleportation, and in Chapter
\ref{chap:qops} we return to look at quantum teleportation from a
completely different angle, as a noisy quantum channel. Finally, in
Part II of the Dissertation we will repeatedly use teleportation as an
elementary operation as part of more sophisticated quantum information
processing operations.  These and many other uses emphasize the role
quantum teleportation has as an exemplar useful for the study of more
complex forms of quantum information processing.

\section{Quantum computation}
\index{quantum computation} \index{quantum circuit}

The theory of quantum computation is an attempt to capture the
essential elements of a theory of quantum information processing in a
single unified theory. I say ``attempt'' because it is not yet clear
that the theory of quantum computation provides a complete account
of the information processing capabilities afforded by quantum
mechanics.

This section describes a single model of quantum computation, the
{\em quantum circuit} model. Other, equivalent, formulations of quantum
computation have also been proposed, but these will not be discussed
in any detail here. Without further ado, here is an outline of the
quantum circuit model of quantum computation:

\begin{enumerate}

\item {\bf Classical resources:} The quantum computer consists of two
parts, a classical part and a quantum part. In principle, there is no
need for the classical part of the computer, but in practice certain
tasks may be made much easier if parts of the computation can be done
classically. For example, many schemes for quantum error correction
are likely to involve classical computations in order to maximize
efficiency. While classical computations can always be done, in
principle, on a quantum computer, it may be more convenient to perform
the calculations on a classical computer.

\item {\bf A suitable state space:} We assume that the quantum part of
the computer consists of some number, $n$, of qubits. The state space
is thus a $2^n$ dimensional complex Hilbert space. Product states of
the form $|x_1,\ldots,x_n\ra$, where $x_i = 0,1$, are known as {\em
computational basis states} of the computer. We sometimes write
$|x\ra$ for a computational basis state, where $x$ is the number whose
binary representation is $x_1\ldots x_n$.

\item {\bf Ability to prepare states in the computational basis:} It is
assumed that any computational basis state $|x_1,\ldots,x_n\ra$ can be
prepared in at most $n$ steps.

\item {\bf Ability to perform quantum gates:} It is assumed that it is
possible to perform the {\em Hadamard gate} and the {\em $\pi/4$} phase
shift gate on any single qubit of the quantum computer. It is assumed
that it is possible to perform the {\em controlled not} gate on any
pair of qubits in the quantum computer. Recall that these gates are
defined in the computational basis as follows:
\begin{itemize}
\item The Hadamard gate:
\beqn
H \equiv \frac{1}{\sqrt 2} \left[ \begin{array}{cc} 1 & 1 \\ 1 & -1
  \end{array} \right]. \eeqn
\item The $\pi/4$ phase shift gate:
\beqn
S \equiv \left[ \begin{array}{cc} 1 & 0 \\ 0 & e^{i\pi/4} \end{array}
\right]. \eeqn
\item The controlled not gate:
\beqn
C \equiv \left[ \begin{array}{cccc} 1 & 0 & 0 & 0 \\ 0 & 1 & 0 & 0 \\
	0 & 0 & 0 & 1 \\ 0 & 0 & 1 & 0 \end{array} \right]. \eeqn
\end{itemize}

\item {\bf Ability to perform measurements in the computational
basis:} Measurements may be performed in the computational basis of
one or more of the qubits in the computer.

\item \label{item:uniformity} {\bf Algorithm to build the quantum
circuit:} Suppose we wish to solve a problem using the quantum circuit
model of computation. As an example, suppose we wish to factor numbers
in the quantum circuit model. Then given an input for the problem, in
this case the number to be factored, there must be a procedure telling
us how to build the quantum circuit to perform the desired
computation. That is, we must have an algorithm which, given the
input, describes how many qubits will be needed to do the computation,
which computational basis state must be prepared, what gates must be
applied during the computation, and when those gates are to be
applied, what measurements are to be performed during the computation,
and a specification of what measurement results are to be regarded as
output from the computation. This requirement -- that the structure of
the quantum circuit be specified by a classical algorithm -- is known
as the {\em uniformity requirement}\index{uniformity requirement for
quantum computation} for quantum computation.  Without imposing this
important requirement, many impossible tasks would become trivial
within the quantum circuit model, or even in the classical circuit
model of computation \cite{Papadimitriou94a}.
 
\end{enumerate}

This model of computation is equivalent to many other models of
computation which have been proposed, in the sense that other models
result in similar resource requirements for the same problems. For
example, one might wonder whether moving to a design based on three
level quantum systems, rather than the two level qubits, would confer
any computational advantage. Of course, although there may be some
slight advantage in using three level quantum systems over two level
systems, any difference will be essentially negligible from the
theoretical point of view. At a less trivial level, the ``quantum
Turing machine'' model of computation, a quantum generalization of the
classical Turing machine model, has been shown to be equivalent to the
model based upon quantum circuits \cite{Yao93a,Bernstein97a}.

In what ways may the quantum circuit model of computation be
criticized? How might it be modified?  Perhaps my sharpest criticism
of the quantum circuit model is that its basis, although expressed in
terms of quantum mechanics, is not yet wholly rooted in fundamental
physical law. The basic assumptions underlying the model are {\em ad
hoc}, and do not seem to have been analyzed in the literature with
respect to fundamental physical law, at least not in any great depth.

For example, it is by no means clear that the basic assumptions
underlying the state space and starting conditions in the quantum
circuit model are justified. Everything is phrased in terms of finite
dimensional state spaces. Might there be anything to be gained by
using systems whose state space is infinite dimensional? What about
the assumption that the starting state of the computer is a
computational basis state? We know that many systems in nature
``prefer'' to sit in highly entangled states of many systems; might it
be possible to exploit this preference to obtain extra computational
power? It might be that having access to certain states allows
particular computations to be done much more easily than if we are
constrained to start in the computational basis. Likewise, if
measurements could be performed outside the computational basis, it
might be possible to harness those measurements to perform tasks
intractable within the quantum circuit model.

It is not my purpose here to do a detailed examination of the physics
underlying the models used for quantum computation, although I believe
that this is a problem well worth considerable time and effort. I wish
merely to raise in your mind the question of the completeness of the
quantum circuit model, and re-emphasize the fundamental point that
information is physical, and in our attempts to formulate models for
information processing we should always attempt to go back to
fundamental physical laws. A very desirable goal for the future is to
use fundamental physics to demonstrate or refute the following modern
version of the Church-Turing thesis (see also \cite{Deutsch85a}):

\index{Church-Turing thesis!modern form}
{\em Any physically reasonable model of computation can be simulated
in the quantum circuit model with at most polynomial overhead in
physical resources.}

At a more practical level, the quantum circuit model will be used in
this Dissertation to provide a language for the description of quantum
information processing tasks. We have already done that, for example,
in the description of quantum teleportation.  It is only in the next
section that the question of the completeness of the quantum circuit
model of quantum information processing will be an important issue.

\section{What quantum measurements may be realized?}

Let us return to the question asked in section \ref{sec:turing}: What
observables may be realized as measurements on a quantum system? In
this section we discuss this problem from a somewhat different point
of view than was done earlier. The point of view we take is that the
quantum circuit model provides an essentially complete account of the
information processing tasks, including measurement, that can be
accomplished within quantum theory.  As noted in the previous section,
that this is a valid assumption has not yet been established beyond
doubt, however, it will allow us to make progress on the question of
determining what measurements may be performed within quantum
mechanics.

\index{halting problem!proof of unsolvability}
We begin with the halting problem, and a proof that the halting
problem is algorithmically unsolvable. The discussion is a little
different to Turing's \cite{Turing36a}, since we allow probabilistic
algorithms, which generalize the deterministic algorithms considered
by Turing. The central outcome of our discussion is the same as
Turing's though: the halting problem may not be solved by any
algorithm, even a probabilistic algorithm.

Turing's key insight was to formalize what he meant by an
algorithm. Essentially, Turing invented the modern concept of a
programming language for his computers. An algorithm to compute a
function is expressed in terms of a program, which takes as input a
number, and outputs a number -- the input and output of the function
computed by the program. Strictly speaking, the functions computed by
programs are {\em partial} functions, since it is possible that for
some inputs a program will fail to ever halt; the function computed by
the program is therefore undefined for that input.

A key point made by Turing is that his programs can be numbered
$0,1,2,\ldots$. There is no need to explicitly give the details of
Turing's model of computation here; it is well covered in computer
science texts such as
\cite{Davis83a} and \cite{Cormen90a}. We need only imagine a computer
running a program in a familiar language such as C or PASCAL, with the
caveat that programs running on the computer, while finite, may make
use of an arbitrarily large amount of scratch memory while running.
We consider a slight generalization of Turing's model in that we
assume that the computer is equipped with a good random number
generator which generates either a zero or one with equal
probability. This random number generator can be called as part of the
algorithm. It is not difficult to see that even in this slightly
generalized model of computation, it is still possible to number the
possible programs, $0,1,2,\ldots$.

We define the halting function in the probabilistic model of
computation as \beqn h(x) \equiv \left\{ \begin{array}{ll} 0 &
\mbox{if program } x \mbox{ halts with probability } < \frac 12 \mbox{
on input } x\\ 1 & \mbox{if program } x \mbox{ halts with probability
} \geq \frac 12 \mbox{ on input } x. \end{array} \right.  \eeqn Is
there an algorithm which computes the halting function?  More
precisely, does there exist an algorithm which can compute the halting
function better than just randomly guessing, that is, with probability
of correctness greater than one half for each input? We will give a
proof by contradiction that such an algorithm cannot exist, by
assuming that such an algorithm does exist, and showing that it leads
to a contradiction.  More formally, for each possible input $x$, the
algorithm, which we shall call HALT, outputs $h(x)$ with probability
greater than one half. We make use of the algorithm for HALT to
construct another program, which we call TURING, which calls HALT as a
subroutine. In pseudocode:

\begin{verbatim}
program TURING(x)
y = HALT(x)
if y = 0 then
   return x and halt
else
   loop forever
\end{verbatim}

We have assumed that HALT is computable, and thus the program TURING
is also computable, and must have an associated program number, say
$t$. What is the value of $h(t)$? Notice that $h(t) = 1$ if and only
if TURING halts on input of $t$ with probability at least one
half. Inspecting the program for TURING, we see that this is true if
and only if HALT$(t) = 0$ with probability at least one half, and
thus, strictly greater than one half, by assumption. Finally, by
definition this last is true if and only if $h(t) =0$. That is, $h(t)
= 1$ if and only if $h(t) = 0$, clearly a contradiction. Thus, our
original assumption must have been wrong: there is no algorithm which
can compute the halting function with success probability greater than
one half for all inputs.

Having demonstrated that there is no algorithm capable of computing
the halting function, even probabilistically, let us return to the
problem of measurements in quantum mechanics.  Suppose that, rather
than wishing to implement a specific type of measurement, we wish to
implement a {\em family} of measurements. We define a family of
measurements to consist of a sequence ${\cal M} = \{ M_1,M_2,\ldots
\}$ of observables, where $M_n$ is an observable on $n$ qubits. Given
$n$, is it possible, in general, to perform a measurement of $M_n$ on
$n$ qubits?

\index{halting observable!family of observables}
The answer is no. In particular, define the {\em halting family}
${\cal M}$ of observables by
\beqn
M_n \equiv \sum_{x=0}^{2^n-1} h(x) |x\ra\la x|. \eeqn The halting
family can not be measured, even approximately, within the quantum
circuit model of computation. To see why not, suppose that it is
possible to measure this family of observables within the quantum
circuit model. We will outline an algorithm for a Turing machine that
will compute the halting function. The algorithm is very simple: given
$x$, it chooses $n$ greater than $\log x$, and then simulates the
quantum circuit used to measure the observable $M_n$ on $n$ qubits,
with the starting state for the circuit chosen to be $|x\ra$. By
carrying out the simulation to a high enough level of
accuracy,
%\footnote{The required accuracy level can be determined by
%repeated iteration of the simulation.}
we can ensure that the value of $h(x)$ can be read out from the output
of the simulation.

We are left to conclude that it is {\em not} possible, in principle,
to measure the halting family of observables within the quantum
circuit model. If we assume that the quantum circuit model provides a
complete description of the class of information processing tasks
which may be performed in quantum mechanics then we are left to
conclude that physical law does not allow measurement of the halting
family of observables.

It is intriguing to consider the consequences if it were possible to
measure the halting observable or, which is effectively the same
thing, the halting family of observables. Perhaps there really exist
in nature quantum processes which can be used to compute functions
which are classically non-computable. It is far-fetched, but not
logically inconsistent, to imagine some type of experiment - perhaps a
scattering experiment - which can be used to evaluate the halting
function.

Recognizing such a process poses some problems.  How could we verify
that a process computes the halting function (or any other
non-computable function)?  Because of the algorithmic unsolvability of
the halting problem, it is not possible to verify directly that the
candidate ``halting process'' does, in fact, computing the halting
function.  Nevertheless, one can imagine inductively verifying that
the process computes the halting function. In principle, one could do
this by running a large number of programs on a computer for a long
time, and checking that all the programs which halt are predicted to
halt by the candidate halting process, and that programs predicted not
to halt by the candidate halting process have not halted.  Given
sufficient empirical evidence of this sort, one could then {\em
postulate} as a new physical law that the process computes the halting
function.

Physically, the most important conclusion we can draw from this
discussion is that there may be significant limitations on the class
of observables which can be realized in quantum mechanics. Related
restrictions apply for unitary dynamics \cite{Nielsen97d}. These
limitations may go considerably beyond the familiar limits of the type
discovered by Heisenberg, although it is not yet clear precisely what
class of measurements is realizable in quantum mechanics. In what
future directions may this line of thinking be taken? The most obvious
is to clarify the extent to which the quantum circuit model of
computation is a complete framework for the description of quantum
information processing. I believe this would be a long and difficult
task, but well worth doing. If such a result could be established,
then one could develop a theory of realizable measurements, along
lines similar to recursive function theory in computer science
\cite{Davis83a,Papadimitriou94a}.

That concludes our discussion of realizable quantum measurements. In
many ways it is a digression from the main stream of the Dissertation,
but it is a digression that reinforces many of the points made in the
main stream, and alerts us to some open problems in fundamental
physics that I would very much like to see solved. Let us now turn to
the more immediately practical topic of experimental quantum
information processing.

%\subsection{Universal quantum gate arrays}

\section{Experimental quantum information processing}

The theory of quantum information processing has progressed very
quickly over the past twenty years. By contrast, experimental progress
has been much slower, despite much ingenuity and effort on the part of
experimentalists.

This section reviews the requirements that must be met in order to do
interesting quantum information processing tasks, and describes in
some detail one of the specific technologies proposed to perform
quantum information processing. The section begins with a discussion
of some of the general principles to be met by quantum information
processors\footnote{\index{quantum information processors}We follow
Steane \cite{Steane97a} in using the general term ``quantum
information processor'' to describe any system that can be used to do
quantum information processing, from the most elementary tasks, up to
full-fledged quantum computation.}. We then discuss in some detail the
approach to quantum computing based upon liquid state nuclear magnetic
resonance\index{NMR}\index{nuclear magnetic resonance}.
% two systems which have yielded the most significant
%experimental results to date: the ion trap quantum information
%processor, and liquid state nuclear magnetic resonance.
The section concludes with an account of the use of nuclear magnetic
resonance to accomplish quantum teleportation.

The specific requirements which must be met by a system which is to do
quantum information processing depend upon the task which the system
is to perform. For example, tasks such as superdense coding require a
high level of control over single qubits, but only a small number of
qubits in order to be accomplished. Optical methods have been used to
successfully implement an impressive variety of quantum information
processing tasks of this high precision-small size type, including
quantum cryptography (\cite{Hughes95a}, and references therein), a
variant of superdense coding
\cite{Mattle96a}, and quantum teleportation
\cite{Boschi98a,Bouwmeester97a}. Given this impressive progress, it
seems likely that optical methods will remain important for quantum
information processing, at least in the short term.

These same optical methods are of little use in their present form for
more general quantum information processing tasks. Purely optical
methods do not appear to scale very well, and with present techniques
it is very difficult to implement the non-linear optical interactions
which are necessary for quantum logic. Physically, in order to achieve
the interactions between photons necessary for quantum logic, there
must be some other medium present to mediate the interaction, and
presently known mediums for this interaction are not especially
efficient. Moreover, it is difficult to store photons in a controlled
fashion for long periods of time. These problems make it seem unlikely
that photons will be the primary basis for large scale quantum
information processors. In the near term, optical methods are likely
to remain an important means for doing small scale investigations of
quantum information processing, and it certainly seems reasonably
likely that optical methods will have {\em some} role to play in other
technologies for quantum information processing.

\index{quantum computation!requirements for} \index{quantum
circuit!requirements for}

What general requirements are desirable in a system which is to be
used for large scale quantum information processing? Obviously, the
requirements to be met depend on the exact model of quantum
information processing which is to be implemented; this is one of the
reasons it is interesting to formulate different but equivalent models
of quantum information processing. If the goal is to implement the
quantum circuit model described in the previous section, then the
following requirements must be met:

\begin{itemize}
\item The system must have a suitable $n$ qubit state space.
\item Ability to prepare the system in computational basis states.
\item Ability to perform an appropriate universal set of
gates on the system, for example, the controlled not, phase shift, and
Hadamard gates.
\item Ability to perform measurements in the computational basis.
\item Precise external control over the system, allowing an arbitrary
sequence of gates and computational basis state measurements to be
performed on the system.
\end{itemize}

Finally, there is one additional requirement not directly related to
the abstract theoretical model for the quantum circuit model, but of
overwhelming practical importance: the ability to cope with noise. The
performance of each of the above tasks will inevitably be imperfect,
and quantum computers must be resilient in the face of such noise. In
particular, the timescale $t_c$ over which the coherent dynamics of
the system takes place (roughly, the longest time required to perform
one of the fundamental logical operations), must be very short
compared to the timescale $t_n$ over which the system's state is
effectively messed up due to the effects of noise. Roughly speaking,
the number of operations which can be done before a quantum computer
becomes useless as a quantum computer is $t_n/t_c$. Thus, the goal is
to find systems which maximize $t_n$ while minimizing the time
required for dynamics. In Chapter \ref{chap:qec} we will investigate
quantum error correcting codes which, it has recently been shown, can
be used to effectively increase $t_n / t_c $ for a quantum system, for
little cost in the time required to do the coherent dynamics.

\subsection{Proposals for quantum information processing}
\index{ion trap} \index{NMR} \index{nuclear magnetic resonance}
\index{cavity QED}

Many proposals have been made for systems capable of functioning as
quantum information processors. Two of these proposals stand out as
they have led to the successful implementation of simple quantum
logical operations, and promise substantially more in the relatively
near future. These proposals are based on the linear ion trap,
originally proposed by Cirac and Zoller \cite{Cirac95a} and further
developed by several groups of researchers
\cite{Steane97a,Monroe97a,Poyatos97a}, and the liquid state nuclear magnetic
resonance (NMR) approach to quantum information processing.
% In this subsection both proposals will be briefly described.
In this subsection we focus on a description of the NMR approach, in
preparation for the next subsection, which describes the results of a
collaboration with Knill and Laflamme to do quantum teleportation in
NMR.  It is also worth noting that a third technology, cavity QED
\cite{Turchette95a}, has been used to implement simple quantum
logic. This technology will not be reviewed here as the task of using
this implementation of quantum logic to do more complex operations is
even more formidable than for the ion trap, or NMR.

Methods for doing quantum information processing using liquid state
NMR were proposed independently at about the same time by Cory, Fahmy
and Havel \cite{Cory97a}, and by Gershenfeld and Chuang
\cite{Gershenfeld97a}.
The scheme has since been applied to do numerous interesting quantum
information processing tasks
\cite{Chuang98a,Chuang98c,Cory97b,Cory98a,Jones98a,Jones98b,Laflamme97a,Nielsen98b}.

The NMR method is unusual in that it makes use of a model of quantum
information processing that is significantly different to the quantum
circuit model of quantum computation. In particular, the computation
is done in a bulk system, at room temperature. Therefore, the initial
state of the system is not a pure computational basis state, but
rather is a thermal mixture of states of the system. Furthermore,
because of the bulk nature of the system it is not possible to do
projective measurements on a single system, but rather, only ensemble
averaged measurements can be made.  Fortunately, both these problems
can be circumvented in our effort to do quantum information
processing.

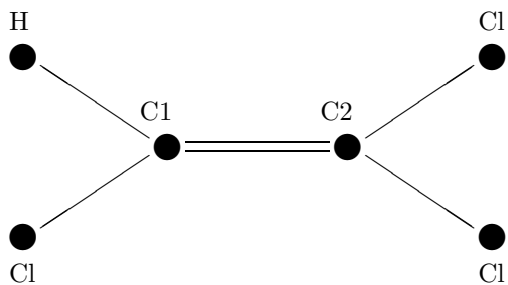
\begin{figure}[h]
\unitlength 1.2cm
\begin{center}
\begin{picture}(6,4)(0,0)
\put(2.7,2.05){\line(1,0){1.6}}
\put(2.7,1.95){\line(1,0){1.6}}
\put(2.5,2){\circle*{0.3}}
\put(2.2,2.3){C1}
\put(4.5,2){\circle*{0.3}}
\put(4.2,2.3){C2}
\put(2.3,2.1){\line(-3,2){1.2}}
\put(0.9,3.0){\circle*{0.3}}
\put(0.75,3.3){H}
\put(4.7,2.1){\line(3,2){1.2}}
\put(6.1,3.0){\circle*{0.3}}
\put(5.95,3.3){Cl}
\put(4.7,1.9){\line(3,-2){1.2}}
\put(6.1,1.0){\circle*{0.3}}
\put(5.95,0.5){Cl}
\put(2.3,1.9){\line(-3,-2){1.2}}
\put(0.9,1.0){\circle*{0.3}}
\put(0.75,0.5){Cl}
\end{picture}
\caption{Schematic representation of the labeled TCE molecule. The two
carbon atoms, C1 and C2, are $^{13}$C isotopes, which a net nuclear
spin of $1/2$.
\label{fig:tce}}
\end{center}
\end{figure}

\index{TCE}
\index{trichloroethylene}
The liquid state NMR approach to quantum information processing makes
use of a large number of molecules dissolved in a solvent such as
chloroform. For example, in experiments done in collaboration with
Knill and Laflamme \cite{Nielsen98b} at the Los Alamos National
Laboratory, the molecule trichloroethylene, or TCE, was used. The
structure of the molecule is shown in figure \ref{fig:tce}.

This molecule consists of two Carbon atoms, double chemically bonded,
a Hydrogen atom, and three Chlorine atoms. The molecules are prepared
in such a way that the Carbon atoms are actually the $^{13}$C isotope,
in order to give us a usable net spin $1/2$ contribution from the
nucleus. In our setup, the Chlorines are not usable because of the
lack of a suitable detector.

The sample is placed in a large, homogeneous, static magnetic field,
oriented in what we shall call the $z$ direction. The field is as
large as can be made with current technology for reasonable cost,
typically in the range of 10 Tesla or so; in our experiments an 11.5
Tesla magnetic field was used. In the liquid state, the molecules in
the sample tumble rapidly around, leading to a situation where
interactions between the molecules can be ignored.

% Hamiltonian

In this limit, the Hamiltonian describing the behaviour of the system
is ($\hbar = 1$)
\be
H = \sum_i \omega_i Z_i + \sum_{ij} J_{ij} Z_i Z_j, \ee where the
second sum is over all pairs of spins. The values of the
frequencies $\omega_i$ and $J_{ij}$ depend on the particular spins
being used; typically, $\omega_i \approx 10^8$-$10^9$ Hz, while for
neighbouring spins $J_{ij} \approx 10^2$-$10^3$ Hz.

In the TCE molecule, the frequencies are as follows: \be \omega_H
\approx 500.133491 \mbox{MHz}; & \,\,\,\, \omega_{C1} \approx
125.772580 \mbox{MHz}; &\,\,\,\, \omega_{C2} \approx \omega_{C1}- 911
\mbox{Hz} \nonumber \\ & & \\ J_{H \, C1} \approx 201 \mbox{Hz}; &
\,\,\,\, J_{C1 \, C2} \approx 103 \mbox{Hz}. & \ee The coupling
frequencies between H and C2, as well as the Chlorines to H, C1 or C2,
are much lower, on the order of ten Hertz for the former, and less
than a Hertz for the latter. These couplings can be effectively
removed by a technique known as \emph{refocusing}, described below,
and will be ignored in what follows. We will also ignore the
Chlorines, as they were not visible in our experiment. Note that the
frequencies of C1 and C2 are not identical; they have slightly
different frequencies, due to the different chemical environments of
the two atoms. This effect is known as the {\em chemical shift}.

% Application of rf pulses

In addition to the uniform magnetic field, it is possible to apply
radio frequency (rf) pulses on resonance to each of the spins in a
direction transverse to the direction of the uniform magnetic field,
that is, in the $x-y$ plane.  In a frame rotating about the $z$ axis
with the spins at respective frequencies $\omega_i$, the Hamiltonian
for this system can therefore be approximated as
\be
H = \sum_{ij} J_{ij} Z_i Z_j + \sum_i P_{i\theta}(t) (\cos \theta
\, X_i+\sin \theta \, Y_i), \ee
where $\theta$ is some phase that may be externally controlled, and
the strength $P_{i\theta}(t)$ of the rf pulse applied to spin $i$ may
also be controlled externally.

Using these external rf fields it is possible to perform single qubit
rotations on individual nuclei in the molecule, by applying a field
tuned to the appropriate resonance frequency. The necessary
interactions happen fast enough that the contribution from the $ZZ$
coupling between spins may be neglected.  For our present purposes, it
is sufficient to consider $\pi/2$ and $\pi$ rotations about the $x, y,
-x$ and $-y$ axes. For example, a $\pi/2$ rotation about the $x$ axis
has the effect
\be
\exp(-i\pi X/4) = \frac{I-iX}{\sqrt 2}. \ee
A $-\pi/2$ rotation about the $y$ axis has the effect
\be
\exp(+i\pi Y/4) = \frac{I+iY}{\sqrt 2}. \ee
A $\pi$ rotation about the $x$ axis has the effect $\exp(-i\pi X/2) =
-iX$. Similar observations may be made about the other possible
rotations.

% j coupling

In the absence of externally applied rf fields, in the rotating frame
the spins evolve according to the Hamiltonian
\be
H = \sum_{ij} J_{ij} Z_i Z_j, \ee where the sum is over all pairs
of interacting spins, $(i,j)$. In many situations, it is desirable to
be able to ``turn off'' one or more of these interactions. A clever
technique known as {\em refocusing} allows this to be done. For
simplicity, we explain how refocusing works in the specific case of
the TCE molecule. Suppose we wished to obtain a $\pi/2$ $ZZ$ coupling
between C1 and C2, with no interaction occurring between C1 and the
Hydrogen.

In order to achieve, this, let $t$ be any length of time. For example,
we might choose $t$ equal to the length of time necessary for a $\pi/2$
coupling between C1 and C2.
% $ t \equiv \pi/(4 J_{\mbox{C1 C2}})$
%be the time necessary for a $\pi/2$ coupling of C1 and C2 to
%occur.
Suppose we cause the following sequence of operations to occur:
\begin{enumerate}
\item Let the system freely evolve for a time $t/2$.
\item Apply a $\pi$ pulse to H, in the $X$ phase.
\item Let the system freely evolve for a time $t/2$.
\end{enumerate}
The total evolution is thus \be & & \exp(-i (t/2) J_{C1\,C2}
Z_{C1} Z_{C2}) \exp(-i (t/2) J_{H\,C1}Z_{H} Z_{C1}) (iX_H) \times
\nonumber \\ & & \exp(-i (t/2) J_{H\,C1} Z_{H} Z_{C1}) \exp(-i
(t/2) J_{C1\,C2} Z_{C1} Z_{C2}). \ee By the anticommutation
relation $XZ+ZX = 0$ we see that \be \exp(-i (t/2) J_{H\,C1}
Z_{H} Z_{C1}) (iX_H) \exp(-i (t/2) J_{H\,C1} Z_{H}Z_{C1}) & = &
\nonumber \\ (iX_H) \exp(+i (t/2) J_{H\,C1} Z_{H} Z_{C1})
\exp(-i(t/2) J_{H\,C1} Z_{H} Z_{C1}) & = & iX_H. \ee Thus, the
total effect of this sequence of operations is \be iX_H \exp(-i t
J_{C1 C2} Z_{C1}Z_{C2}). \ee That is, it is effectively as if a
$ZZ$ coupling of time $t$ between C1 and C2 had occurred, together
with a $\pi$ rotation on the Hydrogen. The interaction between H and
C1 has vanished; we say that it has been refocused.

We will use single qubit rotations and spin-spin couplings to perform
unitary dynamics on our nuclear spins.  Whether this forms a universal
set for quantum computation depends upon the details of the molecule
being considered; see \cite{Gershenfeld97a} for a discussion of this
point. For our much less grandiose purpose of doing quantum
teleportation the interactions available are certainly sufficient to
implement the quantum circuit for teleportation.  The chief difficulty
is perhaps that pulses applied to the two carbon nuclei are applied
non-selectively.  However, standard tricks based upon the chemical
shift can be used to apply selective pulses to C2 \cite{Ernst90a}.

% readout

Liquid state NMR involves bulk systems; typically, on the order of
$10^{15}$ sample molecules occur in the sample being examined. The
signal which is read out from the sample is an {\em ensemble average}
over all those molecules, not a projective measurement which yields a
single result, as in the quantum circuit model. In an NMR machine,
magnetic pick-up coils are used to determine the magnetization in the
$x$-$y$ plane. The signal read-out from the coils is then Fourier
transformed to give a spectrum for the system. The number of
observables whose ensemble average can be directly observed in this
way is thus rather limited. However, by making use of {\em reading
pulses} immediately before the final measurement, it is possible to
greatly extend the range of observables which can be determined. For
example, a $\pi/2$ rotation about the $y$ axis takes a $Z$ operator to
an $X$ operator. Thus, although the ensemble average $\la Z \ra$ for a
single nuclei cannot be directly observed, by applying a $\pi/2$
reading pulse about the $y$ axis immediately before observation, the
value of $\la Z \ra$ before the reading pulse can be inferred from the
observed value of $\la X \ra$ after the reading pulse.

% state preparation and gradients

At room temperature, the initial state of the system is highly
mixed. At temperature $T$, the spins start out in the state
$\exp(-H/kT)$, where $k$ is Boltzmann's constant, $H$ is the
Hamiltonian, and a normalization factor, the partition function, has
been omitted. To a first approximation, the coupling between nuclear
spins may be omitted, and at high temperature the state of the system
is proportional to
\be
\prod_i \exp(-\omega_i Z_i/kT) \approx \prod_i (I-\omega_i
Z_i/kT), \ee which is a mixture of computational basis states.  This
state does not appear to be at all like the pure computational basis
state which is used in the quantum circuit model of quantum
computation. There is a clever idea which allows us to work around
this problem, suggested independently by Cory, Fahmy and Havel
\cite{Cory97a}, and Gershenfeld and Chuang
\cite{Gershenfeld97a}. The idea is to extract a part of the state of
the system which ``looks like'' a pure state. Perhaps the simplest
scheme to illustrate the basic idea is the following method, known as
{\em temporal labeling} \cite{Knill98b}.

Suppose we have a molecule with $n$ nuclei. The idea is to define a
set of unitary operators which permute all the computational basis
states, $|0\ra,\ldots,|2^n-2\ra$ amongst themselves, while leaving the
state $|2^n-1\ra$ alone. We can then perform a series of $2^n-1$
experiments as follows. In each experiment, the corresponding unitary
operator is applied before the experiment begins. At the end, the
experimental results from all $2^n-1$ experiments are averaged. The
net contribution due to the states $|0\ra,\ldots,|2^n-2\ra$ averages
out, leaving a net contribution due only to the state
$|2^n-1\ra$. Thus we have performed a computation with an {\em
effectively} pure state.

To see how this works in more detail, define unitary operators $U_k$,
$0 \leq k \leq 2^n-2$, by $U_k|x\ra \equiv |x+k\ra$ for $1 \leq x \leq
2^n-2$ and $U_k |2^n-1\ra = |2^n-1\ra$, where the addition is done
modulo $2^n-1$. It is straightforward to efficiently implement such
operations using standard quantum gates \cite{Barenco95a,Beckman96a},
so this can be done in NMR. Note then that if $\rho = \sum_x p_x
|x\ra\la x|$ is diagonal in the computational basis then \be \sum_k
U_k \rho U_k^{\dagger} & = & (2^n -1) p_N |N\ra \la N|+ (1-p_N) \sum_{x
\neq N} |x\ra \la x| \\ & = & (2^n p_N-1)|N\ra\la N|+(1-p_N)I, \ee
where $N \equiv 2^n-1$.  Suppose in each of these experiments we
perform the unitary $U_k$, followed by some unitary operation $U$, and
then observe some component of the spin, say $\la X_i \ra$. Summing
over the results observed in each of the $2^n-1$ experiments, we
obtain \beqn \sum_{k=0}^{N-1} \tr(X_i U U_k \rho U_k^{\dagger}
U^{\dagger}) = (2^n p_N-1) \tr(X_i U |N\ra \la N| U^{\dagger}),\ee as
$\tr(X_i I) = 0$. That is, the summed averages behave as if the pure
state $|N\ra\la N|$ had been prepared, the unitary operation $U$
applied to that pure state, and the average of $X_i$ observed. Similar
remarks apply to other observations which may be made in NMR. By
appropriate labeling we can ensure that $p_N$ is the smallest (or
largest) of the the eigenvalues of the initial density operator, in
which case $2^n p_N \neq 1$, unless we are at infinite temperature,
and the ensemble is uniform. Even the small deviation away from
uniformity available at room temperature can be exploited to make the
factor $2^n p_N-1$ appearing in front of the observed average large
enough that this method can be used to obtain effectively pure state
behaviour out of a mixed state system.

This method is known as temporal averaging because it requires that
the experiment be repeated many different times, and the results
summed.  Temporal averaging is only one possible means for performing
state preparation in NMR quantum information processing. It is an
especially easy method to explain, but in the laboratory other methods
may be considerably better. In our experiments, a technique based upon
the use of {\em gradient pulses} was used. The precise details of what
was done are beyond our present scope, but the basic idea may be
explained quite easily.

Essentially what is done is to vary the strength of the magnetic field
applied in the $z$ direction across the sample.  This causes nuclei at
different locations in the sample to rotate around the $z$ axis at
different frequencies.  When applied for the appropriate length of
time, the ensemble averaged values for the $X$ and $Y$ components of
magnetization average to zero.  That is, a gradient pulse applied to a
single spin has the effect of setting the $x$ and $y$ components of
the Bloch vector for the ensemble to zero, while leaving the $z$
component of the Bloch vector untouched.  Cory {\em et al}
\cite{Cory97a} have described how a combination of gradient pulses, rf
pulses, and delays may be combined to prepare effectively pure states,
along similar lines to the temporal labeling method described above.
We will not give further details of this method here.

NMR-based approaches to quantum information processing have many
attractive features. NMR is a well-developed technology, and a
considerable amount of high quality, easy-to-use equipment has been
developed for use off-the-shelf. The noise timescale is typically on
the order of a second, while the time to perform a two qubit gate is
on the order of one to ten milliseconds, giving a best-case estimate
of about one thousand couplings possible, although there is no doubt
that achieving this in a useful computation will be extraordinarily
difficult.  Present experimental work in NMR quantum information
processing usually involves on the order of ten couplings.

With regard to the power of NMR quantum information processing from
the point of view of computational complexity, and in comparison with
the quantum circuit model, I will not essay an opinion here.  A
considerable amount of interesting discussion has taken place on or
closely related to this topic and I refer the reader to, for example,
\cite{Chuang98b,Gershenfeld97a,Knill98b,Knill98c,Schulman98a} for further
discussion. What does seem certain is that NMR provides a powerful
means for conducting interesting investigations into small-scale
quantum information processing. A few qubits may not be much, but it
represents the current best we can do with our quantum information
processors.

\subsection{Experimental demonstration of quantum teleportation using NMR}
\label{subsec:teleport_NMR}
\index{quantum teleportation!experimental implementation}

The ideas of NMR quantum information processing have recently been
exploited to provide an experimental demonstration of quantum
teleportation, in collaboration with Knill and Laflamme
\cite{Nielsen98b}. The essential idea of the scheme is to implement
the quantum circuit for teleportation discussed in section
\ref{sec:teleportation}, using the NMR-based techniques for quantum
information processing discussed in the previous subsection.

\begin{figure}[ht]
\begin{centering}
\unitlength 0.95cm
\begin{picture}(15,3)(-2,0)
\put(-1.3,2.6){data}
\put(-1.29,2.25){(C2)}
\put(-1.45,1.6){ancilla}
\put(-1.26,1.25){(C1)}
\put(-1.4,0.6){target}
\put(-1.22,0.10){(H)}
\put(-0.2,2.45){$|\Psi\rangle$}
\put(0,1.5){$0$}
\put(0,0.4){$0$}
% Labels on Alice and Bob.
\put(-2.16,1.93){{\begin{sideways}\textbf{Alice}\end{sideways}}}
\put(-2.16,0.28){{\begin{sideways}\textbf{Bob}\end{sideways}}}
% Frames around Alice and Bob
\put(-2.0,1.10){\framebox(2.5,1.9){}}
\put(-2.0,-0.15){\framebox(2.5,1.1){}}
\put(0.5,2.5){\line(1,0){3}}
\put(0.5,1.5){\line(1,0){0.5}}
\put(0.5,0.5){\line(1,0){0.5}}
\put(1,0.2){\framebox(2,1.6){entangle}}
\put(3,1.5){\line(1,0){0.5}}
\put(3,0.5){\line(1,0){4.5}}
\put(3.5,1.2){\framebox(3.5,1.6){}}
\put(3.5,2.1){\makebox(3.5,0.5){measure in the}}
\put(3.5,1.4){\makebox(3.5,0.5){Bell basis}}
\put(7,2.5){\line(1,0){2.4}}
\put(7,1.5){\line(1,0){2.4}}
\put(7.5,0.0){\framebox(4,1){}}
\put(7.5,0.5){\makebox(4,0.5){conditional unitary}}
\put(7.5,0.0){\makebox(4,0.5){$U_{00},U_{01},U_{10},U_{11}$}}
\put(9.5,1){\line(0,1){0.4}}
\put(9.5,1.5){\circle{0.2}}
\put(9.5,1.6){\line(0,1){0.8}}
\put(9.5,2.5){\circle{0.2}}
\put(9.6,2.5){\line(1,0){2.4}}
\put(9.6,1.5){\line(1,0){2.4}}
\put(11.5,0.5){\line(1,0){0.5}}
\put(12.3,2.6){classical}
\put(12.5,2.3){data}
\put(12.3,1.6){classical}
\put(12.5,1.3){data}
\put(12.65,0.43){$| \Psi \rangle$}
\end{picture}
\end{centering}
\caption{Schematic of quantum teleportation.  See text for a full description.
\label{fig:schematic}}
\end{figure}
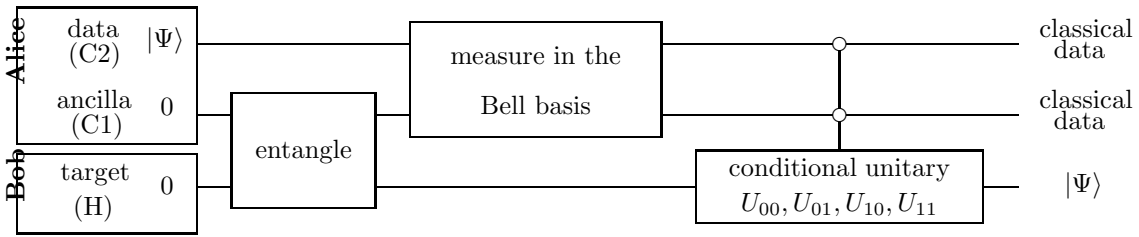

Our implementation of teleportation is performed using liquid state
nuclear magnetic resonance (NMR), applied to an ensemble of molecules
of labeled trichloroethylene TCE, as discussed in the previous
section. To perform teleportation we make use of the Hydrogen nucleus
(H), and the two Carbon 13 nuclei (C1 and C2), teleporting the state
of the second Carbon nucleus to the Hydrogen.  Figure
\ref{fig:schematic} shows a schematic illustration of the
teleportation process we used, based upon the circuit described in
\cite{Brassard98a}, illustrated in figure \ref{fig: Brassard}.  The
circuit has three inputs, which we will refer to as the {\em data}
(C2), {\em ancilla} (C1), and {\em target} (H) qubits. The goal of the
circuit is to teleport the state of the data qubit so that it ends up
on the target qubit.

% starts with an overview of the Brassard et al circuit.

State preparation is done in our experiment using the gradient-pulse
techniques described by Cory {\em et al} \cite{Cory97a}, and phase
cycling \cite{Ernst90a,Grant96a}.  The unitary operations performed
during teleportation may be implemented in a straightforward manner in
NMR, using non-selective rf pulses tuned to the Larmor frequencies of
the nuclear spins, and delays allowing entanglement to form through
the interaction of neighboring nuclei, as described in the previous
section.  Commented pulse sequences for our experiment may be obtained
on the world wide web \cite{Nielsen98c}.
%  Other demonstrations of
%quantum information processing with three qubits using NMR are
%described in \cite{Cory97b,Laflamme97a, Cory98a}, and with two qubits
%in \cite{Chuang98a,Jones98a,Chuang98c,Jones98b}.

% Explain how to implement this circuit in NMR.

An innovation in our experiment was the method used to implement the
Bell basis measurement.  In NMR, the measurement step allows us to
measure the expectation values of $\sigma_x$ and $\sigma_y$ for each
spin, averaged over the ensemble of molecules, rather than performing
a projective measurement in some basis.  For this reason, we must
modify the projective measurement step in the standard description of
teleportation, while still preserving the remarkable teleportation
effect.

We use a procedure inspired by Brassard {\em et al}
\cite{Brassard98a}, who suggested a two-part procedure for performing
the Bell basis measurement. Part one of the procedure is to rotate
from the Bell basis into the computational basis, $|00\ra,|01\ra,
|10\ra,|11\ra$.  We implement this step in NMR by using the natural
spin-spin coupling between the Carbon nuclei, and rf
% we deleted single qubit ... as it is a bit misleading kl 7/1/1998
pulses.  Part two of the procedure is to perform a projective
measurement in the computational basis.  As Brassard {\em et al} point
out, the effect of this two part procedure is equivalent to performing
the Bell basis measurement, and leaving the data and ancilla qubits in
one of the four states, $|00\ra, |01\ra, |10\ra, |11\ra$,
corresponding to the different measurement results.

We cannot directly implement the second step in NMR.  Instead, we
exploit the natural phase decoherence occurring on the Carbon nuclei
to achieve the same effect.  Recall that phase decoherence completely
randomizes the phase information in these nuclei and thus will destroy
coherence between the elements of the above basis.  Its effect on the
state of the Carbon nuclei is to diagonalize the state in the
computational basis, \be \rho & \longrightarrow & |00\ra \la 00| \rho
|00\ra \la 00| + |01\ra \la 01| \rho |01\ra \la 01| +|10\ra \la 10|
\rho |10\ra \la 10| \nonumber \\ & & + |11\ra \la 11| \rho |11\ra \la
11|. \ee As emphasized by Zurek \cite{Zurek91a}, the decoherence
process is indistinguishable from a measurement in the computational
basis for the Carbons accomplished by the environment.  We do not
observe the result of this measurement explicitly, however the state
of the nuclei selected by the decoherence process contains the
measurement result, and therefore we can do the final transformation
conditional on the particular state the environment has selected.  As
in the scheme of Brassard {\em et al}, the final state of the Carbon
nuclei is one of the four states, $|00\ra,|01\ra,|10\ra,|11\ra$,
corresponding to the four possible results of the measurement.

In our experiment, we exploit the natural decoherence properties of
the TCE molecule.  The phase decoherence times ($T_2$) for the C1 and
C2 are approximately $0.4s$ and $0.3s$.  All other $T_2$ and $T_1$
times for all three nuclei are much longer, with a $T_2$ time for the
Hydrogen of approximately $3s$, and relaxation times ($T_1$) of
approximately $20-30s$ for the Carbons, and $5s$ for the Hydrogen.

This implies that for delays on the order of $1s$, we can
approximate the total evolution by exact phase decoherence on the
Carbon nuclei. The total scheme therefore implements a measurement in
the Bell basis, with the result of the measurement stored as classical
data on the Carbon nuclei following the measurement.  We can thus
teleport the information from the Carbon to the Hydrogen and verify
that the information in the final state decays at the Hydrogen rate
and not the Carbon one.

Examining figure \ref{fig:schematic} we see how remarkable
teleportation is from this point of view.  During the stage labeled
``Measure in the Bell basis'' in figure \ref{fig:schematic}, we allow
the C1 and C2 nuclei to decohere and thus be measured by the
environment, destroying all phase information on the data and ancilla
qubits.  Experimentally, the use of multiple refocusing pulses ensures
that the data qubit has effectively not interacted with the target
qubit. Classical intuition therefore tells us that the phase
information about the input state, $|\Psi\rangle$, has been lost
forever. Nevertheless, quantum mechanics predicts that we are still
able to recover the complete system after this decoherence step, by
quantum teleportation.

We implemented this scheme in TCE using a Bruker DRX-500 NMR
spectrometer.  Experimentally, we determined the Larmor and coupling
frequencies for the Hydrogen, C1 and C2 to be: \be \omega_H \approx
500.133491 \mbox{MHz}; & \,\,\,\, \omega_{C1} \approx 125.772580
\mbox{MHz}; &\,\,\,\, \omega_{C2} \approx \omega_{C1}- 911 \mbox{Hz}
\nonumber \\ & & \\ J_{H \, C1} \approx 201 \mbox{Hz}; & \,\,\,\,
J_{C1 \, C2} \approx 103 \mbox{Hz}. & \ee The coupling frequencies
between H and C2, as well as the Chlorines to H, C1 and C2, are much
lower, on the order of ten Hertz for the former, and less than a Hertz
for the latter.  Experimentally, these couplings are suppressed by
multiple refocusings, and will be ignored in the sequel.  Note that
the frequencies of C1 and C2 are not identical; they have slightly
different frequencies, due to the different chemical environments of
the two atoms.

We performed two separate sets of experiments. In one set, the full
teleportation process was executed, making use of a variety of
decoherence delays in place of the measurement. The readout was
performed on the Hydrogen nucleus, and a figure of merit -- the
dynamic fidelity -- was calculated for the teleportation process.  The
dynamic fidelity is a quantity in the range $0$ to $1$ which measures
the combined strength of {\em all} noise processes occurring during
the process, which we will study in detail in Chapter
\ref{chap:distance}\footnote{In the language of that Chapter, we
determined the dynamic fidelity for teleportation of the state I/2.}.
In particular, an dynamic fidelity of $1$ indicates perfect
teleportation, while an dynamic fidelity of $0.25$ indicates complete
randomization of the state.  Perfect {\em classical transmission}
corresponds to an dynamic fidelity of $0.5$, so dynamic fidelities of
greater than $0.5$ indicates that teleportation of some quantum
information is taking place.

The second set of experiments was a control set. In those experiments,
only the state preparation and initial entanglement of H and C1 were
performed, followed by a delay for decoherence on C1 and C2. The
readout was performed in this instance on C2, and once again, a figure
of merit, the dynamic fidelity, was calculated for the entire
process.

% results

\begin{figure}[ht]
\begin{center}
\scalebox{0.8}{\includegraphics{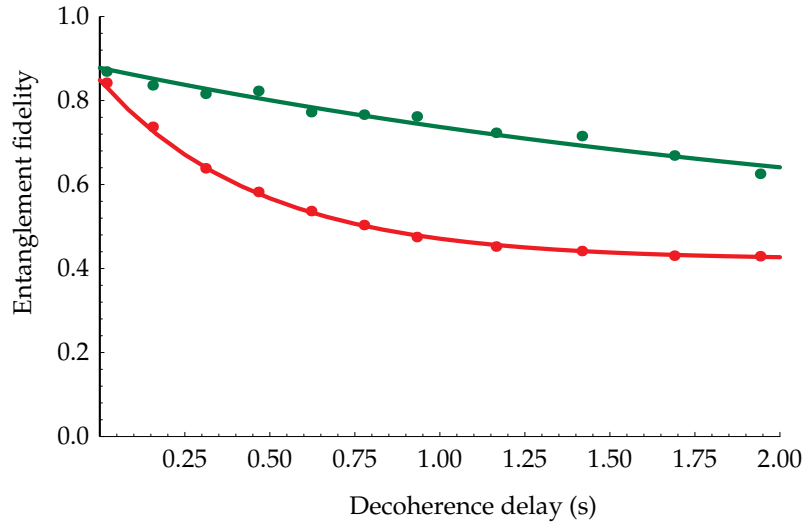}}
\end{center}
%\epsfxsize 4.5in
%\begin{center}
%\epsfbox{2/nature.eps}
%\end{center}
\caption{Dynamic fidelity is plotted as a function of decoherence
time.  The bottom curve is a control run where the information remains
in C2.  The curve shows a decay time of approximately 0.5s.  The top
curve represents the fidelity of the quantum teleportation process.
The decay time is approximately 2.6s.  The information is preserved
for a longer time, corresponding approximately to the combined effects
of decoherence and relaxation for the Hydrogen, confirming the
prediction of teleportation.
\label{fig:tele_fidelity}}
\end{figure}

The results of our experiment are shown in figure
\ref{fig:tele_fidelity}, where the dynamic fidelity is plotted
against the length of the delay which was used for the
decoherence. Errors in our experiment arise from the strong coupling
effect, imperfect calibration of rf pulses, and rf field
inhomogeneities. The estimated uncertainty in our values for
dynamic fidelity are less than $\pm 0.05$.  These uncertainties
are due primarily to rf field inhomogeneity and imperfect calibration
of rf pulses.

In order to determine the dynamic fidelities for the
teleportation and control experiments, we performed {\em quantum
process tomography}.  This procedure, described in detail in section
\ref{sec:quantum_process_tomography}, exploits the linearity of
quantum mechanics to completely characterize a quantum mechanical
process.  In particular, we will show in section
\ref{sec:teleportation_qops} that the linearity of quantum mechanics
implies that the single qubit input and output for the teleportation
process are always related by a linear quantum operation.  By
preparing a complete set of four linearly independent initial states,
we were able to obtain a complete description of the quantum process.
In particular, we used a procedure known as {\em quantum state
tomography} \cite{Vogel89a,Smithey93a} to determine the output states,
and used the linearity of quantum mechanics to extend this to a
description of the complete process.  This description, in turn,
enabled us to calculate the dynamic fidelity for the process, as
described in sections \ref{sec:quantum_process_tomography} and
\ref{sec:dynamic_measures}.

Three elements ought to be noted in figure \ref{fig:tele_fidelity}.
First, for small decoherence delays, the dynamic fidelity for the
teleportation experiments significantly exceeds the value of $0.5$ for
perfect classical transmission of data, indicating that we have
successfully teleported quantum information from C2 to H, with
reasonable fidelity.  Second, it is notable that the dynamic
fidelity decays very quickly for the control experiments as the delay
is increased.  Theoretically, we expect this to be the case, due to a
$T2$ time for C2 of approximately $0.3s$.  Third, the decay of the
dynamic fidelity for the teleportation experiments occurs much
more slowly.  Theoretically, we expect this decay to be due mainly to
the effect of phase decoherence and relaxation on the {\em Hydrogen}.
Our experimental observations are consistent with this prediction, and
provide more support for the claim that the quantum data is being
teleported in this set of experiments.

In conclusion, we have exhibited evidence of successful quantum
teleportation in liquid state NMR. This experiment is not the first
experimental implementation of quantum teleportation, however it is
the first implementation in NMR.  Earlier experiments by Boschi {\em
et al} \cite{Boschi98a} and Bouwmeester {\em et al}
\cite{Bouwmeester97a} used optical methods to achieve quantum
teleportation.  The present NMR-based method illustrates some of the
advantages of using NMR to do elementary quantum information
processing. In particular, the NMR experiment was relatively
straightforward to set up and perform, using off-the-shelf methods,
and could be repeated easily in many laboratories around the world. By
contrast, the optical techniques required much more customized
equipment, and were generally much more difficult to achieve.

\vspace{1cm}
\begin{center}
\fbox{
\parbox{14cm}{
\begin{center}
{\bf Summary of Chapter \ref{chap:fundamentals}: Quantum information:
fundamentals}
\end{center}

\begin{itemize}

\item
{\bf Qubits:} The fundamental unit of quantum information. A two level
quantum system.

\item
{\bf Superdense Coding:} Preshared entanglement can be used to
transmit two classical bits with the transmission of only one qubit.

\item
{\bf Quantum teleportation:} Preshared entanglement can be used to
transmit a qubit with the transmission of two classical bits.

\item
{\bf Quantum circuits:} {\em ad hoc} model with the following features:
\begin{enumerate}
\item \textbf{Classical external control.}
\item {\bf A suitable state space:} $n$ qubits.
\item {\bf Ability to prepare states in the computation basis.}
\item {\bf Dynamics:} An {\em algorithm} for applying {\em quantum
gates} (controlled-not and single qubit unitary gates) and {\em
projective measurements} in the {\em computational basis} to the
system.
\end{enumerate}

%\item
%{\bf Experimental requirements for quantum information processing:}
%\begin{itemize}
%\item Controllable coherent dynamics.
%\end{itemize}

\item
{\bf Experimental implementation of quantum teleportation in NMR}

\end{itemize}

}}
\end{center}

\chapter{Quantum operations}
\label{chap:qops} \index{quantum operations}

%To add: discussion of unital operators, and qops between different
%spaces. Examples on a single qubit are commented out.

%\section{Introduction}

Quantum mechanics describes the {\em dynamics} which can occur in
physical systems. Elementary quantum mechanics texts usually do this
by separating the dynamics into two different types. The first type is
the evolution of a closed quantum mechanical system, which is assumed
to be described by Schr\"odinger's equation. Under such an evolution,
the change in the state $|\psi\rangle$ of a quantum system between two
fixed times is described by a unitary operator $U$ which depends on
those times,
\beqn
|\psi\ra \rightarrow U |\psi\ra. \eeqn
The equivalent map on density operators is given by
\beqn
\rho \rightarrow {\cal E}(\rho) \equiv U \rho U^{\dagger}. \eeqn

\index{measurement}

The second type of dynamics described in basic quantum mechanics texts
is associated with the {\em measurement} of a quantum mechanical
system. The system being measured is no longer a closed system, since
it is interacting with the measuring device. The usual way to describe
such a measurement is the following. Suppose a measurement is
performed which has outcomes labeled by $m$. Then to each outcome $m$
there is associated a projector $P_m$ onto the state space of the
system such that
\beqn
P_m P_n & = & \delta_{mn} P_m \\
\sum_m P_m & = & I. \eeqn
If the state of the system immediately before the measurement was
$\rho$, and the result of the measurement is $m$,
then the state of the system immediately after the measurement is
\beqn
\frac{\evop_m(\rho)}{\tr({\evop_m(\rho)})}, \eeqn
where
\beqn
\evop_m(\rho) \equiv P_m \rho P_m. \eeqn
Moreover, the probability of obtaining this measurement result is
given by
\beqn
p(m) = \tr(\evop_m(\rho)). \eeqn

The maps ${\cal E}$ and ${\cal E}_m$ are examples of {\em quantum
operations}. The theory of quantum operations can be used to describe
a wide class of state changes that may occur in quantum systems. You
may wonder how it is possible to go beyond the usual textbook
description of state changes in terms of unitary transformations and
projective measurements. The key observation is that many state
changes of interest occur in {\em open quantum systems}. The
interaction of the quantum system with an external world allows
dynamics that are neither unitary nor described by the usual model of
projective measurements.

%One example of great interest is that of a quantum system in
%the presence of an environment. No laboratory system is
%ever completely isolated from its environment, and in
%general the interaction of a system with its environment causes
%the dynamics to be non-unitary.
%
%A second example of great interest occurs in the theory of
%quantum measurements. Suppose we wish to perform
%a measurement on some quantum system. It turns out that
%there are many interesting measurements that can only
%be done by first introducing an extra quantum system,
%known as the {\em ancilla}, causing an interaction between
%the system and the ancilla, and then doing a joint measurement
%on the system and ancilla.

To make the idea of quantum operations more concrete, consider the
following example.  Suppose we have a single qubit quantum system, the
{\em principal} system, in a state $\rho$, which is brought into
contact with an environment. We will suppose this environment is also
a single qubit system, which is initially in the state $|0\ra$. For
instance, the principal system might be a nuclear spin in a molecule
being used to do NMR, and the environment a neighbouring spin. Left to
themselves these systems will interact according to some unitary
interaction $U$. For the sake of definiteness we will suppose that $U$
is the controlled not operation, with the principal system the control
qubit, and the environment the data qubit. $U$ can be written \beqn U
= P_0 \otimes I + P_1 \otimes X, \eeqn where the first system is the
principal system, the second system is the environment, and $P_0
\equiv |0\ra\la 0|, P_1 \equiv |1\ra\la 1|$ are projectors.  The state
of the joint system after the interaction is \beqn & & U (\rho \otimes
|0\ra\la 0|) U^{\dagger} = \nonumber \\ & & P_0 \rho P_0 \otimes
|0\ra\la 0| + P_1 \rho P_1 \otimes |1\ra \la 1| + P_0 \rho P_1 \otimes
|0\ra \la 1| + P_1 \rho P_0 \otimes |1\ra \la 0|. \eeqn Tracing out
the environment we obtain the final state of the principal system,
\beqn P_0 \rho P_0 + P_1 \rho P_1. \eeqn That is, the evolution of the
principal system can be described by the map $\evop$, \beqn \rho
\rightarrow \evop(\rho) \equiv P_0 \rho P_0 + P_1 \rho P_1. \eeqn

The map just described is an example of a quantum operation, in which
the quantum state undergoes one single, definite evolution. By
contrast, in the case of a measurement, several different outcomes may
occur, each outcome being associated with a particular classical
measurement value. For example, suppose a principal system consisting
of one qubit is being coupled, once again, to a one qubit
environment. The initial state of the total system is again assumed to
be $\rho \otimes |0\ra\la 0|$, and the coupling is described by a
unitary operator $U$ defined by \be U = \frac{X}{\sqrt 2} \otimes I +
\frac{Y}{\sqrt 2} \otimes X. \ee Following the unitary evolution, a
measurement is done on the environment qubit, in the computational
basis. Note that the state of the system after the unitary evolution
is \be \frac{X \rho X \otimes |0\ra\la 0| + Y\rho Y \otimes |1\ra\la
1| + X \rho Y \otimes |0\ra \la 1| + Y \rho X \otimes |1\ra\la
0|}{2}. \eeqn Conditioned on the result of the measurement, we see by
inspection that the state of the principal system after the
measurement is either $X\rho X$ or $Y \rho Y$, depending upon whether
the measurement result was 0 or 1, with probability $1/2$ for each of
the two possibilities.  Again, these are quantum operations, this time
associated with different measurement outcomes possible in the
process.

The primary purpose of this Chapter is to review the general theory of
quantum operations. In addition to elementary review material, the
Chapter shows how the quantum operations formalism can be used to gain
insight into quantum teleportation, and describes {\em quantum process
tomography}, a general method for the experimental determination of
the dynamics of a quantum system.  The elementary material appearing
here has its origins in earlier work by people such as Hellwig and
Kraus \cite{Hellwig69a,Hellwig70a}, Choi \cite{Choi75a} and Kraus
\cite{Kraus83a}.  The material relating quantum teleportation and the
quantum operations formalism is based upon a collaboration with Caves
\cite{Nielsen97c}, and the work on quantum process tomography is based
upon a collaboration with Chuang \cite{Chuang97a}.  In places the
Chapter contains rather detailed mathematics; upon a first read, these
sections may be read lightly, and returned to later for reference
purposes.

\section{Quantum operations: fundamentals}
\label{sec:qops_fundamentals}
\index{quantum operations}

Suppose we have a quantum system $Q$, initially in an {\em input
state}, $\rho$. We suppose some physical process occurs, which results
in an {\em output state}, $\rho'$.  That output state need not even be
a state of the same system; all we require is that the final state is
uniquely determined by some physical process, starting with the input
state. What requirements must the map ${\cal E}: \rho \rightarrow
\rho'$ satisfy?  We will enumerate a set of axioms which any such map
must satisfy, and then go on to show that any map satisfying these
requirements is physically reasonable.

The formalism we develop shall, ideally, include deterministic quantum
processes, such as unitary evolution and interaction with an
inaccessible environment, as well as measurements, in which a quantum
system undergoes a state change chosen at random, depending on what
measurement outcome occurred.

To cope with the case of measurements, it turns out that it is
extremely convenient to make the convention that the map $\evop : \rho
\rightarrow \rho'$ does not necessarily preserve the trace property of
density operators, that $\tr(\rho) = 1$. Rather, we make the
convention that $\evop$ is to be defined in such a way that
$\tr(\evop(\rho))$ is equal to the probability of the measurement
outcome described by $\evop$ occurring. More concretely, suppose that
we are doing a projective measurement in the computational basis of a
single qubit. Then two quantum operations are used to describe this
process, defined by $\evop_0(\rho) \equiv |0\ra\la 0|\rho|0\ra\la 0|$
and $\evop_1(\rho) \equiv |1\ra\la 1| \rho |1\ra\la 1|$. Notice that
the probabilities of the respective outcomes are correctly given by
$\tr(\evop_0(\rho))$ and $\tr(\evop_1(\rho))$. With this convention
the correctly normalized final quantum state is therefore \be
\frac{\evop(\rho)}{\tr(\evop(\rho))}. \ee

\index{complete quantum operations}
\index{incomplete quantum operations}
\index{quantum operations!complete}
\index{quantum operations!incomplete}
\index{physical quantum operations}
\index{quantum operations!physical}

Thus, generically, we impose a requirement of mathematical
convenience, that $\tr(\evop(\rho))$ be equal to the probability of
the process represented by $\evop$ occurring, when $\rho$ is the
initial state. In the case where the process is deterministic, that
is, no measurement is taking place, this reduces to the requirement
that $\tr(\evop(\rho)) = 1 = \tr(\rho)$. In this case, we say that the
quantum operation is a {\em complete} quantum operation, since on its
own it provides a complete description of the quantum process. On the
other hand, if there is a $\rho$ such that $\tr(\evop(\rho)) < 1$,
then the quantum operation is {\em incomplete}, since on its own it
does not provide a complete description of the processes that may
occur in the system. A {\em physical} quantum operation is one that
satisfies the requirement that probabilities never exceed $1$,
$\tr(\evop(\rho)) \leq 1$.

Next, we impose our first physical requirement on quantum
operations. Suppose the input $\rho$ to the quantum operation is
obtained by randomly selecting the state from an ensemble $\{ p_i,
\rho_i \}$ of quantum states, that is, $\rho = \sum_i p_i
\rho_i$. Then we would expect that the resulting state, ${\cal
E}(\rho)/\tr(\evop(\rho))$ corresponds to a random selection from the
ensemble $\{ p(i|\evop), {\cal E}(\rho_i)/\tr(\evop(\rho_i)) \}$,
where $p(i|\evop)$ is the probability that the state prepared was
$\rho_i$, given that the process represented by $\evop$
occurred. Thus, we demand that \beqn \label{eqtn:Dick_Tracy} {\cal
E}(\rho) = p(\evop) \sum_i p(i|\evop) \frac{{\cal
E}(\rho_i)}{\tr(\evop(\rho_i))}, \eeqn where $p(\evop) =
\tr(\evop(\rho))$ is the probability that $\evop$ occurs on input of
$\rho$. By Bayes' rule, \be p(i|\evop) = p(\evop|i)
\frac{p_i}{p(\evop)} = \frac{\tr(\evop(\rho_i))p_i}{p(\evop)} , \ee so
the equation (\ref{eqtn:Dick_Tracy}) reduces to \be \evop(\sum_i p_i
\rho_i) = \sum_i p_i \evop(\rho_i). \ee That is, we require that
quantum operations be {\em convex linear} on the set of density
operators. Indeed, any convex linear map on density operators can be
uniquely extended to a linear map on Hermitian operators, so we make
this additional requirement, again, as a mathematical convenience.

\index{complete positivity}
\index{complete positivity!example of a positive map not completely
positive}
\index{transpose operation}

Finally, we require that the quantum operation must preserve the
positivity of density operators. This requirement, known as {\em
complete positivity}, means that quantum operations take positive
operators to positive operators. This requirement applies both to
density operators on the system for which the dynamics is occurring,
the {\em principal system}, and also for super-systems of the
principal system.

To illustrate the importance of this point, consider the transpose
operation on a single qubit. By definition, this map transposes the
density operator in the computational basis: \beqn \left[
\begin{array}{cc} a & b \\ c & d \end{array} \right]
\stackrel{T}{\longrightarrow} \left[ \begin{array}{cc} a & c \\ b & d
\end{array} \right]. \eeqn
This map preserves positivity of a single qubit. However, suppose that
qubit is part of a two qubit system initially in the entangled state
\beqn
\frac{|00\ra+|11\ra}{\sqrt 2},\eeqn
and the transpose operation is applied to the first of these two
qubits, while the second qubit is subject to trivial dynamics. Then
the density operator of the system after the dynamics has been applied
is
\beqn
\frac 12 \left[ \begin{array}{cccc} 1 & 0 & 0 & 0 \\ 0 & 0 & 1 & 0 \\
0 & 1 & 0 & 0 \\ 0 & 0 & 0 & 1 \end{array} \right]. \eeqn
It is easy to verify that this operator has eigenvalues $1/2,1/2,1/2$
and $-1/2$, so this is not a valid density operator. Thus, the
transpose operator is an example of a map that preserves the
positivity of operators on the principal system, but does not
continue to preserve positivity when applied to systems which contain
the principal system as a subsystem\footnote{According to Weinberg
\cite{Weinberg95a} there are selection
rules in some system that prohibit, for example, superpositions of
states with different electric charge existing. It is amusing to
speculate that in systems in which such selection rules exist it might
be allowable for systems to undergo dynamics which are not completely
positive, as this would not necessarily lead to density operators
which were not positive, and thus unphysical.}.

\index{quantum operations!definition}

Summarizing, the requirements for a map to be a quantum operation are
as follows:
\begin{enumerate}
\item By definition, $\tr(\evop(\rho))$ is the probability that the process
represented by $\evop$ occurs, when $\rho$ is the initial state.
\item The map ${\cal E}$ is a linear map. The domain of ${\cal E}$ is
the (real vector space) of Hermitian operators on $H_Q$, the input
Hilbert space. The range of ${\cal E}$ is contained in the (real
vector space) of Hermitian operators on $H_Q'$, the output Hilbert space.
\item The map ${\cal E}$ is {\em completely positive}. That is,
suppose we introduce a second system, $R$, with (finite dimensional)
Hilbert space $H_R$. Let ${\cal I}$ denote the identity map on
system $R$. Then the map ${\cal I} \otimes {\cal E}$ takes positive
operators to positive operators.
\end{enumerate}

Surprisingly to me, at least, these requirements are sufficient to
characterize quantum operations. Later, we will show how any map
satisfying these requirements can be physically realized, in a finite
dimensional quantum system. One step along the way to this result is
an elegant representation theorem which relates these abstract
requirements for a quantum operation to an explicit formula:

\index{operator-sum representation}

\begin{theorem} \textbf{(Operator-sum representation)} \cite{Kraus83a}

The map ${\cal E}$ is a quantum operation if and only if
\beqn
{\cal E}(A) = \sum_i E_i A E_i^{\dagger}, \eeqn
for some set of operators $E_i$ which map the input Hilbert space to
the output Hilbert space.
\end{theorem}

The operators $E_i$ appearing in this expression are said to generate
an {\em operator-sum} representation for the quantum operation
$\evop$.

\begin{proof} \cite{Schumacher96a}

Suppose ${\cal E}(A) = \sum_i E_i A E_i^{\dagger}$. ${\cal E}$ is
obviously linear, so to check that ${\cal E}$ is a quantum operation
we need only prove that it is completely positive. Let $A$ be any
positive operator acting on the state space of an extended system,
$RQ$, and let $|\psi\ra$ be some state of that system. Defining
$|\phi_i\ra \equiv (I_R \otimes E_i^{\dagger})|\psi\ra$, we have
\beqn
\la \psi| (I_R \otimes E_i) A (I_R \otimes E_i^{\dagger}) |\psi\ra & =
	& \la \phi_i| A |\phi_i\ra \\
	& \geq & 0, \eeqn
by the positivity of the operator $A$. It follows that
\beqn
\la \psi| ({\cal I} \otimes {\cal E})(A) |\psi\ra =\sum_i \la
	\phi_i|A|\phi_i\ra \geq 0, \eeqn
and thus for any positive operator $A$, the operator $({\cal I} \otimes
{\cal E})(A)$ is also positive, as required. This completes the first
part of the proof.

Suppose next that ${\cal E}$ is a quantum operation. Our aim will be
to find an operator-sum representation for ${\cal E}$. Suppose we
introduce a system, $R$, with the same dimension as the original
quantum system, $Q$. Let $|i_R\ra$ and $|i_Q\ra$ be orthonormal bases
for $R$ and $Q$. It will be convenient to use the same index, $i$, for
these two bases, and this can certainly be done as $R$ and $Q$ have
the same dimensionality. Define a joint state $|\alpha\ra$ of $RQ$ by
\beqn |\alpha\ra \equiv \sum_i |i_R\ra|i_Q\ra. \eeqn The state
$|\alpha\ra$ is, up to a normalization factor, a maximally entangled
state of the systems $R$ and $Q$.  This interpretation of $|\alpha\ra$
as a maximally entangled state may help in understanding the following
construction. Next, we define an operator $\sigma$ on the state space
of $RQ$ by \beqn \label{eqtn:sigma} \sigma \equiv ({\cal I}_R \otimes
{\cal E})(|\alpha\ra \la \alpha|). \eeqn We may think of this as the
result of applying the quantum operation ${\cal E}$ to one half of a
maximally entangled state of the system $RQ$. It is a truly remarkable
fact, which we will now demonstrate, that the operator $\sigma$
completely specifies the quantum operation ${\cal E}$. That is, to
know how ${\cal E}$ acts on an arbitrary state of $Q$, it is
sufficient to know how it acts on a single maximally entangled state
of $Q$ with another system\footnote{It is interesting and enlightening
to contemplate a similar construction for classical systems, based
upon a maximally {\em correlated} state of two classical systems. A
construction of this sort is given at the beginning of Chapter
\ref{chap:distance}.}.

The trick which allows us to recover ${\cal E}$ from $\sigma$ is as
follows. Let $|\psi\ra = \sum_j \psi_j |j_Q\ra$ be any state of system
$Q$. Define a corresponding state $|\tilde \psi\ra$ of system $R$ by
the equation \beqn |\tilde \psi\ra \equiv \sum_j \psi_j^*
|j_R\ra. \eeqn Notice that \beqn \la \tilde \psi| \sigma |\tilde
\psi\ra & = & \la \tilde \psi| \left( \sum_{ij} |i_R\ra \la j_R|
\otimes {\cal E}(|i_Q\ra \la j_Q|) \right) |\tilde \psi \ra \\ & = &
\sum_{ij} \psi_i \psi_j^* {\cal E}(|i_Q\ra \la j_Q|) \\ & = & {\cal
E}(|\psi\ra \la \psi|). \eeqn Let $\sigma = \sum_i |s_i\ra \la s_i|$
be some decomposition of $\sigma$, where the vectors $|s_i\ra$ need
not be normalized. Define a map \beqn E_i(|\psi\ra) \equiv \la \tilde
\psi|s_i\ra. \eeqn A little thought shows that this map is a linear
map, so $E_i$ is a linear operator on the state space of
$Q$. Furthermore, we have \beqn \sum_i E_i|\psi\ra \la \psi|
E_i^{\dagger} & = & \la \tilde \psi|s_i\ra \la s_i| \tilde \psi \ra \\
& = & \la \tilde \psi| \sigma |\tilde \psi\ra \\ & = & {\cal
E}(|\psi\ra \la \psi|). \eeqn Thus \beqn {\cal E}(|\psi\ra \la \psi|)
= \sum_i E_i |\psi\ra \la \psi| E_i{\dagger}, \eeqn for all pure
states, $|\psi\ra$, of $Q$. By linearity it follows that \beqn {\cal
E}(A) = \sum_i E_i A E_i^{\dagger} \eeqn in general.

\end{proof}

This result allows us to give easy proofs that many interesting maps
are quantum operations. For instance, it is clear that the unitary
evolution ${\cal E}(\rho) = U \rho U^{\dagger}$ is a quantum
operation. It is also clear that a measurement is described by a set
of quantum operations ${\cal E}_m(\rho) = P_m \rho P_m$ indexed by the
measurement outcome $m$.

\index{quantum operations!trace map}

Slightly less obviously, we see that the trace map $A \rightarrow
\tr(A)$ is a quantum operation. To see this, let $H_Q$ be any input
Hilbert space, spanned by an orthonormal basis $|1\ra \ldots |d\ra$,
and let $H_Q'$ be a one dimensional output space, spanned by the state
$|0\ra$. Define \beqn {\cal E}(A) \equiv \sum_{i=1}^d |0\ra \la i| A
|i\ra \la 0|, \eeqn so that ${\cal E}$ is a quantum operation, by the
operator-sum representation theorem. Notice that ${\cal E}(A) = \tr(A)
|0\ra \la 0|$, so that, up to the unimportant $|0\ra \la 0|$
multiplier, this quantum operation is identical to the trace function.

\index{quantum operations!partial trace map}

An even more useful result is the observation that the partial trace
is a quantum operation. Suppose we have a joint system $AB$, and wish
to trace out system $B$. Let $|j\ra$ be a basis for system $B$. Define
a linear operator $E_i : H_{AB} \rightarrow H_A$ by
\beqn
E_i \sum_j \lambda_j |a_j\ra |j\ra \equiv \lambda_i |a_i\ra, \eeqn
where $\lambda_j$ are complex numbers, and $|a_j\ra$ are arbitrary
states of system $A$. Define
\beqn
{\cal E}(A) \equiv \sum_i E_i A E_i^{\dagger}. \eeqn
By the operator-sum representation theorem for quantum operations,
this is a quantum operation from system $AB$ to system $A$. Notice that
\beqn
{\cal E}(A \otimes |j\ra \la j'|) = A \delta_{j,j'} = \tr_B(A \otimes
|j\ra \la j'|), \eeqn where $A$ is any Hermitian operator on the state
space of system $A$, and $|j\ra$ and $|j'\ra$ are members of the
orthonormal basis for system $B$. By linearity of ${\cal E}$ and
$\tr_B$, it follows that ${\cal E} = \tr_B$.

In terms of the operator-sum representation, it is easy to
characterize a quantum operation as being complete, incomplete, or
physical. Recall that a quantum operation is complete if
$\tr(\evop(\rho)) = 1$ for all input states $\rho$. Clearly, this is
equivalent to the requirement that $\sum_i E_i^{\dagger} E_i = I$ for
the operators $E_i$ in the operator-sum representation. Similarly, the
property that a quantum operation be incomplete is equivalent to the
condition that $\sum_i E_i^{\dagger} E_i < I$, while the property that
a quantum operation is physical is equivalent to the condition that
$\sum_i E_i^{\dagger} E_i \leq I$.

One reason for our interest in the operator-sum representation is that
it gives us a way of characterizing the dynamics of a system in terms
of {\em intrinsic quantities}. Non-unitary behaviour of quantum system
can only arise because of the action of external systems. The operator
sum representation gives us a way of describing the dynamics of the
principal system, without having to explicitly consider properties of
those external systems; all that we need to know is bundled up into
the operators $E_i$, which act on the Hilbert space of the principal
system alone. Furthermore, we will see soon that many different
interactions with an external system may give rise to the same
dynamics on the principal system. If it is only the dynamics of the
principal system which are of interest then it makes sense to choose a
representation of the dynamics which does not include unimportant
information about other systems.

We can relate the operator-sum representation picture of quantum
operations to the idea of a quantum system interacting with other
systems. We will prove two results. The first result shows how to
determine the operator-sum representation appropriate for a quantum
system interacting in a specified way with other quantum systems. The
second result shows that for {\em any} quantum operation, we can
always find a reasonable {\em model external system and dynamics}
which give rise to that quantum operation. By reasonable, we here mean
that the dynamics must be either a unitary evolution or a projective
measurement.

%\begin{theorem}
%Suppose ${\cal E}$ is a quantum operation with output space identical
%to the input space of another quantum operation ${\cal F}$. Then the
%composition ${\cal F} \circ {\cal E}$ is a quantum operation.
%\end{theorem}
%
%\begin{proof}
%
%This is clear from the definition of quantum
%operations. Alternatively, let $E_i$ be an operator-sum decomposition
%for ${\cal E}$ and $F_j$ an operator-sum decomposition for ${\cal
%F}$. Then
%\beqn
%({\cal F} \circ {\cal E})(A) = {\cal F}(\sum_i E_i A E_i^{\dagger}) \\
%	& = & \sum_{ij} F_j E_i A E_i^{\dagger} F_j^{\dagger}, \eeqn
%which shows that $F_j E_i$ is an operator-sum decomposition
%for ${\cal F} \circ {\cal E}$, and thus ${\cal F} \circ {\cal
%E}$ is a quantum operation.
%
%\end{proof}

\index{environmental models!complete quantum operations}
\index{quantum operations!complete!environmental models for}

Suppose we have a quantum system initially in a state $\rho$. We will
denote this system by the letter $Q$.  Adjoined to $Q$ is another
system which we will refer to variously as the {\em ancilla} or {\em
environment} system, and denote by $E$. We suppose that $Q$ and $E$
are initially independent systems, and that $E$ starts in some
standard state, $\sigma$. The joint state of the system is thus
initially
\begin{eqnarray}
\rho^{QE} = \rho \otimes \sigma. \end{eqnarray}
We suppose that the systems interact according to some unitary
interaction $U$.

After the unitary interaction a measurement may be performed on the
joint system. This measurement is described by projectors $P_m$.  The
case where no measurement is made corresponds to the special case
where there is only a single measurement outcome, $m=0$, which
corresponds to the projector $P_0 \equiv I$.

\begin{figure}[ht]
\begin{center}
\unitlength 1cm
\begin{picture}(8,5)(0,0)
% Q --> Q'
\put(2,3){\framebox(1,1){$Q$}}
\put(3,3.5){\vector(1,0){3}}
\put(6,3){\framebox(1,1){$Q'$}}
%
% initial Q state
\put(1.25,3.5){\makebox(0,0){$\rho^{Q}$}}
%
% purify first stage dynamics
\put(4,0.5){\framebox(1,1){$E$}}
\put(4.4,1.7){\vector(0,1){1.6}}
\put(4.6,3.3){\vector(0,-1){1.6}}
\put(4.8,2.5){\makebox(0,0)[cl]{$P_m U$}}
\end{picture}
\caption{Environmental model for a quantum operation. \label{fig:
basic quantum operation}}
\end{center}
\end{figure}
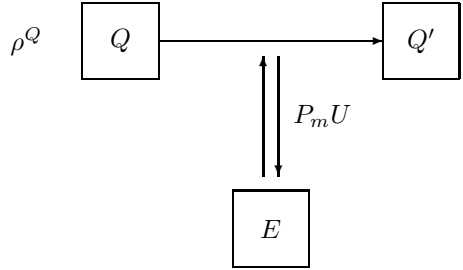

The situation is summarized in figure \ref{fig: basic quantum operation}.
Our aim is to determine the final state of $Q$ as a function of the
initial state, $\rho$. The final state of $QE$ is given by
\begin{eqnarray}
\frac{P_m U (\rho \otimes \sigma) U^{\dagger}P_m}{\tr(P_m U (\rho\otimes\sigma
  ) U^{\dagger} P_m )},
\eeqn
given that measurement outcome $m$ occurred.  Tracing out $E$ we see
that the final state of $Q$ alone is
\beqn \label{eqtn: qop measurement model}
\frac{\tr_E(P_m U (\rho \otimes \sigma) U^{\dagger}P_m)}
	{\tr(P_m U (\rho\otimes\sigma ) U^{\dagger}P_m)}.
\eeqn
This representation of the final state involves the initial state
$\sigma$ of the environment, the interaction $U$ and the measurement
operators $P_m$.  Define a map
\be
\evop_m(\rho) \equiv \tr_E(P_m U (\rho \otimes \sigma)
U^{\dagger}P_m). \ee Note that $\tr(\evop_m(\rho))$ is the probability
of outcome $m$ of the measurement occurring. Let $\sigma = \sum_j q_j
|j\ra\la j|$ be an ensemble decomposition for $\sigma$. Introduce an
orthonormal basis $|k\ra\la k|$ for the system $E$. Note that
\be
\tr(\evop_m(\rho)) & = & \sum_{jk} q_j \tr_E(|k\ra\la k| P_mU(\rho \otimes
|j\ra\la j|) U^{\dagger} P_m |k\ra \la k|) \\
& = & \sum_{jk} E_{jk} \rho E_{jk}^{\dagger}, \eeqn
where
\beqn
E_{jk} \equiv \sqrt{q_j} \la k| P_m U |j\ra. \ee This equation gives
an explicit means for calculating the operators appearing in an
operator-sum representation for $\evop_m$, given that the initial
state $\sigma$ of $E$ is known, and the dynamics between $Q$ and $E$
are known.  Indeed, two examples of this prescription in action were
already given, in the opening section to this Chapter.

We now review a construction converse to this, which shows that for
any quantum operation $\evop$, we can {\em mock up} the dynamics
$\evop$ using an appropriate model. The construction will only be
given for quantum operations mapping the input space to the same
output space, although it is mainly a matter of notation to generalize
the construction to the more general case. In particular, we show that
for any physical quantum operation, $\evop$, there exists a model
environment, $E$, starting in a pure state $|0\ra$, and model dynamics
specified by a unitary operator $U$ and projector $P$ onto $E$ such
that
\be
\evop(\rho) = \tr_E(PU(\rho \otimes |0\ra\la 0|)U^{\dagger} P). \ee
To see this, suppose first that $\evop$ is a complete quantum
operation, with operator-sum representation generated by operators
$E_i$ satisfying the completeness relation $\sum_i E_i^{\dagger} E_i =
I$, so we are only attempting to find an appropriate unitary operator
$U$ to model the dynamics. Let $|i\ra$ be an orthonormal basis set for
$E$, in one-to-one correspondence with the index $i$ for the operators
$E_i$. Define an operator $U$ which has the following action on states
of the form $|\psi\ra|0\ra$,
\be
U|\psi\ra|0\ra \equiv \sum_i E_i |\psi\ra|i\ra. \ee
Note that for arbitrary states $|\psi\ra$ and $|\phi\ra$ of $Q$,
\be
\la \psi|\la 0|U^{\dagger} U |\phi\ra |0\ra = \sum_i \la \psi|
E_i^{\dagger} E_i |\phi\ra = \la \psi|\phi\ra, \ee
by the completeness relation. Thus the operator $U$ can be extended to
a unitary operator acting on the entire state space of the joint
system. It is easy to verify that
\be
\tr_E(U (\rho \otimes |0\ra\la 0|)U^{\dagger}) = \sum_i E_i \rho
E_i^{\dagger}, \ee so this model provides a realization of the quantum
operation $\evop$.

\index{environmental models!incomplete quantum operations}
\index{quantum operations!incomplete!environmental models for}

Incomplete quantum operations can easily be modeled using a
construction along the same lines. Simply introduce an extra operator,
$E_{\infty}$, into the set of operators $E_i$, chosen so that when
summing over the complete set of $i$, including $i =\infty$, one
obtains $\sum_i E_i^{\dagger} E_i = I$. Now repeat the same
construction as before to obtain a unitary operator $U$. Following the
unitary $U$, however, it is necessary to do a projection onto the
states $|i\ra$ where $i \neq \infty$, to remove this operator from the
operator-sum representation of the quantum operation being modeled.

A more interesting generalization of this construction is the case of
a set of physical quantum operations, $\{ \evop_m \}$. Note that if a
set of quantum operations $\{ \evop_m \}$ corresponded to possible
outcomes from a measurement, then the quantum operation $\sum_m
\evop_m$ is complete, since the probabilities of the distinct outcomes
sum to one, $1 = \sum_m p(m) = (\sum_m \evop_m)(\rho)$ for all
possible inputs $\rho$.

Conversely, if we are given a set of physical quantum operations $\{
\evop_m \}$ such that $\sum_m \evop_m$ then it is possible to
construct a {\em measurement model} giving rise to this set of quantum
operations. For each $m$, let $E_{mi}$ be a set of operators
generating an operator-sum representation for $\evop_m$. Introduce an
environmental system, $E$, with an orthonormal basis $|m,i\ra$ in
one-to-one correspondence with the set of indices for the operators
generating the respective operator-sum representations. Analogously to
the earlier construction, define an operator $U$ such that \be
U|\psi\ra|0\ra = \sum_{mi} E_{mi} |\psi\ra |m,i\ra. \ee As before,
this operator may be extended to a unitary operation, because of the
completeness relation $\sum_{mi} E_{mi}^{\dagger} E_{mi} = I$. Next,
define projectors $P_m \equiv \sum_i |m,i\ra \la m,i|$ on the
environmental system, $E$.

Suppose we perform the unitary operation $U$ on the state $\rho
\otimes |0\ra \la 0|$, and follow that up with a measurement on the
environmental system, with the measurement being defined by the
complete set of orthogonal projectors $P_m$. Then it is easy to verify
that the (unnormalized) state of the principal system if the
measurement result $m$ is recorded is $\sum_i E_{mi}\rho
E_{mi}^{\dagger} = \evop_m(\rho)$, with the probability of the outcome
$m$ being given by the trace of $\evop_m(\rho)$, exactly as required.

\subsection{Quantum operations on a single qubit}
\label{subsect:qops_qubit}

\index{qubit!quantum operations on}
\index{quantum operations!on a qubit}

There is a nice geometric way of picturing quantum operations when the
principal system is a single qubit. This method allows one to get an
intuitive feel for the behaviour of quantum operations in terms of
their action on the Bloch sphere. Recall from section \ref{sec:qubits}
that the state of a single qubit can always be written in the Bloch
representation, \beqn \rho = \frac{I + \vec \lambda \cdot \vec
\sigma}{2}, \eeqn where $\vec \lambda$ is a three component real
vector.

In this representation, it turns out that an arbitrary complete quantum
operation is equivalent to a map of the form
\beqn
	\vec \lambda \stackrel{\cal E}{\rightarrow} \vec \lambda' 
		= M \vec \lambda + \vec c,
\label{eqtn: affine map}
\eeqn
where $M$ is a $3 \times 3$ matrix, and $\vec c$ is a constant
vector. This is an {\em affine map}, mapping the Bloch sphere into
itself. Suppose the operators $E_i$ generating the operator-sum
representation for $\evop$ are written in the form
\beqn
	E_i = \alpha_i I + \sum_{k=1}^3 a_{ik} \sigma_k.
\eeqn
Then it is not difficult to check that
\beqn
	M_{jk} & = &  \sum_l \left[ a_{lj} a_{lk}^* + a_{lj}^* a_{lk}
	+ \left( |\alpha_l|^2- \sum_p a_{lp} a_{lp}^* \right) \delta_{jk}   
	\right. \nonumber \\  & & \left. 
	+ i \sum_p \epsilon_{jkp}
	    ( \alpha_l a_{lp}^* - \alpha_l^* a_{lp} ) \right]
\\
	c_k &=& 2i \sum_l \sum_{jp} \epsilon_{jpk} a_{lj} a_{lp}^*,
\eeqn
where we have made use of the completeness relation $\sum_i
E_i^{\dagger} E_i = I$ to simplify the expression for $\vec c$.

The meaning of the affine map equation (\ref{eqtn: affine map}) is
made clearer by considering the polar decomposition \cite{Horn91a} of
the matrix $M$.  Any real matrix $M$ can always be written in the form
\beqn
	M = O S, 
\end{eqnarray}
where $O$ is a real orthogonal matrix with determinant $1$,
representing a proper rotation, and $S$ is a real symmetric
matrix. Viewed this way, equation (\ref{eqtn: affine map}) is just a
deformation of the Bloch sphere along principal axes determined by
$S$, followed by a proper rotation due to $O$, followed by a
displacement due to $\vec c$. 

This picture can be used to obtain simple pictures of quantum
operations on single qubits. For example, unitary operations
correspond to (possibly improper) rotations of the Bloch sphere.

Less trivially, consider the completely decohering quantum operation,
\be \rho \rightarrow \evop(\rho) = P_0 \rho P_0 + P_1 \rho P_1, \ee
for which we introduced an environmental model in the opening section
of this Chapter. Using the above prescription it is easy to see that
the corresponding map on the Bloch sphere is given by \be
(\lambda_x,\lambda_y,\lambda_z) \rightarrow (0,0,\lambda_z). \ee
Geometrically, the Bloch vector is projected along the $z$ axis, and
the $x$ and $y$ components of the the Bloch vector are lost.  This
geometric picture makes it very easy to verify certain facts about
this quantum operation. For example, it is easy to verify that the
quantity $\tr(\rho^2)$ for a single qubit is equal to
$(1+|\lambda|^2)/2$, where $|\lambda|$ is the norm of the Bloch
vector. The projection process above cannot increase the norm of the
Bloch vector, and therefore we can immediately conclude that
$\tr(\rho^2)$ can only ever decrease for the completely decohering
quantum operation. This is but one example of the use of this
geometric picture; once it becomes sufficiently familiar it becomes a
great source of insight about the properties of quantum operations on
a single qubit.

\section{Freedom in the operator-sum representation}
\label{sec:freedom_qops}
\index{operator-sum representation!freedom in}

Consider quantum operations ${\cal E}$ and ${\cal F}$ acting on a
single qubit with the operator-sum representations,
\beqn
{\cal E}(\rho) & = & \frac{\rho}{2} + \frac{Z\rho Z}{2} \\
{\cal F}(\rho) & = & |0\ra\la 0| \rho |0\ra \la 0| + |1\ra \la 1| \rho
|1\ra \la 1|. \eeqn
What is interesting is that these two quantum operations are actually
the same quantum operation. To see this, note that $|0\ra\la 0|
=(I+Z)/2$ and $|1\ra\la 1| = (I-Z)/2$. Thus
\beqn
{\cal F}(\rho) & = & \frac{(I+Z)\rho(I+Z)+(I-Z)\rho(I-Z)}{4} \\
	& = & \frac{I \rho I + Z \rho Z}{2} \\
	& = & {\cal E}(\rho). \eeqn

This freedom in the representation is very interesting. Suppose we
flipped a fair coin, and, depending on the outcome of the coin toss,
applied either the unitary operator $I$ or $Z$ to the quantum
system. This process corresponds to the first operator-sum
representation for ${\cal E}$. The second operator-sum representation
for ${\cal E}$ (labeled ${\cal F}$ above) corresponds to performing a
projective measurement in the $\{|0\ra,|1\ra\}$ basis, with the
outcome of the measurement unknown. These two apparently very
different physical processes give rise to exactly the same system
dynamics.

In this section we study in more detail the question of when two sets
of operators give rise to the same quantum operation. Understanding
this question is important for at least two different reasons. First,
from a physical point of view, understanding the freedom in the
representation gives us more insight into how different physical
processes can give rise to the same system dynamics. Second, in later
chapters we will have occasion to use the characterization we find to
simplify certain constructions. In particular, it will simplify some
of the constructions involving quantum error correction.

To begin, we actually need to answer a different question.  Suppose
$|\psi_i\ra$ is a set of states. We say the set $|\psi_i\ra$ {\em
generates} the operator $A \equiv \sum_i |\psi_i\ra \la \psi_i|$. When
do two sets of states, $|\psi_i\ra$ and $|\phi_j\ra$ generate the same
operator $A$? It turns out that the answer to this question has a
surprising number of interesting and useful consequences, amongst
which is the solution to our problem of determining the freedom in the
operator-sum representation.

\begin{theorem}

The sets $|\psi_i\ra$ and $|\phi_j\ra$ generate the same operator if
and only if
\beqn
|\psi_i\ra = \sum_j u_{ij} |\phi_j\ra, \eeqn
where $u_{ij}$ is a unitary matrix of complex numbers, and we ``pad''
whichever set of states $|\psi_i\ra$ or $|\phi_j\ra$ is smaller with
additional states $0$ so that the two sets have the same number of elements.
\end{theorem}

As an example of the theorem, suppose we have
\beqn
\rho = \frac{3}{4} |0\ra\la 0| + \frac{1}{4} |1\ra\la 1|. \eeqn
Let
\beqn
|a\ra & \equiv & \frac{1}{\sqrt 2} \sqrt{\frac 34} |0\ra + \frac{1}{\sqrt
2} \sqrt{\frac 14}|1\ra \\
|b\ra & \equiv & \frac{1}{\sqrt 2} \sqrt{\frac 34} |0\ra - \frac{1}{\sqrt
2} \sqrt{\frac 14}|1\ra. \eeqn
Then it is easily checked that $\rho = |a\ra \la a|+|b\ra \la b|$.

\begin{proof}

Suppose $|\psi_i\ra = \sum_j u_{ij} |\phi_j\ra$ for some unitary
$u_{ij}$. Then
\beqn
\sum_i |\psi_i\ra\la \psi_i| & = & \sum_{ijk} u_{ij}u_{ik}^*
|\phi_j\ra \la \phi_k| \\
	& = & \sum_{jk} \left( \sum_i u^{\dagger}_{ki} u_{ij} \right)
	|\phi_j\ra \la \phi_k | \\
	& = & \sum_{jk} \delta_{kj} |\phi_j\ra \la \phi_k| \\
	& = & \sum_j |\phi_j\ra \la \phi_j|, \eeqn
which shows that $|\psi_i\ra$ and $|\phi_j\ra$ generate the same
operator.

Conversely, suppose
\beqn
A = \sum_i |\psi_i\ra \la \psi_i| = \sum_j |\phi_j\ra\la
\phi_j|. \eeqn
A little thought shows that for this equation to hold each
$|\psi_i\ra$ can be expressed as a linear combination of the
$|\phi_j\ra$, $|\psi_i\ra = \sum_j c_{ij} |\phi_j\ra$. Thus
\be
\sum_j |\phi_j\ra \la \phi_j| & = & \sum_{j_1 j_2} \left( \sum_i
c_{ij_1} c_{ij_2}^* \right) |\phi_{j_1}\ra \la \phi_{j_2}|, \ee
from which we see that $c$ is unitary, as required.

\end{proof}

This result allows us to characterize the freedom in operator-sum
representations. Suppose $E_j$ and $F_k$ are two sets of operators,
both giving rise to the same quantum operation, $\sum_j E_j A
E_j^{\dagger} = \sum_k F_k A F_k^{\dagger}$ for all $A$. Define
\beqn
|e_j\ra & \equiv & \sum_i |i_R\ra \left(E_j|i_Q\ra \right) \\ |f_k\ra &
\equiv & \sum_i |i_R\ra \left(F_k|i_Q\ra \right). \eeqn Recall, the
earlier definition of $\sigma$, equation (\ref{eqtn:sigma}), from
which it follows that $\sigma = \sum_j |e_j\ra\la e_j| = \sum_k
|f_k\ra \la f_k|$, and thus there exists unitary $u_{jk}$ such that
\beqn
|e_j\ra = \sum_k u_{jk} |f_k\ra. \eeqn
But for arbitrary $|\psi\ra$ we have
\beqn
E_j|\psi\ra & = & \la \tilde \psi|e_j\ra \\
	& = & \sum_k u_{jk} \la \tilde \psi|f_k\ra \\
	& = & \sum_k u_{jk} F_k|\psi\ra. \eeqn
Thus
\beqn
E_j = \sum_k u_{jk} F_k. \eeqn Conversely, supposing $E_j$ and $F_k$
are related by a unitary transformation of the form $E_j = \sum_{jk}
u_{jk} F_k$, simple algebra shows that the quantum operation generated
by the operators $E_j$ is the same as the quantum operation generated
by the operators $F_k$.

Summarizing, we have shown that a quantum operation $\evop$ is
generated in the operator-sum representation by two sets of operators
$E_j$ and $F_k$ if and only if there exists a unitary matrix of
complex numbers $u_{jk}$ such that
\be
E_j = \sum_k u_{jk} F_k, \ee
where it may be necessary to ``pad'' the shorter set of operators with
zero operators to ensure that the matrix $u$ is square.

This result is surprisingly useful. We will use it, for example, in
our study of quantum error correction in Chapter \ref{chap:qec}. In
that Chapter we will see that certain sets operators in the operator
sum representation give more useful information about the quantum
error correction process, and it behooves us to study quantum error
correction from that point of view.  As usual, having multiple ways of
understanding a process gives us much more insight into what is going
on.

\section{Teleportation as a quantum operation}
\label{sec:teleportation_qops}
\index{teleportation!as a quantum operation}

Let's switch gears, and move away from abstract generalities into a
more specific scenario: quantum teleportation. As discussed in section
\ref{sec:teleportation}, quantum teleportation allows us to transmit
an unknown quantum state from one location to another using preshared
entanglement and classical communication. In this section we show how
quantum teleportation can be understood within the quantum operations
formalism. This, in turn, allows us to relate quantum teleportation to
quantum error correction.  The work in this section is based upon a
collaboration with Caves \cite{Nielsen97c}. Some of the ideas were
arrived at independently about the same time by Bennett, DiVincenzo,
Smolin and Wootters \cite{Bennett96a}. I would especially like to
thank Chris Fuchs, who got this work started by suggesting that it
might be valuable to try to understand quantum teleportation in terms
of reversible measurements.

%Initially the composite system is prepared in a state with density 
%operator $\rho \otimes \sigma$, where $\rho$ is an unknown state of 
%the input system $1$, and $\sigma$ is a maximally entangled pure 
%state of systems $2$ and $3$,
%\begin{eqnarray}
%\sigma = {1\over2}\bigl(|\mathord{\uparrow}\mathord{\downarrow}\rangle +
%	 |\mathord{\downarrow}\mathord{\uparrow}\rangle\bigr)
%         \bigl(\langle\mathord{\uparrow}\mathord{\downarrow}|+
%         \langle\mathord{\downarrow}\mathord{\uparrow}|\bigr)\;.
%\end{eqnarray}
%
%Alice's goal is to ``teleport'' the input state $\rho$ to the target 
%system, Bob's system $3$. This is done as follows. Alice performs a 
%measurement on systems $1$ and $2$ in the Bell operator basis 
%\cite{Braunstein92a}, which consists of four entangled states for systems 
%$1$ and $2$,
%\begin{eqnarray}
%|\psi^{\pm}\rangle & = & {1\over\sqrt2}
%   \bigl(|\mathord{\uparrow}\mathord{\downarrow}\rangle \pm
%	 |\mathord{\downarrow}\mathord{\uparrow}\rangle\bigr)\;, \\
%|\phi^{\pm}\rangle & = & {1\over\sqrt2}
%   \bigl(|\mathord{\uparrow}\mathord{\uparrow}\rangle \pm
%	 |\mathord{\downarrow}\mathord{\downarrow}\rangle\bigr)\;. 
%\end{eqnarray}

Recall that teleportation involves a sender, Alice, and a receiver,
Bob.  Suppose Alice has possession of an {\em input system}, which we
label 1, in an unknown input state $\tilde\rho^1$.  To avoid
confusion, we use a superscript to denote the appropriate state space
for a vector or an operator; the reason for the tilde becomes clear
shortly.  Alice might also have access to another system, which we
label $2$.  Bob has access to the {\em target system}, which we label
$3$.  Systems $2$ and $3$ are assumed to be prepared initially in some
standard state $\sigma^{23}$, which is assumed to be uncorrelated with
$\tilde\rho^1$; that is, the initial state of the composite system
consisting of 1, 2, and 3 is
\begin{eqnarray}
\tilde\rho^1\otimes\sigma^{23}\;. \end{eqnarray}
The case where Bob has access to an additional system, labeled $4$, is 
discussed briefly later in this section.

We assume that systems $1$ and $3$ are identical and thus have the same 
state space.  This means that there is a one-to-one linear map from 
the state space of 3 onto the state space of 1.  Though this map is not 
unique, we choose a particular one, thereby setting up a one-to-one 
correspondence between vectors in the state space of 3 and vectors in 
the state space of 1.  We denote this one-to-one correspondence by 
\begin{eqnarray}
|\psi^3\rangle\leftrightarrow|\tilde\psi^1\rangle\;. \end{eqnarray}
The one-to-one correspondence between vectors induces a one-to-one
correspondence between operators on 3 and operators on 1, which we
denote by $A^3\leftrightarrow\tilde A^1$. This correspondence is
given by linearly extending the map $|\psi^3\rangle \langle \phi^3|
\leftrightarrow |\tilde \psi^1\rangle \langle \tilde \phi^1|$ to all
operators on systems $3$ and $1$.  In particular, for each
state $\tilde\rho^1$ of the input system, there is a unique 
counterpart state $\rho^3$ of the target system.

The choice of a correspondence between the state spaces of 1 and 3 is
{\it physically\/} motivated: the correspondence defines what it means
to transport a system unchanged from the location of system 1 to the
location of system 3.  Different procedures for performing this
transportation lead to different correspondences.  For example,
suppose we wish to teleport the state of a spin-$1\over2$ particle
from Los Alamos to Pasadena.  To say what it means to teleport the
state requires a correspondence between the state spaces in Los Alamos
and Pasadena.  We could set up the correspondence by agreeing that the
$z$ axis in each location lies along the local acceleration of gravity
and the $x$ axis along the local magnetic north or by adopting
arbitrary orthogonal axes in the two locations.  Ordinarily we assume
implicitly such a correspondence, as was done earlier in the
Dissertation, and write $\tilde\rho^1=\rho^3=\rho$.  In the present
setting it is advantageous to adopt a notation which more explicitly
distinguishes between states of system 1 and of system 3.

The correspondence can be extended to a one-to-one correspondence between 
the joint state space of 2 and 3 and the joint state space of 1 and 2.  
If $|b^2\rangle|c^3\rangle$ is a product basis for the joint state space 
of 2 and 3, this one-to-one correspondence is given by
\begin{eqnarray}
|\psi^{23}\rangle=\sum_{b,c}\alpha_{bc}|b^2\rangle|c^3\rangle
\leftrightarrow
\sum_{b,c}\alpha_{bc}|\tilde c^1\rangle|b^2\rangle=
|\tilde\psi^{12}\rangle \;. \end{eqnarray}
This correspondence induces a one-to-one correspondence between operators
on the joint state space of 2 and 3 and operators on the joint state
space of 1 and 2.  

The correspondence can be extended further to a one-to-one linear map 
from the state space of the composite system 1, 2, and 3 onto itself:
\begin{eqnarray}
|\psi^{123}\rangle\leftrightarrow
|\tilde\psi^{123}\rangle=
U_{13}|\psi^{123}\rangle\;. \end{eqnarray} 
This map is accomplished by a unitary operator $U_{13}$, which acts on 
product states according to
\begin{eqnarray}
U_{13}|\tilde a^1\rangle|b^2\rangle|c^3\rangle= 
|\tilde c^1\rangle|b^2\rangle|a^3\rangle
\end{eqnarray}
and thus is called the ``swap'' operator because it swaps the states
of systems 1 and 3, while leaving system 2 alone.  The swap operator
clearly satisfies $(U_{13})^2=I^{123}$, that is, $U_{13}^\dagger=U_{13}$.
When extended to operators on the composite system, the correspondence 
becomes
\begin{eqnarray}
A^{123}\leftrightarrow\tilde A^{123}=
U_{13} A^{123} U_{13}^\dagger\;. \end{eqnarray}

Suppose now that Alice performs a measurement on systems $1$ and
$2$. This measurement will be described by a set of quantum operations
$\evop_m$ such that $\sum_m \evop_m$ is a complete quantum operation.
We assume that each ${\cal E}_m$ has an operator-sum representation
generated by operators $\tilde E_{mj}^{12}$ on the systems $1$ and
$2$.
%
%For convenience, we will assume that each $\evop_m$ has but a single
%operator in its operator-sum representation, $\evop_m(\rho) = \tilde E_m^{12}
%\rho (\tilde E_m^{12})^{\dagger}$, where $\tilde E_m^{12}$ is an
%operator on systems $1$ and $2$. The general case is notationally more
%complicated, but no more conceptually difficult than the argument that
%follows, and the conclusions are the same.

If the measurement has outcome $m$, then the unnormalized state of the 
target system $3$ after the measurement is given by 
\begin{equation} \label{cond state of 3}
\hat\rho_m^3 = \mbox{tr}_{12}
\!\left( \sum_j (\tilde E_{mj}^{12}\otimes I^3)
	        (\tilde\rho^1\otimes\sigma^{23}) 
                [(\tilde E_{mj}^{12})^{\dagger}\otimes I^3]
\right) . \end{equation}
where the caret denotes an unnormalized state.  

We now show that $\hat\rho_m^3$ is related to $\rho^3$ by a quantum 
operation, which we denote ${\cal T}_m$.  We first notice that
\begin{eqnarray} \label{sigma eqtn}
\tilde\rho^1\otimes\sigma^{23}=
U_{13}(\tilde\sigma^{12}\otimes\rho^3)U_{13}^{\dagger}\;, \end{eqnarray}
where $\tilde\sigma^{12}$ is the counterpart of $\sigma^{23}$.  
Substituting this into (\ref{cond state of 3}) gives
\begin{eqnarray} \label{eqtn: rho_i again}
\hat\rho_m^3 &=& \mbox{tr}_{12}
\Biggl( \sum_j 
                (\tilde E_{mj}^{12}\otimes I^3) \nonumber \\
&\mbox{}&\hphantom{\mbox{tr}_{12}\Biggl(}\times
                [U_{13}(\tilde\sigma^{12}\otimes\rho^3)U_{13}^{\dagger}]
                [(\tilde E_{mj}^{12})^{\dagger}\otimes I^3]
\Biggr)\;. \end{eqnarray}
The form of this equation allows us to think of $\hat\rho_m^3$ as
arising from the following process.  The composite system begins in
the state $\tilde\sigma^{12}\otimes\rho^3$, in which the joint system
1 and 2 is in the state $\tilde\sigma^{12}$ and system 3 is in the
state $\rho^3$.  After the composite system evolves under the unitary
swap operator, a measurement is performed on the joint system 1 and 2,
and then the joint system 1 and 2 is discarded.  This process ought to
seem highly familiar -- it is the same process we used to generate
selective quantum operations earlier in the Chapter! Of course, it
does not matter that this sequence of events does not literally occur;
what matters is that it is effectively as if this occurred. Next,
we'll explicitly complete the construction of the quantum operation
${\cal E}_m$.  This having been done, the problem of teleportation is
for Bob to reverse the quantum operation ${\cal E}_m$.  If the
reversal can be done, then Bob can recover the state $\rho^3$ from the
output state $\hat\rho_m^3={\cal E}_m(\rho^3)$ of system 3.

We write
\begin{eqnarray}
\tilde\sigma^{12}=\sum_k p_k 
|\tilde s_k^{12}\rangle \langle\tilde s_k^{12}|\;, \end{eqnarray}
where the vectors $|\tilde s_k^{12}\rangle$ make up the complete orthonormal 
set of eigenvectors of $\tilde\sigma^{12}$ in the joint space of $1$ 
and $2$. Furthermore, we let 
$\tilde\Pi_l^{12} = |\tilde P_l^{12}\rangle\langle\tilde P_l^{12}|$ be 
any complete set of orthogonal one-dimensional projectors for the joint 
system $1$ and $2$. Performing the partial trace of 
Eq.~(\ref{eqtn: rho_i again}) in the basis $|\tilde P_l^{12}\rangle$
yields
\begin{eqnarray}
\hat\rho_m^3 & = & 
\sum_{jkl} \Bigl(\sqrt{p_k}\langle\tilde P_l^{12}|
	 (\tilde E_{mj}^{12}\otimes I^3)U_{13}|\tilde s_k^{12}\rangle\Bigr )
\,\rho^3 \nonumber \\
&\mbox{}&\phantom{\sum_{jkl}}\times
	 \Bigl(\sqrt{p_k}\langle\tilde s_k^{12}|U_{13}^{\dagger}
	 [(\tilde E_{mj}^{12})^{\dagger}\otimes I^3]
         |\tilde P_l^{12}\rangle\Bigr)\;.
\end{eqnarray}
Using the single index $n$ to denote the triple $(j,k,l)$ and defining
the system~3 operators
\begin{eqnarray}
B_{mn}^3 
&\equiv&\sqrt{p_k}\langle\tilde P_l^{12}|
        (\tilde E_{mj}^{12}\otimes I^3)U_{13}|\tilde s_k^{12}\rangle 
\nonumber \\
&=&\sqrt{p_k}\langle\tilde P_l^{12}|U_{13}
   (I^1\otimes E_{mj}^{23})|\tilde s_k^{12}\rangle\;,
\end{eqnarray}
we can write the output state of system~3 as
\begin{eqnarray}
\hat\rho_m^3 = \sum_n B_{mn}^3 \rho^3 (B_{mn}^3)^{\dagger} 
\equiv {\cal E}_m(\rho^3)\;. \end{eqnarray}
As we set out to show, $\hat\rho_m^3$ is related to $\rho^3$ by a 
quantum operation ${\cal E}_m$.

%Notice that because of the sums introduced by the partial trace and the 
%orthogonal decomposition of $\tilde\sigma^{12}$, the quantum operations 
%${\cal E}_i$ generally are not ideal even if the measurement on 1 and 2 
%is ideal.   In the next section we explore a case where the quantum
%operations ${\cal E}_i$ are ideal.

\index{quantum teleportation!connection to error correction}

We have shown how to construct a quantum operation explicitly linking
the the input to the teleportation process to the output. The exact
form of the quantum operation depends upon how the teleportation
process is performed. In collaboration with Caves I have used this
description elsewhere to obtain necessary and sufficient conditions
for teleportation, for a subclass of possible teleportation processes
\cite{Nielsen97b}. In the present context, the importance of this
discussion is as an example of how the quantum operations formalism
may be used to obtain explicit representations for interesting quantum
processes. In Chapters \ref{chap:qec} and \ref{chap:capacity} we will
study the problem of quantum error correction, which turns out to be
closely related to teleportation.  The connection is to note that for
Bob to complete the teleportation process, he must perform a complete
quantum operation ${\cal R}_m$ on system 3 such that
\begin{eqnarray}
{\cal R}_m\!\left( 
\frac{{\cal E}_m(\rho^3)}{\mbox{tr}\bigl({\cal E}_m(\rho^3)\bigr)}
\right) = 
\rho^3\;.
\end{eqnarray}
That is, Bob must be able to reverse the quantum operation $\evop_m$,
recovering the original state $\rho^3$. The subject of quantum error
correction is actually the study of when such a reversal is possible;
thus the connection between quantum teleportation and quantum error
correction.

%We have shown that the problem of understanding teleportation can be
%reduced to the problem of understanding how to reverse quantum
%operations.  Given the work that has been done on reversing
%deterministic quantum operations that arise from decoherence, this
%would seem to be a useful insight (see
%\cite{Schumacher96a,Shor95a,Ekert96a,Knill97a,Nielsen96a,%
%Schumacher96b,Shor96a,Steane96a} for a sample of this work).
%
%We have shown how quantum teleportation can be understood in terms
%of the general problem of reversing quantum operations, thereby
%demonstrating the crucial connection between teleportation and the fact 
%that no information about the state to be teleported is gained during 
%the process.  We have used the condition for unitarily reversing an
%ideal quantum operation to characterize completely teleportation schemes
%of the type introduced by Bennett {\it et al.}

\section{Quantum process tomography}
\label{sec:quantum_process_tomography}
\index{quantum process tomography}
\index{quantum state tomography}
\index{process tomography}
\index{state tomography}
\index{tomography}

Suppose an experimentalist wishes to completely characterize the
dynamics of a quantum system. For finite dimensional systems we
explain in this section how this task can be performed with the aid of
the quantum operations formalism, and a process known as {\em quantum
state tomography}
\cite{Raymer94a,Leonhardt96a,Leibfried96a}. The resulting procedure is
called {\em quantum process tomography}, since it gives a method for
completely characterizing a quantum process. The work in this section
is based upon work done in collaboration with Chuang
\cite{Chuang97a}. Similar work was done independently by Poyatos,
Cirac and Zoller \cite{Poyatos97a} at about the same time. Some of
these questions have been considered in a partial manner by other
researchers, including Jones \cite{Jones94a}, Turchette {\em et al}
\cite{Turchette95a} and Mabuchi \cite{Mabuchi96b}.

%Such quantum operations are in a one to one correspondence with the set of
%transformations arising from the joint unitary evolution of the quantum system
%and an initially uncorrelated environment\cite{Kraus83a}.  In other words, the
%quantum operations formalism also describes the master equation and quantum
%Langevin pictures widely used in quantum optics \cite{Louisell,Gardiner91},
%where the system's state change arises from an interaction Hamiltonian between
%the system and its environment\cite{Mabuchi96b}.
%
%The major limitation of this picture is that the formalism assumes that the
%input state $\rho$ is in a product state with the rest of the world.
%Otherwise, if the initial system were to be in a correlated state with part of
%the quantum black box -- for example, if it were entangled with an environment
%shared with the black box -- then Eq.(\ref{eq:rhomapfirst}) would not be
%valid, and the quantum operations description as defined above would not be
%applicable.  For the purposes of this paper, we shall {\em define} a quantum
%black box operation as one satisfying Eq.(\ref{eq:rhomapfirst}).

The experimental procedure may be outlined as follows.  Suppose the
state space of the system has $N$ dimensions; for example, $N=2$ for a
single qubit.  $N^2$ pure quantum states $|\psi_1\ra\la\psi_1|,
\ldots,|\psi_{N^2}\ra \la \psi_{N^2}|$ are prepared, and the output
state ${\cal E}(|\psi_j\ra\la\psi_j|)$ is measured for each input.  In
general, performing such a measurement is not easy, but in recent
years a procedure known as {\em quantum state tomography}
\cite{Raymer94a,Leonhardt96a,Leibfried96a} has been developed which
enables such measurements to be performed.  In principle, the quantum
operation ${\cal E}$ can now be determined by a linear extension of
${\cal E}$ to all states, provided the input operators $|\psi_1\ra\la
\psi_1|,\ldots,|\psi_{N^2}\ra \la \psi_{N^2}|$ form a linearly
independent set.

From a purist's point of view, we are done. In practice, of course, we
would like to have a way of determining a useful representation of
$\evop$ from experimentally available data. In this section we give a
general description of such a method, and an example of how it may be
applied in the single qubit case.

Our goal is to determine a set of operators, $E_i$, generating an
operator-sum representation for $\evop$,
\be \label{eq:eeffect}
\evop(\rho) = \sum_i E_i \rho E_i^{\dagger}. \ee
However, experimental results involve numbers, not operators, which
are a theoretical concept.  To determine the $E_i$ from measurable
parameters, it is convenient to consider an equivalent description of
${\cal E}$ using a {\em fixed} set of operators $\tilde{E}_i$, which
form a basis for the set of operators on the state space, so that
\begin{eqnarray}
		E_i = \sum_m e_{im} \tilde{E}_m
\label{eq:atildedef}
\end{eqnarray}
for some set of complex numbers $e_{im}$.  Eq.(\ref{eq:eeffect}) may
thus be rewritten as
\be 
\label{eqtn: two sided rep}
	{\cal E}(\rho) = \sum_{mn} \tilde{E}_m \rho 
			\tilde{E}_{n}^{\dagger} \chi_{mn}
\,,
\ee
where $\chi_{mn} \equiv \sum_i e_{im} e_{in}^*$ is a matrix which is
positive Hermitian by definition. 
% This shows that ${\cal E}$ can be
%completely described by a complex number matrix, $\chi$, once the set
%of operators $\tilde{E}_i$ has been fixed.

In general, $\chi$ will contain $N^4-N^2$ independent real parameters,
because a general linear map of $N$ by $N$ complex matrices to $N$ by
$N$ matrices is described by $N^4$ independent parameters, but there
are $N^2$ additional constraints due to the fact that $\rho$ remains
Hermitian with trace one; that is, the completeness relation
\begin{eqnarray}
\sum_i E_i^{\dagger} E_i = I, \end{eqnarray}
is satisfied, giving $N^2$ real constraints.
Note that the restriction that the map be a quantum operation does not change
the counting, since by Choi's results \cite{Choi75a}, the set of
quantum operations is just the positive cone in the real vector
space of Hermitian-preserving maps, and the positive cone of a real vector
space has the same dimensionality as the underlying vector space.
%
%The same result may be obtained by realizing that $\chi$ results from the
%restriction of a $N^2$ dimensional (system + environment) unitary transform to
%a $N$ dimensional space (the system).  $N^4$ parameters describe all the
%possible transforms, but of these, $N^2$ describe irrelevant transforms which
%affect only the half space which is traced over (the environment).  Thus,
%$\chi$ is parameterized by $N^4-N^2$ degrees of freedom.
%
% Put in a reference to purification to explain why the env. need only have N
% dimensions? 
%
% END MODIFICATION
%
We will show how to determine $\chi$ experimentally, and then show how an
operator-sum representation of the form (\ref{eq:eeffect}) can be
recovered once the $\chi$ matrix is known.

Let $\rho_j$, $1\leq j \leq N^2$ be a fixed set of linearly
independent basis elements for the space of $N$$\times$$N$ matrices.
A convenient choice is the set of operators $|n\ra\la m|$.
Experimentally, the output state ${\cal E}( | n\ra\la m|)$ may be
obtained by preparing the input states $\ket{n}$, $\ket{m}$,
$\ket{n_+} = (\ket{n}+\ket{m})/\sqrt{2}$, and $\ket{n_-} =
(\ket{n}+i\ket{m})/\sqrt{2}$ and forming linear combinations of ${\cal
E}(\ket{n}\bra{n})$, ${\cal E}(\ket{m}\bra{m})$, ${\cal
E}(\ket{n_+}\bra{n_+})$, and ${\cal E}(\ket{n_-}\bra{n_-})$.  Thus, it
is possible to determine ${\cal E}(\rho_j)$ by state tomography, for
each $\rho_j$.

Furthermore, each ${\cal E}(\rho_j)$ may be expressed as a linear combination
of the basis states,
\be
	{\cal E}(\rho_j)
	= \sum_k \lambda_{jk} \rho_k
\,,
\ee
and since ${\cal E}(\rho_j)$ is known, $\lambda_{jk}$ can be
determined by standard linear algebraic algorithms.
To proceed, we may write
\be
	\tilde{E}_m \rho_j \tilde{E}_n^\dagger = \sum_k \beta^{mn}_{jk} \rho_k
\,,
\label{eq:betadef}
\ee
where $\beta^{mn}_{jk}$ are complex numbers which can be determined by
standard algorithms from linear algebra given the $\tilde{E}_m$
operators and the $\rho_j$ operators.  Combining the last two
expressions we have
\be
	\sum_k \sum_{mn} \chi_{mn} \beta^{mn}_{jk} \rho_k 
	= \sum_k \lambda_{jk}\rho_k
\,.
\ee
{} From independence of the $\rho_k$ it follows that for each $k$,
\be
\label{eqtn: chi condition}
	\sum_{mn} \beta^{mn}_{jk} \chi_{mn} = \lambda_{jk}
\,.
\ee
This relation is a necessary and sufficient condition for the matrix
$\chi$ to give the correct quantum operation ${\cal E}$.  One may
think of $\chi$ and $\lambda$ as vectors, and $\beta$ as a
$N^4$$\times$$N^4$ matrix with columns indexed by $mn$, and rows by
$ij$.  To show how $\chi$ may be obtained, let $\kappa$ be the
generalized inverse for the matrix $\beta$, satisfying the relation
\begin{equation}
	\beta^{mn}_{jk} = \sum_{st,xy} \beta_{jk}^{st} \kappa_{st}^{xy}
		 \beta_{xy}^{mn}
\,.
\end{equation}
Most computer packages for matrix manipulation are capable of finding
such generalized inverses.  We now prove that $\chi$ defined by
\begin{eqnarray}
	\chi_{mn} = \sum_{jk} \kappa_{jk}^{mn} \lambda_{jk}
\label{eqtn:chidefn}
\end{eqnarray}
satisfies the relation (\ref{eqtn: chi condition}).

The difficulty in verifying that $\chi$ defined by
(\ref{eqtn:chidefn}) satisfies (\ref{eqtn: chi condition}) is that, in
general, $\chi$ is not uniquely determined by equation (\ref{eqtn: chi
condition}). For convenience we will rewrite these equations in matrix
form as
\begin{eqnarray}
\label{eqtn: chi cond app}
        \beta \vec \chi & = & \vec \lambda \\
\label{eqtn: chi defn app}
        \vec \chi & \equiv & \kappa \vec \lambda
\,.
\end{eqnarray}
{} From the construction that led to equation (\ref{eqtn: two sided
rep}) we know there exists at least one solution to equation
(\ref{eqtn: chi cond app}), which we shall call $\vec \chi '$. Thus
$\vec \lambda = \beta \vec \chi '$. The generalized inverse satisfies
$\beta \kappa \beta = \beta$.  Premultiplying the definition of $\vec
\chi$ by $\beta$ gives
\begin{eqnarray}
        \beta \vec \chi & = & \beta \kappa \vec \lambda \\
         & = & \beta \kappa \beta \vec \chi ' \\
         & = & \beta \vec \chi ' \\
         & = & \lambda 
\,.
\end{eqnarray}
Thus $\chi$ defined by (\ref{eqtn: chi defn app}) satisfies the
equation (\ref{eqtn: chi cond app}), as we wanted to show.

Having determined $\chi$ one immediately obtains the operator-sum
representation for ${\cal E}$ in the following manner.  Let the
unitary matrix $U^\dagger$ diagonalize $\chi$,
\begin{eqnarray}
	\chi_{mn} = \sum_{xy} U_{mx} d_{x} \delta_{xy} U^*_{ny}
\,. 
\label{eq:chipolar}
\end{eqnarray}
{} From this it can easily be verified that
\begin{eqnarray}
%	E_i = \sqrt{d_i} \sum_j U_{ij} \tilde{E}_j
%
% 12-Apr-97 ILC: NEXT LINE CHANGED (U index swapped)
%
	E_i = \sqrt{d_i} \sum_j U_{ji} \tilde{E}_j
\label{eq:chiopsum}
\end{eqnarray}
gives an operator-sum representation for the quantum operation ${\cal
E}$.  Our algorithm may thus be summarized as follows: $\lambda$ is
experimentally measured, and given $\beta$, determined by a choice of
$\tilde{E}$, we find the desired parameters $\chi$ which completely
describe ${\cal E}$, and which determine a set of operators $E_i$
generating an operator-sum representation for $\evop$.

\subsection{One qubit example}

The above general method can be simplified in the case of a one qubit
operation to provide explicit formulas which may be useful in
experimental contexts, such as the teleportation experiment described
in subsection \ref{subsec:teleport_NMR}.  This simplification is made
possible by choosing the fixed operators $\tilde{E}_i$ to have
commutation properties which conveniently allow the $\chi$ matrix to
be determined by straightforward matrix multiplication.  In the one
qubit case, we use: \beqn \tilde{E}_0 &=& I \label{eq:fixedonebit} \\
\tilde{E}_1 &=& X \\ \tilde{E}_2 &=& -i Y \\ \tilde{E}_3 &=& Z.
\label{eq:fixedonebitend}
\eeqn
There are 12 parameters, specified by $\chi$, which determine an
arbitrary single qubit quantum operation ${\cal E}$.
%; three of these
%describe arbitrary unitary transforms $\exp(i\sum_k r_k\sigma_k)$ on
%the qubit, and nine parameters describe possible correlations
%established with the environment $E$ via $\exp(i\sum_{jk} \gamma_{jk}
%\sigma_j\otimes\sigma^E_k)$.  Two combinations of the nine parameters
%describe physical processes analogous to the $T_1$ and
%%
%% 12-Apr-97 ILC: NEXT LINE CHANGED
%%
%$T_2$ spin-lattice and spin-spin relaxation rates familiar to us from
%%
%classical magnetic spin systems.  However, the dephasing and energy loss rates
%determined by $\chi$ do not simply describe ensemble behavior; rather, $\chi$
%describes the dynamics of a {\em single quantum system}.  Thus, the
%decoherence of a single qubit must be described by {\em more than just two
%parameters}.  {\em Twelve} are needed in general.
These 12 parameters may be measured using four sets of experiments.  As a
specific example, suppose the input states $\ket{0}$, $\ket{1}$,
$\ket{+}=(\ket{0}+\ket{1})/\sqrt{2}$ and $\ket{-} =
(\ket{0}+i\,\ket{1})/\sqrt{2}$ are prepared, and the four matrices
\beqn
	\rho'_1 &=& {\cal E}(\ket{0}\bra{0}) \label{eq:rone}
\\	\rho'_4 &=& {\cal E}(\ket{1}\bra{1})
\\	\rho'_2 &=& {\cal E}(\ket{+}\bra{+}) 
			- i {\cal E}(\ket{-}\bra{-})
			- (1-i)(\rho'_1 + \rho'_4)/2
\\	\rho'_3 &=& {\cal E}(\ket{+}\bra{+}) 
			+ i {\cal E}(\ket{-}\bra{-})
			- (1+i)(\rho'_1 + \rho'_4)/2
\label{eq:rthree}
\eeqn
are determined using state tomography.  These correspond to $\rho'_j = {\cal
E}(\rho_j)$, where
\beqn
	\rho_1 = \left[ \begin{array}{cc} 1 & 0 \\ 0 & 0 \end{array} \right],
\eeqn
$\rho_2 = \rho_1 X$, $\rho_3=X\rho_1$, and $\rho_4 = X
\rho_1X$.  From Eq.(\ref{eq:betadef}) and
Eqs.(\ref{eq:fixedonebit}-\ref{eq:fixedonebitend}) we may determine $\beta$,
and similarly $\rho'_j$ determines $\lambda$.  However, due to the particular
choice of basis, and the Pauli matrix representation of $\tilde{E}_i$, we may
express the $\beta$ matrix as the Kronecker product $\beta = \Lambda\otimes
\Lambda$, where
\beqn
\Lambda = \frac{1}{2} \left[ \begin{array}{cc} I & X \\
	X & -I \end{array} \right],
\label{eq:lambdaone}
\eeqn
so that $\chi$ may be expressed conveniently as
\be
	\chi = \Lambda \left[ \begin{array}{cc} \rho'_1 & {\rho'_2} \\
	{\rho'_3} & {\rho'_4} \end{array} \right]  \Lambda
\,,
\label{eq:chione}
\ee
in terms of block matrices.

Consider a one-qubit black box of unknown dynamics ${\cal E}_1$.  Suppose that
the following four density matrices are obtained from experimental
measurements, performed according to Eqs.(\ref{eq:rone}-\ref{eq:rthree}):
\beqn
	\rho_1' &=& \mattwoc{1}{0}{0}{0}
\\	\rho_2' &=& \mattwoc{0}{\sqrt{1-\gamma}}{0}{0}
\\	\rho_3' &=& \mattwoc{0}{0}{\sqrt{1-\gamma}}{0}
\\	\rho_4' &=& \mattwoc{\gamma}{0}{0}{1-\gamma},
\eeqn
where $\gamma$ is a numerical parameter.  From a independent study of each of
these input-output relations, one could make several important observations:
the ground state $|0\ra$ is left invariant by ${\cal E}_1$, the excited state
$|1\ra$ partially decays to the ground state, and superposition states are
damped.  
%This would suggest that both $T_1$ and $T_2$ processes are at work in
%this black box.

However, let us proceed systematically and determine $\chi$ using this data.
From Eqs.(\ref{eq:lambdaone}-\ref{eq:chione}), we find the $\chi$ matrix for
this process to be
\beqn
	\chi = \frac{1}{4} \left[\begin{array}{cccc}
		(1+\sqrt{1-\gamma})^2 & 0 & 0 & \gamma
	\\	0 & \gamma & -\gamma & 0
	\\	0 & -\gamma & \gamma & 0
	\\	\gamma & 0 & 0 & (1-\sqrt{1-\gamma})^2
	\end{array}\right]
\eeqn
Using Eqs.(\ref{eq:chipolar}-\ref{eq:chiopsum}), we then obtain (after
a little simplification) the operators $E_i$ which generate the
operator-sum representation for this quantum operation,
\beqn
	E_0 &=& \mattwoc{1}{0}{0}{\sqrt{1-\gamma}}
\\	E_1 &=& \mattwoc{0}{\sqrt{\gamma}}{0}{0}.
\eeqn \index{amplitude damping}
These operators define a well-known process called {\em amplitude
damping}.  It can result from a relaxation process with a microscopic
interaction Hamiltonian of the form ${\cal H}_I = \gamma'(\sigma^-
b^\dagger + \sigma^+ b)$, where $\sigma^+$ and $b^\dagger$ are system
and environment creation operators, and $\gamma$ is related to
$\gamma'$ and the interaction time.  
%This description of ${\cal E}_1$
%captures the fact that for a qubit, relaxation is a process which
%cannot be described as a combination of independent $T_1$ or $T_2$
%process.
This process is important, for instance, in quantum error correction,
where one wishes to reverse the effects of noise, because better codes
exist to correct amplitude damping than for general error
processes\cite{Leung97a}.

The dynamics of a two-qubit quantum black box ${\cal E}_2$ pose an
even greater challenge for our understanding.  In this case there are
240 parameters which need to be determined in order to do completely
specify the quantum operation acting on the quantum system! This is
obviously quite a considerable undertaking, however, as for the single
qubit case, it is relatively straightforward to implement a numerical
routine which will automate the calculation, provided experimental
tomography and state preparation are available in the laboratory. We
will not give an example here, as it does not serve the purpose of the
present Chapter, referring the reader instead to \cite{Chuang97a} for
more details.

Until now we have been considering complete quantum operations. In a
situation where quantum measurements may be involved, the
corresponding quantum operations may be incomplete. We now briefly
outline how to determine the quantum operation corresponding to each
measurement outcome in this instance.

Recall that for each measurement outcome, $m$, there is associated a
quantum operation, ${\cal E}_m$. The corresponding state change is
given by
\begin{eqnarray}
	\rho \rightarrow \frac{{\cal E}_m(\rho)}{\mbox{tr}({\cal E}_m(\rho))}
\,,
\end{eqnarray}
where the probability of the measurement outcome occurring is $p_m =
\mbox{tr}({\cal E}_m(\rho))$.  Note that this mapping is nonlinear,
because of this renormalization factor, so the earlier methods do not
apply.

\index{process tomography!for incomplete quantum operations}
\index{quantum process tomography!for incomplete quantum operations}

Despite the possible nonlinearity, the procedure we have described may
be adapted in a straightforward manner to evaluate the quantum
operations describing a measurement.  To determine ${\cal E}_m$ we
proceed exactly as before, except now we must perform the measurement
a large enough number of times that the probability $p_m$ can be
reliably estimated, for example, by using the frequency of occurrence
of outcome $m$. This must be done for each input $\rho_j$ which is to
be used for the tomography procedure.  Note that standard statistical
tools may be used to estimate the accuracy with which the probability
$p_m$ has been determined.  Once the $p_m$ have all been estimated to
some desired accuracy, $\rho'_j$ is determined using state tomography,
allowing us to obtain
\begin{eqnarray}
	{\cal E}_m(\rho_j) = p_m \rho'_j,
\end{eqnarray}
for each input $\rho_j$ which we prepare, since each term on the right hand
side is known. Now we proceed exactly as before to evaluate the quantum
operation ${\cal E}_i$.  

Summing up, we have shown how a useful representation for the dynamics
of a quantum system may be experimentally determined using a
systematic procedure. This procedure of {\em quantum process
tomography} is analogous to the {\em system identification} step
\cite{Ljung87a} performed in classical control theory. Quantum process
tomography opens the way for robust experimental determination of a
wide variety of interesting quantities associated to noisy quantum
processes. As such, I expect it will eventually become an
indispensable tool in the experimental study of quantum information
processing.

\section{The POVM formalism}
\index{POVMs}
\index{generalized measurements}
\index{positive operator valued measures}

One of the main uses of the quantum operations formalism is to
describe the effects of measurement. Quantum operations can be used to
describe both the probability of getting a particular outcome from a
measurement on a quantum system, and also the state change in the
system effected by the measurement.

In many cases, though, the state change in the system being measured
is not particularly interesting, since the system itself is discarded
after the measurement is performed. For example, this is the case for
photons detected by a photodetector, which destroys the photon.

What is still interesting in these examples is the probabilities of
different measurement outcomes. It turns out that the quantum
operations formalism simplifies rather nicely if one is only
interested in the probabilities of different measurement outcomes, and
not also the corresponding state changes. This simplified formalism
has become known for historical reasons as the {\em Positive Operator
Valued Measure} formalism, or POVM formalism for short.

You may ask why we should bother studying a formalism which is a
special case of a more general formalism. The reason is that it
sometimes simplify matters to consider a problem from the point of
view of POVMs. New sources of intuition in quantum information are
to be valued, and the simplicity of the POVM formalism is one such
source of intuition.

Suppose we consider a set of Hermitian operators $M_{m}$ indexed by an
index which we denote $m$, satisfying the conditions
\beqn \label{eqtn: POVM 1}
M_{m} & \geq & 0 \\ \label{eqtn: POVM 2}
\sum_{m} M_{m} & = & I. \eeqn
Consider now a measurement described by quantum operations $\evop_m$
defined by the equations $\evop_m(\rho) = \sqrt{M_m} \rho
\sqrt{M_m}$. Notice that the quantum operation $\sum_m \evop_m$ is a
complete quantum operation by equation (\ref{eqtn: POVM 2}), and that
the probability of outcome $m$ occurring is given by
\beqn \label{eqtn: POVM 3}
p(m) = \tr(\sqrt{M_m} \rho \sqrt{M_m} ) = \tr(M_{m} \rho).
\eeqn
Thus, given a set of operators $M_{m}$ satisfying the conditions
(\ref{eqtn: POVM 1}) and (\ref{eqtn: POVM 2}), it is possible to find
a measurement model such that equation (\ref{eqtn: POVM 3}) correctly
gives the probability of the measurement outcome $m$.

Conversely, suppose a measurement is taking place, which is described
by quantum operations $\evop_{m}$ associated to the measurement
outcomes $m$. Let $E_{mi}$ be a set of operators generating the
quantum operation $\evop_m$. Define $M_m \equiv \sum_i
E_{mi}^{\dagger} E_{mi}$. Note that $M_m \geq 0$, and
\be
\tr(\evop_m(\rho)) & = & \sum_i \tr(E_{mi} \rho E_{mi}^{\dagger}) \\
& = & \tr(M_m \rho), \ee
so $\tr(M_m \rho)$ gives the probability of outcome $m$ occurring in
the measurement. The completeness relation $\sum_{mi} E_{mi}^{\dagger}
E_{mi} = I$ is true if and only if $\sum_m M_m = I$.

These two results suggest the following formal definition.  A {\em
POVM} consists of a set of operators $M_{m}$ satisfying the two
conditions:
\begin{enumerate}
\item ({\em Positivity})
\beqn
M_{m} \geq 0. \eeqn

\item ({\em Completeness})
\beqn
\sum_{m} M_{m} = I. \eeqn
\end{enumerate}
A POVM describes the probabilities of the measurement outcomes via the
rule
\begin{eqnarray}
p(m) = \tr(M_{m} \rho). \end{eqnarray} These three equations -- the
positivity requirement, completeness, and the probability rule --
completely summarize the POVM formalism. Our results imply that any
description of a quantum measurement in terms of quantum operations
gives rise to a unique POVM describing the measurement statistics for
that measurement. We have also shown that given any POVM, there exists
a measurement model whose statistics agree with those predicted by the
POVM.

\section{Beyond quantum operations?}

\index{quantum operations!limitations to the formalism}

Are there interesting quantum systems whose dynamics are not described
by quantum operations? In this section we give a very brief discussion
of this question. A more detailed discussion of some of these issues
has been provided by Royer \cite{Royer96a}. In this section we will
construct an artificial example of a system whose evolution is not
described by a quantum operation, and try to understand the
circumstances under which this is likely to occur.

Suppose a single qubit is prepared in some unknown quantum state,
which we denote $\rho$. The preparation of this qubit involves certain
procedures to be carried out in the laboratory in which the qubit is
prepared.  Suppose that amongst the laboratory degrees of freedom is a
single qubit which, as a side effect of the state preparation
procedure, is left in the state $|0\rangle$ if $\rho$ is a state on
the bottom half of the Bloch sphere, and is left in the state
$|1\rangle$ if $\rho$ is a state on the top half of the Bloch
sphere. That is, the state of the system after preparation is
\beqn
\rho \otimes |0\ra \la 0| \otimes \mbox{other degrees of freedom} \eeqn
if $\rho$ is a state on the bottom half of the Bloch sphere, and
\beqn
\rho \otimes |1\ra \la 1| \otimes \mbox{other degrees of freedom} \eeqn
if $\rho$ is a state on the top half of the Bloch sphere.

Once the state preparation is done, the system begins to interact with
the environment, in this case all the laboratory degrees of
freedom. Suppose the interaction is such that a controlled not is
performed between the principal system and the extra qubit in the
laboratory system. Thus, if the system's Bloch vector was initially in
the bottom half of the Bloch sphere it is left invariant by the
process, while if it was initially in the top half of the Bloch sphere
it is rotated into the bottom half of the Bloch sphere.

Obviously, this process is not an affine map acting on the Bloch
sphere, and therefore, by the results of subsection
\ref{subsect:qops_qubit}, it {\em can not be a quantum operation}. The
lesson to be learned from this discussion is that {\em a quantum
system which interacts with the degrees of freedom used to prepare
that system after the preparation is complete will in general suffer a
dynamics which is not adequately described within the quantum
operations formalism.} This is an important conclusion to have
reached, as it indicates that there are physically reasonable
circumstances under which the quantum operations formalism may not
adequately describe the processes taking place in a quantum
system. This should be kept in mind, for example, in applications of
the quantum process tomography procedure discussed in the previous
section.

For the remainder of this Dissertation we will, however, work within
the quantum operations formalism. It provides a powerful, and
reasonably general tool for describing the dynamics experienced by
quantum systems. Most of all, it provides a means by which concrete
progress can be made on problems related to quantum information
processing. It is an interesting problem for further research to study
quantum information processing beyond the quantum operations
formalism.

\newpage
\vspace{1cm}
\begin{center}
\fbox{\parbox{14cm}{
\begin{center} {\bf Summary of Chapter \ref{chap:qops}:
Quantum operations} \end{center}

\begin{itemize}

\item {\bf Axioms for complete quantum operations:} Linear maps on
density operators which preserve trace, and preserve positivity of
density operators, even when extended in a natural way to larger
systems.

\item {\bf Operator-sum representation for a quantum operation:}
$$
{\cal E}(\rho) = \sum_i E_i \rho E_i^{\dagger}. $$
The quantum operations generated by operators $E_i$ and $F_j$ in the
operator-sum representation are the same if and only if there exists a
unitary matrix $u_{ij}$ such that $E_i = \sum_j u_{ij} F_j$. It may be
necessary to append $0$ operators so that both sets of operators have
the same number of elements.

\item {\bf Environmental models for quantum operations:} A complete 
quantum operation can always be regarded as arising from the unitary
interaction of a system with an initially uncorrelated environment,
and vice versa. Incomplete quantum operations may be treated
similarly, except an additional projective measurement is performed on
the composite of system and environment, with the different outcomes
corresponding to different incomplete quantum operations.

\item {\bf Quantum teleportation:} The input and output states to the
quantum teleportation procedure are related by a set of quantum
operations. The problem of teleportation is to reverse or error
correct those quantum operations.

\item {\bf Quantum process tomography:} A procedure used to completely
characterize the dynamics of a quantum system in the laboratory.

\end{itemize}

}}
\end{center}

\chapter{Entropy and information}
\label{chap:entropy}

{\em Entropy} is a key concept of quantum information theory. It
measures how much uncertainty there is in the state of a physical
system. In this Chapter we review the basic definitions and properties
of entropy in both classical and quantum information theory.  In
places the Chapter contains rather detailed and lengthy mathematical
arguments; upon a first read, these sections may be read lightly, and
returned to later for reference purposes.

%This
%Chapter is meant as an introduction and reference Chapter for the
%reader, to be referred to again in later Chapters. At a first read it
%may be scanned fairly rapidly.

%%%%%%%%%%%%%%%%%%%%%%%%%%%%%%%%%%%%%%%%%%%%%%%%%%%%%%%%%%%%%%%%%%%%%%%%%%%%%
\section{Shannon entropy}
\index{Shannon entropy} \index{entropy!classical}

The key concept of classical information theory is the {\em Shannon
entropy}. Suppose we learn the value of a random variable, $X$. The
Shannon entropy associated with $X$ quantifies how much information we
gain, on average, when we learn the value of $X$. An alternative view
is that the entropy of $X$ measures the amount of {\em uncertainty}
about $X$ before we learn the value of $X$.  These two views are
complementary: we can view the entropy either as a measure of
uncertainty {\em before} we learn the value of $X$, or as a measure of
how much information we have gained {\em after} we learn the value of $X$.

The entropy of a random variable is completely determined by the
probabilities of the different possible values that random variable
takes. For that reason, we will often write the entropy as a function
of a probability distribution, $p_1,\ldots,p_n$. The {\em Shannon
entropy} associated with that probability distribution is defined by
\beqn
H(X) \equiv H(p_1,\ldots,p_n) \equiv -\sum_i p_i \log p_i. 
\eeqn
We will justify this definition shortly.  Note that in the definition
-- and throughout the Dissertation, unless otherwise noted --
logarithms indicated by $\log$ are taken to base two, while $\ln$
indicates a natural logarithm. The reader may wonder what happens when
$p_i = 0$, since $\log 0$ is undefined. Intuitively, an event which
can never occur should not contribute to the entropy, so by convention
we agree that $0 \log 0 \equiv 0$. More formally, note that $\lim_{x
\rightarrow 0} x \log x = 0$, which provides further support for our
intuition, and thus our convention.

Why is the entropy defined in this way? 
%Later in this section there is an
%exercise for the reader which gives an intuitive justification for this
%definition of the entropy, based upon certain ``reasonable'' axioms which you
%might expect a measure of information to have. This intuitive justification is
%reassuring, but it is not the whole story.
In the pedagogical literature, it is common to give an axiomatic
characterization of the entropy, based upon certain intuitive
properties we would expect a measure of information to possess (see,
for example \cite{Cover91a,Csiszar81a} for excellent pedagogical
introductions to information theory which contain such
characterizations). These axioms are then used to deduce the above
formula for entropy. While appealing, there is a better reason than
axiomatics for choosing this definition for entropy. The better reason
for this definition of entropy is that it can be used to quantify the
resources needed to store information.  More concretely, suppose there
is some {\em source} (perhaps a radio antenna) which is producing
information, say in the form of a bit string. Let's consider a very
simple model for a source: we describe it as producing a string
$X_1,X_2,\ldots$ of independent, identically distributed random
variables.  Most real sources don't behave quite that way, but often
it's a good approximation. Shannon asked what minimal physical
resources are required to store the information being produced by the
source, in such a way that at a later time the original source
information can be reconstructed \cite{Shannon48a,Shannon49a}? The
answer to this question turns out to be the entropy, that is, $H(X_1)$
bits are required per source symbol. This result is known as {\em
Shannon's noiseless coding theorem}, and we will prove both classical
and quantum versions of it in Chapter
\ref{chap:data_compress}. \index{Shannon's noiseless coding theorem}
\index{noiseless coding theorem}

\index{information theory!motivation for definitions}
More abstractly, this motivation for the definition of entropy expresses one
of the key philosophies of information theory, both quantum and classical:
{\em fundamental measures of information arise as the answers to fundamental
questions about the quantity of physical resources required to solve some
information processing problem}.

\section{Basic properties of entropy}

\subsection{The binary entropy}
\index{binary entropy}

%Consider a random process which outputs a sequence $X_i$ of independent random
%variables each taking the values $0$ or $1$, with respective probabilities
%$1/8$ and $7/8$.  The entropy of this process is
%\be
%	H(X_i) = -\frac{1}{8} \log \frac{1}{8} - \frac{7}{8} \log\frac{7}{8}
%%
%	\approx 0.544.
%\ee
%According to Shannon's noiseless coding theorem, a sequence of $1000$ bits from
%this source could be faithfully stored by a sequence of approximately
%$544$ bits.  This is intuitively reasonable, since the rare zeros could be
%encoded by something like $00$ and strings of multiple one's, which are
%likely, are encoded as $1n00$, where $n$ is the length of the string, properly
%encoded so that no $00$'s appear in its representation.

\begin{figure}[ht]
%\begin{center}
%\vspace{2in}
\begin{center}
\scalebox{0.4}{\includegraphics{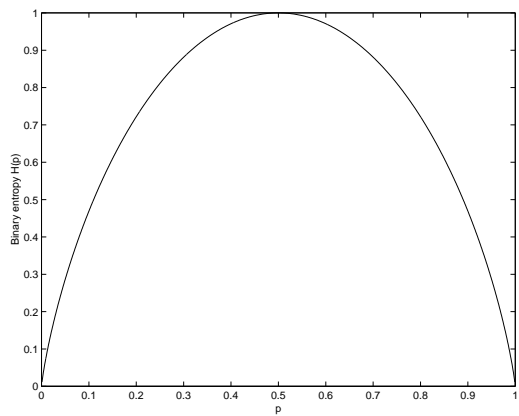}}
\end{center}
%
%	\mbox{\psfig{file=4/hbin-fig.epsf,scale=0.7}}
%\end{center}
\caption{Binary entropy function $H(p)$.}
\label{fig:hbin}
\end{figure}

The entropy of a two outcome random variable is so useful that we will give it
a special name, the {\em binary entropy function}, defined as
\be
	H_{\mbox{bin}}(p) = -p \log p - (1-p) \log (1-p)
\,,
\ee
where $p$ and $1-p$ are the probabilities of the two outcomes. Where context
makes the meaning clear we will write $H(p)$ rather than
$H_{\mbox{bin}}(p)$. Note again that logarithms will be taken to be base two,
unless otherwise stated.  The binary entropy function is plotted in
figure~\ref{fig:hbin}. Notice that $H(p) = H(1-p)$ and that $H(p)$ attains its
maximum value of $1$ at $p = 1/2$.

\subsection{The relative entropy}
\index{relative entropy!classical}

There is a very useful entropy-like measure of the closeness of two
probability distributions, $p(x)$ and $q(x)$, over the same index
set, $x$.  Suppose $p(x)$ and $q(x)$ are two probability
distributions on same index set, $x$. Define the {\em relative entropy}
of $p(x)$ to $q(x)$ by
\beqn
H(p(x) \| q(x)) \equiv \sum_x p(x) \log \frac{p(x)}{q(x)} \equiv
	-H(X) - \sum_x p(x) \log q(x). \eeqn
We define $-0 \log 0 \equiv 0$ and $-p(x) \log 0 \equiv \infty$ if
$p(x) > 0$. 

It is probably not immediately obvious what the relative entropy is good for,
or even why it is a good measure of distance between two distributions. The
following theorem gives some motivation for why it is regarded as being like a
distance measure.

\begin{theorem} \cite{Cover91a}
The relative entropy is non-negative, $H(p(x)\| q(x)) \geq 0$, with equality
if and only if $p(x) = q(x)$ for all $x$.
\end{theorem}

As exemplified here, many of the references for elementary results in
this Chapter will be to the excellent text of Cover and Thomas
\cite{Cover91a} or the review paper of Wehrl \cite{Wehrl78a}, to which
you may refer for historical details.

\begin{proof}

A very useful inequality in information theory is $\log x \ln 2 = \ln
x \leq x -1$, for all positive $x$, with equality if and only if $x =
1$. Here we need to rearrange the result slightly, to $-\log x \ln 2
\geq 1-x$, and then note that
\beqn
H(p(x)\|q(x)) & = & -\sum_x p(x) \log \frac{q(x)}{p(x)} \\
	& \geq & \frac{1}{\ln 2} \sum_x p(x) \left( \frac{q(x)}{p(x)} - 1 \right) \\
	& = & \frac{1}{\ln 2} \sum_x \left( q(x)-p(x) \right) \\
	& = & \frac{1}{\ln 2} (1-1) = 0, \eeqn
which is the desired inequality. The equality conditions are easily deduced by
noting that equality occurs in the second line if and only if $q(x)/p(x) = 1$
for all $x$, that is, the distributions are identical.

\end{proof}

The relative entropy is usually useful, not in itself, but because
other entropic quantities can be regarded as special cases of the
relative entropy. Theorems about the relative entropy then give as
special cases theorems about other entropic quantities. For example,
we can use the non-negativity of the relative entropy to prove the
following fundamental fact about entropies. Suppose $p(x)$ is a
probability distribution for $X$, over $d$ outcomes. Let $q(x) \equiv
1/d$ be the uniform probability distribution over those outcomes. Then
\beqn
H(p(x) \| q(x)) = H(p(x) \| 1/d) = -H(X) - \sum_x p(x) \log (1/d) = \log d -
H(X). \eeqn
From the non-negativity of the relative entropy we see that $\log d - H(X)
\geq 0$, with equality if and only if $X$ is uniformly distributed. This is an
elementary fact, but so important that we restate it formally as a theorem.

\begin{theorem} \cite{Cover91a}
Suppose $p(x)$ is a probability distribution for $X$, on $d$
outcomes. Then $H(X) \leq \log d$, with equality if and only if $p(x)$
is uniformly distributed.
\end{theorem}

We will use this technique -- finding expressions for entropic
quantities in terms of the relative entropy -- often in the study of
both classical and quantum entropies. As another example, it is easily
verified that $H(p(x,y) \| p(x)p(y)) = H(p(x))+H(p(y))-H(p(x,y))$.
From this observation and the non-negativity of the relative entropy,
we see that $H(X,Y) \leq H(X)+H(Y)$, with equality if and only if $X$
and $Y$ are independent random variables.

\subsection{Mutual information and conditional entropy}
\label{subsec:mutual_conditional}
\index{mutual information!classical}
\index{conditional entropy!classical}

Suppose $X$ and $Y$ are two random variables. How is the information
content of $X$ related to the information content of $Y$? In this
subsection we introduce two concepts -- the {\em conditional entropy}
and {\em the mutual information} -- which help answer this
question. The definitions of these concepts which we give are rather
formal, and at times the reader may be confused as to why a particular
quantity -- say, the conditional entropy -- is to be interpreted in
the way we indicate. Keep in mind that the ultimate justification for
these definitions is that they answer resource questions, which will
become clearer in later Chapters. The interpretation given to the
quantities depends on the nature of the resource question being
answered.

We already met the {\em joint entropy} of a pair of random variables
implicitly in the last subsection. For clarity, we now make this
definition formal.  The {\em joint entropy} of $X$ and $Y$ is defined
in the obvious way,
\beqn
	H(X,Y) \equiv -\sum_{x,y} p(x,y) \log p(x,y),
\eeqn
and may be extended in the obvious way to any vector of random
variables.  The joint entropy measures our total uncertainty about the
pair $(X,Y)$.  Suppose we know the value of $Y$, so we have acquired
$H(Y)$ bits of information about the pair, $(X,Y)$. The remaining
uncertainty about the pair $(X,Y)$, is associated with our remaining
lack of knowledge about $X$, even given that we know $Y$.  The {\em
entropy of $X$ conditional on knowing $Y$} is therefore defined by
\beqn
	H(X|Y) \equiv H(X,Y)-H(Y).
\eeqn
The conditional entropy is a measure of how uncertain we are, on
average, about the value of $X$, given that we know the value of
$Y$. 

A second quantity, the {\em mutual information content of $X$ and
$Y$}, measures how much information $X$ and $Y$ have in
common. Suppose we add the information content of $X$, $H(X)$, to the
information content of $Y$. Then all the information in the pair
$(X,Y)$ will have been counted at least once in the sum. Information
which is common to $X$ and $Y$ will have been counted twice in this
sum, while information which is not common will have been counted only
once. Subtracting off the joint information of $(X,Y)$, $H(X,Y)$, we
obtain the common or {\em mutual information} of $X$ and $Y$:
\beqn
	H(X:Y) \equiv H(X)+H(Y)-H(X,Y). 
\eeqn
Notice the useful equality $H(X:Y) = H(X)-H(X|Y)$ relating the
conditional entropy and mutual information.

To get some feeling for how the Shannon entropy behaves, we will
prove some simple relationships between the different entropies.
\begin{theorem}{} \textbf{(Basic properties of entropy)} \cite{Cover91a}
\label{thm:basic_entropy}

\begin{enumerate}

\item $H(X,Y) = H(Y,X)$, $H(X:Y) = H(Y:X)$.
\item $H(Y|X) \geq 0$ and thus $H(X:Y) \leq H(Y)$, with equality if and only
if $Y$ is a function of $X$, $Y = f(X)$.
\item $H(X) \leq H(X,Y)$, with equality if and only if $Y$ is a function of
$X$.
\item {\bf Subadditivity:} $H(X,Y) \leq H(X) + H(Y)$ with equality if and
only if $X$ and $Y$ are independent random
variables. \index{subadditivity!classical} 
\item $H(Y|X) \leq H(Y)$ and thus $H(X:Y) \geq 0$, with equality in each
 if and only if $X$ and $Y$ are independent random variables.
\item {\bf Strong subadditivity:} $H(X,Y,Z)+H(Y) \leq H(X,Y) +
H(Y,Z)$. \index{strong subadditivity!classical}
\end{enumerate}
\end{theorem}

\begin{proof}

\begin{enumerate} 

\item Obvious from the relevant definitions.
\item Since $p(x,y) = p(x)p(y|x)$ we have
\beqn 
H(X,Y) & = & -\sum_{xy} p(x,y) \log p(x)p(y|x) \\
 & = & -\sum_x p(x) \log p(x) -\sum_{xy} p(x,y) \log p(y|x) \\
	 & = & H(X)-\sum_{xy} p(x,y) \log p(y|x). 
\eeqn
Thus $H(Y|X) = -\sum_{xy} p(x,y) \log p(y|x)$. But $-\log p(y|x) \geq 0$, so
$H(Y|X) \geq 0$ with equality if and only if $Y$ is a deterministic function
of $X$.
\item Follows from the previous result.
\item To prove subadditivity and, later, strong subadditivity we use the
fact that $\ln x \leq x-1$ for all positive $x$, with equality if and
only if $x = 1$. This fact is easily proved using calculus. We find
that
\beqn
\sum_{x,y} p(x,y) \ln \frac{p(x) p(y)}{p(x,y)} & \leq &
	\sum_{x,y} p(x,y) \left( \frac{p(x) p(y)}{p(x,y)}-1 \right) \\
	& = & \sum_{x,y} p(x)p(y) - p(x,y) = 1-1 = 0.\eeqn
Subadditivity may easily be recovered by multiplying by a constant
(to change the base of the logarithm to base $2$), and rearranging the
expression. Notice that equality is achieved if and only if
$p(x,y)=p(x)p(y)$ for all $x$ and $y$. That is, the subadditivity
inequality is saturated if and only if $X$ and $Y$ are independent.
\item Follows from subadditivity and the relevant definitions.
\item Strong subadditivity of Shannon entropy follows from the same
	technique as used to prove subadditivity; the difficulty level
	is about the same as that proof. Interestingly, while carrying
	out the proof one notes that the equality conditions for
	strong subadditivity are that $Z \rightarrow Y \rightarrow X$
	forms a Markov chain.
\end{enumerate}

\end{proof}

\begin{figure}[htbp]
\begin{center}
\mbox{\psfig{file=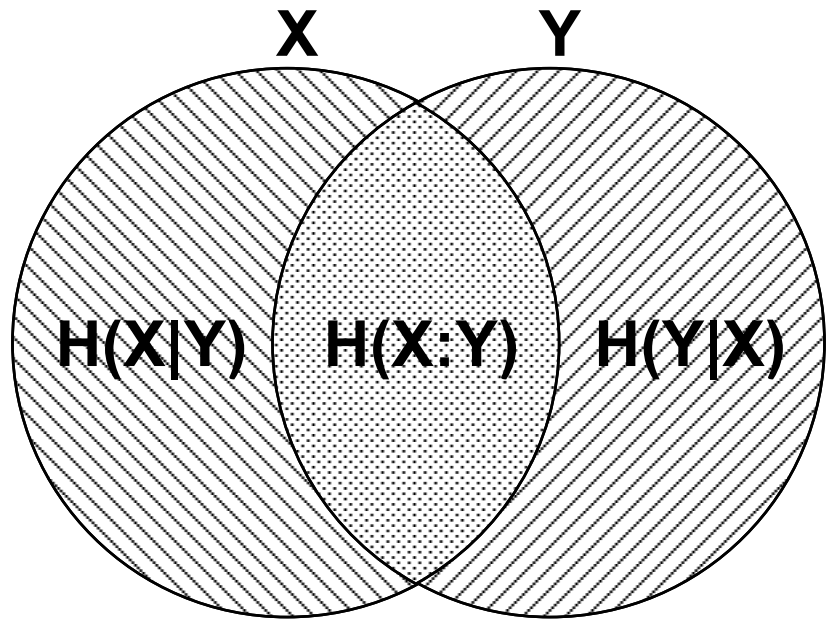,scale=1}}
\end{center}
\caption{Relationships between different entropies. \label{fig: venn}}
\end{figure}

The various relationships between entropies may mostly be deduced from the
``entropy Venn diagram'' shown in figure \ref{fig: venn}. These figures
are not completely reliable as a guide to the properties of entropy, but
they are a useful mnemonic for remembering the various definitions and
properties of entropy.

Intuitively, we expect that the uncertainty about $X$, given that we
know the value of $Y$ and $Z$, is less than our uncertainty about $X$,
given that we only know $Y$. More formally,

\begin{theorem}{}
\textbf{(Conditioning reduces entropy)} \cite{Cover91a}
\beqn
H(X|Y,Z) \leq H(X|Y). \eeqn

\end{theorem}

\begin{proof}

Inserting the relevant definitions, the result is equivalent to
\beqn
H(X,Y,Z)-H(Y,Z) \leq H(X,Y)-H(Y), \eeqn
which is a rearranged version of the strong subadditivity inequality proved
earlier.

\end{proof}

The next result gives a simple, useful formula for the conditional entropy.

\begin{theorem}{}
\textbf{(Chaining for conditional entropies)} \cite{Cover91a}
\label{thm:chain_rule_entropies}

Let $X_1,\ldots,X_n$ and $Y$ be any set of random variables. Then
\beqn
H(X_1,\ldots,X_n |Y) = \sum_{i=1}^n H(X_i | Y,X_1,\ldots,X_{i-1}). \eeqn

\end{theorem}

\begin{proof}

We prove the result for $n = 2$, and then induct on $n$. Using only the
definitions and some simple algebra we have
\beqn
H(X_1,X_2|Y) & = & H(X_1,X_2,Y)-H(Y) \\
	& = & H(X_1,X_2,Y)-H(X_1,Y)+H(X_1,Y)-H(Y) \\
	& = & H(X_2|Y,X_1)+H(X_1|Y), \eeqn
which establishes the result for $n = 2$. Now we assume the result for
general $n$, and show the result holds for $n+1$. Using the already
established $n=2$ case, we have
\beqn
H(X_1,\ldots,X_{n+1} | Y) & = & H(X_2,\ldots,X_n | Y,X_1)+
	H(X_1|Y). \eeqn
Applying the inductive hypothesis to the first term on the right hand
side gives
\beqn
H(X_1,\ldots,X_{n+1} | Y) & = & \sum_{i=2}^{n+1} H(X_i | Y,X_1,\ldots,X_{i-1})+
	H(X_1|Y) \\
	& = & \sum_{i=1}^{n+1} H(X_i |Y,X_1,\ldots,X_{i-1}), \eeqn
so the induction goes through.

\end{proof}

\label{note:subadd}

\index{subadditivity!of mutual information}
Finally, we conclude with a note that will be of interest in Chapter
\ref{chap:capacity} on the quantum channel capacity. In that Chapter
we will be much interested in the subadditivity properties of
quantities like the mutual information. We now note that the mutual
information is not generally subadditive in either or both
entries. For instance, let $X$ and $Y$ be independent identically
distributed random variables taking the values $0$ or $1$ with
probability $1/2$. Let $Z \equiv X+Y$, where the addition is done
modulo two. Then it is easy to see that
\beqn
1 = H(X,Y:Z) \not \leq H(X:Z)+H(Y:Z) = 0+0. \eeqn 
Neither is the mutual information superadditive. For example, suppose
$X_1 = X_2 = Y_1 = Y_2$, and $X_1$ is chosen to have the values $0$ or
$1$ with respective probabilities of one half. Then
\beqn
1 = H(X_1,X_2:Y_1,Y_2) < H(X_1:Y_1)+H(X_2:Y_2) = 1+1 = 2. \eeqn

\subsection{The data processing inequality}
\label{subsec:data_proc}
\index{data processing inequality!classical}

In many applications of interest we perform computations on the
information we have available, but that information is imperfect, as
it has been subjected to noise before it becomes available to us.  A
basic inequality of information theory, the {\em data processing
inequality}, states that the information we have available about a
source of information can only {\em decrease with time}: once
information has been lost, it is gone forever. Making this statement
more precise is the goal of this subsection.

The intuitive notion of {\em information processing} is captured in the idea
of a Markov chain of random variables. Formally, a {\em Markov chain} is a
sequence $X_1,X_2,\ldots$ of random variables such that $X_{n+1}$ is
independent of $X_1,\ldots,X_{n-1}$, given $X_n$. More formally,
\beqn
p(X_{n+1}=x_{n+1} | X_n=x_n,\ldots,X_1=x_1) = p(X_{n+1}=x_{n+1} |
		X_n = x_n). 
\eeqn
When is a Markov chain losing information? The following {\em data processing
inequality} gives an information-theoretic way of answering this question.

\begin{theorem} {\em (Data processing inequality)} \cite{Cover91a}

Suppose $X \rightarrow Y \rightarrow Z$ is a Markov chain. Then
\beqn
	H(X) \geq H(X:Y) \geq H(X:Z). 
\eeqn
Moreover, the first inequality is saturated if and only if, given $Y$, it is
possible to reconstruct $X$.

\end{theorem}

This result is intuitively plausible: it tells us that if a random variable
$X$ is subject to noise, producing $Y$, then further actions on our part
(``data processing'') cannot be used to increase the amount of mutual
information between the output of the process and the original information
$X$.

\begin{proof}

The first inequality was proved in theorem \ref{thm:basic_entropy} on
page \pageref{thm:basic_entropy}. From the definitions we see that
$H(X:Z) \leq H(X:Y)$ is equivalent to $H(X|Y) \leq H(X|Z)$. From the
fact that $X \rightarrow Y \rightarrow Z$ is a Markov chain it is easy
to prove that $Z \rightarrow Y \rightarrow X$ is also a Markov chain,
and thus $H(X|Y) = H(X|Y,Z)$. The problem is thus reduced to proving
that $H(X,Y,Z)-H(Y,Z) = H(X|Y,Z) \leq H(X|Z) = H(X,Z)-H(Z)$. This is
just the strong subadditivity inequality, which we already proved.

Suppose $H(X:Y) < H(X)$. Then it is not possible to reconstruct $X$ from $Y$,
since if $Z$ is the attempted reconstruction based only on knowledge of $Y$,
then $X\rightarrow Y\rightarrow Z$ must be a Markov process, and thus
$H(X)>H(X:Z)$ by the data processing inequality. Thus $X \neq Z$. On the other
hand, if $H(X:Y) = H(X)$, then we have $H(X|Y) = 0$ and thus whenever
$p(X=x,Y=y) > 0$ we have $p(X=x|Y=y) = 1$. That is, if $Y=y$ then we can
infer with certainty that $X$ was equal to $x$, allowing us to reconstruct
$X$. 

\end{proof}

From the definition of Markov chains, it is easy to verify that if $X
\rightarrow Y \rightarrow Z$ is a Markov chain, then so is $Z
\rightarrow Y \rightarrow X$. Thus, as a simple corollary to the data
processing inequality, we see that if $X \rightarrow Y \rightarrow Z$
is a Markov chain, then
\beqn
H(Z:Y) \geq H(Z:X). \eeqn \index{data pipelining inequality!classical}
We will refer to this result as the {\em
data pipelining inequality}. Intuitively, it says that any information
$Z$ share with $X$ must be information which $Z$ also shares with $Y$;
the information is ``pipelined'' from $X$ through $Y$ to $Z$.

%%%%%%%%%%%%%%%%%%%%%%%%%%%%%%%%%%%%%%%%%%%%%%%%%%%%%%%%%%%%%%%%%%%%%%%%%%%%%
\section{Von Neumann entropy}
\label{sec:vne}
\index{von Neumann entropy}
\index{entropy!quantum}

The Shannon entropy measures the uncertainty associated with a classical
probability distribution. Quantum states are described in a similar fashion,
with density operators replacing probability distributions. In this section we
generalize the definition of the Shannon entropy to quantum states.

Von Neumann defined the {\em entropy} of a quantum state $\rho$
by the formula
\beqn
	S(\rho) \equiv -\tr(\rho \log \rho). 
\eeqn
In this formula logarithms are taken to base two, and we define $0
\log 0$ to be equal to zero. If $\lambda_i$ are the eigenvalues of
$\rho$ then von Neumann's definition can be re-expressed
\beqn
S(\rho) = -\sum_i \lambda_i \log \lambda_i. 
\eeqn
For calculations it is usually this last formula which is most
useful. For instance, the completely mixed density operator in a $d$
dimensional space, $I/d$, has entropy $\log d$.

From now on, when we refer to entropy, it will usually be clear from
context whether we mean Shannon or von Neumann entropy. 

\subsection{Quantum relative entropy}
\index{relative entropy!quantum}

As for the Shannon entropy, it is extremely useful to define a quantum
version of the relative entropy.  Suppose $\rho$ and $\sigma$ are
density operators. The {\em relative entropy} of $\rho$ to $\sigma$ is
defined by
\beqn
S(\rho || \sigma) \equiv \tr(\rho \log \rho)-\tr(\rho \log \sigma ).
\eeqn
Conventionally, this is defined to be $+\infty$ if the kernel of
$\sigma$ has non-trivial intersection with the support of $\rho$, and
is finite otherwise. The quantum relative entropy is non-negative, a
result sometimes known as {\em Klein's inequality}.

\begin{theorem} {\em (Klein's inequality)} \cite{Wehrl78a}
\index{Klein's inequality}

The relative entropy is non-negative,
\beqn
	S(\rho || \sigma) \geq 0, 
\eeqn
with equality if and only if $\rho = \sigma$.
\end{theorem}

\begin{proof}

Let $\rho = \sum_i p_i |i\ra \la i|$ and $\sigma = \sum_j q_j |j\ra \la j|$ be
orthogonal decompositions for $\rho$ and $\sigma$. From the definition of the
relative entropy we have
\beqn
S(\rho || \sigma) = \sum_i p_i \log p_i - \sum_i \la i| \rho \log \sigma
	|i\ra. 
\eeqn
We substitute into this equation the equations $\la i|\rho = p_i \la i|$ and
\beqn
\la i| \log \sigma |i\ra = \la i| \left( \sum_j \log(q_j) |j\ra \la j| \right)
	|i\ra = \sum_j \log(q_j) P_{ij}, 
\eeqn
where $P_{ij} \equiv \la i|j\ra \la j| i\ra \geq 0$. Notice that
$P_{ij}$ satisfies the equations $\sum_i P_{ij} = 1$ and $\sum_j
P_{ij} = 1$ (such matrices are called {\em doubly
stochastic}). Substitution gives
\beqn
S(\rho||\sigma) & = & \sum_i p_i \left( \log p_i - \sum_j P_{ij} \log(q_j)
\right). \eeqn
$\log$ is a strictly concave function, so $\sum_j P_{ij} \log q_j \leq
\log r_i$, where $r_i \equiv \sum_j P_{ij} q_j$, with equality of and only
if there exists a value of $j$ for which $P_{ij} = 1$. Thus
\beqn
S(\rho||\sigma) & \geq & \sum_i p_i \log \frac{p_i}{r_i}, \eeqn
with equality if and only if $P_{ij}$ is a permutation matrix. This
has the form of the classical relative entropy, from which we deduce
that
\beqn
S(\rho||\sigma) & \geq & 0, \eeqn with equality if and only if $p_i =
r_i$ for all $i$, and $P_{ij}$ is a permutation matrix. To simplify
the equality conditions further, note that by relabeling the
eigenstates of $\sigma$ if necessary, we can assume that $P_{ij}$ is
the identity matrix, and thus that $\rho$ and $\sigma$ are diagonal in
the same basis. The condition $p_i = r_i$ tells us that the
corresponding eigenvalues of $\rho$ and $\sigma$ are identical, and
thus $\rho = \sigma$ are the equality conditions.

\end{proof}

\subsection{Basic properties of entropy}

The entropy has many interesting and useful properties.
\begin{theorem} {} \cite{Wehrl78a}

\begin{enumerate}
\item The entropy is non-negative. The entropy is zero if
and only if the state is pure.
\item In a $d$ dimensional Hilbert space the entropy
is at most $\log d$. The entropy is equal to $\log d$
if and only if the system is in the completely mixed state
$I/d$.
\item Suppose a composite system $AB$ is in a pure state. Then $S(A) = S(B)$.
\item Suppose $p_i$ are probabilities, and $\rho_i$ are states with
mutually disjoint support. Then
\beqn
S(\sum_i p_i \rho_i) = H(p_i)+ \sum_i p_i S(\rho_i). \eeqn
\item {\bf Joint entropy theorem}: \label{thm:joint_entropy}
\index{joint entropy theorem} Suppose $p_i$ are probabilities, $|i\ra$
are orthogonal states for a system $A$, and $\rho_i$ is any set of
density operators for another system, $B$. Then \beqn S(\sum_i p_i
|i\ra \la i| \otimes \rho_i) = H(p_i) + \sum_i p_i S(\rho_i), \eeqn
where $H(p_i)$ is the Shannon entropy of the distribution $p_i$.
\end{enumerate}	
\end{theorem}

\begin{proof}
\begin{enumerate}
\item Clear from the definition.
\item From the non-negativity of the relative entropy, $0 \leq S(\rho
|| I/d) = -S(\rho)+\log d$, from which the result follows.
\item From the Schmidt decomposition, as discussed in Appendix
\ref{app:mixed}, we know that the eigenvalues of systems
$A$ and system $B$ are the same. The entropy is determined completely
by the eigenvalues, so $S(A) = S(B)$.
\item
Let $\lambda_i^j$ and $|e_i^j\ra$ be the eigenvalues and corresponding
eigenvectors of $\rho_i$. Observe that $p_i \lambda_i^j$ and $|e_i^j\ra$
are the eigenvalues and eigenvectors of $\sum_i p_i \rho_i$, and thus
\beqn
S(\sum_i p_i \rho_i) & = & -\sum_{ij} p_i \lambda_i^j
	\log p_i \lambda_i^j \\
	& = & -\sum_i p_i \log p_i -\sum_i p_i \sum_j \lambda_i^j \log
	\lambda_i^j \\
		& = & H(p_i) + \sum_i p_i S(\rho_i), 
\eeqn
as required.
\item Immediate from the preceding result.
\end{enumerate}
\end{proof}

\index{mutual information!quantum}
\index{conditional entropy!quantum}
By analogy with the Shannon entropies it is possible to define
conditional and mutual von Neumann entropies. We make the definitions:
\beqn
	S(A|B) & \equiv & S(A,B)-S(B) \\
	S(A:B) & \equiv & S(A)+S(B)-S(A,B) \\
 & = & S(A)-S(A|B) = S(B)-S(B|A). 
\eeqn
Some properties of the Shannon entropy fail to hold for the von
Neumann entropy, and this has many interesting consequences for
quantum information theory. For instance, for random variables $X$ and
$Y$, the inequality $H(X) \leq H(X,Y)$ holds. This makes sense: surely
we cannot be more uncertain about the state of $X$ than we are about
the joint state of $X$ and $Y$. This intuition fails for quantum
states.  Consider a system $AB$ of two qubits in the entangled state
$(|00\ra+|11\ra)/\sqrt 2$. This is a pure state, so $S(A,B) = 0$. On
the other hand, system $A$ has density operator $I/2$, and thus has
entropy equal to one. Another way of stating this is that for this
system, the quantity $S(B|A) = S(A,B)-S(A)$ is negative.

Notice that this example involved entanglement. This is a generic
feature: differences between classical and quantum information seem
always to involve either or both of entanglement and the potential
non-orthogonality of quantum states. For example, in Chapter
\ref{chap:ent} we will prove that the negativity of the conditional
entropy always indicates that two systems are entangled, and, indeed,
how negative the conditional entropy is provides a lower bound on how
entangled the two systems are.

\subsection{Measurements and entropy}
\label{subsec:measurements_entropy}

How does the entropy of a quantum system behave when we perform a measurement
on that system? Not surprisingly, the answer to this question depends on the
type of measurement which we perform. Nevertheless, there are some
surprisingly general assertions we can make about how the entropy behaves.

Suppose for example, that an orthogonal measurement described by projectors
$P_i$ is performed on a quantum system, but we never learn the result of the
measurement. If the state of the system before the measurement was $\rho$ then
the state after is given by
\beqn
	\rho' \equiv \sum_i P_i \rho P_i. 
\eeqn
The following result shows that the entropy is never decreased in this
case, and remains the same only when the state is not changed by the
measurement.

\begin{theorem} ({\bf Orthogonal measurements increase entropy})
\cite{Wehrl78a}

Suppose $P_i$ is a complete set of orthogonal projectors and $\rho$ is a
density operator. Then the entropy of the state $\rho' \equiv \sum_i P_i \rho
P_i$ of the system after the measurement is at least as great as the original
entropy,
\beqn
	S(\rho') \geq S(\rho), 
\eeqn
with equality if and only if $\rho = \rho'$.

\end{theorem}

\begin{proof} {\em (Original?)}

The proof is to apply Klein's inequality to $\rho$ and $\rho'$,
\beqn
	0 \leq S(\rho' || \rho) = -S(\rho)-\tr(\rho \log \rho'). 
\eeqn
The result will follow if we can prove that $-\tr(\rho \log \rho') =
S(\rho')$. To do this, we apply the cyclic property of the trace and the
completeness and orthogonality relations for the projectors to obtain
\beqn
-\tr(\rho \log \rho') & = & -\tr( \sum_i P_i \rho \log \rho') \\
		& = & -\tr( \sum_i P_i \rho \log \rho' P_i). 
\eeqn
A little thought shows that $P_i$ commutes with $\rho'$ and thus with $\log
\rho'$, so
\beqn
-\tr(\rho \log \rho') & = & -\tr(\sum_i P_i \rho P_i \log \rho') \\
		& = & -tr(\rho' \log \rho') = S(\rho'). 
\eeqn
This completes the proof.

\end{proof}

\subsection{The entropy of ensembles}
\index{entropy!of an ensemble}

\begin{theorem} \cite{Wehrl78a} \label{thm:entropy_ensemble}

Suppose $\rho = \sum_i p_i \rho_i$, where $p_i$ are some set of probabilities,
and the $\rho_i$ are density operators. Then
\beqn
	S(\rho) \leq H(p_i) + \sum_i p_i S(\rho_i), 
\eeqn
with equality if and only if the states $\rho_i$ have support on orthogonal
subspaces.

\end{theorem}

\begin{proof} % \cite{Wehrl78a}

We begin with the pure state case, $\rho_i = |\psi_i\ra \la \psi_i|$. Let $A$
be a system with the same state space as the $\rho_i$, and introduce a system
$B$ with an orthonormal basis $|i\ra$ corresponding to the index $i$ on the
probabilities $p_i$. Define
\beqn
	|AB\ra \equiv \sum_i \sqrt{p_i} |\psi_i\ra|i\ra. 
\eeqn
Since $|AB\ra$ is a pure state we have
\beqn
	S(B) = S(A) = S(\sum_i p_i |\psi_i\ra \la \psi_i|) = S(\rho). 
\eeqn
Suppose we perform an orthonormal measurement on the system $B$ in the $|i\ra$
basis. After the measurement the state of system $B$ is
\beqn
	B' = \sum_i p_i |i\ra \la i|. 
\eeqn
But orthogonal measurements never decrease entropy, so $S(\rho) = S(B) \leq
S(B') = H(p_i)$. Observing that $S(\rho_i) = 0$ for the pure state case, we
have proved that
\beqn
	S(\rho) \leq H(p_i)+ \sum_i p_i S(\rho_i), 
\eeqn
when the states $\rho_i$ are pure states. Furthermore, equality holds if and
only if $B = B'$, which is easily seen to occur if and only if the
states $|\psi_i\ra$ are orthogonal.

The mixed state case is now easy. Let $\rho_i = \sum_j p^i_j |e^i_j\ra\la
e^i_j|$ be orthonormal decompositions for the states $\rho_i$, so
$\rho=\sum_{ij} p_ip^i_j |e^i_j\ra\la e^i_j|$. Applying the pure state result
and the observation that $\sum_j p^i_j = 1$ for each $i$, we have
\beqn
S(\rho) \leq -\sum_{ij} p_ip^i_j \log(p_ip^i_j) \\
	& = & -\sum_i p_i \log p_i -\sum_i p_i \sum_j  p^i_j \log p^i_j \\
		& = & H(p_i) + \sum_i p_i S(\rho)_i, 
\eeqn
which is the desired result. The equality conditions for the mixed state case
follow immediately from the equality conditions for the pure state case.

\end{proof}

\subsection{Subadditivity}
\label{subsec:subadditivity}
\index{subadditivity!quantum}
\index{triangle inequality}
\index{Araki-Lieb inequality}

Suppose distinct quantum systems, $A$ and $B$, have a joint state
$\rho_{AB}$.  Then the joint entropy for the two systems satisfies the
inequalities
\beqn
S(A,B) & \leq & S(A)+S(B)  \label{eqtn: subadditivity} \\
S(A,B) & \geq & |S(A)-S(B)|. 
\eeqn
The first of these inequalities is known as the {\em subadditivity}
inequality for Von Neumann entropy, and holds with equality if and
only if systems $A$ and $B$ are uncorrelated, that is, $\rho_{AB} =
\rho_A \otimes \rho_B$. The second is called the {\em triangle}
inequality, or sometimes the {\em Araki-Lieb} inequality.

The proof of subadditivity is a simple application of Klein's
inequality, $S(\rho) \leq -\tr(\rho \log \sigma)$. Setting $\rho \equiv
\rho_{AB}$ and $\sigma\equiv \rho_A \otimes \rho_B$, note that
\be-\tr(\rho \log \sigma) & = & -\tr(\rho_{AB} (\log \rho_A + \log
\rho_B)) \\
& = & -\tr(\rho_A \log \rho_A)-\tr(\rho_B \log \rho_B) \\
& = & S(\rho_A)+S(\rho_B).\ee 
Klein's inequality therefore gives $S(\rho_{AB}) \leq
S(\rho_A)+S(\rho_B)$, as desired. The equality conditions $\rho = \sigma$ for
Klein's inequality give equality conditions $\rho_{AB} = \rho_A \otimes
\rho_B$ for subadditivity.

To prove the triangle inequality, let $R$ be a system which purifies
systems $A$ and $B$\footnote{See Appendix \ref{app:mixed} for a review
of purifications. $R$ purifies $A$ and $B$ if the joint state of $RAB$
is pure.}. Applying subadditivity we have
\beqn
	S(R)+S(A) \geq S(A,R). 
\eeqn
Since $ABR$ is in a pure state, $S(A,R) = S(B)$ and $S(R) = S(A,B)$.  The
previous inequality then may be rearranged to give
\beqn
	S(A,B) \geq S(B)-S(A). 
\eeqn
The equality conditions for this inequality are not so easy to
understand. Formally, the equality conditions are that $\rho_{AR}=\rho_A
\otimes \rho_R$. Intuitively, what this means is that $A$ is already as
entangled as it can possibly be with the outside world, given its
existing correlations with system $B$.  Note also that by symmetry
between the systems $A$ and $B$ we also have, $S(A,B) \geq S(A)-S(B)$.
Combining these two inequalities gives the triangle inequality.

\subsection{Concavity of the entropy}
\index{entropy!concavity of}
\index{concavity of the entropy}

The entropy is a {\em concave} function of its inputs. That is, given
real numbers $\lambda_i$ satisfying $\lambda_i \geq 0, \sum_i
\lambda_i = 1$, and corresponding density operators $\rho_i$, the
entropy satisfies the equation:
\beqn \label{eqtn: entropy concave}
	S(\sum_i \lambda_i \rho_i) \geq \sum_i \lambda_i S(\rho_i). 
\eeqn
To understand why this should be so, imagine that the $\lambda_i$s are
probabilities. Then $\sum_i \lambda_i \rho_i$ expresses the state of a
quantum system which is in an unknown state $\rho_i$ with probability
$\lambda_i$. Not surprisingly, our uncertainty about this mixture of
states should be higher than the average uncertainty of the states
$\rho_i$.

Let $A$ have a state space containing the state $\rho_i$, and let $B$ have a
state space with orthonormal basis $|i\ra$. Define the joint state
\beqn
	\rho^{AB} \equiv \sum_i \lambda_i \rho_i \otimes |i\ra \la i|. 
\eeqn
To prove concavity we use the subadditivity of the entropy. Note that
\beqn
S(A) & = & S(\sum_i \lambda_i \rho_i) \\
S(B) & = & S(\sum_i \lambda_i |i\ra \la i|) = H(\lambda_i) \\
	S(AB) & = & H(\lambda_i) + \sum_i \lambda_i S(\rho_i). 
\eeqn
Applying the inequality $S(AB) \leq S(A) + S(B)$ we obtain
\beqn
	\sum_i \lambda_i S(\rho_i) \leq S(\sum_i \lambda_i \rho_i), 
\eeqn
which is the desired concavity result. Note that equality holds if and
only if all the states $\rho_i$ are identical; that is, the entropy is
a strictly concave function\footnote{This observation can be used to
give an elegant proof that the unique maximal entropy state is the
completely mixed state. Let $\rho$ be given, and note that $I/d =
\sum_{\pi} \rho_{\pi} / d!$, where the sum is over all permutations
$\pi$ on $d$ elements, and $\rho_{\pi}$ is obtained from $\rho$ by a
permutation of the basis elements in which $\rho$ is diagonal. The
result follows by strict concavity.} of its inputs.

It's worth pausing here to think about the strategy we've employed in
this proof, and the similar strategy used to prove the triangle
inequality. We introduced an {\em auxiliary system}, $B$, in order to
prove a result about the system $A$. Introducing auxiliary systems is
something often done in quantum information theory, and we'll see this
trick again and again. The intuition behind the introduction of $B$ in
this particular manner is as follows: we want to find a system part of
which is in the state $\sum_i
\lambda_i \rho_i$, where the value of $i$ is not known. System $B$
effectively stores the ``true'' value of $i$: if $A$ were ``truly'' in
state $\rho_i$, the system $B$ would be in state $|i\ra\la i|$, and
observing system $B$ in the $|i\ra$ basis would reveal this
fact. Using auxiliary systems in this way to encode our intuition in a
rigorous way is something of an art, but it is also an essential part
of many proofs in quantum information theory.

%%%%%%%%%%%%%%%%%%%%%%%%%%%%%%%%%%%%%%%%%%%%%%%%%%%%%%%%%%%%%%%%%%%%%%%%%%%%%
\section{Strong subadditivity}
\label{sec:ssa}
\index{strong subadditivity!quantum}
\index{Lieb's theorem}

The subadditivity inequalities proved in the last section for two
quantum systems can be extended to three systems. The basic result is
known as the {\em strong subadditivity} inequality, and it is one of
the most important and useful results in quantum information
theory. Unfortunately, unlike in the classical case, proving the
quantum strong subadditivity inequality appears to be quite
difficult. However, it will be used frequently throughout this
Dissertation, so we give a full proof here.  The result was first
proved by Lieb and Ruskai \cite{Lieb73b}, based upon an earlier result
of Lieb \cite{Lieb73a}.  The proof of Lieb's theorem given here is
adapted from Bhatia \cite{Bhatia97a}, which is an adaptation of a
proof of Simon \cite{Simon79a}.

The {\em strong subadditivity inequality} for von Neumann
entropies states that for a trio of quantum systems, $A,B,C$,
\beqn
S(A,B,C) +S(B)\leq S(A,B)+S(B,C). \eeqn The proof of this inequality
which we give is based upon a deep mathematical result known as {\em
Lieb's theorem}. We begin with a few simple notations and definitions.

Suppose $f(A,B)$ is a real valued function of two matrices, $A$ and
$B$. Then $f$ is said to be {\em jointly concave} in $A$ and $B$ if
for all $0\leq \lambda \leq 1$, \beqn f(\lambda A_1 + (1-\lambda)
A_2,\lambda B_1 + (1-\lambda) B_2) \geq \lambda
f(A_1,B_1)+(1-\lambda)f(A_2,B_2). \eeqn For matrices $A$ and $B$, we
say $A \geq B$ if $A-B$ is a positive matrix. If $A$ is a positive
matrix, and $t$ a real number, then we define $A^t$ as follows. Let $A
= U D U^{\dagger}$, where $U$ is unitary and $D$ is a diagonal matrix
with non-negative entries. Define $D^t$ to be the diagonal matrix with
entries $d_i^t$, where $d_i$ are the diagonal entries in $D$. Define
$A^t \equiv U D^t U^{\dagger}$. Let $A$ be an arbitrary matrix. We
define the {\em norm} of $A$ by \beqn \| A \| \equiv \max_{\la u|u\ra
= 1} | \la u| A |u\ra |. \eeqn In our proof of Lieb's theorem and
strong subadditivity, we will have occasion to use the following
easily verified observations.

\begin{enumerate}
\item \label{ex: lieb positive}
	If $A \leq B$, then $X A X^{\dagger} \leq X B
	X^{\dagger}$ for all matrices $X$.
\item Let $f(A,B)$ be a jointly concave function. Then $f(A,B)$
is concave in $A$, with $B$ held fixed.
It is easy to find a function of two variables that is concave in
each of its inputs, but is not jointly concave.
\item $A \geq 0$ if and only if $A$ is a positive operator.
\item The relation $\geq$ is a partial order on operators -- that is, it is
transitive ($A\geq B$ and $B \geq C$ implies $A \geq C$), asymmetric
($A \geq B$ and $B \geq A$ implies $A = B$), and reflexive ($A \geq
A$).
\item Suppose $A$ has eigenvalues $\lambda_i$. Define $\lambda$ to be
the maximum of the set $|\lambda_i|$. Then:
\begin{enumerate}
\item  $\| A \| \geq \lambda$.
\item When $A$ is Hermitian, $\| A \| = \lambda$.
\item When 
\beqn
A = \left[ \begin{array}{cc} 1 & 0 \\ 1 & 1 \end{array} \right] 
\eeqn
it is easy to verify the $\| A \| = 3/2 > 1 = \lambda$.
\end{enumerate}
\item The eigenvalues of $A$ are the solutions to the characteristic equation
$\det(xI-A) = 0$. For invertible $A$, note that $\det(xI-AB) = 
\det A \det(xI-BA) \det A^{-1} = \det(xI - BA)$, and thus the
eigenvalues of $AB$ and $BA$ are the same. A simple continuity
argument shows that this is generally true in finite dimensions.
\item \label{ex: lieb commutative} Suppose $A$ and $B$ are such that
$AB$ is Hermitian. Then from the previous two observations it follows that
$\| AB \| \leq \| BA \|$.
\item \label{ex: lieb less than 1}
	Suppose $A$ is positive. Then
        $\| A \| \leq 1$ if and only if $A \leq I$.
\item \label{ex: lieb Hilbert-Schmidt}
Let $A$ be a positive matrix. Define a superoperator (linear
operator on matrices) by the equation ${\cal A}(X) \equiv AX$. Then
${\cal A}$ is positive with respect to the Hilbert-Schmidt inner
product. That is, for all $X$, $\tr(X^{\dagger} {\cal A}(X)) \geq
0$. Similarly, the superoperator defined by ${\cal A}(X) \equiv
XA$ is positive with respect to the Hilbert-Schmidt inner product on
matrices.

\end{enumerate}

With these results in hand, we are now in a position to state and
prove Lieb's theorem.

\begin{theorem} {\bf (Lieb's theorem)} \cite{Lieb73a}

Let $X$ be a matrix, and $0 \leq t \leq 1$. Then the function
\beqn
f(A,B) \equiv \tr(X^{\dagger} A^t X B^{1-t}) \eeqn
is jointly concave in positive matrices $A$ and $B$.
\end{theorem}

Lieb's theorem is an easy corollary of the following lemma:
\begin{lemma}
Let $R_1,R_2,S_1,S_2,T_1,T_2$ be positive operators such that
$0=[R_1,R_2]=[S_1,S_2]=[T_1,T_2]$, and
\beqn \label{eqtn: lieb conditions 1}
R_1 & \geq & S_1 + T_1 \\ \label{eqtn: lieb conditions 2}
R_2 & \geq & S_2+T_2. \eeqn
Then for all $0 \leq t \leq 1$,
\beqn \label{eqtn: lieb conc}
R_1^t R_2^{1-t} \geq S_1^t S_2^{1-t}+T_1^t T_2^{1-t} \eeqn
is true as a matrix inequality.
\end{lemma}

\begin{proof} ({\em Adapted from \cite{Bhatia97a}})

We begin by proving the result for $t = 1/2$, and then use this to
establish the result for general $t$.

Let $|x\ra$ and $|y\ra$ be any two vectors. Applying the Cauchy-Schwartz
inequality twice and some straightforward manipulations, we have
\beqn
& & |\la x|(S_1^{1/2}S_2^{1/2}+T_1^{1/2}T_2^{1/2})|y\ra| \nonumber \\
& \leq & |\la x| S_1^{1/2}S_2^{1/2}|y\ra| + |\la x|T_1^{1/2}T_2^{1/2}|y\ra| \\
& \leq & \|S_1^{1/2} |x\ra \| \, \| S_2^{1/2}|y\ra \| + \|T_1^{1/2}
|x\ra \| \, \|T_2^{1/2}|y\ra \| \\
& \leq & \sqrt{ \left( \| S_1^{1/2} |x\ra \|^2 + \| T_1^{1/2} |x\ra \|^2
	\right) \left( \| S_2^{1/2} |y\ra \|^2 + \| T_2^{1/2} |y\ra \|^2
	\right) } \\
& = & \sqrt{ \la x|(S_1 +T_1)|x\ra \la y|(S_2+T_2)|y\ra }. \eeqn
By hypothesis, $S_1+T_1 \leq R_1$ and $S_2+T_2 \leq R_2$, so
\beqn
|\la x|(S_1^{1/2}S_2^{1/2}+T_1^{1/2}T_2^{1/2})|y\ra| & \leq & \sqrt{
\la x| R_1 |x\ra \la y| R_2|y\ra }. \eeqn

Let $|u\ra$ be any unit vector. Then applying the previous result with
$|x\ra \equiv R_1^{-1/2} |u\ra$ and $|y\ra \equiv R_2^{-1/2} |u\ra$ gives
\beqn
& & \la u|R_1^{-1/2} (S_1^{1/2}S_2^{1/2}+T_1^{1/2}T_2^{1/2})
R_2^{-1/2}|u\ra \nonumber \\
& \leq & \sqrt{\la u|R_1^{-1/2} R_1 R_1^{-1/2}|u\ra \la u| R_2^{-1/2}R_2
	R_2^{-1/2}|u\ra} \\
& = & \sqrt{\la u|u\ra \la u|u\ra } = 1. \eeqn
Thus
\beqn \label{eqtn: lieb intermediate}
\| R_1^{-1/2} (S_1^{1/2}S_2^{1/2}+T_1^{1/2}T_2^{1/2}) R_2^{-1/2} \|
\leq 1. \eeqn
Define
\beqn
A & \equiv & R_1^{-1/4} R_2^{-1/4}
 (S_1^{1/2}S_2^{1/2}+T_1^{1/2}T_2^{1/2}) R_2^{-1/2} \\
B & \equiv & R_2^{1/4} R_1^{-1/4}. \eeqn
Note that $AB$ is Hermitian, so by observation
number \ref{ex: lieb commutative} on page \pageref{ex: lieb commutative},
\beqn
& & \| R_1^{-1/4} R_2^{-1/4} (S_1^{1/2}S_2^{1/2}+T_1^{1/2}T_2^{1/2})
	R_2^{-1/4}R_1^{-1/4} \| \nonumber \\
& = & \| AB \| \leq  \| BA \|  \\
 & = & \| R_1^{-1/2} (S_1^{1/2}S_2^{1/2}+T_1^{1/2}T_2^{1/2})
 	R_2^{-1/2} \| \\
 & \leq & 1, \eeqn
where the last inequality is just (\ref{eqtn: lieb intermediate}).
$AB$ is a positive operator, so by observation number
\ref{ex: lieb less than 1} on page \pageref{ex: lieb less than 1} 
and the previous inequality,
\beqn
R_1^{-1/4} R_2^{-1/4} (S_1^{1/2}S_2^{1/2}+T_1^{1/2}T_2^{1/2})
	R_2^{-1/4}R_1^{-1/4} \leq I. \eeqn Finally, by observation
	\ref{ex: lieb positive} on page \pageref{ex: lieb positive},
and the commutativity of $R_1$ and $R_2$,
\beqn
S_1^{1/2}S_2^{1/2}+T_1^{1/2}T_2^{1/2} \leq R_1^{1/2}R_2^{1/2}, \eeqn
which establishes that (\ref{eqtn: lieb conc}) holds for $t = 1/2$.

Let $I$ be the set of all $t$ such that (\ref{eqtn: lieb conc}) holds. By
inspection, we see that $0$ and $1$ are elements of $I$. We now use
the result for $t = 1/2$ to prove the result for general $t$. Suppose
$\mu$ and $\eta$ are elements of $I$, so
\beqn
R_1^{\mu} R_2^{1-\mu} & \geq & S_1^{\mu}
	S_2^{1-\mu}+T_1^{\mu}T_2^{1-\mu} \\ R_1^{\eta} R_2^{1-\eta} &
	\geq & S_1^{\eta} S_2^{1-\eta}+T_1^{\eta}T_2^{1-\eta}. \eeqn
	These inequalities are of the form (\ref{eqtn: lieb conditions
	1}) and (\ref{eqtn: lieb conditions 2}) for which the $t=1/2$
	case has already been proved. Using the $t=1/2$ result we see that
\beqn
\left( R_1^{\mu} R_2^{1-\mu} \right)^{1/2} \left(R_1^{\eta}
	R_2^{1-\eta} \right)^{1/2} \geq \left(S_1^{\mu} S_2^{1-\mu}
	\right)^{1/2} \left( S_1^{\eta} S_2^{1-\eta} \right)^{1/2} +
	\left( T_1^{\mu} T_2^{1-\mu} \right)^{1/2} \left( T_1^{\eta}
	T_2^{1-\eta} \right)^{1/2}. \eeqn
Using the commutativity assumptions $0=[R_1,R_2]=[S_1,S_2]=[T_1,T_2]$,
we see that for $\nu \equiv (\mu+\eta)/2$,
\beqn
R_1^{\nu} R_2^{1-\nu} \geq S_1^{\nu} S_2^{1-\nu} + T_1^{\nu}
T_2^{1-\nu}. \eeqn
Thus whenever $\mu$ and $\eta$ are in $I$, so is $(\mu+\eta)/2$. Since
$0$ and $1$ are in $I$, it is easy to see that any number $x$ between
$0$ and $1$ with a finite binary expansion must be in $I$. Thus $I$ is
dense in $[0,1]$. The result now follows from the continuity in $t$ of
the conclusion, (\ref{eqtn: lieb conc}).

\end{proof}

The proof of Lieb's theorem is a simple application of the lemma. The
main novelty is that the operators in the lemma are chosen to be {\em
superoperators} -- linear maps on operators. These will be chosen in
such a way as to be positive with respect to the Hilbert-Schmidt inner
product $(A,B) \equiv \tr(A^{\dagger} B)$.

\begin{proof} (Lieb's theorem) {\em (Adapted from \cite{Bhatia97a})}

Define
\beqn
{\cal S}_1(X) & \equiv & \lambda A_1 X \\
{\cal S}_2(X) & \equiv & \lambda X B_1 \\
{\cal T}_1(X) & \equiv & (1-\lambda) A_2 X \\
{\cal T}_2(X) & \equiv & (1-\lambda) X B_2 \\
{\cal R}_1 & \equiv & {\cal S}_1+{\cal T}_1 \\
{\cal R}_2 &  \equiv & {\cal S}_2+{\cal T}_2. \eeqn
Observe that ${\cal S}_1$ and ${\cal S}_2$ commute, as do ${\cal T}_1$
and ${\cal T}_2$, and ${\cal R}_1$ and ${\cal R}_2$. Recall
observation \ref{ex: lieb Hilbert-Schmidt} on page \pageref{ex: lieb Hilbert-Schmidt}, that
all these operators are positive with respect to the Hilbert-Schmidt
inner product. By the lemma,
\beqn
{\cal R}_1^t {\cal R}_2^{1-t} \geq {\cal S}_1^t {\cal S}_2^{1-t} +
	{\cal T}_1^t {\cal T}_2^{1-t}. \eeqn
Taking the $X \cdot X$ matrix element of the previous inequality gives
\beqn
 & & \tr \left[ X^{\dagger} \left( \lambda A_1 + (1-\lambda) A_2 \right)^t
	X \left(\lambda B_1+(1-\lambda)B_2 \right)^{1-t} \right] \nonumber \\
 & \geq & \tr \left[ X^{\dagger} (\lambda A_1)^t X (\lambda B_1)^{1-t} \right]
	+ \tr \left[ X^{\dagger} ((1-\lambda)A_2)^t X
	((1-\lambda)B_2)^{1-t} \right] \\
	& = & \lambda \tr(X^{\dagger} A_1^tX B_1^{1-t})+(1-\lambda)
	\tr(X^{\dagger} A_2^t X B_2^{1-t}), \eeqn
which is the desired statement of joint concavity.

\end{proof}

Let $B$ and $C$ be density operators. Recall that the
{\em relative entropy} of $B$ to $C$ is defined by
\beqn
S(B \| C) \equiv -S(B)-\tr(B \log C). \eeqn

\index{relative entropy!quantum!convexity of}
\index{convexity of quantum relative entropy}
\begin{theorem} {\em (Convexity of the relative entropy)} \cite{Wehrl78a}

The relative entropy $S(B\| C)$ is jointly convex in its arguments.
\end{theorem}

\begin{proof} \cite{Bhatia97a}

Define
\beqn
I_t(A,X) \equiv \tr(X^{\dagger} A^t X A^{1-t})-\tr(X^{\dagger} X
A). \eeqn
Note that the first term in this expression is concave in $A$, by
Lieb's theorem, and the second term is linear in $A$. Thus, $I_t(A,X)$
is concave in $A$. Define
\beqn
I(A,X) \equiv \frac{d}{dt} \left|_{t=0} \right. I_t(A,X) =
	\tr(X^{\dagger} (\log A) X A) - \tr(X^{\dagger}X (\log A)
	A). \eeqn
Noting that $I_0(A,X) = 0$ and using the concavity
of $I_t(A,X)$ in $A$ we have
\beqn
I(\lambda A_1+(1-\lambda)A_2,X) & = & \lim_{\Delta \rightarrow 0}
\frac{I_{\Delta}(\lambda A_1 + (1-\lambda)A_2,X)}{\Delta} \\
	& \geq & \lambda \lim_{\Delta \rightarrow 0}
	\frac{I_{\Delta}(A_1,X)}{\Delta} + (1-\lambda) \lim_{\Delta
	\rightarrow 0} \frac{I_{\Delta}(A_2,X)}{\Delta} \\
	& = & \lambda I(A_1,X) + (1-\lambda) I(A_2,X). \eeqn
That is, $I(A,X)$ is a concave function of $A$.

Finally, defining the block matrices
\beqn
A \equiv \left[ \begin{array}{cc} B & 0 \\
	0 & C \end{array} \right], \,\,\,\, X \equiv
	\left[ \begin{array}{cc} 0 & 0 \\ I & 0 \end{array} \right] \eeqn
we can easily verify that $I(A,X) = -S(B\| C)$. The joint convexity of $S(B \| C)$
now follows from the concavity of $I(A,X)$ in $A$.

\end{proof}

Suppose $\rho_{AB}$ is the state of a joint system, $AB$. Recall the
definition of the
{\em conditional entropy} of system $A$ given system $B$,
\beqn
S(A|B) \equiv S(A,B)-S(B). \eeqn

\index{conditional entropy!quantum!concavity of}
\index{concavity of the quantum conditional entropy}
\begin{corollary}
\label{corollary:concavity_conditional}
$S(A|B)$ is concave in $\rho_{AB}$.

\end{corollary}

\begin{proof}

Let $d$ be the dimension of system $A$. Note that
\beqn
S(\rho_{AB} \| \frac{I}{d} \otimes \rho_B) & = & -S(A,B)- \tr(\rho_{AB}
	\log( \frac{I}{d} \otimes \rho_B) ) \\
	& = & -S(A,B)-\tr(\rho_B \log \rho_B) +\log d \\
	& = & -S(A|B) + \log d. \eeqn
Thus $S(A|B)=\log d - S(\rho_{AB} \| I/d \otimes \rho_B)$. The concavity
of $S(A|B)$ follows from the joint convexity of the relative entropy.

\end{proof}

Strong subadditivity can now be proved using the convexity of the
conditional entropy.

\index{strong subadditivity!quantum!proof of}
\begin{theorem} \cite{Lieb73b}

For any trio of quantum systems, $A, B, C$, the inequalities
\beqn
S(A)+S(B) & \leq & S(A,C)+S(B,C) \\
S(A,B,C) +S(B) & \leq & S(A,B)+S(B,C) \eeqn
hold.
\end{theorem}

\begin{proof} \cite{Lieb73b}

The two inequalities are, in fact, equivalent. We will use convexity
of the conditional entropy to prove the first, and show that the
second follows. Define a function of density operators on the system $ABC$,
\beqn
T(\rho_{ABC}) \equiv S(A)+S(B)-S(A,C)-S(B,C) = -S(C|A)-S(C|B). \eeqn
From the concavity of the relative entropy we see that $T(\rho_{ABC})$
is a convex function of $\rho_{ABC}$. Let $\rho_{ABC} = \sum_i p_i
P_i$, where $P_i$ is a pure state of the system $ABC$ and the $p_i$
are probabilities. From the convexity of $T$, $T(\rho_{ABC}) \leq
\sum_i p_i T(P_i)$. But for a pure state, $T(P_i) = 0$, as $S(A,C) =
S(B)$ and $S(B,C) = S(A)$ for a pure state. It follows that
$T(\rho_{ABC}) \leq 0$, and thus
\beqn
S(A)+S(B)-S(A,C)-S(B,C) \leq 0, \eeqn
which is the first inequality.

Finally, to obtain the second inequality, introduce a fourth system,
$R$, purifying the system $ABC$. Then
\beqn
S(R)+S(B) \leq S(R,C)+S(B,C). \eeqn
Since $ABCR$ is a pure state, $S(R)=S(A,B,C)$ and $S(R,C) = S(A,B)$,
so the previous inequality becomes
\beqn
S(A,B,C)+S(B) \leq S(A,B)+S(B,C), \eeqn
as we set out to show.

\end{proof}

Strong subadditivity and the convexity of the relative entropy are
results which have many useful consequences. We will use these results
many, many times throughout the remainder of this Dissertation. For
the time being, it is interesting to note a few elementary
consequences.

First, it is worth emphasizing how remarkable it is that the
inequality $S(A)+S(B) \leq S(A,C)+S(B,C)$ holds. The corresponding
inequality holds also for Shannon entropies, but for quite different
reasons. For Shannon entropies it is true that $H(A)\leq H(A,C)$ and
$H(B) \leq H(B,C)$, so the sum of the two inequalities must
necessarily be true. In the quantum case, it is possible to have
either $S(A) > S(A,C)$ or $S(B) > S(B,C)$, yet somehow nature manages
to conspire in such a way that both these possibilities are not true
simultaneously, in order to ensure that the condition $S(A)+S(B) \leq
S(A,C)+S(B,C)$ is always satisfied. Other ways of rephrasing this are
in terms of conditional entropies and mutual informations,
\beqn \label{eq:rel_ent_in}
0 & \leq & S(C|A) + S(C|B) \\ S(A:B)+S(A:C) & \leq & 2S(A), \eeqn both
of which are also remarkable inequalities, for similar reasons. Note,
however, that the inequality $0 \leq S(A|C) +S(B|C)$, which one might
hope to be true based upon (\ref{eq:rel_ent_in}) is not, as can easily
be seen by choosing $BC$ to be a Bell state in a product state with
system $A$. In part, it is these wonderful facts which brought home to
me how strange and counter-intuitive quantum entropies may be.

%Another useful consequence of strong subadditivity is the concavity of
%the conditional entropy. Suppose $\rho_i^{AB}$ is a collection of states
%of the joint system $AB$, and $p_i$ is a probability distribution. Defining
%$\rho^{AB} \equiv \sum_i p_i \rho^{AB}_i$, we will prove that
%\beqn
%	S(A|B) \geq \sum_i p_i S(A_i | B_i). 
%\eeqn
%That is, the conditional entropy is concave in the state of $AB$.

%To prove this result, introduce an additional system $R$, with an orthonormal
%basis $|i\ra$ in one-to-one correspondence with the index $i$ of the
%state $\rho_i^{AB}$. Define 
%\beqn
%	\rho^{ABC} \equiv \sum_i \rho_i^{AB} \otimes |i\ra \la i|, 
%\eeqn
%and notice that this traces down to the state $\rho^{AB}$ which we defined
%earlier.
%
%Notice that
%\beqn
%	S(A,B,C) = H(p_i) + \sum_i S(A_i,B_i), 
%\eeqn
%and
%\beqn
%	S(A,C) = H(p_i)+\sum_i p_i S(A_i). 
%\eeqn
%Substituting these equations into the strong subadditivity inequality
%\beqn
%	S(A,B,C) + S(A) \leq S(A,B) + S(A,C) 
%\eeqn
%we obtain
%\beqn
%	\sum_i p_i \left[ S(A_i,B_i) -S(A_i) \right] \leq S(A,B)-S(A), 
%\eeqn
%which can be rewritten as
%\beqn
%	\sum_i p_i S(B_i|A_i) \leq S(B|A), 
%\eeqn
%which is the desired concavity.

%Consider the von Neumann conditional entropy, defined for systems $A$
%and $B$ by \beqn S(A|B) \equiv S(A,B) -S(B), \eeqn where $S(\cdot)$ is
%the familiar von Neumann entropy function.

\index{subadditivity!of the quantum conditional entropy}
There is an interesting set of questions related to the subadditivity
properties of quantum conditional entropies. We already saw earlier
that the Shannon mutual information is not subadditive, and thus the
quantum mutual information is not subadditive, either. What about the
subadditivity of the conditional entropy? That is, is it true that
\beqn S(A_1,A_2|B_1,B_2) \leq S(A_1|B_1)+S(A_2|B_2), \eeqn for any
four quantum systems $A_1,A_2,B_1$ and $B_2$? It turns out that this
inequality is correct. To prove this, we apply the strong
subadditivity inequality.

By strong subadditivity,
\beqn
S(A_1,A_2,B_1,B_2)+S(B_1) \leq S(A_1,B_1) + S(A_2,B_1,B_2). \eeqn
Adding $S(B_2)$ to each side of this inequality, we obtain
\beqn
S(A_1,A_2,B_1,B_2) + S(B_1) + S(B_2)
 \leq  S(A_1,B_1)+S(A_2,B_1,B_2)+S(B_2). \eeqn
Applying strong subadditivity to the last two terms of the right hand
side gives
\beqn
 S(A_1,A_2,B_1,B_2)+S(B_1)+S(B_2) 
 \leq  S(A_1,B_1)+S(A_2,B_2)+S(B_1,B_2). \eeqn
Rearranging this inequality gives
\beqn
S(A_1,A_2|B_1,B_2) \leq S(A_1|B_1)+S(A_2|B_2), \eeqn
which is the desired statement of subadditivity of the conditional
entropy.

Two closely related results are the subadditivity of the conditional
entropy in the first and second entries.  These results are attributed
by Ruskai \cite{Ruskai94a} to work of Lieb, which I have not been able
to locate. For example, subadditivity in the first entry, $S(A,B|C)
\leq S(A|C)+S(B|C)$ is trivially seen to be equivalent to strong
subadditivity. Subadditivity in the second entry is slightly more
difficult to prove. We wish to show that $S(A|B,C) \leq
S(A|B)+S(A|C)$. Note that this is equivalent to demonstrating the
inequality
\be
S(A,B,C)+S(B)+S(C) \leq S(A,B)+S(B,C)+S(A,C). \ee To prove this note
that at least one of the inequalities $S(C) \leq S(A,C)$ or $S(B) \leq
S(A,B)$ must be true, as $S(A|B)+S(A|C) \geq 0$. Suppose $S(C) \leq
S(A,C)$. Adding to this inequality the strong subadditivity
inequality, $S(A,B,C)+S(B) \leq S(A,B)+S(B,C)$ gives the result. A
similar proof holds in the case when $S(B) \leq S(A,B)$.

We will return to the subadditivity properties of conditional
information in Chapter \ref{chap:capacity}, in the context of the
quantum channel capacity, where they play a crucial role in
understanding what is going on.

%\index{quantum operations!doubly stochastic}
%\index{quantum operations!unital}
%\index{doubly stochastic quantum operations}
%\index{unital quantum operations}
%
%Another useful consequence of the monotonicity of the relative entropy
%relates to the properties of entropy under quantum operations.  A
%quantum operation ${\cal E}$ is said to be {\em unital} if ${\cal
%E}(I) = I$. A quantum operation is {\em doubly stochastic} if it is
%both unital and complete; in terms of an operator-sum representation
%$\{A_i\}$ this is equivalent to the condition that $\sum_i A_i
%A_i^{\dagger} = \sum_i A_i^{\dagger} A_i = I$.  
%
%Using the monotonicity of the relative entropy it is easy to show that
%a complete quantum operation ${\cal E}$ is entropy non-decreasing if
%and only if it is doubly stochastic.  To see this, note first of all
%that if ${\cal E}$ is not doubly stochastic, then ${\cal E}(I/d) \neq
%I/d$, where $d$ is the dimension of Hilbert space, and therefore
%${\cal E}$ decreases the entropy of the completely mixed state, $I/d$.
%To prove the converse, assume ${\cal E}$ is doubly stochastic and note
%that $S(\rho \| I/d) = -S(\rho)+\log d$.  From the monotonicity of the
%relative entropy and the double stochasticity of ${\cal E}$ it follows
%that
%\be
%-S(\rho)+\log d = S(\rho|I/d) \geq S(\rho' | I/d) = -S(\rho')+\log
%d. \ee
%Thus $S(\rho') \geq S(\rho)$ for doubly stochastic quantum operations.

To finish the Chapter, let us look a little more closely at the
convexity of the relative entropy. Earlier we defined the relative
entropy for density operators, however there is no reason we cannot
extend the definition to any two positive operators, $A$ and $B$,
\beqn
S(A||B) \equiv -S(A) +\tr(A \log B), \eeqn
with the same conventions as before. Following the earlier argument
used to establish the convexity of the relative entropy we
see that the general relative entropy is also convex. This has an
interesting consequence for the case of density operators, although I
am yet to find any practical use for the pretty theorem we will
shortly prove.

Ruskai \cite{Ruskai88a} has pointed out the following interesting
homogeneity relation,
\beqn
S(\alpha A || \beta B) = \alpha S(A||B) +\alpha \tr(A)
\log(\alpha/\beta), \eeqn
which holds for $\alpha,\beta > 0$. Note that when $\alpha =
\beta$ we deduce that $S(\alpha A||\alpha B) = \alpha S(A||B)$.

This is an observation with many interesting consequences.  First, we
see that to prove the double convexity of the relative entropy, it
suffices to prove that \beqn S(A_1+A_2|| B_1 + B_2) \leq
S(A_1||B_1)+S(A_2||B_2). \eeqn We might refer to this inequality as
the ``joint subadditivity'' of the relative entropy; we will see below
that it also follows from joint convexity, so the two statements are
equivalent.  If joint subadditivity were true, then we would have
\beqn S(\lambda A_1+(1-\lambda)A_2 || \lambda B_1+(1-\lambda)B_2) &
\leq & S(\lambda
A_1 || \lambda B_1) + S((1-\lambda)A_2||(1-\lambda)B_2) \nonumber \\
 & & \\
& = &
\lambda S(A_1||B_1)+(1-\lambda)S(A_2||B_2), \eeqn
which is the desired double convexity result. The reason I mention
this is with a viewpoint
to future proofs of strong subadditivity: it may be that it is easier
to try proving the joint subadditivity property of the relative
entropy, rather than attempting the joint convexity directly.

Next we turn this result around and see that the joint subadditivity
follows from joint convexity, that is, 
\beqn
S(\sum_i A_i || \sum_i B_i) \leq \sum_i S(A_i || B_i) \eeqn
is itself a consequence of the double convexity, since if $i$ ranges
over $n$ indices, then
\beqn
S(\sum_i A_i || \sum_i B_i) & = & S(\frac 1n \sum_i (nA_i) || \frac 1n
\sum_i (nB_i)) \\
	& \leq & \sum_i \frac{S(nA_i||nB_i)}{n} \\
	& = & \sum_i S(A_i||B_i). \eeqn

This circle of ideas can be combined to give a new and rather pretty
convexity result for relative entropy. The proof is immediate from
Ruskai's homogeneity relation and the convexity of the relative
entropy:

\index{relative entropy!quantum!strengthened convexity result}
\index{strengthened convexity of quantum relative entropy}
\begin{theorem}
Let $p_i$ and $q_i$ be probability distributions over the same set of
indices. Then \beqn S(\sum_i p_i A_i || \sum_i q_i B_i) \leq \sum_i
p_i S(A_i||B_i) + \sum_i p_i \tr(A_i) \log (p_i/q_i). \eeqn In the
case where the $A_i$ are density operators so $\tr(A_i) = 1$, this
reduces to the remarkable formula \beqn S(\sum_i p_i A_i || \sum_i q_i
B_i) \leq \sum_i p_i S(A_i||B_i) + H(p_i||q_i), \eeqn
where $H(\cdot||\cdot)$ is the Shannon relative entropy.
\end{theorem}

\vspace{1cm}
\begin{center}
\fbox{\parbox{14cm}{
\begin{center} {\bf Summary of Chapter \ref{chap:entropy}:
Entropy and information} \end{center}

\begin{itemize}
\item 
\textbf{Fundamental measures of information arise as the answers to fundamental
questions about the quantity of physical resources required to solve
some information processing problem.}

\item \textbf{Basic definitions:}
\beqn
S(A) & \equiv & -\tr(A \log A) \,\,\,\, \mbox{\em (entropy)} \\
S(A || B) & \equiv & -S(A)+\tr(A \log B) \,\,\,\, \mbox{\em (relative
entropy)} \\
S(A|B) & \equiv & S(A,B) -S(B) \,\,\,\, \mbox{\em (conditional
entropy)} \\ 
S(A:B) & \equiv & S(A)+S(B)-S(A,B) \,\,\,\, \mbox{\em (mutual
information)}. \eeqn

\item \textbf{The relative entropy is jointly convex in its arguments.}

\item {\bf Strong subadditivity:} $S(A,B,C) + S(B) \leq
S(A,B)+S(B,C)$. The other entropy inequalities we discussed are
corollaries of this or the joint convexity of the relative entropy.
\end{itemize}
}}
\end{center}

\chapter{Distance measures for quantum information}
\label{chap:distance} \index{distance measures}

What does it mean to say that information is preserved during some
process? What does it mean to say that two items of information are
similar? A well developed theory of quantum information must provide
useful answers to these questions. Because of the wide variety of
information types in quantum mechanics, inequivalent answers to these
questions are possible, with each answer useful in the context of a
specific class of information processing tasks.

The principle concern of this Chapter is the development of {\em
distance measures} for quantum information. We will be concerned with
two classes of distance measures, {\em static measures}, and {\em
dynamic measures}. Static measures provide a quantitative means of
determining how close two quantum states are, while dynamic measures
provide a quantitative means of determining how well information has
been preserved during a dynamic process. The strategy used in this
Chapter is to begin by developing good static measures of distance,
and then to use those static measures to aid in the development of
good dynamic measures of distance.

Each of the distance measures introduced in this Chapter can be viewed
in two ways.  First, and most important, we inquire as to the {\em
operational meaning} of the distance measures.  That is, we attempt to
find a physical question which leads naturally to that distance
measure.  For example, one of the measures we introduce, the absolute
distance, turns out to be directly related to the ability to
distinguish two quantum states by measurements on those
states. Second, distance measures can be viewed as purely mathematical
constructs, useful for proving facts about the behaviour of quantum
systems.  For example, we will introduce a quantity, the fidelity,
which does not appear to have an especially clear physical
meaning. However, properties of the fidelity can be used to prove
facts of great physical significance, such as the existence of unique
stationary states for certain open quantum systems.

The primary purpose of the Chapter is to serve as review and reference
for basic properties of the distance measures we will consider,
however, it is also contains numerous results which I am not aware of
elsewhere in the literature.  In places the Chapter contains rather
detailed mathematics; upon a first read, these sections may be read
lightly, and returned to later for reference purposes.

\section{Distance measures for classical information}

We begin by studying distance measures for classical information.  In
the classical setting, we will discuss three primary notions of
distance.  Two of these notions will be {\em static measures} of
distance, involving the comparison of two classical probability
distributions. The third notion of distance which we examine is a {\em
dynamical measure} of distance, which is associated with a process.
We won't prove many general results in this section, because we will
prove quantum generalizations of the classical results later in the
Chapter. Indeed, the discussion in this section may appear rather
trivial, and the reader might wonder why we didn't simply skip to the
quantum case. However, the intuitive justification for the distance
measures is easier to grasp in the classical situation, justifying a
separate treatment.

\index{Hamming distance}
What are the objects to be compared in classical information theory?
In some circumstances it is useful to compare strings of bits. For
that purpose, the {\em Hamming distance} is perhaps the most commonly
used measure of distance; it is defined to be the number of places at
which two bit strings are not equal. In this Dissertation we will have
little concern with the actual labeling of bit strings, so notions
such as Hamming distance are of little interest to us.

\index{information source!classical}
By contrast, we will be very concerned with the comparison of
information sources. In classical information theory, an information
source is usually modeled as a random variable, or equivalently, a
probability distribution, over some source alphabet. For example, an
unknown source of English text may be modeled as a sequence of random
variables over the Roman alphabet. Before the text is read, we can
make a fair guess at the relative frequency of the letters that will
appear in the text, and certain correlations among them, such as fact
that occurrences of the pair of letters ``th'' are much more common
than the pair ``zx'' in English text. This characterization of
information sources as probability distributions over some alphabet
causes us to concentrate on the comparison of probability
distributions in our search for measures of distance.

What does it mean to say that two classical probability distributions,
$p_x$ and $q_x$, over the same index set, $x$, are near to one
another?  It is difficult to give an answer to this question which is
obviously the unique ``correct'' answer, so instead we will propose
several different answers, each of which is useful in particular
contexts.

\index{$L_1$ distance!classical}
\index{absolute distance!classical}
The first measure is the $L_1$ distance, defined by the equation
\beqn
D(p_x,q_x) \equiv \sum_x |p_x-q_x|. \eeqn
More usually, we will refer to this as the {\em absolute distance}
between the probability distributions $p_x$ and $q_x$. The absolute
distance is easily seen to be a metric on probability distributions,
so the use of the term ``distance'' is justified. ``Absolute'' refers
to the absolute value signs appearing in the definition.

As an example of the absolute distance, consider probability
distributions on $\{0,1\}$ defined by $p_0 = p, p_1 = 1-p$ and $q_0 =
q, q_1 =1-q$, where $p \leq q$. Then $D(p_x,q_x) = q-p + (1-p-(1-q)) =
2(q-p)$ is the absolute distance.

\index{fidelity!classical}

A second measure of distance between probability distributions, the
{\em fidelity} of the probability distributions $p_x$ and $q_x$ is
defined by
\beqn
F(p_x,q_x) \equiv \sum_x \sqrt{p_x q_x}. \eeqn The fidelity is a quite
different way of measuring distance between probability distributions
than is the absolute distance. To begin with, it is not a metric,
although we will see later that there is a metric which can be
constructed from the fidelity. One way of seeing that the fidelity is
not a metric is to note that when the distributions $p_x$ and $q_x$
are identical, $F(p_x,q_x) = 1$. More generally, we will prove later
that the fidelity is always in the range zero to one, and is equal to
one if and only if the probability distributions are identical.

As an example of the fidelity, consider as before probability
distributions on $0,1$ defined by $p_0 = p, p_1 = 1-p$ and $q_0 = q,
q_1 =1-q$, where $p \leq q$. Then $F(p_x,q_x) = \sqrt{pq}
+\sqrt{(1-p)(1-q)}$.

\index{absolute distance!classical!operational meaning for}
The absolute distance and fidelity are mathematically useful means of
defining the notion of a distance between two probability
distributions.  Do these measures have physically motivated
operational meanings? In the case of the absolute distance, the answer
to this question is yes. In particular, it can be shown that
\cite{Luby96a}
\be
D(p_x,q_x) = 2 \max_S \left( \sum_{x \in S} p_x-\sum_{x \in S} q_x)
\right), \ee
where the maximization is over all subsets $S$ of the index set. The
quantity being maximized is the difference between the probability
that the event $S$ occurs, according to the distribution $p_x$, and
the probability that the event $S$ occurs, according to the
distribution $q_x$. The event $S$ is thus the optimal event to examine
when trying to distinguish the distributions $p_x$ and $q_x$. The
absolute distance governs how well it is possible to make this
distinction, using statistical tools such as the Chernoff Bound
\cite{Papadimitriou94a}.

I am not aware of a similarly clear operational interpretation for the
fidelity. However, in the next section we will see that the fidelity
is a sufficiently useful quantity for mathematical purposes to justify
its study, even without a clear physical interpretation. Moreover, I
can not rule out the possibility that a clear physical interpretation
of the fidelity will be discovered in the future. Finally, it turns
out that there are close connections between the fidelity and the
absolute distance, which allow one to use properties of one quantity to
deduce properties of the other.

%\index{dynamical measures of distance}

The third notion of distance with which we are concerned is a {\em
dynamical measure} of distance. Suppose the random variables $X$ and
$Y$ form a Markov process\footnote{Strictly speaking, any pair of
random variables form a Markov process, but this usage is to get you
in the mood for the less trivial Markov processes in the next
paragraph.}, $X \rightarrow Y$, with values over the same possible
range of values, which we denote by $x$. Then the probability that $Y$
is not equal to $X$, $p(X \neq Y)$, is an obvious, but still important
measure of the degree to which information has been preserved by the
channel.

\label{abs_distance_meaning}
This measure of distance can be recast in the form of the absolute
distance introduced earlier. Imagine that the random variable $X$ is
given to you, and you first make a copy of $X$, creating a new random
variable $\tilde X = X$. The random variable $X$ now undergoes some
Markov dynamics, leaving as the output of the process a random
variable, $Y$. How close is the initial perfectly correlated pair,
$(\tilde X, X)$, to the final pair, $(\tilde X, Y)$? Using the
absolute distance as our measure of distance, we see that the answer
to this question is the absolute distance between the distributions
$p_{x,x'} \equiv p(\tilde X=x,X = x') = \delta_{xx'} p(X=x)$ and
$q_{x,x'} \equiv p(\tilde X=x, Y = x')$, \be D( (\tilde X,X),(X,Y)) &
= & D(p_{x,x'},q_{x,x'}) \\ & = & \sum_{xx'} |\delta_{xx'} p(X=x) -
p(\tilde X=x, Y=x')| \\ & = & \sum_{x \neq x'} p(\tilde X=x,Y=x') +
\sum_x \left( p(X = x)-p(\tilde X =x,Y=x) \right) \\ & = & p(\tilde X
\neq Y) + 1 - p(\tilde X = Y) \\ & = & p(X \neq Y)+p(\tilde X \neq Y)
\\ & = & 2 p(X \neq Y). \ee It is worthwhile to reflect on this
example. The probability of an error occurring during the Markov
process is equal to (half) the absolute distance between $(\tilde
X,X)$ and $(\tilde X,Y)$, which we can regard as a measure of the
extent to which correlation between $X$ and the external world is
destroyed by the dynamics undergone by $X$. A quantum analogue of this
idea will be used later to define a notion of information preservation
through a quantum channel, based on the idea that it is quantum
entanglement, rather than correlation, which is the important thing to
preserve during the channel's dynamics.

The previous example concerned a Markov process containing only two
random variables.  More often, we will be concerned with a multipart
Markov process. For example, imagine that we have a four part Markov
process, $W
\rightarrow X \rightarrow Y \rightarrow Z$. Such a situation arises,
for example, in communications problems: $W$ is an information source
which is encoded using an error correcting code to give a random
variable $X$, before being sent over a noisy communications channel
which has $Y$ as output, before being decoded to yield $Z$. Once again,
the total probability of error, $p(W \neq Z)$ is an important distance
measure for the channel.

\section{How close are two quantum states?}

What does it mean to say that two quantum states, $\rho$ and $\sigma$,
are close together? In this section we review two measures of the
closeness of quantum states, the absolute distance and the fidelity,
both of which generalize the corresponding classical concepts
introduced in the previous section. Furthermore, we introduce two
additional measures of distance, both of which arise naturally from
the fidelity. The section concludes by examining relationships between
the absolute distance and the fidelity.

\subsection{Absolute distance}
\label{subsec:absolute_distance}
\index{absolute distance!quantum}

We begin by defining the {\em absolute distance} between states $\rho$
and $\sigma$, \beqn D(\rho,\sigma) \equiv \| \rho -\sigma \| \equiv
\tr |\rho-\sigma|. \eeqn where $|A| \equiv \sqrt{A^{\dagger}
A}$. Notice that this measure of distance generalizes the classical
absolute distance, in the sense that if $\rho$ and $\sigma$ are
diagonal in the same basis, then the (quantum) absolute distance
between $\rho$ and $\sigma$ is equal to the classical absolute
distance between the eigenvalues of $\rho$ and $\sigma$.

% basic properties: metric; boundedness; invariance properties.
There is a useful alternate formula for the absolute distance, \be
D(\rho,\sigma) = 2 \max_P \tr(P(\rho-\sigma)), \ee where the
maximization may be taken alternately over all projectors, $P$, or
over all positive operators $P \leq I$; the formula is valid in either
case. This formula, which we shortly prove, gives rise to an appealing
interpretation of the absolute distance. Using the identification of
events which may occur as measurement outcomes in a quantum system
with POVM elements -- positive operators $P \leq I$ -- we see that the
absolute distance is equal to twice the difference in probabilities
that an event $P$ may occur, depending on whether the state is $\rho$
or $\sigma$, maximized over all possible events $P$.

We prove this formula for the case where the maximization is over
projectors; the case of positive operators $P \leq I$ follows the same
reasoning. We begin by using the spectral decomposition of
$\rho-\sigma$ to write $\rho-\sigma = Q-S$, where $Q$ and $S$ are
positive operators with disjoint support. Note that $|\rho-\sigma| =
Q+S$, so $D(\rho,\sigma)=\tr(Q)+\tr(S)$. But
$\tr(Q-S)=\tr(\rho-\sigma)=0$, so $\tr(Q)=\tr(S)$, and therefore
$D(\rho,\sigma) = 2\tr(Q)$. Let $P$ be the projector onto the support
of $Q$. Then $2 \tr(P(\rho-\sigma)) = 2 \tr(P(Q-S)) = 2\tr(Q) =
D(\rho,\sigma)$. Conversely, let $P$ be any projector. Then
$2\tr(P(\rho-\sigma)) = 2\tr(P(Q-S)) \leq 2\tr(PQ)
\leq 2\tr(Q)=D(\rho,\sigma)$. This completes the proof. 

Perhaps the most important property of the absolute distance is that
it is a metric on the space of density operators. It is clear that
$D(\rho,\sigma) = 0$ if and only if $\rho=\sigma$, and that
$D(\cdot,\cdot)$ is a symmetric function of its inputs. The triangle
inequality,
\be
D(\rho,\tau) \leq D(\rho,\sigma)+D(\sigma,\tau), \ee
follows from the observation that there exist a projector $P$ such
that
\be
D(\rho,\tau) & = & 2 \tr(P(\rho-\tau)) \\
& = & 2 \tr(P(\rho-\sigma))+2 \tr(P(\sigma-\tau)) \\
& \leq & D(\rho,\sigma)+D(\sigma,\tau). \ee
This completes the proof that the absolute distance is a metric.

% concavity

The same method of proof can be used to show that the absolute
distance is doubly convex in its inputs,
\beqn
D(\sum_i p_i \rho_i, \sum_i p_i \sigma_i) \leq \sum_i p_i
D(\rho_i,\sigma_i). \eeqn
To see this, note that there exist a projector $P$ such that
\be
D(\sum_i p_i \rho_i,\sum_i p_i \sigma_i) & = & 2 \sum_i p_i
\tr(P(\rho_i-\sigma_i)) \\
& \leq & \sum_i p_i D(\rho_i,\sigma_i). \ee
Indeed, it is possible to prove a generalization of double convexity,
using the same line of reasoning. Let $p_i$ and $q_i$ be probability
distributions over the same index set. Then there exists a projector
$P$ such that
\be
D(\sum_i p_i \rho_i,\sum_i q_i \sigma_i) & = & 2\sum_i p_i \tr(P\rho_i)
- 2 \sum_i q_i \tr(P\sigma_i) \\
& = & 2 \sum_i p_i \tr(P(\rho_i-\sigma_i)) + 2 \sum_i
(p_i-q_i)\tr(P\sigma_i) \\
& \leq & \sum_i p_i D(\rho_i,\sigma_i) + 2 D(p_i,q_i), \ee
where $D(p_i,q_i)$ is the absolute distance between the probability
distributions $p_i$ and $q_i$.

% decreasing property

Suppose ${\cal E}$ is a complete quantum operation.  Let $\rho$ and
$\sigma$ be density operators. Then Ruskai \cite{Ruskai94a} has shown
that \beqn D({\cal E}(\rho),{\cal E}(\sigma)) \leq
D(\rho,\sigma). \eeqn That is, physical quantum operations are
contractive maps on the space of density operators. To prove this, use
the spectral decomposition to write $\rho-\sigma =Q-S$, where $Q$ and
$S$ are positive matrices with disjoint support, and let $P$ be a
projector such that $D({\cal E}(\rho),{\cal E}(\sigma)) = 2 \tr(P({\cal
E}(\rho)-{\cal E}(\sigma)))$. Note that \beqn D(\rho,\sigma) & = & \tr
|Q-S| \\ & = & \tr(Q)+\tr(S) \\ & = & \tr({\cal E}(Q))+\tr({\cal
E}(S)) \\ & = & 2 \tr({\cal E}(Q)) \\ & \geq & 2 \tr(P {\cal E}(Q)) \\
& \geq & 2 \tr(P ({\cal E}(Q)-{\cal E}(S))) \\ & = & 2 \tr(P({\cal
E}(\rho)-{\cal E}(\sigma))) \\
& = & D({\cal E}(\rho),{\cal E}(\sigma)), \eeqn which completes the proof.

Contractivity together with double convexity can be used to prove
results about the existence of stationary states for a quantum
operation. Suppose $\evop$ is a quantum operation for which there
exists a fixed density operator $\rho_0$ and a quantum operation
$\evop'$ such that
\beqn
\evop(\rho) = p\rho_0 + (1-p) \evop'(\rho), \eeqn
for some $p$, $0 < p \leq 1$. Physically, this means that with a
certain probability $p$, the input state is thrown out and replaced
with the fixed state $\rho_0$. With probability $1-p$, the operation
$\evop'$ is applied. An important example of a channel of this type is
the much-studied {\em depolarizing channel} for a qubit
\cite{Bennett96a}, which with probability $p$ randomizes the state,
that is, replaces it with the fixed operator $I/2$, and with
probability $1-p$ leaves the state untouched. By the double convexity
of the absolute distance, it follows that
\beqn
D(\evop(\rho),\evop(\sigma)) & \leq & p D(\rho_0,\rho_0) +
(1-p)D(\evop'(\rho),\evop'(\sigma)) \\ & \leq & (1-p) D(\rho,\sigma),
\ee where on the second line we have applied the contractivity of the
absolute distance with respect to physical quantum operations. Thus,
the class of quantum operations which have this form are {\em strictly
contractive}, and it is not difficult to see that they have a unique
fixed point; see the Lemma in Appendix One of \cite{Simmons63a} for a
proof. 

%Contractivity is a remarkable property of quantum operations. It
%implies that every quantum operation has a fixed point, via Brouwer's
%fixed point theorem \cite{Bredon93a} and the convexity of the set of
%density operators. This result, in turn, implies that every Lindblad
%operator must have at least one fixed point (physically, a stationary
%solution of the Lindblad equation), as we saw in Chapter
%\ref{chap:qops} that Lindblad operators are merely quantum operations
%in disguise.

% operational interpretation

We noticed earlier that the absolute distance has an interpretation as
half the maximal difference in probabilities that may arise from a
single measurement result on the two density operators. We now explore
a slightly different way of viewing the operational meaning of the
absolute distance. Suppose $M_m$ is a set of POVM elements describing
a measurement on the quantum system. Let $p_m \equiv \tr(\rho M_m)$
and $q_m \equiv \tr(\sigma M_m)$ be the probabilities associated with
the POVM measurement. Then $D(p_m,q_m) \leq D(\rho,\sigma)$. To see
this, note that \beqn D(p_m,q_m) & = & \sum_m
|\tr(M_m(\rho-\sigma))|. \eeqn Using the spectral theorem we may
decompose $\rho-\sigma = Q-S$, where $Q$ and $S$ are positive
operators with disjoint support. Thus $|\rho-\sigma| = Q+S$, and \beqn
|\tr(M_m(\rho-\sigma))| & = & |\tr(M_m(Q-S))| \\ & \leq & \tr(M_m
(Q+S)) \\ & \leq & \tr(M_m |\rho-\sigma|). \eeqn Thus \beqn D(p_m,q_m)
& \leq & \sum_m \tr(M_m |\rho-\sigma|) \\ & = & \tr(|\rho-\sigma|) \\
& = & D(\rho,\sigma), \eeqn where we have applied the completeness
relation for POVM elements, $\sum_m M_m = I$.

Thus, if two density operators are close in absolute distance, then
any measurement performed on those quantum states will give rise to
probability distributions which are close together in the classical
sense of absolute distance. Conversely, by choosing a measurement
whose POVM elements include projectors onto the support of $Q$ and
$S$, we see that there exist measurements which give rise to
probability distributions such that $D(p_m,q_m) = D(\rho,\sigma)$.

Thus, we have a second interpretation of the absolute distance between
two quantum states, as an achievable upper bound on the absolute
distance between probability distributions arising from measurements
performed on those quantum states.

% Fannes' inequality

\index{Fannes' inequality}

We conclude our survey of elementary properties of the absolute
distance with an elegant result linking the absolute distance to
entropy. This result is known as {\em Fannes' inequality}
\cite{Ohya93a}.  It states that for density operators $\rho$ and
$\sigma$ such that $D(\rho,\sigma) \leq 1/e$, 
\be
|S(\rho)-S(\sigma)| \leq D(\rho,\sigma) \log d + \eta(D(\rho,\sigma)),
\ee
where $d$ is the dimensionality of the underlying Hilbert space, and
$\eta(x) \equiv -x \log x$.

To prove Fannes' inequality we need a simple result relating the
absolute distance between two operators to their eigenvalues.  Let
$r_1 \geq r_2 \geq \ldots \geq r_d$ be the eigenvalues of $\rho$, in
descending order, with corresponding orthonormal eigenvectors
$|e_i\ra$, and $s_1 \geq s_2 \geq \ldots \geq s_d$ be the eigenvalues
of $\sigma$, again in descending order, with corresponding eigenvectors
$|f_i\ra$.  Then decompose $\rho-\sigma = Q-R$, where $Q$ and $R$ are
positive operators with disjoint support.  Defining $T \equiv R+\rho
=Q+\sigma$, we have
\be \label{eqtn:Truman}
D(\rho,\sigma) = \tr(R+Q) = \tr(2T)-\tr(R)-\tr(Q). \ee
Let $t_1 \geq t_2 \geq \ldots \geq t_d$ be the eigenvalues of
$T$. Note that $t_i \geq \max(r_i,s_i)$, so $2t_i \geq
r_i+s_i+|r_i-s_i|$. From equation (\ref{eqtn:Truman}) it follows that
\be \label{eqtn:Cristof}
D(\rho,\sigma) \geq \sum_i |r_i-s_i|, \ee
which is the relation we shall need to prove Fannes' inequality. 

To prove Fannes' inequality we use the inequality \be
|\eta(r)-\eta(s)| \leq \eta(|r-s|), \ee which may be easily verified
by calculus whenever $|r-s| \leq 1/2$.  A little thought shows that
$|r_i-s_i| \leq 1/2$ for all $i$, so \be |S(\rho)-S(\sigma)| & = &
\left| \sum_i (\eta(r_i)-\eta(s_i)) \right|\\ & \leq & \sum_i
\eta(|r_i-s_i|). \ee Setting $\Delta \equiv \sum_i |r_i-s_i|$, we see
that \be |S(\rho)-S(\sigma)| & \leq & \Delta
\eta(|r_i-s_i|/\Delta)+\eta(\Delta) \\ & \leq & \Delta \log d +
\eta(\Delta). \ee But $\Delta \leq D(\rho,\sigma)$ by
(\ref{eqtn:Cristof}), so by the monotonicity of $\eta$ on the interval
$[0,1/e]$, \be |S(\rho)-S(\sigma)| \leq D(\rho,\sigma) \log d +
\eta(D(\rho,\sigma)), \ee whenever $D(\rho,\sigma) \leq 1/e$, which is
Fannes' inequality.  A minor modification to the previous reasoning
shows that for general $D(\rho,\sigma)$, the slightly weaker form of
Fannes' inequality, \be |S(\rho)-S(\sigma)| \leq D(\rho,\sigma) \log d
+ \frac 1e, \ee holds.

\subsection{Fidelity}

\index{fidelity!quantum}

% fidelity

A second measure of distance between two quantum states is the {\em
fidelity}.  This subsection reviews the definition and basic
properties of the fidelity.  At the outset, it is well to mind that
the fidelity is not a true measure of distance, as it is not a metric,
but it does give rise to a metric, which will be reviewed in
the next subsection.

The fidelity of states $\rho$ and $\sigma$ is defined to
be\footnote{The reader ought to be aware that in the literature both
the quantity we call fidelity and its square have been referred to as
the fidelity. Compare also the definition of the dynamic fidelity
given below, in section \ref{sec:dynamic_measures}.}
\cite{Uhlmann76a,Jozsa94c,Fuchs96a}
\beqn \label{eqtn:fidelity}
F(\rho,\sigma) \equiv \tr \sqrt{\rho^{1/2} \sigma \rho^{1/2}}. \eeqn
Note that when $\rho$ and $\sigma$ are diagonal in the same basis,
this reduces to the classical fidelity between the eigenvalues of the
two states.

\index{Uhlmann's formula for fidelity}

There is a useful alternative characterization of the fidelity due to
Uhlmann \cite{Uhlmann76a}. Suppose we denote the quantum system where
our states live by the letter $Q$. Introduce another quantum system,
$R$, which is a copy of $Q$. Then, as discussed in Appendix
\ref{app:mixed}, for any mixed state $\rho$ of $Q$, it is possible to
find a pure state $|\psi\ra$ of $RQ$ such that $|\psi\ra$ extends
$\rho$ in the natural way. We call such a $|\psi\ra$ a {\em
purification} of $\rho$. It can be shown that
\beqn
F(\rho,\sigma) = \max_{|\psi\ra,|\phi\ra} | \la \psi|\phi\ra |,
\eeqn
where the maximization is performed over all purifications $|\psi\ra$
of $\rho$, and $|\phi\ra$ of $\sigma$. We will not prove this formula
here, but instead refer the reader to \cite{Fuchs96a} for an elegant
proof. There are several variants of this formula which are easily
seen to be equivalent. For instance, it is possible to fix any
purification $|\psi\ra$ of $\rho$, and simply maximize over
purifications of $\sigma$. Moreover, purifying the states $\rho$ and
$\sigma$ into the space $RQ$ was not necessary; any space large enough
to contain purifications of both $\rho$ and $\sigma$ will suffice.

Uhlmann's formula does not provide a calculational tool for evaluating
the fidelity, as does equation (\ref{eqtn:fidelity}). However, in many
instances, properties of the fidelity are more easily proved using
Uhlmann's formula than equation (\ref{eqtn:fidelity}).

Uhlmann's formula makes it clear that the fidelity is symmetric in its
inputs, $F(\rho,\sigma) = F(\sigma,\rho)$, and that the fidelity is
bounded between $0$ and $1$, $0 \leq F(\rho,\sigma) \leq 1$. If $\rho
= \sigma$ then it is clear that $F(\rho,\sigma) = 1$, from Uhlmann's
formula. If $\rho \neq \sigma$ then $|\psi\ra \neq |\phi\ra$ for any
purifications $|\psi\ra$ and $|\phi\ra$ of $\rho$ and $\sigma$,
respectively, so $F(\rho,\sigma) < 1$. From equation
(\ref{eqtn:fidelity}) we see that $F(\rho,\sigma)=0$ if and only if
$\rho$ and $\sigma$ have disjoint support.

Summarizing, the fidelity is symmetric in its inputs, $0 \leq
F(\rho,\sigma) \leq 1$, with equality in the first if and only if
$\rho$ and $\sigma$ have orthogonal support, and equality in the
second if and only if $\rho = \sigma$.

\index{fidelity!quantum!explicit formula}

There is a simple instance in which a useful explicit formula for the
fidelity may be given. Suppose we wish to calculate the fidelity
between a pure state $|\psi\ra$ and an arbitrary state, $\rho$. From
equation (\ref{eqtn:fidelity}) we see that
\be
F(|\psi\ra,\rho) & = & \tr \sqrt{\la \psi| \rho|\psi\ra \, |\psi\ra
\la \psi|} \\
& = & \sqrt{\la \psi| \rho |\psi\ra}. \ee
That is, the fidelity is equal to the square root of the overlap
between $|\psi\ra$ and $\rho$. This is an important result which we
will make much use of.

As already noted, the fidelity is not a metric. However, in many other
ways the fidelity closely resembles the absolute distance. The
remainder of this section is used to prove two results about the
fidelity which are analogous to properties already proved of the
absolute distance. These results concern, respectively, a strong
concavity result for the fidelity; and a proof that the fidelity can
not increase under quantum operations.

% fidelity of three operators.

%First, let us show that the fidelity satisfies an inequality similar
%in spirit to the triangle inequality satisfied by all metrics,
%including the absolute distance. Suppose $\rho, \sigma$ and $\tau$ are
%three density operators. We will show that\footnote{It is hard to look
%at this inequality and not be tempted to take logarithms. I have not
%investigated the quantity $-\log F(\rho,\sigma)$ in any detail, but
%this inequality and certain other elementary observations suggest
%that it may well be interesting to do so!}
%\be
%F(\rho,\tau) \geq F(\rho,\sigma) F(\sigma,\tau). \ee
%To see this, let
%$|\phi\ra$ be a purification of $\sigma$, and choose purifications
%$|\psi\ra$ and $|\eta\ra$ of $\rho$ and $\tau$, respectively, such
%that $F(\rho,\sigma) = |\la \psi| \phi\ra|$ and $F(\sigma,\tau) = |\la
%\phi|\eta\ra|$. Then $F(\rho,\tau) \geq |\la \psi|\eta\ra| \geq |\la
%\psi|\phi\ra \la \phi| \eta\ra| = F(\rho,\sigma)F(\sigma,\tau)$, which
%establishes the result.

First, we examine the concavity properties of the fidelity. We will
use the Uhlmann formula for fidelity to prove a strong concavity
property for the fidelity. Let $p_i$ and $q_i$ be probability
distributions over the same index set, and $\rho_i$ and $\sigma_i$
density operators also indexed by the same index set. Then \be
F(\sum_i p_i \rho_i, \sum_i q_i \sigma_i) \geq \sum_i \sqrt{p_i q_i}
F(\rho_i,\sigma_i). \ee To see this, let $|\psi_i\ra$ and $|\phi_i\ra$
be purifications of $\rho_i$ and $\sigma_i$ chosen such that
$F(\rho_i,\sigma_i) = \la \psi_i|\phi_i\ra$. Introduce a system $I$
which has orthonormal basis states $|i\ra$ corresponding to the index
set $i$ for the probability distributions. Define \beqn |\psi\ra &
\equiv & \sum_i \sqrt{p_i} |\psi_i\ra|i\ra \\ |\phi\ra & \equiv &
\sum_i \sqrt{q_i} |\phi_i\ra|i\ra. \eeqn Note that $|\psi\ra$ is a
purification of $\sum_i p_i \rho_i$ and $|\phi\ra$ is a purification
of $\sum_i q_i \sigma_i$, so by Uhlmann's formula, \beqn F(\sum_i p_i
\rho_i,\sum_i q_i \sigma_i) & \geq & |\la \psi|\phi\ra| \\ & = &
\sum_i \sqrt{p_i q_i} \la \psi_i|\phi_i\ra \\ & = & \sum_i \sqrt{p_i
q_i} F(\rho_i,\sigma_i), \eeqn which establishes the result we set out
to prove. We refer to this result as the {\em strong concavity} of the
fidelity.

The strong concavity of the fidelity has a number of useful
consequences. One is the {\em joint concavity} of the fidelity. In
particular, note that if $p_i = q_i$, then strong concavity reduces to
\beqn
F(\sum_i p_i \rho_i,\sum_i p_i \sigma_i) \geq \sum_i p_i
F(\rho_i,\sigma_i). \eeqn
The joint concavity, in turn, implies that the fidelity is concave in
each entry. For example, for each $i$ set $\rho_i = \rho$ for some
fixed $\rho$. Then the joint concavity of the fidelity reduces to
\be
F(\rho,\sum_i p_i \sigma_i) \geq \sum_i p_i F(\rho,\sigma_i), \eeqn
that is, the fidelity is concave in the second entry. By symmetry, the
fidelity is also concave in the first entry.

The second property of the fidelity which we prove is that it is
non-decreasing under complete quantum operations \cite{Barnum96a},
\be
F(\evop(\rho),\evop(\sigma)) \geq F(\rho,\sigma). \ee To prove this,
let $|\psi\ra$ and $|\phi\ra$ be purifications of $\rho$ and $\sigma$
into a joint system $RQ$ such that $F(\rho,\sigma) = |\la
\psi|\phi\ra|$. Introduce a model environment $E$ for the quantum
operation, $\evop$, which starts in a pure state $|0\ra$, and
interacts with the quantum system $Q$ via a unitary interaction
$U$. Note that $U|\psi\ra|0\ra$ is a purification of $\evop(\rho)$,
and $U|\phi\ra|0\ra$ is a purification of $\evop(\sigma)$. By
Uhlmann's formula it follows that
\be
F(\evop(\rho),\evop(\sigma)) & \geq & |\la \psi |\la 0| U^{\dagger} U
|\phi\ra|0\ra | \\
& = & |\la \psi| \phi\ra| \\
& = & F(\rho,\sigma), \ee
establishing the property that we set out to prove.

This completes our discussion of elementary properties of the
fidelity. Note that the fidelity and the absolute distance have many
similar properties, although I am not aware of any simple physical
interpretation of the fidelity. Why do we bother developing both
quantities? We do so because it often helps to have more than one way
of doing things; one obtains new insights from multiple ways of
viewing the same phenomena.  It is also potentially the case that in
the future a powerful property of one of these quantities will be
found that has no natural analogue which applies to the other
quantity. Indeed, I have seen fit to discuss both the fidelity and the
absolute distance in this Chapter because most of the research later
in the Dissertation has been carried out using the fidelity as a tool,
while now it seems to me that the absolute distance has a more
compelling physical interpretation, and is equally powerful
mathematically.

\subsection{Distance measures derived from fidelity}

The fidelity may be used to develop many other useful measures of
distance between density operators.  This subsection develops two
natural measures of distance derived from the fidelity, the {\em
error} and the {\em angle}, and develops some elementary properties of
these measures, most importantly, the fact that the angle is a
metric on the space of density operators.

% The error

\index{error}

Given that the fidelity is bounded between $0$ and $1$, and is equal
to one if and only if the states being compared are equal, the most
obvious candidate for a metric is the function defined by
\beqn
\tilde E(\rho,\sigma) \equiv 1-F(\rho,\sigma). \eeqn
It turns out, however, that it is slightly more convenient to define
the function
\beqn
E(\rho,\sigma) \equiv 1-F(\rho,\sigma)^2,\ee
which we shall refer to
as the {\em error} for $\rho$ and $\sigma$.  Both functions are
metrics on density operators, and have many other nice properties,
however it turns out that the error
function has properties that will be of especial use in the study of
dynamic measures of distance, properties which $\tilde E$ does not
have. The error has numerous
useful properties which it inherits from the fidelity:
\begin{enumerate}
\item $E(\rho,\sigma) = 0$ if and only if $\rho = \sigma$.
\item Symmetry. $E(\rho,\sigma) = E(\sigma,\rho)$.
%\item The triangle
%inequality, $E(\rho,\tau) \leq E(\rho,\sigma)+E(\rho,\tau)$. To see
%this, note that by calculus, $F(\rho,\sigma)^2+F(\sigma,\tau)^2 \leq
%1+F(\rho,\sigma)^2F(\rho,\tau)^2$. Recall that $F(\rho,\sigma)
%F(\sigma,\tau) \leq F(\rho,\tau)$, so that
%$F(\rho,\sigma)^2+F(\sigma,\tau)^2 \leq 1+F(\rho,\tau)^2$, which is easily
%seen to be equivalent to the triangle inequality for error. Together
%with properties 1 and 2 this establishes that the error is a metric on
%the space of density operators.
\item Let $\evop$ be a complete quantum operation. Then
\be
E(\evop(\rho),\evop(\sigma)) \leq E(\rho,\sigma). \ee
\end{enumerate}

%It is a slightly unfortunate fact that, while both the fidelity and
%the fidelity squared have many nice properties, not all those
%properties are shared by both quantities. In the next section, on
%dynamic measures of distance, it will be convenient to make more use
%of the fidelity squared, rather than the fidelity. Anticipating this,
%we pause here to note that the quantity $E(\rho,\sigma) \equiv
%1-F(\rho,\sigma)^2$ is also a metric on density operators, and contractive
%under complete quantum operations. The proofs of these facts are
%identical to the proof for the error.

\index{angle}

The error will assume a significant role in later discussions of
dynamic measures of distance, and throughout the remainder of this
Dissertation.  We now switch to a second measure of distance derived
from the fidelity, the {\em angle}. This measure will play a much
lesser role in the remainder of this Dissertation. It is included here
as a teaser to indicate just one of the wide variety of natural
directions which research into distance measures for quantum states
may take. 

Recall Uhlmann's formula, that the fidelity between two states is
equal to the maximum inner product between purifications of those
states. Recall that in Cartesian geometry the inner product between
two unit vectors has an interpretation as the cosine of the angle
between these states. This suggests that we define the {\em
generalized angle} between states $\rho$ and $\sigma$ by
\beqn
A(\rho,\sigma) \equiv \arccos F(\rho,\sigma). \ee The generalized
angle, which we will usually refer to just as the angle, is a real
number in the range $0$ to $\pi/2$. The angle is also a true distance
measure on density operators. The following is a summary of the
elementary properties of the angle, each of which is immediate from
properties of the fidelity, together with the observations from
calculus that $\arccos$ is a decreasing concave function on the
interval $[0,1]$. Where this is not the case a brief proof is given.
\begin{enumerate}
\item $A(\rho,\sigma) = 0$ if and only if $\rho = \sigma$.
\item Symmetry. $A(\rho,\sigma) = A(\sigma,\rho)$.
\item $A(\rho,\sigma)$ satisfies the triangle inequality, and
therefore is a metric.  This is immediate from Uhlmann's formula, and
the definition of the angle.
%
%
%Recall that $F(\rho,\tau) \geq
%F(\rho,\sigma)F(\sigma,\tau)$. By the decreasing property of the
%$\arccos$ function,
%\be
%A(\rho,\tau) \leq \arccos(F(\rho,\sigma)F(\sigma,\tau)). \ee
%Calculus shows that
%\be
%\arccos ( F(\rho,\sigma) F(\sigma,\tau)) \leq
%A(\rho,\sigma)+A(\sigma,\tau), \ee
%which completes the proof.
\item Let $\evop$ be a complete quantum operation. Then
\be
A(\evop(\rho),\evop(\sigma)) \leq A(\rho,\sigma). \ee
\end{enumerate}

\subsection{Relationships between distance measures}

% Relationships between fidelity and absolute distance

There are several useful relationships between the absolute distance
and the fidelity. 

Consider the absolute distance between two pure states, $|a\ra$ and
$|b\ra$. Introduce orthonormal states $|0\ra$ and $|1\ra$ such that
$|a\ra = |0\ra$ and $|b\ra = \cos \theta |0\ra + \sin \theta
|1\ra$. Notice that $F(|a\ra,|b\ra) = |\cos \theta|$. Furthermore,
\be
D(|a\ra,|b\ra) & = & \tr \left| \left[ \begin{array}{cc} 1-\cos^2
\theta & -\cos \theta \sin \theta \\ - \cos \theta \sin \theta &
\sin^2 \theta \end{array} \right] \right| \\ & = & 2 | \sin \theta |
\\ & = & 2 \sqrt{1-F(|a\ra,|b\ra)^2} = 2 \sqrt{E(|a\ra,|b\ra)}. \eeqn
Let $\rho$ and $\sigma$ be any two quantum states, and let $|\psi\ra$
and $|\phi\ra$ be purifications chosen such that $F(\rho,\sigma) =
|\la \psi|\phi\ra| = F(|\psi\ra,|\phi\ra)$. Recalling that absolute
distance is non-increasing under the partial trace, we see that
\be
D(\rho,\sigma) & \leq & D(|\psi\ra,|\phi\ra) \\ & = & 2
\sqrt{1-F(\rho,\sigma)^2} = 2\sqrt{E(\rho,\sigma)}. \eeqn Thus, if the
error between two states is small, it will follow that the states are
also close in absolute distance. The converse is also true, at least
when one of the two states is a pure state, which will be sufficient
for the applications we shall consider. Let $|\psi\ra$ be a pure
state, and $\sigma$ an arbitrary state. Then
\be
D(|\psi\ra,\sigma) & = & 2 \max_P \tr(P(|\psi\ra \la \psi|-\sigma)) \\
& \geq & 2 \tr(|\psi\ra\la \psi|(|\psi\ra\la \psi|-\sigma)) \\
& = & 2(1-F(|\psi\ra,\sigma)^2). \ee
Restating these bounds in term of the error, we see that
\be \label{eqtn:distance_vs_fidelity}
2 E(|\psi\ra,\sigma) \leq D(|\psi\ra,\sigma)) \leq
2\sqrt{E(|\psi\ra,\sigma)}. \ee The implication of this relation is
that when one of the inputs is a pure state, and the other state is
arbitrary, the absolute distance and the error are equally good
measures of closeness for quantum states, at least in terms of their
limiting behaviours. Part II of this Dissertation is largely concerned
with such limiting behaviours; in such instances this relation implies
that it does not matter whether the error or the absolute distance is
used as a measure of distance, since any result about one will imply a
qualitatively similar result about the other.

\section{Dynamic measures of information preservation}
\label{sec:dynamic_measures}

%\index{dynamic measures!quantum}

This section uses the static measures of distance discussed in
previous sections to develop several measures of how well a quantum
operation preserves information.  A major concern of this Dissertation
is the transmission of entangled states through quantum channels, so
we will focus on measures related to this problem.
%, touching briefly on
%other dynamic measures in the final section of this Chapter.

\begin{figure}
\begin{center}
\unitlength 1cm
\begin{picture}(8,4)(0,2.5)
% Q --> Q'
\put(2,3){\framebox(1,1){$Q$}}
\put(3,3.5){\vector(1,0){3}}
\put(6,3){\framebox(1,1){$Q'$}}
%
% add R plus initial RQ state
\put(2,5){\framebox(1,1){$R$}}
\put(2.4,4.10){\line(0,1){.05}} \put(2.6,4.10){\line(0,1){.05}}
\put(2.4,4.25){\line(0,1){.05}}  \put(2.6,4.25){\line(0,1){.05}}
\put(2.4,4.40){\line(0,1){.05}}  \put(2.6,4.40){\line(0,1){.05}}
\put(2.4,4.55){\line(0,1){.05}}  \put(2.6,4.55){\line(0,1){.05}}
\put(2.4,4.70){\line(0,1){.05}}  \put(2.6,4.70){\line(0,1){.05}}
\put(2.4,4.85){\line(0,1){.05}}  \put(2.6,4.85){\line(0,1){.05}}
\put(1,4.5){\makebox(0,0){$\left | RQ \right \rangle$}}
\end{picture}
\end{center}
\caption{The $RQ$ picture of a quantum channel. The initial state of
$RQ$ is a pure state. \label{fig:RQ}}
\end{figure}
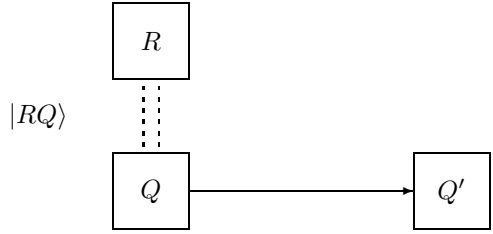

We will primarily be interested in the following model scenario. A
quantum system, $Q$, is prepared in a state $\rho$. The state of $Q$
is entangled in some way with the external world. We represent
this entanglement by introducing a fictitious system $R$, such that
the joint state of $RQ$ is a pure state. It turns out that all results
that we prove do not depend in any way on how this purification is
performed, so we may as well suppose that this is an arbitrary
entanglement with the outside world. The system $Q$ is then subjected
to a dynamics described by a quantum operation, ${\cal E}$. The basic
situation is illustrated in figure \ref{fig:RQ}.

\index{dynamic distance}

How well is the entanglement between $R$ and $Q$ preserved by the
quantum operation ${\cal E}$? We investigate two ways of quantifying
this.  The first measure we refer to as the {\em dynamic distance}. It
is defined by the expression
\beqn
D(\rho,{\cal E}) \equiv D(RQ, R'Q'), \eeqn where the use
of a prime indicates the state of a system after the quantum operation
has been applied, and the absence of a prime indicates the state of a
system before the quantum operation has been applied. Note that this
expression depends only upon $\rho$ and $\evop$, and not upon the
details of the purification $RQ$. To see this, we use the fact, proved
in Appendix \ref{app:mixed}, that any two purifications $R_1Q_1$ and
$R_2Q_2$ of $\rho$ are related by a unitary operation, $U$, {\em that
acts upon $R$ alone}, $R_2Q_2 = U (R_1Q_1) U^{\dagger}$. Thus
\be
D(R_2Q_2,R_2'Q_2') & = & \tr|R_2Q_2-R_2'Q_2'|\\
& = & \tr(U |R_1Q_1-R_1'Q_1'|U^{\dagger}) \\
& = &  \tr|R_1Q_1-R_1'Q_1'| \\
& = & D(R_1Q_1,R_1'Q_2'), \ee
which establishes the result. Notice that the dynamic distance
provides a measure of how well the entanglement between $Q$ and $R$ is
preserved by the process, with values close to zero indicating that
the entanglement has been very well preserved, and larger values
indicating that it has not been so well preserved.

\index{dynamic fidelity}
\index{entanglement fidelity}

The second measure\footnote{Historically, this measure was introduced
earlier than the dynamic distance. It was introduced by Schumacher
\cite{Schumacher96a}.} quantifying how well the entanglement is
preserved is known as the {\em entanglement fidelity}, although we
will usually refer to it just as the {\em fidelity}, or the {\em
dynamic fidelity}.  As for the dynamic distance, the dynamic fidelity
is defined for a {\em process}, specified by a quantum operation
${\cal E}$ acting on some initial state, $\rho$.  We denote it by
$F(\rho,{\cal E})$.
% The
%concerns motivating the definition of the dynamic fidelity are
%twofold:
%\begin{enumerate}
%
%\item $F(\rho,{\cal E})$ measures how well the {\em state},
%$\rho$, is preserved by the operation ${\cal E}$. An dynamic fidelity
%close to one indicates that the process preserves the state well.
%
%\item $F(\rho,{\cal E})$ measures how well the {\em entanglement} of $\rho$
%with other systems is preserved by the operation ${\cal E}$. An
%dynamic fidelity close to one indicates the process preserves the
%entanglement well.
%
%\end{enumerate}
%Conversely, a fidelity close to zero indicates that the state or its
%entanglement were not well preserved by the operation ${\cal E}$.

Formally, the dynamic fidelity is defined by
\begin{eqnarray}
F(\rho,{\cal E}) & \equiv & F(RQ, R'Q')^2 \\
 & = & \frac{\langle RQ| ({\cal I}_R \otimes {\cal E})
	(|RQ\rangle \langle RQ|) |RQ \rangle}{\tr ({\cal I}_R \otimes
 \evop)(|RQ\ra\la RQ|)},
\end{eqnarray}
where the quantity appearing on the right hand side is the static
fidelity between the initial and final states of $RQ$, {\em
squared}. Thus, the dynamic fidelity provides a measure of how well
the entanglement between $R$ and $Q$ is preserved by the process
$\evop$, with values close to $1$ indicating that the entanglement has
been well preserved, and values close to $0$ indicating that most of
the entanglement has been destroyed. The choice of whether to use the
static fidelity squared or the static fidelity is essentially
arbitrary; the present definition seems to result in slightly more
attractive mathematical properties. Recall that the fidelity is not
known to have a clear physical interpretation, so in this instance it
is legitimate for us to make a decision based upon mathematical
elegance, rather than physical necessity. To go with the dynamic
fidelity, we define the {\em dynamic error}\index{dynamic error} in a
manner analogous to the earlier definition of the static error, \be
E(\rho,\evop) \equiv 1-F(\rho,\evop). \ee Note that there is no
square appearing on the right hand side, since the dynamic fidelity
is already a fidelity squared. Thus, \be E(\rho,\evop) = E(RQ,R'Q'),
\ee where the $E(\cdot,\cdot)$ appearing on the right hand side is the
static error introduced in the previous section.

Note further that the dynamic fidelity and dynamic error do not depend
upon the particular purification $RQ$ of $Q$ that is chosen, but only
upon the initial state $Q$, and the quantum operation $\evop$
\cite{Schumacher96a}. The proof of this is as for the proof of the
dynamic distance, which was adapted from \cite{Schumacher96a}. 

Suppose $E_i$ is a set of operation elements for a quantum operation
$\evop$. Then
\beqn \label{eqtn:Dolittle}
F(\rho,\evop) = \la RQ|R'Q'|RQ\ra = \frac{\sum_i |\la RQ| E_i
|RQ\ra|^2}{\tr(\evop(\rho))}. \eeqn Note that
\beqn
\la RQ| E_i |RQ\ra & = &\sum_{jk} \sqrt{p_j p_k} \la j|k\ra \la
j|E_i|k\ra \\ & = & \sum_j p_j \la j| E_i |j\ra \\ & = & \tr(E_i
\rho). \eeqn Combining this expression with equation
(\ref{eqtn:Dolittle}) we obtain the useful computational formula
\cite{Schumacher96a,Nielsen98a}
\beqn \label{eqtn:comp_fidelity}
F(\rho,\evop) = \frac{\sum_i |\tr(\rho
E_i)|^2}{\tr(\evop(\rho))}. \eeqn This expression simplifies for
trace-preserving quantum operations since the denominator is one. The
dynamic fidelity has the following properties, with corresponding
properties of the dynamic error obvious corollaries:
\begin{enumerate}
\item $0 \leq F(\rho,{\cal E}) \leq 1$ \cite{Schumacher96a}. Follows
from properties of the static fidelity.

\item The dynamic fidelity is convex in the density operator input, and linear
in the quantum operation input, for complete quantum operations
\cite{Schumacher96a,Barnum98a}. The linearity is immediate from the
definition of the dynamic fidelity. The convexity may be proved in
many ways; one simple technique is to use equation
(\ref{eqtn:comp_fidelity}) to show that the function $f(x) \equiv
F(x\rho_1+(1-x)\rho_2,\evop)$ has a non-negative second
derivative. Elementary calculus shows that
\be
f''(x) = \sum_i | \tr((\rho_1-\rho_2)E_i)|^2 \geq 0, \ee
as required.

\item Unfortunately, the dynamic fidelity is not jointly convex
in the density operator and the quantum operation. To see this,
consider the following example on a space spanned by orthonormal
states $|0\ra, |1\ra, |2\ra, |3\ra$. Let $P_0,P_1,P_2,P_3$ denote
projections onto the corresponding one dimensional subspaces, and
$P_{12}$ and $P_{34}$ projections onto the corresponding two
dimensional subspaces. Define complete quantum operations $\evop_1$
and $\evop_2$ by $\evop_1(\rho) \equiv P_1 \rho P_1 + P_2 \rho P_2 +
P_{34} \rho P_{34}$ and $\evop_2(\rho) \equiv P_{12} \rho P_{12} + P_3
\rho P_3 + P_4 \rho P_4$. Then
\beqn
F\left( \frac{I}{4}, \frac{\evop_1}{2} + \frac{\evop_2}{2} \right) =
\frac{3}{8} \not \leq \frac{1}{4} = \frac 12 F\left(
\frac{P_{12}}{2},\evop_1 \right) + \frac 12 F\left(
\frac{P_{34}}{2},\evop_2 \right), \ee
so this is the desired counterexample.

\item The dynamic fidelity is a lower bound on the static fidelity
squared between the input and output to the process \cite{Schumacher96a},
\begin{eqnarray}
F(\rho,{\cal E}) \leq \left[ F(\rho,{\cal E}(\rho)/
\tr(\evop(\rho))) \right]^2. \end{eqnarray} The proof is an elementary
application of the non-decreasing property of the static fidelity
under partial trace, $F(\rho,\evop) = F(RQ,R'Q')^2 \leq
F(Q,Q')^2$. The intuitive meaning of the result is obviously a
desirable property: it is harder to preserve a state plus entanglement
with the outside world than it is to merely preserve the state alone.

\item For pure state inputs, the dynamic fidelity is equal to the
static fidelity squared between input and output, \be
F(|\psi\ra,\evop) = F(|\psi\ra,\evop(|\psi\ra\la
\psi|)/\tr(\evop(|\psi\ra\la\psi|)))^2. \eeqn This is immediate from
the observation that the state $|\psi\ra$ is a purification of itself,
and the definition.

\item Using convexity and the result that the dynamic fidelity is a
lower bound on the static fidelity squared, we see that if $\{ p_i,
\rho_i \}$ is an ensemble of states generating $\rho$ then
\be
F(\rho,\evop) \leq \sum_i p_i F(\rho_i, \evop(\rho_i))^2, \ee
for a complete quantum operation $\evop$.

\item \label{item:TaDa}
$F(\rho,{\cal E}) = 1$ if and only if for all pure states
$|\psi\rangle$ lying in the support of $\rho$,
\begin{eqnarray}
{\cal E}(|\psi\rangle \langle \psi|) = |\psi\rangle \langle \psi|.
\end{eqnarray}
Suppose $F(\rho,\evop) = 1$, and $|\psi\ra$ is a pure state in the
support of $\rho$. Define $p \equiv \la \psi|\rho|\psi\ra > 0$ and
$\sigma$ to be a density operator such that $(1-p)\sigma =
\rho-p|\psi\ra\la \psi|$. Then by convexity,
\be
1 = F(\rho,\evop) \leq p F(|\psi\ra,\evop) +(1-p), \ee and thus
$F(|\psi\ra,\evop)=1$, establishing the result one way. The other way
is a straightforward application of the definition of the dynamic
fidelity.  This result was proved in \cite{Schumacher96b}, via a
different technique, based upon the next property in this list.

\item The following result, due to Knill and Laflamme \cite{Knill97a},
is essentially a strengthening of the previous result. Suppose that
$\langle \psi|{\cal E}(|\psi\rangle \langle \psi |) |\psi\rangle
\geq 1 - \eta$ for all $|\psi\rangle$ in the support of $\rho$, for
some $\eta$. Then
$F(\rho, {\cal E}) \geq 1 - (3\eta/2).$ 
\end{enumerate}

Earlier we derived equation (\ref{eqtn:distance_vs_fidelity}) which
related the absolute distance and the fidelity in the static case,
when one of the inputs was a pure state. This set of inequalities
immediately implies a corresponding set of inequalities for the
dynamic distance and dynamic error,
\be
2 E(\rho,\evop) \leq D(\rho,\evop) \leq 2\sqrt{E(\rho,\evop)}. \ee This
is a very useful result. First, it tells us that the dynamic distance
and dynamic error are essentially equivalent as measures of how well
entanglement is preserved when it undergoes a process, that is, the
dynamic distance is small if and only if the dynamic error is
small. Second, the result gives us alternate means for investigating
quantum channels: if the dynamic error, for instance, is proving to be
difficult to use, one can switch to the dynamic distance in an effort
to simplify the analysis. If successful, the results obtained using
the dynamic distance can then be translated back in terms of the
dynamic error. Two measures are better than one.

\subsection{Continuity relations}

\index{continuity properties of distance measures}

What continuity properties are possessed by the dynamic measures of
information preservation? Naturally, we expect that if the input to a
quantum process is perturbed slightly, then the distance measures
associated with that process should only change by a small amount. In
this section we give a bound on the extent to which this is true.

We will call such relations {\em continuity relations}, although this
is perhaps a slightly dubious coinage. After all, the distance
measures being investigated are defined in terms of the self-same
metrics with respect to which we are investigating their continuity
properties. 
By the triangle inequality and the non-increasing property of the
absolute distance under quantum operations, \be D(\rho_1,\evop) & = &
D(R_1Q_1,R_1'Q_1') \\ & \leq &
D(R_1Q_1,R_2Q_2)+D(R_2Q_2,R_2'Q_2')+D(R_2'Q_2',R_1'Q_1') \\ & \leq &
2D(R_1Q_1,R_2Q_2)+D(\rho_2,\evop). \ee Minimizing over purifications
of $\rho_1$ and $\rho_2$ we obtain \be D(\rho_1,\evop) \leq
D(\rho_2,\evop) + 4\sqrt{E(\rho_1,\rho_2)} . \eeqn Thus, if $\rho_1$
is close to $\rho_2$, as measured by the fidelity, then
$D(\rho_1,\evop)$ and $D(\rho_2,\evop)$ must also be close together.

\subsection{Chaining quantum errors}

\index{chaining properties of distance measures}

Suppose we have a composite quantum process generated by quantum
operations $\evop_1$ and $\evop_2$,
\beqn
\rho \stackrel{\evop_1}{\longrightarrow} \rho'
	\stackrel{\evop_2}{\longrightarrow} \rho''.
\eeqn
Is there a way of relating the absolute distance associated with the
complete two-part process with the absolute distance associated with
the component processes? We will show that
\be
D(\rho,\evop_2 \circ \evop_1) \leq D(\rho,\evop_1) + D(\rho',\evop_2). \ee
Thus, in order for there to be little error in the composite process
$\evop_2 \circ \evop_2$, it suffices that there be little error caused
by the process $\evop_1$ or $\evop_2$.

To see this, introduce a system $R_2'$ which purifies the system
$R'Q'$. Then notice that
\beqn
D(R'Q',R''Q'') \leq D(R'R_2'Q',R''R_2''Q'')  = D(\rho',\evop_2). \eeqn
Thus
\be
D(\rho,\evop_2 \circ \evop_1) & = & D(RQ,R''Q'')  \\
	& \leq & D(RQ,R'Q') +D(R'Q',R''Q'') \\
	& \leq & D(\rho,\evop_1) + D(\rho',\evop_2), \ee
as we set out to demonstrate.

%In exactly the same way, we can prove the following result for the
%dynamic error,
%\be
%E(\rho,\evop_2 \circ \evop_1) \leq E(\rho,\evop_1) +
%E(\rho',\evop_2). \ee
%This result can be given an obvious translation in terms of the
%dynamic fidelity. A somewhat less obvious translation arises from
%the result for static fidelity, $F(\rho,\tau) \geq F(\rho,\sigma) F(\sigma,\tau)$:
%\beqn
%F(\rho,\evop_2 \circ \evop_1) & = & F(RQ,R''Q'')^2 \\
%	& \geq & F(RQ,R'Q')^2 F(R'Q',R''Q'')^2 \\
%	& \geq & F(\rho,\evop_1) F(R'R_2'Q',R''R_2''Q'')^2 \\
%	& = & F(\rho,\evop_1) F(\rho',\evop_2). \ee

This result about the chaining behaviour of errors will not be used
much later in the Dissertation. Nevertheless, it is very important
conceptually. Essentially, it tells us in a quantitative way that if
we want to ensure that a complicated quantum process is is carried out
well, then it is sufficient to ensure that each step of that process
is carried out reliably. 
%At present we have shown this only for the
%state preservation process; later we will prove the result for any
%sequence of quantum gates.

\section{Alternative view of the dynamic measures}

In the previous section we presented several dynamic measures of
quantum information preservation based upon the $RQ$ picture of a
quantum process. In this picture, the state of a quantum system, $Q$,
is first purified into a fictitious quantum system, $R$. $R$ is used
to represent the possibility that $Q$ is entangled with another
system. 

In this section we will give a more obviously physical account of the
dynamic measures of information. To avoid repetition, this account
will be phrased entirely in terms of the dynamic error; identical
arguments apply to the dynamic distance and dynamic fidelity.  The
account is based upon \cite{Nielsen96c}.

The scenario we wish to consider is that of a system which is part of
another, possibly much larger, system. For example, we may be
interested in the performance of a single qubit memory element in a
large quantum computer.  One could argue that what should be done is
to look at the fidelity of the total system -- qubit plus
computer. However, in general, quantum computers can be very large
systems compared to the subsystem whose performance as a memory
element we wish to analyze, and inclusion of the entire state and
dynamics of the quantum computer would make the analysis enormously
complicated. 

Given that we do not wish to analyze the complete dynamics of the
total system, the natural thing to do is to define a quantity which
captures the worst-case error possible in the system. We define
\begin{eqnarray}
E_1(\rho, {\cal E}) \equiv \max_{\tilde \rho, {\cal E}'} E \Bigl (
	 (\evop' \otimes {\cal I}_Q ) (\tilde \rho), ({\cal E}' \otimes
	 {\cal E})(\tilde \rho) \Bigr ), \end{eqnarray} where the
	 maximization is over all extensions $\tilde \rho$ of $\rho$
	 to larger systems $RQ$, and all possible complete quantum
	 operations ${\cal E}'$ that could occur on $R$. $E_1$ is a
	 measure of how well the subsystem plus its entanglement with
	 the remainder of the system is stored. Note especially that
	 the initial state $RQ$ is not necessarily a pure state; it
	 can be any extension of $\rho$ whatsoever.

We maximize over all possible extensions and dynamics for the
remainder of the system in order to obtain the {\em worst possible
value} the error could have, regardless of the actual state or
dynamics of the remainder of the system, $R$. The advantage of this is
that this quantity depends only on the part of the computer, $Q$,
under consideration, not on the detailed dynamics and state of the
entire computer.

A second, related, quantity is also a useful measure of how well a
system plus entanglement is stored. It will turn out that this
quantity is equal to $E_1$. Define
\begin{eqnarray}
E_2(\rho,{\cal E}) := \max_{\tilde \rho} E \Bigl ( \tilde \rho,({\cal
	I}_R \otimes {\cal E})(\tilde \rho) \Bigr ). \end{eqnarray}
	The motivation for this quantity is similar to that for $E_1$,
	except now we assume that $R$ is subject to the identity
	dynamics ${\cal I}$, instead of maximizing over all possible
	dynamics ${\cal E}'$ for $R$.

To see that $E_1$ and $E_2$ are equal, note that
\begin{eqnarray}
E_1(\rho, {\cal E}) \geq E_2(\rho,{\cal E}), \end{eqnarray}
since the maximization in $E_1$ clearly includes all the values being maximized
over for $E_2$. To see the reverse inequality, notice that
\begin{eqnarray}
E \Bigl ( \tilde \rho, ({\cal I}_R \otimes \evop)(\tilde \rho) \Bigr ) \geq
	 E \Bigl ( ({\cal E}' \otimes {\cal I}_Q)(\tilde \rho),
	 ({\cal E}' \otimes {\cal E})(\tilde \rho) \Bigr ), \end{eqnarray}
by the non-increasing property of the error under quantum operations, and thus
\begin{eqnarray}
E_2(\rho,{\cal E}) \leq E_1(\rho,{\cal E}). \end{eqnarray}
It follows that
\begin{eqnarray}
E_1(\rho,{\cal E}) = E_2(\rho,{\cal E}). \end{eqnarray}

A similar argument can be used to show that $E(\rho,\evop) =
E_2(\rho,\evop)$. First, note that $E_2(\rho,\evop) \geq
E(\rho,\evop)$, by choosing the initial extension $RQ$ to be a
purification of $Q$. Second, $E_2(\rho,\evop) \leq E(\rho,\evop)$, by
the non-increasing property of the error under partial traces, which
completes the proof.

It follows that \be E(\rho,\evop) = \max_{\rho,\evop'} E \left(
(\evop' \otimes {\cal I}_Q)(\tilde \rho), (\evop' \otimes
\evop)(\tilde \rho) \right). \ee This expression brings home the
operational meaning of the dynamic error in a way that is, perhaps,
somewhat more compelling than the original abstract definition in terms
of purifications, because it emphasizes the dynamic error as a
quantity which arises as a \emph{worst-case} scenario in contexts
where preservation of entanglement may be important.

\vspace{1cm}
\begin{center}
\fbox{
\parbox{14cm}{
\begin{center}
{\bf Summary of Chapter \ref{chap:distance}: Distance measures for
quantum information}
\end{center}

\begin{itemize}

\item {\bf Absolute distance:} $D(\rho,\sigma) \equiv
\tr|\rho-\sigma|$. Doubly convex metric on density operators,
contractive under quantum operations.
\item {\bf Fidelity:} 
$$F(\rho,\sigma) \equiv \tr\sqrt{\rho^{1/2}\sigma
\rho^{1/2}} = \max_{|\psi\ra,|\phi\ra} |\la \psi|\phi\ra|.$$
Strongly concave, $F(\sum_i p_i \rho_i,\sum_i q_i \sigma_i) \geq
\sum_i \sqrt{p_iq_i} F(\rho_i,\sigma_i)$.
%\item {\bf Error:} $E(\rho,\sigma) = 1-F(\rho,\sigma)^2$. Doubly convex
%metric on density operators, contractive under quantum operations.

\item \textbf{Dynamic fidelity and dynamic distance:} $F(\rho,\evop)$
and $D(\rho,\evop)$. Measure how well entanglement is preserved during
a quantum mechanical process, starting with the state $\rho$ of a
system $Q$, which is assumed to be entangled with another quantum
system, $R$, and applying the quantum operation $\evop$ to system $Q$.

\item {\bf Chaining of errors:} $E(\rho,\evop_2\circ\evop_1) \leq
E(\rho,\evop_1)+E(\rho',\evop_2)$, and similarly for the dynamic
distance.
\end{itemize}

}}
\end{center}

\part{Bounds on quantum information transmission}

\chapter{Quantum communication complexity}
\label{chap:qcomm} \index{communication complexity}
\index{communication complexity!quantum}
\index{quantum communication complexity}
%
% To be done:
% Revise the coherent communication complexity results.
% add general bounds: space bound, normalizer bound, classical bound.

Suppose a number of widely separated parties wish to perform a
distributed computation. Each of the parties has access to some part
of the data which is to be used as input to the computation. However,
no party has access to all of the data, so in general no party can
complete the computation on their own. The {\em communication
complexity} of a problem is defined to be the minimal communication
cost incurred in performing the distributed computation.  The
classical theory of communication complexity was initiated by Yao
\cite{Yao79a}, and has since blossomed into a dynamic field of
research, as may be seen by consulting one of the excellent surveys of
the field that have been written, such as \cite{Kushilevitz96a}.
Recently, Yao \cite{Yao93a} has initiated the study of {\em quantum
communication complexity}, in which quantum resources may be used to
assist in the performance of a distributed computation. 

% purpose of the Chapter: review qcomm.

The purpose of this Chapter is to develop some elementary results in
quantum communication complexity. We will explore several different
models for quantum communication complexity, in which different types
of quantum resources may be used for communication, and with different
computational goals.  The models may be divided into two broad
classes. The first class is concerned with the {\em quantum
communication complexity of classical functions}.  The problems in
this class concern the computation of classical functions, but with
quantum resources allowed to assist in the computation. We will
examine several variants of this class, differentiated by the nature
of the quantum resources used.  For example, Yao \cite{Yao93a}
considered the computation of a classical function assisted by the
ability of the computing parties to communicate using qubits. An
important variant of this model was introduced by Cleve and Buhrman
\cite{Cleve97a}, who considered the computation of a classical
function in which the communication is carried out using classical
bits, but in which an arbitrary pre-shared entanglement is allowed.
Other variants within this class will be mentioned during this
Chapter.

\index{communication complexity!coherent quantum}

The second class of models concern {\em coherent quantum communication
complexity}.  In this class, the problems involve the distributed
computation of a {\em quantum function}, such as a joint unitary
operation performed by Alice and Bob on their qubits.  To my
knowledge, this class of problems has not been discussed prior to this
Dissertation.

The structure of the Chapter is as follows.  Section \ref{sec:Holevo}
reviews the work of other researchers on the Holevo bound, an
important result in quantum information theory, and the keystone of
much of the later work in this chapter.  Section
\ref{sec:cap_class_comm} presents a complete, original solution to a
basic problem in quantum information theory: what quantum resources
are required to transmit classical information from one location to
another, in the absence of noise?  This result is used in section
\ref{sec:IP} to give an original lower bound on the quantum
communication complexity of an interesting distributed computation,
known as the {\em inner product problem}. The work reported in
sections \ref{sec:cap_class_comm} and \ref{sec:IP} is the result of a
collaboration with Cleve, van Dam, and Tapp \cite{Cleve97b}.  Section
\ref{sec:cqcc} reports the first results on coherent quantum
communication complexity.  I demonstrate a lower bound on the coherent
quantum communication complexity of an important unitary operation,
the quantum Fourier transform.  Furthermore, a new and seemingly quite
powerful general technique for proving lower bounds to the coherent
quantum communication complexity is proved, and applied to several
problems in coherent quantum communication complexity.  Some of the
results in subsection \ref{subsec:cqcc_lower_bound} were inspired by a
conversation with Manny Knill.  Section \ref{sec:unified_qcc} outlines
an original formalism which can be used to unify coherent quantum
communication complexity with the quantum communication complexity for
computing a classical function. The Chapter concludes in section
\ref{sec:qcc_future} with a survey of some open problems in quantum
communication complexity, and suggestions for future research
directions.

My especial thanks to Richard Cleve for the many enjoyable and
stimulating discussions about quantum communication complexity which
stirred my interest in the field, and provoked many of the thoughts
reported in this Chapter.

\section{The Holevo bound}
\label{sec:Holevo}
\index{Holevo bound}

We begin with a review of what is historically perhaps the first major
result in quantum information theory, the {\em Holevo bound}
\cite{Holevo73a}.  This result will be the basis for our later results
in quantum communication complexity.  The line of proof used here
follows Schumacher, Westmoreland, and Wootters \cite{Schumacher96c}.

The setting is a game to be played by two fictitious protagonists,
Alice and Bob.  Alice is in possession of a classical source producing
symbols $X = 1,\ldots,n$ according to a corresponding probability
distribution $p_1,\ldots,p_n$. The aim of the game is for Alice to
convey the value of $X$ to Bob. However, for some reason, Alice can't
give $X$ directly to Bob. Rather, she prepares the quantum state
$\rho_X$, where $\rho_X$ is chosen from some fixed set
$\rho_1,\ldots,\rho_n$ of quantum states. She then gives that state to
Bob, whose task it is to determine the value of $X$, as best he can.

Suppose Bob performs a measurement on the quantum system he has been
given, with measurement result $Y$. A measure of how much information
he has gained about $X$ is the mutual information $H(X:Y)$ discussed
in Chapter \ref{chap:entropy}. By the data processing inequality we
know that Bob can infer $X$ from $Y$ if and only if $H(X:Y) = H(X)$,
and that in general $H(X:Y) \leq H(X)$. More generally, it is true
that the closer $H(X:Y)$ is to $H(X)$, the better Bob can do at
inferring $X$ from the observed value of $Y$. Bob's goal, therefore,
is to choose a measurement which maximizes $H(X:Y)$, bringing it as
close as possible to $H(X)$.

The Holevo bound states that:
\beqn
H(X:Y) \leq S(\rho) - \sum_x p_x S(\rho_x), \eeqn
where $\rho \equiv \sum_x p_x \rho_x$.  Thus, the Holevo bound is an
upper bound on the mutual information between Alice's classical data,
$X$, and the result of Bob's measurement, $Y$.  This bound holds {\em
for any} measurement Bob may choose to do.  

Before we proceed to the details of the proof, it is useful to note a
few elementary formulas concerning the probabilities of various
events. Suppose Bob does a measurement whose statistics are described
by POVM elements $M_y$, corresponding to the different possible values
which $Y$ may take. Then the probability that $Y = y$, given that the
state $\rho_x$ was prepared is given by \beqn p(Y=y | X=x) = \tr(M_y
\rho_x). \eeqn Thus $p(X=x,Y=y) = \tr(M_y \rho_x) p_x$, from which we
can calculate $H(X,Y), H(X), H(Y)$ and thus $H(X:Y)$.

Our proof of Holevo's bound is not quite the most direct possible,
however the route we take allows us to prove several facts that will
be useful later in the Chapter.  Our proof of Holevo's bound makes use
of the following result:

\index{partial trace property of $\chi$}

\begin{theorem} {\bf (Partial trace property of $\chi$)}
\cite{Schumacher96c} \index{partial trace property of $\chi$}
\index{Holevo $\chi$} \index{$\chi$}

Suppose states $\rho_x$ of a system $A$ are prepared, with respective
probabilities $p_x$. Define the Holevo $\chi$ quantity for system $A$,
\beqn
\chi_A \equiv S(\rho) - \sum_x p_x S(\rho_x), \eeqn
where $\rho \equiv \sum_x p_x \rho_x$.
Suppose a quantum system consists of two parts, $A$
and $B$, and $\{ p_x, \rho_x \}$ is an ensemble of states for the
joint system $AB$. Then
\beqn
\chi_{A} \leq \chi_{AB}, \eeqn
where $\chi_{AB}$ and $\chi_A$ are the natural $\chi$ quantities
associated with the ensemble $\{p_x,\rho_x\}$ for systems $AB$ and
$A$, respectively.
\end{theorem}

\begin{proof} \cite{Schumacher96c}

Introduce a system, $P$, with an orthonormal basis $|x\ra$ of states
with index $x$ corresponding to the index of the states
$\rho_x$. Suppose the initial state of $PAB$ is \beqn \rho^{PAB}
\equiv \sum_x p_x |x\ra \la x| \otimes \rho_x^{AB}. \eeqn Applying the
joint entropy theorem on page \pageref{thm:joint_entropy} and doing a
little algebra, we see that $\chi_{AB} = -S(P:A,B) $ and $\chi_A =
-S(P:A)$. The result now follows from the observation that $S(P:A,B)
\leq S(P:A)$, which is a restatement of strong subadditivity.

\end{proof}

It is straightforward to generalize this result to the case where a
complete quantum operation, $\evop$, replaces the partial trace in the
above theorem. Suppose the system of interest is labeled $Q$, and let
$\rho_x'', \rho''$ be the states obtained from $\rho_x, \rho$ by
applying ${\cal E}$, and let $\chi$ and $\chi''$ be the corresponding
Holevo $\chi$ quantities before and after application of the quantum
operation ${\cal E}$. It is a simple corollary of the previous result
that $\chi'' \leq \chi$. To see this, introduce a model environment
$E$ for the quantum operation ${\cal E}$. Suppose $U^{QE}$ is the
model interaction giving rise to the operation ${\cal E}$, and $|0\ra$
is the initial state of $E$. Let $\rho_x'$ be the {\em joint states}
of $QE$ after the model unitary operator has been applied, with
$\rho'$ and $\chi'$ defined in the obvious way. From the unitary
invariance of the entropy, $\chi' = \chi$. But the states $\rho_x''$
may be obtained from the states $\rho_x'$ by tracing out $E$, so by
the previous theorem, $\chi'' \leq \chi$, as we set out to prove. We
state this as a theorem generalizing the previous theorem:

\index{non-increasing property of $\chi$}

\begin{theorem} \label{thm:holevo} 
{\bf (Non-increasing property of $\chi$ under complete quantum operations)}
\cite{Schumacher96c}

The Holevo $\chi$ quantity can not be increased under complete quantum
operations.
\end{theorem}

The proof of the Holevo bound is to combine the partial trace property
of $\chi$ with a beautiful construction involving four quantum
systems, which we shall label $P,Q,M$ and $E$. The interesting thing
about the proof is that none of these systems need be associated with
the ``reality'' of the problem at all; that is, these systems need not
be directly related to the preparer, quantum system or observer
appearing in the statement of Holevo's theorem; recall the use of
similar constructions to prove entropic results in Chapter
\ref{chap:entropy}. The reason we can do this is because Holevo's
theorem is an inequality between entropic quantities which do not
depend on particular realizations for their meaning.

$P$ is to be thought of as the ``preparation'' system. It has an
orthonormal basis $|x\ra$ whose elements correspond to possible
preparations $\rho_x$ for the quantum system, $Q$.  $M$ and $E$ start
out in standard pure states, which we will label $|0\ra$ for both
systems. The initial state of the total system is assumed to be \beqn
\rho^{PQME} = \sum_x p_x |x\ra\la x| \otimes \rho_x \otimes |0\ra\la
0| \otimes |0\ra \la 0|.  \eeqn The intuition behind this construction
is that system $P$ represents Alice, the preparer, who knows the value
of $x$, and depending on this value prepares an appropriate state for
system $Q$. System $M$ represents Bob's measuring apparatus, which
records the result of the measurement, and $E$ represents an
additional ``environmental record'' \cite{Zurek91a} of this
measurement. Formally, this measurement process is represented by a
unitary dynamics on the system $PQME$ defined by the equation \beqn U
|PQ\ra|0\ra|0\ra \equiv \sum_y \left( I^P \otimes \sqrt{M_y}
\right)|PQ\ra |y\ra|y\ra, \eeqn where $|PQ\ra$ is any pure state of
$PQ$. As we saw in Chapter \ref{chap:qops} the unitarity of the
dynamics defined by this equation is ensured by the completeness
relation $\sum_y M_y = I$.  The state of the system after this
evolution is \beqn \rho^{PQME{'}} = \sum_{x,y_1,y_2} p_x |x\ra \la x|
\otimes \left( \sqrt{M_{y_1}} \rho_x \sqrt{M_{y_2}} \right) \otimes
|y_1\ra \la y_2| \otimes |y_1\ra \la y_2|. \eeqn

Let $\chi_M'$ be the Holevo $\chi$ quantity associated with system $M$
after the interaction. The respective states of system $M$ after the
interaction, are given by
\be
M_x' & = & \sum_y p(y|x) |y\ra \la y| \\
M' & = & \sum_y p(y) |y\ra \la y|, \ee
so $\chi_M' = H(Y)-H(Y|X) = H(X:Y)$. By the partial trace property,
$H(X:Y) = \chi_M' \leq \chi_{QME}'$. But the interaction of $Q, M$ and
$E$ was unitary, so $\chi_{QME}' = \chi_{QME} = \chi_Q$. Putting it
all together, we see that
\be
H(X:Y) \leq \chi_Q = S(\rho) -\sum_x p_x S(\rho_x), \ee
which is the Holevo bound.

Holevo's bound is a keystone in the proof of many results in quantum
information theory. The remainder of this section samples some of the
well-known uses to which the Holevo bound may be put, in order to
sharpen your intuition about the uses of this result.  Recall from
page \pageref{thm:entropy_ensemble} that \beqn S(\rho) \leq H(X) +
\sum_x p_x S(\rho_x), \eeqn with equality if and only if the states
$\rho_x$ have orthogonal support. Suppose that the states $\rho_x$ do
not have orthogonal support. Then Holevo's theorem allows us to
conclude that $H(X:Y)$ is always strictly less than $H(X)$. This is
just our intuitive notion that if the states prepared by Alice are not
orthogonal, then it is not possible for Bob to determine with
certainty which state Alice prepared.

A more concrete example may be useful. Suppose Alice prepares a single
qubit in one of two quantum states. Which state she prepares is
determined by a fair coin toss.  If the coin toss yields heads, then
Alice prepares the state $|0\ra$, and if the coin toss yields tails,
then Alice prepares the state $\cos \theta |0\ra + \sin \theta |1\ra$,
where $\theta$ is some real parameter. In the $|0\ra, |1\ra$ basis it
follows that $\rho$ can be written \beqn \rho = \frac{1}{2} \left[
\begin{array}{cc} 1 & 0 \\ 0 & 0 \end{array} \right] + \frac{1}{2}
\left[ \begin{array}{cc} \cos^2 \theta & \cos \theta \sin \theta \\
\cos \theta \sin \theta & \sin^2 \theta \end{array} \right]. \eeqn A
simple calculation shows that the eigenvalues of $\rho$ are $(1\pm
\cos \theta)/2$. This allows us to calculate the Holevo bound, which
in this case of pure state signals, is just the entropy of $\rho$.

\begin{figure}[htbp]
\begin{center}\mbox{\epsfig{file=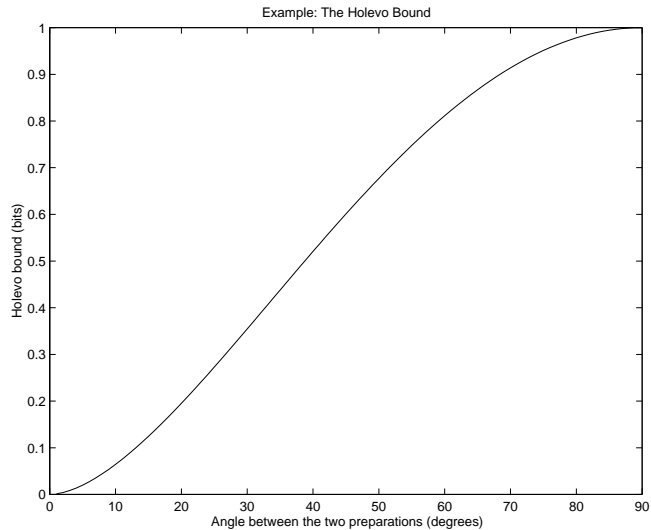,scale=0.5}}
\end{center}
\caption{Plot of the Holevo bound when the states $|0\ra$ and $\cos
\theta |0\ra + \sin \theta |1\ra$ are prepared with equal
probability. Notice that the Holevo bound reaches a maximum when the
angle between the two states is $\theta = \pi/2$, corresponding to
orthogonal states. At this point only it is possible for Bob to
determine with certainty which state Alice prepared. \label{fig: Holevo
example}}
\end{figure}

A plot of the Holevo bound for this example is shown in figure
\ref{fig: Holevo example}. Notice that the Holevo bound is maximized
at one bit, when $\theta$ is 90 degrees, corresponding to orthogonal
states.  At this point it is possible for Bob to determine with surety
which state Alice prepared. For other values of the $\theta$, the
Holevo bound is strictly less than one bit, and it is impossible for
Bob to determine with surety which state Alice prepared, since $H(X)$,
Alice's preparation entropy, is equal to one bit.

This can be quantified more precisely by making use of the {\em Fano
inequality}.  The Fano inequality is a result of classical information
theory, proved in the box on page
\pageref{box:Fano}, which provides a connection between the loss of
mutual information and the likelihood that an error is made in
inference.  Suppose Bob makes a guess $\tilde X = f(Y)$ as to which
state Alice prepared, based on the outcome of his measurement, $Y$. In
the general case, according to the Fano inequality, \beqn
H_{\rm{bin}}(p(\tilde X \neq X)) + p(\tilde X \neq X)) \log(|X|-1)
\geq H(X|Y) = H(X)-H(X:Y). \eeqn Combining this result with the Holevo
bound allows us to place bounds on how well Bob may infer the value of
$X$. Heuristically, the smaller $\chi$ is, the harder it is for Bob to
determine which state Alice prepared.  More precisely, we have \be
S(\rho)-\sum_x p_x S(\rho_x) + H_{\rm{bin}}(p(\tilde X \neq X)) +
p(\tilde X \neq X)) \log(|X|-1) \geq H(X). \eeqn We can use this
equation to numerically set a lower bound on the probability of Bob
making an error in inference of Alice's original state.  For instance,
in the present example, where Alice prepares $|0\ra$ with probability
one half and $\cos \theta |0\ra + \sin \theta |1\ra$ with probability
one half, the inequality reduces to \be H_{\rm{bin}}(p(\tilde X \neq
X)) \geq 1-H\left(\frac{1+\cos \theta}{2}\right). \ee A little thought
shows that when $\theta \neq 90$ degrees, we must have $p(\tilde X \neq
X) > 0$; moreover, the further away $\theta$ is from $90$ degrees, the
larger the probability of an error in inference is constrained to be
by this inequality.

%
% Box on Fano's inequality
%

\begin{center}
\fbox{ \label{box:Fano}
\parbox[t]{6in}{
\renewcommand{\baselinestretch}{0.5} \normalsize
\centerline{\bf Box \ref{chap:qcomm}$.1$: Fano's inequality}
\index{Fano inequality!classical}

Suppose we wish to infer the value of an unknown random variable, $X$,
based on knowledge of another random variable, $Y$. Intuitively, we
expect that the conditional entropy $H(X|Y)$ limits how well we may
perform this inference. The {\em Fano inequality} \cite{Cover91a}
provides a useful bound on how well we may infer $X$, given $Y$.

Suppose $\tilde X = f(Y)$ is some function of $Y$ which we are using
as our best guess for $X$. Let $p_e \equiv p(\tilde X \neq X)$ be the
probability that this guess is incorrect. Then the Fano inequality
states that \beqn H_{\rm{bin}}(p_e)+p_e \log(|X|-1) \geq H(X|Y),
\eeqn where $H_{\rm{bin}}$ is the binary entropy and $|X|$ is the
number of values $X$ may assume. Examining the inequality, what it
tells us is that if $H(X|Y)$ is large, then the probability of making
an error in inference, $p_e$, must also be large.

To prove the Fano inequality, define an ``error'' random variable, $E
\equiv 1$ if $X \neq \tilde X$, and $E \equiv 0$ if $X = \tilde X$.
Notice that $H(E) = H_{\rm{bin}}(p_e)$. Using the chain rule for
entropies proved on page \pageref{thm:chain_rule_entropies}, we have
$H(E,X|Y) = H(E|X,Y)+H(X|Y)$. But $E$ is completely determined once
$X$ and $Y$ are known, so $H(E|X,Y) = 0$ and thus $H(E,X|Y) =
H(X|Y)$. Applying the chain rule for entropies in a different fashion,
we obtain $H(E,X|Y) = H(X|E,Y) +H(E|Y)$. Conditioning reduces entropy,
so $H(E|Y) \leq H(E) = H_{\rm{bin}}(p_e)$. Finally, \beqn H(X|E,Y) &
= & p(E=0)H(X|E=0,Y)+p(E=1) H(X|E=1,Y) \\ & \leq & p(E=0) \times 0 +
p_e \log(|X|-1), \eeqn where $H(X|E=1,Y) \leq \log(|X|-1)$ follows
from the fact that when $E=1$, $X \neq Y$, and $X$ can assume at most
$|X|-1$ values, bounding its entropy, and thus its conditional entropy
by $\log(|X|-1)$.  Summarizing, we have \beqn H(X|Y) = H(E,X|Y) \leq
H_{\rm{bin}}(p_e)+p_e \log(|X|-1), \eeqn which establishes the Fano
inequality.

%\end{singlespace}
}
}
\end{center}

%\begin{figure}[htbp]
%\vspace{3cm}
%\caption{A lower bound on the probability of Bob making an error in
%inferring whether Alice prepared the state $|0\ra$ or 
%$\cos \theta |0\ra + \sin \theta |1\ra$. Notice that this bound
%decreases to zero as $\theta$ gets close to $\pi/2$, where the
%states may be distinguished without probability of error.
%\label{fig: Holevo p_e}}
%\end{figure}

\section{Capacity theorem for qubit communication}
\label{sec:cap_class_comm}

\index{capacity!theorem for noiseless qubit communication}

We can use the Holevo bound to analyze the following two-party game.
Alice is in possession of $n$ bits which she would like to transmit to
Bob. To achieve this, she is allowed to send qubits to Bob, and Bob
may send qubits to Alice, with no other form of communication allowed.
How many qubits must Alice and Bob use in order to successfully
transmit the $n$ bits from Alice to Bob?  The following {\em capacity
theorem} provides a complete answer to this question:

\begin{theorem} {\bf (Capacity theorem for communication using qubits)}

Suppose that Alice possesses $n$ bits of information, and wants to
convey this information to Bob.  Suppose that Alice and Bob possess no
prior entanglement but qubit communication in either direction is
allowed.  Let $n_{AB}$ be the number of qubits Alice sends to Bob, and
$n_{BA}$ the number of qubits Bob sends to Alice.  Then, Bob can
acquire the $n$ bits if and only if the following inequalities are
satisfied:
\begin{eqnarray}
n_{AB}, n_{BA} & \geq & 0 \\
n_{AB} & \geq & \lceil n/2 \rceil \label{eqtn:app1}\\ 
n_{AB}+n_{BA} & \geq & n \label{eqtn:app2}.
\end{eqnarray}
Moreover, the necessity of the condition $n_{AB} \geq \lceil n/2
\rceil$ remains valid even if pre-shared entanglement is allowed.  More
generally, Bob can acquire $m$ bits of mutual information with respect
to Alice's $n$ bits if and only if the above equations hold with $m$
substituted for $n$.
\end{theorem}

\begin{figure}[htbp]
\setlength{\unitlength}{1cm}
\begin{center}
\begin{picture}(4.5,5)
\put(1,1){\vector(0,1){4}}
\put(1,1){\vector(1,0){4}}
\put(0.2,4.5){$n_{AB}$}
\put(4.5,0.2){$n_{BA}$}
\put(0.5,4){$n$}
\put(0,2.5){$\lceil n/2 \rceil$}
\put(2.2,0.4){$\lfloor n/2 \rfloor $}
\thicklines
%\linethickness{0.99mm}
\put(1,4){\line(1,-1){1.5}}
%\linethickness{0.5mm}
\put(2.5,2.5){\vector(1,0){2.5}}
%\put(1,4){\vector(0,1){1}}
\linethickness{0.2mm}
\put(1,4){\dashbox{0.2}(4,0){}}
\thinlines
\put(4.84,4){\vector(1,0){0.16}}
\put(2.3,3.5){Capacity region}
\end{picture}
\end{center}
\vspace{-7mm}
\caption{Capacity region to send $n$ bits from Alice to Bob.
$n_{AB}$ is the number of qubits Alice sends to Bob,
and $n_{BA}$ is the number of qubits Bob sends to Alice. The
dashed line indicates the bottom of the classical capacity region.
\label{fig: capacity}}
\end{figure}
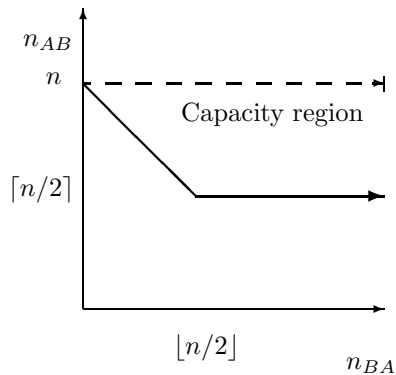

Graphically, the capacity region for the above communication problem
is shown in figure~\ref{fig: capacity}. Note the difference with the
classical result for communication with bits, where the capacity
region is given by the equation $n_{AB} \geq n$; that is, classically,
communication from Bob to Alice does not help.

\begin{proof}

%{\bf Proof of capacity theorem}

The sufficiency of equations (\ref{eqtn:app1}) and (\ref{eqtn:app2})
follows from the superdense coding technique discussed in section
\ref{sec:superdense}.  Sufficiency in the case $n_{AB} \geq n$ is
obvious, so we suppose $n_{AB} < n$.  Bob prepares $n - n_{AB} \leq
n_{BA}$ maximally entangled pairs of qubits, and sends one qubit of
each pair to Alice, who can use them in conjunction with sending $n -
n_{AB} \le n_{AB}$ qubits to Bob to transmit $2(n - n_{AB})$ bits to
Bob, using superdense coding.  Alice uses her remaining allotment of
$n_{AB}-(n-n_{AB}) = 2n_{AB} - n \geq 0$ qubits to transmit the
remaining $2n_{AB} - n$ bits in the obvious way.

The proof that equations (\ref{eqtn:app1}) and (\ref{eqtn:app2}) are
necessary follows from an application of Holevo's theorem and the
non-increasing property of the Holevo $\chi$ under partial traces, as
discussed in the previous section. The details are as follows.

Let $X$ be Alice's $n$ bits of information, which we assume is
uniformly distributed over $\{0,1\}^n$.  Without loss of generality,
it can be assumed that the protocol between Alice and Bob is of the
following form.  For any value $(x_1,\ldots,x_n)$ of $X$, Alice begins
with a set of qubits in state $\ket{x_1,\ldots,x_n}\ket{0,\ldots,0}$
and Bob begins with a set of qubits in a standard state
$\ket{0,\ldots,0}$.  The protocol consists of a sequence of steps,
where at each step one of the following four processes takes place.
\begin{enumerate}
\item
Alice performs a unitary operation on the qubits in her possession.
\item
Bob performs a unitary operation on the qubits in his possession.
\item
Alice sends a qubit to Bob.
\item
Bob sends a qubit to Alice.
\end{enumerate}
After these steps, Bob performs a measurement on the qubits in his
possession, which has outcome $Y$.  Bob's goal is to maximize the
mutual information between $Y$ and Alice's input, $X$.  You may be
wondering about the possibility of a protocol in which Bob starts with
a mixed state, or in which non-unitary operations are allowed.  Note
that using the techniques of Chapter \ref{chap:qops} any non-unitary
operation can be simulated by a unitary operation, with the
introduction of an appropriate environmental model, and by the
purification procedure discussed in Appendix \ref{app:mixed} it is
possible to simulate any protocol in which Bob starts with a mixed
state by a protocol in which Bob starts with a pure state.

Let $\rho_i^X$ be the density operator of the set of qubits that are
in Bob's possession after $i$ steps in the protocol have been
executed, and $\rho_i \equiv \sum_x p_x \rho_i^x$ be the density
operator of Bob's system after $i$ steps, averaged over all possible
inputs, $x$.  Due to Holevo's bound, it suffices to upper bound the
final value of $\chi(\rho_i^X)$.  We consider the evolution of
$\chi(\rho_i^X)$ and $S(\rho_i)$.  Initially, $\chi(\rho_0^X) =
S(\rho_0) = 0$, since Bob begins in a state independent of $X$.  Now,
consider how $\chi(\rho_i^X)$ and $S(\rho_i)$ change for each of the
four processes above.
\begin{enumerate}
\item
Alice performing a unitary operation on her qubits does not affect
$\rho_i^X$ and hence has no effect on $\chi(\rho_i^X)$ or $S(\rho_i)$.
\item
It is easy to verify that $\chi$ and $S$ are invariant under unitary
transformations, so Bob performing a unitary on his qubits does not
affect $\chi(\rho_i^X)$ and $S(\rho_i)$, either.
\item
Alice sends a qubit to Bob. Let $B$ denote Bob's qubits after $i$
steps and $Q$ denote the qubit that Alice sends to Bob at step number
$i+1$.  By the subadditivity inequality discussed in subsection
\ref{subsec:subadditivity} and the fact that, for a single qubit $Q$,
$S(Q) \leq 1$, $S(B,Q) \le S(B)+S(Q) \le S(B) + 1$.  Also, by the
triangle inequality discussed in subsection
\ref{subsec:subadditivity}, $S(B,Q) \ge S(B) - S(Q) \geq S(B) - 1$.  It
follows that $S(\rho_{i+1}) \le S(\rho_i) + 1$ and thus
\begin{eqnarray}
\chi(\rho_{i+1}^X) 
& = & S(\rho_{i+1}) - \sum_x p_x S(\rho_{i+1}^x) \nonumber \\
& \le & (S(\rho_i)+1) - \sum_x p_x (S(\rho_x)-1) \nonumber \\
& = & \chi(\rho_i^X) + 2.
\end{eqnarray}
\item Bob sends a qubit to Alice.  In this case, $\rho_{i+1}^X$ is
$\rho_i^X$ with one qubit traced out.  We saw in the previous section
that $\chi$ does not increase under partial trace, so
$\chi(\rho_{i+1}^X) \le \chi(\rho_i^X)$.  Note also that
$S(\rho_{i+1}) \le S(\rho_i) + 1$ for this process, by the triangle
inequality.
\end{enumerate}

Now, since $\chi(\rho_i^X)$ can only increase when Alice sends a qubit
to Bob and by at most 2, equation (\ref{eqtn:app1}) follows from the
Holevo bound.  Also, since $S(\rho_i)$ can only increase when one
party sends a qubit to the other and by at most 1, equation
(\ref{eqtn:app2}) follows from the observation that $S(\rho_i)$ is an
upper bound on the Holevo $\chi$, and the Holevo bound.  Finally, note
that even if pre-shared entanglement is allowed, so $S(\rho_i)$ may
start out greater than zero, $\chi(\rho_i)$ is still zero at the start
of the protocol, and thus the reasoning leading to the constraint
$n_{AB} \geq \lceil n/2 \rceil$ still holds. This completes the proof.

\end{proof}

%Finally, notice that even if Alice and Bob pre-share a physical
%resource, Bob's initial $\chi$ is still $0$, and thus
%$I(X:Y) \leq \chi_F \leq 2n_{AB}$.
%  Thus, even in the presence of a pre-shared physical
% resource, the bound $n_{AB} \geq \lceil n/2 \rceil $ must still be
% satisfied if Alice is to communicate $n$ bits to Bob. On the other
% hand, the proof that $n_{AB} + n_{BA} \geq n$ fails in the presence of
%a pre-shared entanglement. The reason is because in this case we have
%$S(\rho) > 0$, which causes the inequality $S(\rho_F) \leq
%n_{AB}+n_{BA}$ to fail. This is reflected by the fact that if Alice
%and Bob pre-share $\lceil n/2 \rceil$ EPR pairs, then it is possible
%for Alice to transmit her $n$ bits to Bob using superdense coding,
%with $n_{BA} = 0$. It would be interesting to understand in more depth
%how different physical resources may be used to save on the
%communication requirements for this problem.
%
%{\bf QED}

Our main interest in this capacity theorem is as a step along the way
to proving results about quantum communication complexity. However, it
is interesting to briefly consider other directions in which this work
could be taken.

First, let us consider what the essential difference is between the
classical and the quantum resources required to perform the task under
consideration. Recall that when Alice sends a qubit to Bob, we showed
that $\Delta \chi \leq 2$. To prove this, we used the triangle
inequality $S(B,Q) \geq S(B) - S(Q)$ to show that $S(\rho_x') \geq
S(\rho_x)-1$. However, in the absence of entanglement between $Q$ and
$B$, we will show in Chapter \ref{chap:ent} that the stronger
inequality $S(B,Q) \geq S(B)$ is true.  We deduce that in the absence
of entanglement, $\Delta \chi \leq 1$, from which the familiar
classical lower bound $n_{AB} \geq n$ emerges.

Second, note that we have assumed noiseless transmission of quantum
information between Alice and Bob.  What are the resource requirements
if there is noise in the channel between Alice and Bob?  Another
interesting path for generalization is to consider the many-party
version of the problem. What quantum resources are required to
accomplish communication of classical information amongst a network of
$k$ users?  Finally, we may ask for a precise characterization of what
quantum resources are required to transmit $n$ bits of information in
the presence of a pre-shared entanglement between Alice and
Bob. Answering this question in full generality may give new insight
into the meaning of entanglement, and suggests a means for defining
measures of entanglement for quantum systems consisting of two or more
components.

\section{Communication complexity of the inner product}
\label{sec:IP}

\index{inner product}
\index{communication complexity!of the inner product}
\index{quantum communication complexity!of the inner product}

We now have the tools we need to investigate several interesting
problems in quantum communication complexity. We begin with the
communication complexity of the {\it inner product modulo two (IP)}
function:
\begin{eqnarray}
\mbox{\it IP\/}(x,y) &=& (x_1 \cdot y_1 + x_2 \cdot y_2 + \cdots + 
x_n \cdot y_n) (\bmod 2).
\label{IP}
\end{eqnarray}
The communication complexity of the IP function is fairly well
understood in the classical models of communication complexity.  For
worst-case inputs and deterministic protocols guaranteed to give the
correct answer, the communication complexity is $n$. For randomized
protocols (with either an independent or a shared random source),
uniformly distributed or worst-case inputs, and with error probability
${1 \over 2} - \delta$ required, the communication complexity is $n -
O(\log (1 / \delta))$ \cite{Chor88a} (see also
\cite{Kushilevitz96a}).
% I deleted the following footnote, realizing that this is the only
% place in the Dissertation where I make use of the big $O$ notation!
%
%\footnote{If you are unfamiliar with the big $O$
%and big $\Omega$ notation used by computer scientists to quantity
%resources, I recommend that you read Appendix \ref{app:asymptotic} for
%a brief introduction.}

Yao \cite{Yao93a} has introduced a model of quantum communication
complexity in which Alice and Bob start of with uncorrelated systems,
and are allowed to communicate using qubits, in order to compute some
joint classical function.  The cost, or communication complexity, in
this model, is defined to be the minimum number of qubits which must
be communicated in order to compute the classical function.  Kremer
\cite{Kremer95a}, using a proof methodology which he attributes to
Yao, demonstrated that in this model the communication complexity of
IP is asymptotically linear in $n$, whenever the required correctness
probability is $1 - \epsilon$ for a constant $0 \leq \epsilon < {1
\over 2}$.

In this section, we consider the communication complexity of IP in two
models different to that introduced by Yao: with prior entanglement
and qubit communication; and with prior entanglement and classical bit
communication.  In both models, an arbitrary prior entanglement may be
set up, at no cost to the protocol. As far as is presently known, the
proof methodology of the lower bound in the qubit communication model
without prior entanglement \cite{Kremer95a} does not carry over to
either of these two models.  Nevertheless, we show linear lower bounds
in these models.

To state our lower bounds more precisely, we introduce the following
notation.  Let $f : \{0,1\}^n \times \{0,1\}^n \rightarrow \{0,1\}$ be
a communication problem, and $0 \le \epsilon < \frac 12$.  Let
$Q_{\epsilon}^{\ast}(f)$ denote the communication complexity of $f$
in terms of {\em qubits}, where quantum entanglement is available and
the requirement is that Bob determines the correct answer with
probability at least $1 - \epsilon$; the $\ast$ superscript is
intended to highlight the fact that prior entanglement is available.
Also, let $C_{\epsilon}^{\ast}(f)$ denote the corresponding
communication complexity of $f$ in the scenario where the
communication is in terms of {\em bits}; again, quantum entanglement
is available and Bob is required to determine the correct answer with
probability at least $1 - \epsilon$.  When $\epsilon = 0$, we
refer to the protocols as {\em exact}, and, when $\epsilon > 0$, we
refer to them as {\em bounded-error} protocols.  With this notation,
the results we will prove in this section may be summarized:
\begin{eqnarray}
Q_0^{\ast}(\mbox{IP})  & = & \lceil n/2 \rceil \label{eqtn:Q} \\
Q_{\epsilon}^{\ast}(\mbox{IP}) & \ge & \frac 12(1 - 2 \epsilon)^2 n - \frac 12 \label{eqtn:Qe} \\
C_0^{\ast}(\mbox{IP})  & = & n \label{eqtn:C} \\
C_{\epsilon}^{\ast}(\mbox{IP}) & \ge & \max(\frac 12(1 - 2 \epsilon)^2,(1 - 2 \epsilon)^4) n - \frac 12 \label{eqtn:Ce} 
\end{eqnarray}
Note that all the lower bounds are linear in $n$ whenever $\epsilon$
is held constant.  Also, these results subsume the lower bounds in
\cite{Kremer95a}, since the qubit model defined by Yao \cite{Yao93a}
differs from the bounded-error qubit model defined above only in that
it does not permit a prior entanglement.

The lower bound proofs employ a novel kind of ``quantum'' reduction
between protocols, which reduces the problem of communicating, say,
$n$ bits of information to the IP problem.  It is noteworthy that
there does not appear to be a similar classical reduction between the
two problems.  This reduction is particularly remarkable since quantum
information theory subsumes classical information theory, and
therefore our results also represent new proofs of nontrivial lower
bounds on the {\em classical} communication complexity of IP. It is
intriguing that we are able to prove such lower bounds using a quantum
mechanical methodology fundamentally different from previous methods
used for proving classical lower bounds.

\subsection{Converting exact protocols into clean form}

\index{clean protocols for quantum communication}

We begin by showing how to reduce a general protocol for computing a
function $f(x,y)$ into a special type of protocol which we call a {\em
clean protocol}.  A clean protocol is a special kind of qubit protocol
inspired by the general spirit of the reversible computing paradigm
\cite{Lecerf63a,Bennett73a}, in a quantum setting.  In particular, a
clean protocol is set up so that none of the qubits involved in the
protocol changes, except for one, which contains the answer, $f(x,y)$.

In general, the initial state of a qubit protocol is of the form 
\begin{equation}
\underbrace{\ket{y_1,\ldots,y_n} \otimes \ket{0,\ldots,0} \otimes |\Phi
}%
_{\mbox{\small Bob's qubits}}
\underbrace{\!{}_{BA}\rangle \otimes \ket{x_1,\ldots,x_n} \otimes
\ket{0,\ldots,0}}%
_{\mbox{\small Alice's qubits}},
\end{equation}
where $\ket{\Phi_{\!BA}}$ is the state of the entangled qubits shared
by Alice and Bob, and the $\ket{0,\ldots,0}$ states can be regarded as
work space for the protocol.  At each turn, one party performs some
unitary operation on all the qubits in their possession and then sends
a subset of these qubits to the other party.  Note that, due to the
communication, the set of qubits possessed by each party varies during
the execution of the protocol.
 
We say that a protocol which exactly computes a function $f(x,y)$ is
{\em clean\/} if, when executed on the initial state
\begin{equation}
\ket{z} \otimes \ket{y_1,\ldots,y_n} \otimes \ket{0,\ldots,0} \otimes \ket{\Phi_{BA}}  
\otimes \ket{x_1,\ldots,x_n} \otimes \ket{0,\ldots,0}, 
\label{eqtn:ip_initial}
\end{equation}
the protocol results in the final state
\begin{equation}
\ket{z + f(x,y)} \otimes \ket{y_1,\ldots,y_n}  
\otimes \ket{0,\ldots,0} \otimes \ket{\Phi_{BA}} \otimes \ket{x_1, \ldots, x_n} \otimes \ket{0,\ldots,0}
\label{eqtn:ip_final}
\end{equation}
(where the addition is mod 2).  The input, the work qubits, and the
initial entangled qubits will typically change states during the
execution of the protocol, but they are reset to their initial values
at the end of the protocol.

We will show how to transform a general protocol into a clean
protocol. This transformation comes at a cost, which we quantity using
the following notation. If a qubit protocol consists of $m_1$ qubits
from Alice to Bob and $m_2$ qubits from Bob to Alice then we refer to
the protocol as an $(m_1,m_2)$-qubit protocol.

The following argument shows it is always possible to transform an
exact $(m_1,m_2)$-qubit protocol to compute $f(x,y)$ into a clean
$(m_1+m_2,m_1+m_2)$-qubit protocol that computes the same function.
First, the protocol for $f$ is run once, creating a state of the form
\be |z\ra|f(x,y)\ra \otimes |\Phi_{BA}')\ra , \ee where
$|\Phi_{BA}'\ra$ is some extra ``garbage'' state of the joint system,
$BA$, which will depend on $x_1,\ldots,x_n,y_1,\ldots,y_n$.  We now
apply a controlled not gate with $|f(x,y)\ra$ as the control and
$|z\ra$ as the ancilla, to create the state \be |z+f(x,y)\ra|f(x,y)\ra
\otimes |\Phi_{BA}'\ra. \ee Finally, note that all the steps in the protocol
to compute $f$ were reversible.  We now run the original protocol in
reverse, putting the system in the desired state, \be |z+f(x,y)\ra
\otimes |y_1,\ldots,y_n\ra \otimes |0,\ldots,0\ra \otimes
|\Phi_{BA}\ra \otimes |x_1, \ldots, x_n\ra \otimes |0,\ldots,0\ra. \ee
Note that, for each qubit that Bob sends to Alice when the protocol is
run forwards, Alice sends the qubit to Bob when run in the backwards
direction.  Therefore, we have constructed a $(m_1+m_2,m_1+m_2)$-qubit
protocol that maps state (\ref{eqtn:ip_initial}) to state
(\ref{eqtn:ip_final}).

\subsection{Reduction from the communication problem}

We now show how to transform a clean $(m_1+m_2,m_1+m_2)$-qubit
protocol that exactly computes IP for inputs of size $n$, to an
$(m_1+m_2,m_1+m_2)$-qubit protocol that transmits $n$ bits of
information from Alice to Bob.  This is accomplished in four stages:
\begin{enumerate}
\item
Bob initializes his qubits as indicated in equation (\ref{eqtn:ip_initial})
with $z = 1$ and $y_1 = \cdots = y_n = 0$, while Alice prepares the
state $|x_1,\ldots,x_n\ra$.
\item
Bob performs a Hadamard gate on each of his first $n+1$ qubits.
\item 
Alice and Bob execute the clean protocol for the inner product function.
\item
Bob again performs a Hadamard gate on each of his first 
$n+1$ qubits.
\end{enumerate}
Let $\ket{B_i}$ denote the state of Bob's first $n+1$ qubits after the
$i^{\mbox{\scriptsize th}}$ stage.  Recalling the definition of the
Hadamard gate from page \pageref{defn:Hadamard}, $H|0\ra =
(|0\ra+|1\ra)/\sqrt 2$ and $H|1\ra = (|0\ra-|1\ra)/\sqrt 2$, we see
that
\begin{eqnarray}
\ket{B_1} &=& \ket{1} \otimes \ket{0,\ldots,0}  \label{eqtn:st1} \\
\ket{B_2} &=& {\textstyle{1 \over \sqrt{2^{n+1}}}}
\sum_{a,b_1,\ldots,b_n \in \{0,1\}}
(-1)^a\ket{a} \otimes \ket{b_1,\ldots,b_n} \\
\ket{B_3} &=& {\textstyle{1 \over \sqrt{2^{n+1}}}}
\sum_{a,b_1,\ldots,b_n \in \{0,1\}} 
(-1)^a\ket{a + b_1 x_1 + \cdots + b_n x_n} \otimes \ket{b_1,\ldots,b_n} \nonumber \\
&=& {\textstyle{1 \over \sqrt{2^{n+1}}}}
\sum_{c,b_1,\ldots,b_n \in \{0,1\}} 
(-1)^{c + b_1 \cdot x_1 + \cdots + b_n \cdot x_n}
\ket{c} \otimes \ket{b_1,\ldots,b_n} \label{eqtn:st3} \\
\ket{B_4} &=& \ket{1} \otimes \ket{x_1,\ldots,x_n}, \label{eqtn:st4}
\end{eqnarray}
where, in equation (\ref{eqtn:st3}), the substitution $c = a + b_1 x_1
+ \cdots + b_n x_n$ has been made, and arithmetic over bits is taken
modulo 2.  The above transformation was inspired by \cite{Terhal97a}; 
see also \cite{Bernstein97a}.

Since the above protocol conveys $n$ bits of information (namely,
$x_1,\ldots,x_n$) from Alice to Bob, by the capacity theorem of
section \ref{sec:cap_class_comm}, we have $m_1+m_2 \ge n/2$.  Since
this protocol can be constructed from an arbitrary exact
$(m_1,m_2)$-qubit protocol for IP, this establishes the lower bound of
equation (\ref{eqtn:Q}).  That this bound is achievable follows
immediately from the superdense coding technique; Alice need merely
send all $n$ of her bits to Bob using $\lceil n/2 \rceil$ qubits and
superdense coding. Bob can then calculate the inner product.  This
completes the proof of equation (\ref{eqtn:Q}).

The approximate result (\ref{eqtn:Qe}) follows in a straightforward
fashion by running essentially the same argument, and using the Fano
inequality to bound the probability of error for the IP protocol.  We
will not go through the details here; they may be found in
\cite{Cleve97b}.

Note that, classically, the reduction used here to prove the
communication complexity lower bound is not possible.  For example, if
a clean protocol for IP is executed in any classical context, it can
never yield more than one bit of information to Bob, whereas, in this
quantum context, it yields $n$ bits of information to Bob.

\subsection{Lower bounds for bit protocols}

We now use the just-proved exact {\em quantum bit} communication
complexity for IP to prove an exact {\em classical bit} communication
complexity for IP, in the presence of pre-shared entanglement,
equation (\ref{eqtn:C}).

Using quantum teleportation it is straightforward to simulate any
$m$-qubit protocol by a $2m$-bit classical protocol, and appropriate
pre-shared entanglement.  Also, if the communication pattern in an
$m$-bit protocol is such that an even number of bits is always sent
during each party's turn then it can be simulated by an $m/2$-qubit
protocol by superdense coding \cite{Bennett92c} (which also employs
EPR pairs).  However, this latter simulation technique cannot, in
general, be applied directly, especially for protocols where the
parties take turns sending single bits.

We can nevertheless obtain a slightly weaker simulation of bit
protocols by qubit protocols for IP that is sufficient for our
purposes.  The result is that, given any $m$-bit protocol for
$\mbox{IP}_n$ (that is, IP instances of size $n$), one can construct
an $m$-qubit protocol for $\mbox{IP}_{2n}$.  This is accomplished by
interleaving two executions of the bit protocol for $\mbox{IP}_n$ to
compute two independent instances of inner products of size $n$.  We
make two observations.  First, by taking the sum (mod 2) of the two
results, one obtains an inner product of size $2n$.  Second, due to
the interleaving, an even number of bits is sent at each turn, so that
the above superdense coding technique can be applied, yielding a
$(2m)/2 = m$-qubit protocol for $\mbox{IP}_{2n}$.  Now, equation
(\ref{eqtn:Q}) implies $m \ge n$, which establishes the lower bound of
equation (\ref{eqtn:C}).  The achievability of this lower bound
follows from the obvious protocol: Alice sends her classical data to
Bob, establishing equality in equation (\ref{eqtn:C}).  

To obtain the lower bound (\ref{eqtn:Ce}), suppose we apply the same
proof technique as above to any $m$-bit protocol computing
$\mbox{IP}_n$ with probability $1-\epsilon$.  We obtain an $m$-qubit
protocol which computes $\mbox{IP}_{2n}$ with probability
$(1-\epsilon)^2 = 1-2\epsilon(1-\epsilon)$. Applying equation
(\ref{eqtn:Qe}), with $2n$ replacing $n$ and $2\epsilon(1-\epsilon)$
replacing $\epsilon$, we find that $m \geq (1-2\epsilon)^4 n -
1/2$. For $\epsilon > 1/2 - \sqrt 2 / 4 \approx 0.146$ a better bound
is obtained by noting that $C_{\epsilon}^* \geq Q_{\epsilon}^*$ is
always true, since quantum bits can always be used in the place of
bits, and applying equation (\ref{eqtn:Qe}).  This establishes
equation (\ref{eqtn:Ce}).

\section{Coherent quantum communication complexity}
\label{sec:cqcc}

\index{communication complexity!coherent quantum}

In the previous section we considered the distributed computation of a
classical function using quantum resources. Analogous questions can be
asked about the distributed computation of a quantum function using
quantum resources, a field of investigation which we will call {\em
coherent quantum communication complexity}, or more usually just {\em
coherent communication complexity}.

In this section we develop some elementary lower bounds on the
coherent communication complexity.  The original inspiration for this
investigation was the following problem, which we shall call FT, for
{\em Fourier transform}.  Suppose Alice is in possession of $n$
qubits, Bob is in possession of $n$ qubits, and they wish to perform
the quantum Fourier transform
\cite{Shor94a,Coppersmith94a,Deutsch94a}. How many qubits must be
communicated between Alice and Bob if they are to achieve this goal?

We will prove a lower bound of $n$ qubits for this problem, using a
method inspired by that used to prove the IP lower bound.  We then
prove a much more general lower bound, which applies to any unitary
operator.  This general lower bound is then used to give an alternate
proof that the quantum Fourier transform has a coherent communication
complexity of at least $n$ qubits. The section closes with some
general remarks about further directions for exploration in the field
of coherent quantum communication complexity.

\subsection{Coherent communication complexity of the quantum Fourier
transform}

It is clear that the computation of FT -- and, indeed, of any unitary
transform -- can be done using $2n$ qubits of communication: Alice
sends her $n$ qubits to Bob, who performs the quantum Fourier
transform on all $2n$ qubits. Bob then sends the $n$ qubits which
initially belonged to Alice back to Alice, completing the quantum
Fourier transform. We will show that this is essentially the best
procedure that can be achieved, to within a constant factor.  Our
general strategy will be to use a technique similar to that used for
the inner product function, transforming a protocol for computing the
quantum Fourier transform into a communications protocol.

It has been shown by Danielson and Lanczos (see \cite{Press89a} for a
discussion) that the Fourier transform on $m$ qubits has the following
effect: \be \label{eqtn:IQFT} |x_1, \ldots, x_m\ra \longrightarrow
\nonumber \\ \left( |0\ra + e^{2\pi i x_m} |1\ra \right) \otimes
\left( |0\ra + e^{2\pi i 0.x_{m-1} x_m} |1\ra \right) \otimes \ldots
\otimes \left( |0\ra + e^{2\pi i 0.x_1 \ldots x_m} |1\ra \right), \ee
where $x_1,\ldots,x_m$ is any set of $m$ bits. Throughout this section
normalization factors are omitted.  This decomposition, discovered by
Danielson and Lanczos in 1942, has been rediscovered many times since;
in the quantum context it has been rediscovered by Griffiths and Niu
\cite{Griffiths96a} and, somewhat later, but independently, by Cleve
{\em et al} \cite{Cleve98a}.

The strategy we use is to turn a protocol for computing the quantum
Fourier transform into a method for classical communication between
Alice and Bob. Suppose Alice has a string $x_1,\ldots,x_n$ of classical
bits which she wishes to transmit to Bob. The following protocol
achieves this. Alice prepares a system of $n$ qubits in the state
\beqn \label{eqtn:Alice_start} |x_1,\ldots,x_n\ra, \ee
while Bob prepares a system of $n$ qubits in the all $|0\ra$ state.

Alice and Bob now jointly apply the $2n$-qubit quantum Fourier
transform to their system, resulting in the state \be & & \left(
|0\ra+|1\ra \right)^{\otimes n} \otimes \left( |0\ra + e^{2\pi i
0.x_n}|1\ra \right) \otimes \left( |0\ra + e^{2\pi i
0.x_{n-1}x_n}|1\ra \right) \otimes \ldots \otimes \left( |0\ra +
e^{2\pi i 0.x_1 \ldots x_n}|1\ra \right). \nonumber \\ & & { } 
\ee Bob now performs an
$n$-qubit inverse quantum Fourier transform on his $n$ qubits,
resulting in a final state for Bob of $|x_1,\ldots,x_n\ra$, from which
he can simply read off the values of $x_1,\ldots,x_n$ which were
originally in Alice's possession.

Thus any procedure for performing the quantum Fourier transform
immediately yields a procedure for communicating $n$ classical bits of
information from Alice to Bob, for the same cost.  The results of
section \ref{sec:cap_class_comm} imply a lower bound on the coherent
communication complexity of the quantum Fourier transform of $n$
qubits. We conclude that the coherent communication complexity of the
quantum Fourier transform is in the range $n$ to $2n$ qubits.

So far we have considered the coherent communication complexity of the
quantum Fourier transform in the case where the quantum Fourier
transform must be done exactly. What if we are willing to allow an
{\em approximate} performance of the quantum Fourier transform?
Suppose we are attempting to perform the unitary operation $U$, but
the protocol instead performs a quantum operation $\evop$. Define the
{\em absolute distance} for the protocol by \beqn D \equiv D(U,\evop)
\equiv \min_{|\psi\ra} D(U|\psi\ra,\evop(|\psi\ra\la \psi|)), \ee
where the minimization is over all pure states $|\psi\ra$, and the
function $D(\cdot,\cdot)$ appearing on the right hand side is the
absolute distance of Chapter \ref{chap:distance}.  The absolute
distance for the protocol is a measure of how well the protocol
computes the quantum Fourier transform.

Suppose we want a protocol such that $D < \epsilon$.  Intuitively, any
such protocol must involve nearly $n$ qubits, with the allowed
deviation determined by the magnitude of $\epsilon$.

Suppose $q$ qubits are sent during the protocol for the approximate
quantum Fourier transform. Suppose we substitute this approximate
quantum Fourier transform into the bit communication protocol used
earlier to obtain the exact lower bound.  Since only $q$ qubits are
sent, we have an upper bound on the final Holevo $\chi$ quantity of
Bob's system of $q$ bits. Combining the Fano inequality and the Holevo
bound, this implies that the minimal probability of Bob making an
error in his inference, $p_e$, satisfies \be (n-q) \leq h(p_e)+p_e
n. \ee But from page \pageref{abs_distance_meaning} we know that $p_e
\leq D$, and assuming that $\epsilon < 1/2$, it follows that $h(p_e)
\leq h(D)$, so \be q \geq n(1-\epsilon)-h(\epsilon). \ee In the case
when $\epsilon = 0$ this reduces to $q \geq n$, as expected, and more
generally gives a lower bound on the number of qubits required to
achieve a specified accuracy in the coherent communication complexity.

%Finally, we will outline a procedure for improving this bound
%somewhat.  Thus far, we have made use of the difficulty of sending
%classical information from Alice to Bob. However, the protocol we have
%described may be used to send quantum information form Alice to
%Bob. Suppose Alice starts with the state $|y_1,\ldots,y_n\ra$, which
%she Fourier transforms to obtain the state
%(\ref{eqtn:Alice_start}). Then she and Bob use the inverse Fourier
%transform protocol as before, with Bob's final state being
%$|y_1,\ldots,y_n\ra$. Each step of this procedure is unitary, so the
%procedure preserves superpositions of states of the form
%$|y_1,\ldots,y_n\ra$ as well. That is, the procedure can be used to
%transmit $n$ qubits of quantum information from Alice to Bob.last
%
%Now, it might seem obvious that $n$ qubits {\em from Alice to Bob} is
%the minimal number of qubits which must be used to communicate $n$
%qubits of information from Alice to Bob! Nevertheless, we must prove
%this, since it is possible that Bob's capability of sending
%information to Alice might be able to short-circuit this
%expectation. We do not have the tools to give a simple direct proof
%here, although the tools of Chapters
%\ref{chap:qec} and \ref{chap:capacity} can be used to do so. In any
%case, there is little to be gained from such a result, since I do not
%believe that it will lead to a sharp bound for the coherent complexity
%of the quantum Fourier transform. My conjecture, which I have not been
%able to prove, is that a full $2n$ qubits of communication are
%required to compute the quantum Fourier transform.

It is somewhat disappointing that there remains the gap between $n$
and $2n$ for the coherent communication complexity of the quantum
Fourier transform.  I have not been able to close this gap.  I
conjecture that $2n$ is the actual coherent communication complexity
of the quantum Fourier transform; the proof or refutation of this
conjecture is an interesting problem for future work.

\subsection{A general lower bound}
\label{subsec:cqcc_lower_bound}

Let $U$ be a general unitary operator on a joint system, $AB$. Suppose
the action on $A$ is on $m$ qubits, and on $B$ is on $n$ qubits.  Note
that $U$ can always be written in the form \beqn U = \sum_i A_i
\otimes B_i, \ee where $A_i$ and $B_i$ are non-zero operators on the
systems $A$ and $B$, respectively. The {\em Schmidt number} of $U$,
$\Sch(U)$, is defined to be the minimal number of operators $A_i$
(equivalently $B_i$) required in any such decomposition of $U$.

A general lower bound on the coherent communication complexity of $U$
is \be \label{eqtn:cqcc_bound} Q_0(U) \geq \lceil \log_4 \Sch(U)
\rceil. \ee We will prove this result shortly.  It provides a general
technique for proving lower bounds on the coherent communication
complexity of a given unitary operator.  In order to make use of the
bound, we must have a means of determining the Schmidt number of a
given unitary operator.  Fortunately, such a means is provided by the
Schmidt decomposition, described in Appendix \ref{app:mixed}.

Recall that the space of operators on the joint system $AB$ can be
regard as a Hilbert space formed from the tensor product of the
Hilbert space of operators on system $A$ with the Hilbert space of
operators on system $B$.  Any convenient inner product may be used to
turn the vector spaces of operators on systems $A$ and $B$ into
Hilbert spaces; we will use the trace inner product, $(A_1,A_2) \equiv
\tr(A_1^{\dagger}A_2)$ and $(B_1,B_2)\equiv
\tr(B_1^{\dagger}B_2)$. This inner product, in turn, gives rise to a
Schmidt decomposition for vectors (that is, operators) on the joint
space $AB$, \be U = \sum_i A_i \otimes B_i, \eeqn where the sets $A_i$
and $B_i$ are guaranteed to be orthogonal with respect to the trace
inner product.  Moreover, properties of the Schmidt decomposition
guarantee that this decomposition contains the minimal number of
operators possible.

In practice, finding the Schmidt decomposition of an operator may be
done in a straightforward manner, along the lines outlined in Appendix
\ref{app:mixed}, and we will not recap that method here in the slightly
different operator language.  It is, however, instructive to look at a
couple of examples of the use of the Schmidt decomposition.  First, we
give the operator-Schmidt decomposition for the controlled not
operation, \be C = |0\ra\la 0| \otimes I + |1\ra \la 1| \otimes X. \ee
Notice that this neatly encapsulates the verbal description often used
for this process -- if the control qubit is zero, the data qubit is
left alone, while if the control qubit is one, the data qubit is
flipped, while still emphasizing that this process is undertaken {\em
coherently}, a point that may not always be clear from verbal
descriptions of the controlled not gate.

To prove the lower bound (\ref{eqtn:cqcc_bound}), consider a general
protocol for computing $U$ using qubit communication between Alice and
Bob. Let $U_s$ be the total unitary operation performed after $s$
steps in the protocol.  Notice that $U_0 = I$ has $\Sch(U_0) =
1$. Once again, using the techniques of Chapter \ref{chap:qops}, we
may introduce work bits to ensure that, without loss of generality, we
may restrict ourselves to consideration of unitary operations.  Note
that the operations which may be performed during the protocol are of
four types:

\begin{enumerate}
\item Alice does a unitary operation on her qubits.
\item Bob does a unitary operation on his qubits.
\item Alice sends a qubit to Bob.
\item Bob sends a qubit to Alice.
\end{enumerate}

Clearly, $\Sch(U_s) = \Sch(U_{s+1})$ if steps 1 or 2 are carried
out. Suppose $U_s = \sum_i A_{s,i} \otimes B_{s,i}$ is a minimal
decomposition of $U_s$. Suppose Alice sends a qubit $Q$ to Bob,
leaving Alice with a system $A'$. Note that \be A_{s,i} = \sum_{j=1}^4
A'_{s,i,j} \otimes Q_{s,i,j}, \eeqn for some set of four operators
$A'_{s,i,j}$ on $A'$ and $Q_{s,i,j}$ on $Q$. Thus, \be U_{s+1} =
\sum_{i,j=1\ldots 4} A'_{s,i,j} \otimes Q_{s,i,j} \otimes B_{s,i}, \ee
from which we deduce that
\be
\Sch(U_{s+1}) \leq 4 \, \Sch(U_s). \ee

Similarly, if Bob sends a qubit to Alice then $\Sch(U_{s+1}) \leq
4\, \Sch(U_s)$.  Putting these observations together, if Alice and Bob
employ $q$ qubits of communication to compute $U$, then we must have
$\Sch(U) \leq 4^q$, from which we have the general lower bound,
(\ref{eqtn:cqcc_bound}).

%For a fixed $U$, is the bound (\ref{eqtn:cqcc_bound}) tight, or within
%a constant factor of being tight?

A simple application of this lower bound is to the communication
complexity of the swap operation. Suppose Alice and Bob each have $n$
qubits, which they wish to swap.  The unitary operator implementing
this swap has Schmidt decomposition \be U = \sum_{ij} |i\ra\la j|
\otimes |j\ra \la i|, \ee where the sum is over all computational
basis states $|i\ra$ and $|j\ra$. Thus $\Sch(U) = 4^n$, from which it
follows that the coherent communication complexity is at least $n$
qubits.  The obvious method to achieve such a swap is for Alice to
send her $n$ qubits to Bob, and for Bob to send his $n$ qubits to
Alice, for a total cost of $2n$ qubits.  Once again, as for the
quantum Fourier transform, we are within a factor of two of knowing
the exact quantum communication complexity.  In actual fact, it is
straightforward to adapt the methods used to prove the capacity
theorem of section \ref{sec:cap_class_comm} to show that at least $2n$
qubits of quantum communication must be employed to perform the swap
operation; this is easy, and the proof will be omitted. Nevertheless,
this example of the swap operation provides a simple example where the
general lower bound (\ref{eqtn:cqcc_bound}) provides useful
information.

A second, less trivial application of (\ref{eqtn:cqcc_bound}) is to
the quantum Fourier transform.  We could explicitly work out the
Schmidt decomposition for the quantum Fourier transform by using the
procedure in Appendix \ref{app:mixed}. However, in this case we can
fortuitously note that the quantum Fourier transform on $2n$ qubits
can be written \be U & = & \sum_{x_1,\ldots,x_n,y_1,\ldots,y_n} \left[
\left( |0\ra + e^{2\pi i 0.y_n}|1\ra \right) \ldots \left( |0\ra +
e^{2\pi i 0.y_1\ldots y_n}|1\ra \right) \la x_1,\ldots,x_n| \right]
\otimes \nonumber \\ & & \left[ \left( |0\ra + e^{2 \pi i 0.x_n
y_1\ldots y_n}|1\ra\right) \ldots \left( |0\ra + e^{2\pi i 0.x_1\ldots
x_n y_1 \ldots y_n} |1\ra \right) \la y_1,\ldots,y_n| \right]. \ee
Simple algebra verifies that, as written, this is already the
operator Schmidt decomposition for the quantum Fourier transform.  It
follows that $\mbox{Sch}(U) = 2^{2n} = 4^n$, and thus, by
(\ref{eqtn:cqcc_bound}), the coherent quantum communication complexity
of the quantum Fourier transform is at least $n$ qubits.

Admittedly, this is a result which we were able to prove earlier, by
different means, however it is interesting to see that this result can
be obtained as a special case of a more general result.  To what other
problems might it be possible to apply this general technique?
Unfortunately, there do not seem to be many interesting unitary
operations known, for quantum computation.  One problem of some
interest would be to investigate the quantum communication complexity
of the Fourier transform over an arbitrary Abelian group rather than
the group of integers modulo $2^n$, as we have been considering.
Nevertheless, this is a somewhat artificial problem. Less artificial
is the iteration used by Grover \cite{Grover96a} in his search
algorithm; unfortunately, it is easy to see that this can be done
using two qubits of communication, making the problem rather trivial.

Another problem, recently suggested to me by Raymond Laflamme, is that
of evaluating the difficulty of performing quantum error correction in
a distributed fashion.  This is a problem which is potentially of great
interest in schemes for distributed quantum computation, such as that
suggested by Cirac, Zoller, Kimble and Mabuchi \cite{Cirac97a}.
Laflamme has also asked me whether the above proof techniques can be
adapted to a different model of distributed computation in which the
allowed operation is not qubit communication between the parties, but
rather quantum gates which may be performed jointly by the parties.
The answer is that yes, these techniques may be adapted in a
straightforward manner; a detailed working out of these developments
will appear elsewhere.

\section{A unified model for communication complexity}
\label{sec:unified_qcc}
\index{communication complexity!unified model for}
\index{quantum communication complexity!unified model for}

We have considered two broadly different classes of models for
communication complexity -- a class involving the computation of
classical functions, using quantum resources, and a class involving
the computation of quantum functions, using quantum resources.  In
this section a formalism is briefly outlined which has both these
classes of models as special limiting cases.

An obvious means of generalizing coherent communication complexity is
to consider the communication complexity for an arbitrary {\em quantum
operation}.  For example, suppose $\evop$ is a complete quantum
operation acting, jointly, on two systems, $A$ and $B$.  What is the
minimal number of qubits which must be communicated between $A$ and
$B$ if the quantum operation $\evop$ is to be implemented exactly?
Another very interesting case is the performance of collective
measurements on the system; suppose we have a set $\evop_0, \evop_1$
of incomplete quantum operations which represent a measurement.  How
much quantum communication must be performed between Alice and Bob if
they are to be able to perform that measurement?

For definiteness, we will study the case when $\evop$ is a complete
quantum operation on $AB$, which is to be implemented exactly, there
is no preshared resource existing between Alice and Bob, and
communication is to be carried out using qubits alone.  Each of the
choices implied in the previous sentence could be varied to provide
problems of considerable interest, but we will restrict ourselves to a
single problem.  Furthermore, we will not consider the very
interesting problem of {\em families} of communication problems, which
imply additional {\em uniformity requirements} for communication
protocols, along the lines sketched in item \ref{item:uniformity} on
page
\pageref{item:uniformity} for the quantum circuit model of quantum
computation.  We denote the communication complexity in this model by
$Q_0(\evop)$, the minimal number of qubits that must be communicated
in order to compute the quantum operation $\evop$ {\em exactly}.

Let $\evop_1$ and $\evop_2$ be two complete quantum operations on
$AB$.  We say $\evop_1 \leq \evop_2$, read {\em $\evop_1$ can be
reduced to $\evop_2$}, if there exist complete quantum operations
$\evop_A, \evop_A'$ on system $A$ and $\evop_B,\evop_B'$ on system $B$
such that \be \evop_1 = (\evop_A' \otimes \evop_B') \circ \evop_2
\circ (\evop_A \otimes \evop_B). \ee It is easily verified that $\leq$
is a partial order on the set of quantum operations.  It is clear that
$Q_0$ preserves this order, $Q_0(\evop_1) \leq Q_0(\evop_2)$, since to
perform $\evop_1$, all we need do is perform $\evop_A$ on system $A$,
$\evop_B$ on system $B$, then $\evop_2$, and finish by applying
$\evop_A'$ on system $A$ followed by $\evop_B'$ on system $B$, for a
total cost the same as the communication cost to compute $\evop_2$.

This result, incidentally, is a special case of a more general {\em
triangle inequality for communication complexity} \index{triangle
inequality!for communication complexity}.  This is the obvious
statement that if $\evop_1$ and $\evop_2$ are complete quantum
operations then 
\be
Q_0(\evop_2 \circ \evop_1) \leq Q_0(\evop_2) + Q_0(\evop_1). \ee

Let $f : \{0,1\}^n \times \{0,1\}^n \rightarrow \{0,1\}$ be any
classical function. Define a complete quantum operation $\evop_f$
which has as input $n$ qubits from system $A$, $n$ qubits from system
$B$, and as output, $1$ qubit system $A$, by the condition that
$\evop_f \equiv \evop_1 \circ {\cal D}$, where ${\cal D}$ completes
decoheres the system in the computational basis, and \be
\evop_1(|x\ra\la x| \otimes |y\ra \la y|) \equiv |f(x,y)\ra \la
f(x,y)|. \ee That is, $\evop_f$ is a quantum operation which takes the
state $|x\ra|y\ra$ as input, and outputs $|f(x,y)\ra$, while
destroying all coherences between computational basis states.

Let $\evop$ be any quantum operation on $AB$ which we would naturally
think of as computing $f$.  That is, we require that, on input of the
state $|x,y\ra$, the quantum operation $\evop$ should output
$|f(x,y)\ra$ on a fixed one of Alice's qubits.  Let ${\cal D}$ be the
operation which decoheres all Alice's qubits and all Bob's qubits.
The operation of decoherence in the computational basis can be
performed locally by both Alice and Bob, so $\evop \circ {\cal D} \leq
\evop$.  Furthermore, recall from Chapter \ref{chap:qops} that the
partial trace is a complete quantum operation.  This can certainly be
done locally: we are merely ignoring a system!  Thus, $\evop_f \leq
\evop \circ {\cal D} \leq \evop$, from which it follows that
$Q_0(\evop_f)) \leq Q_0(\evop)$.  It follows that if we wish to
calculate the communication complexity of the classical function $f$,
it suffices to calculate the communication complexity of the quantum
operation $\evop_f$.

Aesthetically, this is a pleasing result; it allows us to connect the
communication complexity of the classical function $f$ to the
communication complexity of a single quantum operation.  Thus, the
general question of the communication complexity of a quantum
operation contains both the coherent communication complexity, and the
quantum communication complexity of a classical function as special
cases.

%In fact, we can go even
%go further. Suppose $f$ is a permutation.  Suppose we have a $q$ qubit
%protocol which computes the quantum operation $\evop_f$. As per usual,
%we may introduce extra work bits for Alice and Bob which they can use
%to implement the protocol. Thus, the protocol can be regarded as a
%unitary protocol which takes the state
%\be
%|x_1,\ldots,x_n\ra|0,\ldots,0\ra_A \otimes |0,\ldots,0\ra_B |y_1,\ldots,y_n\ra
%|0,\ldots,0\ra, \ee
%and produces the output state
%\be |f(x,y)_1, \ldots, f(x,y)_n\ra |\psi_{AB}\ra
%|f(x,y)_{n+1},\ldots,f(x,y)_{2n}\ra, \ee
%where $|\psi_{AB}\ra$ is the final state of the work bits.  Suppose now
%we copy the classical result of the computation to an extra register
%using 

\section{Conclusion}
\label{sec:qcc_future}

Distributed classical computation is still only incompletely
understood; how much more true this is in the quantum case!  What are
some of the interesting open problems and directions for research in
the study of distributed quantum computation? In this section I
enumerate a few of the problems which I believe are particularly
interesting and important:

\begin{enumerate}

\item Find a general simulation technique for the entanglement
assisted classical communication model introduced by Cleve and Buhrman
\cite{Cleve97a} in the qubit communication model introduced by Yao
\cite{Yao93a}.

\item How is the coherent communication complexity affected by the
presence of one or more of the following resources: pre-shared
entanglement, pre-shared classical correlation, or classical
communication?

\item What are the coherent communication complexities for some more,
truly quantum, operations, beyond the quantum Fourier transform?

\item How are results on quantum communication complexity affected by
the presence of noise in the communications channel?

\item Can notions of quantum communication complexity be used to
define measures of entanglement in multipartite quantum systems?  In
Chapter \ref{chap:ent} we will study quantitative measures of the the
entanglement between two quantum systems.  These measures are based
upon {\em resource problems}.  Quantum communication complexity is a
natural source of such resource problems, and it is possible that one
of these resource problems may be used to provide a good measure of
entanglement, perhaps even for systems consisting of more than two
parts, a major bugbear of present efforts to study entanglement.
\end{enumerate}

This is just a sample of the sorts of questions which naturally arise
out of consideration of distributed quantum computation.  Judging from
the rapid progress over the past eighteen months, I expect that the
field of quantum communication complexity will be one of the major
areas of significant development in quantum information theory over
the next few years.  This progress, in turn, should help stimulate
other parts of the field with new insights into the nature of quantum
information.

\vspace{1cm}
\begin{center}
\fbox{
\parbox{14cm}{
\begin{center}
{\bf Summary of Chapter \ref{chap:qcomm}: Quantum communication
complexity}
\end{center}

\begin{itemize}

\item {\bf Capacity theorem for communication using qubits:}
Suppose that Alice possesses $n$ bits of information, and wants to
convey this information to Bob.  Suppose that Alice and Bob possess no
prior entanglement but qubit communication in either direction is
allowed.  Let $n_{AB}$ be the number of qubits Alice sends to Bob, and
$n_{BA}$ the number of qubits Bob sends to Alice.  Then, Bob can
acquire the $n$ bits if and only if the following inequalities are
satisfied:
\begin{eqnarray}
n_{AB}, n_{BA} & \geq & 0 \\
n_{AB} & \geq & \lceil n/2 \rceil \\ 
n_{AB}+n_{BA} & \geq & n.
\end{eqnarray}

\item {\bf Entanglement-assisted communication complexity:} The number
of bits of classical information that must be communicated between
Alice and Bob if they are to compute a given (classical) function,
given that they may preshare an arbitrary entanglement. In order to
compute the inner product, modulo 2, of an $n$ bit string belonging to
Alice, and an $n$ bit string belonging to Bob, requires precisely $n$
bits of classical communication.

\item {\bf Coherent quantum communication complexity:} How many qubits
need to be communicated between Alice and Bob if they are to compute a
quantum function, that is, some family of quantum operations? For the
$2n$ qubit quantum Fourier transform, $n$ qubits belonging to Alice
and $n$ qubits belonging to Bob, an $n$ qubit lower bound can be
proved.  Furthermore, to do an approximate Fourier transform a
distance $D < 1/2$ from the exact Fourier transform, at least
$n(1-D)-h(D)$ qubits must be sent, where $h(\cdot)$ is the binary
entropy.

\item {\bf Lower bound on the coherent communication complexity:}
\be
Q_0(U) \geq \lceil \log_4 \Sch(U) \rceil . \ee

\end{itemize}

}}
\end{center}

\chapter{Quantum data compression}
\label{chap:data_compress}

%\section{Quantum data compression}

\index{quantum data compression}
\index{data compression}
\index{Schumacher's theorem}

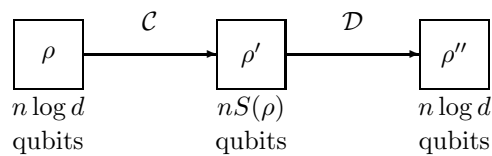
\begin{figure}
\begin{center}
\unitlength 0.9cm
\begin{picture}(9,2.5)(0,-0.2)
\put(1,1){\framebox(1,1){$\rho$}}
\put(0,0.2){\makebox(3,1){$n\log d$}}
\put(0,-0.3){\makebox(3,1){qubits}}
\put(2,1.5){\vector(1,0){2}}
\put(2.5,1.5){\makebox(1,1){${\cal C}$}}
\put(4,1){\framebox(1,1){$\rho'$}}
\put(3,0.2){\makebox(3,1){$n S(\rho)$}}
\put(3,-0.3){\makebox(3,1){qubits}}
\put(5,1.5){\vector(1,0){2}}
\put(5.5,1.5){\makebox(1,1){${\cal D}$}}
\put(7,1){\framebox(1,1){$\rho''$}}
\put(6,0.2){\makebox(3,1){$n\log d$}}
\put(6,-0.3){\makebox(3,1){qubits}}
\end{picture}
\end{center}
\caption{Quantum data compression. The compression operation
${\cal C}$ compresses a quantum source $\rho$ stored in $n \log d$ qubits
into $n S(\rho)$ qubits. The source is accurately recovered via the
decompression operation ${\cal D}$.
\label{fig: compression}}
\end{figure}

The storage of states produced by a quantum source using the fewest
possible resources is a fundamental problem of quantum information
theory.  Schumacher and co-workers
\cite{Schumacher95a,Jozsa94a,Barnum96a} have shown that a quantum
source $\rho$ produced by picking from an ensemble of quantum states
$\{ p_i, |\psi_i\rangle \}$ may be compressed so that it requires only
$S(\rho)$ qubits per source state for reliable storage. Barnum, Fuchs,
Jozsa and Schumacher \cite{Barnum96a} have shown that $S(\rho)$ qubits
per source state is the minimal resources required for reliable
storage.  The basic idea of quantum data compression is illustrated in
figure \ref{fig: compression}.  A quantum source $\rho$ on $d$ qubits
is used $n$ times.  A compression operation, ${\cal C}$ is used to
compress that source into roughly $n S(\rho)$ qubits.  At some later
time, a decompression operation, ${\cal D}$ is used to recover the
original state produced by the source, with high fidelity.

This Chapter addresses the compression of a quantum source producing
states $\rho$ which are entangled with another, inaccessible quantum
system, using the tools introduced in Chapter \ref{chap:distance},
especially the {\em dynamic fidelity}. As motivation for the use of
the dynamic fidelity, we might imagine that we are trying to compress
part of the memory of a quantum computer, and that we wish to recover
the entanglement with the rest of the quantum computer at some later
time.  This approach based upon the dynamic fidelity is quite
different to the work of Schumacher and collaborators, who used a
different measure of reliability, to be discussed below.  An advantage
of the present approach is that several of the proofs appear more
natural than in the approach pioneered by Schumacher
\cite{Schumacher95a}.  In particular, Schumacher did not find a simple
proof that $S(\rho)$ is the minimal resources required to do quantum
data compression. Schumacher \cite{Schumacher95a} and later Schumacher
and Jozsa \cite{Jozsa94a} gave incomplete proofs of this result that
did not consider the most general possible decoding schemes. The proof
was later completed by Barnum {\em et al} \cite{Barnum96a}, using an
ingenious but rather complicated argument. The techniques introduced
in this Chapter lead to simple and direct proofs.

Using these techniques we also study the problem of {\em universal
quantum data compression}.  In classical information theory the
existence of universal compression algorithms, such as the well-known
Lempel-Ziv \cite{Ziv78a} algorithm, is an important and useful fact,
exploited in many widely available programs and devices, such as the
UNIX {\em compress} program. A universal compression algorithm is one
which can compress a large class of sources, not just a single
source. We study the limits to universal quantum data compression, and
exhibit a quantum scheme which is universal with respect to a large
set of sources.

The Chapter is organized as follows.  Section \ref{sect: Schmidt}
defines the {\em Schmidt number} of a pure state of a composite
system, and studies some of the properties of the Schmidt number.
Many of the results in the Chapter are simple counting arguments based
on properties of the Schmidt number. Section \ref{sect: typical
subspaces} reviews some basic facts about typical subspaces of a
quantum source.  With these tools in hand, section \ref{sect:
compression} proves the quantum data compression theorem. Section
\ref{sect: side channel} extends this result to the case of a quantum
channel with a side channel for classical information. 
Section \ref{sect: universal with} studies universal quantum data
compression.   Section \ref{sect: compress_future} concludes the
Chapter with a discussion of future directions.  The work reported in
the Chapter is largely my original work, with the exception of Section
\ref{sect: typical subspaces}, which is based upon the work of
Schumacher and Jozsa.

\section{Schmidt numbers}
\label{sect: Schmidt}

\index{Schmidt bases}
\index{Schmidt number}
\index{Schmidt decomposition}

This section reintroduces an extremely useful tool for proving
properties about entangled systems: the {\em Schmidt number} of an
entangled pure state.  This tool was also used, in a different and
less central guise, in Chapter \ref{chap:qcomm}, to prove results
about quantum communication complexity. All of the results in this
section are rather elementary, yet they play a crucial role in our
proof of the quantum data compression theorem.

For convenience, we restate the {\em Schmidt decomposition} theorem,
an extremely useful structural theorem for pure states of composite
quantum systems, proved in Appendix \ref{app:mixed}.  Suppose
$|AB\rangle$ is a pure state for some joint system $AB$. Then there
exists a {\em Schmidt decomposition} for $|AB\rangle$,
\begin{eqnarray}
|AB\rangle = \sum_i \sqrt{p_i} |i^A \rangle |i^B \rangle
\end{eqnarray} where $|i^A\rangle$ is an orthonormal basis for $A$ and
$|i^B\rangle$ is an orthonormal basis for $B$ and $p_i \geq 0, \sum_i
p_i = 1$.  Recall also that these bases are identical to the bases in which
the reduced density operators on $A$ and $B$ are diagonal, since
\begin{eqnarray}
A & \equiv & \mbox{tr}_B(|AB\rangle \langle AB|) = \sum_i p_i |i^A\rangle
	\langle i^A| \\
B & \equiv & \mbox{tr}_A(|AB\rangle \langle AB|) = \sum_i p_i |i^B\rangle
	\langle i^B|. \end{eqnarray}

We define the {\em Schmidt number} of $|AB\rangle$ to be the
number of non-zero $p_i$ in the Schmidt decomposition of
$|AB\rangle$.  An equivalent and rather useful way of defining
the Schmidt number is using the concept of {\em support}. Given a
diagonalizable operator $D$ the {\em support} $\mbox{supp}(D)$ of that
operator is defined to be the vector space spanned by those
eigenvectors of $D$ whose corresponding eigenvalue is non-zero.
Clearly, the Schmidt number is equal to the dimension of the support
of $A$, which is also equal to the dimension of the support of
$B$,
\begin{eqnarray}
\mbox{Sch}(|AB\rangle) \equiv \dim \mbox{supp}(A) =
	\dim \mbox{supp}(B). \end{eqnarray} 
The Schmidt number has two especially useful properties
under operations:
\begin{enumerate}

\item Suppose ${\cal C} : Q \rightarrow Q'$ is any quantum operation
mapping density operators on the space $Q$ to density operators on the
space $Q'$. Let $RQ$ be a given density operator on the composite
system $RQ$. Let ${\cal I}$ be the identity operation on $R$. Then the
state of the system after the action of ${\cal C}$,
\begin{eqnarray}
R'Q' \equiv \frac{({\cal I} \otimes {\cal C})(RQ)}
	{\mbox{tr}\left[ ({\cal I}\otimes{\cal C})(RQ)\right]}, \end{eqnarray}
can be written in the form
\begin{eqnarray}
R'Q'= \sum_j q_j |R'Q'_j\rangle \langle R'Q'_j|, \end{eqnarray}
where $q_j \geq 0, \sum_j q_j = 1$, the $|R'Q'_j\ra$ form an
orthonormal set, and
\begin{eqnarray} \label{eqtn:Schmidt_1}
\mbox{Sch}(|R'Q'_j\rangle) \leq \mbox{dim}(Q'). \end{eqnarray}

\item Let ${\cal D}: Q' \rightarrow Q''$ be any quantum operation
mapping density operators on the space $Q'$ to density operators on
the space $Q''$. Let $R'Q'$ be any state of $R'Q'$. Then
\begin{eqnarray}
R''Q'' \equiv \frac{({\cal I} \otimes {\cal D})(R'Q')}{\mbox{tr}
	({\cal I} \otimes {\cal D})(R'Q')}
\end{eqnarray}
can be written in the form
\begin{eqnarray}
R''Q'' = \sum_k s_k |R''Q''_k \rangle \langle R''Q''_k|, \end{eqnarray}
where $s_k \geq 0, \sum_k s_k = 1$, the $|R''Q''_k\rangle$ are pure
states, not necessarily orthonormal, and
\begin{eqnarray} \label{eqtn:Schmidt_2}
\mbox{Sch}(|R''Q''_k\rangle) \leq \mbox{dim}(Q'). \end{eqnarray}

\end{enumerate}

Both properties follow immediately from the definition of the Schmidt
number.  Heuristically, the first property is just the obvious fact
that the output ensemble from a quantum operation can not have a
Schmidt number higher than the dimension of the output space. The
second result is only slightly less obvious, stating that a quantum
operation can not increase the Schmidt number of elements in an
ensemble. Actually, it is clear that stronger results than 2 are true,
and rather interesting, although we will not need such a result
here. For completeness, we describe one such result:

Let ${\cal D}: Q'\rightarrow Q''$ be any quantum operation, and $R'Q' =
|R'Q'\rangle\langle R'Q'|$ be any pure state of $R'Q'$. Defining $R''Q''$
as before, it follows that $R''Q''$ can be written in the form
\begin{eqnarray}
R''Q'' = \sum_k s_k |R''Q''_k\rangle \langle R''Q''_k|,
\end{eqnarray}
where $s_k \geq 0$, $\sum_k s_k = 1$ and
\begin{eqnarray}
\mbox{Sch} (|R''Q''_k\ra) \leq \mbox{Sch} (|R'Q'\ra).
\end{eqnarray}
Property 2, above, is clearly a consequence of this result, which is
also immediate from the definition of the Schmidt number.

\section{Typical subspaces}
\label{sect: typical subspaces}

\index{typical subspace}
\index{typical sequence}

The notion of a {\em typical subspace} was introduced by Schumacher
\cite{Schumacher95a} and Jozsa and Schumacher \cite{Jozsa94a} as the
quantum analogue of an important notion in classical information
theory, that of a typical sequence. They proved the following two
results, which we will refer to jointly as the {\em typical subspace
theorem}.  Both parts of the typical subspace theorem follow easily
from the weak law of large numbers, proved in the box on page
\pageref{box:lln}.

\begin{theorem}  {\bf (Typical subspace theorem)} \cite{Schumacher95a,Jozsa94a}

\begin{enumerate}

\item Fix a quantum state $\rho$ in a state space of $d$
qubits\footnote{The restriction to qubit systems is not necessary, but
it may help make the discussion more concrete.}.
Let $\epsilon > 0$ be given. For all $n$ sufficiently large there
exists a projector $P^n_{\epsilon}$ onto a space of at most
$2^{n(S(\rho)+\epsilon)}$ dimensions such that
\begin{eqnarray} \label{eqtn:typical_1}
\mbox{tr}(\rho^{\otimes n} P^n_{\epsilon}) & > &1-\epsilon \\ 
\label{eqtn:typical_2} \left[ \rho^{\otimes n},P^n_{\epsilon} \right] & = & 0.
\end{eqnarray}

\item Fix $\rho$. Let $\epsilon > 0$ be given. Let $P_n$ be any
sequence of projectors such that $P_n$ projects onto a space
of at most $2^{nR}$ dimensions. Then
\begin{eqnarray}
\mbox{tr}(\rho^{\otimes n}P_n) \leq 2^{-n(S(\rho)-R-\epsilon)} + \epsilon
\end{eqnarray}
for all sufficiently large values of $n$.
\end{enumerate}

\end{theorem}

\begin{proof} \cite{Schumacher95a,Jozsa94a}

For completeness, we outline the construction of the projector
$P^n_{\epsilon}$ onto the typical subspace. Suppose $\rho$ has
orthogonal decomposition,
\begin{eqnarray}
\rho^{\otimes n} = \sum_x p_x |x\rangle \langle x|. \end{eqnarray}
Then $\rho^{\otimes n}$ has orthogonal decomposition
\begin{eqnarray}
\rho = \sum_{\bf x} p_{\bf x} |{\bf x}\rangle \langle {\bf x}|,
\end{eqnarray}
where the sum is over all sequences ${\bf x} = x_1,\ldots,x_n$,
$p_{\bf x} \equiv p_{x_1} p_{x_2} \ldots p_{x_n}$ and
$|{\bf x}\rangle = |x_1\rangle|x_2\rangle \ldots |x_n\rangle$. We say
a sequence ${\bf x}$ is $\epsilon$-typical if
\begin{eqnarray}
2^{-n(S(\rho)+\epsilon)} \leq p_{\bf x} \leq 2^{-n(S(\rho)-\epsilon)}.
\end{eqnarray}
Intuitively, the sequence ${\bf x}$ can be thought of as the sequence
of outputs produced by a classical source producing independent random
variables, identically distributed according to $p_x$.  Using the law
of large numbers and taking the logarithm of the above definition, we
see that in the limit as $n$ goes to infinity, a typical sequence
occurs with probability going to one. Furthermore, since the sum of a
set of probabilities is at most one, and
\begin{eqnarray}
p_{\bf x} \geq 2^{-n(S(\rho)+\epsilon)}, \end{eqnarray} we see that
there at most $2^{n(S(\rho)+\epsilon)}$ $\epsilon$-typical sequences.
We now define
\begin{eqnarray}
P^n_{\epsilon} \equiv \sum_{\epsilon-\mbox{typical } {\bf x}} |{\bf
x}\rangle \langle {\bf x}| \end{eqnarray} to be the projector onto the
$\epsilon$-typical subspace.  Note that \be \tr(\rho^{\otimes n}
P^n_{\epsilon}) = \sum_{\epsilon-\mbox{typical } {\bf x}} p_{\bf x},
\ee so by the law of large numbers, for sufficiently large $n$,
$\tr(\rho^{\otimes n} P_{\epsilon}^n ) > 1-\delta$, for any $\delta >
0$. Setting $\delta \equiv \epsilon$ proves equation
(\ref{eqtn:typical_1}).  Furthermore, by definition $P^n_{\epsilon}$
is diagonal in the same basis as $\rho^{\otimes n}$, and thus commutes
with $\rho^{\otimes n}$, equation (\ref{eqtn:typical_2}).

The second property now follows from the identity
\begin{eqnarray}
\mbox{tr}(\rho^{\otimes n} P_n) = \mbox{tr} (\rho^{\otimes n} P^n_{\epsilon}
	P_n) + \mbox{tr}(\rho^{\otimes n}(I-P^n_{\epsilon})P_n)
\end{eqnarray}
and the observations that
\begin{eqnarray}
\mbox{tr}(\rho^{\otimes n} P^n_{\epsilon} P_n) \leq 2^{-n(S(\rho)-\epsilon)}
	2^{nR}, \end{eqnarray}
and
\begin{eqnarray}
\mbox{tr}(\rho^{\otimes n}(I-P^n_{\epsilon})P_n) & \leq &
	\mbox{tr}(\rho^{\otimes n}(I-P^n_{\epsilon})) \\
	& \leq & \epsilon. \end{eqnarray}
\end{proof}

\newpage
\begin{center}
\fbox{
\parbox[t]{6in}{
\renewcommand{\baselinestretch}{0.5} \normalsize
%\begin{singlespace}
\centerline{\bf The law of large numbers} \label{box:lln}

\index{law of large numbers}
\index{weak law of large numbers}

Suppose we repeat an experiment a large number of times, each time
measuring the value of some parameter, $X$. We label the results of
the experiments $X_1,X_2,\ldots$. Assuming that the results of the
experiments are independent, we intuitively expect that the value of
the estimator \beqn S_n \equiv \sum_{i=1}^n \frac{X_i}{n} \eeqn of the
average $\ex(X)$, should approach $\ex(X)$ as $n\rightarrow\infty$.
The {\em law of large numbers} \cite{Grimmett92a} is a rigorous
statement of this intuition.

{\bf Theorem} (Law of large numbers)
Suppose $X_1,X_2,\ldots$ are independent and identically distributed
random variables, with finite first and second moments,
$ |\ex(X_1)| < \infty $ and $\ex(X_1^2) < \infty$.
Then for any $\epsilon > 0$,
$p(|S_n - \ex(X)|>\epsilon) \rightarrow 0$
as $n \rightarrow \infty$.

{\bf Proof:}

To begin we assume that $\ex(X_i) = 0$. We will discuss what
happens when $\ex(X_i) \neq 0$ upon completion of the proof.
Since the random variables are independent
with mean zero, it follows that $\ex(X_i X_j) = \ex(X_i) \ex(X_j) = 0$
when $i \neq j$, and thus
\beqn
\ex( S_n^2) = \frac{\sum_{i,j=1}^n \ex(X_i X_j)}{n^2} = 
 \frac{\sum_{i=1}^n \ex(X_i^2)}{n^2} = \frac{\ex(X_1^2)}{n},
 \label{eqtn:cchan: lln one} \eeqn
where the final equality follows from the fact
that $X_1,\ldots,X_n$ are identically distributed. By the same token,
from the definition of the expectation we have
\beqn
\ex(S_n^2) & = & \int dP \, S_n^2, \eeqn
where $dP$ is the underlying probability measure. It is clear that either
$|S_n| \leq \epsilon$ or $|S_n| > \epsilon$, so we can split this
integral into two pieces, and then drop one of these pieces, observing
that it is non-negative,
\beqn
\ex(S_n^2) & = & \int_{|S_n| \leq \epsilon} dP \, S_n^2 +
        \int_{|S_n| > \epsilon} dP \, S_n^2
  \, \, \geq \, \int_{|S_n| > \epsilon} dP \, S_n^2. \eeqn
In the region of integration $S_n^2 > \epsilon^2$, and thus
\beqn
\ex(S_n^2) & \geq & \epsilon^2 \int_{|S_n| > \epsilon} dP \,\, =
 \, \epsilon^2 p(|S_n| > \epsilon). \eeqn
Comparing this inequality with (\ref{eqtn:cchan: lln one}) we
see that
\beqn
p(|S_n| > \epsilon) \leq \frac{\ex(X_1^2)}{n \epsilon^2}. \eeqn
Letting $n \rightarrow \infty$ completes the proof.
In the case when $\ex(X_1) \neq 0$, it is easy to obtain the
result, by defining $Y_i \equiv X_i - \ex(X_1)$. The $Y_i$ are
a sequence of independent, identically distributed random variables with
$\ex(Y_1) = 0$ and $\ex(Y_1^2)<\infty$. The result follows from
the earlier reasoning.

\qed

%\end{singlespace}
}
}
\end{center}

\section{Quantum data compression theorem}
\label{sect: compression}

\index{quantum data compression}
\index{information source!quantum}

We now have all the tools necessary to prove the quantum data
compression theorem. To understand the result, we first need to make
more formal the notions of quantum sources, and block encoding and
decoding.

An {\em i.i.d. (independent, identically distributed) quantum source}
$\{ H, \rho \}$ consists of a Hilbert space $H$ and a density operator
$\rho$ on that Hilbert space. The $n$-{\em blocked} source is the pair
$\{ H^{\otimes n}, \rho^{\otimes n} \}$. A {\em compression scheme of
rate} $R$ for a source $\{ H, \rho \}$ is a sequence $\{ {\cal C}^n,
{\cal D}^n \}$ of quantum operations such that the {\em encoding
operation} ${\cal C}^n$ maps the n-blocked source space $H^{\otimes
n}$ into a Hilbert space $H_c^n$ of dimension $2^{nR}$, and the {\em
decoding operation} ${\cal D}^n$ maps the $2^{nR}$ dimensional Hilbert
space $H_c^n$ back into the source Hilbert space $H^{\otimes n}$.

Our criterion for whether or not the compression-decompression
procedure has been successfully accomplished is whether or not the
dynamic fidelity for the total procedure is close to one.  As we saw
in Chapter \ref{chap:distance}, a dynamic fidelity close to one is
equivalent to the requirement that the source and any entanglement it
has with other systems has been well preserved by the process.

More precisely, we say that a compression scheme is {\em reliable} if
\begin{eqnarray}
\lim_{n \rightarrow \infty} F(\rho^{\otimes n}, {\cal D}^n \circ {\cal C}^n) = 1.
\end{eqnarray}
A compression scheme is said to be {\em weakly unreliable} if it is
not reliable. A compression scheme is said to be {\em strongly
unreliable} if
\begin{eqnarray}
\lim_{n \rightarrow \infty} F(\rho^{\otimes n}, {\cal D}^n \circ {\cal C}^n) = 0.
\end{eqnarray}
Clearly a  strongly unreliable compression scheme is also weakly unreliable.

\begin{theorem} {\bf (Quantum entanglement compression theorem)}

Let $\{ H, \rho \}$ be a quantum source.
\begin{enumerate}

\item (Achievability)

If $R > S(\rho)$ then there exists a reliable compression scheme 
of rate $R$ for the source $\{ H, \rho \}$.

\item (Weak converse).

If $R < S(\rho)$ then all compression schemes are weakly unreliable.

\item (Strong converse)

If $R < S(\rho)$ then all compression schemes are strongly unreliable.

\end{enumerate}

\end{theorem}

Obviously the strong converse implies the weak converse. Both results
are stated here because we will give a proof of the weak converse
which is independent of the proof of the strong converse.

\begin{proof}

{\em Proof of achievability}

The compression scheme used to prove achievability is exactly the same
as that used by Jozsa and Schumacher \cite{Jozsa94a}, although the
analysis is made slightly different by the use of dynamic fidelity as
the reliability criterion.

Let $\epsilon > 0$ be such that $S(\rho) + \epsilon \leq R$.  Define
$P^n_{\epsilon}$ to be the projector onto the $\epsilon$-typical
subspace, and use $T^n_{\epsilon}$ to denote the $\epsilon$-typical
subspace.  By the typical subspace theorem, for all $n$ sufficiently
large,
\begin{eqnarray}
\mbox{tr}(\rho^{\otimes n} P^n_{\epsilon}) \geq 1 - \epsilon, \end{eqnarray}
and
\begin{eqnarray}
\dim(T^n_{\epsilon}) \leq 2^{nR}. \end{eqnarray} Let $H^n_c$ be any
$2^{nR}$ dimensional Hilbert space containing $T^n_{\epsilon}$.  The
encoding is done in the following fashion. First a measurement is
made, described by the complete set of orthogonal projectors
$P^n_{\epsilon}, I-P^n_{\epsilon}$, with corresponding outcomes we
will call $1$ and $0$. If outcome $0$ occurs nothing more is done and
the state is left in the typical subspace. If outcome $1$ occurs then
we replace the state of the system with some standard state
$|0\rangle$ chosen from the typical subspace.  It follows that the
encoding is a map
\begin{eqnarray}
{\cal C}^n : H^{\otimes n} \rightarrow H^n_c \end{eqnarray}
and has the operator-sum representation
\begin{eqnarray}
{\cal C}^n(\sigma) \equiv P^n_{\epsilon} \rho P^n_{\epsilon} + \sum_i A_i \sigma A_i^{\dagger},
\end{eqnarray}
where
\begin{eqnarray}
A_i = |0 \rangle \langle i| \end{eqnarray}
and $|i\rangle$ is an orthonormal basis for the orthocomplement of the typical subspace.

The decoding operation
\begin{eqnarray}
{\cal D}^n : H^n_c \rightarrow H^{\otimes n} \end{eqnarray} is just
the identity on $H^n_c$, ${\cal D}^n(\sigma) = \sigma$.  With these
definitions for the encoding and decoding it follows that
\begin{eqnarray}
F(\rho^{\otimes n}, {\cal D}^n \circ {\cal C}^n) & = &
	|\mbox{tr}(\rho^{\otimes n} P^n_{\epsilon})|^2 + \sum_i
	|\mbox{tr}( \rho^{\otimes n} A_i)|^2 \\ & \geq &
	|\mbox{tr}\rho^{\otimes n} P^n_{\epsilon})|^2 \\ & \geq &
	|1-\epsilon|^2 \geq 1- 2\epsilon, \end{eqnarray} where the
	last line follows from the theorem of typical subspaces.  But
	$\epsilon$ can be made arbitrarily small and thus it follows
	that there exists a reliable compression scheme $\{ {\cal
	C}^n, {\cal D}^n \}$ of rate $R$ whenever $S(\rho) < R$.

{\em Proof of the weak converse}

In Chapter \ref{chap:capacity}, on page
\pageref{lemma:entropy-fidelity}, we prove the following result, known
as the {\em entropy-fidelity} inequality. For any $\rho$ and complete
quantum operations ${\cal C}$ and ${\cal D}$,
\begin{eqnarray} \label{eqtn: entropy-fidelity inequality}
S(\rho) \leq I(\rho,{\cal C}) + 2 + 4 (1-F(\rho,{\cal D} \circ {\cal C})) \log d, \end{eqnarray}
where $d$ is the dimension of the Hilbert space of $\rho$ and $I(\rho,{\cal C})$
is the {\em coherent information} defined by
\begin{eqnarray}
I(\rho,{\cal C}) \equiv S({\cal C}(\rho)) - S(\rho,{\cal C}),
\end{eqnarray} and $S(\rho,{\cal C})$ is a non-negative quantity known
as the {\em entropy exchange}. From the non-negativity of the entropy
exchange it follows that $I(\rho,{\cal C}) \leq S({\cal C}(\rho))$ and
thus from the entropy-fidelity inequality
\begin{eqnarray} \label{eqtn: Ahristopher Fuchs}
S(\rho) \leq S({\cal C}(\rho)) + 2 +4(1-F(\rho,{\cal D} \circ {\cal C})) \log d. \end{eqnarray}

Applying (\ref{eqtn: Ahristopher Fuchs}) to $\rho^{\otimes n}, {\cal C}^n, {\cal D}^n$
and noting that $S({\cal C}^n(\rho^{\otimes n})) \leq \log 2^{nR} = nR$ we see that
\begin{eqnarray}
n S(\rho) \leq nR + 2 + 4 n (1-F(\rho,{\cal D}^n \circ {\cal C}^n))  \log d, \end{eqnarray}
where $d$ is the dimension of the source space $H$. Dividing by $n$ and taking the limit as
$n \rightarrow \infty$ we see that
\begin{eqnarray}
S(\rho) \leq R + \lim_{n \rightarrow \infty} (1-F(\rho,{\cal D}^n \circ {\cal C}^n)) \log d
\end{eqnarray}
(when the limit exists). Thus, for reliable transmission we obtain
\begin{eqnarray}
S(\rho) \leq R. \end{eqnarray}
It follows that if $R < S(\rho)$ then all compression schemes must be 
weakly unreliable.

{\em Proof of the strong converse}

Suppose $|RQ\rangle$ purifies $\rho$. Then taking $n$ copies of $RQ$,
$|RQ_n \rangle \equiv |RQ\rangle^{\otimes n}$ purifies
$\rho^{\otimes n}$. Define
\begin{eqnarray}
\rho^{RQ}_n & \equiv & |RQ_n\rangle \langle RQ_n| \\
\rho^{RQ'}_n & \equiv & ({\cal I} \otimes {\cal C}^n)(\rho^{RQ}_n) \\
\rho^{RQ''}_n & \equiv & ({\cal I} \otimes {\cal D}^n)(\rho^{RQ'}_n),
\end{eqnarray}
where ${\cal I}$ is the identity operation on $R^{\otimes n}$.

From properties 1 and 2 of the Schmidt number enumerated on page
\pageref{eqtn:Schmidt_1} we see that $\rho^{RQ''}_n$ can be written in
the form
\begin{eqnarray}
\rho^{RQ''}_n = \sum_i p_i |\phi^{RQ''}_i\rangle \langle \phi^{RQ''}_i|, \end{eqnarray}
where $p_i \geq 0, \sum_i p_i = 1$ and
\begin{eqnarray}
\mbox{Sch}(|\phi^{RQ''}_i\rangle) \leq 2^{nR}. \end{eqnarray}
By definition we have
\begin{eqnarray} 
F(\rho^{\otimes n}, {\cal D}^n \circ {\cal C}^n) =
	\langle RQ_n| \rho^{RQ''}_n |RQ_n\rangle, \end{eqnarray}
and thus
\begin{eqnarray} \label{eqtn: Groucho}
F(\rho^{\otimes n}, {\cal D}^n \circ {\cal C}^n) =
  \sum_i p_i |\langle RQ_n |\phi_i^{RQ''}\rangle|^2. \end{eqnarray}
To bound the dynamic fidelity we examine the individual terms in this equation.
Note that
\begin{eqnarray}
| \langle RQ_n|\phi_i^{RQ''}\rangle|^2 = \mbox{tr}\left[ \rho^{RQ}_n
	\sigma_i^{RQ''} \right], \end{eqnarray}
where $\sigma_i^{RQ''} \equiv |\phi_i^{RQ''} \rangle \langle \phi_i^{RQ''}|$.
Let $P_i$ be the projector onto the support of $\sigma_i^{Q''}$. Notice that
\begin{eqnarray}
\sigma_i^{Q''} \leq (I^R \otimes P_i),
\end{eqnarray}
as an operator inequality, and thus
\begin{eqnarray}
\mbox{tr} \left[ \rho^{RQ} \sigma_i^{RQ''} \right] \leq
	\mbox{tr} \left[ \rho^Q_n P_i \right]. \end{eqnarray}
But
\begin{eqnarray}
\mbox{dim}(P_i) & = & \mbox{Sch} (|\phi_i^{RQ''}\rangle) \leq 2^{nR}
\end{eqnarray}
and by the second part of the typical subspace theorem it follows that
for sufficiently large $n$,
\begin{eqnarray}
\mbox{tr} \left[ \rho^Q_n P_i \right] \leq 2^{-n(S(\rho)-R-\epsilon)}. \end{eqnarray}
Putting it all together we have
\begin{eqnarray}
| \langle RQ_n|\phi_i^{RQ''}\rangle|^2 \leq 2^{-n(S(\rho)-R-\epsilon)}+\epsilon,
\end{eqnarray}
for sufficiently large $n$. Note, incidentally, that how large $n$ needs to be
depends only on $\epsilon$ and $R$, and not on $i$. Inserting this
equation into (\ref{eqtn: Groucho}) gives the result
\begin{eqnarray}
F(\rho^{\otimes n},{\cal D}^n \circ {\cal C}^n) \leq
	2^{-n(S(\rho)-R-\epsilon)}+\epsilon \end{eqnarray}
for all $\epsilon > 0$, for sufficiently large $n$. It follows that if $R < S(\rho)$ then
\begin{eqnarray}
F(\rho^{\otimes n},{\cal D}^n \circ{\cal C}^n) \rightarrow 0
\end{eqnarray}
as $n \rightarrow \infty$, which is what we set out to prove.

\end{proof}

Note that the proof of the weak converse presented here is more
difficult than that of the strong converse, since the proof makes
implicit use of the strong subadditivity inequality for entropies via
the entropy fidelity inequality, (\ref{eqtn: entropy-fidelity
inequality}), which is proved in Chapter
\ref{chap:capacity}. Nevertheless, the proofs of both converse
theorems appear quite natural compared to the proof found in
\cite{Barnum96a}. The proof of the strong converse, especially,
depends only upon elementary facts.  The same proof of the weak
converse was obtained independently by Allahverdyan and Saakian
\cite{Allahverdyan97a}. In the next section similar ideas will be used
to prove a stronger version of the weak converse than was proved by
Allahverdyan and Saakian.

%We have proved the quantum data compression theorem for i.i.d. sources.
%In fact, the only property of the sources used in the proof was
%the asymptotic equipartition property. Appendix \ref{app: general sources}
%discusses the case of more general sources for which the asymptotic
%equipartition property and thus the quantum data compression theorem holds.

\section{Quantum data compression with a classical side channel}
\label{sect: side channel}

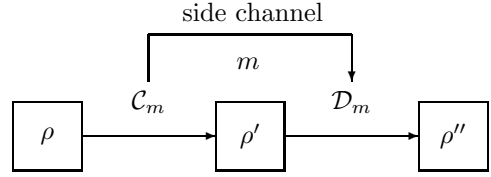
\begin{figure}[htbp]
\begin{center}
\unitlength 0.9cm
\begin{picture}(9,3)(0,0.5)
%\begin{picture}(9,3)(0,-0.2)
\put(1,1){\framebox(1,1){$\rho$}}
%\put(0,0.2){\makebox(3,1){$n\log d$}}
%\put(0,-0.3){\makebox(3,1){qubits}}
\put(2,1.5){\vector(1,0){2}}
\put(2.5,1.5){\makebox(1,1){${\cal C}_m$}}
\put(4,1){\framebox(1,1){$\rho'$}}
%\put(3,0.2){\makebox(3,1){$n S(\rho)$}}
%\put(3,-0.3){\makebox(3,1){qubits}}
\put(5,1.5){\vector(1,0){2}}
\put(5.5,1.5){\makebox(1,1){${\cal D}_m$}}
\put(7,1){\framebox(1,1){$\rho''$}}
%\put(6,0.2){\makebox(3,1){$n\log d$}}
%\put(6,-0.3){\makebox(3,1){qubits}}
% now we add the classical side channel
\put(3,2.3){\line(0,1){0.7}}
\put(3,3){\line(1,0){3}}
\put(4.3,2.5){$m$}
\put(3.5,3.2){\mbox{side channel}}
\put(6,3){\vector(0,-1){0.7}}
\end{picture}
\end{center}
\caption{Quantum data compression with a classical side channel. Results of
measurements made during compression are made available to assist
during decompression.
\label{fig: side channel}}
\end{figure}

In general, measurements are performed during compression of a quantum
source. These measurements yield classical information: the measurement
outcome. Suppose a {\em classical side-channel} is available, as shown in
figure \ref{fig: side channel}, so that the outcomes of the measurements
performed during compression are available to assist in decompression.

Intuitively, it seems likely that such a classical side channel cannot assist
in the compression of the quantum information. The reason is because any
measurement performed on the quantum system cannot obtain information
about that state without causing some irreversible disturbance to the
state. Thus, for the storage to be reliable it is necessary that the classical
side channel contain no information about the source. Therefore, it seems
unlikely that a classical side channel can assist in compression. 

Such intuitive arguments are at best heuristic guides. Consider that
quantum teleportation, review in Chapter \ref{chap:fundamentals},
involves the use of a classical side channel which is necessary for
the recovery of the quantum state of interest. Yet the information in
that side channel contains no information about the state which is
being teleported. This section gives a rigorous proof that the use of
a classical side channel does not decrease the minimal storage
requirements for a quantum state.

A {\em compression scheme of rate $R$ with side information} for a
source $\{ H, \rho \}$ is a sequence $\{ {\cal C}^n_i, {\cal D}^n_i
\}$ of quantum operations such that the {\em encoding operations}
${\cal C}^n_i$ map the n-blocked source space $H^{\otimes n}$ into a
Hilbert space $H_c^n$ of dimension $2^{nR}$, $\sum_i {\cal C}^n_i$ is a
complete quantum operation, and the {\em decoding operation} ${\cal
D}^n_i$ is a complete quantum operation that maps the $2^{nR}$
dimensional Hilbert space $H_c^n$ back into the source Hilbert space
$H^{\otimes n}$. 

Such a compression scheme is {\em reliable} if
\begin{eqnarray}
\lim_{n \rightarrow \infty} F(\rho^{\otimes n}, \sum_i 
	{\cal D}^n_i \circ {\cal C}^n_i) = 1,
\end{eqnarray}
and is said to be {\em weakly unreliable} if it is not reliable. Such
a compression scheme is said to be {\em strongly unreliable} if
\begin{eqnarray}
\lim_{n \rightarrow \infty} F(\rho^{\otimes n}, 
	\sum_i {\cal D}^n_i \circ {\cal C}^n_i) = 0.
\end{eqnarray}
Clearly a  strongly unreliable compression scheme is also weakly unreliable.

\begin{theorem} {\bf (Quantum entanglement compression with a classical side-channel)}

Let $\{ H, \rho \}$ be a quantum source. Then:
\begin{enumerate}

\item (Weak converse).

If $R < S(\rho)$ then all compression schemes with side information
are weakly unreliable.

\item (Strong converse)

If $R < S(\rho)$ then all compression schemes with side
information are strongly unreliable.

\end{enumerate}
\end{theorem}

Achievability need not be considered, since the compression scheme in
the last section already achieves the best possible rate of
compression without the need for classical side information. Once
again the strong converse implies the weak converse, and we outline
independent proofs of the two results.

\begin{proof}

{\em Outline proof of the weak converse}:

We use the ``generalized entropy-fidelity'' lemma from section
\ref{subsec: upper bounds on observed channel}. This result states
that for any set of operations ${\cal C}_i, {\cal D}_i$ such that
$\sum_i {\cal C}_i$ and ${\cal D}_i$ are trace-preserving,
\begin{eqnarray}
S(\rho) & \leq & \sum_i \mbox{tr}({\cal C}_i(\rho))I(\rho,{\cal C}_i) + 2 + \nonumber \\
	& & 4(1-F(\rho,\sum_m {\cal D}_i \circ {\cal C}_i)) \log d.
\end{eqnarray}
Applying this result with $\rho = \rho^{\otimes n}$, $d = d^n$, ${\cal C}_i = {\cal C}^n_i$,
and ${\cal D}_i = {\cal D}^n_i$ gives the result by the same reasoning used earlier in
the proof of the weak converse without a classical side channel.

{\em Outline proof of the strong converse}:

Define
\begin{eqnarray}
\rho^{RQ''}_n \equiv \left[ {\cal I} \otimes \left( \sum_i {\cal D}_i \circ {\cal C}_i \right)
	\right] (\rho^{RQ}_n).
\end{eqnarray}
The proof of the strong converse in the presence of a classical side
channel now proceeds exactly as that for the strong converse without a
classical side channel.

\end{proof}

\section{Universal data compression with a classical side channel}
\label{sect: universal with}

In order to do quantum data compression by the means we have described
it is necessary to know the source density operator $\rho$ in order to
construct the projector onto the typical subspace of $\rho$. An
elegant quantum circuit which does essentially this has been described
by Cleve and DiVincenzo \cite{Cleve96a}, demonstrating the possibility
of doing quantum data compression on a quantum computer.

In order for such methods for data compression to work it is necessary
to know the source density operator. Many popular algorithms for data
compression on classical computers work in a very different way,
succeeding for all source distributions within some large class of
possible source distribution.  Essentially, they do this by sampling
from the source to build up a good knowledge of the nature of the
source distribution, and use that knowledge to perform encoding in an
appropriate manner.  For example, Lempel-Ziv coding \cite{Ziv78a},
used in popular programs such as the UNIX {\em compress}, compresses
all stationary ergodic sources to the limit allowed by Shannon's
noiseless channel coding theorem. Obviously, such algorithms are
highly desirable whenever one does not know the source distribution
{\em a priori}.

Clearly, it is desirable to have analogous universal {\em quantum}
data compression algorithms. At first sight this may appear hopeless:
classically, universal data compression works because the compressor
obtains information about a source as the source is sampled, which
asymptotically allows efficient compression. Quantum mechanically, we
know that we cannot obtain information about a source without
disturbing it, so it would seem difficult to perform universal quantum
data compression.  Despite, this difficulty, we will see in this
section that it is possible to perform a useful form of universal
quantum data compression, making use of an auxiliary classical side
channel.  We begin with a simple example.

Suppose $\rho_1$ and $\rho_2$ are density operators such that
$S(\rho_1) < S(\rho_2)$. Suppose $P_i^n$ is the projector onto the
$\epsilon$-typical subspace of $\rho_i^{\otimes n}$ and define
\begin{eqnarray}
Q_i^n \equiv I^{\otimes n} - P_i^n,
\end{eqnarray}
the projector onto the complement of the $\epsilon$-typical subspace of
$\rho_i^{\otimes n}$. The data compression stage is done as follows:
\begin{enumerate}

\item Perform the measurement defined by the projectors
$P_1^n$ and $Q_1^n$. Label the corresponding measurement
result $M_1$. $M_1=0$ if $P_1^n$ occurred, and $M_1=1$ if
$Q_1^n$ occurred.

\item If $M_1 = 1$, then perform the measurement defined by the projectors
$P_2^n$ and $Q_2^n$. Label the corresponding measurement
result $M_2$. $M_2=0$ if $P_2^n$ occurred, and $M_2=1$ if
$Q_2^n$ occurred. If $M_1 = 0$ then define $M_2 = 1$.

\item If $M_1 = 1$ and $M_2 = 1$ then set $M_3 \equiv 0$.
\end{enumerate}
Let $r$ be the minimal value such that $M_r = 0$. Suppose $r=3$.  Then
we don't send anything; we give up.  If $r = 1$ then we send (a
unitary transformation of) the typical subspace of $\rho_1$, which
asymptotically requires $S(\rho_1)$ qubits per use of the source.  If
$r = 2$ then we send (a unitary transformation of) the typical
subspace of $\rho_2$, which asymptotically requires $S(\rho_2)$ qubits
per use of the source.  The decoding operation is merely to map the
state of the transmitted qubits back into the appropriate typical
subspace for $\rho_1$ or $\rho_2$, depending on whether $r = 1$ or $r=
2$.  If $r = 3$, then no decoding is performed; we give up.

If the source is $\rho_1$, then the dynamic fidelity for this
compression scheme satisfies \be F(\rho_1^{\otimes n}, {\cal D}^n
\circ {\cal C}) \geq |\tr(P^n_1 \rho_1^{\otimes n})|^2, \ee which, as
we saw earlier, asymptotically tends to $1$ as $n \rightarrow \infty$.
Much more interesting is the case when $\rho_2$ is the source.  In
this case the dynamic fidelity satisfies 
\be F(\rho_2^{\otimes n},{\cal D}^n \circ {\cal C}^n) \geq |\tr(P^n_2
Q^n_1 \rho_2^{\otimes n})|^2. \ee

Note, however, that
\be
|\tr(P^n_2 P^n_1 \rho_2^{\otimes n})| \leq
\tr(P^n_1 | \rho_2^{\otimes n} P^n_2 |). \ee
But the largest eigenvalue of $|\rho_2^{\otimes n}P^n_2|$ is less than
the largest eigenvalue of $\rho_2^{\otimes n}$, so by the theorem of
typical subspaces
\be
|\tr(P^n_2 P^n_1 \rho_2^{\otimes n})| \leq 2^{n(S(\rho_1)+\epsilon)}
2^{-n(S(\rho_2)-\epsilon)} = 2^{n(S(\rho_1)-S(\rho_2)+2\epsilon)}, \ee
which tends to zero as $n \rightarrow \infty$ and we allow $\epsilon$
to tend to zero. Thus
\be
F(\rho_2^{\otimes n},{\cal D}^n \circ {\cal C}^n) & \geq & |\tr(P^n_2
Q^n_1 \rho_2^{\otimes n})|^2 \\
& \geq & |\tr(P^n_2 \rho_2^{\otimes n})-\tr(P^n_2 P^n_1
\rho_2^{\otimes n})|^2 \\
& \geq & |1-\epsilon-2^{-n(S(\rho_1)-S(\rho_2)+2\epsilon)}|^2, \ee
from which we deduce that this compression scheme is asymptotically
reliable as $n \rightarrow \infty$.

We have demonstrated a compression scheme which compresses both
$\rho_1$ and $\rho_2$. Notice that the measurement result $r$
identifies, with high probability, whether the input density operator
was $\rho_1$ or $\rho_2$.

More generally, suppose $\rho_1,\ldots,\rho_M$ is a set of density
operators which is {\em entropy distinct}, that is, the density
operators have distinct Von Neumann entropies. We will show that there
is a compression scheme which is optimal with respect to all of these
sources, by a straightforward generalization of the previous
construction.  In particular, order the density operators such that
\be
S(\rho_1) < S(\rho_2) < \ldots < S(\rho_M). \ee
Then perform the following procedure:
\begin{enumerate}
\item Starting with $i = 1$, perform the measurement defined by the
projectors $P_i^n$ and $Q_i^n$. Label the corresponding measurement
result $M_i$; $M_i=0$ if $P_i^n$ occurred, and $M_i=1$ if $Q_i^n$
occurred.  Repeat, incrementing $i$ until $M_i = 0$ for some value of
$i$, and then proceed to the next step.  If $M_i \neq 0$ for all $i$,
then set $M_{M+1} = 0$, and proceed to the next step.
\item Let $r$ be the minimal value such that $M_r = 0$. If $r = M+1$
then give up.  Otherwise, compress the typical subspace of $\rho_r$
into $S(\rho_r)$ qubits per use of the source.
\item To decompress, if $r = M+1$, then do nothing.  Otherwise, do the
appropriate unitary transform on the $S(\rho_r)$ qubits per source
symbol to move them back to the typical subspace of $\rho_r$.
\end{enumerate}

%To see that this works, define a quantum operation ($i \geq j$)
%\be
%\evop_{ij}(\sigma) \equiv P^n_i Q^n_{i-1} \ldots Q^n_j \sigma Q^n_j
%\ldots Q^n_{i-1} P^n_i. \ee
%Note that
%\be
%F(\rho_i^{\otimes n},{\cal D}^n \circ {\cal C}^n) & \geq & p(i)
%F(\rho_i^{\otimes n},\evop_{i1}), \ee
%where $p(i)$ is the probability of obtaining 

Note that the dynamic fidelity of this compression scheme if the
source is $\rho_i$ is lower bounded by \be F(\rho_i^{\otimes n},{\cal
D}^n \circ {\cal C}^n) \geq |\tr(P^n_i Q^n_{i-1} \ldots Q^n_1
\rho_i^{\otimes n})|^2. \ee Note, however, that for any set $1 \leq
i_1 \leq \ldots \leq i_k \leq i-1$, \be \left| \rho_i^{\otimes n}
P^n_{i_1} \ldots P^n_{i_k} P^n_i \right| \leq
2^{-n(S(\rho_i)-\epsilon)} P^n_i, \ee and therefore, for sufficiently
large $n$ and small $\epsilon$, \be F(\rho_i^{\otimes n},{\cal D}^n
\circ {\cal C}^n) \geq |1-\epsilon-2^i \epsilon|^2 \\ & \geq &
|1-(2^m+1)\epsilon|^2. \ee Letting $\epsilon \rightarrow 0$ we obtain
the result.

We have shown how to compress a finite set of entropy-distinct
sources.  What about more general sources?  In section \ref{subsec:
dense subset} the following surprising result is proved: there exists
a countable set $\pi = \{ \rho_1,\rho_2,\ldots \}$ of source density
operators which is both entropy distinct, and dense in the set of all
source density operators.

Using previous results we can construct a compression scheme which is
optimal with respect to all the source density operators in
$\pi$. This is not universal data compression, but it is data
compression which works for a set of sources dense in the set of all
possible sources. Recall that a set $X$ is said to be dense in $Y$ if
every point in $Y$ can be arbitrarily well approximated by a point in
$X$.

\begin{figure}
\begin{eqnarray*}
\begin{array}{|c|cccc|} \hline & & & & \\
	\mbox{Block} & \{ \rho_1 \} & \{ \rho_1,\rho_2 \}
	& \{ \rho_1,\rho_2,\rho_3 \} & \ldots \\ 
	\mbox{length} & & & & \\ \hline & & & & \\ 
	1 & \underline{({\cal C}^1_1,{\cal D}^1_1)} & ({\cal C}^1_2,{\cal D}^1_2)
	   & ({\cal C}^1_3,{\cal D}^1_3) & \ldots \\ & & & & \\
	2 & \underline{({\cal C}^2_1,{\cal D}^2_1)} & ({\cal C}^2_2,{\cal D}^2_2)
	   & ({\cal C}^2_3,{\cal D}^2_3) & \ldots \\  & & & & \\
	3 & \underline{({\cal C}^3_1,{\cal D}^3_1)} & ({\cal C}^3_2,{\cal D}^3_2)
	   & ({\cal C}^3_3,{\cal D}^3_3) & \ldots \\ & & & & \\
	\ldots & \underline{(\ldots,\ldots)} & (\ldots,\ldots) & (\ldots,\ldots) & \ldots \\
	& & & & \\
	n(2) & ({\cal C}^1_1,{\cal D}^1_1) & \underline{({\cal C}^1_2,{\cal D}^1_2)}
	   & ({\cal C}^1_3,{\cal D}^1_3) & \ldots \\ & & & & \\
	n(2)+1 & ({\cal C}^1_1,{\cal D}^1_1) & \underline{({\cal C}^1_2,{\cal D}^1_2)}
	   & ({\cal C}^1_3,{\cal D}^1_3) & \ldots \\ & & & & \\
	\ldots & (\ldots,\ldots) & \underline{(\ldots,\ldots)} & (\ldots,\ldots) & \ldots \\
	& & & & \\
	n(3) & ({\cal C}^1_1,{\cal D}^1_1) & ({\cal C}^1_2,{\cal D}^1_2)
	   & \underline{({\cal C}^1_3,{\cal D}^1_3)} & \ldots \\  & & & & \\
	n(3)+1 & ({\cal C}^1_1,{\cal D}^1_1) & ({\cal C}^1_2,{\cal D}^1_2)
	   & \underline{({\cal C}^1_3,{\cal D}^1_3)} & \ldots \\ & & & & \\
	\ldots & \ldots & \ldots & \ldots & \ldots \\ \hline
\end{array}
\end{eqnarray*}
\caption{\label{fig: extension to infinity} Construction of a data
compression scheme which works on all elements in $\pi$. The vertical
columns indicate a compression scheme that is optimal with respect to
the set of source density operators at the head of the column. For
example, the second column contains a compression scheme which is
optimal with respect to both $\rho_1$ and $\rho_2$. From this table
the underlined elements are used to construct a new compression scheme
which is optimal with respect to all elements of $\pi$.}
\end{figure}

Fix $m$. Let $({\cal C}^n_m,{\cal D}^n_m)$ be a
compression scheme which is optimal with respect to the first $m$ elements
of $\pi$, namely $\rho_1,\ldots,\rho_m$.  We will use this sequence
of compression schemes to construct a compression scheme
which works on every element in $\pi$. The essential
elements of  the construction are illustrated in figure \ref{fig: extension to infinity}.

For each $m$, let $n(m)$ be such that the compression scheme $({\cal
C}^n_m,{\cal D}^n_m)$ gives fidelities greater than $1 - 1/m$ for all
$n \geq n(m)$. (It is clear that $n(m)$ can be chosen in such a way
that it increases with $m$). Now for $i$ in the range $n(m)$ to
$n(m+1)-1$, define
\begin{eqnarray}
{\cal C}^i & \equiv & {\cal C}^i_m \\
{\cal D}^i & \equiv & {\cal D}^i_m.
\end{eqnarray}
This defines a compression scheme $({\cal C}^i,{\cal D}^i)$ which is
optimal for all source density operators in $\pi$, by construction. It
is an interesting open problem to determine whether or not
asymptotically reliable compression is possible for all
i.i.d. sources.

%Finally, suppose $\rho$ is a density operator not in $\pi$. Choose
%a density operator $\sigma$ in $\pi$ such that $F(\rho,\sigma) \geq
%2^{-\Delta}$, where $\Delta > 0$ is arbitrary.  Then, for sufficiently
%large $n$ we have, by equation (\ref{eqtn:John_Lennon}),
%\be
%F(\rho^{\otimes n}, {\cal D}^n \circ {\cal C}^n) \geq F(\rho^{\otimes
%n},\sigma^{\otimes n})^4 F(\sigma^{\otimes n},{\cal D}^n \circ {\cal
%C}^n) \geq (1-\epsilon) 2^{-4 n \Delta}, \ee
%for any $\epsilon > 0$.  But $\Delta$ was arbitrary, so what this
%result implies is that, as $n \rightarrow \infty$, the rate of decay
%of dynamic fidelity is zero, for all i.i.d. sources.  This does not,
%unfortunately, imply that this compression scheme is asympotically
%reliable for all i.i.d. sources, however it is the next-best-thing.

\subsection{A dense subset of density operators with distinct entropies}
\label{subsec: dense subset}

This subsection proves a technical theorem which is needed in our work
on universal data compression. As such, it is rather ancillary to the
main line of thought in the Chapter, however I find the result to be
interesting and surprising in its own right.

We begin by defining some terms. A collection of density operators is
{\em entropy distinct} if no two density operators in the collection
have the same entropy.  A collection $\Lambda$ of density operators is
said to be {\em dense} in the set of all density operators if for any
$\epsilon > 0$ and density operator $\rho$, there exists
$\rho_{\Lambda} \in \Lambda$ such that $D(\rho,\rho_{\Lambda}) <
\epsilon$, where $D(\cdot,\cdot)$ is the absolute distance studied in
Chapter \ref{chap:distance}.  

\begin{theorem}

There exists a countable set $\{ \rho_1,\rho_2,\ldots \}$ of density
operators which is both entropy distinct, and dense in the set of all
density operators.

\end{theorem}

We will begin the proof of the theorem with some definitions and a
lemma.  Recall from subsection \ref{subsec:absolute_distance} that
the entropy function is continuous with respect to the absolute
distance.  Moreover, the set of density operators is easily seen to be
compact with respect to the absolute distance.

Suppose $\epsilon > 0$. Then an {\em $\epsilon$-net} on the set of
density operators is a collection $\Lambda_{\epsilon}$ such that given
any density operator $\rho$, there exists $\rho_{\Lambda} \in \Lambda$
such that
\begin{eqnarray}
D(\rho_{\Lambda},\rho) < \epsilon. \end{eqnarray}

\begin{lemma}

Let $\epsilon > 0$ be given. Then there exists a finite entropy
distinct $\epsilon$-net $\{ \rho_1,\ldots,\rho_n \}$.

\end{lemma}

\begin{proof}

Since the set of density operators is compact, there exists a finite
$\epsilon/2$-net $\{ \sigma_1,\ldots,\sigma_n \}$. Suppose two or more
of these density operators have the same entropy. For example, suppose
$S(\sigma_1) = S(\sigma_2)$. Then we set $\rho_1 \equiv \sigma_1$ and
perturb $\sigma_2$ by a small amount, for example
\begin{eqnarray}
\rho_2 \equiv p \sigma_2 + (1-p) \frac{I}{d} \end{eqnarray} where $p
\approx 1$ is chosen in such a way that $S(\rho_2) \neq S(\sigma_2)$,
but $D(\sigma_2,\rho_2) < \epsilon/2$. (Note that if $\sigma_2 = I/d$
then it may be necessary to perturb $\sigma_2$ using a different state,
say a pure state).  It is easy to see that by perturbing all the
density operators $\sigma_i$ by a distance less than $\epsilon/2$ it
is possible to ensure that the resulting perturbed set $\rho_i$ is
entropy distinct. Moreover, the resulting set is an $\epsilon$-net,
since
\begin{eqnarray}
D(\rho,\rho_i) \leq D(\rho,\sigma_i) + D(\sigma_i,\rho_i) <
	\frac{\epsilon}{2} + \frac{\epsilon}{2}, \end{eqnarray} where
	$i$ has been chosen such that $D(\rho,\sigma_i) < \epsilon/2$.
	This completes the proof of the lemma.

\end{proof}

\begin{proof} (\textbf{Main theorem})

Returning to the main theorem, for each $n = 1,2,\ldots$, let
$\Lambda_n$ be a finite entropy distinct $1/(2n)$-net. We will use
these nets to construct an increasing sequence $\Lambda_n'$ of
$1/n$-nets. Set $\Lambda_1' \equiv \Lambda_1$.  Given $\Lambda_n'$ we
construct $\Lambda_{n+1}'$ as follows. By perturbing each element in
$\Lambda_{n+1}$ by a distance at most $1/[2(n+1)]$ to create a perturbed
set $\Lambda^p_{n+1}$ it is possible to ensure that the resulting
union
\begin{eqnarray}
\Lambda_{n+1}' \equiv \Lambda_n' \cup \Lambda_{n+1}^p
\end{eqnarray}
is entropy distinct. Observing that $\Lambda_{n+1}^p$ is a $1/(n+1)$-net we
see that $\Lambda_n'$ is a finite, entropy distinct $1/n$-net such that
\begin{eqnarray}
\Lambda_1' \subseteq \Lambda_2' \subseteq \Lambda_3' \subseteq \ldots.
\end{eqnarray}
Define
\begin{eqnarray}
\Lambda \equiv \bigcup_{i=1}^{\infty} \Lambda_i'. \end{eqnarray} It is
now easy to see that $\Lambda$ is a countable, entropy distinct subset
of the set of density operators. Furthermore, $\Lambda$ is dense,
since given $\rho$ and $\epsilon > 0$, choose $n$ such that
$\frac{1}{n} < \epsilon$, and then $\rho' \in \Lambda_N' \subseteq
\Lambda$ such that
\begin{eqnarray}
d(\rho,\rho') < \frac 1n < \epsilon. \end{eqnarray}
This completes the proof of the theorem.

\end{proof}

Notice, incidentally, that there was very little about the entropy
function that was used in the proof. All one needs is a function which
varies sufficiently near any point in the space of interest in order
for the result to hold. It is not even necessary that the function be
continuous. Similarly, it is obviously possible to extend the result
well beyond the choice of the absolute distance and the set of density
operators. We will not investigate such generalizations here.

\section{Conclusion}
\label{sect: compress_future}

This Chapter has presented simplified proofs of the quantum data
compression theorem, and studied universal quantum data compression.
How might we extend the work further? There are several natural
questions we might ask:
\begin{itemize}

\item What is largest class of sources for which
an analogue of the typical subspace theorem holds? Any such source
will necessarily satisfy the quantum data compression theorem.

\item Is reliable data compression possible for all i.i.d. sources?

\item Can universal data compression (or similar schemes, such as the
compression scheme described in this Chapter) be implemented
efficiently using a quantum circuit?

\end{itemize}
The ultimate utility of quantum data compression depends upon whether
large scale quantum computers are ever built.  At present, this
eventuality appears to be quite far off. Nevertheless, I am hopeful
that such large scale devices will one day be built, and that it may
even be found useful to implement data compression in those devices.
In any case, it is certainly true that studies of quantum data
compression give insights into quantum information that allow us to
make progress in other, perhaps more immediately practical, areas of
quantum information theory.

\vspace{1cm}
\begin{center}
\fbox{
\parbox{14cm}{
\begin{center}
{\bf Summary of Chapter \ref{chap:data_compress}: Quantum
data compression}
\end{center}

\begin{itemize}

\item \textbf{Quantum data compression:} We have shown that
compression can be performed in such a way that the entanglement of
the source with another system can be recovered with high
fidelity.This is in contrast to Schumacher's original theorem, which
used a weaker measure of fidelity, the ensemble fidelity, as a measure
of reliability.

\item \textbf{A classical side channel does not decrease the minimal
resources required for reliable storage of quantum information.}

%\item \textbf{Universal data compression is not possible for quantum data
%without a classical side channel.}

\item \textbf{Universal quantum compression scheme is possible for a
set of sources dense in the set of all (i.i.d.) quantum sources.}

\end{itemize}

}}
\end{center}

\chapter{Entanglement}
\label{chap:ent}

% add the Hydrogen stuff back in?
% measurement of entanglement?

What is entanglement? This is a question we've skirted, until
now. We've seen that entanglement is a useful resource which can be
used to assist in the performance of quantum information processing
tasks. It would be useful to have a more precise way of quantifying
what we mean by entanglement, and understanding what it can be used to
do.  The purpose of this Chapter is to begin to develop such an
understanding.

The most familiar example of an entangled system which we've met is
the maximally entangled state of a two qubit system,
\begin{eqnarray}
|\psi\rangle = \frac{1}{\sqrt 2} \left( |00\ra+|11\ra
\right). \end{eqnarray} As we have seen in earlier chapters, this
state has many remarkable properties which do not appear to have
classical analogues.  In this Chapter we will explore in greater
detail what it means for two quantum systems to be entangled.

A great deal of work has been done studying the properties of
entanglement, and I will not attempt to list all of this work here.
It has been widely suggested (see, for example, \cite{Jozsa97a}) that
entanglement is of central importance in giving quantum information
processing systems an edge over classical information processing
systems, although other authors \cite{QCC97a} have argued that
entanglement is at most part of the story in explaining this
difference in computational power.  Nevertheless, despite extensive
work entanglement remains a poorly understood phenomenon.

I regret to say that I have not succeeded in understanding
entanglement as deeply as I would have liked.  It seems to be a
difficult subject; this is reflected also in the relatively slow
progress that has been made in the literature, despite much ingenuity
on the part of many researchers.  This Chapter is mainly concerned
with reviewing a simple quantitative tool which has been developed to
study entanglement, the {\em entanglement of formation}, and proving
several new properties of this tool.  The most significant new result
is a relationship between the entanglement of formation and negative
conditional quantum entropies.  In Chapter \ref{chap:capacity} this
result will be used to help gain insight into the quantum channel
capacity.

The Chapter is structured as follows.  In section
\ref{sec:pure_state_entanglement} we study entanglement for pure
states of a two part composite system.  Section
\ref{sec:mixed_state_entanglement} develops many elementary properties
of entanglement for arbitrary quantum states, including mixed states.
Section \ref{sec:entanglement_examples} steps back from the abstract
properties of entanglement studied in the preceding section, and
examines some simple examples which may be used to build intuition
about entanglement.  Along the way, we note an amusing fact about
entanglement which makes clear that it must play a rather subtle role
in quantum information processing: evidence that there may be quantum
computations which cannot be simulated efficiently on a classical
computer, but in which no two qubits ever become entangled.
%Section \ref{sec:dynamic_entanglement} studies the \emph{dynamic
%entanglement}, a quantity which measures the amount of entanglement
%processed by a quantum channel. In Chapter
%\ref{chap:capacity} this quantity will be used to obtain an upper
%bound on the quantum channel capacity.  
Section \ref{sec:future_entanglement} concludes the Chapter with an
overview of some of the future directions which could be taken to the
study of entanglement in quantum systems.

Section \ref{sec:pure_state_entanglement} reviews the work of other
researchers on pure state entanglement.  Section
\ref{sec:mixed_state_entanglement} is based partially upon other
people's work, however the main result of the section, the
entropy-entanglement inequality, is my own work.  The remainder of the
Chapter is largely my own original work.  I would like to thank Dorit
Aharonov and Bill Wootters for several fun conversations and some
great papers that got me more deeply interested in the subject, and
convinced me that understanding this mysterious stuff we call
entanglement is one of the central problems of quantum information
theory.  Their thoughts and words helped motivate much of the work
here.

\section{Pure state entanglement}
\label{sec:pure_state_entanglement}

What does it mean for a composite system to be entangled?  In the case
of pure states, an answer to this question has been given by Popescu
and Rohrlich \cite{Popescu97a} for the case of a two part composite
system. Suppose $|AB\ra$ is a pure state of a two part composite
system, $AB$. Let us assume that $E(A:B)$ is some measure of the
entanglement between systems $A$ and $B$.  Suppose we demand that the
following properties hold for $E$:
\begin{enumerate}
\item $E$ is a function of the state $|AB\ra$ alone.
\item $E$ is continuous with respect to the Hilbert space distance for
pure states of $AB$.
\item Suppose there are classical parties Alice and Bob who can
manipulate systems $A$ and $B$, respectively. Suppose they can perform
operations on their own systems, and communicate classical
information. The entanglement $E(A:B)$ between them can not be
increased by such operations.
\item The entanglement is additive.  That is, suppose Alice and Bob
jointly possess a number of systems, $(AB)_1, (AB)_2,
\ldots, (AB)_n$.  The systems $A_i$ and $B_i$ may be entangled, but we
assume that the total state of the system is a product state of the
pairs $(AB)_i$.  Then the total entanglement between Alice and Bob is
the sum of the entanglement in each subsystem, $E(A:B) = \sum_i
E(A_i:B_i)$.
\end{enumerate}
Popescu and Rohrlich showed that, up to an undetermined overall
constant factor, the entanglement associated to a pure state is then
$E(A:B) = S(A)$, where $S(A)$ is the von Neumann entropy of subsystem
$A$.  Note that $S(A) = S(B)$ as $A$ and $B$ are in a pure state.

The method used by Popescu and Rohrlich may be briefly described as
follows.  Bennett {\em et al} \cite{Bennett96b,Bennett96a} had earlier
considered the problem of {\em formation} for an entangled
state. Specifically, suppose Alice and Bob want to create $n$ copies
of a pure state $|AB\ra$, using only local operations and classical
communication.  The {\em entanglement of formation} for $|AB\ra$ is
defined to be the maximal number, $c$, such that if Alice and Bob are
provided with $\lfloor cn \rfloor$ Bell states, then as $n \rightarrow
\infty$ they can create those $n$ copies of $|AB\ra$ with
asymptotically good fidelity.  Bennett {\em et al} show that the
entanglement of formation is equal to $S(A)$. 

There is a converse process to formation, which is the {\em
distillation} of Bell states.  Suppose Alice and Bob are supplied with
$n$ copies of the state $|AB\ra$.  The {\em entanglement of
distillation} for $|AB\ra$ is defined to be the maximal number, $c$,
such that as $n \rightarrow \infty$, Alice and Bob can produce
$\lfloor cn \rfloor$ Bell states with asymptotically good fidelity,
using local operations and classical communication.  Bennett {\em et
al} show that the entanglement of distillation for the pure state
$|AB\ra$ is equal to the entanglement of formation, $S(A)$.

Popescu and Rohrlich's argument is to suppose we have $n$ copies of
the state $|AB\ra$.  This is then transformed by local operations and
classical communication into $\lfloor n S(A) \rfloor$ approximate Bell
states. Let $B_e$ be the entanglement of a Bell state.  Then we see
from continuity and additivity that $nE(A:B) \geq nS(A) B_e$ must hold
approximately. But we may transform the $\lfloor n S(A) \rfloor$ Bell
states back into $n$ copies of $|AB\ra$, again by local operations and
classical communication, so $n S(A) B_e \geq nE(A:B)$ must also be
approximately true.  Letting $n \rightarrow \infty$ we see that
$E(A:B) = S(A) B_e$ holds exactly.  But $B_e$ is a constant which does
not depend on the state $|AB\ra$, so we see that, up to an overall
proportionality factor, the entanglement of a pure state is uniquely
determined by the above axioms to be given by $E(A:B) = S(A)$.

The obvious reaction at this point is to think ``That's terrific!''
After all, we understand the von Neumann entropy pretty well, so it
seems as though we are well on our way to understanding entanglement
in general.

Unfortunately, though, it seems to be fairly difficult to even {\em
define} entanglement in a more general context than two-part composite
systems which are in pure states.  Plausible operational motivations
for definitions of entanglement in more general scenarios are not
difficult to generate, however researchers have had limited success in
calculating with these definitions.  In the next section we will study
some of the best developed tools for understanding the entanglement
present in an arbitrary state of two quantum systems.

\section{Mixed state entanglement}
\label{sec:mixed_state_entanglement}

The {\em entanglement of formation} \cite{Bennett96a} of a state,
$\rho$, of a composite system $AB$, is defined to be the minimum
number of Bell states which must be shared between $A$ and $B$ if they
are to be able to form the state $\rho$ using only local operations
and classical communication.

More precisely, suppose we have a family of protocols, one protocol
for each positive integer $n$, such that in the $n$th protocol,
parties $A$ and $B$ start out sharing one half each of $\lfloor cn
\rfloor$ Bell states, for some $c \geq 0$, and the protocols only
involve local operations on systems $A$ and $B$, and classical
communication between $A$ and $B$. The family of protocols is said to
be a good entanglement forming protocol for the state $\rho$ if the
$n$th protocol produces the state $\rho^{\otimes n}$ with asymptotic
fidelity approaching one as $n$ approaches infinity.

The entanglement of formation has been extensively studied by Wootters
and collaborators \cite{Bennett96a,Hill97a,Wootters98a}. Modulo a
problem to be discussed below, they have shown that the entanglement
of formation between systems $A$ and $B$ is given by the expression
\beqn \label{eqtn:e_of_f}
{\cal F}(A:B) = \min \sum_x p_x S(A_x), \eeqn
where $\{ p_x, AB_x\}$ is an ensemble of pure states generating the
state $AB$. The minimum in the definition of the entanglement of
formation is over all pure state ensembles generating the state $AB$.

%\begin{figure}[ht]
%\setlength{\unitlength}{2000sp}
%\begin{picture}(3024,6024)(1189,-7100)
%\thinlines
%\put(1201,-1561){\framebox(1500,1200){$A_1$}}
%\put(1201,-3961){\framebox(1500,1200){$A_2$}}
%\put(1201,-6361){\framebox(1500,1200){$A_3$}}
%\put(2700,-1561){\framebox(1500,1200){$B_1$}}
%\put(2700,-3961){\framebox(1500,1200){$B_2$}}
%\put(2700,-6361){\framebox(1500,1200){$B_3$}}
%\put(2550,-4750){$\otimes$}
%\put(2550,-2250){$\otimes$}
%\put(2650,-7080){$\vdots$}
%\end{picture}
%\caption{The entanglement of formation is, by definition, additive
%across uncorrelated systems.\label{fig:entanglement_additive}.}
%\end{figure}

The possible problem with equation (\ref{eqtn:e_of_f}) is whether the
quantity appearing within it is {\em additive}\footnote{This problem
was pointed out by Sandu Popescu.}. Recall that the operational
definition of the entanglement of formation was an asymptotic
definition, expressed in terms of the creation of a large number of
copies of the state $AB$. Strictly speaking, what is shown in
\cite{Bennett96a} is that \beqn {\cal F}_o(A:B) = \limsup_{n
\rightarrow \infty} \frac{{\cal F}(A_1\ldots A_n: B_1\ldots B_n)}{n},
\eeqn where ${\cal F}_o$ is the entanglement of formation,
operationally defined, the quantity ${\cal F}$ appearing on the right hand
side is as defined in equation (\ref{eqtn:e_of_f}), and the system
$A_1\ldots A_nB_1\ldots B_n$ is a tensor product of $n$ copies of
$AB$. The expression ${\cal F}_o(A:B)$ would be equal to ${\cal
F}(A:B)$ if it could be shown that the quantity ${\cal F}(A:B)$
defined by equation (\ref{eqtn:e_of_f}) is {\em additive} in the sense
that \beqn {\cal F}(A_1\ldots A_n:B_1\ldots B_n) = {\cal F}(A_1:B_1)+
\ldots + {\cal F}(A_n:B_n). \eeqn 
%where $A_1\ldots A_n B_1 \ldots B_n$
%is $n$ uncorrelated copies of the original state $AB$, in an obvious
%notation.

So we have two definitions of the entanglement of formation: an
operational definition, based upon the number of Bell states it takes
to form the state in question, and an explicit formula, equation
(\ref{eqtn:e_of_f}). It is believed but not yet known that these two
definitions are the same. Let ${\cal F}_o(A:B)$ denote the operational
definition of entanglement, and ${\cal F}(A:B)$, the definition based upon
the formula (\ref{eqtn:e_of_f}). Then
\beqn
{\cal F}_o(A:B) = \limsup_{n \rightarrow \infty} \frac{{\cal F}(A_1,\ldots,A_n:
B_1,\ldots,B_n)}{n}. \eeqn
From subadditivity of the entropy and equation (\ref{eqtn:e_of_f}) it
is clear that
\beqn
{\cal F}_o(A:B) \leq \limsup_{n \rightarrow \infty}
\frac{{\cal F}(A_1:B_1)+\ldots+{\cal F}(A_n:B_n)}{n} = {\cal F}(A:B). \eeqn

Note that \beqn {\cal F}_o(A:B) & = & \limsup_{n \rightarrow \infty}
\frac{\min \sum_x p_x S((A_1\ldots A_n)_x)}{n} \\ & = & \limsup_{n
\rightarrow \infty} \frac{\min -\sum_x p_x S((A_1\ldots A_n)_x|
(B_1\ldots B_n)_x)}{n}, \eeqn where the minimum is taken over all
ensembles $\{p_x, (A_1\ldots A_nB_1\ldots B_n)_x \}$ generating
$A_1\ldots A_nB_1\ldots B_n$.  From the subadditivity of the
conditional entropy, proved in section (\ref{sec:ssa}), we see that
\beqn {\cal F}_o(A:B) & \geq & \min -\sum_x p_x S(A_x|B_x), \eeqn
where now the minimum is taken over all ensembles $\{p_x AB_x\}$
generating $AB$. From the concavity of the conditional entropy, it
follows that \beqn {\cal F}_o(A:B) \geq -S(A|B). \eeqn By concavity of
the entropy we also have ${\cal F}(A:B) \leq S(B)$.  Combining this
with the previous equation gives \beqn {\cal F}(A:B)-S(A,B) \leq {\cal
F}_o(A:B) \leq {\cal F}(A:B). \eeqn Thus for states which are nearly
pure, the equation (\ref{eqtn:e_of_f}) is guaranteed to be pretty
close to the operational definition for the entanglement of formation.

In \cite{Wootters98a} it is stated that numerical tests provide
evidence for the conjecture that the expression given in equation
(\ref{eqtn:e_of_f}) is additive, and thus is the correct operational
formula for the entanglement of formation. From now on we will assume
that (\ref{eqtn:e_of_f}) is indeed the correct formula for the
(operational) entanglement of formation.

%The following are some elementary properties of the entanglement of
%formation:
%\begin{enumerate}
%\item ${\cal F}(A:B) \leq \min(S(A),S(B))$.
%\item ${\cal F}(A:B) = \min \sum_x p_x S(A_x)$, where this time the
%minimization is over {\em all} ensembles $\{p_x, AB_x\}$ generating
%$AB$, not just pure state ensembles. The result is immediate from the
%concavity of the entropy.
%\item ${\cal F}(A:B) \leq {\cal F}(A:B,C)$. Clearly, if we can form the state
%$ABC$ by local operations and classical communication between $A$ and
%$BC$, it must be possible to form $AB$: just carry out the procedure
%to form $ABC$, and then throw away system $C$. In terms of the
%entropic definition,
%\beqn
%{\cal F}(A:B,C) = \sum_x p_x S(A_x), \eeqn
%for some pure state ensemble $\{ p_x, ABC_x \}$ for $ABC$. This
%induces a natural mixed state ensemble $\{p_x, AB_x \}$ for $AB$, so
%from the previous item we see that
%\beqn
%{\cal F}(A:B) \leq \sum_x p_x S(A_x) = {\cal F}(A:B,C), \eeqn
%as required.
%\end{enumerate}

% e of a: defn and motivation

% Remote control view of entanglement

%There is an interesting and useful {\em remote control} view which can
%be taken to the entanglement of an ensemble. Suppose Alice and Bob are
%in joint possession of a general quantum state, $AB$. Suppose this
%system is {\em purified} (see Appendix \ref{app:mixed} for an
%introduction to purification) by a third system, $C$, which we shall
%suppose belongs to Charlie. 
%
%Charlie performs a measurement on that system

% e of f and the conditional entropy

The entanglement has many simple, useful properties:
\begin{theorem} \textbf{(Elementary properties of entanglement)}

\begin{enumerate}
\item {\bf Symmetry:}
\be
%Q(p_x; A:B_x) = Q(p_x; B:A_x); \,\,\,\, 
{\cal F}(A:B) = {\cal F}(B:A).
%\,\,\,\, {\cal A}(A:B) = {\cal A}(B:A). 
\ee
%\item {\bf Ordering properties of entanglement:}
%\be
%{\cal F}(A:B) \leq Q(p_x; A:B_x) \leq {\cal A}(A:B). \ee
\item {\bf The entanglement is entropy-bounded:}
\be
{\cal F}(A:B) \leq \min(S(A),S(B)). \ee
\item {\bf The entropy-entanglement inequality:}
\be \label{eqtn:entropy-entanglement}
{\cal F}(A:B) \geq -S(A|B); \,\,\,\, {\cal F}(A:B) \geq -S(B|A). \ee
%\item {\bf The assistance-formation inequality:}
%\be
%{\cal A}(A:B) - {\cal F}(A:B) \leq S(A:B). \ee
\item {\bf More systems means more entanglement:}
\be
{\cal F}(A:B) \leq {\cal F}(A:B,C). \ee
\item {\bf Adding uncorrelated systems does not change the
entanglement:}

Suppose the system $AB$ is in a product state with system $C$. Then
\beqn
{\cal F}(A:B,C) = {\cal F}(A:B). \eeqn
\end{enumerate}

\end{theorem}

\begin{proof}

The symmetry %and ordering 
property is obvious from the definition. 
%Note that the ordering properties may be combined with
%later inequalities in a trivial way to deduce inequalities that do not
%appear explicitly on this list. 
As already noted, the entropy-boundedness is an immediate consequence
of the definition of ${\cal F}(A:B)$, and the concavity of the entropy.
Also as noted, the entropy-entanglement inequality ${\cal F}(A:B) \geq
-S(A|B)$ follows from the concavity of the conditional entropy.

%The assistance-formation inequality ${\cal A}(A:B)-{\cal F}(A:B) \leq S(A:B)$
%is immediate from the entropy-entanglement inequality and the
%entropy-boundedness of the entanglement of assistance.  This property
%of entanglement was first brought to my attention by Ashish Thapliyal
%\cite{DiVincenzo98b}.

We give two proofs that more systems means more entanglement. The
first proof is from the operational definition of entanglement.  If we
have enough singlet pairs to make $n$ good copies of $A:B,C$ then by
throwing away all $n$ copies of $C$ we obtain $n$ good copies of
$AB$. The result follows. The result also follows easily from the
entropic definition of entanglement. Suppose \beqn {\cal F}(A:B,C)=\sum_i p_i
S(\mbox{tr}_{BC}(|\psi_i\ra\la \psi_i|)). \eeqn Define \beqn \rho_i
\equiv \mbox{tr}_C(|\psi_i\ra\la\psi_i|), \eeqn and let $\{
\lambda^i_j,|\psi^i_j\ra\}$ be any ensemble for $\rho_i$.  From the
concavity of entropy it follows that \beqn \sum_{ij} p_i
\lambda^i_j S(\mbox{tr}_B(|\psi^i_j\ra\la \psi^i_j|) & \leq & \sum_i
p_i S(\mbox{tr}_B(\rho_i)) \\ & = & \sum_i p_i
S(\mbox{tr}_{BC}(|\psi_i\ra\la \psi_i|)) \\ & = & {\cal F}(A:B,C). \eeqn This
establishes the result.

To prove that adding uncorrelated systems does not change the
entanglement, note first that ${\cal F}(A:B,C) \geq {\cal
F}(A:B)$. Suppose $\{ p_i; |AB_i\ra \}$ is an ensemble for $AB$ such
that \beqn {\cal F}(A:B) = \sum_i p_i S(A_i). \eeqn Suppose $C =
\sum_j \lambda_j |j\ra\la j|$ is an eigenensemble decomposition for
system $C$. Then it is clear that $\{ p_i \lambda_j; |AB_i \ra
\otimes |j\ra \}$ is an ensemble for $ABC$, and thus \beqn {\cal F}(A:B,C)
&\leq & \sum_{ij} p_i \lambda_j S(\mbox{tr}_{BC}(|AB_i\ra\la AB_i|
\otimes |j\ra\la j|)) \\ & = & \sum_i p_i
S(\mbox{tr}_{B}(|AB_i\ra\la AB_i|)) \\ & = & {\cal F}(A:B).\eeqn Thus
${\cal F}(A:B,C) = {\cal F}(A:B)$.  
%It would be interesting to know
%whether the condition that $AB$ be in a product state with $C$ can be
%replaced in the above theorem with the requirement that the system be
%separable with respect to the $AB:C$ division.

\end{proof}

% relating accessible information to E of F

The following simple example illustrates the importance of the
entanglement for fundamental operational problems.

Suppose Alice has a composite of two systems, $A$ and $B$, in her
possession. She prepares the joint system $AB$ in the state $AB_x$,
according to some probability distribution $p_x$. She then gives the
system $B$ to Bob, whose task it is to determine as much information
as possible about the value of $x$. Let $I$ denote the maximum
possible mutual information Bob can obtain about $x$ by performing
operations on $B$. Applying the Holevo bound, as proved in section
\ref{sec:Holevo}, we see that \be I \leq S(B)-\sum_x p_x S(B_x) =
S(B)\sum_x p_x S(B_x). \ee Combining this with the observation that
${\cal F}(A:B) \leq \sum_x p_x S(B_x)$, we see that \be I \leq
S(B)-{\cal F}(A:B). \ee A similar line of reasoning shows that \be I
\leq S(A)-{\cal F}(A:B). \ee That is, the amount of information which
Bob can obtain about the preparation of system $AB$ is bounded by a
quantity determined by the entanglement existing between those
systems.

Developing such general theoretical connections between the
entanglement and other quantities of practical importance is one of
the great open problems in the study of entanglement.  We will
indicate a few more such connections in the following Chapters, but it
is my hunch that many more, and deeper, connections can be found
between the entanglement and other aspects of quantum information
theory.  It would be particularly useful to be able to connect the
computational power of quantum computers, either for distributed
computation, as in Chapter \ref{chap:qcomm}, or for straight
computation, to measures of entanglement.

\section{Entanglement: Examples}
\label{sec:entanglement_examples}

In this section we will look at some simple examples where
entanglement arises naturally.  These examples will give us a feel for
how entanglement behaves in real physical systems.  As a bonus we will
obtain some clues as to how entanglement enters into quantum
computation.

%Many physical systems exhibit interesting phase transitions, such as
%the transition of H$_2$O from water to ice, the superconducting properties
%acquired by many materials at sufficiently low temperatures, and the
%recently observed formation of a Bose-Einstein condensate in a dilute
%atomic gas at low temperatures. Many phase transitions exhibiting
%interesting quantum
%effects are associated with the formation of entanglement in the system
%of interest. For example, the Cooper pairs formed in a superconducting metal
%are pairs of electrons in the entangled spin singlet state. This spin singlet
%behaves like a single spin zero particle, which is responsible for
%many of the important properties of superconductors.
%
%This paper studies the question of whether there is a phase transition
%directly associated with the entanglement present in a quantum system. For
%many systems of interest we find such a phase transition. In the
%simplest cases, the system exhibits no entanglement above a certain
%critical temperature, and entanglement below that temperature.
%Our investigations demonstrate a rich
%set of behaviours associated with these ``entanglement phase transitions'',
%and suggest ways of experimentally probing these behaviours.
%
%\begin{itemize}
%
%\item What static constraints does the entanglement satisfy?
%
%\item What processes create and destroy entanglement? What constraints do
%these processes satisfy?
%
%\item How may entanglement be transferred from one system to another?
%
%\end{itemize}

In general the entanglement of formation is a difficult quantity to
evaluate, and no general prescription is known.  However, for a
bipartite system of two spins, Wootters \cite{Wootters98a} has proved
a conjecture of Hill and Wootters \cite{Hill97a}, which gives an
explicit prescription for evaluating the entanglement.  This
prescription is somewhat involved, but is straightforward in
principle, and quite simple to implement on a computer. Hill and
Wootters introduce the {\em magic basis}, \index{magic basis}
\begin{eqnarray} 
|a\rangle & = & \frac{1}{\sqrt 2} \left( |00 \rangle +
	| 11 \rangle \right) \\
|b\rangle & = & \frac{i}{\sqrt 2} \left( | 00 \rangle -
	| 11 \rangle \right) \\
|c\rangle & = & \frac{i}{\sqrt 2} \left( | 01 \rangle +
	| 10 \rangle \right) \\
|d\rangle & = & \frac{1}{\sqrt 2} \left( | 01 \rangle -
	| 10 \rangle \right). 
\end{eqnarray}
For any density operator $\rho$ of the two spin system they define
\begin{eqnarray}
R \equiv \sqrt{ \sqrt \rho \rho^* \sqrt \rho }, \end{eqnarray}
where $\rho^*$ is the complex conjugate of $\rho$ when $\rho$ is
expressed in the magic basis. Defining $\lambda$ to be the
largest eigenvalue of $R$, they define the {\em concurrence} of
$\rho$ by
\begin{eqnarray}
c(\rho) \equiv \max(0,2 \lambda - \mbox{tr}(R)). \end{eqnarray}
The entanglement of $\rho$ is then
\begin{eqnarray}
{\cal F}(\rho) = H\left(\frac 12 + \frac 12 \sqrt{1-c^2}\right), \end{eqnarray}
where $H(x) \equiv -x \log_2 x - (1-x) \log_2 (1-x)$ is the binary
Shannon entropy function.

Suppose we have an ensemble of two spin systems in equilibrium
at temperature $T$. The state of this system is given by
\begin{eqnarray}
\rho = \frac{\exp(-H/kT)}{Z}, \end{eqnarray}
where $k$ is Boltzmann's constant, $H$ is the Hamiltonian for the two
spin system, $Z$ is the partition function, and we use $k = \hbar =
1$,
\begin{eqnarray}
Z \equiv \mbox{tr}(\exp(-H/T). \end{eqnarray}

Consider a system of two spins, labeled $A$ and $B$, described by
the Hamiltonian
\begin{eqnarray}
H = \frac{a}{2} (\sigma_z^A + \sigma_z^B) + \frac{b}{4}
	\vec \sigma_A \cdot \vec \sigma_B 
\end{eqnarray}
$a$ and $b$ are real constants characterizing the internal
energies of the two spins and the strength of the
coupling between them, respectively.

%It is clear from the operational definition of entanglement that the
%entanglement is an {\em additive} quantity. That is
%\begin{eqnarray}
%{\cal F}(\rho^{A_1 B_1} \otimes \ldots \otimes \rho^{A_n B_n}) =
%	{\cal F}(\rho^{A_1 B_1}) + \ldots + {\cal F}(\rho^{A_n B_n}). \end{eqnarray}
Suppose then that we have a large number of two qubit molecules in
thermodynamic equilibrium. Assuming that the intermolecular
interactions are essentially negligible, so that the total system is
in a product state, $\rho \otimes \ldots \otimes \rho$, it follows
from the additive property of entanglement that the total entanglement
present in the system is $N$ times the entanglement present in a
single molecule, where $N$ is the total number of molecules present in
the system.

At thermal equilibrium the state of the system depends only on
$H/T$, and this can be written in the form
\begin{eqnarray}
\frac{H}{T} = \frac{ \frac 12 (\sigma_z^A + \sigma_z^b) +
	\frac 1 4 \frac b a \vec \sigma_A \cdot \vec \sigma_B
	}{\frac{T}{a}}. \end{eqnarray}
From this form we
see that one of the three parameters in the problem ($a, b$ and
$T$) can be eliminated by using the rescaled temperature
$T/a$ and rescaled coupling strength $b/a$. This can be
accomplished by setting $a=1$, and using $b$ and $T$, as before,
which is what we do for the remainder of this section.

%We can imagine this system as being one containing a large number of
%two particle composite systems, which we will refer to as molecules
%for the sake of convenience, with each molecule described by the
%Hamiltonian above. Furthermore, we assume that the different molecules
%are independent.  The goal of this section is to study the behaviour
%of the entanglement of formation associated with the molecules in thee
%ensemble.
 
Written out in the magic basis the Hamiltonian of the system becomes
\begin{eqnarray}
H  & = &  \left[ \begin{array}{cccc}
	\frac{b}{4} & i & 0 & 0 \\
	-i & \frac{b}{4} & 0 & 0 \\
	0 & 0 & \frac{b}{4} & 0 \\
	0 & 0 & 0 & \frac{-3b}{4} \end{array}
	\right]. \end{eqnarray}
From this form we find the energy eigenstates and corresponding
eigenvalues,
\begin{eqnarray}
|E_1\rangle & = & |d\rangle \\
|E_2\rangle & = & \frac{|a\rangle+i|b\rangle}{\sqrt 2} = |11 \ra \\
|E_3\rangle & = & |c\rangle \\
|E_4\rangle & = &  \frac{|a\rangle-i|b\rangle}{\sqrt 2} = |00\ra \\
E_1 & = & -\frac{3b}{4} \\
E_2 & = & \frac{b}{4} - 1\\
E_3 & = & \frac b 4\\
E_4 & = & \frac b4 + 1
\end{eqnarray}
For now we will assume that $b \geq 0$ so that $E_1\leq E_2\leq E_3
\leq E_4$, and return to the general case later. Given the spectrum
and eigenstates it is straightforward to calculate the state of the
system at temperature $T$,
\begin{eqnarray}
\rho(T) & = & \frac{1}{1+e^{b/T}+2 \cosh(1/T)} \times \nonumber \\
	& & \left[
	\begin{array}{cccc} \cosh(1/T) & -i \sinh(1/T) & 0 & 0 \\
	i \sinh(1/T) & \cosh(1/T) & 0 & 0 \\
	0 & 0 & 1 & 0 \\
	0 & 0 & 0 & e^{b/T} \end{array} \right] \end{eqnarray}
It is now straightforward to calculate $R$ and to see that
\begin{eqnarray}
2 \lambda - \mbox{tr}(R) =  \frac{e^{b/T}-3}{1+e^{b/T}+2\cosh(1/T)},
\end{eqnarray}
from which the concurrence and entanglement follow immediately.

Note that the entanglement is non-zero for $0 < T < T_e$ , where
\begin{eqnarray}
T_e \equiv \frac{b}{\ln 3}. \end{eqnarray} Clearly the entanglement
vanishes for $T \geq T_e$.  We will refer to $T_e$ as the {\em
critical temperature}, by analogy with the physics of phase
transitions -- at the critical temperature, a qualitative change in
the system takes place; where before, no entanglement was present in
the system, now there is.  Note, however, that in this example at
least this change is \emph{not} associated with the presence of
long-range order in the system, as is the case in a true phase
transition.  The concurrence is given by the expression
\begin{eqnarray} \label{eqnt: example concurrence}
c(b,T) = \left\{ \begin{array}{ll} \frac{e^{b/T}-3}{1+e^{b/T}+2\cosh(1/T)}
	& \mbox{if } T \leq T_e \\
	0 & \mbox{if } T > T_e. \end{array} \right. 
\end{eqnarray}
Since the entanglement and the concurrence are monotonically related,
our study will be focused on the concurrence, since it is easier to
deal with.

\begin{figure}[ht]
\begin{center}
{\mbox{\psfig{file=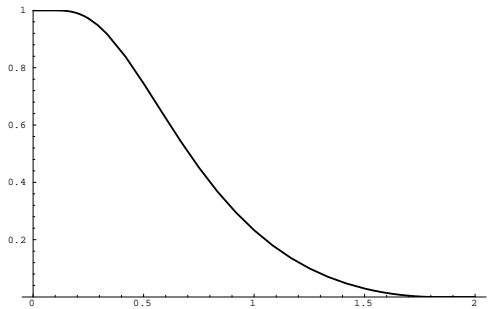,width=2.5in}}}
\end{center}
\vspace{-0.3cm}
\caption{Entanglement $(E)$ plotted as a function of temperature
$(T)$ for a coupling strength $b=2$ in the regime where the coupling
dominates.
\label{fig: strong coupling}}
\end{figure}

Consider now the case where $b > 1$, which we will refer to as the
\emph{strong coupling} regime. Writing the entanglement as a function
of coupling strength $b$ and temperature $T$, ${\cal F}(b,T)$, it is
clear that as $T \rightarrow 0$ we have ${\cal F}(b,T) \rightarrow 1$,
since the ground state is the maximally entangled spin singlet state,
$|d\rangle$. Furthermore, it is easy to see that for temperatures less
than the critical temperature,
\begin{eqnarray}
\frac{\partial c}{\partial(1/T)}&=&\frac{2e^{b/T}}{(1+e^{b/T}+2\cosh(1/T))^2}
	\times \left[ 2b + 3\sinh(1/T)e^{-b/T} +
	b\cosh(1/T)-\sinh(1/T) \right]. \nonumber \\
& & {  }
\end{eqnarray}
For $b \geq 1$ we have $b\cosh(1/T) \geq \cosh(1/T) > \sinh(1/T)$,
from which it follows that $c(b,T)$ and thus ${\cal F}(b,T)$ is a {\em
decreasing} function of temperature. It is also easy to verify that as
$T \rightarrow 0$ the rate of change of ${\cal F}(b,T)$ with $T$ goes to
zero.

Figure \ref{fig: strong coupling} shows the entanglement plotted as a
function of temperature in the strong coupling regime. As expected we
see that the entanglement at $T = 0$ is $1$, and then decreases
monotonically to zero at $T_e$. For the value of the coupling strength
chosen for this plot, $b=2$, the value of the critical temperature is
$T_e = 2/\ln 3 \approx 1.82$.

\begin{figure}[ht]
\vspace{-0.3cm}
\begin{center}
{\mbox{\psfig{file=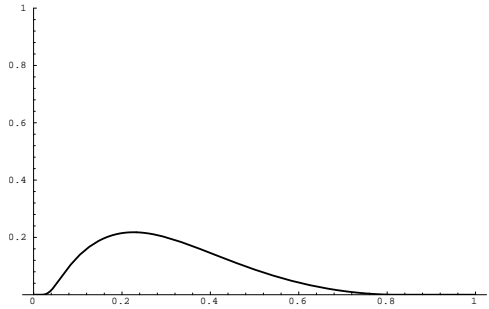,width=2.5in}}}
\end{center}
\vspace{-0.3cm}
\caption{Entanglement $(E)$ plotted as a function of temperature
$(T)$ for a weak coupling strength, $b=0.9$. 
\label{fig: weak coupling}}
\end{figure}

Consider next the case where $b < 1$, the \emph{weak coupling
regime}. Once again we have
\begin{eqnarray}
\frac{\partial c}{\partial(1/T)}&=&
	\frac{2e^{b/T}}{(1+e^{b/T}+2\cosh(1/T))^2} \times \left[ 2b +
	3\sinh(1/T)e^{-b/T} + b\cosh(1/T)-\sinh(1/T)
	\right]. \nonumber \\
& & { }
\end{eqnarray}
However, it is not difficult to see by inspection that for
sufficiently small $T$ the $-\sinh(1/T)$ term in the derivative
dominates, and thus ${\partial c}/{\partial (1/T)} < 0$, from which it
follows that there is a regime in which the entanglement {\em
increases} with temperature.

Figure \ref{fig: weak coupling} demonstrates this behaviour
graphically.  The figure shows the entanglement plotted as a function
of temperature in the weak coupling regime. As expected we see that
the entanglement at $T = 0$ is $0$. It increases to about $0.2$ near
$T = 0.2$, and then decreases to zero at $T_e$. For the value of the
coupling strength chosen for this plot, $b=0.9$, the value of the
critical temperature is $T_e = 0.9/\ln 3 \approx 0.82$.

It is not difficult to gain some intuitive feeling for why the
entanglement initially increases with temperature in this case. For
weak coupling the ground state of the system is the unentangled
product state $|E_2\rangle = |11\rangle$.  Thus, as $T \rightarrow 0$
the entanglement of the system goes to zero.  For small temperatures
we expect some of the population of the system to be in the first
excited state, $|E_1\rangle = |d\rangle$, the maximally entangled spin
singlet. The state of the system is thus largely a mixture of an
unentangled state with a maximally entangled state, resulting in a
small amount of entanglement.

\begin{figure}[ht]
\vspace{-0.3cm}
\begin{center}
{\mbox{\psfig{file=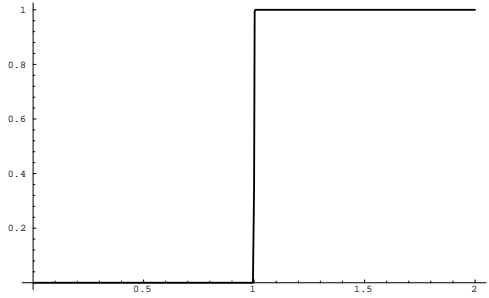,width=2.5in}}}
\end{center}
\vspace{-0.3cm}
\caption{Entanglement $(E)$ plotted as a function of the
coupling strength $(b)$ at zero temperature.
\label{fig: zerotemp}}
\end{figure}

Consider now the behaviour of the entanglement as a function of the
coupling strength, $b$. For fixed finite temperature $T$ we see that there
is a ``critical value'' of the coupling,
\begin{eqnarray}
b_e = T \ln 3. \end{eqnarray} For values of the coupling above this
strength the system exhibits entanglement, while for values of the
coupling below this strength no entanglement exists in the system. 
%Of
%course, calling $b_e$ a critical value is merely analogy, since in
%most physical systems the coupling strength is a constant. 
Taking derivatives we find
\begin{eqnarray}
\frac{\partial c}{\partial b} = \frac{2e^{b/T}}{T (1+e^{b/T}+2\cosh(1/T))^2}
	(2+\cosh(1/T)), \end{eqnarray}
which is always positive, so the entanglement {\em increases} as
a function of $b$.

For $T = 0$ the preceding analysis does not apply, since the
exponential terms in the denominator of the concurrence diverge as $T
\rightarrow 0$.  At $T = 0$ the system is in the ground state, which
is the maximally entangled spin singlet state $|d\rangle$ for $b > 1$,
and is the unentangled state $|11\rangle$ for $b < 1$. Thus at $T = 0$
we expect a sharp jump in the entanglement as a function of $b$ from
zero to one, at $b = 1$. Figure \ref{fig: zerotemp} shows the
entanglement as a function of coupling strength at zero
temperature. We clearly see such a jump in the entanglement, which is
is zero below $b=1$, and one above $b=1$.

\begin{figure}[ht]
\vspace{-0.3cm}
\begin{center}
{\mbox{\psfig{file=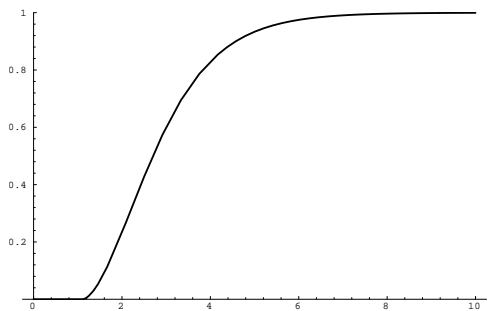,width=2.5in}}}
\end{center}
\vspace{-0.3cm}
\caption{Entanglement $(E)$ plotted as a function of the
coupling strength $(b)$ at a finite temperature, $T=1$.
\label{fig: finitetemp}}
\end{figure}

Figure \ref{fig: finitetemp} shows the entanglement as a function of
coupling strength at finite temperature, in this case $T=1$.  The
entanglement remains zero out to $b=1$ at which point it suddenly
begins increasing, eventually rising up to approach one for large
values of the coupling.

\begin{figure}[ht]
\vspace{-0.3cm}
\begin{center}
{\mbox{\psfig{file=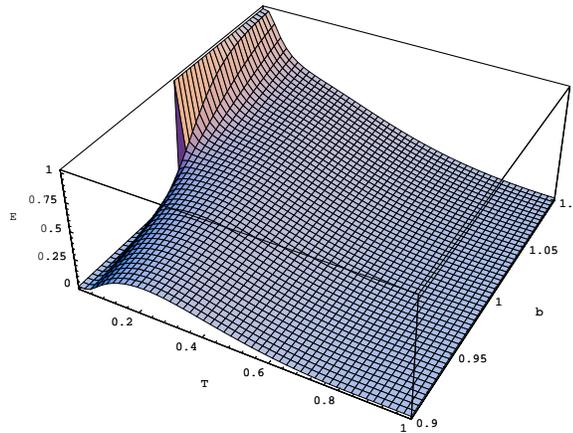,width=3in}}}
\end{center}
\vspace{-0.3cm}
\caption{Entanglement $(E)$ plotted as a function of temperature
$(T)$ and the coupling strength $(b)$. This figure
shows the behaviour for values of the
coupling strength near the crossover at $b = 1$. \label{fig: closeup}}
\end{figure}

Figure \ref{fig: closeup} shows the entanglement as a function of both
the coupling strength and the temperature. The parameters have been
chosen so that $b$ is near the value one, where most of the
interesting behaviour in this model occurs. The differing behaviour of
the entanglement for $b < 1$ and $b > 1$ is clearly visible on this
diagram.

\begin{figure}[ht]
\begin{center}
{\mbox{\psfig{file=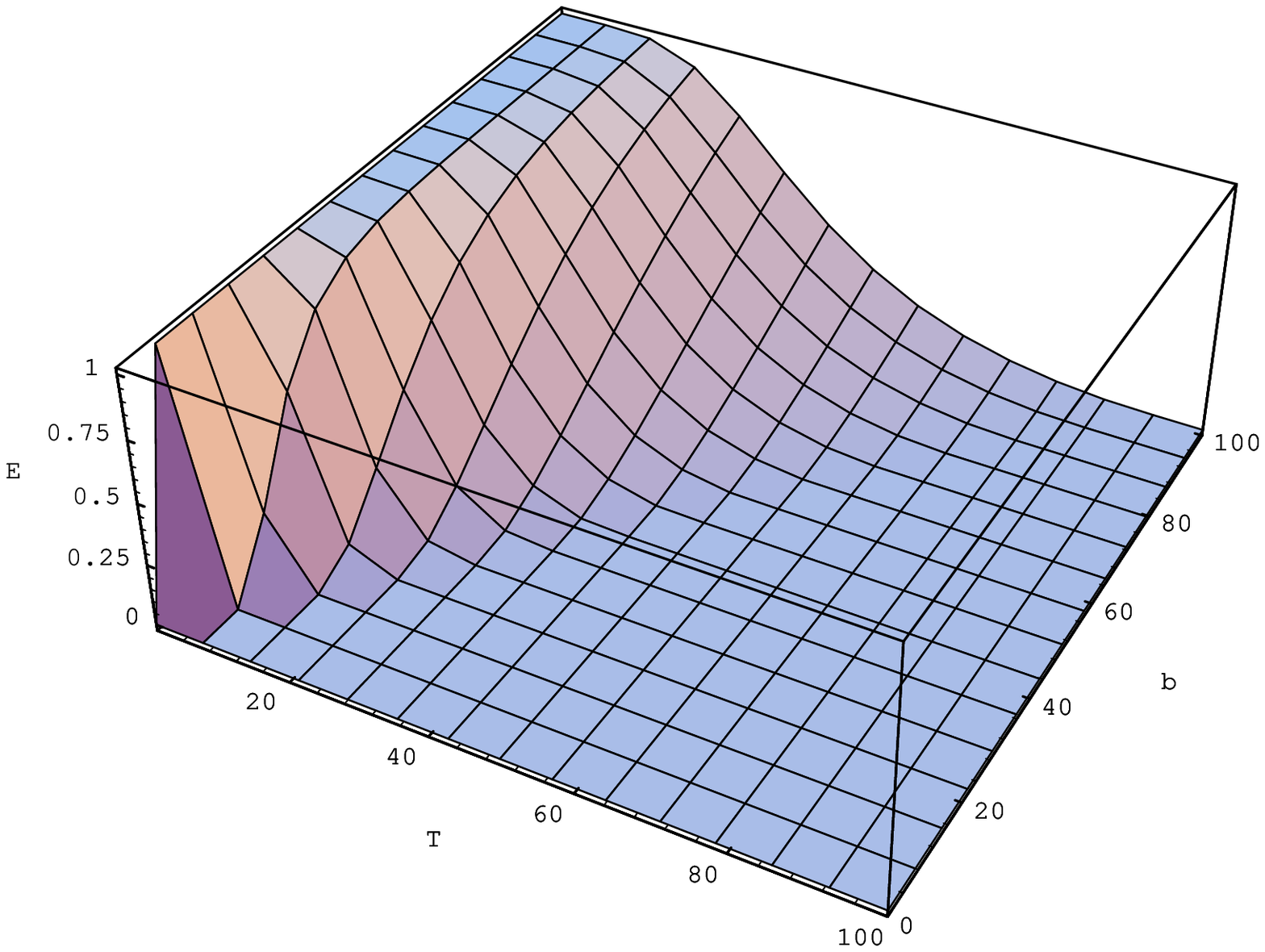,width=2.5in}}}
\end{center}
\caption{Entanglement $(E)$ plotted as a function of temperature
$(T)$ and the coupling strength $(b)$. This figure
shows the behaviour for large values of the
coupling strength, $b \gg 1$. \label{fig: longshot}}
\end{figure}

Figure \ref{fig: longshot} shows the entanglement as a function of
coupling strength and temperature, this time for very large values of
the coupling and temperature. One sees plainly from this figure that
as the coupling strength is increased, the critical temperature also
increases.

%By analogy with the heat capacity we can define the {\em entanglement
%capacity} for the system,
%\begin{eqnarray}
%E_C \equiv \frac{dE}{dT}. \end{eqnarray}

The results in this section imply that properties associated with the
entanglement of composite quantum systems are not derivable completely
within the formalism of ordinary statistical mechanics.  If that were
the case then the properties of entanglement would be determined
completely by the partition function,
\begin{eqnarray}
Z \equiv \mbox{tr}(\exp(-H/kT)). \end{eqnarray} Consider a two spin
system with the same energies as the spin system we have been
considering, but whose energy eigenstates are unentangled product
states of the two systems. It is clear that the partition functions
for these two systems are the same, since they have the same energies,
and thus the two systems are identical from the point of view of
statistical mechanics. It is also clear that the entanglement of such
a system is always zero, since its density operator is diagonal in a
product basis for the system.

It seems to me that one of the major directions in which research into
entanglement can be taken is to develop a theory of statistical
mechanics and thermodynamics which adequately accounts for
entanglement\footnote{Hideo Mabuchi suggested the elegant term
\emph{thermodynamics of entanglement} for such a subject to me in early
1996.  We had independently been having similar thoughts about the
thermodynamics and statistical mechanics of entanglement.} in a
natural fashion.  It may be interesting, for example, to study
transport properties of entanglement in non-equilibrium systems, or to
investigate whether there is a connection between entanglement and
quantum phase transitions.

A rather different direction to take the study of entanglement is in
the study of the power of quantum computation.  Making use of the
results in this section, I will note one amusing result in this
connection, associated with the behaviour of entanglement in the
context of NMR quantum computing.  For definiteness, we will consider
the scenario defined by Knill and Laflamme \cite{Knill98d}, in which
they consider the power of ``one bit of quantum information''.
Specifically, they consider a scenario in which one qubit in the state
$|0\ra$ is available, together with $n$ qubits in the completely mixed
state $I^{\otimes n}/2^n$.  This can be achieved, effectively, in NMR,
by making use of gradient-pulse techniques to eliminate polarization
of $n$ qubits of the molecule, and to create a pseudo-pure state on
$1$ qubit. We will denote the (true) initial state of the pseudo-pure
qubit as $\rho$.

Consider a two qubit system in the completely mixed state $I^{\otimes
2}/4$.  Note that $2\lambda - \tr(R) = -1/2$ for this state.  It
follows from continuity that there exists $\epsilon > 0$ such that for
all states within an absolute distance $\epsilon$ of $I/4$, the
entanglement of formation is zero. Note that for sufficiently high
temperatures, the state $\rho \otimes I/2$ is within an absolute
distance $\epsilon$ of $I^{\otimes 2}/4$.  Moreover, suppose ${\cal
E}$ is a doubly stochastic quantum operation\footnote{A doubly
stochastic quantum operation is a complete quantum operation which
preserves the completely mixed state, $\evop(I) = I$.} on two qubits.
Then by the contractivity property of absolute distance under quantum
operations and the fact that $I/4$ remains fixed under the doubly
stochastic operation ${\cal E}$, it follows that the state ${\cal
E}(\rho \otimes I/2)$ is also within a distance $\epsilon$ of $I/4$,
and thus has zero entanglement of formation.

Next, suppose $i$ and $j$ are two qubits in the Knill-Laflamme
scenario.  Let $S$ be the unitary operator which swaps the pseudo-pure
qubit with qubit $i$.  Note that the state of qubits $i$ and $j$ after
performing a unitary operation $U$ on the $n+1$ spins is given by
\beqn \rho_{ij}' = \tr_{E}\left( \frac{U (I^{\otimes n}\otimes \rho)
U^{\dagger}}{2^n} \right), \ee where $E$ consists of all qubits except
qubits $i$ and $j$. This can be rewritten as \be \label{eqtn:Dunsany}
\rho_{ij}' = \tr_E\left( \frac{U S (I^{\otimes n-1} \otimes \rho_i
\otimes I_j )S^{\dagger} U^{\dagger}}{2^{n}} \right), \ee where
$\rho_i$ indicates the state $\rho$, but on the $i$th qubit, and $I_j$
is the identity operator on the $j$th qubit.  It follows from the
results on quantum operations in Chapter \ref{chap:qops} that \be
\rho_{ij}' = {\cal E}(\rho \otimes I/2), \ee for some complete quantum
operation ${\cal E}$. Moreover, as may be explicitly verified using
equation (\ref{eqtn:Dunsany}), $I \otimes I$ is left invariant under
${\cal E}$, so ${\cal E}$ is doubly stochastic.

It follows that, in the Knill-Laflamme model of computation, with
unitary operations, for sufficiently high starting temperatures, there
can never be any pairwise entanglement between qubits, as measured by
the entanglement of formation.

This is an interesting observation, because Knill and Laflamme
\cite{Knill98d} have found an example of simulation problems which can
be solved efficiently within this model of computation which have no
known efficient classical solution.  Suppose no efficient classical
solution is possible.  Then this would be an example of a problem
where quantum computers give an advantage of computational power over
classical computers, yet there is never any entanglement existing
between any pair of qubits. Perhaps there is entanglement in this
algorithm between subsystems larger than single qubits; I do not know.
However, this example does suggest that the statement ``entanglement
is responsible for the power of quantum computation'' needs to be
explored in much greater depth if it is to be made into a precise
statement about the difference in computational power between quantum
and classical computation.  In any case, it seems as though
exploration of the connection between entanglement and the power of
quantum computation is any area of research that deserves considerable
effort over the next few years, in order to determine to what extent
entanglement is a necessary condition for quantum computers to exhibit
greater computational power than classical computers.

\section{Conclusion}
\label{sec:future_entanglement}
\index{phase transitions and entanglement}

In this Chapter we have discussed the elementary properties of
entanglement, and made some tentative steps towards understanding its
relationship with other aspects of quantum information processing.
Unfortunately, I have not yet been able to integrate the subject of
entanglement with other aspects of quantum information theory as
smoothly as is desirable.  Doing so is a very interesting challenge,
which I intend to work on as part of my future research.  Some
questions which I intend to address include:

\begin{enumerate}

\item What is it about entanglement that makes it useful for quantum
information processing?

%\item Is it possible to efficiently simulate a quantum computer whose
%state remains separable at all times?

\item What characteristics can be used to understand entanglement in
composite systems consisting of three or more components?  It is
tempting to conjure up real valued measures of entanglement, just as
has been done for the two-system case.  I hypothesize that it may be
useful to make use of more sophisticated algebraic techniques.  An
analogy may be helpful here.  In algebraic topology \cite{Bredon93a},
great progress in the study of topological spaces is made by
associating algebraic constructs to topological spaces.  By studying
the relatively simple algebraic constructs, properties of the much
more complicated topological spaces can be deduced.  A similar
situation may obtain in the study of entanglement.
%
%At present there is little agreement over what measure(s) of
%entanglement should be used for composite systems consisting of three
%or more components. In the absence of good measures it is difficult to
%
%Suppose we have a three qubit system in the GHZ state
%\beqn
%|\psi\ra = \frac{|0_A0_B0_C\ra+|1_A1_B1_C\ra}{\sqrt 2}. \eeqn
%We see that
%\beqn
%\rho_{AB} = \rho_{AC} = \rho_{BC} = \frac 12 |0\ra \la 0| \otimes |0\ra \la 0|
%+ \frac 12 |1\ra \la 1| \otimes |1\ra \la 1|. \eeqn That is, each pair
%of qubits is separable. Yet the GHZ state itself is highly entangled.
%
%An analogous situation arises in classical information theory.
%Consider three random variables, $X$, $Y$ and $Z$. $X$ and $Y$ are
%independent and identically distributed random variables, each taking
%the values zero and one with probabilities half. $Z$ is defined to be
%the sum, modulo two, of $X$ and $Y$, $Z \equiv X \oplus Y$. It is easy
%to verify that each pair of random variables, $X$-$Y$, $Y$-$Z$ and
%$X$-$Z$ are independent random variables. That is, we have $H(X:Y) =
%H(X:Z)=H(Y:Z) = 0$. However, it is also clear that the three random
%variables as a whole are correlated, since given any two of the random
%variables, the value of the third can be determined. For example,
%given $X$ and $Z$, $Y$ can be determined, since $Y = X\oplus Z$.

\item How may the entanglement of formation be computed for composite
systems whose components have more than two dimensions?

\item Continuous quantum phase transitions \cite{Sondhi97a} occur at
zero temperature in quantum systems, as some parameter in the
Hamiltonian is varied.  For systems with a non-degenerate ground
state, the long range order associated with this phase transition must
be associated with correlations arising out of entanglement in the
ground state.  It would be interesting to study such zero-temperature
phase transitions from the point of view of quantum information theory.

%\item What other systems exhibit an entanglement phase
%transition, and what physical effects are associated with the
%transitions in those systems? We expect entanglement phase
%transitions to be the generic case for systems of physical
%interest. Clearly there are many, many
%systems where entanglement phase transitions could occur.
%Examples include:
%\begin{itemize}
%
%\item An atom coupled to a single mode of the electromagnetic field.
%
%\item An ion in an ion trap.
%
%\item Qubits in a quantum computer.
%
%\end{itemize}

\item What are the experimental signatures of entanglement?  What are
the simplest, most physically meaningful tests for the presence or
absence of entanglement in a quantum system?

%Entanglement phase transitions should be experimentally
%verified. Doing so means finding a system in which the
%phase transition occurs within an accessible parameter range,
%and where there is an observable signature associated with
%the transition.

%\item Is the entanglement phase transition related to other
%phase transitions of interest? Many interesting phase transitions
%occur in systems whose ground state is known to be highly
%entangled, and this entanglement is often related to those
%properties which make the phase transition interesting. It would
%be of interest to derive some relationships between those
%phase transitions, and entanglement phase transitions.

\item Can the presence of entanglement be related to other interesting
physical phenomena?  For example, in superconductivity, Cooper pairs
form as a result of phonon exchange between electrons in a metal.  For
ordinary superconductors these pairs are assumed to be in the
(entangled) spin singlet state.  It would be interesting to know
whether necessary conditions phrased in terms of entanglement can be
found for the superconducting phase transition.

\end{enumerate}

The study of entanglement suggests many other problems; this is only a
tiny sample.  I expect that this study will yield a rich and deep
structure that will give us insight into both quantum information, and
also into naturally occurring physical systems.

\vspace{1cm}
\begin{center}
\fbox{\parbox{14cm}{
\begin{center} {\bf Summary of Chapter \ref{chap:ent}:
Entanglement} \end{center}

\begin{itemize}
\item For pure states of $AB$, the entanglement is essentially unique,
$E(A:B) = S(A) = S(B)$.

\item \textbf{The entropy-entanglement inequality:}
$$
{\cal F}(A:B) \geq -S(A|B). $$

\item An efficient quantum algorithm with no known efficient classical
analogue has been found by Knill and Laflamme \cite{Knill98d}, in
which there may be no entanglement between any pair of two qubits, at
any stage of the algorithm.

\end{itemize}
}}
\end{center}

\chapter{Error correction and Maxwell's demon}
\label{chap:qec}

% overview of Chapter

%% entropy exchange
%%% measure of noise
%%% 
%% quantum Fano
%% data processing and coherent information
%% information-theoretic approach to qec
%%% example of error correction; use book.
%%% info theoretic conditions
%% other inequalities
%%
%% qec and Maxwell's demon
%%%% example of a Maxwell's demon. Szilard's engine
%%%% Landauer's principle
%%%% qec as a Maxwell's demon.

Large scale quantum information processing will be enormously
sensitive to the effects of noise on quantum systems. Shor
\cite{Shor95a} and Steane \cite{Steane96a} have introduced methods for
doing quantum error correction in order to preserve quantum
information in the presence of noise. These methods have been
developed much further by a large number of researchers, notably
Gottesman \cite{Gottesman96a} and Calderbank {\em et al}
\cite{Calderbank97a}, who developed a powerful framework for the study
of quantum codes, and by Aharonov and Ben-Or \cite{Aharonov97a},
Gottesman \cite{Gottesman97a,Gottesman98a}, Kitaev \cite{Kitaev97a},
Knill, Laflamme and Zurek \cite{Knill98c,Knill98a}, Preskill
\cite{Preskill98b}, and Shor \cite{Shor96a}, who developed methods for
performing quantum information processing in the presence of noise.

In this Chapter we study quantum error correction from an
information-theoretic point of view. Information-theoretic necessary
and sufficient conditions for doing quantum error correction are
formulated, and the information-theoretic point of view is used to
study quantum error correction as a thermodynamic process, analogous
to Maxwell's famous Demon \cite{Bennett87a}, an information processing
system which {\em apparently} violates the second law of
thermodynamics.  The material in this Chapter also serves as the basis
for work in the next Chapter, on the quantum channel capacity.

Throughout the Chapter we will make heavy use of two constructions
introduced earlier in this Dissertation in the study of quantum
systems. First, as in Chapter \ref{chap:distance}, we assume that
quantum system of interest, $Q$, has been purified by a second system,
$R$, before any dynamics has occurred.  The system $R$ is assumed to
undergo the trivial dynamics during any quantum process on the system
$Q$. Moreover, quantum operations are modeled by a unitary
interaction of $Q$ with an environment, $E$, which is assumed to
initially be in a pure state. As shown in Chapter \ref{chap:qops}, it
is always possible to introduce such a model for any complete quantum
operation. For incomplete quantum operations the unitary operation on
$QE$ is followed by a projection on the system $E$. We refer to this
picture of quantum operations as the $RQE$ picture of quantum
operations.

The Chapter is structured as follows.  Section
\ref{sec:entropy_exchange} introduces the \emph{entropy exchange}, a
tool for quantifying the effects of noise in a quantum system.
Section \ref{sec:quantum_Fano} introduces the \emph{quantum Fano
inequality}, a quantum analogue to the classical Fano inequality
proved in section \ref{sec:Holevo}.  Section
\ref{sec:q_data_processing} introduces the \emph{coherent
information}, a quantitative measure of the amount of quantum
information transmitted through a quantum channel.  Section
\ref{sec:q_data_processing} also proves the \emph{quantum data
processing inequality}, a quantum analogue of the classical data
processing inequality proved in subsection \ref{subsec:data_proc}.
Section \ref{sec:error_correction} reviews the basic concepts of
quantum error correction.  Section \ref{sec:info_conditions} uses the
quantum Fano and data processing inequalities to obtain
information-theoretic necessary and sufficient conditions for quantum
error correction.  Section \ref{sec:other_inequalities} proves several
information theoretic inequalities for quantum channels.  Section
\ref{sec:Maxwell's_demon} formulates quantum error correction as a
type of \emph{Maxwell's demon} -- a famous system proposed by Maxwell
last century that was apparently able to violate the second law of
thermodynamics by making observations upon a system. We do a
thermodynamic analysis of quantum error correction, and show that
there is no possibility of using quantum error correction to violate
the second law.  A consequence of our analysis, however, is that it is
possible to do quantum error correction in a thermodynamically
efficient manner.  Section \ref{sec:qec_conclusion} concludes the
Chapter.

Sections \ref{sec:entropy_exchange}, \ref{sec:quantum_Fano} and
\ref{sec:error_correction} are largely reviews of background material.
The remaining sections of the Chapter report original work, based upon
collaborations with with Schumacher \cite{Schumacher96b}, with Caves
\cite{Nielsen97b}, with Barnum and Schumacher \cite{Barnum98a}, and
with Caves, Schumacher, and Barnum \cite{Nielsen98a}. Section
\ref{sec:other_inequalities} has not appeared elsewhere, and is an
original contribution.  I am especially grateful to Ben Schumacher for
the many enjoyable discussions we have had about quantum information
theory.

\section{Entropy exchange}
\label{sec:entropy_exchange}
\index{entropy exchange}

How much noise does a quantum operation cause when applied to a
particular state, $\rho$, of a quantum system, $Q$?  One measure of
this is the extent to which the state, $RQ$, initially pure, becomes
mixed as a result of the quantum operation. To this end, following
Schumacher \cite{Schumacher96a}, we define the {\em entropy exchange}
of the operation $\evop$ upon input of $\rho$ by
\beqn
S_e \equiv S(\rho,{\cal E}) \equiv S(\rho^{RQ'}) = S(E'), \eeqn where the
equality of the entropy exchange with $S(E')$ follows from the purity
of the total state $R'Q'E'$. Thus, the entropy exchange can be
regarded as the amount of entropy introduced into an initially pure
environment as a result of the quantum operation $\evop$. We use the
notation $S_e$ for the entropy exchange in situations where the
arguments $\rho$ and $\evop$ are implied, and the notation
$S(\rho,\evop)$ otherwise.

Note that the entropy exchange does not depend upon the way in which
the initial state of $Q$, $\rho$, is purified into $RQ$. The reason is
because any two purifications of $Q$ into $RQ$ are related by a
unitary operation on the system $R$, as shown in appendix
\ref{app:mixed}. This unitary operation commutes with the action of
the quantum operation on $RQ$, and thus the two final states of $R'Q'$
induced by the two different purifications are related by an overall
unitary transformation which does not affect the entropy of $R'Q'$,
giving rise to the same value for the entropy exchange.  Furthermore,
it follows from these results that $S(E')$ does not depend upon the
particular model for $\evop$ which is used, provided the model starts
with $E$ in a pure state.

A useful explicit formula \cite{Schumacher96a} for the entropy
exchange can be given, based upon the operator-sum representation for
quantum operations. Suppose a complete quantum operation $\evop$ has
the operator-sum representation $\evop(\rho) = \sum_i E_i \rho
E_i^{\dagger}$. Then, as shown in section \ref{sec:qops_fundamentals},
a unitary model implementing this quantum operation is given by
defining a unitary operator $U$ on $QE$ such that \beqn U|\psi\ra
|0\ra = \sum_i E_i |\psi\ra |i\ra, \eeqn where $|0\ra$ is the initial
state of the environment, and $|i\ra$ is an orthonormal basis for the
environment. Note that the state $E'$ after application of $\evop$ is
given in this model by \beqn E' = \sum_{i,j} \tr(E_i \rho
E_j^{\dagger}) |i\ra\la j|. \eeqn That is, $\tr(E_i \rho
E_j^{\dagger})$ are the matrix elements of $E'$ in the $|i\ra$
basis. Schumacher \cite{Schumacher96a} suggests defining a matrix $W$
whose elements are given by \beqn W_{ij} \equiv \tr(E_i \rho
E_j^{\dagger} ), \eeqn that is, $W$ is the matrix of $E'$, in an
appropriate basis. This formula applies only for complete quantum
operations.  In the case of incomplete quantum operations a similar
argument shows that the matrix elements of $E'$ are contained in the
matrix $W$ defined by \be \label{eqtn:W_defn} W_{ij} \equiv
\frac{\tr(E_i \rho E_j^{\dagger} )}{\tr(\evop(\rho))}. \eeqn This
gives rise to the useful calculational formula \be S(\rho,\evop) =
S(W) \equiv -\tr(W \log W). \ee

Recall from Chapter \ref{chap:qops} that a quantum operation may have
many different operator-sum representations. In particular, sets of
operators $E_i$ and $F_j$ generate the same quantum operation if and
only if $F_j = \sum_j u_{ji} E_i$, where $u$ is a unitary matrix of
complex numbers, and it may be necessary to append $0$ operators to
the sets $E_i$ or $F_j$ so that the matrix $u$ is a square matrix.

\index{canonical representation for a quantum operation}

$W$ contains matrix elements of the environmental density operator,
and thus is a positive matrix, which may be diagonalized by a unitary
matrix, $v$, $D = v W v^{\dagger}$, where $D$ is a diagonal matrix with
non-negative entries. Define operators $F_j$ by the equation
\beqn
F_j = \sum_i v_{ji} E_i, \eeqn
so the operators $F_j$ give rise to the same quantum operation in the
operator-sum representation. This representation of $\evop$ gives rise
to a $W$ matrix,
\beqn
\tilde W_{kl} & = & \frac{\tr(F_k \rho F_l^{\dagger})}{\tr(\evop(\rho))} \\
	& = & \sum_{mn} v_{km} v_{ln}^* W_{mn} \\ & = & D_{kl}. \ee
Thus, there is a set of operators $F_j$ with respect to which
the $W$ matrix for the system is diagonal, with non-negative
entries.  Any set of operators $F_j$ giving rise to an
operator-sum representation for $\evop$, and for which the
matrix $W$ is diagonal is said to be a {\em canonical
representation} for $\evop$ with respect to the input
$\rho$. We will see later that canonical representations
turn out to have a special significance for quantum error
correction.

Many properties of the entropy exchange follow easily from properties
of the entropy discussed in Chapter \ref{chap:entropy}. For example,
working in a canonical representation for a complete quantum
operation, $\evop$, on a $d$-dimensional space, we see immediately
that $S(I/d,\evop) = 0$ if and only if $\evop$ is a unitary quantum
operation.  Therefore, $S(I/d,\evop)$ quantifies the extent to which
incoherent quantum noise may occur on the system as a whole.  A second
example is that when $\evop$ is restricted to be a complete quantum
operation, the matrix $W$ is easily seen to be convex-linear in
$\rho$, and the state $R'Q'$ is convex-linear in $\evop$. From the
concavity of the von Neumann entropy it follows that $S(\rho,\evop)$
is concave in $\rho$ and $\evop$. Since the system $RQ$ can always be
chosen to be at most $d^2$ dimensional, where $d$ is the dimension of
$Q$, it follows that the entropy exchange is bounded above by $2\log
d$. Other properties will be derived as needed later in this Chapter,
and in the next Chapter.

\section{Quantum Fano inequality}
\label{sec:quantum_Fano}
\index{quantum Fano inequality}
\index{Fano inequality!quantum analogue to}

Intuitively, if an entanglement $RQ$ is subject to noise which results
in it becoming mixed, then the fidelity of the final state $R'Q'$ with
the initial state $RQ$ cannot be perfect. Moreover, the greater the
mixing, the worse the fidelity. In section \ref{sec:Holevo} an
analogous situation arose in the study of classical channels, where
the uncertainty $H(X|Y)$ about the input of a channel, $X$, given the
output, $Y$, was related to the probability of being able to recover
the state of $X$ from $Y$ by the Fano inequality. Schumacher
\cite{Schumacher96a} has proved a very useful analogue of the
classical Fano inequality, the {\em quantum Fano inequality}, which
relates the entropy exchange and the dynamic fidelity:
\begin{eqnarray} \label{eqtn: quantum Fano}
S(\rho,{\cal E}) \leq h(F(\rho,{\cal E})) + (1-F(\rho,{\cal E}))
  \log_2 (d^2-1),
\end{eqnarray}
where $h(x)$ is the binary Shannon entropy.  Inspection of this
inequality reveals its intuitive meaning: if the entropy exchange for
a process is large, then the dynamic fidelity for the process must
necessarily be small, indicating that the entanglement between $R$ and
$Q$ has not been well preserved.  It will be useful to note for our
later work that $0 \leq h(x) \leq 1$ and $\log (d^2-1) \leq 2
\log d$, so from the quantum Fano inequality,
\begin{eqnarray} \label{eqtn: quantum Fano 2}
S(\rho,{\cal E}) \leq 1 + 2(1-F(\rho,{\cal E}))
  \log d.
\end{eqnarray}

To prove the quantum Fano inequality, consider
an orthonormal set of $d^2$ basis states, $|\psi_i\rangle$, for the system
$RQ$. This basis set is chosen so $|\psi_1\rangle = |RQ\rangle$. If we
form the quantities $p_i \equiv \langle \psi_i| (R'Q') | \psi_i \rangle$,
then from the results of subsection \ref{subsec:measurements_entropy}
it follows that
\begin{eqnarray}
S(R'Q') \leq H(p_1,\ldots,p_{d^2}), \end{eqnarray}
where $H(p_i)$ is the Shannon information of the set $p_i$.
Elementary algebra shows that
\begin{eqnarray}
\label{eqtn: entropy grouping}
H(p_1,\ldots,p_{d^2}) & = & h(p_1) \nonumber \\
& & + (1-p_1)H\left(
\frac{p_2}{1-p_1},\ldots,\frac{p_{d^2}}{1-p_1}\right).
\end{eqnarray}
Combining this with the observation that
$H(\frac{p_2}{1-p_1},\ldots,\frac{p_{d^2}}{1-p_1}) \leq \log (d^2-1)$ and
$p_1 = F(\rho,{\cal E})$ by definition of the dynamic fidelity gives,
\begin{eqnarray}
S(\rho,{\cal E}) \leq h(F(\rho,{\cal E})) + (1-F(\rho,{\cal E}))
  \log (d^2-1),
\end{eqnarray}
which is the quantum Fano inequality.

The quantum Fano inequality has been proved using the dynamic fidelity
as a measure of how well information is preserved when it is passed
through a quantum channel. It is possible to give an alternative
formulation of the quantum Fano inequality based upon the dynamic
distance.  The simplest such statement to prove is \be S(\rho,\evop)
\leq \frac{1}{e} + D(\rho,\evop) \log d. \eeqn Note that the intuitive
meaning of this inequality is essentially the same as for the quantum
Fano inequality based upon the dynamic fidelity: a large value for the
entropy exchange implies that the dynamic distance for the process
must be quite large, indicating that entanglement has not been well
preserved. To prove this inequality, we make use of {\em Fannes'
inequality}, which we proved in subsection
\ref{subsec:absolute_distance}. Fannes' inequality states that for two
density operators $A$ and $B$, \be |S(A)-S(B)| \leq \frac{1}{e} +
D(A,B) \log d. \eeqn Thus \be S(\rho,\evop) & = & |S(R'Q')-S(RQ)| \\ &
\leq & \frac{1}{e}+D(R'Q',RQ) \log d \\ & = &
\frac{1}{e}+D(\rho,\evop) \log d, \eeqn which completes the proof.
This inequality may easily be strengthened by making use of the
stronger form of Fannes' inequality which we proved in subsection
\ref{subsec:absolute_distance}, at the cost of some loss in clarity.

In our work we will make use of the quantum Fano inequality based upon
the dynamic fidelity, rather than the dynamic distance.  Nevertheless,
it is useful to keep in mind that an alternate formulation of the
quantum Fano inequality is available, and may be potentially useful
for some applications.

\section{The quantum data processing inequality}
\label{sec:q_data_processing}
\index{data processing inequality!quantum}
\index{quantum data processing inequality}
\index{coherent information}

In subsection \ref{subsec:data_proc} we reviewed
a classical result about Markov processes known as the {\em data
processing inequality}. Recall that the data processing inequality
states that for a Markov process $X \rightarrow Y \rightarrow Z$,
\beqn
H(X) \geq H(X:Y) \geq H(X:Z), \eeqn
with equality in the first stage if and only the random variable
$X$ can be recovered from $Y$ with probability one.

There is a quantum analogue to the data processing inequality, which
Schumacher and I proved in \cite{Schumacher96b}.  Suppose a two stage
quantum process occurs, described by quantum operations $\evop_1$ and
$\evop_2$,
\begin{eqnarray} 
\rho \stackrel{\evop_1}{\longrightarrow} \rho' 
	\stackrel{\evop_2}{\longrightarrow}
	\rho''. \end{eqnarray}
We define the quantum {\em coherent information} by
\beqn
I(\rho,{\cal E}) \equiv S(\rho') - S(\rho,\evop). \eeqn This quantity,
coherent information, is intended to play a role in quantum
information theory analogous to the role played by the mutual
information $H(X:Y)$ in classical information theory. It is not
immediately apparent that the coherent information is the correct
quantum analogue of the mutual information, and we will spend some
time over the next two Chapters in an attempt to justify this claim.

Part of the reason for taking the coherent information seriously as a
quantity like mutual information is that later in the section we prove
that it satisfies the following {\em quantum data processing
inequality},
\begin{eqnarray} \label{eqtn:quantum_data_processing}
S(\rho) \geq I(\rho,{\cal E}_1) \geq I(\rho,{\cal E}_2 \circ {\cal E}_1),
\end{eqnarray}
with equality in the first inequality if and only if it is possible to
reverse the operation $\evop_1$, in a sense to be described
below. Comparison with the classical data processing inequality shows
that the coherent information plays a role in the quantum data
processing inequality identical to the role played by the mutual
information in the classical data processing inequality.

%There is another easily stated reason for considering the coherent
%information to be a quantum analogue of the mutual information.
%We saw above that the entropy exchange and dynamic fidelity are
%related by a quantum Fano inequality, in a similar way to the
%classical Fano inequality which relates the conditional entropy
%$H(X|Y)$, and the probability of making an error in inferring the
%value of $X$ from the output of a classical channel $X \rightarrow
%Y$. Based on the observation the the classical mutual information
%satisfies the relationship $H(X:Y) = H(Y)-H(Y|X)$, we might therefore
%guess that a quantum analogue of the mutual information ought to have
%the form $S(\rho')-S(\rho,\evop)$. In the next Chapter, we will see
%that this line of thought even has an operational consequence, in the
%form of a bound on the quantum channel capacity.

Such a heuristic argument can not be regarded as any sort of a
rigorous justification for the view that the coherent information is
the correct quantum analogue of the classical mutual information. Such
a justification ought to come as a consequence of the role played by
the coherent information in questions related to the quantum channel
capacity. This question will be the topic of the next Chapter.

Let us return to the proof of the quantum data processing inequality.
This result is proved using four systems: $R, Q, E_1$ and $E_2$.  $R$
and $Q$ are used in their familiar roles from Chapter
\ref{chap:distance}. $E_1$ and $E_2$ are systems initially in pure
states, chosen such that a unitary interaction between $Q$ and $E_1$
generates the dynamics ${\cal E}_1$, and a unitary interaction between
$Q$ and $E_2$ generates the dynamics ${\cal E}_2$.  The proof of the
first stage of the quantum data processing inequality is to apply the
subadditivity inequality $S(R'E_1') \leq S(R') + S(E_1')$ to obtain
\begin{eqnarray}
I(\rho,{\cal E}_1) & = & S({\cal E}_1(\rho)) - S(\rho,{\cal E}_1) \\
 & = & S(Q') - S(E_1') \\
 & = & S(R'E_1')-S(E_1') \\
 & \leq & S(R') = S(R) = S(Q) = S(\rho).
\end{eqnarray}

The proof of the second part of the data processing inequality is to apply
the strong subadditivity inequality,
\begin{eqnarray} \label{eqtn:strong_subadditivity_data_proc}
S(R''E_1''E_2'') + S(E_1'') \leq S(R''E_1'')
	+ S(E_1''E_2'').
\end{eqnarray}
{}From purity of the total state of $R''Q''E_1''E_2''$ it follows that
\begin{eqnarray}
S(R''E_1''E_2'') = S(Q''). \end{eqnarray}
Neither of the systems $R$ or $E_1$ are involved in the second stage of
the dynamics in which $Q$ and $E_2$ interact unitarily. Thus, their state
does not change during this stage: $R''E_1'' = R'E_1'$.
But from the purity of $RQE_1$ after the first stage of the dynamics,
\begin{eqnarray}
S(R''E_1'') = S(R'E_1') = S(Q').
\end{eqnarray}
The remaining two terms in the subadditivity inequality are now recognized
as entropy exchanges,
\begin{eqnarray} 
S(E_1'') = S(E_1') = S(\rho,{\cal E}_1), \\
S(E_1''E_2'') = S(\rho,{\cal E}_2 \circ {\cal E}_1). 
\end{eqnarray}
Making these substitutions into the inequality obtained from
strong subadditivity (\ref{eqtn:strong_subadditivity_data_proc})
yields
\begin{eqnarray}
S(Q'') + S(\rho,{\cal E}_1) \leq S(Q') + S(\rho,
	{\cal E}_2 \circ {\cal E}_1), 
\end{eqnarray}
which can be rewritten as the second stage of the data processing
inequality,
\begin{eqnarray}
I(\rho,{\cal E}_1) \geq I(\rho,{\cal E}_2 \circ {\cal E}_1).
\end{eqnarray}
This concludes the proof of the quantum data processing inequality.

The data processing inequality will be invaluable in our study of
quantum error correction, and the quantum channel capacity.  To
understand why it is important, consider a somewhat analogous
statement, the Second Law of Thermodynamics.  The constraint that the
entropy of a closed system can never decrease is tremendously useful
in thermodynamics.  In a somewhat similar fashion, we have obtained
here a quantity (the coherent information) which is non-increasing
under arbitrary quantum operations.  I expect that this non-increasing
property will have many uses beyond even those to which we put it in
this Dissertation.

We conclude the section by noting for future reference that the first
part of the data processing inequality need not hold when ${\cal E}_1$
is not trace-preserving. The reason for this is that it is no longer
necessarily the case that $R' = R$, and thus it may not be possible to
make the identification $S(R') = S(R)$. For example, suppose we have a
three dimensional state space with orthonormal states $|1\rangle$,
$|2\rangle$ and $|3\rangle$.  Let $P_{12}$ be the projector onto the
two dimensional subspace spanned by $|1\rangle$ and $|2\rangle$, and
$P_3$ the projector onto the subspace spanned by $|3\rangle$. Let
$\rho = \frac{p}{2} P_{12} + (1-p)P_3$, where $0 < p < 1$, and ${\cal
E}(\rho) = P_{12} \rho P_{12}$. Then by choosing $p$ small enough we
can make $S(\rho) \approx 0$, but $I(\rho,{\cal E}) = 1$, so we have
an example of a non trace-preserving operation which does not obey the
data processing inequality.

\section{Quantum error correction}
\label{sec:error_correction}
\index{quantum error correction}
\index{error correction}

Noise is a great bane of information processing systems.  Whenever
possible we build our systems to avoid noise completely, and where
that is not possible, we try to protect against the effects of
noise. For example, components in modern computers are extremely
reliable, with a failure rate typically below one error in $10^{17}$
operations. For most practical purposes we can act as if computer
components are completely noiseless. On the other hand, many systems
in widespread use do suffer from a substantial noise problem. Modems
and CD players make use of error correcting codes to protect against
the effects of noise. The details of the techniques used to protect
against noise in practice are sometimes rather complicated, but the
basic principles are easily understood. The key idea is that if we
wish to protect a message against the effects of noise, then we should
{\em encode} the message by adding some redundant information to the
message.  That way, even if some of the information in the encoded
message is corrupted by noise, there will be enough redundancy in the
encoded message that it is possible to recover or {\em decode} the
message so that all the information in the original message is
recovered.

\begin{figure}[htbp]
\begin{center}
\setlength{\unitlength}{1cm}
\begin{picture}(6,3)
\put(0.5,0.5){0}
\put(0.5,2.5){1}
\put(5.5,0.5){0}
\put(5.5,2.5){1}
\put(0.8,0.5){\vector(1,0){4.4}}
\put(2.7,0.2){$1-p$}
\put(0.8,2.5){\vector(1,0){4.4}}
\put(2.7,2.8){$1-p$}
\put(1.0,0.78){\vector(3,1){4}}
\put(2.0,0.9){$p$}
\put(1.0,2.12){\vector(3,-1){4}}
\put(2.0,2.1){$p$}
\end{picture}
%\mbox{\psfig{file=figures/qinfo/binary-symmetric-channel.epsf,scale=150}}
% The figure commented out is by Ike.
\end{center}
\caption{Binary symmetric channel.}
\label{fig:qec:bsc}
\end{figure}
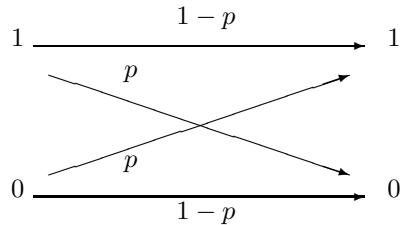

For example, suppose we wish to send a bit from one location to
another through a noisy communications channel. Suppose that the
effect of the noise in the channel is to flip the bit being
transmitted with probability $p > 0$; with probability $1-p$ the bit
is transmitted without error. This is known as the binary symmetric
channel (see figure~\ref{fig:qec:bsc}).  A simple means of protecting
the bit against the effects of noise is to replace the bit we wish to
protect with three copies of itself: \beqn 0 & \rightarrow & 000 \\ 1
& \rightarrow & 111.  \eeqn We now send all three bits through the
channel. At the receiver's end of the channel three bits are output,
and the receiver has to decide what the value of the original bit
was. Suppose $001$ was output from the channel. Provided the
probability $p$ of a bit flip is not too high, it is very likely that
the third bit was flipped by the channel, and that $0$ was the bit
that was sent.

This type of decoding is called {\em majority voting}, since the decoded
output from the channel is whatever value, $0$ or $1$, appears more times in
the actual channel output.  Majority voting fails if two or more of the bits
sent through the channel were flipped, and succeeds otherwise.  The
probability that two or more of the bits is flipped is $3p^2(1-p)+p^3$, so
the probability of error is $p_e \equiv 3p^2-2p^3$. Without encoding the
probability of an error was $p$, so the code improves matters if $p_e < p$,
which occurs whenever $p < 1/2$.

The type of code we have described is called a {\em repetition code}, since
we encode the message to be sent by repeating it a number of times. A
similar technique has been used for millenia as a part of everyday
conversation: if we're having difficulty understanding someone's spoken
language, perhaps because they have a foreign accent, we ask them to repeat
what they're saying. We may not catch all the words either time, but we can
put the iterations together to produce a coherent message.

Many interesting and clever techniques have been developed in the
theory of classical error correcting codes; unfortunately these
techniques are beyond the scope of this Dissertation. However, the key
idea is always to encode messages by adding enough redundancy that the
original message is recoverable after noise has acted on the encoded
message.  How much redundancy needs to be added depends on how severe
the noise in the channel is.

To protect quantum states against the effects of noise we would like
to have {\em quantum error correcting codes}. This section is a review
of the elementary theory of quantum error correcting codes.  In the
next section we will re-examine quantum error correcting codes from an
information-theoretic viewpoint.  

There are some important differences between classical information and
quantum information that require new ideas to be introduced to make
quantum error correcting codes possible:
\begin{itemize}
\item {\em No cloning}: One might try to implement the repetition
code quantum mechanically by duplicating the quantum state three or more
times. This is forbidden by the no cloning theorem
\cite{Dieks82a,Wootters82a}. Even if cloning were possible, it would
not be possible to measure and compare the three quantum states output
from the channel.

\item {\em Errors are continuous}: A continuum of different errors may
occur on a single qubit.  Determining which error occurred in order to
correct it would appear to require infinite precision, and therefore
infinite resources.

\item {\em Measurement destroys quantum information}: In
classical error correction we observe the output from the channel, and
decide what decoding procedure to adopt. Observation in quantum mechanics
generally destroys the quantum state under observation, and makes recovery
impossible.
\end{itemize}

Suppose we send qubits through a channel which leaves the qubits
untouched with probability $1-p$, and flips the qubits with
probability $p$. That is, with probability $p$ the state $|\psi\ra$ is
taken to the state $X|\psi\ra$. This channel is called the {\em bit
flip channel}\index{bit flip channel}, and we will now show how to
protect qubits against the effects of noise from this channel.

Suppose we encode the single qubit state $a|0\ra+b|1\ra$ in three qubits as
$a|000\ra+b|111\ra$. A convenient way to write this encoding is\index{bit
flip code}
\beqn 
	|0\ra & \rightarrow & |0_L\ra \equiv |000\ra \\
	|1\ra & \rightarrow & |1_L\ra \equiv |111\ra, 
\eeqn
where it is understood that superpositions of basis states are taken
to corresponding superpositions of encoded states.  The notation
$|0_L\ra$ and $|1_R\ra$ indicate that these are the {\em logical} zero
and one\index{logical zero and one} states, not the {\em physical}
zero and one states.

Suppose the initial state $a|0\ra+b|1\ra$ has been perfectly encoded. Each
of the three qubits is passed through an independent copy of the bit flip
channel. Suppose a bit flip occurred on one or fewer of the qubits. There is
a simple two stage {\em error correction} procedure which can be used to
recover the correct quantum state in this case:
\begin{enumerate}
\item ({\em Error detection} or {\em syndrome diagnosis}). We
perform a measurement which tells us what error, if any, occurred on
the quantum state. The measurement result is called the {\em error
syndrome}. For the bit flip channel there are four error syndromes,
corresponding to the four projection operators
\beqn
P_0 = |000\ra\la 000| +|111\ra \la 111| & \mbox{no error} \\
P_1 = |100\ra\la 100| +|011\ra \la 011| & \mbox{bit flip on qubit one} \\
P_2 = |010\ra\la 010| +|101\ra \la 101| & \mbox{bit flip on qubit two} \\
P_3 = |001\ra\la 001| +|110\ra \la 110| & \mbox{bit flip on qubit three}.
\eeqn
Suppose, for example, that a bit flip occurs on qubit one, so the
corrupted state is $a|100\ra+b|011\ra$. Notice that $\la \psi|P_1|\psi
\ra = 1$ in this case, so the outcome of the measurement result (the
error syndrome) is certainly $1$. Notice, furthermore, that syndrome
measurement does not cause any change to the state: it is
$a|100\ra+b|011\ra$ both before and after syndrome measurement.

\item ({\em Recovery}). We use the value of the error syndrome to tell
us what procedure can be used to recover the initial state. For
example, if the error syndrome was $1$, indicating a bit flip on the
first qubit, then we flip that qubit again, recovering the original
state $a|000\ra+b|111\ra$ with perfect accuracy. The four possible
error syndromes and the recovery procedure in each case are: $0$ (no
error) -- do nothing; $1$ (bit flip on first qubit) -- flip the first
qubit again; $2$ (bit flip on second qubit) -- flip the second qubit
again; $3$ (bit flip on third qubit) -- flip the third qubit again. In
each case it is easy to see that the original state is recovered with
perfect accuracy for each value of the error syndrome.
\end{enumerate}
This error correction procedure works perfectly, provided bit flips occur on
one or fewer of the three qubits. This occurs with probability
$(1-p)^3+3p(1-p)^2 = 1-3p^2+2p^3$. The probability of an error remaining
uncorrected is therefore $3p^2-2p^3$, just as for the repetition code we
studied earlier. Once again, provided $p < 1/2$ the encoding and decoding
improve the reliability of storage of the quantum state.

In some ways this error analysis is inadequate. The problem is that
not all errors and states in quantum mechanics are created equal:
quantum states live in a continuous space, so it is possible for some
errors to corrupt a state by a tiny amount, while others mess it up
completely. An extreme example is provided by the bit flip ``error''
$X$, which does not affect the state $(|0\ra+|1\ra)/\sqrt 2$ at all,
but flips the $|0\ra$ state so it becomes a $|1\ra$. In the former
case we would not be worried about a bit flip error occurring, while
in the latter case we would obviously be very worried.

To address this problem we make use of the \emph{fidelity} quantity
introduced in Chapter \ref{chap:distance}.  Recall that the fidelity
between a pure and a mixed state is given by
\beqn
	F(|\psi\ra,\rho) \equiv \la \psi| \rho |\psi\ra. 
\eeqn
The object of quantum error correction is to increase the fidelity
with which quantum information is stored.  By the results of Chapter
\ref{chap:distance}, if we can perform computations with a high enough
fidelity, then the measurement results output from the computation
will be sufficiently close in distribution to the desired distribution
to consider the computation successful.

Let's compare the {\em minimum} fidelity achieved by the three qubit
bit flip code with the minimum fidelity achieved without error
correction.  Suppose the quantum state of interest is
$|\psi\ra$. Without using the error correcting code the state of the
qubit after being sent through the channel is \beqn \rho = (1-p)
|\psi\ra\la \psi| + p X |\psi\ra \la \psi| X.  \eeqn The fidelity is
given by \beqn F = \la \psi|\rho|\psi\ra = (1-p) + p \la \psi| X |\psi
\ra \la \psi| X |\psi\ra.  \eeqn The second term on the right hand
side is non-negative. When $|\psi\ra = |0\ra$ the second term is zero
so we see that the minimum fidelity is $F = 1-p$. Suppose the three
qubit error correcting code is used to protect the state $|\psi\ra =
a|0_L\ra + b|1_L\ra$. The quantum state after the channel and error
correction is \beqn \rho = \left[(1-p)^3 +3p(1-p)^2 \right] |\psi\ra
\la \psi| + \ldots \eeqn The included term is all the contributions
from the correctable errors -- no error at all, and a bit flip on a
single qubit. The omitted terms are the contributions from bit flips
on two or three qubits. The omitted terms are non-negative, so the
fidelity we calculate will be a {\em lower bound} on the true
fidelity. We see that $F = \la \psi| \rho|\psi\ra \geq
(1-p)^3+3p(1-p)^2$. That is, the fidelity is at least
$1-3p^2+2p^3$. We see that the fidelity of storage for the quantum
state is improved provided $p < 1/2$, which is the conclusion we came
to earlier based on a cruder analysis.

The bit flip code is interesting, but it does not seem to go beyond
classical error correcting codes in any significant manner.  A more
interesting noisy quantum channel is the {\em phase flip} error model
for a single qubit. In this error model the qubit is left alone with
probability $1-p$, and the relative phase of the $|0\ra$ and $|1\ra$
states is flipped. More precisely, the phase flip operator\index{phase
flip operator} $Z$ (sometimes called the Pauli sigma $z$
operator\index{Pauli sigma matrices} $\sigma_z$) is applied to the
qubit with probability $p > 0$. The action of the phase flip $Z$ is
defined by $Z|0\ra \equiv |0\ra, Z|1\ra
\equiv -|1\ra$. Thus the state $a|0\ra +b|1\ra$ is taken to the state
$a|0\ra-b|1\ra$ under the phase flip. The reason this is called a phase flip
is that the relative phase of the $|0\ra$ and $|1\ra$ states is flipped by
the action of the phase flip operator $Z$.

There is no classical equivalent to the phase flip channel, since
classical channels don't have any property equivalent to
phase. However, there is an easy way to turn the phase flip channel
into a bit flip channel. Suppose we apply the Hadamard gate
immediately before and after the action of the phase flip channel.  If
the phase flip channel left the state alone then the additional
Hadamard gates cancel out (since $H^2 = I$) and can be ignored.  If
the phase flip $Z$ occurs, then the action with the Hadamard gates
taken into account is $H Z H = X$, which is the bit flip.

Quantum error correction for the phase flip channel can therefore be
accomplished by encoding in three qubits as for the bit flip channel
and then applying a Hadamard gate to each qubit to complete the
encoding for the phase flip channel.  The phase flip channel then acts
independently on each qubit.  Finally, we error correct by applying a
Hadamard gate to each qubit and then applying the usual error
correction procedure for the bit flip code.

The encoded $|0\ra$ and $|1\ra$ for the three qubit phase flip code are thus
\beqn
	|0_L\ra & = & H|0\ra H|0\ra H|0\ra = |+\ra|+\ra|+\ra \\
	|1_L\ra & = & H|1\ra H|1\ra H|1\ra = |-\ra|-\ra|-\ra, 
\eeqn
where $|+\ra \equiv H|0\ra = (|0\ra+|1\ra)/\sqrt 2$ and $|-\ra \equiv H|1\ra
= (|0\ra-|1\ra)/\sqrt 2$.

Obviously this code for the phase flip channel has the same
characteristics as the earlier code for the bit flip channel.  In
particular, the minimum fidelity for this code is the same as that for
the three qubit bit flip code, and we have the same criteria for the
code producing an improvement over the case with no error correction.
We say that these two channels are {\em unitarily equivalent}, since
there is a unitary operator $U$ (in this case the Hadamard gate) such
that the action of one channel is the same as the other, provided the
first channel is preceded by $U$ and followed by $U^{\dagger}$. These
operations may be trivially incorporated into the encoding and error
correction operations. 
%For general unitary operators these ideas are
%worked out in a problem at the end of the Chapter.

\begin{figure}[htbp]
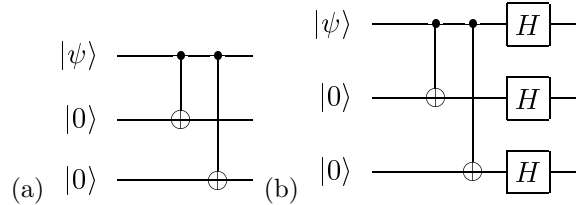

\begin{center}
(a)
\mbox{\psfig{file=encode-bit-flip-circuit.epsf,scale=1}}
(b)
\mbox{\psfig{file=encode-phase-flip-circuit.epsf,scale=1}}
\end{center}
\caption{Encoding circuits for the (a) bit flip and (b) phase flip
codes.  
\label{fig: three qubit encoding}}
\end{figure}

So far we've talked about encoding and error correction in the
abstract.  How can these operations be performed in practice?  Quantum
circuits for encoding the three qubit bit flip and phase flip code are
shown in figure \ref{fig: three qubit encoding}.  To see that the bit
flip encoding circuit works, just note that the state $|000\ra$ is
left unchanged by the circuit, while the state $|100\ra$ is taken to
the state $|111\ra$ by the circuit.  The phase flip encoding circuit
is exactly the same, except we apply an extra Hadamard gate at the end
of the encoding, as expected.  The simplicity of design in these
circuits is a general feature of many of the quantum error correcting
codes which have been proposed \cite{Gottesman97a}, however it is by
no means always the case that quantum error correction can be
performed efficiently by means of a quantum circuit.  One drawback of
the information-theoretic approach to quantum error correction which
we take later in the Chapter is that it does not seem to provide many
clues about how to efficiently perform encodings and decodings.

%%%%%%%%%%%%%%%%%%%%%%%%%%%%%%%%%%%%%%%%%%%%%%%%%%%%%%%%%%%%%%%%%%%%%%%%%%%%%
\subsection{Shor's code}
\index{Shor's quantum error correcting code}

There is a simple quantum code which can protect against the effects
of {\em any} error, provided the error only affects a single
qubit. The code is known as the {\em Shor code}, after its inventor
\cite{Shor95a}.  The code is a combination of the three qubit phase
flip and bit flip codes. We first encode the qubit using the phase
flip code: $|0\ra \rightarrow |+++\ra, |1\ra \rightarrow
|---\ra$. Next, we encode each of these three qubits using the bit
flip code: $|+\ra$ is encoded as $(|000\ra+|111\ra)\sqrt 2$ and
$|-\ra$ is encoded as $(|000\ra-|111\ra)\sqrt 2$.  The result is a
nine qubit code, with codewords given by:
\beqn
|0\ra \rightarrow |0_L\ra & \equiv & \frac{
	(|000\ra+|111\ra)(|000\ra+|111\ra)(|000\ra+|111\ra)}
	{2 \sqrt 2} \\
|1\ra \rightarrow |1_L\ra & \equiv & \frac{
	(|000\ra-|111\ra)(|000\ra-|111\ra)(|000\ra-|111\ra).}
	{2 \sqrt 2}.
\eeqn

The quantum circuit encoding the Shor code is shown in Fig.~\ref{fig:
Shor code}. As described above, the first part of the circuit encodes
the qubit using the three qubit phase flip code; comparison with
Fig.~\ref{fig: three qubit encoding} (b) shows that the circuits are
identical. The second part of the circuit encodes each of these three
qubits using the bit flip code. To do this three copies of the bit
flip code encoding circuit (Fig.~\ref{fig: three qubit encoding} (a))
is used.  This method of encoding using a hierarchy of levels in this
way is known as {\em concatenation}\cite{Knill96a}.
% It's a great trick
%for obtaining new codes from old, and we will use it later to prove
%some important results about quantum error correction.

The Shor code is able to protect against phase flip and bit flip errors on
any qubit. To see this, suppose a bit flip occurs on the first qubit. As for
the bit flip code, we perform a measurement comparing the two qubits, and
find that they are different. We conclude that an error occurred on the
first or second qubit.  As before, we do not actually measure the first and
second qubit, which would destroy the coherence between them, rather, we
merely compare them. Next we, compare the second and third qubit. We find
that they are the same, so it could not have been the second qubit which
flipped. We conclude that the first qubit must have flipped, and recover
from the error by flipping the first qubit again, back to its original
state.

In a similar way we can cope with a phase flip on the first qubit. We
do this by comparing the {\em sign} of the first block of three qubits
with the sign of the second block of three qubits. The phase flip on
the first qubit caused the sign in the first block to be flipped, so
we find that these signs are different. Next, we compare the sign of
the second block of three qubits with the sign of the third block of
three qubits. We find that these are the same, and conclude that the
phase must have flipped in the first block of three qubits. We recover
from this by flipping the sign in the first block of three qubits
again, back to its original value.  

Note that this procedure also allows us to recover when both a bit
flip and a phase flip occur, simply by performing both procedures.

The bit and / or phase flip errors are not the only errors which the
Shor code can protect against. In fact, the Shor code can protect
against an {\em arbitrary} error, provided it only affects a single
qubit. The error could even be so drastic as to remove the qubit
entirely and replace it with complete garbage! The interesting thing
is, no additional work needs to be done in order to protect against
arbitrary errors -- the procedure already described works just fine.
An outline proof is as follows.

Suppose an arbitrary error occurs on the first qubit, described by a
set of operators $\{ E_i \}$ in some operator-sum representation (see
Chapter \ref{chap:qops}). Each $E_i$ is a single qubit operator, and
thus can be expanded as a linear combination of the identity, $I$, the
bit flip, $X$, the phase flip, $Z$, and the combined bit and phase
flip, $XZ$: \beqn E_i = e_{i0} I + e_{i1} X + e_{i2} Z + e_{i3} X Z.
\eeqn After the noise has acted, the code is in a mixture of states,
$E_i |\psi\ra$, each of which is a superposition of the states that
would have resulted if nothing had occurred (the $I$ term in the
expression for $E_i$), if a bit flip had occurred (the $X$ term), if a
phase flip had occurred (the $Z$ term), or if both a bit and phase
flip occurred (the $XZ$ term).  The quantum measurement used to
perform error detection causes these four possible outcomes to
decohere. Thus, we have a mixture of states of the form $|\psi\ra,
X|\psi\ra, Z|\psi\ra, XZ|\psi\ra$. However, we have already proved
that it is possible to recover the original state of the system given
such a mixture, so the error correction procedure works correctly.
% It is
%interesting to note that this decoherence only obtains information
%about what type of error occurred, it does not result in information
%being obtained about the state being protected.  

\section{Information-theoretic conditions for error correction}
\label{sec:info_conditions}
\index{information-theoretic conditions for quantum error correction}

There is an elegant set of information-theoretic conditions for
quantum error correction. Suppose first that $\evop$ is a complete
quantum operation, and $\rho$ is some input state. We will say that
$\evop$ is {\em perfectly reversible} upon input of $\rho$ if there
exists a complete quantum operation ${\cal R}$ such that
\be \label{eqtn:qec_conditions}
F(\rho,{\cal R} \circ {\cal E}) = 1. \ee
From item \ref{item:TaDa} on page \pageref{item:TaDa}, it follows that
a quantum operation is perfectly reversible if and only if for
every state $|\psi\ra$ in the support of $\rho$,
\be \label{eqtn:qec_conditions2}
({\cal R} \circ {\cal E})(|\psi\ra \la \psi|) = |\psi\ra \la \psi|. \ee

We may connect the notion of perfect reversibility with the quantum
error correcting codes which we studied in the previous section.
Specifically, a quantum error correcting code was a {\em subspace}
spanned by codewords in some larger Hilbert space.  To be resilient
against the noise induced by some quantum operation, $\evop$, it is
necessary that the quantum operation $\evop$ be reversible on the
subspace spanned by the codewords.  Letting $P$ be the projector onto
that subspace, and $d$ be the dimensionality, we see that the noise
process $\evop$ is correctable if and only if the operation $\evop$ is
perfectly reversible upon input of the density operator $P/d$.

The information-theoretic condition for a complete quantum operation
$\evop$ to be perfectly reversible upon input of $\rho$ is that the
first inequality in the quantum data processing inequality be
satisfied with equality, \be S(\rho) = I(\rho,\evop) =
S(\rho')-S(\rho,\evop). \ee To prove necessity, suppose that $\evop$
is perfectly reversible upon input of $\rho$. From the second stage of
the quantum data processing inequality it follows that \be
S(\rho)-S(\rho,\evop) \geq S(\rho'') -S(\rho,{\cal R} \circ
\evop). \ee From the reversibility requirement it follows that $\rho''
= \rho$. Furthermore, from the quantum Fano inequality, (\ref{eqtn:
quantum Fano}), and the reversibility requirement
(\ref{eqtn:qec_conditions}) it follows that $S(\rho,{\cal R} \circ
\evop) = 0$. Thus the second stage of the quantum data processing
inequality may be rewritten \be S(\rho')-S(\rho,\evop) \geq
S(\rho). \ee Combining this with the first part of the quantum data
processing inequality, $S(\rho) \geq S(\rho')-S(\rho,\evop)$, we
deduce that \be S(\rho') = S(\rho)-S(\rho,\evop), \ee for any $\evop$
which is reversible upon input of $\rho$.

Next, we will give a constructive proof that satisfaction of the
condition
\be
S(\rho) = S(\rho')-S(\rho,\evop) \ee
implies that the quantum operation $\evop$ is reversible upon input of
$\rho$. Noting that $S(\rho) = S(Q)=S(R) = S(R')$, $S(\rho') = S(Q') =
S(R'E')$ and $S(\rho,\evop) = S(E')$, we see that
\be
S(R')+S(E') = S(R'E'). \eeqn Recall from subsection
\ref{subsec:subadditivity} that this is equivalent to the condition
that $R'E' = R' \otimes E'$. Suppose that the initial state of $Q$ is
$\sum_i p_i |i\ra \la i|$, and that we purify this state into $RQ$ as
$|RQ\ra = \sum_i \sqrt{p_i} |i\ra|i\ra$. Note that $R' = R = \sum_i
p_i |i\ra \la i|$. Furthermore, suppose that $E' = \sum_j q_j |j\ra
\la j|$ for some orthonormal set $|j\ra$, so that
\be
R'E' = \sum_{ij} p_i q_j |i\ra \la i | \otimes |j\ra \la j|. \ee
Next, we use the Schmidt decomposition to write the total state of
$R'Q'E'$ after the quantum operation has been applied, as
\beqn
|R'Q'\ra = \sum_{ij} \sqrt{p_i q_j} |i\ra |i,j\ra |j\ra, \eeqn
where $|i,j\ra$ is some orthonormal set of states in system
$Q$. Define projectors $P_j$ by
\beqn
P_j \equiv \sum_i |i,j\ra \la i,j|. \eeqn
The idea of the restoration operation is to first perform a
measurement described by the projectors $P_j$, which reveals the state
$|j\ra$ of the environment, and then conditional on the measurement
result do a unitary rotation $U_j$ which satisfies the equation
\be
U_j |i,j\ra = |i\ra. \ee
That is, the restoration operation is given by
\beqn
{\cal R}(\sigma) \equiv \sum_j U_j P_j \sigma P_j U_j^{\dagger}. \eeqn
The projectors $P_j$ are orthogonal, by
the orthogonality of the states $|i,j\ra$, but may not be
complete. If this is the case, then to ensure that the quantum
operation ${\cal R}$ is complete, it is necessary to add an extra
projector $\tilde P \equiv I-\sum_j P_j$ to the set of projectors to
make the operation complete.

Finally, note that the state of the system $RQE$ after the reversal
operation is given by
\be
\sum_j U_j P_j |\psi'\ra \la \psi'| P_j U_j^{\dagger} & = &
\sum_j \sum_{i_1 i_2} \sqrt{p_{i_1} p_{i_2}} q_j |i_1\ra \la
i_2| \otimes (U_j |\psi_{i_1j}\ra \la \psi_{i_2 j}|
U_j^{\dagger}) \otimes |j\ra \la j| \\
& = & \sum_j \sqrt{p_{i_1}p_{i_2}} |i_1\ra \la i_2| \otimes |i_1\ra
\la i_2| \otimes E', \eeqn
from which we see that $R''Q'' = RQ$, and thus $F(\rho,{\cal R} \circ
\evop) = 1$, that is, the operation $\evop$ is perfectly reversible
upon input of the state $\rho$, as we desired to show.

This completes the proof of the information-theoretic reversibility
conditions for complete quantum operations.  Some intuition about the
result may be obtained by imagining that $Q$ is a memory element in a
quantum computer, $R$ is the remainder of the quantum computer, and
$E$ is an environment whose interaction with $Q$ causes noise.  The
information-theoretic reversibility condition may be stated as
follows: the state of the environment, $E'$, after the interaction,
should not be correlated with the state of the remainder of the
quantum computer, $R'$, after the interaction between $Q$ and $E$.
That is, the environment does not learn anything about the rest of the
quantum computer through interacting with $Q$.

We have discussed information-theoretic conditions for the
perfect reversibility of complete quantum operations. It is possible
to give a similar characterization of perfect reversibility for
incomplete quantum operations which generalize these conditions. 
What does it mean for an incomplete quantum operation $\evop$ to be
perfectly reversible? As before, we will take the criterion for
reversibility to be the requirement that there exist a complete
quantum operation ${\cal R}$ such that
\be
F(\rho, {\cal R} \circ \evop) = 1. \ee
As for the case of complete quantum operations, it is not difficult
to show that this is equivalent to the condition that
\be
\frac{({\cal R} \circ \evop)(|\psi\ra\la\psi |)}{\tr(({\cal R} \circ
\evop)(|\psi\ra \la \psi|))} =  |\psi\ra \la \psi|, \ee
for all $|\psi\ra$ in the support of $\rho$.

Necessary and sufficient conditions for an arbitrary quantum operation
$\evop$ to be perfectly reversible are as follows:
\cite{Nielsen97b,Nielsen98a}
\begin{enumerate}
\item There exists a constant $c > 0$ such that for all states
$|\psi\ra$ in the support of $\rho$, $\tr(\evop(|\psi\ra\la \psi|)) =
c$.
\item $S(\rho) = S(\rho')-S(\rho,\evop)$.
\end{enumerate}
The first requirement can be given an elegant information-theoretic
interpretation.  We saw in Chapter \ref{chap:qops} that incomplete
quantum operations are associated with measurements on quantum
systems.  Suppose we think of $\evop$ as a possible outcome that can
occur as the result of a measurement on $Q$.  Then the first
requirement is just the condition that this measurement result occurs
with the same probability, regardless of which state in the support of
$\rho$ is prepared.  Because of this uniformity, it follows that no
information about the identity of the state is revealed to the
observer through the measurement.  The second requirement can be
interpreted exactly as before.

To see necessity of condition 1, note that for all states $\sigma$ whose
support is contained within the support of $\rho$,
\beqn
({\cal R} \circ \evop) (\sigma) = \tr( \evop(\sigma)) \sigma. \eeqn The
linearity of the left hand side implies that the right hand side must
also be linear, and therefore $\tr(\evop(\sigma)) = c$, for some
constant $c$, for all states whose support is contained within the
support of $\rho$.

To see the necessity of condition 2, we make use of the already proved
necessity of condition 1.  Let $c$ be the constant value of
$\tr(\evop(\rho))$ from condition 1. Let $E_i$ be a set of operators
giving an operator-sum representation for $\evop$.  Let $P$ be a
projection onto the support of $\rho$, and $Q \equiv I-P$ the
projection onto the orthocomplement of the support, that is, the
kernel of $\rho$.  Define \be \tilde \evop(\sigma) \equiv \frac{E_i P
\sigma P E_i^{\dagger}}{c}+Q\sigma Q. \ee Note that $\tilde \evop$ is
a complete quantum operation such that $\tilde \evop(\sigma) =
\evop(\sigma) / \tr(\evop(\sigma))$ for all states $\sigma$ such that
the support of $\sigma$ lies within the support of $\rho$.  That is,
the action of $\tilde \evop$ within the support of $\rho$ is identical
to the action of $\evop$.  It follows that reversibility of $\evop$
upon input of $\rho$ is equivalent to the reversibility of $\tilde
\evop$ upon input of $\rho$, and therefore \be S(\rho) =
S(\rho')-S(\rho,\tilde \evop). \ee But $S(\rho,\tilde \evop) =
S(\rho,\evop)$, since $\tilde \evop$ and $\evop$ act in the same way
within the support of $\rho$, from which we conclude \be S(\rho) =
S(\rho')-S(\rho,\evop), \ee as required.

To prove that these two conditions are sufficient for reversibility,
construct the quantum operation $\tilde \evop$ as above.  As before,
$\tilde \evop$ and $\evop$ have identical actions upon states in the
support of $\rho$, and therefore $S(\rho)=S(\rho')-S(\rho,\tilde
\evop)$.  But $\tilde \evop$ is a complete quantum operation, and
therefore there exists a reversing operation ${\cal R}$ for $\tilde
\evop$, as constructed earlier.  Since $\tilde \evop$ and $\evop$ have
the same action on states in the support of $\rho$, it follows that
${\cal R}$ is a reversing operation for $\evop$ as well.  This
completes the proof of the information-theoretic conditions for
reversibility of a quantum operation.

Verifying the information-theoretic conditions for a specific quantum
error-correcting code may not be completely trivial.  It can certainly
be done; for example, for the Shor code presented in the last section,
however, it is usually much easier to verify that a given quantum
error-correcting code works by algebraic techniques
\cite{Bennett96a,Knill97a,Gottesman97a,Calderbank97a}.  The true
benefit of the information-theoretic approach to quantum
error-correction lies in the insight it gives into other problems,
such as the thermodynamics of quantum error-correction, to be
discussed later in this Chapter, and the quantum channel capacity, to
be discussed in the next Chapter.

\section{Information-theoretic inequalities for quantum processes}
\label{sec:other_inequalities}

The data processing inequality is an interesting inequality with a
practical use: the study of quantum error correcting codes. However,
it is possible to prove many other related inequalities. In this
section I will tabulate {\em all} the inequalities that can be proved
for a two-part quantum process, using subadditivity and strong
subadditivity.

First, consider a quantum process with a single stage, described by a
complete quantum operation $\evop$.  \beqn \rho
\stackrel{\evop}{\longrightarrow} \rho' \eeqn Three systems are
involved in this process, $R$, $Q$ and $E$. Applying all possible
permutations of the subadditivity inequality to $R'Q'E'$ yields three
non-trivial inequalities: \beqn S(R'Q') & \leq & S(R')+S(Q') \\
S(R'E') & \leq & S(R')+S(E') \\ S(Q'E') & \leq & S(Q')+S(E').  \eeqn
In terms of system quantities alone these inequalities are easily
rewritten \beqn S(\rho,\evop) & \leq & S(\rho)+S(\rho') \\ S(\rho') &
\leq & S(\rho) + S(\rho,\evop). \\ S(\rho) & \leq &
S(\rho')+S(\rho,\evop). \eeqn The first inequality puts an upper bound
on the entropy exchange in terms of the input and output
entropies. The second inequality is familiar as the first stage of the
data processing inequality, slightly rewritten.  The third inequality
can be rewritten in a form that will be especially useful in the study
of the thermodynamics of quantum error-correction, \beqn
\label{eqtn:second_law} \Delta S + S(\rho,\evop) \geq 0,\eeqn where
$\Delta S \equiv S(\rho')-S(\rho)$ is the difference between output
and input entropies for the process.  What this inequality tells us is
that the total entropy change associated to the process is positive,
where both the system $Q$ and the environment $E$ are included in the
entropic accounting.
% This viewpoint will be useful
%later in our study of the relationship of error correction
%to Maxwell's demon. 

It is easily checked that applying all possible permutations of
subadditivity and strong subadditivity to the joint system $R'Q'E'$
yields no further inequalities. Note that in all three cases the
equality conditions for saturation of the inequality are obvious from
the usual equality conditions for subadditivity. In particular,
equality holds in (\ref{eqtn:second_law}) if and only if
\begin{equation} 
\label{eqtn: second law equality}
Q'E' = Q' \otimes E'
\end{equation}
These equality conditions are useful in our later analysis of 
thermodynamically efficient error correction.

Consider next the case of a two stage quantum process, \beqn \rho
\stackrel{\evop_1}{\longrightarrow} \rho'
\stackrel{\evop_2}{\longrightarrow} \rho''. \eeqn This process
involves four systems, $R, Q, E_1$ and $E_2$, where $E_1$ and $E_2$
are the environments associated with the complete quantum operations
$\evop_1$ and $\evop_2$, respectively. Consider the state
$R''Q''E_1''E_2''$. We wish to apply all possible permutations of the
subadditivity and strong subadditivity inequalities to this state.

When we do this, we discover an interesting fact.  The entropies of
all possible subsystems of $R''Q'',E_1''E_2''$ can be expressed in
terms of things we already understand, such as entropy exchanges, with
one exception.  That exception is the quantity $S(Q''E_1'') =
S(R''E_2'')$, which we will refer to as the \emph{correlation
entropy}\index{correlation entropy}, since it is a measure of the
correlation existing between the environment causing noise in the
first part of the dynamics, $E_1$, and the final state of the quantum
system, \be C(\rho,\evop_1,\evop_2) \equiv S(Q''E_1''). \ee To
calculate the value of the correlation entropy, let $\{ E_i^1 \}$ and
$\{ E_i^2 \}$ be sets of operators generating operator-sum
representations for $\evop_1$ and $\evop_2$, respectively. Applying
the usual unitary model for quantum operations, we see that the final
state of $Q''E_1''E_2''$ is 
\be \sum_{i_1i_2j_1j_2} E_{i_2}^2
E_{i_1}^1 \rho (E_{j_1}^1)^{\dagger} (E_{j_2}^2)^{\dagger} \otimes 
|i_1\ra \la j_1| \otimes |i_2\ra \la j_2|. \ee
Introducing an orthonormal basis $|k\ra$ for system $Q$, we see that
the matrix elements of $Q''E_1''$ are given by the matrix $U$ defined
by
\be
U_{ik,jl} \sum_m \la k|E_m^2 E_i^1 \rho (E_i^1)^{\dagger}
(E_m^2)^{\dagger} |l\ra, \ee
and therefore $C(\rho,\evop_1,\evop_2) = S(U)$.  We will not have any
occasion to calculate correlation entropies, however it is useful to
know that such an explicit formula exists which could be used to
calculate such quantities if the need arises.  Note also that the
matrix $U$ picks up a normalization factor of $1/\tr((\evop_2 \circ
\evop_1)(\rho))$ in the case where $\evop_2$ or $\evop_1$ is incomplete.

Let's begin by enumerating in a table all the inequalities which can
be obtained from subadditivity. To keep track of which inequalities we
have evaluated, we write $(X:Y)$, where $X$ and $Y$ are (different)
subsystems of $R''Q''E_1''E_2''$.  Alongside these entries we write
the corresponding entropy inequality, $S(X,Y) \leq S(X)+S(Y)$, in
terms of appropriate system quantities:

\vspace{0.6cm}
\begin{center}
\fbox{
\parbox{14cm}
{
\begin{center} {\bf Entropy Inequalities: Subadditivity} \end{center}
$$
\begin{array}{||c|rcl||}  \hline
(R'':Q'') & S(\rho,\evop_2 \circ \evop_1) & \leq & S(\rho)+S(\rho'')
\\
(R'':E_1'') & S(\rho') & \leq & S(\rho)+S(\rho,\evop_1) \\
(R'':E_2'') & C(\rho,\evop_1,\evop_2) & \leq & S(\rho)+S(\rho',\evop_2) \\
(Q'':E_1'') & C(\rho,\evop_1,\evop_2) & \leq & S(\rho'')+S(\rho,\evop_1) \\
(Q'':E_2'') & S(\rho') & \leq & S(\rho'')+S(\rho',\evop_2) \\
(E_1'':E_2'') & S(\rho,\evop_2 \circ \evop_1) & \leq &
S(\rho,\evop_1)+S(\rho',\evop_2) \\ \hline
(R''Q'':E_1'') & S(\rho',\evop_2) & \leq & S(\rho,\evop_2\circ\evop_1)+S(\rho,\evop_1) \\
(R''Q'':E_2'') & S(\rho,\evop_1) & \leq & S(\rho,\evop_2\circ\evop_1)+S(\rho',\evop_2) \\
(R''E_1'':Q'') & S(\rho',\evop_2) & \leq & S(\rho')+S(\rho'') \\
(R''E_1'':E_2'') & S(\rho'') & \leq & S(\rho')+S(\rho',\evop_2) \\
(R''E_2'':Q'') & S(\rho,\evop_1) & \leq & C(\rho,\evop_1,\evop_2)+S(\rho'') \\
(R''E_2'':E_1'') & S(\rho'') & \leq & C(\rho,\evop_1,\evop_2)+S(\rho,\evop_1) \\
(Q''E_1'':R'') & S(\rho',\evop_2) & \leq & C(\rho,\evop_1,\evop_2)+S(\rho) \\
(Q''E_1'':E_2'') & S(\rho) & \leq & C(\rho,\evop_1,\evop_2)+S(\rho',\evop_2) \\
(Q''E_2'':R'') & S(\rho,\evop_1) & \leq & S(\rho')+S(\rho) \\
(Q''E_2'':E_1'') & S(\rho) & \leq & S(\rho')+S(\rho,\evop_1) \\
(E_1''E_2'':R'') & S(\rho'') & \leq & S(\rho,\evop_2\circ \evop_1)+S(\rho) \\
(E_1''E_2'':Q'') & S(\rho) & \leq & S(\rho,\evop_2\circ\evop_1)+S(\rho'') \\ \hline
\end{array}
$$

}}
\end{center}

Next, we construct a table containing all the inequalities obtainable
directly from the strong subadditivity inequality, $S(X,Y,Z) +S(Y)
\leq S(X,Y)+S(Y,Z)$.  At the start of each row we write $(X:Y:Z)$ to
indicate to which three subsystems of $R''Q''E_1''E_2''$ the strong
subadditivity inequality is being applied:

\vspace{0.6cm}
\begin{center}
\fbox{
\parbox{14cm}
{
\begin{center} {\bf Entropy Inequalities: Strong Subadditivity} \end{center}
$$
\begin{array}{||c|rcl||}  \hline
(R'':Q'':E_1'') & S(\rho',\evop_2)+S(\rho'') & \leq & S(\rho,\evop_2\circ\evop_1)+C(\rho,\evop_1,\evop_2) \\
(Q'':E_1'':R'') & S(\rho',\evop_2)+S(\rho,\evop_1) & \leq & C(\rho,\evop_1,\evop_2)+S(\rho') \\
(E_1'':R'':Q'') & S(\rho',\evop_2)+S(\rho) & \leq & S(\rho'')+S(\rho,\evop_2\circ\evop_1) \\ \hline
(R'':Q'':E_2'') & S(\rho,\evop_1)+S(\rho'') & \leq & S(\rho,\evop_2\circ\evop_1)+S(\rho') \\
(Q'':E_2'':R'') & S(\rho,\evop_1)+S(\rho',\evop_2) & \leq & S(\rho')+C(\rho,\evop_1,\evop_2) \\
(E_2'':R'':Q'') & S(\rho,\evop_1)+S(\rho) & \leq & C(\rho,\evop_1,\evop_2)+S(\rho,\evop_2\circ\evop_1) \\ \hline
(R'':E_1'':E_2'') & S(\rho'')+S(\rho,\evop_1) & \leq & S(\rho')+S(\rho,\evop_2\circ\evop_1) \\
(E_1'':E_2'':R'') & S(\rho'')+S(\rho',\evop_2) & \leq & S(\rho,\evop_2\circ\evop_1)+C(\rho,\evop_1,\evop_2) \\
(E_2'':R'':E_1'') & S(\rho'')+S(\rho) & \leq & C(\rho,\evop_1,\evop_2)+S(\rho') \\ \hline
(Q'':E_1'':E_2'') & S(\rho)+S(\rho,\evop_1) & \leq & C(\rho,\evop_1,\evop_2)+S(\rho,\evop_2\circ\evop_1) \\
(E_1'':E_2'':Q'') & S(\rho)+S(\rho',\evop_2) & \leq & S(\rho,\evop_2\circ\evop_1)+S(\rho') \\
(E_2'':Q'':E_1'') & S(\rho)+S(\rho'') & \leq &
S(\rho')+C(\rho,\evop_1,\evop_2) \\ \hline
\end{array}
$$
}}
\end{center}

It is, perhaps, slightly unfortunate that we do not make use of this
plethora of entropy inequalities.  Certainly, it is interesting to
peruse these tables of entropy inequalities, attempting to discern the
significance of each of these results.  I hope that some of them may
have a role to play in future research into quantum information theory.

\section{Quantum error correction and Maxwell's demon}
\label{sec:Maxwell's_demon}
\index{Maxwell's demon}

Error correction may decrease the entropy of a quantum system, so it
is natural to inquire about the thermodynamic efficiency of this
process.  In this section we discuss the question of the entropy cost
of error correction and show that error correction can be regarded as
a sort of {\em refrigeration}, wherein information about the system
dynamics, obtained through measurement, is used to keep the system
cool.  Indeed, the method of operation of an error correction scheme
is very similar to that of a famous old paradox of thermodynamics, the
{\em Maxwell demon} paradox \cite{Bennett87a} introduced by Maxwell
last century, and the methods we will use to analyze the
thermodynamics of quantum error correction are based upon those used
by Bennett \cite{Bennett82a,Bennett87a} to resolve the paradox.

\subsection{Error-correction by a ``Maxwell demon''}

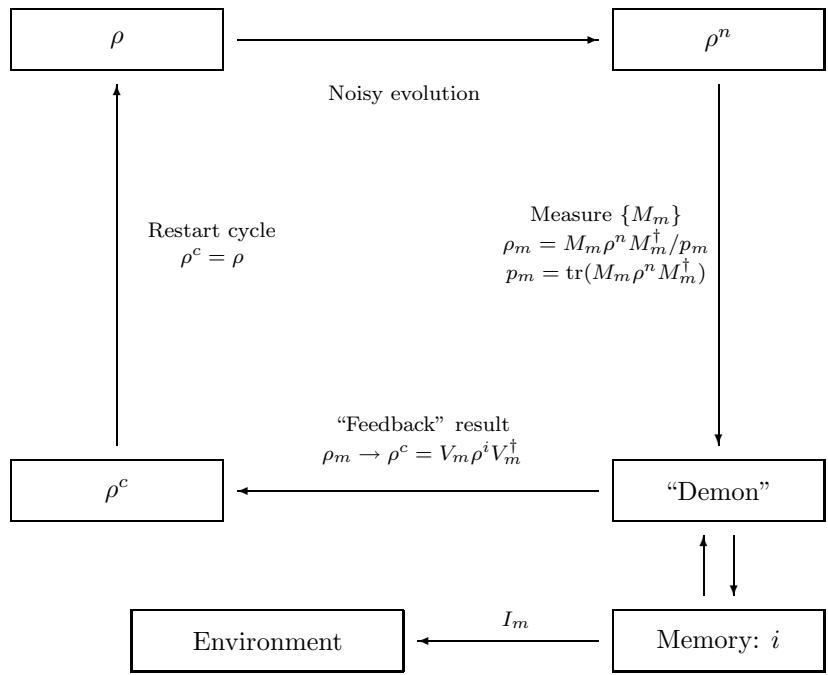
\begin{figure}
\begin{center}
\unitlength 0.04cm
\begin{picture}(280,250)(0,-50)

% draw four boxes
\put(0,160){\framebox(70,20){$\rho$}}
\put(200,160){\framebox(70,20){$\rho^n$}}
\put(200,10){\framebox(70,20){``Demon''}}
\put(0,10){\framebox(70,20){$\rho^c$}}
\put(200,-40){\framebox(70,20){Memory: $i$}}
\put(40,-40){\framebox(90,20){Environment}}
 
% connect boxes with arrows
\put(195,20){\vector(-1,0){120}}
\put(75,170){\vector(1,0){120}}
\put(235,155){\vector(0,-1){120}}
\put(35,35){\vector(0,1){120}}
\put(195,-30){\vector(-1,0){60}}
\put(230,-15){\vector(0,1){20}}
\put(240,5){\vector(0,-1){20}}

% labels
{\footnotesize
\put(100,150){\begin{tabular}{c}
         Noisy evolution % \\ $(\Delta S \geq 0) $
        \end{tabular} }
\put(158,100){\begin{tabular}{c}
   	Measure $\{ M_m \}$ \\
 	$\rho_m = M_m \rho^n M_m^{\dagger}/p_m$ \\
	$p_m = \mbox{tr}(M_m \rho^n M_m^{\dagger})$
        \end{tabular} }
\put(98,35){\begin{tabular}{c}
        ``Feedback'' result \\
        $\rho_m \rightarrow \rho^c= V_m\rho^i V_m^{\dagger}$
        \end{tabular}}
\put(40,100){\begin{tabular}{c}
	Restart cycle \\ $\rho^c=\rho$
	\end{tabular}}
\put(163,-25){$I_m$}
}

%\put(53,-80){Figure~1. Error correction cycle.}
 
\end{picture}
\caption{Error correction cycle. \label{fig: cycle}}
\end{center}
\end{figure}

Consider the error-correction ``cycle'' depicted in figure
\ref{fig: cycle}.  The cycle 
can be decomposed into four stages: 

\begin{enumerate}

\item The system, starting in a state $\rho$, is subjected to a noisy
quantum evolution that takes it to a state $\rho^n$.  We denote the 
change in entropy of the system during this stage by $\Delta S$.  In 
typical scenarios for error correction, we are interested in cases where
$\Delta S \geq 0$, though this is not necessary.

\item  A ``demon'' performs a measurement on the state $\rho^n$.  We
will suppose that the measurement can be described by quantum
operations $\evop_m(\rho) = M_m \rho M_m^{\dagger}$.  As shown in
section \ref{sec:info_conditions}, the error detection stage of
quantum error correction can always be performed in such a way. The
probability that the demon obtains result $m$ is
\begin{equation}
p_m=\mbox{tr}(M_m\rho^n M_m^\dagger)\;,
\end{equation}
and the state of the system conditioned on result $m$ is
\begin{equation}
\rho_m=M_m\rho^n M_m^\dagger/p_m\;.
\end{equation}

\item  The demon ``feeds back'' the result $m$ of the measurement as
a unitary operation $V_m$ that creates a final system state 
\begin{equation}
\rho^c_m=
V_m\rho_m V_m^\dagger=
V_m M_m\rho^nM_m^\dagger V_m^\dagger/p_m\;,  
\end{equation}
In the case of error correction this final state is the ``corrected''
state.  The state of the system, averaged over all possible
measurement outcomes, is given by
\be
\rho_c \equiv \sum_m p_m \rho^c_m. \ee

\item The cycle is restarted.  In order that this actually be a cycle
and that it be a successful error correction, we must have $\rho^c =\rho$.

\end{enumerate}

\noindent
The second and third stages are the ``error-correction'' stages.  The 
idea of error correction is to restore the original state of the system 
during these stages.  In this section we show that the reduction in 
the system entropy during the error-correction stages comes at the
expense of entropy production in the environment, which is at least
as large as the entropy reduction.

\index{Landauer limit}
\index{algorithmic information}

To investigate the balance between the entropy reduction of the system
and entropy production in the environment, we adopt what Caves
\cite{Caves94a} has termed the ``inside view'' of the demon.  The
``outside view'' of the demon regards it as a specific physical
system.  By contrast, the only aspect of the Demon relevant from the
``inside view'' are its properties as an information processing
system; it appears to itself as a set of decohered classical bits
stored in some memory. After stage~3 the only record of the
measurement result~$m$ is the record in the demon's memory.  To reset
its memory for the next cycle, the demon must erase its record of the
measurement result.  Associated with this erasure is a thermodynamic
cost, the {\it Landauer erasure cost} \cite{Landauer61a}, which
corresponds to an entropy increase in the environment.  The erasure
cost of information is equivalent to the thermodynamic cost of
entropy, when entropy and information are measured in the same units,
conveniently chosen to be bits.  Bennett \cite{Bennett82a} used the
idea of an erasure cost to resolve the paradox of Maxwell demons, and
Zurek \cite{Zurek89b} and later Caves \cite{Caves90a} showed that a
correct entropic accounting from the ``inside view'' can be obtained
by quantifying the amount of information in a measurement record by
the algorithmic information content $I_m$ of the record.  Algorithmic
information is the information content of the most compressed form of
the record, quantified as the length of the shortest program that can
be used to generate the record on a universal computer.  We show here
that the average thermodynamic cost of the demon's measurement record
is at least as great as the entropy reduction achieved by error
correction.

In a particular error-correction cycle where the demon obtains 
measurement result~$i$, the total thermodynamic cost of the 
error-correction stages is $I_m+\Delta S^c$, where
\begin{equation}
\Delta S^c\equiv S(\rho^c)-S(\rho^n)
\end{equation}
is the change in the system entropy in the error-correction stages.
Note that the stage change in the quantum system results in a change
in entropy of $\Delta S^c$, not $S(\rho^c_m)-S(\rho^n)$, because the
result of the measurement record is erased after the error-correction
stage, leaving the quantum system in the state $\rho^c \equiv \sum_m p_m
\rho^c_m$. What is of interest to us is the average thermodynamic
cost,
\begin{equation}
\sum_mp_mI_m +\Delta S^c\;,
\end{equation}
where the average is taken over the probabilities for the measurement
results.  To bound this average thermodynamic cost, we now proceed through
a chain of three inequalities.  

The first inequality is a strict consequence of algorithmic information 
theory: the average algorithmic information of the measurement records 
is not less than the Shannon information for the probabilities $p_m$,
that is,
\begin{equation}
\sum_m p_mI_m\ge H( p_m)=-\sum_mp_m\log p_m\;.
\end{equation}
{}Furthermore, Schack \cite{Schack94b} has shown that any universal computer
can be modified to make a new universal computer that has programs for
all the raw measurement records which are at most one bit longer than 
optimal code words for the measurement records.  On such a modified
universal computer, the average algorithmic information for the measurement 
records is within one bit of the Shannon information $H$.

To obtain the second and third inequalities, notice that the corrected
state $\rho^c$ can be written as
\begin{equation}
\rho^c=
\sum_m p_mV_m\rho_mV_m^\dagger=
\sum_m V_mM_m\rho^n M_m^\dagger V_m^\dagger\equiv
{\cal R}(\rho^n)
\;,
\end{equation}
where ${\cal R}$ is the deterministic reversal operation for the 
error-correction stages.  The operators $V_mM_m$ make up an operator-sum
decomposition for the reversal operation.  The probabilities $p_m$ are 
the diagonal elements of the $W$ matrix for this decomposition,
\begin{equation}
p_m=
\mbox{tr}(M_m\rho^nM_m^\dagger)=
\mbox{tr}(V_mM_m\rho^nM_m^\dagger V_m^\dagger)
\;.
\end{equation}
From the results of subsection \ref{subsec:measurements_entropy} we see that
\begin{equation}
H(p_m)\ge S(\rho^n,{\cal R}),
\end{equation}
with equality if and only if the operators $V_mM_m$ are a canonical
decomposition of ${\cal R}$ with respect to $\rho^n$.  We stress that
different measurements and conditional unitaries at stages 2 and 3 lead 
to the same reversal operation, but may yield quite different amounts 
of Shannon information.

The third inequality is obtained by applying the 
inequality~(\ref{eqtn:second_law}) to ${\cal R}$ and $\rho^n$:
\begin{equation} 
\label{eqtn: entropy 2}
S(\rho^n,{\cal R}) + \Delta S^c \geq 0\;. 
\end{equation}
This inequality is automatically satisfied with equality if ${\cal R}$
error corrects $\evop$.  To see this, recall the data processing
inequality gives $S(\rho) \geq S(\rho^n)-S(\rho,\evop) \geq
S(\rho^c)-S(\rho,{\cal R}\circ\evop)$. But the error correcting
property implies that these inequality hold with equality,
$\rho^c=\rho$, and $S(\rho,{\cal R}\circ \evop) = 0$.  Therefore we
have $S(\rho^n)-S(\rho,\evop)=S(\rho^c)$. But $S(\rho,{\cal R}\circ
\evop) = 0$, from which we deduce that $S(\rho,\evop) = S(\rho^n,{\cal
R})$, and therefore $SS(\rho^n,{\cal R})+\Delta S^c = 0$ for
error-correction. 

Combining the three inequalities, we see that the total entropy
produced during the error-correction process is greater than or equal
to zero:
\begin{equation} 
\label{eqtn: second law ok}
\sum_mp_mI_m+\Delta S^c\ge
H(p_m)+\Delta S^c\ge
S(\rho^n,{\cal R})+\Delta S^cc\ge
0\;.
\end{equation}
Stated another way, this result means that the total entropy change 
around the cycle is at least as great as the initial change in entropy 
$\Delta S$, which is caused by the first stage of the dynamics.  The 
error-correction stage can be regarded as a kind of refrigerator, 
similar to a Maxwell demon, achieving a reduction in system entropy 
at the expense of an increase in the entropy of the environment due to 
the erasure of the demon's measurement record.  

How then does this error-correction demon differ from an ordinary Maxwell 
demon?  An obvious difference is that the error-correction demon doesn't
extract the work that is available in the first step of the cycle as 
the system entropy increases under the noisy quantum evolution.  A subtler,
yet more important difference lies in the ways the two demons return the 
system to a standard state, so that the whole process can be a cycle.  For
the error-correction demon, it is the error-correction steps that reset
the system to a standard state, which is then acted on by the noisy
quantum evolution.  For an ordinary Maxwell demon, the noisy quantum 
evolution restores the system to a standard state, typically thermodynamic 
equilibrium, starting from different input states representing the different 
measurement outcomes.

Can this error correction be done in a thermodynamically efficient manner?
Is there a strategy for error correction that achieves equality in the 
Second Law inequality~(\ref{eqtn: second law ok})?  The answer is yes, 
and we give such a strategy here.  The proof of the Second Law 
inequality~(\ref{eqtn: second law ok}) uses three inequalities, 
$\sum_mp_mI_m\ge H$, $H\geq S_e$, and $S_e \geq -\Delta S$.  To achieve 
thermodynamically efficient error correction, it is necessary and sufficient 
that the equality conditions in these three inequalities be achieved.  

We have already noted that Schack has shown that the first inequality,
$\sum_mp_mI_m\ge H(p_m)$, can be saturated to within one bit by using 
a universal computer that is designed to take advantage of optimal coding 
of the raw measurement records~$i$.  On such a universal computer the 
average amount of space needed to store the programs for the measurement
records---that is, the encoded measurement records---is within one bit of 
the Shannon information $H$.  Moreover, it is possible to reduce this one
bit asymptotically to zero by the use of block coding and reversible 
computation.  The demon stores the results of its measurements using an
optimal code for a source with probabilities $p_m$.  Thus the demon stores
an encoded list of measurement results.   Immediately before performing a 
measurement, the demon decodes the list of measurement results using 
reversible computation.  It performs the measurement, appends the result 
to its list, and then re-encodes the enlarged list using optimal block 
coding done by reversible computation.  In the asymptotic limit of large 
blocks, the average length of the compressed list of measurement results 
becomes arbitrarily close to $H(p_m)$ per measurement result.

The second inequality, $H(p_m)\ge S(\rho^n,{\cal R})$, can be
saturated by letting the measurement operators $M_m$ and conditional
unitaries $V_m$ be those defined by the canonical decomposition of the
reversal operation ${\cal R}$.  It should be noted that the optimal
method of encoding the measurement records depends on the
probabilities $p_m$, which in turn are ultimately determined by the
initial state $\rho$.  Thus the type of encoding needed to efficiently
store the measurement record generally depends on the initial state
$\rho$.  The probabilities $p_m$ of the measurement results cannot
depend on the initial state, $\rho$, by the results of section
\ref{sec:info_conditions}.  It follows that for some states with
support in the coding space, this error correction scheme is not
thermodynamically efficient.

The third inequality, $S({\cal R},\rho^n)\ge
-\bigl(S(\rho^c)-S(\rho^n)\bigr)$, is satisfied, as we have already
seen, by any error-correction procedure that corrects errors
perfectly.  
%Indeed, in Sec.~\ref{sect: algebraic discussion} we showed
%that the entropy exchange associated with any reversing operation is
%equal to the entropy reduction achieved by the reversing operation
%(see Eq.~(\ref{eqtn: entropy change equals entropy exchange})).  An
%alternative demonstration that perfect error correction achieves the
%equality $S=-\Delta S^c$ begins by noting that at the end of the
%error-correction process $RQ$ must be in a pure state--the initial
%state---and therefore the overall state must be a product
%$\rho^{RQ''}\otimes\rho^{EA''}$ (recall that $E$ is the environment
%for the noise stage, while $A$ is the ancilla for the reversal stage).
%Thus the condition $\rho^{QA''}=\rho^{Q''}\otimes\rho^{A''}$ certainly
%holds.  This is the equality condition~(\ref{eqtn: second law
%equality}) for the triangle inequality, applied to the reversal
%operation.  Hence we have $S_e=-\Delta S^c$ for the reversal
%operation, and we conclude that any successful error-correction
%procedure automatically achieves equality in Eq.~(\ref{eqtn: entropy
%2}).  
It would be interesting to see whether equality can be achieved in
inequality (\ref{eqtn: entropy 2}) by error-correction schemes that do
not correct errors perfectly.

\subsection{Discussion}

Zurek \cite{Zurek89a}, Milburn \cite{Milburn96a}, and Lloyd
\cite{Lloyd96a} have analyzed examples of quantum Maxwell demons,
though not in the context of error correction.  Lloyd notes that
``creation of new information'' in a quantum measurement is an
additional source of inefficiency in his scheme, which involves
measuring $Z$ for a spin in a static magnetic field applied along the
$z$ axis, in order to extract energy from it.  If the spin is measured
in the ``wrong'' basis -- for example, if it is initially in a pure
state not an eigenstate of $Z$ -- the measurement fails to extract all
the available free energy of the spin, because of the disturbance to
the system state induced by the measurement.  In the case of error
correction, something similar happens, but it is not disturbance to
the system that is the source of the inefficiency.  Instead, if the
ancilla involved in the reversal decoheres in the wrong basis -- that
is, the measurement performed by the demon is not the one defined by
the canonical decomposition of the reversal operation -- then the
Landauer erasure cost is greater than the efficient minimum $S_e$.
This can be thought of as creation of new information, due to
disturbance of the ancilla, but the change in the system state is
independent of the basis in which the ancilla decoheres.

Error correction can be accomplished in ways other than that depicted
in figure \ref{fig: cycle}.  The ``inside view'' of the preceding
subsection, in which the demon makes a measurement described by some
decomposition of the reversal operation, arises when the demon is
decohered by an environment, the particular measurement being defined
by the basis in which the environment decoheres.  If the demon is
isolated from everything except the system and is initially in a pure
state, then its entropy gain is $S_e = -\Delta S$ for the
error-correction process.  One can restart the error-correction cycle
by discarding the demon and bringing up a new demon, the result being
an increase in the environment's entropy by the demon's entropy $S_e$.
This way of performing error correction, which does not involve any
measurement records, is equivalent to the ``outside view'' of the
demon's operation.

The ``inside view'' of the demon's operation, we stress again, arises
if the demon's memory is decohered by interaction with an
environment, the measurement record thus becoming ``classical
information.''  In this case the demon has the entropy $H(p_m)$
of the measurement record, not just the entropy $S_e$.  Once this
decoherence is taken into account, the different decompositions of the
reversal operation, corresponding to different measurements,
constitute operationally different ways of reversing things, rather
than just different interpretations of the same overall interaction.
Keeping in mind the variety of decompositions of the reversal
operation might lead one to consider a greater variety of experimental
realizations, some of which may be easier to perform than others.  As
we emphasize above, a reversal in which the decohered measurement
results correspond to a canonical decomposition of the reversal
operation is the reversal method that is most efficient
thermodynamically.

\section{Conclusion}
\label{sec:qec_conclusion}

In this Chapter we have shown that information-theoretic tools can be
a powerful tool to understand quantum noise, and quantum error
correction.  Information-theoretic necessary and sufficient conditions
for quantum error correction have been formulated, and a thermodynamic
analysis of quantum error correction performed, which shows that
quantum error correction functions as a kind of ``quantum Maxwell's
demon'', for reducing the entropy of a quantum system, through
observation and feedback.

\vspace{1cm}
\begin{center}
\fbox{\parbox{14cm}{
\begin{center} {\bf Summary of Chapter \ref{chap:qec}: Error correction and Maxwell's demon} \end{center}

\begin{itemize}

\item {\bf Entropy exchange:}
Measure of noise a quantum process induces in a state.
$$
S_e \equiv S(\rho,{\cal E}) \equiv S(\rho^{RQ'}). $$

\item {\bf Quantum Fano inequality:}
A large entropy exchange implies a low dynamic fidelity.
$$
S(\rho,{\cal E}) \leq h(F(\rho,\evop)) + (1-F(\rho,\evop))
	\log(d^2-1). $$

\item {\bf Coherent information:}
Quantum analogue of the mutual information.
$$
I(\rho,\evop) \equiv S(\evop(\rho))-S(\rho,\evop). $$

\item {\bf Data processing inequality:}
$$
S(\rho) \geq I(\rho,\evop_1) \geq I(\rho,\evop_2 \circ \evop_1).$$
Equality is satisfied in the first inequality if and only if it is possible
to perfectly error correct $\evop_1$ on the subspace supporting $\rho$.

\item {\bf Error correction as a Maxwell's demon:}
Extracting classical information about a quantum system, we can reduce
its entropy, at the cost of having to {\em erase} the classical measurement
results. This is the thermodynamic cost of quantum error correction; there is
always a way for doing quantum error correction in a thermodynamically
efficient way.

\end{itemize}

}}
\end{center}

\chapter{The quantum channel capacity}
\label{chap:capacity}

\index{channel capacity}
\index{quantum channel capacity}
\index{noisy channel coding theorem}

A central result of Shannon's classical theory of information 
\cite{Shannon48a,Shannon49a,Cover91a} is
the {\em noisy channel coding theorem}. This result provides an {\em
effective procedure} for determining the {\em capacity} of a noisy
channel - the maximum rate at which classical information can be
reliably transmitted through the channel.

This Chapter has two goals. The first goal is to develop general
techniques for proving upper bounds on the capacity of a noisy quantum
channel, which are applied to several different classes of quantum
noisy channel problems. Second, I point out some of the essentially
new features that quantum mechanics introduces into the noisy channel
problem, which make it more difficult than the classical noisy channel
problem.  It is worth emphasizing at this point that this Chapter does
\emph{not} provide an effective procedure for calculating the capacity
of a quantum channel, or even for calculating bounds on the channel
capacity, except in very simple cases.  What it represents is progress
on understanding the quantum channel capacity from the point of view
of the von Neumann entropy and related tools.  The Chapter is based
upon work done in collaboration with Schumacher \cite{Schumacher96b},
and with Barnum and Schumacher \cite{Barnum98a}.  As the work was
being carried out, independent work on the problem was being done by
Lloyd \cite{Lloyd97a}, Bennett {\em et al} \cite{Bennett96a}, and Shor
and Smolin \cite{Shor96a}.  Additional work done since that time will
be pointed out within the Chapter.  The Chapter reports original work;
there is little review material in the Chapter.

The Chapter is organized as follows. In section \ref{sect: channels}
we give a basic introduction to the problem of the noisy quantum
channel, and explain the key concepts.  Section \ref{sect: classical}
shows how the classical noisy channel coding theorem can be put into
the quantum language, and explains why the capacities that arise in
this context are not directly useful for applications such as quantum
computing. Section \ref{sect: coherent information} discusses the {\em
coherent information} introduced in the previous Chapter as an
analogue to the concept of {\em mutual information} in classical
information theory. Many new results about the coherent information
are proved, and we show that quantum entanglement allows the coherent
information to have properties which have no classical analogue. These
properties are critical to understanding what is essentially quantum
about the quantum noisy channel coding problem.  Section \ref{sect:
noisy coding revisited} brings us back to noisy channel coding, and
formally sets up the class of noisy channel coding problems we
consider. Section \ref{sect: upper bounds} proves a variety of upper
bounds on the capacity of a noisy quantum channel, depending on what
class of coding schemes one is willing to allow. This is followed in
section \ref{sect: discussion} by a discussion of the achievability of
these upper bounds and of other work on channel capacity.  Section
\ref{sect: observed channel} formulates the new problem of a noisy
quantum channel with measurement, allowing classical information about
the environment to be obtained by measurement, and then used during
the decoding process. Upper bounds on the corresponding channel
capacity are proved. Finally, section \ref{sect: conc} concludes with
a summary of our results, a discussion of the new features which
quantum mechanics adds to the problem of the noisy channel, and
suggestions for further research.

\section{Noisy channel coding}
\label{sect: channels}

\begin{figure}
\begin{center}
\scalebox{0.8}{\includegraphics{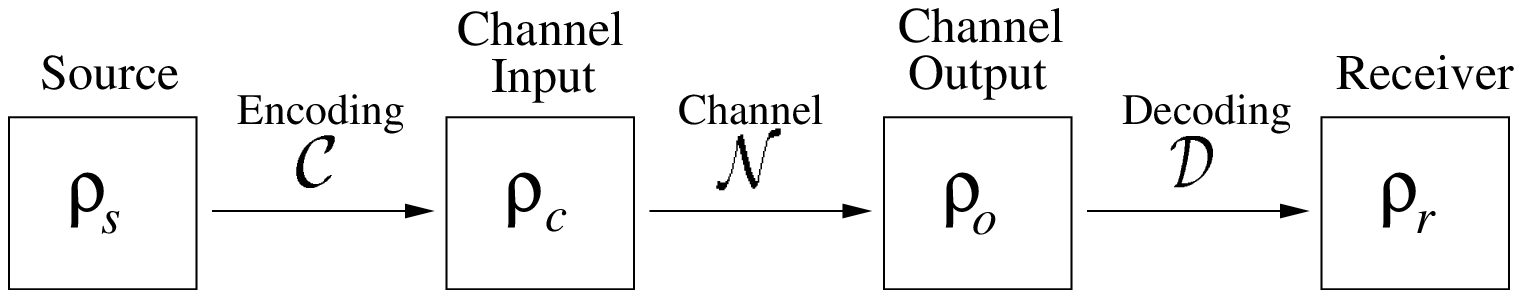}}
\end{center}
\caption{\label{fig: channel0}
The noisy quantum channel, together with encodings and decodings.}
\end{figure}

The problem of noisy channel coding will be outlined in this section.
Precise definitions of the concepts used will be given in later
sections. The procedure is illustrated in figure \ref{fig: channel0}.

There is a {\em quantum source} emitting unknown quantum states, which
we wish to transmit through the channel to some
receiver. Unfortunately, the channel is usually subject to noise,
which prevents it from transmitting states with high fidelity. For
example, an optical fiber suffers losses during transmission. Another
important example of a noisy quantum channel is the memory of a
quantum computer. There the idea is to transmit quantum states {\em in
time}. The effect of transmitting a state from time $t_1$ to $t_2$ can
be described as a noisy quantum channel. Quantum teleportation can
also be described as a noisy quantum channel whenever there are
imperfections in the teleportation process, as shown in section
\ref{sec:teleportation_qops}. 
 
The idea of noisy channel coding is to encode the quantum state
emitted by the source, $\rho_s$, which one wishes to transmit, using
some {\em encoding operation}, which we denote ${\cal C}$. The encoded
state is then sent through the channel, whose operation we denote by
${\cal N}$. The output state of the channel is then {\em decoded}
using some {\em decoding operation}, ${\cal D}$.  The objective is for
the decoded state to match with high fidelity the state emitted by the
source.  As in the classical theory, we consider the fidelity of large
blocks of material produced by repeated emission from the source, and
allow the encoding and decoding to operate on these blocks.  A channel
is said to transmit a source reliably if a sequence of block-coding
and block-decoding procedures can be found that approaches perfect
fidelity in the limit of large block size.

Shannon's classical noisy coding theorem is proved for {\em discrete
memoryless channels}. Discrete means that the channel only has a
finite number of input and output states. By analogy we define a
discrete quantum channel to be one which has a finite number of
Hilbert space dimensions. In the classical case, memoryless means that
the output of the channel is independent of the past, conditioned on
knowing the state of the source. Quantum mechanically we take this to
mean that the output of the channel is completely determined by the
encoded state of the source, and is not affected by the previous
history of the source.

Phrased in the language of quantum operations, we assume that there is
a quantum operation, ${\cal N}$, describing the dynamics of the
channel.  The input $\rho_i$ of the channel is related to the output
$\rho_o$ by the equation
\begin{eqnarray}
\rho_i \rightarrow \rho_o = {\cal N}(\rho_i). \end{eqnarray}
For the majority of this Chapter we assume, as in the previous
equation, that the operation describing the action of the channel is a
complete quantum operation. This corresponds to the physical
assumption that no classical information about the state of the system
or its environment is obtained by an external classical observer.
However, in section \ref{sect: observed channel} we go beyond this to
consider the case of a noisy channel which is being observed by some
classical observer, which will cause us to make use of incomplete
quantum operations.

What then is the {\em capacity} of such a discrete memoryless quantum
channel - the highest rate at which information can be reliably
transmitted through the channel?  The goal of a {\em channel capacity
theorem} is to provide a procedure to answer this question. This
procedure must be an {\em effective procedure}, that is, an explicit
algorithm to evaluate the channel capacity.  Such a theorem comes in
two parts. One part proves an upper bound on the rate at which
information can be reliably transmitted through the channel.  The
other part demonstrates that there are coding and decoding schemes
which attain this bound, which is therefore the channel capacity.  We
do not prove such a channel capacity theorem in this Chapter. We do,
however, derive bounds on the rate at which information can be sent
through a noisy quantum channel.

Before we proceed to the more technical sections of the Chapter, it is
useful to settle on a few notational conventions.  Generically we
denote quantum operations by ${\cal E}$ and the dimension of the
quantum system $Q$ by $d$.  ${\cal N}$ is used to denote noisy quantum
channels, which are also quantum operations.  We work in the $RQE$
picture of quantum operations, as in the previous chapter.  A prime
always denotes a {\em normalized} state. For instance,
\begin{eqnarray}
R'Q' = \frac{({\cal I}_R \otimes {\cal E})(RQ)}
	{\mbox{tr}(({\cal I}_R \otimes {\cal E})(RQ))}.
\end{eqnarray}
Other notational conventions will be introduced as we proceed further.

\section{Classical noisy channels in a quantum setting}
\label{sect: classical}

In this section we show how classical noisy channels can be formulated
in terms of quantum mechanics. We begin by reviewing the formulation in
terms of classical information theory.

A classical noisy channel is described in terms of distinguishable
channel states, which we label by $x$. If the input to the channel is
symbol $x$ then the output is symbol $y$ with probability $p_{y|x}$. 
The channel is assumed to act independently on each input. For
each $x$, the probability sum rule $\sum_y p_{y|x} = 1$ is
satisfied. These {\em conditional probabilities} $p_{y|x}$ completely
describe the classical noisy channel.

Suppose the input to the channel, $x$, is represented by some
classical random variable, $X$, and the output by a random variable
$Y$.  Shannon showed that the capacity of a noisy classical channel is
given by the expression
\begin{eqnarray}
C_S = \max_{p(x)} H(X:Y), \end{eqnarray} where $H(X:Y)$ is the Shannon
mutual information between $X$ and $Y$, as defined in subsection
\ref{subsec:mutual_conditional}, and the maximum is taken over all possible
distributions $p(x)$ for the channel input, $X$. Notice that although
this is not an explicit expression for the channel capacity in terms
of the conditional probabilities $p_{x|y}$, the maximization can
easily be performed using well known techniques from numerical
mathematics. That is, Shannon's result provides an effective procedure
for computing the capacity of a noisy classical channel.

All these results may be re-expressed in terms of quantum mechanics.
We suppose the channel has some preferred orthonormal basis,
$|x\rangle$, of signal states. For convenience we assume the set of input
states, $|x\rangle$, is the same as the set of output states, $|y\rangle$,
of the channel, although
more general schemes are possible. For the purpose of illustration the
present level of generality suffices. A classical input random variable,
$X$, corresponds to an input density operator for the quantum
channel,
\begin{eqnarray}
\rho_X \equiv \sum_x p(x) |x\rangle \langle x|. \end{eqnarray}
The statistics of $X$ are recoverable by measuring $\rho_X$ in the
$|x\rangle$ basis.
Defining operators $E_{xy}$ by
\begin{eqnarray}
E_{xy} \equiv |y\rangle \langle x|, \end{eqnarray}
we find that the channel operation defined by
\begin{eqnarray}
{\cal N}(\rho) \equiv \sum_{xy} p_{y|x} E_{xy} \rho E_{xy}^{\dagger}.
\end{eqnarray}
is a trace-preserving quantum operation,
and that
\begin{eqnarray}
{\cal N}(\rho_X) = \rho_Y = \sum_y p(y) |y\rangle \langle y|,
\end{eqnarray}
where $\rho_Y$ is the density operator corresponding to the random variable
$Y$ that would have been obtained from $X$ given a classical channel
with probabilities $p_{y|x}$. This gives a quantum mechanical
formalism for describing classical sources and channels. It is interesting to
see what form the mutual information and channel capacity take in the
quantum formalism.

Notice that
\begin{eqnarray}
H(X) & = & S(\rho_X) \\
H(Y) & = & S(\rho_Y) = S({\cal N}(\rho_X)).
\end{eqnarray}
Next we compute the entropy exchange associated with the channel
operating on input $\rho_X$, by computing the $W$ matrix given by
equation (\ref{eqtn:W_defn}).  The $W$ matrix corresponding to the
channel with input $\rho_X$ has entries
\begin{eqnarray}
W_{(xy) (x'y')} = \delta_{x,x'} \delta_{y,y'} p(x) p(y|x), \end{eqnarray}
But the joint distribution of $(X,Y)$ satisfies
$p(x) p(y|x) = p(x,y)$. Thus $W$ is diagonal with eigenvalues $p(x,y)$,
so the entropy exchange is given by
\begin{eqnarray}
S(\rho_X,{\cal N}) = H(X,Y). \end{eqnarray}
It follows that
\begin{eqnarray}
H(X:Y) = S(\rho_X)+S({\cal N}(\rho_X))-S(\rho_X,{\cal N}),
\end{eqnarray}
and thus the Shannon capacity $C_S$ of the classical channel is given in
the quantum formalism by
\begin{eqnarray}
C_S = \max_{\rho_X} \left[ S(\rho_X) + S({\cal
N}(\rho_X))-S(\rho_X,{\cal N}) \right] ,
\end{eqnarray}
where the maximization is over all input states for the channel, $\rho_X$,
which are diagonal in the $|x\rangle$ basis.

The problem we have been considering is that of transmitting a
discrete set of orthogonal states (the states $|x\rangle$) through the
channel. In many quantum applications one is not only interested in
transmitting a discrete set of states, but rather the entanglement of
a quantum source with another system.  For this purpose we will use
the \emph{dynamic fidelity} of Chapter \ref{chap:distance} as a figure
of merit for how reliable transmission is.  The capacity in this
scenario is defined to be the highest rate at which quantum source can
be transmitted through a noisy quantum channel, in the sense of having
asymptotically high dynamic fidelity. It is easy to see, and we will
show explicitly later on, that this cannot be done by considering the
transmission of a set of orthogonal pure states alone.  That is, the
transmission of entanglement is a much more stringent condition than
the transmission of classical information which we have been
considering here, and consequently, the channel capacity for
transmission of quantum entanglement -- the main subject of this
Chapter -- may in general be somewhat lower than the channel capacity
for transmission of classical information.

\section{Coherent information}
\label{sect: coherent information}
\index{coherent information}

In this section we investigate in more detail the {\em coherent
information}, defined in section \ref{sec:q_data_processing}, where it
was suggested that the coherent information plays a role in quantum
information theory analogous to the role played by mutual information
in classical information theory; that is, suppose we consider a process
defined by an input $\rho$, and output $\rho'$, with the process
described by a quantum operation, ${\cal E}$,
\begin{eqnarray} \label{eqtn: quantum process}
\rho \stackrel{{\cal E}}{\rightarrow} \rho' = {\cal E}(\rho). \end{eqnarray}
I assert that the coherent information, defined by
\begin{eqnarray}
I(\rho,{\cal E}) \equiv S \left(
	\frac{{\cal E}(\rho)}{\mbox{tr}({\cal E}(\rho))} \right) -
	S(\rho,{\cal E}),
\end{eqnarray}
plays a role in quantum information theory analogous to that played by
the mutual information $H(X:Y)$ in classical information theory, where
$X$ is the input to a classical channel, and $Y$ is the output from
that channel. Heuristic arguments for why this is so were given in the
previous Chapter.  Of course, the true justification for regarding the
coherent information as the quantum analogue of the mutual information
is its success as the quantity appearing in results on channel
capacity, as discussed in later sections. This is the appropriate
motivation for all definitions in information theory, whether
classical or quantum: their success at quantifying the resources
needed to perform some interesting physical task, not some abstract
mathematical motivation.

Subsection \ref{subsec:I_properties} studies in detail the
properties of the coherent information. In particular, we prove
several results related to convexity that are useful both as
calculational aids, and also for proving later results.  Subsection
\ref{subsec:fidelity_lemma} proves the \emph{entropy-fidelity} lemma
that glues together many of our later proofs of upper bounds on the
channel capacity. Finally, subsections
\ref{subsec:example_1} and \ref{subsec:example_2} describe
two important ways the behaviour of the coherent information differs
from the behaviour of the mutual information when quantum entanglement
is allowed.

\subsection{Properties of coherent information}
\label{subsec:I_properties}

The set of quantum operations forms a positive cone, that is, if
${\cal E}_i$ is a collection of quantum operations and $\lambda_i$ is
a set of non-negative numbers then $\sum_i \lambda_i {\cal E}_i$ is
also a quantum operation. In this section we prove two very useful
properties of the coherent information. First, it is easy to see that
for any quantum operation ${\cal E}$ and non-negative $\lambda$,
\begin{eqnarray}
I(\rho,\lambda {\cal E}) = I(\rho,{\cal E}). \end{eqnarray}
This follows immediately from the definition of the coherent information.
A slightly more difficult property to prove is the following.

\begin{theorem} \textbf{(convexity theorem for coherent information)}

Suppose ${\cal E}_i$ are quantum operations. Then
\begin{eqnarray} \label{eqtn: abstract subadditivity}
I(\rho,\sum_i {\cal E}_i) \leq \frac{\sum_i \mbox{tr}({\cal E}_i(\rho))
	I(\rho,{\cal E}_i)}{\mbox{tr}(\sum_i {\cal E}_i(\rho))}.
\end{eqnarray}

\end{theorem}

This result will be extremely useful in  our later work. An important
and immediate corollary is the following:

\begin{corollary}

If a complete quantum operation, ${\cal E} = \sum_i p_i {\cal E}_i$ is
a convex sum ($p_i \geq 0, \sum_i p_i = 1$) of complete quantum
operations ${\cal E}_i$, then the coherent information is convex,
\begin{eqnarray}
I(\rho,\sum_i p_i {\cal E}_i) \leq \sum_i p_i I(\rho,{\cal E}_i).
\end{eqnarray}

\end{corollary}

The proof of the corollary is immediate from the theorem.

\begin{proof} \textbf{(convexity theorem for coherent information)}

The theorem follows from the concavity of the {\em conditional}
entropy, Corollary
\ref{corollary:concavity_conditional}, on page
\pageref{corollary:concavity_conditional}.  By definition
\begin{eqnarray}
I(\rho,{\cal E}) = S(Q')-S(R'Q') = -S(R' | Q').
\end{eqnarray}
The theorem follows immediately from the concavity of the conditional
entropy.

\end{proof}

The following lemma, from \cite{Marcus92a}, is extremely useful in
computing the maxima of convex functions on convex sets.  Later in
this Chapter we will be interested in the computation of such maxima.

\begin{lemma}

Suppose $f$ is a continuous convex function on a compact,
convex set, $S$. Then there is an extremal point at which $f$
attains its global maximum.

\end{lemma}

The proof is obvious. The reason for our interest in the proof is
because for fixed $\rho$ and complete quantum operations ${\cal E}$,
the coherent information $I(\rho,{\cal E})$ is a convex, continuous
function of the operation ${\cal E}$, as just shown. The set of
trace-preserving quantum operations forms a compact, convex set, and
thus by the convexity lemma, $I(\rho,{\cal E})$ attains its maximum
for a quantum operation ${\cal E}$ which is extremal in the set of all
trace-preserving quantum operations.

%Choi \cite{Choi75a} has proved that any extremal point in the set of
%trace-preserving quantum operations has a set of operation elements
%$\{ A_i \}$ such that
%
%\begin{enumerate}
%
%\item There are at most $d$ elements $A_i$. This is to be contrasted
%with the general situation, where there may be up to $d^2$ elements.
%
%\item The $A_i$ are linearly independent.
%
%\end{enumerate}
%
%This result provides a considerable saving in the class of quantum
%operations that must be optimized over in order to numerically
%calculate expressions of the form (\ref{eqtn: general
%bound}). 
%Unfortunately, this only takes us part of the way towards
%proving that the expressions (\ref{eqtn: general bound}) and
%(\ref{eqtn: unitary bound}) are identically equal, or, alternatively,
%it suggests a starting point for a search for counterexamples to the
%proposition that the two quantities are equal. If the extremal points
%of the set of quantum operations were the unitary operations we would
%be done. However that is not the case, as the above theorem shows.

A further useful result concerns the additivity of coherent
information,
\begin{theorem} \textbf{(additivity for independent channels)}

Suppose ${\cal E}_1,\ldots,{\cal E}_n$ are quantum operations and
$\rho_1,\ldots,\rho_n$ are density operators. Then
\begin{eqnarray}
I(\rho_1 \otimes \ldots \otimes \rho_n,{\cal E}_1 \otimes \ldots {\cal E}_n)
	= \sum_i I(\rho_i,{\cal E}_i). \end{eqnarray}

\end{theorem}

The proof is immediate from the additivity property of entropies
for product states.

\subsection{The entropy-fidelity lemma}
\label{subsec:fidelity_lemma}
\index{entropy-fidelity lemma}

The following lemma is the glue which holds together much of our later
work on proving upper bounds to channel capacities. In this section we
will prove the lemma only for the special case of complete quantum
operations. A similar but more complicated result is true for general
quantum operations, and will be given in section \ref{sect: observed
channel}.

\begin{lemma} \textbf{(entropy-fidelity lemma)}
\label{lemma:entropy-fidelity}

Suppose ${\cal E}$ is a complete quantum operation, and $\rho$ is some
quantum state. Then for all complete quantum operations ${\cal D}$,
\begin{eqnarray} \label{eqtn: fidelity lemma}
S(\rho) \leq I(\rho,{\cal E}) + 2
	+ 4 (1-F(\rho,{\cal D} \circ {\cal E})) \log d.
\end{eqnarray}

\end{lemma}

This lemma is extremely useful in obtaining proofs of bounds on the
channel capacity. In order for the dynamic fidelity to be close to
one, the quantity appearing on the right hand side must be close to
zero. This shows that the entropy of $\rho$ cannot greatly exceed the
coherent information $I(\rho,{\cal E})$ if the dynamic fidelity of the
total process -- $\evop$ followed by ${\cal D}$ -- is to be close to
one.

\begin{proof}

To prove the lemma, notice that by the second part of
the data processing inequality, (\ref{eqtn:quantum_data_processing}),
\begin{eqnarray}
S(\rho)-I(\rho,{\cal E}) & \leq & S(\rho) - S(({\cal D} \circ
	{\cal E})(\rho)) +
	S(\rho,{\cal D} \circ {\cal E}). \nonumber \\
& & \end{eqnarray}
Applying inequality (\ref{eqtn:second_law}) gives
\begin{eqnarray}
S(\rho)-S(({\cal D} \circ {\cal E})(\rho))
    \leq S(\rho,{\cal D} \circ {\cal E}), \end{eqnarray}
and combining the previous two inequalities gives
\begin{eqnarray}
S(\rho)-I(\rho,{\cal E}) & \leq & 2 S(\rho,{\cal D} \circ {\cal E}) \\
 & \leq & \label{eqtn: stronger bounds}
	2 h(F(\rho,{\cal D} \circ {\cal E})) + 2(1-F(\rho,{\cal D} \circ {\cal E})) \log (d^2-1),
\end{eqnarray}
where the second step follows from the quantum Fano inequality,
(\ref{eqtn: quantum Fano}). But the binary Shannon entropy $h$ is bounded
above by $1$ and $\log (d^2-1) \leq 2 \log d$, so
\begin{eqnarray}
S(\rho) \leq I(\rho,{\cal E})+2+4 (1-F(\rho,{\cal D} \circ {\cal E}))
	\log d. \end{eqnarray}
This completes the proof.

\end{proof}

The inequality in the statement of the entropy-fidelity lemma is
strong enough to prove the asymptotic bounds of most interest in our
later work. The somewhat stronger inequality (\ref{eqtn: stronger
bounds}) is also useful when proving one-shot results, that is, when
no block coding is being used.  We will not make any use of it in this
Chapter.

\subsection{Quantum characteristics of the coherent information I}
\label{subsec:example_1}
\index{data pipelining inequality}

There are at least two important respects in which the coherent
information behaves differently from the classical mutual
information. In this subsection and the next we will explain what
these differences are.

Classically, suppose we have a Markov process,
\begin{eqnarray}
X \rightarrow Y \rightarrow Z. \end{eqnarray}
Intuitively we expect that
\begin{eqnarray}
H(X:Z) \leq H(Y:Z), \end{eqnarray} and, indeed, in subsection
\ref{subsec:data_proc} we proved this ``data pipelining inequality'', based
on the definition of the mutual information. The idea is that any
information about $X$ that reaches $Z$ must go through $Y,$ and thus
will also be information that $Z$ has about $Y$.  However, the quantum
mechanical analogue of this result fails to hold.   We shall see that
the reason it fails is due to quantum entanglement.

{\bf Example 1:}

Suppose we have a two-part quantum process described by quantum 
operations
${\cal E}_1$ and ${\cal E}_2$.
\begin{eqnarray}
\rho \rightarrow {\cal E}_1 (\rho) \rightarrow 
({\cal E}_2 \circ {\cal E}_1)
	(\rho). \end{eqnarray}
Then, in general
\begin{eqnarray} \label{eqtn: example 1}
I(\rho,{\cal E}_2 \circ {\cal E}_1) \not\leq 
I({\cal E}_1(\rho),{\cal E}_2).
\end{eqnarray}
An explicit example showing that this is the case will be given below.
It is not possible to prove \emph{any} general inequality of this sort
for the coherent information - examples may be found where a $<,>$ or
$=$ sign could occur in the last equation.  We will now show how the
purely quantum mechanical effect of entanglement is responsible for
this property of coherent information.

Observe first that the truth of the equation
\begin{eqnarray} \label{eqtn: qm subadditivity}
I(\rho,{\cal E}_2 \circ {\cal E}_1) \leq
I({\cal E}_1(\rho),{\cal E}_2),
\end{eqnarray}
is equivalent to 
\begin{eqnarray}
S({\cal E}_1(\rho),{\cal E}_2) \leq S(\rho,{\cal E}_2 \circ {\cal E}_1).
\end{eqnarray}
This last equation makes it easy to see why 
(\ref{eqtn: qm subadditivity}) may fail. It is because the entropy of the
joint environment for processes ${\cal E}_1$ and ${\cal E}_2$ (the
quantity on the right-hand side) may be less than the entropy of
the environment for process ${\cal E}_2$ alone (the quantity on the left).
This is a property peculiar to quantum mechanics, which is caused
by entanglement; there is no classical analogue.  In particular, the
entropy-entanglement inequality on page
\pageref{eqtn:entropy-entanglement} showed that the entanglement
between $E_1''$ and $E_2''$ satisfies \be {\cal F}(E_1'':E_2'') \geq
S(E_2'')-S(E_1'',E_2'') = S(\evop_1(\rho),\evop_2) -
S(\rho,\evop_2\circ\evop), \ee demonstrating that entanglement between
$E_1''$ and $E_2''$ must exist in order that (\ref{eqtn: qm
subadditivity}) be violated.

An explicit example where this is the case will now be given. For
convenience we will do so in the language of coding and channel
operations, since this is the language that will be most convenient
later. ${\cal E}_1$ is to be identified with the coding operation,
${\cal C}$, and ${\cal E}_2$ is to be identified with the channel
operation, ${\cal N}$.

Suppose we have a four dimensional state space. We will suppose we
have an orthonormal basis $|1\rangle,|2\rangle,|3\rangle,|4\rangle$,
and that $P_{12}$ is the projector onto the space spanned by
$|1\rangle$ and $|2\rangle$, and $P_{34}$ is the projector onto the
space spanned by $|3\rangle$ and $|4\rangle$. Let $U$ be a unitary
operator defined by
\begin{eqnarray}
U \equiv |3\rangle \langle 1 | + |4\rangle \langle 2| + 
|1\rangle \langle 3|
	+ |2\rangle \langle 4| . \end{eqnarray}
The channel operation is defined by
\begin{eqnarray}
{\cal N}(\rho) \equiv P_{12} \rho P_{12} + U^{\dagger} 
P_{34} \rho P_{34} U,
\end{eqnarray}
and we use an encoding defined by
\begin{eqnarray}
{\cal C}(\rho) \equiv \frac{1}{2} P_{12}\rho P_{12} + 
\frac{1}{2} U P_{12} \rho
	P_{12} U^{\dagger} + P_{34} \rho P_{34}.
\end{eqnarray}
It is easily checked that for any state $\rho$ whose support lies
wholly in the space spanned by $|1\rangle$ and $|2\rangle$,
\begin{eqnarray}
({\cal N} \circ {\cal C})(\rho) = \rho.
\end{eqnarray}
It follows that
\begin{eqnarray}
I(\rho,{\cal N} \circ {\cal C}) = S(\rho).
\end{eqnarray}
It is also easy to verify that
\begin{eqnarray}
I({\cal C}(\rho),{\cal N}) = 2 S(\rho) - 1.
\end{eqnarray}
Thus there exist states $\rho$ such that
\begin{eqnarray}
I(\rho,{\cal N} \circ {\cal C}) > I({\cal C}(\rho),{\cal N}), \end{eqnarray}
providing an example of (\ref{eqtn: example 1}).

\subsection{Quantum characteristics of the coherent information II}
\label{subsec:example_2}
\index{subadditivity}

\begin{figure}
\begin{center}
\scalebox{0.8}{\includegraphics{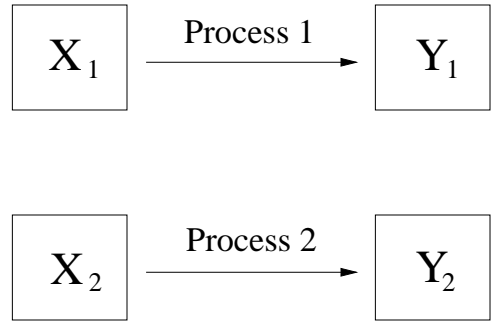}}
\end{center}
\caption{\label{fig: subadditivity}
Dual classical channels operating on inputs $X_1$ and $X_2$
produce outputs $Y_1$ and $Y_2$.}
\end{figure}

The second important difference between coherent information and
classical mutual information is related to the property known classically
as {\em subadditivity of mutual information}. Suppose we have several
independent channels operating. Figure \ref{fig: subadditivity} shows
the case of two channels.

These channels are numbered $1,\ldots,n$ and take as inputs random
variables $X_1,\ldots,X_n$. The channels might be separated spatially,
as shown in the figure, or in time. The channels are assumed to act
independently on their respective inputs, and produce outputs
$Y_1,\ldots,Y_n$. It is not difficult to show that \cite{Cover91a}
(c.f. page \pageref{note:subadd})
\begin{eqnarray} \label{eqtn: classical subadditivity}
H(X_1,\ldots,X_n : Y_1,\ldots,Y_n) \leq \sum_i H(X_i : Y_i).
\end{eqnarray}
This property is known as the {\em subadditivity} of mutual information.
It is used, for example, in proofs of the weak converse to Shannon's
noisy channel coding theorem. We will show that the corresponding
quantum statement about coherent information fails to hold.

{\bf Example 2:}
There exists a quantum operation ${\cal E}$ and
a density operator $\rho_{12}$ such that
\begin{eqnarray} \label{eqtn: example 2}
I(\rho_{12},{\cal E} \otimes {\cal E}) \not \leq 
	I(\rho_1,{\cal E}) + I(\rho_2,{\cal E}), \end{eqnarray}
where $\rho_1 \equiv \mbox{tr}_2(\rho_{12})$ and
$\rho_2 \equiv \mbox{tr}_1(\rho_{12})$ are the usual 
reduced density operators
for systems $1$ and $2$.

An example of (\ref{eqtn: example 2}) is 
the following. Suppose system
$1$ consists of two qubits, $A$ and $B$. System 
$2$ consists of two more
qubits, $C$ and $D$. As the initial state we choose
\begin{eqnarray}
\rho_{12} = \frac{I_A}{2} \otimes |BD\rangle 
\langle BD| \otimes
	\frac{I_C}{2}, \end{eqnarray}
where $|BD\rangle$ is a Bell state shared 
between systems $B$ and
$D$.

The action of the channel on $A$ and $B$ is as follows: it sets bit $B$
to some standard state, $|0\rangle$, and allows $A$ through unchanged.
This is achieved by swapping the state of $B$ out into the environment.
Formally,
\begin{eqnarray}
{\cal E}(\rho_{AB}) = \rho_A \otimes |0 \rangle \langle 0 |.
\end{eqnarray}
The same channel is now set to act on systems $C$ and $D$:
\begin{eqnarray}
{\cal E}(\rho_{CD}) = \rho_C \otimes |0 \rangle \langle 0 |.
\end{eqnarray}
A straightforward though slightly tedious 
calculation shows that with this
channel setup
\begin{eqnarray}
I(\rho_1,{\cal E}) = I(\rho_2,{\cal E}) = 0, \end{eqnarray}
and
\begin{eqnarray}
I(\rho_{12},{\cal E} \otimes {\cal E}) = 2. \end{eqnarray}
Thus this setup provides an example of the violation of subadditivity
for the coherent information, (\ref{eqtn: example 2}).

\section{Noisy channel coding revisited}
\label{sect: noisy coding revisited}

\begin{figure}[ht]
\begin{center}
\scalebox{0.8}{\includegraphics{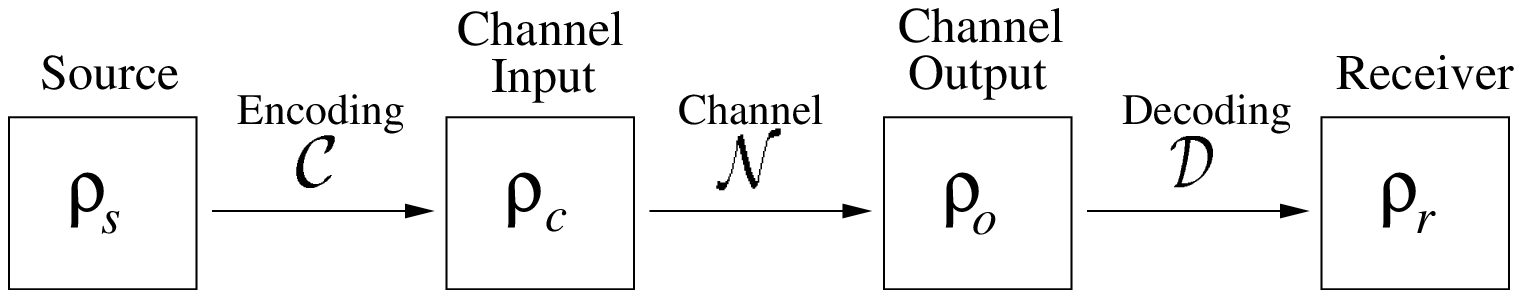}}
\end{center}
\caption{\label{fig: channel1}
The noisy quantum channel, together with encodings and decodings.}
\end{figure}

In this section we return to noisy channel coding. Recall the basic
procedure for noisy channel coding, as illustrated in figure
\ref{fig: channel1}.

Suppose a quantum source has output $\rho_s$. A quantum operation,
which we shall denote ${\cal C}$, is used to {\em encode} the source
source, giving the input state to the channel, $\rho_i \equiv {\cal
C}(\rho)$. The encoded state is used as input to the noisy channel,
giving a channel output $\rho_o \equiv {\cal N}(\rho_i)$. Finally, a
decoding quantum operation, ${\cal D}$, is used to decode the output
of the channel, giving a {\em received state}, $\rho_r \equiv {\cal
D}(\rho_o)$. The goal of noisy channel coding is to find out what
source states can be sent with high dynamic fidelity. That is, we want
to know for what states $\rho_s$ can encoding and decoding operations
be found such that
\begin{eqnarray}
F(\rho_s,{\cal D} \circ {\cal N} \circ {\cal C}) \approx 1. 
\end{eqnarray}
Typically, it is the entropy of a state which determines whether it
can be sent with high dynamic fidelity.  If large blocks of source
states with entropy $R$ per use of the source can be sent through
equally large blocks of channel with high dynamic fidelity, we say the
channel is \emph{transmitting at the rate} $R.$

Shannon's noisy channel coding theorem is an example of a
{\em channel capacity} theorem. Such theorems come in two parts:
\begin{enumerate}

\item First an {\em upper bound} is placed on the rate at which information
can be sent reliably through the channel. This upper bound should be
expressible entirely in terms of channel quantities.

\item Second it is proved that a reliable
scheme for encoding and decoding exists which
comes arbitrarily close to {\em attaining} the upper bound found in 1.

\end{enumerate}
This maximum rate at which information can be reliably sent through the
channel is known as the {\em channel capacity}.

In this Dissertation we consider only the first of these tasks, the
placing of upper bounds on the rate at which quantum information can
be reliably sent through a noisy quantum channel, with high dynamic
fidelity the criterion for successful transmission of quantum
information. That is, we place bounds on the entropy of the source
states, $\rho_s$, that can be reliably sent through such a channel.

The results we will prove are analogous to the weak converse of the
classical noisy coding theorem, but cannot be considered true
converses, since we do not prove that our bounds can be achieved by
a coding scheme.  Thus our results cannot be considered to be a
channel capacity theorem, although if attainability of the upper
bounds we prove could be shown, then a true channel capacity theorem
would result.  I do consider the bounds to be likely candidates for
the quantum channel capacity.

\subsection{Mathematical formulation of noisy channel coding}
\label{subsec: other stuff}

Up to this point the procedure for doing noisy channel coding has been
discussed in broad outline, but we have not made all of our
definitions mathematically precise. This subsection gives a
mathematically precise formulation for the most important concepts
appearing in our work on noisy channel coding.

Define a {\em quantum source}\index{quantum source}, $\Sigma \equiv
(H_s,\Upsilon)$ to consist of a Hilbert space $H_s$ and a sequence
$\Upsilon = [\rho_s^1,\rho_s^2,...,\rho_s^n,...]$ where $\rho_s^1$ is
a density operator on $H_s$, $\rho_s^2$ a density operator on $H_s
\otimes H_s$, and $\rho_s^n$ a density operator on $H_s^{\otimes n},$
etc...  Using, for example, ``${\rm tr}_{34}$'' to denote the partial
trace over the third and fourth copies of $H_s,$ we require as part of
our definition of a quantum source that for all $j$ and all $n > j,$
\begin{eqnarray}
{\rm tr}_{j+1,...,n}(\rho_s^n) = \rho_s^j,
\end{eqnarray}
that is, that density operators in the sequence be consistent with
each other in the sense that earlier ones be derivable from later ones
by an appropriate partial trace.  The $n$-th density operator is meant
to represent the state of $n$ emissions from the source, normally
thought of as taking $n$ units of time.  (We could have used a single
density operator on a countably infinite tensor product of spaces
$H_s,$ but we wish to avoid the technical issues associated with such
products.)  We will define the {\em entropy} of a general source
$\Sigma$ as
\begin{eqnarray}
S(\Sigma) \equiv \lim_{n \rightarrow \infty} \frac{ S(\rho_s^n)}{n},
\end{eqnarray} when this limit exists. 

A special case of this general definition of quantum source
is the i.i.d. source $(H_s,[\rho_s, \rho_s \otimes \rho_s,...,
\rho_s^{\otimes n},...]),$ for some fixed $\rho_s.$ Such a source
corresponds to the classical notion of an {\em independent, identically
distributed} classical source, thus the term i.i.d.  The entropy 
of this source is simply $S(\rho_s).$

A discrete memoryless channel\index{discrete memoryless quantum
channel}, $(H_c,{\cal N})$ consists of a finite-dimensional Hilbert
space, $H_c$, and a trace-preserving quantum operation ${\cal N}$. The
{\em $n$th extension} of that channel is given by the pair
$(H_s^{\otimes n},{\cal N}^{\otimes n})$. The memoryless nature of the
channel is reflected in the fact that the operation performed on the
$n$ copies of the channel system is a tensor product of independent
single-system operations.

Define an {\em $n$-code} from $H_s$ into $H_c$ to consist of a
trace-preserving quantum operation, ${\cal C}$, from $H_s^{\otimes n}$
to $H_c^{\otimes n}$, and a trace-preserving quantum operation ${\cal
D}$ from $H_c^{\otimes n}$ to $H_s^{\otimes n}$. We will refer to
${\cal C}$ as the {\em encoding} and ${\cal D}$ as the {\em decoding}.

The {\em total coding operation} ${\cal T}$ is given by
\begin{eqnarray}
{\cal T} \equiv {\cal D} \circ {\cal N}^{\otimes n} \circ {\cal C}.
\end{eqnarray}
The measure of success we will use for the total procedure
is the {\em total dynamic fidelity},
\begin{eqnarray}
F(\rho_s^n,{\cal T}). \end{eqnarray}

In practice we will frequently abuse notation, usually by omitting
explicit mention of the Hilbert spaces $H_s$ and $H_c$. Note also that
in principle the channel could have different input and output Hilbert
spaces. To ease notational clutter we will not consider that case
here, but all the results we prove go through without change.

Given a source state $\rho_s$ and a channel ${\cal N}$, the goal of
noisy channel coding is to find an encoding ${\cal C}$ and a decoding
${\cal D}$ such that $F(\rho_s,{\cal T})$ is close to one; that is,
$\rho_s$ and its entanglement is transmitted almost perfectly. In
general this is not possible to do. However, Shannon showed in the
classical context that by considering blocks of output from the
source, and performing block encoding and decoding it is possible to
considerably expand the class of source states $\rho_s$ for which this
is possible. The quantum mechanical version of this procedure is to
find a sequence of $n$-codes, $({\cal C}_n,{\cal D}_n)$ such that as
$n \rightarrow \infty$, the measure of success $F(\rho_s^n,{\cal
T}_n)$ approaches one, where ${\cal T}_n = {\cal D}_n \circ {\cal
N}^{\otimes n} \circ {\cal C}_n$.  (We will sometimes refer to such a
sequence as a {\em coding scheme}.)

Suppose such a sequence of codes exists for a given source $\Sigma.$
In this case the channel is said to transmit $\rho_s$ reliably.  We
also say that the channel can transmit reliably at a {\em rate} $R =
S(\Sigma).$ (Note that this definition does not require that the
channel be able to transmit reliably {\em any} source with entropy
less than or equal to $R$; that is a different potential definition of
what it means for a channel to transmit reliably at rate $R$; in the
contexts considered in this Chapter, it has been shown elsewhere
\cite{Barnum98b} that the two to turn out to be equivalent, that is if
a channel can transmit some source with entropy $R$, it can transmit
any source with that entropy.)

A noisy channel coding theorem would enable one to determine, for any
source and channel, whether or not the source can be transmitted
reliably on that channel at a given rate.  Classically, this is
determined by comparing the Shannon entropy of the source to the
capacity of the channel.  If the entropy of the input distribution is
greater than the capacity, the source cannot be transmitted reliably.
If the entropy is less than the capacity, it can.  The conjunction of
these two statements is the noisy channel coding theorem.  (The case
of $H$ precisely equal to $C$ requires separate consideration;
sometimes reliable transmission is achievable, and sometimes not.)  We
expect that in quantum mechanics, the entropy $S(\Sigma)$ of the
source will play a role analogous to the Shannon entropy, and the
coherent information will play a role analogous to the mutual
information.  A channel will be able to transmit reliably any source
with von Neumann entropy less than the capacity; furthermore, {\it no}
source with entropy greater than the capacity will be reliably
transmissible. The first part of this would constitute a quantum noisy
channel coding theorem; the second, a ``weak converse'' of the
theorem.  A ``strong converse'' would require not just that no source
with entropy greater than the capacity can be reliably transmitted,
that is transmitted with asymptotic fidelity approaching unity, but
would require that all such sources have asymptotic fidelity of
transmission approaching zero.

\section{Upper bounds on the channel capacity}
\label{sect: upper bounds}

In this section we investigate a variety of upper bounds on the
capacity of a noisy quantum channel.

\subsection{Unitary encodings}
\label{subsec: unitary encodings}

This subsection will be concerned with the case where the encoding,
${\cal C}$, is unitary.

For this subsection only we define
\begin{eqnarray}
C_n \equiv \max_{\rho} I(\rho,{\cal N}^{\otimes n}), \end{eqnarray}
where the maximization is over all inputs $\rho$ to $n$ copies of the channel.
The bound on the channel capacity proved in this section is defined by
\begin{eqnarray} \label{eqtn: unitary bound}
C \equiv \lim_{n \rightarrow \infty} \frac{C_n}{n}.
\end{eqnarray}
It is not immediately obvious that this limit exists. To see that it
does, notice that $C_n \leq n \log d$ and $C_m + C_n \leq C_{m+n}$ and
apply the following lemma. Notice that $C = C({\cal N})$ is a function
of the noisy channel only.

\begin{lemma}

\label{lemma:limits}
Suppose $c_1,c_2,\ldots$ is a nonnegative sequence such that $c_n \leq
kn$ for some $k \geq 0$, and
\begin{eqnarray} \label{eqtn: abstract superadditivity}
c_m + c_n \leq c_{m+n}, \end{eqnarray}
for all $m$ and $n$. Then
\begin{eqnarray} \label{eqtn: lim exists}
\lim_{n \rightarrow \infty} \frac{c_n}{n} \end{eqnarray}
exists and is finite.

\end{lemma}

\begin{proof}

Define
\begin{eqnarray}
c \equiv \limsup_n \frac{c_n}{n}. \end{eqnarray}
This always exists and is finite, 
since $c_n \leq kn$ for some $k \geq 0$.
Fix $\epsilon > 0$ and choose $n$ sufficiently large that
\begin{eqnarray}
\frac{c_n}{n} > c - \epsilon. \end{eqnarray}
Suppose $m$ is any integer strictly greater than $\max(n,n/\epsilon)$.
Then by (\ref{eqtn: abstract superadditivity}),
\begin{eqnarray} \label{eqtn: intermediate appendix}
\frac{c_m}{m} \geq \frac{c_n}{n} 
\frac{n}{m} \left( 1 + \frac{c_{m-n}}{c_n}
	\right). \end{eqnarray}
Using the fact that $l c_n \leq c_{ln},$ (an immediate consequence 
of (\ref{eqtn: abstract superadditivity})) with $l = \lfloor 
\frac{m}{n} \rfloor- 1$ gives
\begin{eqnarray}
\frac{c_{m-n}}{c_n} & \geq & \lfloor \frac{m}{n} \rfloor - 1 \\
 & \geq & \frac{m}{n} - 2, \end{eqnarray}
where $\lfloor x \rfloor$ 
is the integer immediately below $x$. Plugging the
last inequality into (\ref{eqtn: intermediate appendix}) gives
\begin{eqnarray}
\frac{c_m}{m} \geq \frac{c_n}{n} \left( 1 - \frac{n}{m} \right).
\end{eqnarray}
But $-n/m > -\epsilon$ and $c_n/n \geq c - \epsilon$, so
\begin{eqnarray}
\frac{c_m}{m} \geq (c-\epsilon)(1-\epsilon). \end{eqnarray}
This equation holds for all sufficiently large $m$, and thus
\begin{eqnarray}
\liminf_n \frac{c_n}{n} \geq (c-\epsilon)(1-\epsilon). \end{eqnarray}
But $\epsilon$ was an arbitrary number greater than $0$, so letting
$\epsilon \rightarrow 0$ we see that
\begin{eqnarray}
\liminf_n \frac{c_n}{n} \geq c = \limsup_n \frac{c_n}{n}. 
\end{eqnarray}
It follows that $\lim_n c_n/n$ exists, as claimed.

\end{proof}

The following theorem places a limit on the entropy of a source
which can be sent through a quantum channel.

\begin{theorem}  \textbf{(Upper bound on the capacity with unitary encodings)}

Suppose we consider a source $\Sigma= (H_s, [..\rho_s^n...])$ and a
sequence of unitary encodings ${\cal U}_n$ for the source. Suppose
further that there exists a sequence of decodings, ${\cal D}_n$ such
that
\begin{eqnarray}
\lim_{n \rightarrow \infty} F(\rho_s^n,{\cal D}_n \circ {\cal N}^{\otimes n} 
	\circ {\cal U}_n) = 1.
\end{eqnarray}
Then
\begin{eqnarray}
\limsup_{n \rightarrow \infty} \frac{ S(\rho_s^n)}{n} \leq C.
\end{eqnarray}

\end{theorem}

\begin{proof}

What this theorem tells us is that we cannot reliably transmit more
than $C$ qubits of information per use of the channel.  When the
source entropy exists, it tells us we cannot transmit sources with
entropy greater than $C$; when the entropy of the source is not
defined, it still rules out transmission of sources for which the
limsup in the expression (which is always defined) is too large.  For
unitary ${\cal U}_n$ we have
\begin{eqnarray}
I(\rho_s,{\cal N}^{\otimes n} \circ {\cal U}_n) =
	 I({\cal U}_n(\rho_s),{\cal N}^{\otimes n}),
\end{eqnarray}
and thus
\begin{eqnarray}
I(\rho_s,{\cal N}^{\otimes n} \circ {\cal U}_n) \leq C_n. \end{eqnarray}
By (\ref{eqtn: fidelity lemma}) with ${\cal E} \equiv {\cal N}^{\otimes n}
\circ {\cal U}_n$, and the fact that $I(\rho^{\otimes n},{\cal N}^{\otimes n}) \le 
\max_{\rho^n} I(\rho^n,{\cal N}^{\otimes n}) \equiv C_n$ it now follows that
\begin{eqnarray}
\frac{S(\rho_s^n)}{n} & \leq & \frac{C_n}{n} + \frac{2}{n} + \nonumber \\
 & & 4 (1-F(\rho_s^n,{\cal D}^n \circ {\cal N}^n \circ {\cal U}^n))
	\log d. \end{eqnarray} (Note that $d$ here is the dimension 
of a single copy of the source Hilbert space, so that we have inserted
$d^n$ for the overall dimension $d$ of (\ref{eqtn: fidelity lemma})). 
Taking $\limsup$s on both sides of the equation completes the proof of the
theorem.

\end{proof}

It is extremely useful to study this result at length, since the basic
techniques employed to prove the bound are the same as those that
appear in a more elaborate guise later in the Chapter.  It is
particularly instructive to see how this result differs from the
classical result. In particular, what features of quantum mechanics
necessitate a change in the proof methods used to obtain the classical
bound?

Suppose the quantum analogue of the classical subadditivity of mutual
information were true, namely
\begin{eqnarray}
I(\rho^n,{\cal N}^{\otimes n}) \leq \sum_{i=1}^n I(\rho^n_i,{\cal N}),
\end{eqnarray}
where $\rho^n$ is any density operator that can be used as input to $n$
copies of the channel, and $\rho^n_i$ is the density operator obtained
by tracing out all but channel number $i$. Then 
it would follow easily from the definition that $C_n = C_1$ for all
$n$, and thus 
\begin{eqnarray}
C = C_1 = \max_{\rho} I(\rho,{\cal N}). \end{eqnarray} This expression
is exactly analogous to the classical expression for channel capacity
as a maximum over input distributions of the mutual information
between channel input and output. If this were truly a bound on the
quantum channel capacity then it would allow easy numerical
evaluations of bounds on the channel capacity, as the maximization
involved is easy to do numerically, and the coherent information is
not difficult to evaluate.

Unfortunately, it is not possible to assume that the quantum
mechanical coherent information is subadditive, as shown by example
(\ref{eqtn: example 2}), and thus in general it is possible that
\begin{eqnarray}
C > C_1. \end{eqnarray} In fact, the results of Shor and Smolin
\cite{Shor96a} demonstrate the existence of channels for which the
above strict inequality holds.  In order to evaluate the bound $C$
which we have derived it is thus necessary to take the limit in
(\ref{eqtn: unitary bound}). To numerically evaluate this limit
directly is certainly not a trivial task, in general. The result we
have presented, that (\ref{eqtn: unitary bound}) is an upper bound on
channel capacity is an important theoretical result, that may aid in
the development of effective numerical procedures for obtaining
general bounds. But it does not yet constitute an effective procedure.

\subsection{General encodings}
\label{sect: general encodings}

We will now consider the case where something more general than a
unitary encoding is allowed. In principle, it is always possible to
perform a non-unitary encoding, ${\cal C}$, by introducing an extra
ancilla system, performing a joint unitary on the source plus ancilla,
and then discarding the ancilla.

We define
\begin{eqnarray}
C^n \equiv \max_{\rho,{\cal C}} I(\rho,{\cal N}^n\circ {\cal C}),
\end{eqnarray}
where the maximization is over all inputs $\rho$ to the encoding operation,
${\cal C}$, which in turn maps to $n$ copies of the channel,
\begin{eqnarray} 
{\cal N}^n \equiv {\cal N} \otimes \ldots \otimes {\cal N}; \,\,\,\,
	n \mbox{times}, \end{eqnarray}
The bound on the channel capacity proved in this section is defined by
\begin{eqnarray} \label{eqtn: general bound}
C({\cal N}) \equiv \lim_{n \rightarrow \infty} \frac{C_n}{n}.
\end{eqnarray}
Once again, to prove that this limit exists one applies the lemma
proved on page \pageref{lemma:limits}.

To prove that this quantity is a bound on the channel capacity, one
applies almost exactly the same reasoning as in the preceding
subsection.  The result is:

\begin{theorem} \textbf{(General bound on the channel capacity)}

Suppose we consider a source $\Sigma= (H_s, [..\rho_s^n...])$ and a
sequence of unitary encodings ${\cal U}_n$ for the source. Suppose
further that there exists a sequence of decodings, ${\cal D}_n$ such
that
\begin{eqnarray}
\lim_{n \rightarrow \infty} F(\rho_s^n,{\cal D}^n \circ {\cal N}^{\otimes n} 
	\circ {\cal C}^n) = 1.
\end{eqnarray}
Then
\begin{eqnarray}
\limsup_{n \rightarrow \infty} \frac{ S(\rho_s^n)}{n} \leq C.
\end{eqnarray}

\end{theorem}

\begin{proof}

Again, this result places an upper bound on the rate at which
information can be reliably transmitted through a noisy quantum
channel. The proof is very similar to the earlier proof of a bound for
unitary encodings. One simply applies (\ref{eqtn: fidelity lemma})
with ${\cal E} = {\cal N}^{\otimes n} \circ {\cal C}^n$ and ${\cal D}
= {\cal D}^n,$ again invoking the fact that $I(\rho^{\otimes n},{\cal
N}^{\otimes n}) \le \max_{\rho^n} I(\rho^n,{\cal N}^{\otimes n})$, and
chooses ${\cal C}_n$ to be the coding that maximizes this expression, to
give:
\begin{eqnarray}
\frac{S(\rho_s^n)}{n} \leq \frac{C_n}{n} &+& \frac{2}{n} + \nonumber \\
	&&4(1-F(\rho_s^n,{\cal D}^n \circ {\cal N}^{\otimes n} \circ 
	{\cal C}^n)) \log d. \end{eqnarray}
Taking $\limsup$s on both sides of the equation completes the proof.

\end{proof}

It is instructive to see why the proof fails when the maximization is
done over channel input states alone, rather than over all source
states and encoding schemes. The basic idea is that there may exist
source states, $\rho_s$, and encoding schemes ${\cal C}$, for which
\begin{eqnarray} \label{eqtn: strange inequality}
I(\rho,{\cal N} \circ {\cal C}) > I({\cal C}(\rho),{\cal N}).
\end{eqnarray}
It is clear that the existence of such a scheme would cause the line
of proof suggested above to fail.   Moreover, as we saw in subsection
\ref{subsec:example_1}, it is possible for exactly this situation to
occur, due to quantum entanglement. 

Having proved that $C({\cal N})$ is an upper bound on the channel
capacity, let us now investigate some of the properties of this
bound. First of all we will examine the range over which $C$ can
vary. Note that
\begin{eqnarray}
0 \leq C_n \leq n \log d, \end{eqnarray}
since if $\rho$ is pure then $I(\rho,{\cal N}^n \circ {\cal C}) = 0$
for any encoding ${\cal C}$, and for all $\rho$ and ${\cal C}$,
$I(\rho,{\cal N}^n \circ {\cal C}) \leq \log d^n = n \log d$, since the
channel output has $d^n$ dimensions. It follows that
\begin{eqnarray}
0 \leq C({\cal N}) \leq \log d. \end{eqnarray} This parallels the
classical result, which states that the channel capacity varies
between $0$ and $\log s$, where $s$ is the number of channel
symbols. The upper bound on the classical capacity is attained if and
only if the classical channel is noiseless.

In the case when ${\cal N}$ takes a constant value,
\begin{eqnarray}
{\cal N}(\rho) = \sigma, \end{eqnarray} for all channel inputs $\rho$,
it is not difficult to verify that $C({\cal N}) = 0$. This is
consistent with the obvious fact that the capacity for coherent
quantum information of such a channel is zero.

When is the upper bound, $C({\cal N}) = \log d$ attained? Suppose
the channel is unitary, ${\cal N}(\rho) = U \rho U^{\dagger}$. Encoding
the source $\rho_s = I/d \otimes \ldots \otimes I/d$
using the identity encoding, we see that
$I(\rho_s,{\cal N}^n \circ {\cal C}) = \log d$, and thus $C_n \geq n \log d$,
and thus $C({\cal N}) \geq \log d$. But the reverse inequality also holds
as remarked earlier, and thus
\begin{eqnarray}
C({\cal N}) = \log d, \end{eqnarray}
if ${\cal N}$ is a unitary channel.

It is also of interest to consider what happens when channels ${\cal N}_1$
and ${\cal N}_2$ are composed, forming a joint channel,
${\cal N} = {\cal N}_2 \circ {\cal N}_1$. From the data processing
inequality it follows that
\begin{eqnarray}
C({\cal N}_1) \geq C({\cal N}). \end{eqnarray}
It is clear by repeated application of the data-processing inequality
that this result also holds if we compose more than two
channels together, and even holds if we allow intermediate decoding and
re-encoding stages. Classically, channel capacities also behave in
this way: the capacity of
a channel made by composing two (or more) channels together is no greater
than the capacity of the first part of the channel alone.

Although (\ref{eqtn: example 1}) might seem to suggest otherwise, 
in fact
\begin{eqnarray}
C({\cal N}_2) \geq C({\cal N}). \end{eqnarray}
For let us suppose that ${\cal C}$ is the encoding which achieves
the channel capacity $C({\cal N}),$ so that the total operation 
is ${\cal D} \circ {\cal N} \circ {\cal C} \equiv {\cal D} \circ {\cal N}_2 \circ {\cal N}_1 \circ {\cal C}.$   As our encoding for the channel 
${\cal N}_2$, we may use ${\cal N}_1 \circ {\cal C}$ and decode with
${\cal D},$ hence achieving
precisely the same total operation.

\subsection{Other encoding protocols}
\label{subsec: other encoding protocols}

So far we have considered two allowed classes of encodings: encodings
where a general unitary operation can be performed on a block of
qubits, and encodings where a general trace-preserving quantum
operation can be performed on a block of qubits. If large-scale
quantum computation ever becomes feasible it may be realistic to
consider encoding protocols of this sort. However, for present-day
applications of quantum communication such as quantum cryptography and
teleportation only a much more restricted class of encodings is
possible.  In this section we will describe several plausible classes.

We will begin by considering a toy example which is meant to
illustrate the basic techniques which will be used later.
It is the class involving local unitary operations \index{local unitary operations} only. 
We will refer to this class as $U$-$L$. It consists of the set of
operations ${\cal C}$ which can be written in the form
\begin{eqnarray} \label{eqtn: local unitary encodings}
{\cal C}(\rho) = (U_1 \otimes \ldots \otimes U_n) \rho (U_1^{\dagger} \otimes
	\ldots U_n^{\dagger}), \end{eqnarray}
where $U_1,\ldots,U_n$ are local unitary operations on systems $1$ through
$n$.
Another possibility is the class $L$ of encodings involving
local operations only, i.e. operations of the form:
\begin{eqnarray} \label{eqtn: local encodings}
\sum_{i_1,...i_N} (A_{i_1} \otimes B_{i_2} \otimes \cdots
\otimes Z_{i_N}) \rho \nonumber \\
(A_{i_1}^\dagger \otimes B_{1_2}^\dagger \otimes \cdots
\otimes Z_{i_N}^\dagger).
\end{eqnarray}
In other words, the overall operation has a tensor product form
${\cal A}_1 \otimes {\cal A}_2 \otimes \cdots \otimes {\cal A}_m$.

A more realistic class is $1$-$L$ -- encoding by local
operations with one way classical communication. The idea is that the encoder
is allowed to do encoding by performing arbitrary quantum operations
on individual members (typically, a single qubit) of the strings
of qubits emitted by a source. This is not unrealistic with present
day technology for manipulating single qubits. Such operations could
include arbitrary unitary rotations, and also generalized measurements. After
the qubit is encoded, the results of any measurements done during the
encoding may be used to assist in the encoding of later qubits. This
is what we mean by one way communication - the results of the measurement
can only be used to assist in the encoding of later qubits, not earlier
qubits.

Another possible class is $2$-$L$ - encoding by local operations with
two-way classical communication. These may arise in a situation where
there are many identical channels operating side by side in
space. Once again it is assumed that the encoder can perform arbitrary
local operations, only this time two way classical communication is
allowed when performing the encoding.

For any class of encodings $\Lambda$ arguments analogous to 
those used above for general and for unitary block coding, ensure
that the capacity 
\begin{eqnarray}
C_{\Lambda}({\cal N}) \equiv \lim_{n \rightarrow \infty} \frac{C_{\Lambda}^n}{n},
\end{eqnarray}
where
\begin{eqnarray}
C^n_{\Lambda} \equiv \max_{\rho,{\cal C 2}\Lambda} I(\rho,{\cal N}^n \circ {\cal C}),
\end{eqnarray}
is an upper bound to the rate at which information can be
reliably transmitted using encodings in $\Lambda$.  Thus we 
have expressions for $C_{U}, C_{L}, C_{1-L},$ and $C_{2-L},$
which provide upper bounds on the rate of quantum information
transmission for these types of encodings.

An interesting and important question is whether there are
closed-form
characterizations of the sets of quantum operations corresponding to
particular types of encodings schemes such as $1$-$L$ and $2$-$L$. For
example, in the cases of $U$-$L$ and $L$ there are explicit forms 
(\ref{eqtn: local unitary encodings},\ref{eqtn: local encodings}) for the classes of encodings
allowed.  For $1$-$L$ we believe the operations take the form:
\begin{eqnarray}
\sum_{i_1,...i_N} (A_{i_1} &\otimes& B_{i_1,i_2} \otimes \cdots
\otimes Z_{i_1,i_2,...i_N}) \rho \nonumber \\
&&(A_{i_1}^\dagger \otimes B_{i_1,i_2}^\dagger \otimes \cdots
\otimes Z_{i_1,i_2,...i_N}^\dagger).
\end{eqnarray}
It would be valuable to limit the range of the indices in this
expression.  This is likely to be related to the number of rounds
of classical communication which are involved in an operation.  Since
communication is one-way, it is likely this is bounded.  It would also be useful to find a similar expression
for $2$-$L$ encodings.  One possibility is:
\begin{eqnarray}
\sum_{i} (A_{i} \otimes B_{i} \otimes \cdots
\otimes Z_{i}) \rho (A_{i}^\dagger \otimes B_{i}^\dagger \otimes \cdots
\otimes Z_{i}^\dagger).
\end{eqnarray}  However, although all $2$-$L$ operations involving
a finite number of rounds of communication can certainly be put in
this form, I do not presently see whether all operations expressible
in this form should be realizable with local operations and two-way
classical communication.

Such closed-form expressions would aid in numerical maximizations like
that performed in calculating of bounds on the channel capacity. In
order to perform such maximizations it would be necessary that the
closed form expressions be {\em bounded} in size (hence the interest
in limiting the range of indices above).

The classes we have described in this subsection are certainly not the
only realistic classes of encodings. Many more classes may be
considered, and in specific applications this may well be of great
interest. What we have done is illustrated a general technique for
obtaining {\em bounds} on the channel capacity for different classes
of encodings. A major difference between classical information theory
and quantum information theory is the greater interest in the quantum
case in studying different classes of encodings. Classically it is, in
principle, easy to perform an arbitrary encoding and decoding
operation using a look-up table. However, quantum mechanically this is
far from being the case, so there is correspondingly more interest in
studying the channel capacities that may result from considering
different classes of encodings.

Here we have not addressed the attainability of the bounds we have
described. To qualify as true quantum capacities one must exhibit
explicit coding and decoding schemes which allow the bounds described
in this section to be achieved. The development of general proofs
showing that this can be done or counterexamples showing that it
cannot is a major remaining goal of quantum information theory.

\section{Discussion}
\label{sect: discussion}

What then can be said about the status of the quantum coherent noisy
channel coding theorem in the light of comments made in the preceding
sections?  While we have established upper bounds, we have not proved
achievability.  Lloyd \cite{Lloyd97a} also proposed the maximum of the
coherent information as the channel capacity, although initially
without considering the difficulties engendered by the failure of
subadditivity.  He argued that this capacity was an achievable upper
bound on transmission rate.  The methods by which we derived our upper
bound are quite different from those employed by Lloyd; I hope
comparison of the two approaches will prove illuminating.  The
fidelity criterion he used, average pure-state fidelity for the
uniform ensemble over the typical subspace, is different from the
criterion used here, and although I think it is likely that they lead
to the same capacity asymptotically, I am not aware of results that
imply this.  In Lloyd's work, although the encoding scheme is not
explicitly written out, it appears to be restricted to projection onto
the typical subspace followed by a unitary.  
%It will be shown
%elsewhere that such a restriction is justified \cite{Barnum98b}, which
%establishes that (\ref{eqtn: unitary bound}) is actually truly an
%upper bound on the channel capacity.
%think this is unlikely.  
However, one can still make progress towards a proof that the novel
expression, (\ref{eqtn: general bound}), which we have shown bounds
the channel capacity, is in fact the true capacity of a noisy quantum
channel for sending coherent quantum information.  If we accept
Lloyd's claim that his expression for the channel capacity is correct
for the case when only restricted encodings are allowed, then it is
possible to use the following four-stage construction to show that
(\ref{eqtn: general bound}) is a correct expression for the capacity
for transmission through a noisy quantum channel; i.e. that in
addition to being an upper bound as shown in section \ref{sect: upper
bounds}, it is also achievable.
\begin{figure}[ht]
\begin{center}
\scalebox{0.8}{\includegraphics{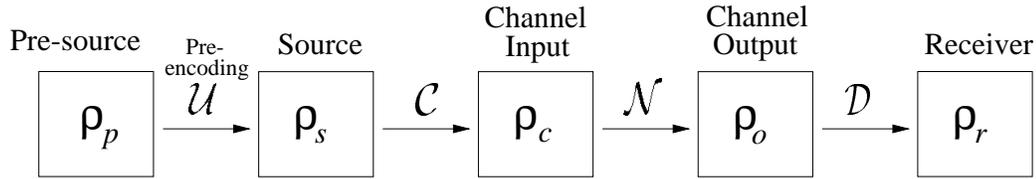}}
\end{center}
\caption{\label{fig: channel3}
Noisy quantum channel with an extra stage, a restricted {\em pre-encoding},
${\cal P}$.}
\end{figure}

For a fixed block size, $n$, one finds an encoding, ${\cal C}_n$, for
which the maximum in
\begin{eqnarray}
C_n \equiv \max_{{\cal C}_n,\rho_s} I(\rho_s,{\cal C}_n),
\end{eqnarray}
is achieved. One then regards the composition ${\cal N}^{\otimes n}
\circ {\cal C}_n$ as a single noisy quantum channel, and applies the
achievability result on restricted encodings to the joint channel
${\cal N}^{\otimes n} \circ {\cal C}_n$ to achieve an even longer $mn$
block coding scheme with high dynamic fidelity. This gives a joint
coding scheme ${\cal P}_{mn} \circ {\cal C}_n^{\otimes m}$ which for
sufficiently large blocks $m$ and $n$ can come arbitrarily close to
achieving the channel capacity (\ref{eqtn: general bound}).

An important open question is whether (\ref{eqtn: general bound}) is
equal to (\ref{eqtn: unitary bound}). It is clear that the former
expression is at least as large as the latter.  Work in progress
\cite{Barnum98b} shows that this is, in fact, the case.

Thus, I think it likely that the expression (\ref{eqtn: unitary
bound}) will turn out to be the maximum achievable rate of reliable
transmission through a quantum channel.  But this is still not quite
as satisfactory as the classical expression for the capacity, because
of the difficulty of evaluating the limit involved.  At a minimum, we
would like to know enough about the rate of convergence of $C_n$ to
its limit to be able to accurately estimate the error in a numerical
calculation of capacity, giving an effective procedure for calculating
the capacity to any desired degree of accuracy.

\section{Channels with a classical observer}
\label{sect: observed channel}
\index{observed channel}

In this section we consider a more general version of the quantum
noisy channel coding problem than has been considered in any previous
work. Suppose that in addition to a noisy interaction with the
environment there is also a classical observer who is able to perform
a measurement. This measurement may be on the channel or the
environment of the channel, or possibly on both.

\begin{figure}[ht]
\begin{center}
\scalebox{0.8}{\includegraphics{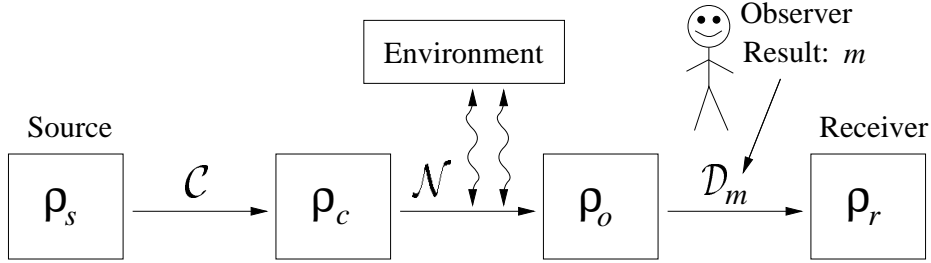}}
\end{center}
\caption{ \label{fig: channel4}
Noisy quantum channel with a classical observer.}
\end{figure}

The result of the measurement is then sent to the decoder, who may use the
result to assist in the decoding. We will assume to begin that this
transmission of classical information is done noiselessly, although it
is also interesting to consider what happens when the classical
transmission also involves
noise. It can be shown \cite{Kraus83a} that the state received by the
decoder is again related to the state $\rho$ used as input to the
channel by a quantum operation ${\cal N}_m$, where $m$ is the
measurement result recorded by the classical observer,
\begin{eqnarray}
\rho \rightarrow \frac{{\cal N}_m(\rho)}{\mbox{tr}({\cal N}_m(\rho))}.
\end{eqnarray}
The basic situation is illustrated in figure \ref{fig: channel4}. The idea
is that by giving the decoder access to classical information about
the environment responsible for noise in the channel it may be possible
to improve the capacity of that channel, by allowing the decoder
to choose different decodings ${\cal D}_m$ depending on the measurement
result $m$.

A simple example which illustrates that this can be the case will now
be given. Suppose have a two-level system in a state $\rho$ and
an initially uncorrelated four-level environment initially in
the maximally mixed state $I/4$, so the total state of the joint system is
\begin{eqnarray}
\rho \otimes \frac{I}{4}.
\end{eqnarray}
Suppose we fix an orthonormal basis $|1\rangle,|2\rangle,|3\rangle,
|4\rangle$ for the environment. We assume that a unitary interaction
between the system and environment takes place, given by the
unitary operator
\begin{eqnarray}
U & = & I \otimes |1\rangle \langle 1| + X \otimes |2\rangle
	\langle 2| + Y \otimes |3\rangle \langle 3| +
	Z \otimes |4 \rangle \langle 4|.
\end{eqnarray}
The output of the channel is thus
\begin{eqnarray}
\rho \rightarrow {\cal N}(\rho) \equiv \mbox{tr}_E(U \rho \otimes
	\frac{I}{4} U^{\dagger} ).
\end{eqnarray}
The quantum operation ${\cal N}$ can be given two particularly
useful forms,
\begin{eqnarray}
{\cal N}(\rho) & = & \frac{1}{4} \left( I\rho I + X \rho X
	+ Y \rho Y + Z \rho Z \right) \\
	& = & \frac{I}{2}.
\end{eqnarray}
It is not difficult to show from the second form that
\begin{eqnarray}
C({\cal N}) = 0.
\end{eqnarray}
Suppose now that an observer is introduced, who is allowed to perform
a measurement on the environment. We will suppose this measurement is
a Von Neumann measurement in the $|1\rangle,|2\rangle,|3\rangle,|4\rangle$
basis, and yields a corresponding measurement result, $m = 1,2,3,4$.
Then the quantum operations corresponding to these four measurement outcomes
are
\begin{eqnarray}
{\cal N}_1(\rho) & = & \frac 14 \rho \\
{\cal N}_2(\rho) & = & \frac 14 X \rho X \\
{\cal N}_3(\rho) & = & \frac 14 Y \rho Y \\
{\cal N}_4(\rho) & = & \frac 14 Z \rho Z.
\end{eqnarray}
Each of these is unitary, up to a constant multiplying factor, so the
corresponding channel capacities are
\begin{eqnarray}
C_m = 1. \end{eqnarray}
Thus $0 = C < C_m = 1$ for each result $m$. Clearly this is consistent
with the fact that ${\cal N} = \sum_m {\cal N}_m$ (cf. (\ref{eqtn: abstract subadditivity})).

This result is particularly clear in the context of teleportation.  In
section \ref{sec:teleportation_qops} we showed that the problem of
teleportation can be understood precisely as the problem of a quantum
noisy channel with an auxiliary classical channel. In the original
single qubit teleportation scheme described in section
\ref{sec:teleportation} \cite{Bennett93a} there are four quantum
operations relating the state Alice wishes to teleport, to the state
Bob receives, corresponding to each of the four measurement
results. In that scheme it happens that those four operations are the
${\cal N}_m$ we have described above. Furthermore in the absence of
the classical channel, that is, when Alice does not send the result of
her measurement to Bob, the channel is described by the single
operation ${\cal N}$. Clearly, in order that causality be preserved we
expect that $C = 0$. On the other hand, in order that teleportation be
able to occur we should expect that $C_m = 1$, as was shown
above. Teleportation understood in this way as a noisy channel with a
classical side channel offers a particularly elegant way of seeing
that the transmission of quantum information may sometimes be greatly
improved by making use of classical information.

The remainder of this section is organized into two
subsections. Subsection \ref{subsec: upper bounds on observed
channel} proves bounds on the capacity of an observed channel. These
results require nontrivial extension of the techniques developed
earlier for proving bounds on the capacity of an unobserved
channel. Subsection (\ref{subsec: relationship to unobserved
channel}) relates work done on the observed channel to the work done
earlier in the Chapter on the unobserved channel.

\subsection{Upper bounds on channel capacity}
\label{subsec: upper bounds on observed channel}

As for the unobserved channel we will now prove several results
bounding the channel capacity of an observed channel. We begin with
the key lemma that will be used to prove bounds.

The following lemma generalizes the entropy-fidelity lemma on page
\pageref{subsec:fidelity_lemma} for quantum operations, which was the
foundation of our earlier proofs of upper bounds on the quantum
channel capacity:

\begin{lemma} \textbf{(generalized entropy-fidelity lemma for operations)}

Suppose ${\cal E}_m$ are a set of quantum operations such that $\sum_m
{\cal E}_m$ is a complete quantum operation. Suppose further that
${\cal D}_m$ is a complete quantum operation for each $m$.  Then
\begin{eqnarray}
\label{eqtn: fidelity lemma for general operations}
S(\rho) & \leq & \sum_m \mbox{tr}({\cal E}_m(\rho)) I(\rho,{\cal E}_m)
	+ 2 + \nonumber \\ & & 4 (1-F(\rho,{\cal T})) \log d,
\end{eqnarray}
where
\begin{eqnarray}
{\cal T} \equiv \sum_m {\cal D}_m \circ {\cal E}_m.
\end{eqnarray}

\end{lemma}

\begin{proof}

By the second step of the data processing inequality,
(\ref{eqtn:quantum_data_processing}), $I(\rho,{\cal E}_m) \geq
I(\rho,{\cal D}_m \circ {\cal E}_m)$ for each $m$, and noting also
that by the completeness of ${\cal D}_m$,
$\mbox{tr}({\cal E}_m(\rho)) =
\mbox{tr}(({\cal D}_m \circ {\cal E}_m)(\rho))$, we obtain
\begin{eqnarray}
S(\rho) & \leq & S(\rho) + \sum_m \left[ \mbox{tr}({\cal E}_m(\rho))
	I(\rho,{\cal E}_m) - \right. \nonumber \\
	& & \left. \mbox{tr}(({\cal D}_m \circ {\cal E}_m)(\rho))
	I(\rho,{\cal D}_m \circ {\cal E}_m) \right]. \end{eqnarray}
Applying now the convexity theorem for coherent information,
\begin{eqnarray}
-\sum_m \mbox{tr}(({\cal D}_m \circ {\cal E}_m)(\rho)) I(\rho,{\cal D}_m
	\circ {\cal E}_m) \leq -I(\rho,{\cal T}).
\end{eqnarray}
we obtain
\begin{eqnarray}
S(\rho) & \leq & \sum_m \mbox{tr}({\cal E}_m(\rho)) I(\rho,{\cal E}_m)
	+ S(\rho) - I(\rho,{\cal T}).
\end{eqnarray}
But ${\cal T} = \sum_m {\cal D}_m \circ {\cal E}_m$ is trace-preserving since
${\cal D}_m$ is trace-preserving and $\sum_m {\cal E}_m$ is
trace-preserving, and thus by (\ref{eqtn:second_law}),
\begin{eqnarray}
S(\rho) - I(\rho,{\cal T}) & = &
	S(\rho) - S({\cal T}(\rho)) + S(\rho,{\cal T}) \\
	& \leq & 2 S(\rho,{\cal T}).
\end{eqnarray}
Finally, an application of the quantum Fano inequality
(\ref{eqtn: quantum Fano}) along with the observations that the entropy
function $h$ appearing in that inequality is bounded above by one,
and $\log (d^2-1) \leq 2 \log d$ gives
\begin{eqnarray}
S(\rho) & \leq & \sum_m \mbox{tr}(({\cal D}_m \circ {\cal E}_m)(\rho)) 
	I(\rho,{\cal D}_m \circ {\cal E}_m) + 2 + \nonumber \\
 & & 	4(1-F(\rho,{\cal T})) \log d,
\end{eqnarray}
as we set out to prove.

\end{proof}

If we define 
\begin{eqnarray}
\label{eqtn: observed capacity}
C(\{ {\cal N}_m \}) =  \limsup_{n \rightarrow \infty}
	\max_{{\cal C}^n, \rho} \nonumber \\
  \sum_{m_1,...m_n}  \mbox{tr}(({\cal C}^n \circ 
	{\cal N}_{m_1} \otimes \cdots \otimes {\cal N}_{m_n} )(\rho))  
	\frac{I(\rho, {\cal N}_{m_1} \otimes \cdots \otimes {\cal N}_{m_n}\circ {\cal C}^n)}{n},
\end{eqnarray}
we may use (\ref{eqtn: fidelity lemma for general operations}) to easily 
prove that
$C(\{ {\cal N}_m \})$ is an upper bound on the rate of reliable
transmission through an observed channel, in precisely the same way we
earlier used (\ref{eqtn: fidelity lemma}) to prove bounds for
unobserved channels.

We may derive the same bound in another fashion if we associate
observed channels with complete quantum operations -- unobserved
channels -- in the following fashion suggested by examples in
\cite{Bennett97a}.  To an observed channel $\{ {\cal N}_m\}$ we
associate a single complete operation ${\cal M}$ from ${\cal H}$ to
the larger Hilbert space ${\cal H} \otimes {\cal R}$.  The operation
is specified by:
\begin{eqnarray}
\label{eqtn: equivmap}
{\cal M}(\rho) = \sum_m {\cal N}_m(\rho) \otimes |m\rangle \langle m|.
\end{eqnarray}
This map is an ``all-quantum'' version of the observed channel.  The
classical information about which $m$ occurred appears in the
``register'' Hilbert space ${\cal R}$ encoded in orthogonal states.
Since our upper bound to the capacity of an unobserved channel applies
also to channels with output Hilbert spaces of different
dimensionality than the input space, they apply to this map as well.
It is easily verified that the coherent information for the map ${\cal
M}$ acting on $\rho$ is the same as the average coherent information
for the observed channel ${\cal N}_m$ acting on $\rho$, which appears
in (\ref{eqtn: fidelity lemma for general operations}) and in the
quantity (\ref{eqtn: observed capacity}).  To show this, define $p_m =
{\rm tr} ({\cal N}_m(Q))$.  Then $Q' = {\cal M}(Q)$ is given by
(\ref{eqtn: equivmap}), so that
\begin{eqnarray}
S(Q') = H(p_m) + \sum_m p_m S\left(\frac{{\cal N}_m (Q)}{p_m}\right)
\end{eqnarray} by the grouping property (\ref{eqtn: entropy grouping}) of
Shannon entropy, which applies since the density matrices ${\cal N}_m
(Q) \otimes |m \rangle \langle m|$ are mutually orthogonal.
Similarly,
\begin{eqnarray}
R'Q' = ({\cal I} \otimes \sum_m {\cal N}_m^*)(RQ),
\end{eqnarray}
 where by definition ${\cal N}_m^*(\rho) = 
{\cal N}_m (\rho) \otimes |m \rangle \langle m|.$  By linearity
this is equal to $\sum_m ({\cal I} \otimes {\cal N}_m)(RQ)
\otimes |m \rangle \langle m|.$   Applying the orthogonality and
grouping argument again, and noting that ${\rm tr} (({\cal I} \otimes
{\cal N}_m)(RQ)) = {\rm tr}({\cal N}_m (Q)) = p_m,$
we get that
\begin{eqnarray}
S(R'Q') = H(p_m) + \sum_m p_m S\left(\frac{({\cal I} \otimes
{\cal N}_m) (RQ)}{p_m}\right).
\end{eqnarray}
Hence the coherent information
for ${\cal M}$ becomes
\begin{eqnarray}
\sum_m p_m \left[S\left(\frac{{\cal N}_m (Q)}{p_m}\right) -
S\left(\frac{({\cal I} \otimes {\cal N}_m) (RQ)}{p_m}\right)\right],
\end{eqnarray}
which is precisely the average coherent information for $\{{\cal
N}_m\}.$ So an application of the bound (\ref{eqtn: general bound}) on
the rate of transmission through the unobserved channel ${\cal M}$
yields the bound (\ref{eqtn: observed capacity}), if one accepts the
intuitively obvious claim that ${\cal M}$ and $\{ {\cal N}_m\}$ are
equivalent with respect to transmission of quantum information.

Bennett {\em et al} \cite{Bennett97a} derive capacities for three
simple channels which may be viewed as taking the form (\ref{eqtn:
equivmap}).  The {\em quantum erasure channel} takes the input state
to a fixed state orthogonal to the input state with probability
$\epsilon$; otherwise, it transmits the state undisturbed.  An
equivalent observed channel would with probability $\epsilon$ replace
the input state with a standard pure state $|0\rangle \langle 0|$
within the input subspace, and also provide classical information as
to whether this replacement has occurred or not. The {\em phase
erasure channel} randomizes the phase of a qubit (or, in our context
of multidimensional input space, diagonalizes the density operator in
a fixed basis) with probability delta, and otherwise transmits the
state undisturbed; it also supplies classical information as to which
of these alternatives occurred.  The {\em mixed erasure/phase-erasure
channel} may either erase or phase-erase, with exclusive probabilities
$\epsilon$ and $\delta$.  Bennett {\em et al} note that the capacity
$1 - 2 \epsilon$ of the erasure channel is in fact the one-shot
maximal coherent information.  We have verified that the capacities
they derive for the phase-erasure channel ($1-\delta$) and the mixed
erasure/phase-erasure channel ($1 - 2\epsilon - \delta$) are the same
as the one-shot maximal average coherent information for the
corresponding observed channels, lending some additional support to
the view that the bounds we have derived here are in fact the
capacities.

\subsection{Relationship to unobserved channel}
\label{subsec: relationship to unobserved channel}

Suppose we have an observed channel which is described by operations
$\{ {\cal N}_m \}$. There are various natural physical ways these
operations can be associated with a channel described by the operation
\begin{eqnarray}
{\cal N} = \sum_m {\cal N}_m. \end{eqnarray}

One physically natural way this association may be made is the
following. Suppose a system is sent through a noisy quantum channel.
During the time and possibly after the system has traversed the
channel, various measurements may be performed, possibly on the
system, and possibly on the environment giving rise to the noise. We
will label the collective results of these measurements by a single
index, $m$.  As discussed earlier, with each $m$ is associated a
quantum operation, ${\cal N}_m$, which describes the state change
undergone by a system passing through the channel, given that result
$m$ occurs. (If the measurement involves the system, and not just the
environment, there is no guarantee that ${\cal N} \equiv \sum_m {\cal
N}_m = {\cal N}_0,$ where ${\cal N}_0$ is the noise due to the channel
without measurement.)

There is a particularly important special case of the above
scenario. Suppose the system is sent through a channel, and interacts
with an environment. The action of this channel is described by the
complete quantum operation ${\cal N}_0$.  After the environment has
interacted with the system, measurements are performed on the {\em
environment alone}.  Averaging over all possible measurement outcomes,
this does not disturb the state of the system, i.e. ${\cal N} \equiv
\sum_m {\cal N}_m = {\cal N}_0$.

We will now show that observing the environment of the channel never
{\em decreases} the bound we have obtained on the channel
capacity. This is certainly a property which we would expect the
channel capacity to have: observing the environment and then sending
the result of the observation on to be used in decoding should not
decrease the channel capacity, since the decoder can always simply
ignore the result of the observation.

Recall the expressions for the bound on the capacity of the unobserved channel,
\begin{eqnarray}
C({\cal N}) = \limsup_{n \rightarrow \infty}
	 \max_{{\cal C}^n,\rho} \frac{I(\rho,{\cal N}^{\otimes n} \circ {\cal C}^n)}{n},
\end{eqnarray}
and the observed channel,
\begin{eqnarray}
C(\{ {\cal N}_m \}) = \limsup_{n \rightarrow \infty} \max_{{\cal C}^n,
\rho} \nonumber \\
\sum_{m_1,...m_n} \mbox{tr}(({\cal C}^n \circ {\cal N}_{m_1} \otimes \cdots \otimes {\cal N}_{m_n} )(\rho)) 
\frac{I(\rho, {\cal N}_{m_1} \otimes \cdots \otimes {\cal N}_{m_n}\circ {\cal C}^n)}{n}.
\end{eqnarray}
%C(\{ {\cal N}_m \}) & = & \limsup_{n \rightarrow \infty}
%	\max_{{\cal C}^n, \rho} \nonumber \\
% & & \sum_m  \mbox{tr}(({\cal C}^n \circ 
%	{\cal N}^{\otimes n})(\rho)) \nonumber \\
%	 \frac{I(\rho,{\cal N}^{\otimes n} \circ {\cal C}^n)}{n}.
%\end{eqnarray}
But the convexity theorem for coherent information implies that
\begin{eqnarray}
	\sum_{m_1,...,m_n} \mbox{tr}(({\cal C}^n \circ 
	{\cal N}_{m_1}\otimes \cdots \otimes {\cal N}_{m_n})(\rho))
\frac{I(\rho,{\cal N}_{m_1}\otimes \cdots \otimes {\cal N}_{m_n} \circ {\cal C}^n)}{n} \nonumber \\
	\leq
    \frac{I(\rho,{\cal N} \circ {\cal C}^n)}{n}, 
\end{eqnarray}
and thus
\begin{eqnarray}
C({\cal N}) \leq C(\{ {\cal N}_m \}). 
\end{eqnarray}

To see that this inequality may sometimes be strict, return to the
example considered earlier in the context of teleportation. In that case it
is not difficult to verify that
\begin{eqnarray}
0 = C({\cal N}) < C(\{ {\cal N}_m \}) = 1. 
\end{eqnarray}

What these results show is that our bounds on the channel capacity are
never made any worse by observing the environment, but sometimes they
can be made considerably better. This is a property that we certainly
expect the quantum channel capacity to have, and we take as an
encouraging sign that the bounds we have proved in this Chapter are in
fact achievable, that is, the true capacities.

\subsection{Discussion}
\label{subsec: discussion unobserved channel}

All the questions asked about the bounds on channel capacity for an
unobserved channel can be asked again for the observed channel:
questions about achievability of bounds, the differences in power
achievable by different classes of encodings and decodings, and so
on. We will not address those problems here, beyond noting that they
are important problems which need to be addressed by future research.

Many new twists on the problem of the quantum noisy channel arise when
an observer of the environment is allowed. For example, one might
consider the situation where the classical channel connecting the
observer to the decoder is noisy. What then are the resources required
to transmit coherent quantum information?

It might also be interesting to prove results relating the
classical and quantum resources that are required to perform a certain
task. For example, in teleportation it can be shown that one requires
not only the quantum channel, but also two bits of classical information,
in order to transmit coherent quantum information with perfect
reliability \cite{Bennett93a}.

\section{Conclusion}
\label{sect: conc}

In this Chapter we have shown that different information transmission
problems may result in different channel capacities for the same noisy
quantum channel.  We have developed some general techniques for
proving upper bounds on the amount of information that may be
transmitted reliably through a noisy quantum channel.

Perhaps the most interesting thing about the quantum noisy channel
problem is to discover what is new and essentially {\em quantum} about
the problem. The following list summarizes what I believe are the
essentially new features:
\begin{enumerate}

\item The insight that there are many essentially different information
transmission problems in quantum mechanics, all of them of interest
depending on the application. These span a spectrum between two extremes:

\begin{itemize}

\item The transmission of a discrete set of mutually orthogonal
quantum states through the channel. Such problems are problems of
transmitting classical information through a noisy quantum channel.

\item The transmission of entire subspaces of quantum states through
the channel, keeping entanglement intact. This is likely to be of
interest in applications such as quantum computation, cryptography and
teleportation where superpositions of quantum states are crucial. Such
problems are problems of transmitting coherent quantum information
through a noisy quantum channel.

\end{itemize}

Both these cases are important for specific applications. For each
case, there is great interest in considering different classes of
allowed encodings and decodings. For example, it may be that encoding
and decoding can only be done using local operations and one-way
classical communication. This may give rise to a different channel
capacity than occurs if we allow non-local encoding and decoding. Thus
there are different noisy channel problems depending on what class of
encodings and decodings is allowed.

\item The use of quantum entanglement to construct examples where the
quantum analogue of the classical equation $H(X:Z) \leq H(Y:Z)$ for a
Markov process $X \rightarrow Y \rightarrow Z$, fails to hold
(compare equation  (\ref{eqtn: example 1})).

\item The use of quantum entanglement to construct examples where the
subadditivity property of mutual information,
\begin{eqnarray}
H(X_1,\ldots,X_n : Y_1,\ldots,Y_n) \leq \sum_i H(X_i : Y_i),
\end{eqnarray}
fails to hold (compare equation (\ref{eqtn: example 2})).

\end{enumerate}

There are many more interesting open problems associated with the
noisy channel problem than have been addressed here.  The following is
a sample of those problems which I believe to be particularly
important:
\begin{enumerate}

\item The development of good numerical algorithms for determining the
different channel capacities. If the expressions for channel
capacities involve limits like those in the upper bounds in this
Chapter, it will also be important either to evaluate those limits
analytically, or to know the rate of convergence to those limits to
aid in evaluating them numerically.

\item Estimation of channel capacities for realistic channels. This
work could certainly be done theoretically and perhaps also
experimentally, using the technique of quantum process tomography
discussed in section \ref{sec:quantum_process_tomography}. An
interesting problem is to analyze how stable the determination of
channel capacities is with respect to experimental error.

\item As suggested in subsection \ref{subsec: other encoding
protocols} it would be interesting to see what channel capacities are
attainable for different classes of allowable encodings, for example,
encodings where the encoder is only allowed to do local operations and
one-way classical communication, or encodings where the encoder is
allowed to do local operations and two-way classical communication.
We have seen how to prove bounds on the channel capacity in these
cases; whether these bounds are attainable is unknown.

\item The development of rigorous general techniques for proving
attainability of channel capacities, which may be applied to different
classes of allowed encodings and decodings.

\item Finding the capacity of a noisy quantum channel for classical
information; considerable progress on this problem has already been
made \cite{Holevo96a,Schumacher97a}, however much remains to be
done. A related problem arises in the context of {\em superdense
coding}, where one half of an EPR pair can be used to send two bits of
classical information. It would be interesting to know to what extent
this performance is degraded if the pair of qubits shard between
sender and receiver is not an EPR pair, but rather the sharing is done
using a noisy quantum channel, leading to a decrease in the number of
classical bits that can be sent. Given a noisy quantum channel, what
is the maximum amount of classical information that can be sent in
this way?

\item All work done thus far has been for discrete channels, that is,
channels with finite dimensional state spaces. It is an important and
non-trivial problem to extend these results to channels with infinite
dimensional state spaces.

\end{enumerate}

There are many other ways the classical results on noisy channels have
been extended - considering channels with {\em feedback}, developing
{\em rate-distortion} theory and so on. Each of these could give rise
to highly interesting work on noisy quantum channels. It is also to be
expected that interesting new questions will arise as experimental
efforts in the field of quantum information develop further. My own
chief interest to us is to develop a still clearer understanding of
the essential differences between the quantum noisy channel and the
classical noisy channel problem, and to provide an effective procedure
for evaluating and achieving the quantum channel capacity.

\newpage
\vspace{1cm}
\begin{center}
\fbox{
\parbox{14cm}{
\begin{center}
{\bf Summary of Chapter \ref{chap:capacity}: The quantum channel capacity}
\end{center}

\begin{itemize}

\item {\bf The quantum channel capacity:} The maximum rate at which
quantum information can be sent through a noisy quantum channel.

\item \textbf{Coherent information:}
$$
I(\rho,\evop) \equiv S(\rho')-S(\rho,\evop).$$
Behaves in many ways as a quantum analogue to the classical mutual
information. 

\item \textbf{Failure of the data pipelining inequality for the
coherent information:}
$$
I(\rho,{\cal E}_2 \circ {\cal E}_1) \not\leq 
I({\cal E}_1(\rho),{\cal E}_2).
$$
The extent to which this inequality is violated is a lower bound on
the entanglement between $E_1''$ and $E_2''$.

\item \textbf{Failure of subadditivity for the coherent information:}
$$
I(\rho_{12},{\cal E} \otimes {\cal E}) \not \leq 
	I(\rho_1,{\cal E}) + I(\rho_2,{\cal E}). $$

\item \textbf{General bound on quantum channel capacity:}
$$
C({\cal N}) \leq \lim_{n \rightarrow \infty} \max_{\rho^n,{\cal C}^n}
\frac{I(\rho^n,{\cal N}^{\otimes n} \circ {\cal C}^n}{n}. $$
It can be shown \cite{Barnum98b} that the maximization over encodings
${\cal C}^n$ is unnecessary, and can be removed.

\item \textbf{The observed quantum channel:} By performing
measurements upon the environment of a quantum system we may be able
to use the result of the measurement to increase the quantum channel
capacity. 

\end{itemize}

}}
\end{center}

\part{Conclusion}

\chapter{Conclusion}
\label{chap:conc}

% summary of the Dissertation
%% summary of achievements and open problems
%% major achievement has been development of numerous new bounds on
%%%%%%%%% qinfo proc, and new methods for proving bounds.
%% Examples:
%%%% Q data compression: New proof of the q data compression theorem.
%%%%%%%%% universal q data compression.
%%%%%%%%% correlated q data comp.
%%%% Q com complexity:
%%%%%%%%% Capacity theorem for dist computation. ASk about it in the
%%%%%%%%% presence of noise.
%%%%%%%%% Proof that n bits is optimal for inner product. 
%%%%%%%%% General bounds on comm comp.
%%%% Measures of entanglement:
%%%%%%%%% many, many new bounds. Related all this stuff to classical
%%%%%%%%% info, Holevo, and the coherent info.
%%%% Q error correction: info theoretic approach to q error
%%%%%%%%% correction.
%%%%%%%%%bounds on the q channel capacity. Developed new techniques
%%%%%%%%%and upper bounds, but did not find an expression for the q
%%%%%%%%%channel capacity. Still to be done. One thing done was to
%%%%%%%%%make it clear how many complex the problem is... give examples.
%%%%%%%%%Also, got a nice discussion of the thermodynamics of quantum
%%%%%%%%%error correction.
%%%%% More than anything else this has been a Dissertation about the
%%%%% development of tools: qops, entropy, and fidelity.

%open problems
%%
%% Ultimate power of qinfo processing
%%%%% P vs NP. PSPACE result.
%%%%% NPI. What we need in order for qinfo to be interesting.
%% Pseudo-randomness, and statistical physics. Phase transitions.

This concluding Chapter briefly surveys some future directions in
quantum information theory. We begin by summarizing the results of the
Dissertation, with an emphasis on novel results, and new problems
suggested by this research.  We conclude by broadening our net to look
at the wider field of quantum information, suggesting some possibly
profitable directions for future research.

\section{Summary of the Dissertation}

We begin by reviewing what has been achieved in the Dissertation, and
open problems arising directly as a result of this research. The major
achievement of the Dissertation is the discovery of numerous bounds on
our ability to do quantum information processing, and the development
of new techniques for proving such bounds.

% part I

Part I of the Dissertation reviewed some of the tools necessary to
make progress in quantum information theory, especially, the quantum
operations formalism, the properties of entropy in quantum mechanics,
and distance measures for quantum information. In many ways it was
gaining a good understanding of these tools that I expect to be the
most useful aspect of doing the research that led to this Dissertation
over the long term, although the research results in Part I are of a
fairly diffuse nature; the primary purpose of Part I is pedagogical.

% part II

Part II of the Dissertation consisted largely of original research on
problems in quantum information theory, focusing on the proof of
bounds to what tasks are possible in quantum information theory, using
the tools introduced in Part I of the Dissertation.

%%%% Q com complexity:
%%%%%%%%% Capacity theorem for dist computation. ASk about it in the
%%%%%%%%% presence of noise.
%%%%%%%%% Proof that n bits is optimal for inner product. 
%%%%%%%%% General bounds on comm comp.

In Chapter \ref{chap:qcomm} we investigated quantum communication
complexity, the study of the communication requirements involved in
distributed quantum computation. Holevo's theorem was used to prove a
new capacity theorem which encapsulates the limits to communication of
classical information between two parties when a two-way noiseless
quantum channel is available for use between the parties. A
generalization of this result to the case of channels with noise would
be of great interest. The capacity theorem was then applied to prove
that the availability of a noiseless quantum channel does not assist
in the calculation of the inner product of two bit strings, when one
of those bit strings belongs to one party, and the other bit string
belongs to a second party. This is a significant result, as it tells
us that there are problems in communication complexity for which
quantum mechanics provides no advantage over the classical result.
The Chapter also contained the first results in coherent quantum
communication complexity, which deals with the communication
requirements incurred when computing a \emph{unitary operator} in a
distributed fashion.  We were able to show that the computation of the
quantum Fourier transform over $2n$ qubits, $n$ of them belonging to
one party, and $n$ to a second party, requires at least $n$ qubits of
communication between those parties.  A general lower bound on the
coherent communication complexity was proved, which we may expect to
be of great assistance in future investigations of the coherent
communication complexity.  Finally, the beginnings of a framework
which unifies previous work on quantum communication complexity was
sketched.  Most importantly, it includes as special cases both the
coherent communication complexity, and the communication complexity of
a classical function, using quantum resources.

% quantum data compression

In Chapter \ref{chap:data_compress} we studied the compression of
information from a quantum source. A new proof was given of the
quantum data compression theorem, which gives an {\em operational
interpretation} of the von Neumann entropy $S(\rho)$ as the minimal
number of qubits with which it is possible to reliably store a quantum
source described by the density operator $\rho$. The new techniques
introduced during the proof were then applied to the problem of
\emph{universal quantum data compression}, a hoped-for technique which
provides a means for compressing quantum information even in the
absence of knowledge about a quantum source's characteristics.  We
constructed a method for performing a potentially useful form of
universal quantum data compression on large class of quantum sources,
although it remains to find an efficient quantum algorithm for
implementing this procedure for universal quantum data compression.

%%%% Measures of entanglement:
%%%%%%%%% many, many new bounds. Related all this stuff to classical
%%%%%%%%% info, Holevo, and the coherent info.

Chapter \ref{chap:ent} focused on the central problem of providing
good quantitative measures of entanglement. Entanglement appears to be
a central resource in most quantum information processing tasks known
to date. This Chapter focused on one particular measure of
entanglement, the {\em entanglement of formation}, a measure of how
many Bell states it takes to create a particular entangled state.  The
most important result of the Chapter was a relationship between the
entanglement of formation and the quantum conditional entropy, $E(A:B)
\geq -S(A|B)$.  In particular, this shows that the puzzling phenomena
of negative quantum conditional entropies are always associated with
the presence of entanglement in a quantum system.  
%We also defined the
%\emph{dynamic entanglement}, a measure of the entanglement processed
%by a quantum channel, and developed numerous elementary properties of
%the dynamic entanglement. 

%%%% Q error correction: info theoretic approach to q error
%%%%%%%%% correction.
%%%%%%%%%bounds on the q channel capacity. Developed new techniques
%%%%%%%%%and upper bounds, but did not find an expression for the q
%%%%%%%%%channel capacity. Still to be done. One thing done was to
%%%%%%%%%make it clear how many complex the problem is... give examples.
%%%%%%%%%Also, got a nice discussion of the thermodynamics of quantum
%%%%%%%%%error correction.
%%%%% More than anything else this has been a Dissertation about the
%%%%% development of tools: qops, entropy, and fidelity.

Chapter \ref{chap:qec} introduced the basic notions of quantum error
correction.  After reviewing the basic ideas of quantum error
correction using the Shor nine qubit code, we developed
information-theoretic necessary and sufficient conditions for quantum
error correction, using a quantum analogue of the classical data
processing inequality.  Next, we analyzed the thermodynamic cost of
quantum error correction, showing that quantum error correction
schemes function as a kind of Maxwell's demon, in which information is
extracted from a system in order to lower its entropy.  We were able to
show that quantum error correction can be performed in a
thermodynamically efficient manner.

Chapter \ref{chap:capacity} studied the problem of the quantum channel
capacity.  The channel capacity measures how much quantum information
can be sent through a noisy quantum channel.  We developed a bound on
this quantity based upon the coherent information, and explained some
of the outstanding problems related to the channel capacity. The
problem of the \emph{observed quantum channel} was introduced, and we
explained how it can be reformulated in purely quantum terms.  Upper
bounds on the channel capacity of an observed quantum channel were
proved, again using the coherent information.

% problems

What are the most important outstanding problems arising directly from
the dissertation?  Perhaps the most immediately fruitful areas for
future research are quantum communication complexity and the
understanding of entanglement.  

With regard to quantum communication complexity, the general lower
bound technique for quantum communication complexity which I proved in
Chapter \ref{chap:qcomm} can doubtless be generalized, and applied to
many interesting problems.  It would, for example, be useful to
understand the communication costs in performing quantum Fourier
transforms over groups more general than the integers modulo $2^n$, as
we considered, or to consider the communication costs incurred when
doing quantum error correction.  More generally, little is presently
known about the relationship between quantum and classical
communication complexity.  Developing these connections more deeply is
an obvious area for further work.

The study of entanglement is a second area that I believe will yield
rich results over the next few years.  Among many possible avenues of
research, I am especially interested in pursuing the behaviour of
entanglement in systems in thermodynamic equilibrium, and trying to
understand how the entanglement behaves near some of the phase
transitions which may occur in such systems.

The other topics addressed by the Dissertation -- quantum data
compression, quantum error correction, and the channel capacity --
also suggest many interesting problems.  Developing a more complete
understanding of universal quantum data compression is a worthwhile
goal, and may be of practical importance in the future.  I am also
actively investigating the problem of data compression of correlated
quantum sources in an attempt to find a quantum analogue of the
Slepian-Wolf theorem of classical information theory \cite{Cover91a};
these results are incomplete, and were not included in the
Dissertation.  With regard to error correction and the channel
capacity, perhaps the most outstanding problem is to better understand
the quantum channel capacity, and to develop a general procedure for
evaluating it, analogous to Shannon's noisy channel coding theorem.
This is a fascinating, albeit apparently quite a difficult problem,
whose solution I expect will give us great new insight into quantum
information.

% summarizing

Summarizing, in this Dissertation I have discovered many new limits to
the ability to perform information processing within quantum
mechanics.  Many of these limitations are concerned with multiple
parties, where some bound is placed on their communication
requirements.  These result in practical limits on the ability of two
parties to compress quantum data, to communicate classical data using
quantum resources, to compute the inner product function and the
quantum Fourier transform, to perform quantum error correction, and to
send quantum information through a noisy quantum channel.  More
general theoretical results have been proved which give general though
not always practically applicable bounds on the ability to perform
distributed quantum computations, and on the ability to send quantum
information through a noisy quantum channel.  Moreover, new tools and
techniques have been developed while solving these problems that will
be of great use in further investigations of quantum information
theory.  Perhaps most importantly, though, many interesting new
problems have been raised.  We now turn to look more broadly at the
problems facing quantum information theory at the present time.

\section{Open problems in quantum information}

In the last section we reviewed the achievements of this Dissertation,
and some of the open problems arising directly from this research. In
this section we discuss in a broader setting some of the challenges
facing quantum information theory. Several simply stated problems may
be identified as especially important:
\begin{enumerate}
\item Develop computationally interesting new applications of quantum
information.
\item What are the ultimate achievable limits to quantum information
processing?  Several subproblems may be identified:
\begin{itemize}
\item What class of problems may be solved efficiently on a quantum
computer? How does this class compare to the class of problems
efficiently soluble on a classical computer?
\item What resources are required to do distributed quantum
computation?
\item In a multi-party situation where not all parties trust one
another, what resources are required to do information processing
tasks with a reasonable level of security?
\end{itemize}
\item What technologies are needed to implement quantum information
processing? Does it have economically practical applications?
\item Can quantum information shed new light on the problems of
fundamental physics?
\item Can other physical theories be used to do information processing
tasks beyond the quantum computational model?
\end{enumerate}

One of the fun and exciting things about quantum information is that
we are still at the point where simple, fundamental questions like
these can be asked, without the answers being known.  Precious little
is known about the computational power of quantum mechanics.
Remedying this situation offers many interesting challenges.

A discussion of directions to take in solving the listed problems
would be enormous.  Instead, I will focus on three directions which I
believe are especially promising as first steps if we are to solve
these problems.

The first of these directions is inspired directly by the results in
this Dissertation.  In subsection \ref{subsec:unifying} I will sketch
out a formalism which can be used to unify several of the disparate
approaches to quantum information theory which have been developed.
In particular, results such as the Holevo theorem which relate to
sending classical resources through quantum channels, appear to have
little relationship to questions of manipulating quantum information
in quantum channels.  I sketch out the beginnings of a formalism
which can unite these approaches to quantum information theory.

The second direction which we look at is the so-called ``decoherence
program'', which aims to explain how classical physics arises as the
limit to quantum physics.  Throughout this Dissertation, we have
assumed that there are two fundamental units of information, one
classical (the ``bit''), one quantum (the ``qubit'').  It appears as
though Nature does not respect such a duality at the fundamental
level.  Rather, classical information arises as the limit of quantum
information under certain special circumstances.  Understanding how
this occurs in more detail is the goal of the decoherence program.  In
subsection \ref{subsec:decoherence} I ask whether quantum information
theory can information this program.

The third direction which we will examine is whether there are
interesting cross-disciplinary advances to be made between quantum
information theory and statistical physics.  This is the most
speculative and the most sketchy of all the proposals made here.
Nevertheless, I feel it is a direction well worth exploring.

\subsection{A unifying picture for quantum information}
\label{subsec:unifying}
\index{unifying picture for quantum information}

% The need for a unifying formalism.
% Example: Schumacher and Westmoreland.
% Properties which a unifying formalism ought to have.
%%% describe an ensemble of mixed states produced by a
%%%      classic source.
%%% describe an state which is actually entangled with another system.
%%% described the classical msment results, allowing for non-ideal
%%%      msmnts.
%%% Describe a closed universe: the total state should be pure at all
%%%      times.
%%% Describe the emergence of a classical world.
%%% Describe multi-part processes.
%%% Explain how classical information arises as a limit to quantum
%%%      information. 
%
% we can do this.
%%% explanation of the formalism.
%%% Schumacher and Westmoreland as an example.

This Dissertation has explored many parts of quantum information
theory.  Despite considerable effort on my part, I did not find it
possible to present the results of all chapters within a single,
unified picture.  Compare, for example, Chapter
\ref{chap:data_compress}, on quantum data compression, with Section
\ref{sec:Holevo}, on the Holevo bound.  Although similar tools are
used in each instance, it is not immediately apparent that the
approaches to both problems fit within a single, unified approach to
quantum information theory. Nowhere is this lack of a unified approach
more apparent than in the existence of two different pictures, one to
handle the problem of sending classical information, using quantum
resources, the other to handle the problem of sending quantum
information using quantum resources.

This is a problem which exists throughout the published literature on
quantum information.  Recently, an interesting paper by Schumacher and
Westmoreland \cite{Schumacher98a} has appeared which demonstrates a
link between the two approaches to quantum information theory.
Broadly speaking, Schumacher and Westmoreland draw our attention to a
connection between a quantity based on the Holevo $\chi$ which
measures how much {\em classical information} can be sent, with {\em
complete privacy}, through a quantum channel, and the {\em coherent
information}, which we have seen is related to how much {\em quantum
information} can be sent through a channel.

This work has stimulated me to think about whether a unified picture
for the description of quantum information might be found.  I believe
I have found the beginnings of such a picture, which I will sketch in
this section.  Nevertheless, considerable work remains to be done
before the new picture can be considered complete.  A partial
statement of the goals of such a picture is that it ought to be able
to achieve {\em all} of the following in an integrated manner:
\begin{enumerate}
\item Describe an ensemble of (potentially mixed) states being
produced by a classical source.
\item Describe a source which represents an entanglement with a
reference system.
\item Describe (potentially multi-part) dynamical processes.
\item Describe the classical results of (possibly non-ideal)
measurements.
\item Describe a closed universe in which the state of the total
system is pure at all times.
\item Describe how classical information theory and the classical
world arise as a limit of quantum information theory and the quantum
world.
\end{enumerate}

This seems like a rather lengthy list of features to require in a
single picture!  The picture I describe will seem somewhat
complicated.  Nevertheless, it offers a remarkably simple way for
rederiving results such as the Schumacher and Westmoreland result.

Suppose a classical source is producing quantum states $\rho_x$
according to some probability distribution $p_x$.  Define $\rho \equiv
\sum_x p_x \rho_x$.  The following construction enables us to describe
the classical source, together with an entangled source producing the
state $\rho$, all within the one formalism.

Let $Q \equiv Q_1$ be the quantum system under consideration.  Let
$Q_2$ be a copy of system $Q$, and let the states $|Q_1Q_2\ra_x$ be
purifications of the states $\rho_x$.  Under some circumstances it
may be advantageous to require additional properties of the
purifications $|Q_1Q_2\ra_x$\footnote{For example, we might enlarge
$Q_2$, and required that the states $|Q_1Q_2\ra_x$ be mutually
orthogonal. Or we might require that the reduced states on system
$Q_2$, $Q_{2x}$, are equal to the original states $\rho_x$ of
$Q_1$. Other possibilities may also be useful.}, but we will not need
such additional properties here.

We also introduce two systems, $P_1$ and $P_2$, each of which has an
orthonormal basis $|x\ra$ of states in one-to-one correspondence with
the outputs which may be produced by the classical source.  Defining
$P \equiv P_1$, we will refer to the system $P$ as the {\em
preparation system}, since it will be used to encode information about
which of the state $\rho_x$ has been prepared.

Notationally, we will write states of the system $P_1P_2Q_1Q_2$ in the
order $P_1P_2Q_1Q_2$, unless otherwise noted.  The state of the system
representing both the classical source $\{ p_x, \rho_x \}$ and the
entangled source $\rho$ is given by
\be
|P_1P_2Q_1Q_2\ra = \sum_x \sqrt{p_x} |x\ra|x\ra|Q_1Q_2\ra_x. \ee
Note that requirement number 5 is met -- this state is pure.

Note that \be PQ = \sum_x p_x |x\ra \la x| \otimes \rho_x. \ee
Intuitively, the system $PQ$ therefore describes a classical system
$P$ which is in one of the orthonormal states $|x\ra$, with respective
probabilities $p_x$, and which prepares a corresponding state of
system $Q$, $\rho_x$.  Compare this with the similar construction in
the proof of Holevo's bound, section \ref{sec:Holevo}.  That is,
requirement 1 is met.

To see that requirement 2 is met, define the \emph{reference system}
$R \equiv P_1P_2Q_2$.  Notice that $RQ$ is a purification of the
system $Q$, which starts in the state $Q = \sum_x p_x \rho_x$.

Requirements 3 and 4 require that we introduce additional
systems. Suppose $\evop$ is a complete quantum operation that may
occur on the system.  This quantum operation may arise, potentially,
as the result of a measurement described by quantum operations
$\evop_m$, $\evop = \sum_m \evop_m$.  Suppose $\evop_m$ has an
operator-sum representation generated by operators $E_{mi}$.
Introduce a system $M$ with an orthonormal basis $|m\ra$, a system $I$
with an orthonormal basis $|i\ra$, and a system $E$ with an
orthonormal basis $|m,i\ra$.  Let $|0_M\ra, |0_I\ra, |0_E\ra$ be
standard pure states of the respective systems $M, I$ and $E$.  Define
a unitary operator $U$ on $QMIE$ which has the action 
\be
U|\psi\ra|0\ra|0\ra|0\ra \equiv \sum_{mi} E_{mi} |\psi\ra
|m\ra|i\ra|m,i\ra. \ee
Intuitively, the system $M$ plays the part of a ``measuring
apparatus'', which records the result of the measurement.  The system
$I$ represents information which is lost when the measurement is
performed.  Finally, the system $E$ can be thought of as an
``environment'', which decoheres the measuring apparatus, in the sense
of Zurek \cite{Zurek91a}.

In the language used to prove the Holevo bound in section
\ref{sec:Holevo}, $M$ plays the same role here as it does there -- it
stores the result of the measurement.  The joint system $IE$ plays the
same role as the system $E$ did in section \ref{sec:Holevo}.  Finally,
the combined system $MIE$ plays the role of a model environment for
the operation $\evop = \sum_m \evop_m$, just as was used throughout 
the Dissertation.

Thus, this formalism encompasses all the constructions contained in
this Dissertation, from constructions in which there is a classical
source of information, as in the Holevo bound, through to results such
as the data processing inequality, which deal with the transmission of
entanglement through a quantum channel.  Indeed, part of the great
attraction of this formalism is the number of results which it gives
you \emph{automatically}.  In section \ref{sec:other_inequalities} we
were able to obtain numerous entropy inequalities, simply by applying
the subadditivity and strong subadditivity inequality in a mindless
fashion.  In a similar way, one can obtain the Holevo bound, the data
processing inequality, the result of Schumacher and Westmoreland
linking the Holevo bound and coherent information
\cite{Schumacher98a}, and many other results, all free-of-charge, as
automatic consequences of the formalism and a few powerful tools such
as strong subadditivity.  For this reason, I believe that this
formalism will be a powerful tool to aid in answering new questions
about the connection between classical and quantum information.

\subsection{Classical physics and the decoherence program}
\label{subsec:decoherence}
\index{classical physics}
\index{decoherence}

% Zurek's work.
%% interaction with an environment.
%% the decoherence program.
%% much is understood in broad outline.
%% several key problems:
%%% How systems emerge.
%%% In what sense is the measuring process irreversible?
%%% what does it mean to say that system A knows something about
%%%    system B?

Since the foundation of quantum mechanics there has been intense
interest in understanding how the classical world we see in our
everyday life arises out of the underlying quantum reality.
This effort was particularly intense in the early years, with
researchers such as Bohr \cite{Bohr28a} stressing general principles
which linked quantum and classical physics, and researchers such as
Mott \cite{Mott29a} who did detailed investigations of specific
phenomena in order to explain how the classical world arises from the
quantum reality.

Since those early days there has been a continual effort to understand
the connection between quantum reality and the classical world.
Unfortunately, this work suffered considerably because of a lack of
experimental progress at the level of single quantum systems.  With a
few notable exceptions, much of the work in this area got bogged down
in a morass of theory and philosophy, with only a few notable pieces
of science emerging from sixty-odd years of work. 

Over the past twenty years or so there has been tremendous
experimental progress at the level of single quantum systems.  To cite
but one important example, there was the development in the mid 1980s
of the {\em quantum jumps} technique, a technique for doing projective
measurements on a single ion in an ion trap
\cite{Nagourney86a,Sauter86a,Bergquist86a}, based upon a Technical
Report written by Dehmelt in the mid 1970s, but never published.

This experimental Renaissance in studying foundational issues in
quantum mechanics has been matched by a similar theoretical
Renaissance.  A particularly broad program, sometimes known under the
rubric of the ``decoherence program'' has been advanced by Zurek and
other researchers, starting in the early 1980s
\cite{Zurek81a,Zurek82a}.  Reviews of this material may be found in
\cite{Zurek98b,Zurek91a}.  The decoherence program is an attempt to
give a detailed explanation for how classical behaviour arises from
quantum reality.  Much of this problem is now well understood, at
least in outline, but some fundamental problems remain.

How may the tools of quantum information be brought to bear upon the
decoherence program?  I do not know how to answers this question, but
I do have a number of problems which I would like solved:
\begin{enumerate}
\item What does it mean to say that system $A$ ``objectively knows''
something about system $B$?  Answering this question in a quantitative
fashion ought to give us a much better handle on the old problem of
determining, from first principles, what physical systems function as
measuring devices, in the sense of inducing a collapse of the state
vector.
\item Zurek \cite{Zurek93a} has proposed what he refers to as the
``predictability sieve''.  This is a proposal intended to solve the
following fundamental problem: given a unitary interaction between a
quantum system and a measuring device, there are many possible
``collapse'' rules for the quantum system consistent with the
non-selective unitary interaction between system and measuring device.
Zurek gives a prescription for determining which of the collapse rules
consistent with the unitary interaction is actually taking place.
This prescription uses the von Neumann entropy to determine the ``most
classical'' set of quantum states possessed by the measuring
apparatus.  Unfortunately, the motivation for using the von Neumann
entropy in this context has never been physically clear.  It would be
interesting to approach the predictability sieve from the point of
view of quantum information theory, and to ask what is the appropriate
quantity for measuring the ``classicallity'' of a set of states of
the measuring device.
%Classically, I am not entirely
%sure how to go about answering the question, but I believe a
%combination of Shannon's theory and algorithmic information theory
%\cite{Solomonoff64a,Solomonoff64b,Kolmogorov65a,Chaitin66a}
%immediately suggests several plausible solutions to the problem.  In
%the quantum case, I have vastly fewer .
\item Zurek \cite{Zurek98b} has asked why it is that a composite,
``system'' structure exists in nature?   This structure is crucial to
the success of the decoherence program, but it has never been
explained why it is found in nature.  It is not obvious that this
question is immediately related to quantum information theory,
however, it is as well to keep what is perhaps the most significant unsolved
problem of the decoherence program in mind while pursuing an
information-theoretic approach to decoherence.
\end{enumerate}

\subsection{Quantum information and statistical physics}

An observation that has recurred in my thoughts for several years now
is that statistical physics and the theory of computation seem like
two different approaches to a very similar problem.  In both cases we
are trying to determine the long-time behaviour of a dynamical system
whose constituent parts behave according to simple rules.

This observation is, apparently, not without foundation in fact, for I
have recently learned that there are, in fact, deep connections which
can be made between computation and statistical physics.  In
particular, it has been shown that certain problems in statistical
physics are NP-complete -- a computer science term for a class of
problems that is believed to be intractable, at least within classical
computational models \cite{Freedman98b}.  The class of NP-complete
problems is tremendously important, containing as it does many of the
most important problems in computer science.  All NP-complete problems
are, essentially, equivalent in terms of computational difficulty.

It is interesting to ask, then, whether similar results hold relating
difficult problems for quantum computers to problems in quantum
statistical mechanics?  How difficult is it to predict the long-time
behaviour of a quantum system such as a spin glass?  What are the
transport properties of entanglement in such a system? Can we relate
the existence of phase transitions in quantum statistical mechanics to
entanglement between the constituent parts?  Answering questions such
as these has the potential benefit of not just illuminating aspects of
quantum information theory, but also other areas of physics.

%An area that I believe is ripe for exploration is the intersection of
%computer science with statistical mechanics.  This is an area of
%research that has received some considerable attention over the past
%twenty years (see, for instance, the influential paper of Baskaran
%{\em et al} \cite{Baskaran86a}), however I believe there are
%significant problems that remain untouched.

%Conventional quantum statistical mechanics is based upon the partition
%function. This function is completely determined by the energies of
%the system under consideration, and does not depend at all on whether
%the corresponding 

\section{Concluding thought}

% quinfo gives us a _long-term goal_.
% promise to build the world's most powerful computational devices,
% perhaps build machines with powers of comprehension that dwarf our
% own, illuminate and stimulate thoughts about fundamental physics.

Quantum information, and more generally, the physics of information,
offers tremendous opportunities.  We may be able to harness the laws
of physics to perform fantastic computations impossible within the
classical laws.  It may even be that computing devices harnessing the
full power of physics will be able to achieve a fuller comprehension
of the world than our classically-limited minds.  The physics of
information stimulates us to ask new questions about the foundations
of the physical world we live in.  The exploration of these
possibilities is a deep and beautiful problem, full of challenges to
stimulate and awe the mind that contemplates them, the hand that
realizes them, and any being that makes practical use of their
eventual fruits.

\appendix
\chapter{Purifications and the Schmidt decomposition}
\label{app:mixed}

\index{purification}
\index{Schmidt decomposition}

Composite quantum systems are used throughout this Dissertation. In
order to get a better grasp of the properties of composite systems we
need tools to understand the states of composite quantum systems. Two
of the most useful tools for doing this are the {\em Schmidt
decomposition}, and {\em purifications}. In this Appendix we will
review both these tools, and try to give a flavour of their power.
The first part of the Appendix gives a review of these results in
their standard form; the second part of the Appendix gives a new
generalization of the Schmidt decomposition for which, unfortunately,
I have not yet been able to find any interesting applications.

\begin{theorem} {(\bf Schmidt decomposition)} \cite{Schmidt06a}

Suppose $|AB\ra$ is a pure state of a composite system, $AB$.
Then there exists orthonormal states $|i_A\ra$ for system
$A$, and orthonormal states $|i_B\ra$ of system $B$ such that
\beqn
|AB\ra = \sum_i \lambda_i |i_A\ra |i_B\ra, \eeqn where $\lambda_i$ are
positive real numbers satisfying $\sum_i \lambda_i^2 = 1$.

\end{theorem}

This innocuous looking theorem is tremendously useful. As a taste
of its power, consider the following consequence: let $|AB\ra$
be a pure state of a composite system, $AB$. Then the eigenvalues of
$A$ and $B$ are identical, namely $\lambda_i^2$ for both
density operators. Many important properties of quantum systems 
are completely determined by the eigenvalues of the system. For
a pure state of a composite system such properties will therefore
be the same for both systems.  As an example, consider that the von
Neumann entropy of a state is completely determined by the eigenvalues
of that state.  Therefore, for a pure state $|AB\ra$ of system $AB$,
the von Neumann entropy of systems $A$ and $B$ are the same.

\begin{proof}

Let $A$ be the state of system $A$ when system $B$ is traced
out, $A \equiv \tr_B(|AB\ra\la AB|)$. Let
\beqn
A = \sum_i p_i |i_A\ra \la i_A| \eeqn
be an orthonormal decomposition for system $A$. Then there
exist vectors $|\psi_i^B\ra$ in the state space of system $B$
such that
\beqn
|\psi\ra = \sum_i |i_A\ra|\psi^i_B\ra. \eeqn
But we know that $A = \tr_B(|AB\ra \la AB|)$, from which
we deduce $\la \psi_i^B|\psi_j^B\ra = \delta_{ij} p_i$. Thus, we can
find orthonormal $|i_B\ra$ such that $|\psi_i^B\ra = \sqrt{p_i} |i_B\ra$,
and thus
\beqn
|AB\ra = \sum_i \sqrt{p_i} |i_A\ra |i_B\ra. \eeqn
Setting $\lambda_i \equiv \sqrt{p_i}$ and noting that
$\sum_i \lambda_i^2 = \sum_i p_i = 1$ completes the proof.

\end{proof}

The bases $|i_A\ra$ and $|i_B\ra$ are called the {\em Schmidt
bases}\index{Schmidt bases} for $A$ and $B$, respectively, and the
number of non-zero values $\lambda_i$ is called the {\em Schmidt
number}\index{Schmidt number} for the state $|\psi\ra$. The Schmidt
number is a tremendously important property of a composite quantum
system. To get some idea of why this is the case, consider the
following obvious but important property: the Schmidt number is
preserved under unitary transformations on system $A$ or system $B$
alone. To see this, notice that if $\sum_i \lambda_i |i_A\ra|i_B\ra$
is the Schmidt decomposition for $|\psi\ra$ then $\sum_i \lambda_i
(U|i_A\ra) |i_B\ra$ is the Schmidt decomposition for $U|\psi\ra$,
where $U$ is a unitary operator acting on system $A$ alone. Invariance
properties of this type are very useful in Chapter
\ref{chap:data_compress}.

Another useful technique in quantum information theory is {\em
purification}. Suppose we are given a state $A$ of a quantum system
$A$. It is possible to introduce another system, which will call $R$,
and {\em define} a {\em pure state} $|AR\ra$ for the joint system $AR$
such that $A = \tr_R(|AR\ra\la AR|)$. That is, the pure state $|AR\ra$
reduces to $A$ when we look at system $A$ alone.  This is a purely
mathematical procedure, known as {\em purification}, which allows us
to associate pure states with mixed states. For this reason we call
system $R$ a {\em reference} system: it is a fictitious system,
without a direct physical significance.

To prove that purification can be done for {\em any} mixed state, we
will construct a system $R$ and purification $|AR\ra$ corresponding to
the state $A$.  Suppose $A$ has spectral decomposition, $A = \sum_i
p_i |i\ra \la i|$.  In order to purify $A$ we introduce a system
$R$ which has the same state space as system $A$, and define a pure
state for the combined system \beqn |AR\ra \equiv \sum_i
\sqrt{p_i} |i\ra |i\ra. \eeqn We now calculate the reduced density
operator for system $A$ associated with $|AR\ra$: \beqn
\tr_R(|AR\ra\la AR|) & = & \sum_{ij} \sqrt{p_ip_j} |i\ra
\la j| \tr(|i\ra \la j|) \\ & = & \sum_{ij} \sqrt{p_i p_j} |i\ra \la
j| \delta_{ij} \\ & = & \sum_i p_i |i\ra \la i| \\ & = & A. \eeqn
Thus $|AR\ra$ is a valid purification of $A$.

Notice the close relationship of the Schmidt decomposition to purification:
the procedure used to purify a mixed state is to define a pure state whose
Schmidt basis is just the basis in which the mixed state is diagonal.
A related observation is that the Schmidt decomposition may be used to
obtain a classification of purifications of the state $A$.  First of
all, note that the Schmidt decomposition implies that if $|AB\ra$ is a
purification of $A$ then $B$ must contain at least as many dimensions
as $A$ has support dimensions.  Suppose $|AB\ra$ and $|AB'\ra$ are two
purifications of $A$. Then by the Schmidt decomposition
\be
|AB\ra & = & \sum_i \sqrt{p_i} |i_A\ra |i_B\ra \\
|AB'\ra & = & \sum_i \sqrt{p_i} |i_A\ra |i_B'\ra, \ee
where $p_i$ are the eigenvalues of $A$, $|i_A\ra$ the corresponding
eigenvectors, and $|i_B\ra$ and $|i_B'\ra$ are each a set of
orthonormal vectors in system $B$.  Since $|i_B\ra$ and $|i_B'\ra$ are
both orthonormal sets, it follows that there exists a unitary operator
$U_B$ on system $B$ such that $U_B|i_B\ra = |i_B'\ra$, and therefore
\be
|AB\ra = U_B |AB'\ra. \ee
Conversely, if $|AB\ra$ is a purification of $A$ and $U_B$ is a
unitary operator on $B$ then $|AB'\ra$ defined by $|AB'\ra \equiv U_B
|AB\ra$ is easily verified to be a purification of $A$.  Thus $|AB\ra$
and $|AB'\ra$ are purifications of $A$ if and only if there exists a
unitary operator on $B$ relating the two states.

\index{Schmidt decomposition!for mixed states}
\index{mixed state Schmidt decomposition}

To conclude the Appendix, we present a new generalization of the
Schmidt decomposition to mixed states of a two-part composite system.
I have not found any applications of this result, which is why this is
an Appendix, and not a Chapter, however I am hopeful that in the
future it may prove useful.

Suppose $\rho$ is any state of a composite system $AB$. For
convenience we assume that $A$ and $B$ have the same number of
dimensions; if this is not true then it can be made true by appending
extra dimensions to whichever system has fewer dimensions. Suppose
$\rho = \sum_k |k\ra \la k|$, where $|k\ra$ is an orthogonal set, with
the eigenvalues $\la k | k\ra$ of $\rho$ absorbed into the
normalizations of the states $|k\ra$. Similarly, suppose $A =
\sum_i |i\ra \la i|$ and $B = \sum_j |j\ra \la j|$, where $|i\ra$
and $|j\ra$ are orthogonal sets, with the eigenvalues of $A$ and
$B$ absorbed into the normalizations of $|i\ra$ and $|j\ra$.

Our goal here is to take $A$ and $B$ as given density operators for
systems $A$ and $B$, and to derive a set of algebraic constraints on
the matrices $a^k$, in order that $\rho$ be an allowed density
operator for the system $AB$, which when system $A$ is traced out
reduces to $B$, and when system $B$ is traced out, reduces to $A$. The
result we obtain gives as a special case the Schmidt decomposition for
pure states.  The key to doing this is to relate the sets $|i\ra,
|j\ra$ and $|k\ra$.  Note first that each $|k\ra$ must be a linear
combination of the $|i\ra$s and the $|j\ra$s, \beqn |k\ra = \sum_{ij}
a^k_{ij} |i\ra|j\ra. \eeqn

Tracing out system 
$B$ from the expression $\rho = \sum_k |k\ra \la k|$ we see that
\beqn
\sum_i |i\ra \la i| = \sum_k \sum_{i i' j} q_j a^k_{ij} (a^k_{i' j})^* |i\ra 
\la i'|, \eeqn
where $q_j \equiv \la j| j\ra$ are the eigenvalues of $B$. Defining $Q$ to
be a diagonal matrix with entries $q_j$, we can rewrite the previous equation as
\beqn
\sum_k a^k Q (a^k)^{\dagger} = I. \eeqn
Tracing out system $A$ from the expression $\rho = \sum_k |k\ra \la k|$
we deduce that
\beqn
\sum_j |j\ra \la j| = \sum_k \sum_{i j j'} p_i a^k_{ij} (a^k_{i j'})^* |j\ra 
\la j'|, \eeqn
where $p_i \equiv \la i|i\ra$ are the eigenvalues of $A$. Defining
$P$ to be a diagonal matrix with entries $p_i$, we see that
\beqn
\sum_k (a^k)^{\dagger} P a^k = I. \eeqn
Furthermore, we have the orthogonality relation
\beqn
\la k| k'\ra = \sum_{ij} (a^k_{ij})^* a^{k'}_{ij} p_i q_j = 
\tr((a^k)^{\dagger} P a^{k'} Q ). \eeqn
The trace condition $\tr(\rho) = 1$ is now seen to be equivalent to
$\sum_k \tr((a^k)^{\dagger} P a^k Q) = 1$. However, since
$\sum_k (a^k)^{\dagger} P a^k = I$, the trace condition is equivalent to
$\tr(Q) = I$. This is true by assumption, since $\tr(B) = 1$, so the
trace condition gives nothing new. We have proved the following theorem:

\begin{theorem}  \textbf{(Mixed state Schmidt decomposition)}

 Let $A$ and $B$ be given density operators on systems
$A$ and $B$. Let $P$ be a matrix whose diagonal entries are the 
eigenvalues of $A$, and $Q$ a matrix whose diagonal entries are the 
eigenvalues of $\rho_B$. Then $\rho$ is a density operator of system
$AB$, consistent with $A$ and $B$, if and only if $\rho$ has
orthogonal decomposition
\beqn
\rho = \sum_k |k\ra \la k|, \eeqn
where the $|k\ra$ are defined by
\beqn
|k\ra = \sum_{ij} a^k_{ij} |i\ra |j\ra, \eeqn
and the $a^k$ are complex matrices satisfying the conditions
\beqn
\sum_k a^k Q (a^k)^{\dagger} & = & I \\
\sum_k (a^k)^{\dagger} P a^k & = & I \\
\tr((a^k)^{\dagger} P a^{k'} Q) & = & \la k|k' \ra. \eeqn
\end{theorem}

This result provides a complete characterization of the possible
states $\rho$ of $AB$, in terms of the density operators $A$ and $B$.
Let's see what it tells us in the pure state case. In this case, there
is a single non-trivial value of $k$. We deduce that $a Q a^{\dagger}
= I$ and $a^{\dagger} P a = I$.  Polar decompose $a$ as $a = uh$, for
some unitary $u$ and Hermitian $h$. Note that $ h Q h = h u^{\dagger}
P u h = I$, from which we see that $\det h \neq 0$ and $\det P \neq
0$, and thus $P = u Q u^{\dagger} $ and $h = Q^{-1/2}$. Since $P$ and
$Q$ are diagonal, $u$ must be a permutation matrix. Relabeling the
basis states, we have $P = Q$ and $u=I$. We then have \beqn |k\ra & =
& \sum_{ij} h_{ij} |i\ra|j\ra \\ & = & \sum_{i} \frac{1}{\sqrt{p_i}}
|i^A\ra|i^B\ra. \eeqn The pure state Schmidt decomposition follows by
renormalizing the states$|i^A\ra$ and $|i^B\ra$, which at present
satisfy $\la i^A|i^A\ra = \la i^B|i^b\ra = p_i = q_i$.

\addcontentsline{toc}{chapter}{{\bf Bibliography}}
%\bibliography{bib}

\begin{thebibliography}{100}

\bibitem{Abrams97a}
D.~S. Abrams and S.~Lloyd.
\newblock Simulation of many-body {F}ermi systems on a quantum computer.
\newblock {\em Phys.~Rev.~Lett.}, 79(13):2586--2589, 1997.
\newblock quant-ph/9703054.

\bibitem{Aharonov97a}
D.~Aharonov and M.~Ben-Or.
\newblock Fault tolerant computation with constant error.
\newblock In {\em Proceedings of the Twenty-Ninth Annual ACM Symposium on the
  Theory of Computing}, pages 176--188, 1997.

\bibitem{Allahverdyan97a}
A.~E. Allahverdyan and D.~B. Saakian.
\newblock Converse coding theorems for quantums source and noisy channel.
\newblock {\em \mbox{arXive} e-print quant-ph/9702034}, 1997.

\bibitem{Anderson72a}
P.~W. Anderson.
\newblock More is different.
\newblock {\em Science}, 177(4047):393--396, 1972.

\bibitem{Araki60a}
H.~Araki and M.~M. Yanase.
\newblock Measurement of quantum mechanical operators.
\newblock {\em Phys. Rev.}, 120:622, 1960.

\bibitem{Bardeen57a}
J.~Bardeen, L.~N. Cooper, and J.~R. Schrieffer.
\newblock Theory of superconductivity.
\newblock {\em Physical Review}, 108(5):1175--1204, 1957.

\bibitem{Barenco95a}
A.~Barenco, C.~H. Bennett, R.~Cleve, D.~P. DiVincenzo, N.~Margolus, P.~Shor,
  T.~Sleator, J.~Smolin, and H.~Weinfurter.
\newblock Elementary gates for quantum computation.
\newblock {\em Physical Review A}, 52:3457--3467, 1995.
\newblock quant-ph/9503016.

\bibitem{Barnum96a}
H.~Barnum, C.~M. Caves, C.~A. Fuchs, R.~Jozsa, and B.~Schumacher.
\newblock Noncommuting mixed states cannot be broadcast.
\newblock {\em Physical Review Letters}, 76(15):2828--2821, 1996.
\newblock quant-ph/9511010.

\bibitem{Barnum98b}
H.~Barnum, E.~Knill, and M.~A. Nielsen.
\newblock On quantum fidelities and channel capacities.
\newblock 1998.
\newblock quant-ph/9809010.

\bibitem{Barnum98a}
H.~Barnum, M.~A. Nielsen, and B.~W. Schumacher.
\newblock Information transmission through a noisy quantum channel.
\newblock {\em Phys. Rev. A}, 57:4153, 1998.

\bibitem{Beckman96a}
D.~Beckman, A.~N. Chari, S.~Devabhaktuni, and J.~Preskill.
\newblock Efficient networks for quantum factoring.
\newblock {\em Physical Review A}, 54(2):1034, 1996.
\newblock \mbox{arXive} E-print quant-ph/9602016.

\bibitem{Bell89a}
J.~S. Bell.
\newblock {\em Speakable and Unspeakable in Quantum Mechanics: Collected Papers
  on Quantum Mechanics}.
\newblock Cambridge University Press, Cambridge, 1989.

\bibitem{Benioff80a}
P.~Benioff.
\newblock The computer as a physical systems: A microscopic quantum mechanical
  hamiltonian model of computers as represented by turing machines.
\newblock {\em J. Stat. Phys.}, 22(5):563--591, 1980.

\bibitem{Bennett73a}
C.~H. Bennett.
\newblock Logical reversibility of computation.
\newblock {\em IBM Journal of Research and Development}, 17(6):525--32, 1973.

\bibitem{Bennett87a}
C.~H. Bennett.
\newblock Demons, engines and the second law.
\newblock {\em Scientific American}, 295(5):108, 1987.

\bibitem{Bennett84a}
C.~H. Bennett and G.~Brassard.
\newblock Quantum cryptography: {P}ublic key distribution and coin tossing.
\newblock In {\em Proceedings of IEEE International Conference on Computers,
  Systems and Signal Processing}, pages 175--179, New York, 1984. IEEE.
\newblock Bangalore, India, December 1984.

\bibitem{Bennett82a}
C.~H. Bennett, G.~Brassard, S.~Breidbart, and S.~Wiesner.
\newblock Quantum cryptography, or unforgeable subway tokens.
\newblock In D.~Chaum, R.~L. Rivest, and A.~T. Sherman, editors, {\em Advances
  in Cryptology: Proceedings of Crypto 82}, pages 267--275, New York, 1982.
  Plenum Press.

\bibitem{Bennett93a}
C.~H. Bennett, G.~Brassard, C.~Cr\'epeau, R.~Jozsa, A.~Peres, and W.~Wootters.
\newblock Teleporting an unknown quantum state via dual classical and {EPR}
  channels.
\newblock {\em Phys. Rev. Lett.}, 70:1895--1899, 1993.

\bibitem{Bennett92a}
C.~H. Bennett, G.~Brassard, and N.~D. Mermin.
\newblock Quantum cryptography without {B}ell's theorem.
\newblock {\em Physical Review Letters}, 68(5):557--559, 1992.

\bibitem{Bennett96b}
C.~H. Bennett, G.~Brassard, S.~Popescu, B.~Schumacher, J.~A. Smolin, and W.~K.
  Wootters.
\newblock Purification of noisy entanglement and faithful teleportation via
  noisy channels.
\newblock {\em Physical Review Letters}, 76:722, 1996.
\newblock quant-ph/9511027.

\bibitem{Bennett97a}
C.~H. Bennett, D.~P. DiVincenzo, and J.~A. Smolin.
\newblock Capacities of quantum erasure channels.
\newblock {\em Physical Review Letters}, 78(16):3217--3220, 1997.
\newblock quant-ph/9701015.

\bibitem{Bennett96a}
C.~H. Bennett, D.~P. DiVincenzo, J.~A. Smolin, and W.~K. Wootters.
\newblock Mixed state entanglement and quantum error correction.
\newblock {\em Phys. Rev. A}, 54:3824, 1996.
\newblock quant-ph/9604024.

\bibitem{Bennett92c}
C.~H. Bennett and S.~J. Wiesner.
\newblock Communication via one- and two-particle operators on
  {E}instein-{P}odolsky-{R}osen states.
\newblock {\em Physical Review Letters}, 69(20):2881--2884, 1992.

\bibitem{Bergquist86a}
J.~C. Bergquist, R.~G. Hulet, W.~M. Itano, and D.~J. Wineland.
\newblock Observation of quantum jumps in a single atom.
\newblock {\em Physical Review Letters}, 57(14):1699--1702, 1986.

\bibitem{Bernstein97a}
E.~Bernstein and U.~Vazirani.
\newblock Quantum complexity theory.
\newblock {\em SIAM J. Comp.}, 26(5):1411--1473, 1997.
\newblock quant-ph/9701001.

\bibitem{Bhatia97a}
R.~Bhatia.
\newblock {\em Matrix analysis}.
\newblock Springer-Verlag, New York, 1997.

\bibitem{Bohr28a}
N.~Bohr.
\newblock The quantum postulate and the recent development of atomic theory.
\newblock {\em Nature}, 121:580--590, 1928.
\newblock Reprinted in \cite{Wheeler83a}.

\bibitem{Boschi98a}
D.~Boschi, S.~Branca, F.~De Martini, L.~Hardy, and S.~Popescu.
\newblock Experimental realization of teleporting an unknown pure quantum state
  via dual classical and {E}instein-{P}odolski-{R}osen channels.
\newblock {\em Phys. Rev. Lett.}, 80:1121--1125, 1998.
\newblock quant-ph/9710013.

\bibitem{Bouwmeester97a}
D.~Bouwmeester, J.~W. Pan, K.~Mattle, M.~Eibl, H.~Weinfurter, and A.~Zeilinger.
\newblock Experimental quantum teleportation.
\newblock {\em Nature}, 390(6660):575--579, 1997.

\bibitem{Brassard96b}
G.~Brassard.
\newblock Teleportation as a quantum computation.
\newblock In T.~Toffoli, M.~Biafore, and J.~Leao, editors, {\em PhysComp 96},
  pages 48--50, Cambridge MA, 1996. New England Complex Systsems Institute.

\bibitem{Brassard98a}
G.~Brassard, S.~Braunstein, and R.~Cleve.
\newblock {\em To appear in Physica D}, 1998.

\bibitem{Bredon93a}
G.~E. Bredon.
\newblock {\em Topology and geometry}.
\newblock Springer-Verlag, New York, 1993.

\bibitem{Calderbank97a}
A.~R. Calderbank, E.~M. Rains, P.~W. Shor, and N.~J.~A. Sloane.
\newblock Quantum error correction and orthogonal geometry.
\newblock {\em Phys. Rev. Lett.}, 78:405--8, 1997.

\bibitem{Caves90a}
C.~M. Caves.
\newblock Quantitative limits on the ability of a {Maxwell} demon to extract
  work from heat.
\newblock {\em Physical Review Letters}, 64(18):2111--2114, 1990.

\bibitem{Caves94a}
C.~M. Caves.
\newblock Information, entropy, and chaos.
\newblock In J.~J. Halliwell, J.~Perez-Mercader, and W.~H. Zurek, editors, {\em
  Physical Origins of Time Asymmetry}. Cambridge University Press, Cambridge,
  1994.

\bibitem{Choi75a}
M.-D. Choi.
\newblock Completely positive linear maps on complex matrices.
\newblock {\em Linear Algebra and Its Applications}, 10:285--290, 1975.

\bibitem{Chor88a}
B.~Chor and O.~Goldreich.
\newblock Unbiased bits from weak sources of randomness and probabilistic
  communication complexity.
\newblock {\em SIAM J. on Comput.}, 17(2):230--261, 1988.

\bibitem{Chuang98a}
I.~L. Chuang, N.~Gershenfeld, and M.~Kubinec.
\newblock Experimental implementation of fast quantum searching.
\newblock {\em Physical Review Letters}, 18(15):3408--3411, 1998.

\bibitem{Chuang98b}
I.~L. Chuang, N.~Gershenfeld, M.~G. Kubinec, and D.~W. Leung.
\newblock Bulk quantum computation with nuclear-magnetic-resonance: theory and
  experiment.
\newblock {\em Proc. Roy. Soc. Lond. A}, 454(1969):447--467, 1998.

\bibitem{Chuang97a}
I.~L. Chuang and M.~A. Nielsen.
\newblock Prescription for experimental determination of the dynamics of a
  quantum black box.
\newblock {\em J. Mod. Opt.}, 44(11-12):2455--2467, 1997.
\newblock quant-ph/9610001.

\bibitem{Chuang98c}
I.~L. Chuang, L.~M.~K. Vandersypen, X.~L. Zhou, D.~W. Leung, and S.~Lloyd.
\newblock Experimental realization of a quantum algorithm.
\newblock {\em Nature}, 393(6681):143--146, 1998.

\bibitem{Cirac95a}
J.~I. Cirac and P.~Zoller.
\newblock Quantum computations with cold trapped ions.
\newblock {\em Phys. Rev. Lett.}, 74:4091, 1995.

\bibitem{Cirac97a}
J.~I. Cirac, P.~Zoller, H.~J. Kimble, and H.~Mabuchi.
\newblock Quantum state transfer and entanglement distribution among distant
  nodes in a quantum network.
\newblock {\em Phys. Rev. Lett.}, 78(16):3221--3224, 1997.
\newblock quant-ph/9611017.

\bibitem{Cleve97a}
R.~Cleve and H.~Buhrman.
\newblock Substituting quantum entanglement for communication.
\newblock {\em Physical Review A}, 56(2):1201--1204, 1997.
\newblock quant-ph/9704026.

\bibitem{Cleve96a}
R.~Cleve and D.~P. DiVincenzo.
\newblock Schumacher's quantum data compression as a quantum computation.
\newblock {\em Phys. Rev. A}, 54:2636, 1996.
\newblock quant-ph/9603009.

\bibitem{Cleve98a}
R.~Cleve, A.~Ekert, C.~Macciavello, and M.~Mosca.
\newblock Quantum algorithms revisited.
\newblock {\em Proc. Roy. Soc. A}, 454(1969):339--354, 1998.

\bibitem{Cleve97b}
R.~Cleve, W.~van Dam, M.~A. Nielsen, and A.~Tapp.
\newblock Quantum entanglement and the communication complexity of the inner
  product function.
\newblock {\em \mbox{arXive} e-print quant-ph/9708019}, 1997.

\bibitem{Coppersmith94a}
D.~Coppersmith.
\newblock An approximate {Fourier} transform useful in quantum factoring.
\newblock {\em IBM Research Report RC 19642}, 1994.

\bibitem{Cormen90a}
T.~H. Cormen, C.~E. Leiserson, and R.~L. Rivest.
\newblock {\em Introduction to Algorithms}.
\newblock The MIT Press, Cambridge, Mass., 1990.

\bibitem{Intel98a}
Intel Corporation.
\newblock Moore's law: Changing the {P}{C} platform for another 20 years.
\newblock {\em Online at
  http://developer.intel.com/solutions/archive/issue2/focus.htm}, 1998.

\bibitem{Cory97a}
D.~G. Cory, A.~F. Fahmy, and T.~F. Havel.
\newblock Ensemble quantum computing by {N}{M}{R} spectroscopy.
\newblock {\em Proc. Natl. Acad. Sci. USA}, 94:1634--1639, 1997.

\bibitem{Cory98a}
D.~G. Cory, W.~Mass, M.~Price, E.~Knill, R.~Laflamme, W.~H. Zurek, T.~F. Havel,
  and S.~S. Somaroo.
\newblock Experimental quantum error correction.
\newblock {\em \mbox{arXive} e-print quant-ph/9802018}, 1998.

\bibitem{Cory97b}
D.~G. Cory, M.~D. Price, and T.~F. Havel.
\newblock Nuclear magnetic resonance spectroscopy: An experimentally accessible
  paradigm for quantum computing.
\newblock {\em \mbox{arXive} e-print quant-ph/9709001}, 1997.

\bibitem{Cover91a}
T.~M. Cover and J.~A. Thomas.
\newblock {\em Elements of Information Theory}.
\newblock John Wiley and Sons, New York, 1991.

\bibitem{Csiszar81a}
I.~Csisz\'ar and J.~K\"orner.
\newblock {\em Information Theory: Coding Theorems for Discrete Memoryless
  Systems}.
\newblock Academic Press, New York, 1981.

\bibitem{Davis65a}
M.~D. Davis.
\newblock {\em The Undecidable}.
\newblock Raven Press, Hewlett, New York, 1965.

\bibitem{Davis83a}
M.~D. Davis and E.~J. Weyuker.
\newblock {\em Computability, Complexity, and Languages}.
\newblock Academic Press, New York, 1983.

\bibitem{Deutsch85a}
D.~Deutsch.
\newblock Quantum theory, the {C}hurch-{T}uring {P}rinciple and the universal
  quantum computer.
\newblock {\em Proc. R. Soc. Lond. A}, 400:97, 1985.

\bibitem{Deutsch94a}
D.~Deutsch.
\newblock Notes on the quantum {Fourier} transform.
\newblock {\em Unpublished}, 1994.

\bibitem{Dieks82a}
D.~Dieks.
\newblock Communication by {EPR} devices.
\newblock {\em Physics Letters A}, 92(6):271--272, 1982.

\bibitem{Dirac58a}
P.~A.~M. Dirac.
\newblock {\em The principles of quantum mechanics 4th ed.}
\newblock Oxford University Press, Oxford, 1958.

\bibitem{Drexler92a}
K.~E. Drexler.
\newblock {\em Nanosystems: Molecular Machinery, Manufacturig, and
  Computation}.
\newblock John Wiley and Sons, New York, 1992.

\bibitem{Ekert96a}
A.~Ekert and R.~Jozsa.
\newblock Quantum computation and {Shor's} factoring algorithm.
\newblock {\em Rev. Mod. Phys.}, 68:1, 1996.

\bibitem{Ernst90a}
R.~Ernst, G.~Bodenhausen, and A.~Wokaun.
\newblock {\em Principles of Nuclear Magnetic Resonance in One and Two
  Dimension}.
\newblock Oxford University Press, Oxford, 1990.

\bibitem{Feynman65d}
R.~P. Feynman.
\newblock {\em The character of physical law}.
\newblock MIT Press, Cambridge, MA, 1965.

\bibitem{Feynman82a}
R.~P. Feynman.
\newblock Simulating physics with computers.
\newblock {\em Int. J. Theor. Phys.}, 21:467, 1982.

\bibitem{Freedman98b}
M.~H. Freedman.
\newblock P/np, and the quantum field computer.
\newblock {\em Proc. Natl. Acad. Sci. USA}, 95:98--101, 1998.

\bibitem{Fuchs96a}
C.~A. Fuchs.
\newblock {\em Distinguishability and Accessible Information in Quantum
  Theory}.
\newblock PhD thesis, The University of New Mexico, Albuquerque, NM, 1996.
\newblock Also \mbox{arXive} e-print quant-ph/9601020.

\bibitem{Gershenfeld97a}
N.~Gershenfeld and I.~L. Chuang.
\newblock Bulk spin resonance quantum computation.
\newblock {\em Science}, 275:350, 1997.

\bibitem{Gottesman96a}
D.~Gottesman.
\newblock Class of quantum error-correcting codes saturating the quantum
  {Hamming} bound.
\newblock {\em Phys. Rev. A}, 54:1862, 1996.

\bibitem{Gottesman97a}
D.~Gottesman.
\newblock {\em Stabilizer Codes and Quantum Error Correction}.
\newblock PhD thesis, California Institute of Technology, Pasadena, CA, 1997.

\bibitem{Gottesman98a}
D.~Gottesman.
\newblock Theory of fault-tolerant quantum computation.
\newblock {\em Physical Review A}, 57(1):127--137, 1998.
\newblock quant-ph/9702029.

\bibitem{Grace97a}
E.~S. Grace.
\newblock {\em Biotechnology unzipped: Promises and realities}.
\newblock National Academy Press, 1997.

\bibitem{Grant96a}
D.~M. Grant and R.~K. Harris, editors.
\newblock {\em Encyclopedia of nuclear magnetic resonance}.
\newblock John Wiley, New York, 1996.

\bibitem{Griffiths96a}
R.~B. Griffiths and C.-S. Niu.
\newblock Semiclassical {Fourier} transform for quantum computation.
\newblock {\em Physical Review Letters}, 76(17):3228--3231, 1996.
\newblock quant-ph/9511007.

\bibitem{Grimmett92a}
G.~R. Grimmett and D.~R. Stirzaker.
\newblock {\em Probability and Random Processes}.
\newblock Clarendon Press, Oxford, 1992.

\bibitem{Grover96a}
Lov~K. Grover.
\newblock A fast quantum mechanical algorithm for database search.
\newblock In {\em 28th ACM Symposium on Theory of Computation}, page 212, New
  York, 1996. Association for Computing Machinery.

\bibitem{Hawking76a}
S.~W. Hawking.
\newblock Breakdown of predictability in gravitational collapse.
\newblock {\em Phys.~Rev.~D}, 14:2460, 1976.

\bibitem{Hellwig69a}
K.-E. Hellwig and K.~Kraus.
\newblock Pure operations and measurements.
\newblock {\em Commun. Math. Phys.}, 11:214--220, 1969.

\bibitem{Hellwig70a}
K.-E. Hellwig and K.~Kraus.
\newblock Operations and measurements. {II}.
\newblock {\em Commun. Math. Phys.}, 16:142--147, 1970.

\bibitem{Helstrom76a}
Carl~W. Helstrom.
\newblock {\em Quantum Detection and Estimation Theory}.
\newblock Mathematics in Science and Engineering, vol.\ 123. Academic Press,
  New York, 1976.

\bibitem{Hill97a}
S.~Hill and W.~K. Wootters.
\newblock Entanglement of a pair of quantum bits.
\newblock {\em Phys. Rev. Lett.}, 78(26):5022--5025, 1997.
\newblock quant-ph/9703041.

\bibitem{Holevo73a}
A.~S. Holevo.
\newblock Statistical problems in quantum physics.
\newblock In Gisiro Maruyama and Jurii~V. Prokhorov, editors, {\em Proceedings
  of the Second Japan--USSR Symposium on Probability Theory}, pages 104--119,
  Berlin, 1973. Springer-Verlag.
\newblock Lecture Notes in Mathematics, vol.\ 330.

\bibitem{Holevo82a}
A.~S. Holevo.
\newblock {\em Probabilistic and Statistical Aspects of Quantum Theory}.
\newblock North-Holland Series in Statistics and Probability, vol.\ 1.
  North-Holland, Amsterdam, 1982.

\bibitem{Holevo96a}
A.~S. Holevo.
\newblock The capacity of quantum channel with general signal states.
\newblock {\em \mbox{arXive} e-print quant-ph/9611023}, 1996.

\bibitem{Horn91a}
R.~A. Horn and C.~R. Johnson.
\newblock {\em Topics in matrix analysis}.
\newblock Cambridge University Press, Cambridge, 1991.

\bibitem{Huelga97a}
S.~F Huelga, C.~Macchiavello, T.~Pellizzari, A.~K. Ekert, M.~B. Plenio, and
  J.~I. Cirac.
\newblock On the improvement of frequency stardards with quantum entanglement.
\newblock {\em Phys.~Rev.~Lett.}, 79(20):3865--3868, 1997.
\newblock quant-ph/9707014.

\bibitem{Hughes95a}
R.~J. Hughes, D.~M Alde, P.~Dyer, G.~G. Luther, G.~L. Morgan, and M.~Schauer.
\newblock Quantum cryptography.
\newblock {\em Contemp. Phys.}, 36(3):149--163, 1995.
\newblock quant-ph/9504002.

\bibitem{Jones98a}
J.~A. Jones and M.~Mosca.
\newblock Implementation of a quantum algorithm to solve {Deutsch's} problem on
  a nuclear magnetic resonance quantum computer.
\newblock {\em \mbox{arXive} e-print quant-ph/9801027}, 1998.

\bibitem{Jones98b}
J.~A. Jones, M.~Mosca, and R.~H. Hansen.
\newblock Implementation of a quantum search algorithm on a nuclear magnetic
  resonance quantum computer.
\newblock {\em Nature}, 393(6683):344, 1998.
\newblock quant-ph/9805069.

\bibitem{Jones94a}
K.~R.~W. Jones.
\newblock Fundamental limits upon the measurement of state vectors.
\newblock {\em Phys. Rev. A}, 50:3682--3699, 1994.

\bibitem{Jozsa94c}
R.~Jozsa.
\newblock Fidelity for mixed quantum states.
\newblock {\em J. Mod. Opt.}, 41:2315--2323, 1994.

\bibitem{Jozsa97a}
R.~Jozsa.
\newblock Entanglement and quantum computation.
\newblock {\em \mbox{arXive} e-print quant-ph/9707034}, 1997.

\bibitem{Jozsa94a}
R.~Jozsa and B.~Schumacher.
\newblock A new proof of the quantum noiseless coding theorem.
\newblock {\em J. Mod. Opt.}, 41:2343--2349, 1994.

\bibitem{Kitaev97a}
A.~Y. Kitaev.
\newblock Quantum error correction with imperfect gates.
\newblock In A.~S.~Holevo O.~Hirota and C.~M. Caves, editors, {\em Quantum
  Communication, Computing, and Measurement}, pages 181--188, New York, 1997.
  Plenum Press.

\bibitem{Knill98b}
E.~Knill, I.~Chuang, and R.~Laflamme.
\newblock Effective pure states for bulk quantum computation.
\newblock {\em Phys. Rev. A}, 57(5):3348--3363, 1998.
\newblock quant-ph/9706053.

\bibitem{Knill96a}
E.~Knill and R.~Laflamme.
\newblock Concatenated quantum codes.
\newblock {\em \mbox{arXive} e-print quant-ph/9608012}, 1996.

\bibitem{Knill97a}
E.~Knill and R.~Laflamme.
\newblock A theory of quantum error-correcting codes.
\newblock {\em Phys. Rev. A}, 55:900, 1997.
\newblock quant-ph/9604034.

\bibitem{Knill98d}
E.~Knill and R.~Laflamme.
\newblock On the power of one bit of quantum information.
\newblock {\em \mbox{arXive} e-print quant-ph/9802037}, 1998.

\bibitem{Knill98c}
E.~Knill, R.~Laflamme, and W.~H. Zurek.
\newblock Resilient quantum computation.
\newblock {\em Science}, 279(5349):342--345, 1998.
\newblock quant-ph/9702058.

\bibitem{Knill98a}
E.~Knill, R.~Laflamme, and W.~H. Zurek.
\newblock Resilient quantum computation: error models and thresholds.
\newblock {\em Proc. Roy. Soc. A}, 454(1969):365--384, 1998.
\newblock quant-ph/9702058.

\bibitem{Kraus83a}
Karl Kraus.
\newblock {\em States, Effects, and Operations:\ Fundamental Notions of Quantum
  Theory}.
\newblock Lecture Notes in Physics, Vol.\ 190. Springer-Verlag, Berlin, 1983.

\bibitem{Kremer95a}
I.~Kremer.
\newblock Quantum communication.
\newblock Master's thesis, The Hebrew University of Jerusalem, 1995.

\bibitem{Kushilevitz96a}
E.~Kushilevitz and N.~Nisan.
\newblock {\em Communication Complexity}.
\newblock Cambridge University Press, Cambridge, 1996.

\bibitem{Laflamme97a}
R.~Laflamme, E.~Knill, W.~H. Zurek, P.~Catasti, and S.~V.~S. Mariappan.
\newblock {NMR} {GHZ}.
\newblock {\em \mbox{arXive} e-print quant-ph/9709025}, 1997.

\bibitem{Landauer61a}
R.~Landauer.
\newblock Irreversibility and heat generation in the computing process.
\newblock {\em IBM J. Res. Dev.}, 5:183, 1961.

\bibitem{Landauer86a}
R.~Landauer.
\newblock Computation and physics: {Wheeler's} meaning circuit?
\newblock {\em Found. Phys.}, 16(6), 1986.

\bibitem{Landauer91a}
R.~Landauer.
\newblock Information is physical.
\newblock {\em Physics Today}, 44(5):22--29, 1991.

\bibitem{Langer97a}
J.~S. Langer.
\newblock Nonequilibrium physics.
\newblock In V.~L. Fitch, D.~R. Marlow, and M.~A.~E. Dementi, editors, {\em
  Critical problems in physics}, pages 281--306. Princeton University Press,
  Princeton, New Jersey, 1997.

\bibitem{Lecerf63a}
Y.~Lecerf.
\newblock Machines de {Turing} r\'eversibles.
\newblock {\em Comptes Rendus}, 257:2597--2600, 1963.

\bibitem{Leibfried96a}
D.~Leibfried, D.~M. Meekhof, B.~E. King, C.~Monroe, W.~M. Itano, and D.~J.
  Wineland.
\newblock Experimental-determination of the motional quantum state of a trapped
  atom.
\newblock {\em Physical Review Letters}, 77(21):4281--5285, 1996.

\bibitem{Leonhardt96a}
U.~Leonhardt.
\newblock Discrete {Wigner} function and quantum-state tomography.
\newblock {\em Physical Review A}, 53(5):2998--3013, 1996.

\bibitem{Leung97a}
D.~W. Leung, M.~A. Nielsen, I.~L. Chuang, and Y.~Yamamoto.
\newblock Approximate quantum error correction can lead to better codes.
\newblock {\em Phys. Rev. A}, 56:2567--2573, 1997.

\bibitem{Lieb73a}
E.~H. Lieb.
\newblock Convex trace functions and the {W}igner-{Y}anase-{D}yson conjecture.
\newblock {\em Advances in Mathematics}, 11:267--288, 1973.

\bibitem{Lieb73b}
E.~H. Lieb and M.~B. Ruskai.
\newblock Proof of the strong subadditivity of quantum mechanical entropy.
\newblock {\em J. Math. Phys.}, 14:1938--1941, 1973.

\bibitem{Ljung87a}
L.~Ljung.
\newblock {\em System identification: Theory for the user}.
\newblock Prentice-Hall, Upper Saddle River, 1987.

\bibitem{Lloyd96a}
S.~Lloyd.
\newblock Universal quantum simulators.
\newblock {\em Science}, 273:1073, 1996.

\bibitem{Lloyd97a}
S.~Lloyd.
\newblock The capacity of the noisy quantum channel.
\newblock {\em Phys. Rev. A}, 56:1613, 1997.

\bibitem{Lloyd97b}
S.~Lloyd.
\newblock Quantum controllers for quantum systems.
\newblock {\em \mbox{arXive} e-print quant-ph/9703042}, 1997.

\bibitem{Luby96a}
M.~Luby.
\newblock {\em Pseudorandomness and cryptographic applications}.
\newblock Princeton University Press, Princeton, New Jersey, 1996.

\bibitem{Mabuchi96b}
H.~Mabuchi.
\newblock Dynamical identification of open quantum-systems.
\newblock {\em Quant. Semi. Opt.}, 8(6):1103--1108, 1996.

\bibitem{Mabuchi98a}
H.~Mabuchi.
\newblock Standard quantum limits for broadband position measurement.
\newblock {\em \mbox{arXive} e-print quant-ph/9801039}, 1998.

\bibitem{Marcus92a}
M.~Marcus and H.~Minc.
\newblock {\em A Survey of Matrix Theory and Matrix Inequalities}.
\newblock Dover, New York, 1992.

\bibitem{Mattle96a}
K.~Mattle, H.~Weinfurter, P.~G. Kwiat, and A.~Zeilinger.
\newblock Dense coding in experimental quantum communication.
\newblock {\em Physical Review Letters}, 76(25):4656--4659, 1996.

\bibitem{McLuhan64a}
M.~McLuhan.
\newblock {\em Understanding media: The extension of man}.
\newblock McGraw-Hill, New York, 1964.

\bibitem{Milburn96a}
G.~J. Milburn.
\newblock A quantum mechanical {Maxwell's} demon.
\newblock Unpublished, 1996.

\bibitem{Monroe97a}
C.~Monroe, D.~Leibfried, B.~E. King, D.~M. Meekhof, W.~M. Itano, and D.~J.
  Wineland.
\newblock Simplified quantum logic with trapped ions.
\newblock {\em Physical Review A}, 55(4):R2489--R2491, 1997.

\bibitem{Mott29a}
N.~F. Mott.
\newblock The wave mechanics of $\alpha$-ray tracks.
\newblock {\em Proc. Roy. Soc. Lond. A}, 126:79--84, 1929.
\newblock Reprinted in \cite{Wheeler83a}.

\bibitem{Nagourney86a}
W.~Nagourney, J.~Sandberg, and H.~Dehmelt.
\newblock Shelved optical electron amplifier: observation of quantum jumps.
\newblock {\em Physical Review Letters}, 56(26):2797--2799, 1986.

\bibitem{Newell76a}
A.~Newell and H.~Simon.
\newblock Computer science as empirical enquiry: Symbols and search.
\newblock {\em Communications of the Association for Computing Machinery},
  19:113--126, 1976.

\bibitem{Nielsen96c}
M.~A. Nielsen.
\newblock The entanglement fidelity and quantum error correction.
\newblock {\em \mbox{arXive} e-print quant-ph/9606012}, 1996.

\bibitem{Nielsen97d}
M.~A. Nielsen.
\newblock Computable functions, quantum measurements, and quantum dynamics.
\newblock {\em Phys. Rev. Lett.}, 79(15):2915--2918, 1997.

\bibitem{Nielsen97b}
M.~A. Nielsen and C.~M. Caves.
\newblock Reversible quantum operations and their application to teleportation.
\newblock {\em Phys. Rev. A}, 55(4):2547--2556, 1997.

\bibitem{Nielsen98a}
M.~A. Nielsen, C.~M. Caves, B.~Schumacher, and H.~Barnum.
\newblock Information-theoretic approach to quantum error correction and
  reversible measurement.
\newblock {\em Proc. Roy. Soc. A}, 454(1969):277--304, 1998.

\bibitem{Nielsen97c}
M.~A. Nielsen and I~L. Chuang.
\newblock Programmable quantum gate arrays.
\newblock {\em Phys. Rev. Lett.}, 79(2):321--324, 1997.

\bibitem{Nielsen98c}
M.~A. Nielsen, E.~Knill, and R.~Laflamme.
\newblock {\em http://wwwcas.phys.unm.edu/~mnielsen/nmr/index.html}, 1998.

\bibitem{Nielsen98b}
M.~A. Nielsen, E.~Knill, and R.~Laflamme.
\newblock Experimental quantum teleportation by {NMR}.
\newblock {\em In preparation}, 1998.

\bibitem{Ohya93a}
M.~Ohya and D.~Petz.
\newblock {\em Quantum entropy and its use}.
\newblock Springer-Verlag, Berlin, 1993.

\bibitem{Papadimitriou94a}
C.~M. Papadimitriou.
\newblock {\em Computational complexity}.
\newblock Addison-Wesley, Reading, Massachusetts, 1994.

\bibitem{Peres93a}
A.~Peres.
\newblock {\em Quantum Theory: Concepts and Methods}.
\newblock Kluwer Academic, Dordrecht, 1993.

\bibitem{Popescu97a}
S.~Popescu and D.~Rohrlich.
\newblock Thermodynamics and the measure of entanglement.
\newblock {\em Phys. Rev. A}, 56(5):R3319--R3321, 1997.

\bibitem{Poyatos97a}
J.~F. Poyatos, J.~I. Cirac, and P.~Zoller.
\newblock Complete characterization of a quantum process: the two-bit quantum
  gate.
\newblock {\em Phys. Rev. Lett.}, 78(2):390--393, 1997.

\bibitem{Preskill97a}
J.~Preskill.
\newblock Fault-tolerant quantum computation.
\newblock {\em \mbox{arXive} e-print quant-ph/9712048}, 1997.

\bibitem{Preskill98b}
J.~Preskill.
\newblock Reliable quantum computers.
\newblock {\em Proc. Roy. Soc. A: Math., Phys. and Eng.}, 454(1969):385--410,
  1998.

\bibitem{Press89a}
W.~H. Press, S.~A. Teukolsky, B.~P. Flannery, and W.~T. Vetterling.
\newblock {\em Numerical recipes in Fortran}.
\newblock Cambridge University Press, Cambridge, 1989.

\bibitem{Raymer94a}
M.~G. Raymer, M.~Beck, and D.~F. McAlister.
\newblock Complex wave-field reconstruction using phase-space tomography.
\newblock {\em Physical Review Letters}, 72(8):1137--1140, 1994.

\bibitem{Royer96a}
A.~Royer.
\newblock Reduced dynamics with initial correlations, and time-dependent
  environment and {Hamiltonians}.
\newblock {\em Physical Review Letters}, 77(16):3272--3275, 1996.

\bibitem{Ruskai88a}
M.~B. Ruskai.
\newblock Extremal properties of relative entropy in quantum statistical
  mechanics.
\newblock {\em Rep. Math. Phys.}, 26(1):143--150, 1988.

\bibitem{Ruskai94a}
M.~B. Ruskai.
\newblock Beyond strong subadditivity: improved bounds on the contraction of
  generalized relative entropy.
\newblock {\em Rev. Math. Phys.}, 6(5A):1147--1161, 1994.

\bibitem{Sauter86a}
T.~Sauter, W.~Neuhauser, R.~Blatt, and P.~E. Toschek.
\newblock Observation of quantum jumps.
\newblock {\em Physical Review Letters}, 57(14):1696--1698, 1986.

\bibitem{Schack94b}
R.~Schack.
\newblock Algorithmic information and simplicity in statistical physics.
\newblock {\em \mbox{arXive} e-print hep-th/9409022}, 1994.

\bibitem{Schmidt06a}
E.~Schmidt.
\newblock Zur theorie der linearen und nichtlinearen integralgleighungen.
\newblock {\em Math. Annalen.}, 63:433--476, 1906.

\bibitem{Schulman98a}
L.~J. Schulman and U.~Vazirani.
\newblock Scalable {NMR} quantum computation.
\newblock {\em \mbox{arXive} e-print quant-ph/9804060}, 1998.

\bibitem{Schumacher95a}
B.~Schumacher.
\newblock Quantum coding.
\newblock {\em Phys. Rev. A}, 51:2738--2747, 1995.

\bibitem{Schumacher97a}
B.~Schumacher and M.~D. Westmoreland.
\newblock Sending classical information via noisy quantum channels.
\newblock {\em Physical Review A}, 56(1):131--138, 1997.

\bibitem{Schumacher98a}
B.~Schumacher and M.~D. Westmoreland.
\newblock Quantum privacy and quantum coherence.
\newblock {\em Phys. Rev. Lett.}, 80(25):5695--5697, 1998.

\bibitem{Schumacher96a}
B.~W. Schumacher.
\newblock Sending entanglement through noisy quantum channels.
\newblock {\em Phys. Rev. A}, 54:2614, 1996.

\bibitem{Schumacher96b}
B.~W. Schumacher and M.~A. Nielsen.
\newblock Quantum data processing and error correction.
\newblock {\em Physical Review A}, 54(4):2629, 1996.

\bibitem{Schumacher96c}
B.~W. Schumacher, M.~Westmoreland, and W.~K. Wootters.
\newblock Limitation on the amount of acessible information in a quantum
  channel.
\newblock {\em Phys. Rev. Lett.}, 76:3453, 1996.

\bibitem{Shannon48a}
C.~E. Shannon.
\newblock A mathematical theory of communication.
\newblock {\em Bell System Tech. J.}, 27:379--423, 623--656, 1948.

\bibitem{Shannon49b}
C.~E. Shannon.
\newblock Communication theory of secrecy systems.
\newblock {\em Bell Sys. Tech. J.}, 28:656--715, 1949.

\bibitem{Shannon49a}
C.~E. Shannon and W.~Weaver.
\newblock {\em The Mathematical Theory of Communication}.
\newblock University of Illinois Press, Urbana, 1949.

\bibitem{Shor94a}
P.~W. Shor.
\newblock Algorithms for quantum computation: discrete logarithms and
  factoring.
\newblock In {\em Proceedings, 35th Annual Symposium on Fundamentals of
  Computer Science}, Los Alamitos, 1994. IEEE Press.

\bibitem{Shor95a}
P.~W. Shor.
\newblock Scheme for reducing decoherence in quantum memory.
\newblock {\em Phys. Rev. A}, 52:2493, 1995.

\bibitem{Shor97a}
P.~W. Shor.
\newblock Polynomial-time algorithms for prime factorization and discrete
  logarithms on a quantum computer.
\newblock {\em SIAM J. Comp.}, 26(5):1484--1509, 1997.

\bibitem{Shor96a}
P.~W. Shor and J.~A. Smolin.
\newblock Quantum error-correcting codes need not completely reveal the error
  syndrome.
\newblock {\em \mbox{arXive} e-print quant-ph/9604006}, 1996.

\bibitem{Simmons63a}
G.~F. Simmons.
\newblock {\em Introduction to Topology and Modern Analysis}.
\newblock McGraw-Hill, Auckland, 1963.

\bibitem{Simon79a}
B.~Simon.
\newblock {\em Trace ideals and their applications}.
\newblock Cambridge University Press, Cambridge, 1979.

\bibitem{Smithey93a}
D.~T. Smithey, M.~Beck, M.~G. Raymer, and A.~Faridani.
\newblock Measurement of the {Wigner} distribution and the density-matrix of a
  light mode using optical homodyne tomography: application to squeezed states
  and the vacuum.
\newblock {\em Phys. Re. Lett.}, 70(9):1244--1247, 1993.

\bibitem{Sondhi97a}
S.~L. Sondhi, S.~M. Girvin, J.~P. Carini, and D.~Shahar.
\newblock Continuous quantum phase transitions.
\newblock {\em Rev. Mod. Phys.}, 69(1):315--333, 1997.

\bibitem{Steane97a}
A.~Steane.
\newblock The ion-trap quantum information processor.
\newblock {\em App. Phys. B -- Lasers and Optics}, 64(6):623--642, 1997.

\bibitem{Steane96a}
A.~M. Steane.
\newblock Error correcting codes in quantum theory.
\newblock {\em Physical Review Letters}, 77:793, 1996.

\bibitem{Terhal97a}
B.~M. Terhal and J.~A. Smolin.
\newblock Superfast quantum algorithms for coin weighing and binary search
  problems.
\newblock {\em \mbox{arXive} e-print quant-ph/9705041}, 1997.

\bibitem{QCC97a}
{The Quantum Computation Collective}.
\newblock What makes quantum computers powerful?
\newblock 1997.

\bibitem{Thorne97a}
K.~S. Thorne.
\newblock Gravitational waves: A new window onto the universe.
\newblock In V.~L. Fitch, D.~R. Marlow, and M.~A.~E. Dementi, editors, {\em
  Critical problems in physics}, pages 167--197. Princeton University Press,
  Princeton, New Jersey, 1997.

\bibitem{Turchette95a}
Q.~A. Turchette, C.~J. Hood, W.~Lange, H.~Mabuchi, and H.~J. Kimble.
\newblock Measurement of conditional phase shifts for quantum logic.
\newblock {\em Phys. Rev. Lett.}, 75:4710, 1995.

\bibitem{Turing36a}
A.~M. Turing.
\newblock On computable numbers, with an application to the
  {Entscheidungsproblem}.
\newblock {\em Proc. Lond. Math. Soc. 2 (reprinted in
  \protect\cite{Davis65a})}, 42:230, 1936.

\bibitem{Uhlmann76a}
A.~Uhlmann.
\newblock The `transition probability' in the state space of a $\,^*$-algebra.
\newblock {\em Reports on Mathematical Physics}, 9:273--279, 1976.

\bibitem{vanGelder98a}
T.~van Gelder.
\newblock The dynamical hypothesis in cognitive science.
\newblock {\em To appear in Brain and Behavioral Sciences}, 1998.

\bibitem{Viola98a}
L.~Viola and S.~Lloyd.
\newblock Dynamical suppression of decoherence in two-state quantum systems.
\newblock {\em \mbox{arXive} e-print quant-ph/9803057}, 1998.

\bibitem{Vogel89a}
K.~Vogel and H.~Risken.
\newblock Determination of quasiprobability distributions in terms of
  probability distributions for the rotated quadrature phase.
\newblock {\em Phys. Rev. A}, 40(12):7113--7120, 1989.

\bibitem{Wehrl78a}
A.~Wehrl.
\newblock General properties of entropy.
\newblock {\em Rev. Mod. Phys.}, 50:221, 1978.

\bibitem{Weinberg93a}
S.~Weinberg.
\newblock {\em Dreams of a Final Theory}.
\newblock Pantheon Books, 1993.

\bibitem{Weinberg95a}
S.~Weinberg.
\newblock {\em The Quantum Theory of Fields I}.
\newblock Cambridge University Press, Cambridge, 1995.

\bibitem{Westmoreland98a}
M.~D. Westmoreland and B.~Schumacher.
\newblock Quantum entanglement and the non-existence of superluminal signals.
\newblock {\em \mbox{arXive} e-print quant-ph/9801014}, 1998.

\bibitem{Wheeler86a}
J.~A. Wheeler.
\newblock Physics as meaning circuit.
\newblock In G.~E. Moore and M.~O. Scully, editors, {\em Frontiers of
  nonequilibrium statistical physics}, volume 135 of {\em NATO ASI Series B}.
  Plenum Press, New York, 1986.

\bibitem{Wheeler83a}
J.~A. Wheeler and W.~H. Zurek.
\newblock {\em Quantum theory and measurement}.
\newblock Princeton University Press, Princeton NJ, 1983.

\bibitem{Wiesner69a}
S.~Wiesner.
\newblock Unpublished manuscript, circa 1969, appeared as
  \protect\cite{Wiesner83a}.

\bibitem{Wiesner83a}
S.~Wiesner.
\newblock Conjugate coding.
\newblock {\em SIGACT News}, 15:77, 1983.

\bibitem{Wigner52a}
E.~P. Wigner.
\newblock Die messung quantenmechanischer operatoren.
\newblock {\em Z. Phys.}, 133:101, 1952.

\bibitem{Wilczek97a}
F.~Wilczek.
\newblock The future of particle physics as a natural science.
\newblock In V.~L. Fitch, D.~R. Marlow, and M.~A.~E. Dementi, editors, {\em
  Critical problems in physics}, pages 281--306. Princeton University Press,
  Princeton, New Jersey, 1997.

\bibitem{Williams98a}
C.~P. Williams and S.~H. Clearwater.
\newblock {\em Explorations in quantum computing}.
\newblock Springer-Verlag, New York, 1998.

\bibitem{Witten97a}
E.~Witten.
\newblock Vistas in theoretical physics.
\newblock In V.~L. Fitch, D.~R. Marlow, and M.~A.~E. Dementi, editors, {\em
  Critical problems in physics}, pages 271--278. Princeton University Press,
  Princeton, New Jersey, 1997.

\bibitem{Wolfram94a}
S.~Wolfram.
\newblock {\em Cellular automata and complexity}.
\newblock Addison-Wesley, Redwood City, CA, 1994.

\bibitem{Wootters98a}
W.~K. Wootters.
\newblock Entanglement of formation of an arbitrary state of two qubits.
\newblock {\em Phys. Rev. Lett.}, 80(10):2245--2248, 1998.

\bibitem{Wootters82a}
W.~K. Wootters and W.~H. Zurek.
\newblock A single quantum cannot be cloned.
\newblock {\em Nature}, 299:802--803, 1982.

\bibitem{Yao79a}
A.~C. Yao.
\newblock Some complexity questions related to distributed computing.
\newblock {\em Proc.\ of the 11th Ann.\ ACM Symp.\ on Theory of Computing},
  pages 209--213, 1979.

\bibitem{Yao93a}
A.~C. Yao.
\newblock Quantum circuit complexity.
\newblock {\em Proc.\ of the 34th Ann.\ IEEE Symp.\ on Foundations of Computer
  Science}, pages 352--361, 1993.

\bibitem{Yariv89a}
A.~Yariv.
\newblock {\em Quantum electronics}.
\newblock John Wiley and Sons, New York, 1989.

\bibitem{Zalka98a}
C.~Zalka.
\newblock Simulating quantum systems on a quantum computer.
\newblock {\em Proc. Roy. Soc.}, 454(1969):313--322, 1998.

\bibitem{Ziv78a}
J.~Ziv and A.~Lempel.
\newblock Compression of individual sequences by variable rate coding.
\newblock {\em IEEE Trans. Inform. Theory}, IT-24:530, 1978.

\bibitem{Zurek81a}
W.~H. Zurek.
\newblock Pointer basis of quantum apparatus: into what mixture does the wave
  packet collapse?
\newblock {\em Phys. Rev. D}, 24(6):1516--1525, 1981.

\bibitem{Zurek82a}
W.~H. Zurek.
\newblock Environment-induced super-selection rules.
\newblock {\em Phys. Rev. D}, 26(8):1862--1880, 1982.

\bibitem{Zurek89b}
W.~H. Zurek.
\newblock Algorithmic randomness and physical entropy.
\newblock {\em Phys. Rev. A}, 40:4731, 1989.

\bibitem{Zurek89a}
W.~H. Zurek.
\newblock Thermodynamic cost of computation, algorithmic complexity and the
  information metric.
\newblock {\em Nature}, 341:119, 1989.

\bibitem{Zurek91a}
W.~H. Zurek.
\newblock Decoherence and the transition from quantum to classical.
\newblock {\em Phys. Tod.}, 44(10):36--44, 1991.

\bibitem{Zurek93a}
W.~H. Zurek.
\newblock Preferred states, predictability, classicality and the
  environment-induced decoherence.
\newblock {\em Prog. Theor. Phys.}, 89(2):281--312, 1993.

\bibitem{Zurek98b}
W.~H. Zurek.
\newblock Decoherence, einselection, and the existential interpretation (the
  rough guide).
\newblock {\em \mbox{arXive} e-print quant-ph/9805065}, 1998.

\end{thebibliography}

\twocolumn
\addcontentsline{toc}{chapter}{{\bf Index}}
\begin{theindex}

  \item $L_1$ distance
    \subitem classical, 90
  \item $\chi$, 112

  \indexspace

  \item absolute distance
    \subitem classical, 90
      \subsubitem operational meaning for, 91
    \subitem quantum, 92
  \item algorithmic information, 182
  \item amplitude damping, 61
  \item angle, 99
  \item Araki-Lieb inequality, 78

  \indexspace

  \item Bell basis, 22
  \item binary entropy, 67
  \item bit flip channel, 170
  \item bit flip code, 171
  \item Bloch sphere, 20
  \item Bloch vector, 20

  \indexspace

  \item canonical representation for a quantum operation, 165
  \item capacity
    \subitem theorem for noiseless qubit communication, 116
  \item cavity QED, 31
  \item chaining properties of distance measures, 104
  \item channel capacity, 188
  \item Church-Turing thesis, 6
    \subitem modern form, 27
  \item classical physics, 224
  \item clean protocols for quantum communication, 119
  \item coherent information, 167, 192
  \item communication complexity, 110
    \subitem coherent quantum, 110, 122
    \subitem of the inner product, 118
    \subitem quantum, 110
    \subitem unified model for, 126
  \item complete positivity, 44
    \subitem example of a positive map not completely positive, 44
  \item complete quantum operations, 43
  \item computational basis states, 20
  \item concavity of the entropy, 79
  \item concavity of the quantum conditional entropy, 84
  \item conditional entropy
    \subitem classical, 69
    \subitem quantum, 76
      \subsubitem concavity of, 84
  \item continuity properties of distance measures, 104
  \item controlled not gate, 24
  \item convexity of quantum relative entropy, 84
  \item correlation entropy, 180

  \indexspace

  \item data compression, 130
  \item data pipelining inequality, 194
    \subitem classical, 73
  \item data processing inequality
    \subitem classical, 72
    \subitem quantum, 167
  \item decoherence, 224
  \item discrete memoryless quantum channel, 199
  \item distance measures, 89
  \item dynamic distance, 101
  \item dynamic error, 102
  \item dynamic fidelity, 101

  \indexspace

  \item entanglement fidelity, 101
  \item entropy
    \subitem classical, 66
    \subitem concavity of, 79
    \subitem of an ensemble, 77
    \subitem quantum, 73
  \item entropy exchange, 164
  \item entropy-fidelity lemma, 194
  \item environmental models
    \subitem complete quantum operations, 48
    \subitem incomplete quantum operations, 49
  \item error, 98
  \item error correction, 169

  \indexspace

  \item Fannes' inequality, 95
  \item Fano inequality
    \subitem classical, 115
    \subitem quantum analogue to, 166
  \item fidelity
    \subitem classical, 90
    \subitem quantum, 96
      \subsubitem explicit formula, 97
  \item fundamental physics, 13

  \indexspace

  \item generalized measurements, 62

  \indexspace

  \item Hadamard gate, 21
  \item halting observable, 7
    \subitem family of observables, 29
  \item halting problem, 7
    \subitem proof of unsolvability, 28
  \item Hamming distance, 90
  \item Holevo $\chi$, 112
  \item Holevo bound, 111

  \indexspace

  \item identity gate, 21
  \item incomplete quantum operations, 43
  \item information source
    \subitem classical, 90
    \subitem quantum, 136
  \item information theory
    \subitem motivation for definitions, 67
  \item information-theoretic conditions for quantum error correction, 
		175
  \item inner product, 118
  \item ion trap, 31

  \indexspace

  \item joint entropy theorem, 75

  \indexspace

  \item Klein's inequality, 74

  \indexspace

  \item Landauer limit, 17, 182
  \item law of large numbers, 135
  \item Lieb's theorem, 80
  \item local unitary operations, 205
  \item logical zero and one, 171

  \indexspace

  \item magic basis, 153
  \item Maxwell's demon, 181
  \item meaning circuits, 15
  \item measurement, 41
  \item mixed state Schmidt decomposition, 229
  \item Moore's law, 16
    \subitem quantum corollary to, 17
  \item mutual information
    \subitem classical, 69
    \subitem quantum, 76

  \indexspace

  \item NMR, 30, 31
  \item noiseless coding theorem, 67
  \item noisy channel coding theorem, 188
  \item non-increasing property of $\chi$, 112
  \item not gate, 21
  \item nuclear magnetic resonance, 30, 31

  \indexspace

  \item observed channel, 207
  \item operator-sum representation, 45
    \subitem freedom in, 51

  \indexspace

  \item partial trace property of $\chi$, 112
  \item Pauli sigma matrices, 172
  \item Pauli sigmas matrices, 20
  \item phase flip operator, 21, 172
  \item phase shift operator, 21
  \item phase transitions and entanglement, 161
  \item physical quantum operations, 43
  \item physics of information, 3
  \item positive operator valued measures, 62
  \item POVMs, 62
  \item process tomography, 57
    \subitem for incomplete quantum operations, 61
  \item purification, 227

  \indexspace

  \item quantum channel capacity, 188
  \item quantum circuit, 26
    \subitem for teleportation, 24
    \subitem requirements for, 31
  \item quantum communication complexity, 110
    \subitem of the inner product, 118
    \subitem unified model for, 126
  \item quantum computation, 26
    \subitem requirements for, 31
  \item quantum data compression, 130, 136
  \item quantum data processing inequality, 167
  \item quantum error correction, 169
  \item quantum Fano inequality, 166
  \item quantum gates, 21
  \item quantum information, 8
  \item quantum information processors, 30
  \item quantum operations, 41, 43
    \subitem complete, 43
      \subsubitem environmental models for, 48
    \subitem definition, 44
    \subitem incomplete, 43
      \subsubitem environmental models for, 49
    \subitem limitations to the formalism, 63
    \subitem on a qubit, 50
    \subitem partial trace map, 47
    \subitem physical, 43
    \subitem trace map, 47
  \item quantum process tomography, 57
    \subitem for incomplete quantum operations, 61
  \item quantum source, 198
  \item quantum state tomography, 57
  \item quantum teleportation
    \subitem connection to error correction, 56
    \subitem experimental implementation, 36
  \item qubit, 19
    \subitem quantum operations on, 50

  \indexspace

  \item relative entropy
    \subitem classical, 68
    \subitem quantum, 74
      \subsubitem convexity of, 84
      \subsubitem strengthened convexity result, 88
  \item reversible computation, 17

  \indexspace

  \item Schmidt bases, 131, 228
  \item Schmidt decomposition, 131, 227
    \subitem for mixed states, 229
  \item Schmidt number, 131, 228
  \item Schumacher's theorem, 130
  \item Shannon entropy, 66
  \item Shannon's noiseless coding theorem, 67
  \item Shor's quantum error correcting code, 173
  \item sigma matrices, 20
  \item state tomography, 57
  \item strengthened convexity of quantum relative entropy, 88
  \item strong subadditivity
    \subitem classical, 70
    \subitem quantum, 80
      \subsubitem proof of, 85
  \item subadditivity, 196
    \subitem classical, 70
    \subitem of mutual information, 72
    \subitem of the quantum conditional entropy, 86
    \subitem quantum, 78
  \item superdense coding, 21

  \indexspace

  \item TCE, 32
  \item teleportation, 23
    \subitem as a quantum operation, 54
  \item tomography, 57
  \item transpose operation, 44
  \item triangle inequality, 78
    \subitem for communication complexity, 127
  \item trichloroethylene, 32
  \item typical sequence, 133
  \item typical subspace, 133

  \indexspace

  \item Uhlmann's formula for fidelity, 96
  \item uniformity requirement for quantum computation, 27
  \item unifying picture for quantum information, 222
  \item universal set of gates, 21

  \indexspace

  \item von Neumann entropy, 73

  \indexspace

  \item weak law of large numbers, 135

\end{theindex}

\end{document}